\documentstyle[11pt,epsfig]{article}
\renewcommand{\theequation}{\arabic{section}.\arabic{equation}}

\begin{document}

\newcommand{\nc}{\newcommand}

\def\muchbigger{>>}
\newcommand\sfrac[2]{\!\! \begin{array}{c} \frac{#1}{#2} \end{array} \!\!}  
\newcommand{\La}{{\mathcal L}} 
\newcommand{\mIm}{\,\mbox{\small $\Im$m\,}} 
\newcommand{\eRe}{\,\mbox{\small $\Re$e\,}} 
\newcommand{\be}{\begin{eqnarray}} 
\newcommand{\ee}{\end{eqnarray}} 
\newcommand{\nn}{\nonumber} 
\newcommand{\dgs}{d^{\gamma}{\!\scriptstyle (}s{\scriptstyle )}\,} 
\newcommand{\dzs}{d^{Z}{\!\scriptstyle (}s{\scriptstyle )}\,} 
\newcommand{\plr}{\stackrel{\leftrightarrow}{\partial}{}\!\!} 

\newcommand{\<}{\langle\,}
\renewcommand{\>}{\,\rangle}
\nc{\lsp}{\;\;\;\;\;\;\;\;}
\nc{\beq}{\begin{equation}}   \nc{\eeq}{\end{equation}}
\nc{\bea}{\begin{eqnarray}}   \nc{\eea}{\end{eqnarray}}
\nc{\baa}{\begin{array}}      \nc{\eaa}{\end{array}}
\nc{\bit}{\begin{itemize}}    \nc{\eit}{\end{itemize}}
\nc{\ben}{\begin{enumerate}}  \nc{\een}{\end{enumerate}}
\nc{\bce}{\begin{center}}     \nc{\ece}{\end{center}}
\nc{\ra}{\rightarrow}
\nc{\non}{\nonumber}
\nc{\barx}{\bar{x}}
\nc{\pbarn}{\rm pb}
\nc{\fmbarn}{\rm fb}
\nc{\re}{\hbox {Re}}
\nc{\mev}{\hbox {MeV}} \nc{\gev}{\;\hbox {GeV}} \nc{\tev}{\;\hbox {TeV}}
\def\gesim{\lower0.5ex\hbox{$\:\buildrel >\over\sim\:$}} 
\def\lesim{\lower0.5ex\hbox{$\:\buildrel <\over\sim\:$}} 

\nc{\ttbar}{t\bar{t}}         \nc{\bbbar}{b\bar{b}}
\nc{\twbdec}{t\to W^+ b}
\nc{\tbwbdec}{\bar{t}\to W^- \bar{b}}
\nc{\epem}{e^+e^-}            \nc{\eett}{\epem \to \ttbar}
\nc{\sigeett}{\sigma_{e\bar{e}\to\ttbar}}  
\nc{\tbar}{\bar{t}}           \nc{\bbar}{\bar{b}}
\nc{\mt}{m_t}                 \nc{\mts}{m_t^2}
\nc{\lp}{\ell^+}              \nc{\lm}{\ell^-}
\nc{\epsl}{\epsilon_L}        \nc{\leff}{L\epsilon_{\ell\ell}}
\nc{\sig}{\frac{1}{\sigma}\frac{d\sigma}{d\phi}}
\nc{\et}{E_T}
\nc{\proc}{p \bar{p} \ra \ttbar +X \ra l^+ l^- + X \;/\; l^\pm + X}

\def\tbar {\overline{t}}
\def\beq {\begin{equation}}
\def\eeq {\end{equation}}
\def\be {\begin{equation}}
\def\ee {\end{equation}}
\def\barr{\begin{array}}
\def\earr{\end{array}}
\def\bea{\begin{eqnarray}}
\def\eea{\end{eqnarray}}
\def\bmath{\begin{displaymath}}
\def\emath{\end{displaymath}}
\def\bq{\begin{quote}}
\def\eq{\end{quote}}
\def\oas{$O(\alpha_s)$}
\def\Tr{\mbox{$\mbox{\rm Tr}\;$}}
\def\g5{\gamma_5}
\def\as{\alpha_s}
\def\real{\mathop{\mbox{\rm Re}}\nolimits}
\def\imag{\mathop{\mbox{\rm Im}}\nolimits}
\def\gE{\gamma_{\scriptscriptstyle E}}
\def\mz{M_{\scriptscriptstyle Z}}
\def\gz{\Gamma_{\scriptscriptstyle Z}}
\def\gf{g_{\scriptscriptstyle F}}
\def\NC{N_{\scriptscriptstyle C}}
\def\st{\sin\theta}
\def\s2t{\sin^2\kern-2pt\theta}
\def\su{\sigma_{\scriptscriptstyle U}}
\def\sl{\sigma_{\scriptscriptstyle L}}
\def\sf{\sigma_{\scriptscriptstyle F}}
\def\stot{\sigma_{\scriptscriptstyle T}}
\def\sfb{\sigma_{\scriptscriptstyle F/0}}
\def\sfl{\sigma_{\scriptscriptstyle F/1}}
\def\Li{\mbox{$\mbox{\rm Li}_2$}}
\def\Frac#1#2{\mbox{$\textstyle{#1\over#2}$}}
\def\half{\Frac{1}{2}}
\def\eps{\varepsilon}
\def\nn{\nonumber\\}
\def\sw{s_{\scriptscriptstyle W}}
\def\tw{\theta_{\scriptscriptstyle W}}
\def\MZ{M_{\scriptscriptstyle Z}}
\def\GZ{\Gamma_{\scriptscriptstyle Z}}
\def\Slash#1{\mbox{$\not{\hspace{-1.03mm}#1}$}}
\def\bm#1{\mbox{$\boldmath{#1}$}}
\def\vecprod#1#2{\mbox{\(\bm{#1}\hspace{-1mm}\cdot\hspace{-1mm}\bm{#2}\)}}
\def\direc{\bm{\hat{p}}}
\def\ndirec{\bm{\hat{n}}}
\def\I#1{{I}_{#1}}
\def\S#1{{S}_{#1}}
\def\J#1{{J}_{#1}}
\def\T#1{{T}_{#1}}
\def\II#1{{\tilde{I}}_{#1}}
\def\SS#1{{\tilde{S}}_{#1}}
\def\JJ#1{{\tilde{J}}_{#1}}
\def\uint{\displaystyle\int dy\,dz\:\:}
\def\pint{\displaystyle\int{dy\,dz\over\sqrt{(1-y)^2-\xi}}\:\:}
\def\jint{\displaystyle\int{dy\,dz \over (1-y)^2-\xi}\:\:}
\def\tint{\displaystyle\int{dy\,dz\over\{(1-y)^2-\xi\}^{3/2}}\:\:}
\def\sxi{\mbox{$\sqrt\xi$}}
\def\Deltab#1{{\bf\bar{\rm\Delta}}^{#1}}
\def\Dsigmap{\Delta\sigma^{(+)}}
\def\Dsigmam{\Delta\sigma^{(-)}}
\def\lapp{\stackrel{<}{~}}
\def\gapp{\stackrel{>}{~}}

\def \ni {\noindent}
\def \tm {\times}
\def \o  {\overline}
\def \ra {\rightarrow}
\def \f  {\frac}
\def \non {\nonumber}

\def \lc {{\cal L}}
\def \ba {\begin{array}}
\def \ea {\end{array}}
\def \be {\begin{equation}}
\def \ee {\end{equation}}
\def \bea{\begin{eqnarray}}
\def \eea{\end{eqnarray}}

\def \idt {{\rm Im}\,d_t}
\def \rdt {{\rm Re}\,d_t}

\newcommand{\sls}[1]{\mbox{$\slash\!\!\!\!\!{#1}$}}
\newcommand{\slstwo}[1]{\mbox{$\:\slash\!\!\!\!{#1}$}}
\newcommand{\slsthree}[1]{\mbox{$\:\slash\:\!\!\!{#1}$}}
\newcommand{\dt}{ \mbox{ $\!\!\cdot\!\!$ } }
\newcommand{\abs}[1]{  \mbox{ $\left|{#1}\right|$ }  }
\newcommand{\ps}   {\mbox{$\rlap /\;\!\!p      $}}
\newcommand{\ks}   {\mbox{$\rlap /\;\!\!k      $}}
\newcommand{\qs}   {\mbox{$\rlap /\;\!\!q      $}}
\newcommand{\xp}[1]{ \mbox{ $\mbox{\large e}^{#1}$ } }

\def \G{\Gamma}

\def \a{$\alpha$}
\def \bt{\beta_t}
\def \bg{\beta_{\gamma}}
\def \bz{\beta_z}
\def \eps{\epsilon}
\def \om {\omega}
\def \sg  {\sigma}
\def \la{\lambda}
\def \lab {\o {\lambda}}
\def \lae {\lambda_e}
\def \leb {\lambda_{\o e}}
\def \lg {\lambda_{\g}}
\def \ll {\lambda_l}
\def \lt {\lambda_t}
\def \ltb {\lambda_{\o t}}
\def \tl{\theta_l}
\def \tht{\theta_t}
\def \th0{\theta_0}

\def \ttg{$t \o t (\g)$}

\def \l  {Lagrangian}
\def \sm {SM}
\def \g {\gamma}
\def \gz {\g^0}
\def \gf {\g^5}
\def \gg {\g \g}

\def \tbar {$\o t$}
\def \bbar {$\o b$}
\def \ttbar{$t \o t$}
\def \ffbar {$f \o f$}
\def \tltrbar {$t_L \o t_R$}
\def \trtlbar {$t_R \o t_L$}
\def \tltlbar {$t_L \o t_L$}
\def \trtrbar {$t_R \o t_R$}
\def \ep {$e^+ e^-$}
\def \eett {$e^+ e^- \ra t \o t$}
\def \ggtt {$\g \g \ra t \o t$}

\def \ach {$A_{ch}$}
\def \afb {$A_{fb}$}
\def \achfb {$A_{ch}^{fb}$}

\newcommand{\square}{\sqcap\mbox{\hglue-.9em}\sqcup}

\newcommand{\mmtbc}{\mbox{ $\mu^+\mu^-\to t\bar c$}}
\newcommand{\mpmm}{\mu^+\mu^-}

\newcommand{\etal}{{\it et al}.}

\newcommand{\lsim}{\mathrel{\lower4pt\hbox{$\sim$}}
\hskip-12.5pt\raise1.6pt\hbox{$<$}\;}

\newcommand{\gsim}{\mathrel{\lower4pt\hbox{$\sim$}}
\hskip-12.5pt\raise1.6pt\hbox{$>$}\;}

\def\downrightarrow{{\mathrel{\lower7.5pt\hbox{$\rightarrow$}}
\hskip-11.2pt\lower4.9pt\hbox{$\vrule height .25truecm$}\;}}

\def\dra#1#2{\mathop{\mathop{\mathrel{\hbox{$#1$}}}\limits_\downrightarrow}
\nolimits_{\hbox{$\ \ #2$}}}

\def\downlongrightarrow{{\mathrel{\lower7.5pt\hbox{$\longrightarrow$}}
\hskip-18.2pt\lower4.9pt\hbox{$\vrule height .25truecm$}\;}}

\def\dlra#1#2{\mathop{\mathop{\mathrel{\hbox{$#1$}}}\limits_\downlongrightarrow}
\nolimits_{\hbox{\hglue16pt$#2$}}}

\def\uprightarrow{{\mathrel{\raise.5pt\hbox{$\rightarrow$}}
\hskip-12.1pt\lower4.pt\hbox{$\vrule height .25truecm$}\;}}

\def\ura#1#2{\mathop{\mathop{\mathrel{\hbox{$#1$}}}\limits^\uprightarrow}
\nolimits^{\raise-3pt\hbox{$\ \ #2$}}}

\def\uplongrightarrow{{\mathrel{\raise.5pt\hbox{$\longrightarrow$}}
\hskip-18.2pt\lower4.pt\hbox{$\vrule height .25truecm$}\;}}

\def\ulra#1#2{\mathop{\mathop{\mathrel{\hbox{$#1$}}}\limits^\uplongrightarrow}
\nolimits^{\raise-3pt\hbox{\hglue16pt$ #2$}}}

\def \f{\frac}
\def \n{\noindent}
\def \ra{\rightarrow}
\def \g{\gamma}
\def\wtZ{{\widetilde Z}}
\def\what{\underbar{\hspace{.5in}}}
\def\beq{\begin{equation}}
\def\eeq{\end{equation}}
\def\beqa{\begin{eqnarray}}
\def\eeqa{\end{eqnarray}}
\def\alphatwos{\alpha^2_s}
\def \alphaX {\tilde\alpha}


\begin{flushright}
{\bf AMES-HET 00-03}\\
{\bf BNL-HET-00/7}\\
{\bf ROME1-1294/00}
\end{flushright}

\vspace*{0.5 in}

\renewcommand{\thefootnote}{\fnsymbol{footnote}}
\begin{center}
{\large\bf CP Violation in Top Physics
\footnote{A 2-up version of this postscript file may be obtained 
at the url http://thy.phy.bnl.gov/$\sim$soni/topreview.html}}

\vspace*{0.5 in}

David Atwood$^a$, Shaouly Bar-Shalom$^b$, Gad Eilam$^c$ and Amarjit Soni$^d$
\vspace*{0.5 in}

$a$) Department of Physics and Astronomy, 
Iowa State University, Ames, Iowa 50011, 
USA\footnote{Email: atwood@iastate.edu}.
\\
$b$) INFN, Sezione di Roma and Dept. of Physics, 
University of Roma I, La Sapienza, Roma, 
Italy\footnote{Email: Shaouly.BarShalom@roma1.infn.it}\footnote{
Much of the work in this review was done while 
S.~Bar-Shalom was at
Physics Department, University of California, Riverside, CA, USA}.
\\
$c$) Physics Department, Technion-Institute of Technology, Haifa 32000, 
Israel\footnote{Email: eilam@physics.technion.ac.il}.
\\ 
$d$) Physics Department, Brookhaven National Laboratory, Upton, NY 11973, 
USA\footnote{Email: soni@bnl.gov}.

\vspace*{0.5 in}

{\bf Abstract}\\
\end{center}

CP violation in top physics is reviewed. The Standard Model has negligible
effects, consequently CP violation searches involving the top quark may
constitute the best way to look for physics beyond the Standard Model.
Non-standard sources of CP violation due to an extended Higgs sector with
and without natural flavor conservation and supersymmetric theories are
discussed. Experimental feasibility of detecting CP violation effects in
top quark production and decays in high energy $e^+ e^-$, $\gamma \gamma$,
$\mu^+ \mu^-$, $pp$ and $p \bar p$ colliders are surveyed.  Searches for
the electric, electro-weak and the chromo-electric dipole moments of the
top quark in $e^+e^- \to t \bar t$ and in $p p \to t \bar t X$ are
descibed. In addition, other mechanisms that appear promising for
experiments, e.g., tree-level CP violation in $e^+ e^- \rightarrow t \bar
t h, t \bar t Z, t \bar t \nu_e \bar\nu_e$ and in the top decay $t \to b
\tau \nu_{\tau}$ and CP violation driven by $s$-channel Higgs exchanges in
$p p,~\gamma \gamma,~\mu^+ \mu^- \to t \bar t$ etc., are also discussed.

\setcounter{footnote}{0}
\renewcommand{\thefootnote}{\arabic{footnote}}

\newpage

\tableofcontents

\newpage
\section{Introduction \label{sec1}}
\setcounter{equation}{0}

Violations of the CP (Charge Conjugation combined with Parity) symmetry
are of great interest in particle physics especially since its origin is
still unclear.  Better understanding of this (so far) rare phenomenon can
lead to new physics which may explain both the origin of mass and the
preponderance of matter over anti-matter in the present universe. Indeed,
reactions that violate CP are such a scarce resource that in over thirty
years the only confirmed examples of CP violation are those found in the
decay of the $K_L$-meson\footnote{For excellent recent books on CP 
violation see ref.~\cite{branco}}.

The first experimental observation of CP violation was in 1964 by
Christenson, Cronin, Fitch and Turlay \cite{prl13138} who observed a
non-vanishing rate for the decay $K_L\to2\pi$ \cite{prd541},

\beq
Br(K_L\to2\pi) = 3.00 \pm 0.04 \times10^{-3} \label{eqintro1}~.
\eeq

\ni Since the dominant decay of $K_L$ is to a $3\pi$ state of
CP $= -1$, the above decay to a manifestly CP $= +1$ state clearly violates
this symmetry.

Another example of CP violation which is well established in $K_L$ is
the difference between $\Gamma(K_L\to\ell^+\nu_\ell \pi^-)$ and 
$\Gamma(K_L\to \ell^-\nu_\ell \pi^+)$, $\ell=e,\mu$ \cite{prd541}:

\beq
\frac{\Gamma(K_L\to\ell^+\pi^-\nu) - \Gamma(K_L\to\ell^-\pi^+\bar
\nu)}{\Gamma (K_L \to\ell^+\pi^-\nu) + \Gamma(K_L \to \ell^-\pi^+
\bar\nu)} =(3.27\pm 0.012) \times10^{-3} \label{eqintro2} ~.
\eeq

\ni All of these observations of CP violation in the $K_L$ system can be
explained by the CP violation in the mixing of the neutral $K$ mesons. Thus

\beqa
| K_L \rangle & = & \left((1+\epsilon) |K^0\rangle - (1-\epsilon) | \bar
K^0\rangle \right) / \sqrt{2 (1+|\epsilon|^2)} 
\label{eqintro3} ~, \\
| K_S \rangle & = & \left( (1+\epsilon) |K^0\rangle + (1-\epsilon) |\bar
K^0\rangle \right) / \sqrt{2 (1+|\epsilon|^2)} \label{eqintro4}~,
\eeqa

\ni where the experimental value of $\epsilon$ is \cite{prd541}:

\beq 
|\epsilon| = (2.263\pm 0.023)\times 10^{-3} \quad ; \quad \arg(\epsilon)
= 43.49 \pm 0.08 \label{eqintro5} ~.
\eeq

\ni As is well known, 
this mixing can be accommodated in the Standard Model (SM) with
three generations where the CP violation originates through a phase in
the Cabibbo Kobayashi Maskawa (CKM) \cite{ptp49652} matrix as will be
discussed in some detail in section \ref{ssec31}.

The SM further predicts that there is an additional CP violation in
$K_L\to\pi\pi$ parameterized by the quantity $\epsilon^\prime$. The
prediction is that $\epsilon^\prime/\epsilon = {\cal O}(10^{-4})$; the
theoretical difficulties in determining the hadronic matrix element
prevent us from making 
a more precise estimate \cite{pl389749}. Experimentally
$\Re{\rm e}(\epsilon^\prime/\epsilon)$ may be measured via \cite{rmp651113}:

\beq
\Re{\rm e} (\epsilon^\prime/\epsilon) \simeq \frac{1}{6} \left[ 1-
\frac{|\eta_{00}|^2}{|\eta_{+-}|^2} \right] \label{reepspr} ~,
\eeq

\noindent where

\beq 
\eta_{ij} = \frac{\langle\pi^i\pi^j | H_W | K_L \rangle}{\langle
\pi^i\pi^j | H_W|K_S \rangle} \label{etaij} ~,
\eeq

\ni and $H_W$ is the relevant weak interaction Hamiltonian.

After some two decades of intensive efforts, new and quite dramatic 
experimental developments have recently taken place that we 
would now briefly like to mention. First of all, let us recall 
that a few years ago   
the CERN experiment NA31 gave the result \cite{plb317233}:

\beq
\Re{\rm e} (\epsilon^\prime/\epsilon) 
= (23 \pm 6.5)\times 10^{-4} \label{na31} ~,
\eeq

\noindent appreciably different from zero. On the other hand, the Fermilab
experiment E731 found it completely consistent with zero \cite{prl701203}:

\beq
\Re{\rm e} (\epsilon^\prime/\epsilon) = 
(7.4 \pm 6)\times10^{-4} \label{e731} ~.
\eeq

\noindent For the past many years
improved experiments have been underway, at CERN (experiment
NA48), and at FNAL (KTEV) with an expected accuracy of about
${\cal O}(10^{-4})$,
KTEV has recently announced their new results on $\epsilon^\prime/\epsilon$, 
based upon analysis of 20\% data collected so far \cite{E832}: 

\beq
\Re{\rm e} (\epsilon^\prime/\epsilon) = 
(28 \pm 4.1)\times10^{-4} \label{ktevnew} ~.
\eeq

\noindent
Combining with \cite{plb317233} and \cite{prl701203} one now finds:

\beq
\Re{\rm e} (\epsilon^\prime/\epsilon) = 
(21.8 \pm 3)\times10^{-4} \label{finalnew} ~,
\eeq

\noindent thus conclusively establishing 
that $\epsilon^\prime/\epsilon \neq 0$. Such a non-vanishing value
formally lays to rest the phenomenological superweak model
\cite{superweak} of CP
violation as it unambiguously predicts $\epsilon^\prime/\epsilon=0$.
However, unless the computational challenges presented by strong
interactions can be overcome, it is unlikely that the measured value of
$\epsilon^\prime/\epsilon$ would confirm or refute the SM in any
reliable fashion.

Experiments involving $B$-mesons are more likely to have a
quantitative bearing on the SM\null. Just as the SM indicates that the
natural size of CP asymmetries in $K$ physics is
${\cal O}(10^{-3}$--$10^{-4})$, it also strongly suggests that
the effects in the
$B$ system are much bigger; in many cases CP asymmetries are expected
to be tens of percents. This expectation renders the $B$ system ideal
for a precise extraction of the CKM phase and, indeed for a thorough
quantitative test of the SM through a detailed
study of the unitarity triangle \cite{utriangle}. The asymmetric and
symmetric $B$ factories currently in the early stages of running
 at KEK, SLAC and
Cornell and hadron machines,
 should have a very important role to play in confronting the
experimental results with the detailed predictions of the SM.
A recent CDF result \cite{hepex9903002} for CP violation in $B^0 \to 
J/\Psi K_S$, though crude at the moment, indicates that CP violation 
may indeed be large in the $b$ system.

Experiments at FNAL have decisively \cite{topmass1} demonstrated 
that the mass of the top
quark is extremely large, i.e., the D0 and CDF average is now 
$m_t \sim 174$ GeV \cite{topmass2}. 
This has some very
important consequences. First the top rapidly undergoes two-body weak
decay: $t\to b+W$, with a time scale of about $10^{-24}$ sec., which is
shorter by an order of magnitude 
than the typical QCD time scale necessary for hadronic
bound states to be formed \cite{plb181p157}. 
Thus, unlike the other five quarks, the top
does not form hadrons. It means that the dynamics of top production and
decay does not get masked by the complications of non-perturbative,
bound state
physics, i.e., the ``brown muck''. All of the CP violation phenomena
relevant to the top are therefore of the ``direct'' type.

We should think of the top  quark as an elementary fermion. For
example, it therefore is sensible to ask for its dipole moment
\cite{prd45p2405,plb279p389,prl69p33,npb408p286}. Unlike
the other quarks the spin of the top quark becomes an extremely
important observable. Indeed the decays of the top quark become very
effective analyzers of its spin, 
see e.g., \cite{prl69p410,prd45p124,prl69p2754}.

The SM predicts, however, that CP-violating effects in
$t$-physics are very small. This is primarily due to the fact that its
large mass in comparison to the other quarks renders the 
Glashow-Iliopoulos-Maiani (GIM) \cite{gim}
cancellation particularly effective \cite{prd44p1473,plb268p106}. 
This being the
case, what then is the motivation for the study of CP violation in the
top quark system?

There are two related reasons why one might expect to find such
effects.
First of all there is another important example of CP violation which
the SM fails to explain, namely the excess of matter over
anti-matter in the universe. It was shown by Sakharov
\cite{adsakharov} in 1967 that CP violation is one of the necessary conditions 
for baryon number asymmetry to appear in the early universe; 
baryon -- anti-baryon asymmetry can be 
dynamically generated at early stages after the big bang even if the 
universe was ``born'' symmetric, provided that: ($i$) C and CP are violated, 
($ii$) there are baryon 
number violating interactions and ($iii$) there is a deviation from 
thermal equilibrium.        
The basic idea is that, if CP is violated, then baryons and anti-baryons
interact with different rates at some point in the early
universe. However, the CP violation due to the SM appears too weak to 
drive such an asymmetry \cite{anrnps4327}. 
In many cases, extensions of the SM such
as the Two Higgs Doublet Model (2HDM) or the SUperSYmmetric (SUSY) 
extensions of the
SM are able to supply the CP violation required to
produce such a baryon asymmetry in the early universe. In fact in some models
\cite{npb373p453,ptp98p1325} it is precisely the couplings of the top quark to
CP-violating phases in beyond the SM physics which drive baryogenesis.
Thus, the study of CP violation in top quark interactions 
in the laboratory could shed light on these primordial processes.

The second motivation for investigating CP violation in top
quark physics is that in many extensions of the SM, CP
violation in the top quark can be particularly large. Indeed, because
the SM contribution to CP violation in the top quark is so small, any
observation of such effects would be a clear evidence of 
physics beyond the SM\null. 
 The argument here parallels the search for the
weak neutral current, in the 1970's, by looking for parity violation in
deep-inelastic-scattering. The point is that the existing
theory of the time, namely QED, could not cause parity violation in
deep-inelastic-scattering\null. 
Such an effect became an unambiguous signature for the
existence of the weak neutral current.

Since various extensions of the SM entail new CP-violating phase(s), we
should seek the optimal strategies for searching each type of new
phase. 
In this context we first recall, what has been emphasized on the preceding
page, that the $b$-quark is very
sensitive to the CKM phase of the SM\null. Existing literature has
revealed that top physics is very sensitive to several different types
of new phases. Upcoming high energy colliders of the next decade
can therefore serve as excellent laboratories for searching for new physics in 
top quark systems, and in particular, for studying CP-violating effects 
associated with those new CP-odd phases.
The upgraded Tevatron $p \bar p$ collider (runs 2 and 3) at Fermilab 
which will be able to produce about $10^4 - 10^5$ $t \bar t$/year, 
the CERN $pp$ Large Hadron Collider (LHC) will produce about 
$10^7 - 10^8$ $t \bar t$/year, and about $10^5 - 10^6$ $t \bar t$/year 
are expected at a future $e^+e^-$ Next 
Linear Collider (NLC).
 
A CP-odd phase due to an extended neutral Higgs sector or 
vertex corrections arising in other extensions of the SM can endow the top
quark with a large dipole moment form factor. 
Such an effect could be
detected both at an $e^+e^-$ collider, such as the NLC or hadron colliders
such as the LHC.  
A CP-violating phase in the neutral Higgs sector
also causes large CP asymmetries in the reactions $e^+e^-\to t\bar tH^0$
and $e^+ e^-\to t\bar t\nu_e\bar\nu_e$, both of which should be a prime
target for the NLC\null. Moreover, CP violation in the neutral Higgs sector 
and in supersymmetry can have interesting effects in single top 
production at the upgraded 
Tevatron $p \bar p$ collider at Fermilab and in $t \bar t$ pair production at 
the LHC.
The transverse polarization of the $\tau$ in the three-body top decay 
$t\to b\tau\nu$ is extremely sensitive to a new phase from a charged Higgs
sector in Multi-Higgs Doublets Models (MHDM's). 
Finally, CP-odd phases in SUSY models have also 
interesting effects in 
Partial Rate Asymmetry (PRA) in $t\to W^+ b$
versus $\bar t\to W^- \bar b$. These processes and others 
will be discussed in the subsequent chapters.

We will not consider CP violation phenomena in which 
the top quark is virtual rather than an external particle. 
It suffices to recapitulate that, in the SM, CP violation 
is often dominated by the virtual top quarks in the loops.
Let us also comment that the discovery of the top 
with the measurement of $m_t$, and the progress in determination of the CKM 
matrix elements as well as considerable progress 
in theory, has influenced our understanding 
of CP violation in K and B physics within the SM \cite{hepph9704376} 
and beyond the SM \cite{hepph9701231}. 
Furthermore, for the Electric Dipole Moments (EDM's) of the electron and 
the neutron \cite{hepph9701357}, the virtual top quark also 
plays a crucial role
in extended Higgs sector scenarios.
\pagebreak

\section{General discussion \label{sec2}}
\setcounter{equation}{0}

\subsection{Definitions of discrete symmetries C, P \& T \label{ssec21}}

Let us now review the definitions of the discrete symmetries $C,P$ and $T$ 
and recall
a few basic facts concerning their manifestation in relativistic
quantum field theory. 

Under the parity transformation, $P$, the spatial coordinate axes are
reversed, i.e., $P \vec x = - \vec x$. 
Thus for an ingoing particle, $X$, in a specific momentum and spin state
$|X;\vec P, \vec S\rangle_{\rm in}$, the action of parity is to reverse the
momentum, leaving the spin fixed as angular momentum is an axial 
vector defined by a cross product. Hence

\beq
P |X; \vec P, \vec S\rangle_{\rm in} \to 
|X; -\vec P,\vec S\rangle_{\rm in} ~.
\label{eq2.1.1}
\eeq

\ni Under the Wigner definition of time reversal, $T$, the sign
of both momenta and spins are reversed, and also, due to the anti-unitary 
nature of $T$, 
$|\rangle_{\rm in}$ and $|\rangle_{\rm out}$ states are interchanged. 
Thus

\beq
T |X;\vec P, \vec S \rangle_{\rm in} \to 
|X; -\vec P,-\vec S\rangle_{\rm out} ~.
\label{eq2.1.2} 
\eeq

\ni Under Charge conjugation, $C$, each particle is replaced by its
anti-particle, and so 

\beq
C |X;\vec P, \vec S \rangle_{\rm in} \to 
|\bar X; \vec P,\vec S\rangle_{\rm in} ~,
\label{eq2.1.3} 
\eeq

\noindent where $\bar X$ means that all charges and other 
additive quantum numbers are 
reversed. 

It can be shown that local relativistic quantum field theories with the 
usual spin-statistics relations are
invariant under the combined action of all three of these symmetries
where \cite{cpt}:

\beq
CPT |X; \vec P, \vec S\rangle_{\rm in} = 
|\bar X; \vec P,-\vec S\rangle_{\rm out} ~.
\label{eq2.1.4} 
\eeq

\ni Thus, such a theory violates $T$ if and only if it violates
CP, where 

\beq
CP |X; \vec P, \vec S\rangle_{\rm in} \to 
|\bar X; -\vec P,\vec S\rangle_{\rm in} ~,
\label{eq2.1.5}
\eeq

\ni and in this sense CP and $T$ violation are equivalent.

Other well known consequences of the CPT theorem are that masses of
particles and anti-particles are the same, $m_X=m_{\bar X}$, and the
total widths of particles and anti-particles are also equal,
$\Gamma_X=\Gamma_{\bar X}$. Note that it does not follow from CPT that
decay rates to specific final states are the same. In fact, partial
width differences, i.e., a non-zero value of 

\beq 
\Delta (X\to A) \equiv \Gamma (X\to A) - \Gamma (\bar X\to\bar A) ~,
\label{eq2.1.6}
\eeq

\ni is a form of CP violation that we will discuss in more detail in
section \ref{ssec23}. 
Clearly it follows from $\Gamma_X=\Gamma_{\bar X}$ that 

\beq 
\sum_A \Delta (X\to A) = 0 ~, \label{eq2.1.7}
\eeq

\ni where the sum is over all possible final states. The relationship
between $\Delta(X\to A)$ and the other final states which compensate
for it will also be discussed in detail in section \ref{ssec23}.

In this report we are largely concerned with the violation of discrete
symmetries in decay and 
scattering experiments. We, therefore, need to consider the
implementation of CP and $T$ on the $S$-matrix.

For $C$ and $P$ this is straightforward. Consider the initial
state of $n_i$ particles 

\beq
|i\rangle = |\vec P_a,\vec P_b,\dots, \vec S_a,\vec S_b,\dots \rangle_{\rm
in}~,  \label{eq2.1.8}
\eeq

\ni and the final state of $n_f$ particles 

\beq 
|f\rangle = |\vec P_1,\vec P_2,\dots,\vec S_1,\vec S_2,\dots
\rangle_{\rm out} \label{eq2.1.9} ~,
\eeq

\ni so that the $S$-matrix element is 

\beq 
\langle f| S |i\rangle = S_{fi} ~. \label{eq2.1.10}
\eeq

\ni The transformations $P$ and $C$ follow from the single particle
transformation

\beqa
|f_P\rangle & = & |-\vec P_1,-\vec P_2,\dots,\vec S_1,\vec S_2,\dots 
\rangle_{\rm out} ~, \label{eq2.1.11} \\
|f_C\rangle & = & |\vec P_{\bar 1}, \vec P_{\bar 2},\dots, \vec S_{\bar
1},\vec S_{\bar 2},\dots \rangle_{\rm out} ~, \label{eq2.1.12}
\eeqa

\ni and likewise for $|i_P\rangle$, $|i_C\rangle$. The transformation of
the $S$-matrix element under these symmetries is thus

\beqa
S_{fi} \stackrel{P}{\longrightarrow} S_{f_Pi_P} ~, \label{eq2.1.13} \\
S_{fi} \stackrel{C}{\longrightarrow} S_{f_Ci_C} ~. \label{eq2.1.14}
\eeqa

\ni and thus

\beqa
S_{fi} \stackrel{CP}{\longrightarrow} S_{f_{CP}i_{CP}} ~.
\eeqa

\ni The nature of time reversal, however, requires the interchange of
$|\rangle_{\rm in}$ and $|\rangle_{\rm out}$ states so that the effect
on the $S$-matrix will be anti-unitary. Thus

\beqa
|f_T\rangle & = & |-\vec P_1, -\vec P_2,\dots,-\vec S_1,- \vec S_2,\dots
\rangle_{\rm in} ~, \label{eq2.1.15} \\
|i_T\rangle & = & |-\vec P_a,-\vec P_b,\dots,-\vec S_a,-\vec S_b,\dots
\rangle_{\rm out} ~, \label{eq2.1.16} 
\eeqa

\ni such that

\beq
S_{fi} \stackrel{T}{\longrightarrow} S_{i_T f_T} ~. \label{eq2.1.17}
\eeq

\ni Needless to say, because of the interchange of initial and final
states, accelerator based experiments seldom test $T$ directly.

These symmetries, as they are defined in $S$-matrix theory are
fundamental in that if the Lagrangian and the vacuum states respect
$C$, $P$ or $T$, then the corresponding symmetry of the $S$-matrix will
apply. We will also find it useful to consider the symmetry $T_N$ - 
``naive'' time reversal - for which this is not true.
The definition of $T_N$ is to apply $T$ to the initial and final
states without interchanging them

\beq
S_{fi} \stackrel{T_N}{\longrightarrow} S_{f_T i_T} ~. \label{eq2.1.18} 
\eeq

\ni Thus, $T_N$ is a ``symmetry'' which can be tested in accelerator based
scattering experiments, but, as we shall see in the following section, it only
corresponds to ``true'' time reversal ($T$) operation at tree-level in
perturbation theory. It is nonetheless useful in categorizing the
various modes of CP violation.

\subsection{CP-violating observables: categorizing according to $T_N$ 
\label{ssec22}} 

It is useful to divide CP-violating observables into two categories
(see e.g., \cite{prd45p2405,prl69p2754,npb338p53,mpla10p627}), 
those that are even under ``naive'' time reversal
($T_N$) and those that are odd. 
Recall that $T_N$ is defined as a transformation which reverses the momenta 
and spins of all particles 
without the interchange of the initial and final states.
This contrasts with true time reversal, $T$, in that under $T$ initial and
final states are also interchanged. 

The symmetry $T_N$ is not a fundamental symmetry like $C$, $P$ and $T$
since the $S$-matrix under $T_N$ need not follow from
the transformation properties of the Lagrangian. 
Nevertheless, it is a useful tool for categorization and as
we shall
presently show, observables which are CP-odd and $T_N$-odd, i.e., are
CP$T_N$-even, may assume non-zero expectation values in the absence of
Final State Interaction (FSI) effects. In particular, tree-level
processes in perturbation theory may lead to non-vanishing 
expectation values for
these operators. On the other hand, CP-odd $T_N$-even (i.e.,
CP$T_N$-odd) operators may only assume non-zero expectation values if
such FSI effects are present giving a non-trivial phase to the Feynman
amplitude. Such a phase, often called a strong phase or absorptive phase, 
may arise in
several ways. It may be present in a loop diagram if the internal
particle(s) can be on-shell. An interesting variation of this, which we
will consider in section~\ref{ssec24}, is in
 the propagator of an unstable
particle where the strong phase is the phase of the Breit-Wigner
amplitude. Non-perturbative rescattering of final state particles can
also give a strong phase though this is more of interest in CP studies
in $B$ and $K$ physics.

Indeed, $T_N$ is useful in understanding when CP-conserving
observables depend on an absorptive phase. In particular, a CP-even
$T_N$-odd observable (typically a CP-even triple product correlation of
momenta and/or spins) will only assume an expectation value if FSI effects are
present. Thus, for instance, if one is looking for a CP-odd, $T_N$-odd
effect, one must have data both on the process of interest and its CP
conjugate (if they are different) in order to distinguish from the 
possible background of CP-even
$T_N$-odd effects (see e.g., \cite{jarlskog2}).

In order to understand the role of $T_N$, let us consider the unitarity
relations of the $S$-matrix (implied by conservation of probabilities). 
Following a derivation analogous to the optical theorem 
\cite{itzykson_cpt} we write the $S$-matrix in terms of the
scattering amplitude ${\cal T}$

\beq
S=1+i{\cal T} ~, \label{eq2.2.1}
\eeq

\ni where, for a given transition $i\to f$, ${\cal T}$ is related to the 
``reduced scattering amplitude'', $\tau$,  by

\beq
<f|{\cal T}|i> = (2 \pi)^4 \delta^4 (p_f-p_i) <f|\tau|i> ~. \label{eq2.2.2}
\eeq 

\ni Substituting Eq.~\ref{eq2.2.1} into the unitarity relation
$S^\dagger S =1$ we obtain

\beq
{\cal T}_{fi} - {\cal T}^\ast_{if} = i\sum_n {\cal T}^\ast_{nf} {\cal T}_{ni} ~, \label{eq2.2.3}
\eeq

\ni where we denote $\langle a|{\cal T}|b\rangle \equiv {\cal T}_{ab}$. In terms of $\tau$
this becomes

\beq
\tau_{fi}-\tau^\ast_{if} = i(2\pi)^4\sum_n \delta^4 (p_n-p_i)
\tau^\ast_{nf}\tau_{ni} ~. \label{eq2.2.4}
\eeq

\ni Let us now assume that there are no rescattering effects and that (to
the order of approximation considered) $i$ and $f$ are stable states
so that $\tau_{ii} = \tau_{ff} =0$. Thus, for each possible intermediate
state $n$, the rhs of Eq.~\ref{eq2.2.4} vanishes. Therefore, in the absence
of rescattering, $\tau$ is hermitian

\beq 
\tau_{if} = \tau^\ast_{fi} ~. \label{eq2.2.5}
\eeq

\ni Now, if $\tau$ is CP invariant, then by the CPT theorem 
it is also $T$ invariant. Thus

\beq
\langle f|\tau|i\rangle  = \langle i_T|\tau |f_T\rangle = \langle
f_T|\tau |i_T\rangle^\ast ~, \label{eq2.2.6}
\eeq

\ni and therefore

\beq   
\left| <f|\tau|i> \right|^2 =  \left| <f_T| \tau |i_T>\right|^2
\label{eq2.2.7}~. 
\eeq 

\ni In fact this equation means precisely that the modulus of 
$\langle f|\tau | i\rangle$ is invariant under $T_N$. 
Since, in the absence of rescattering,
the expectation value of any
operator depends only on $|\langle f | \tau | i\rangle|$, Eq.~\ref{eq2.2.7}
implies that if CP is conserved, then 
only $T_N$-even 
operators can have a non-vanishing expectation value.

What we have shown therefore is that in the absence of rescattering 
effects (i.e., $\Im{\rm m}(\tau)=0$, in which case the requirement 
of CPT invariance leads effectively to 
conservation of the scattering amplitude under CP$T_N$) and 
in the absence of CP violation, 
$T_N$-odd observables have zero expectation value.
Thus, if such a $T_N$-odd observable ${\cal O}$ has a non-zero
expectation value, either CP is violated and ${\cal O}$ is CP-odd
 (i.e., CP$T_N$-even) or 
there are rescattering effects present 
(implying ${\rm CPT} \neq {\rm CP}T_N$) and
${\cal O}$ is CP-even (i.e., CP$T_N$-odd).
Conversely, let us suppose ${\cal O}$ is CP-odd and $T_N$-even (i.e.,
CP$T_N$-odd). Again Eq.~\ref{eq2.2.7} implies that this operator can
only assume a non-zero expectation value if rescattering effects are
present. These properties of the operators are summarized in 
Table \ref{tnprop}. 

\begin{table}
\begin{center}
\begin{tabular}{c|c|c} \hline 
$T_N$ & CP-violating & CP-conserving \\ \hline
even & Y & N \\ \hline
odd & N & Y \\ \hline 
\end{tabular} 
\end{center}
\caption[dummy]{\emph{Transformation properties under $T_N$ and CP and 
presence or absence of final state interactions (FSI). Here 
Y $\equiv$ FSI present and N $\equiv$ FSI absent.}}\label{tnprop}
\end{table}

From Table \ref{tnprop} we see that 
another consequence of Eq.~\ref{eq2.2.7} is that a $T_N$-odd signal
is
only a definite signal for 
$T$ violation and hence of CP violation in the absence of
rescattering effects. To confirm the CP-even or CP-odd nature of such a
reaction one must therefore compare data from $i\to f$ with the charge
conjugate channel $\bar i \to \bar f$ 
to explicitly verify CP violation or else rule out
rescattering effects in some other way.

Recall from the definition of time reversal that the spatial components
of vectors representing momenta and spins are reversed. Thus, an
observable is $T_N$-odd if it is proportional to a term of the form
$\epsilon (v_1,v_2,v_3,v_4)$, where $v_i$ are 4-vectors representing spins
or momenta of initial and final state particles and $\epsilon$ is 
the Levi-Civita tensor. Consequently, $T_N$-odd
signals can only be observed in reactions where there are at least four
independent momenta or spins that can be measured.

There are two important
 venues for  the investigation of CP violation that 
we will deal with extensively. 
The first
one is when CP nonconservation appears in decays of a particle 
and the second, is to search for scattering processes that can give rise 
to CP violation. 
The latter consist of two different possibilities: either
 the CP-violating effect is due to the subsequent decay of the particle
which is produced in the scattering process or the CP nonconservation
is driven by an intrinsic property of the scattering mechanism itself.

An observable which is CP-odd and $T_N$-even, thus requiring an absorptive
phase (as was shown above), and which is widely  used in  
the case where the CP effect appears in decays of
a particle is called PRA (Partial Rate Asymmetry). This observable is
non-vanishing when a particle $A$ decays to a state $B$ with
a partial width $\Gamma (A\to B)$ whereas the partial width of the
conjugate process, i.e., $\Gamma (\bar{A}\to \bar{B})$ is different
from $\Gamma (A\to B)$. Thus, defining

\beq
\alpha_{PRA} \equiv \frac{\Gamma (A\to B) - 
\Gamma (\bar{A}\to \bar{B})}{\Gamma (A\to B) + \Gamma
(\bar{A}\to \bar{B})} \label{pra11}~, 
\eeq 

\ni it is easy to see that $\alpha_{PRA}$ is odd under CP and CP$T_N$. For
$\alpha_{PRA}$ to receive non-vanishing contributions, at least two amplitudes
with different (CP-even) absorptive phases as well as with different CP-odd
phases must contribute to $A\to B$. To see this explicitly, let us define
${\cal M}_1$ and ${\cal M}_2$ to be the two possible amplitudes
contributing to $A\to B$

\beqa
&& {\cal M} \equiv {\cal M}(A\to B) = |{\cal M}_1|e^{i\varphi_1}e^{i\delta_1}
+ |{\cal M}_2|e^{i\varphi_2}e^{i\delta_2} \ , \nonumber\\ 
&& \bar{{\cal M}} \equiv {\cal M}(\bar{A}\to \bar{B}) = |{\cal M}_1|e^{-i
\varphi_1}e^{i\delta_1}+ |{\cal M}_2|e^{-i\varphi_2}e^{i\delta_2} ~,
\eeqa

\noindent where $\varphi_i$ are CP-odd phases that change sign as one 
goes from $A\to B$ to $\bar{A}\to \bar{B}$, and $\delta_i$ are CP-even
phases that can arise due to FSI. It is then easy to see that

\beq
|{\cal M}|^2 - |\bar{{\cal M}}|^2 = - 4 |{\cal M}_1||{\cal M}_2|\sin(\delta_1
- \delta_2)\sin(\varphi_1 - \varphi_2)~. 
\eeq 

\noindent Clearly $\alpha_{PRA}$, being proportional to
$(|{\cal M}|^2 - |\bar{{\cal M}}|^2)$, will vanish if the two amplitudes 
do not have a relative absorptive phase, i.e., $\delta_1 -\delta_2 \neq 0$
as well as a relative CP-odd phase, i.e., $\varphi_1-\varphi_2\ne0$. 
Of course
$\alpha_{PRA}$ does not violate CPT \cite{weinberg_cpt} 
as the requirement of CPT applies only to total widths $\Gamma
(A) = \Gamma (\bar{A})$. 
This equivalence (i.e., $\Gamma(A)=\Gamma(\bar A)$) 
is due to the fact that the absorptive phase of $\Gamma
(A\to B)$ emanates from rescattering through an on-shell intermediate state
$C$ and vice versa, i.e., the absorptive phase for $\Gamma(A\to C)$ will
emanate from the on-shell intermediate state $B$. This fact is 
an example of the well-known ``CP-CPT connection''
\cite{npb352p367,prd43p2909,prd43p151}. The states $B$ and $C$ are
referred to as compensating processes. Of course, one can in general 
have this compensation act between several final states.

It is sometimes useful to define a slightly different asymmetry which also
requires dealing with partial rates. This asymmetry is called 
the Partially Integrated Rate Asymmetry (PIRA) and is defined as

\beq
\alpha_{PIRA} \equiv \frac{\Gamma_{PI} (A\to B) - \Gamma_{PI} (\bar{A}\to
\bar{B})}{\Gamma_{PI}  (A\to B) + \Gamma_{PI} (\bar{A}\to \bar{B})} ~,
\eeq 

\noindent where $\Gamma_{PI}$ is the partially integrated width for $A\to
B$ obtained by integrating only part of the full kinematic range 
of phase-space. Often such asymmetries can be larger than $\alpha_{PRA}$
since the portion of the final states not included in the integral 
may themselves
be the compensating process. For example, in \cite{prl70p1364}
it was shown that detecting CP violation effects in the 
process $t\to b \tau
\nu_{\tau}$ through $\alpha_{PIRA}$ is more efficient than 
through $\alpha_{PRA}$
as the former is driven by tree-level diagrams and the latter by 1-loop 
diagrams.

A related observable which is also CP-odd and $T_N$-even is the energy
asymmetry 

\beq
\alpha_E \equiv \frac{<E_{i}> - <E_{\bar{i}}>}{<E_{i}> + <E_{\bar{i}}>} ~,
\eeq 

\noindent where $<E_i>$ is the average energy of a particle $i$ in a decay
of the ``parent'' particle 
and $<E_{\bar i}>$ is the average energy of the 
corresponding anti-particle $\bar i$ 
in the decay of the conjugate state of the ``parent'' particle. 
Such an asymmetry becomes relevant when
the decay involves three or more particles in the final state and may be
regarded as a weighted PRA\null.

A further generalization of the above constructions of  CP-odd $T_N$-even
observables is by 
considering combinations of dot products (thus being even under $T_N$)
of measurable momenta or spin vectors. Examples of such CP-odd $T_N$-even
observables will be given in the following chapters.

As an example of the various types of operators discussed above let us
consider the reaction 

\beq
e^-(p_e)+e^+(\bar p_e) \to t(p_t, s_t) + \bar t (\bar
p_t, \bar s_t) ~,
\eeq

\ni where $p_i$ are 4-momenta and $s_i$ are spins. Clearly, no
$T_N$-odd observable can be constructed without the  
Levi-Civita tensor, $\epsilon$, and since the momenta satisfy $p_e+\bar p_e =
p_t +\bar p_t$ one needs to use the spins of the $t$ and $\bar t$ to
construct such observables. Indeed no non-trivial CP-odd observable
can be constructed without knowing the spins either, so top polarimetry
is essential for the study of CP violation in this reaction (see
section \ref{ssec28}).

As an example of a $T_N$-even CP-odd operator consider in the c.m. 
frame of the $e^+e^-$: 

\beq 
{\cal O}_1 = (p_e-\bar p_e) \cdot (s_t-\bar s_t) = - (\vec P_e -
\vec{\bar P}_e) \cdot (\vec S_t - \vec{\bar S}_t) ~. 
\eeq

\ni Clearly this is $T_N$-even and also $C$-even. It is $P$-odd since
$\vec S_t,\vec{\bar S}_t$ are axial vectors while 
$\vec P_e,\vec{\bar P}_e$ are
polar vectors. Thus, since ${\cal O}_1$ is CP-odd and CP$T_N$-odd, 
it will have an expectation value if both CP
violation and rescattering effects are present. Consider now the
operator

\beq
{\cal O}_2 = \epsilon^{\alpha\beta\gamma\delta} p_{e\alpha} \bar
p_{e\beta} s_{t\gamma} \bar s_{t\delta} ~ \propto ~ (\vec P_e-\vec{\bar
P}_e) \cdot (\vec S_t \times \vec{\bar S}_t) ~. 
\eeq

\ni Clearly ${\cal O}_2$ is $T_N$-odd and $C$-even but also
$P$-odd. This CP-odd observable can therefore give a signal of
CP violation without the necessity of rescattering effects.

Finally the operator

\beq
{\cal O}_3 = \epsilon^{\alpha\beta\gamma\delta} p_{e\alpha} \bar
p_{e\beta} (s_t-\bar s_t)_\gamma (p_t +\bar p_t)_\delta ~ \propto ~ (\vec
P_e - \vec{\bar P}_e) \cdot \left[ (\vec S_t-\vec{\bar S}_t)\times (\vec P_t +
\vec{\bar P}_t) \right] ~, 
\eeq

\ni is $T_N$-odd, $P$-even and $C$-even and so provides a signal of 
CP-conserving rescattering effects. Similarly many more operators can
be constructed with the symmetries of the above. Some consideration of
the physics to be tested for is helpful in selecting the operator most
useful in measuring possible CP-violating effects. 
We will consider that in more detail in section \ref{ssec26}.

\subsection{Partial rate asymmetries and the CP-CPT connection 
\label{ssec23}}

As mentioned in the previous section, one of the most interesting 
and widely studied observable for testing CP
violation is the difference between
the partial width of a reaction, $\Gamma(A\to B)$, from that of the
conjugate reaction, $\Gamma(\bar A\to \bar B)$. Thus, if $\Gamma(A\to
B)\ne \Gamma(\bar A\to \bar B)$, then CP is violated in the decay. In
practice, it is better to work with the corresponding
dimensionless ratio $\alpha_{\rm PRA}$ in Eq.~\ref{pra11}, 
called the PRA. 
Its use, specifically in the context of heavy quarks and the
SM, first appeared in \cite{prl43p242}.

Since the CPT theorem demands that particle and anti-particle 
have identical life-times (or total widths), that theorem imposes
important restrictions on the form of CP-violating PRA's; these were
first recognized by G\'{e}rard and Hou \cite{prd43p2909}.

In perturbative calculations, which lead to PRA's, if all the diagrams
are systematically included, the requirement of CPT - that the total rate
and its conjugate be identical - should be manifest order by order. The
compensating processes should also be evident as the internal states of
loop diagrams. When simplifications are used in such calculations the
constraint of the CPT theorem provides  an important consistency check.
Furthermore, these restrictions can be used to 
greatly facilitate the calculations of the PRA for a compensating process.

A general formalism for maintaining the CPT constrains in calculating
PRA's was given in \cite{prd43p151}. In particular, it was shown that if one
defines a partial width difference, 
into a particular final state, $I$, as

\beq
\Delta_I \equiv \Gamma(P\to I) - \Gamma(\bar P\to \bar I) ~, \label{pr2}
\eeq

\noindent where $P$ and $\bar P$ are the decaying particle and its
anti-particle, respectively, then equality of total widths
$\Gamma(P)=\Gamma(\bar P)$ implies

\beq
\Delta_I = -\sum_{J\ne I} \Delta_J ~.
\eeq

\ni More specifically, in perturbation theory one can write at a given
order 

\beq
\Delta_I = \sum_{J\ne I} \Delta_I (J) \label{pr3}
\eeq

\noindent and

\beq
\Delta_I(J) = - \Delta_J(I)~. \label{pr4}
\eeq

\noindent Here $\Delta_I(J)$ denotes a contribution to the partial
width difference into the final state $I$ which is driven by the final
state $J$, and conversely, $\Delta_J(I)$ is the contribution to the
particle width difference into the final state $J$ being driven by the
final state $I$. Due to Eq.~\ref{pr4}, 
summing up the partial width differences over all the
final states one gets

\beq
\sum_K \Delta_K =0 \label{pr5} ~,
\eeq

\noindent where $K$ runs over all final states, $I$, $J$,\dots Thus the
requirements of CPT are automatically satisfied. Two important
conclusions that can also be drawn are:

\begin{description}
\item (a)  Rescattering of a state on to itself cannot give rise to
PRA. This follows immediately from Eq.~\ref{pr4} by setting $J=I$.

\item (b) The knowledge of the PRA into some final state $I$ that is
driven by the absorptive cut across a final state $J$, can be used to
deduce the PRA into the final state $J$ arising over a corresponding
cut across the state $I$. As mentioned before, 
such two processes are often called
``compensating processes''.

\end{description}

\begin{figure}
 \begin{center}
  \leavevmode
\epsfig{file=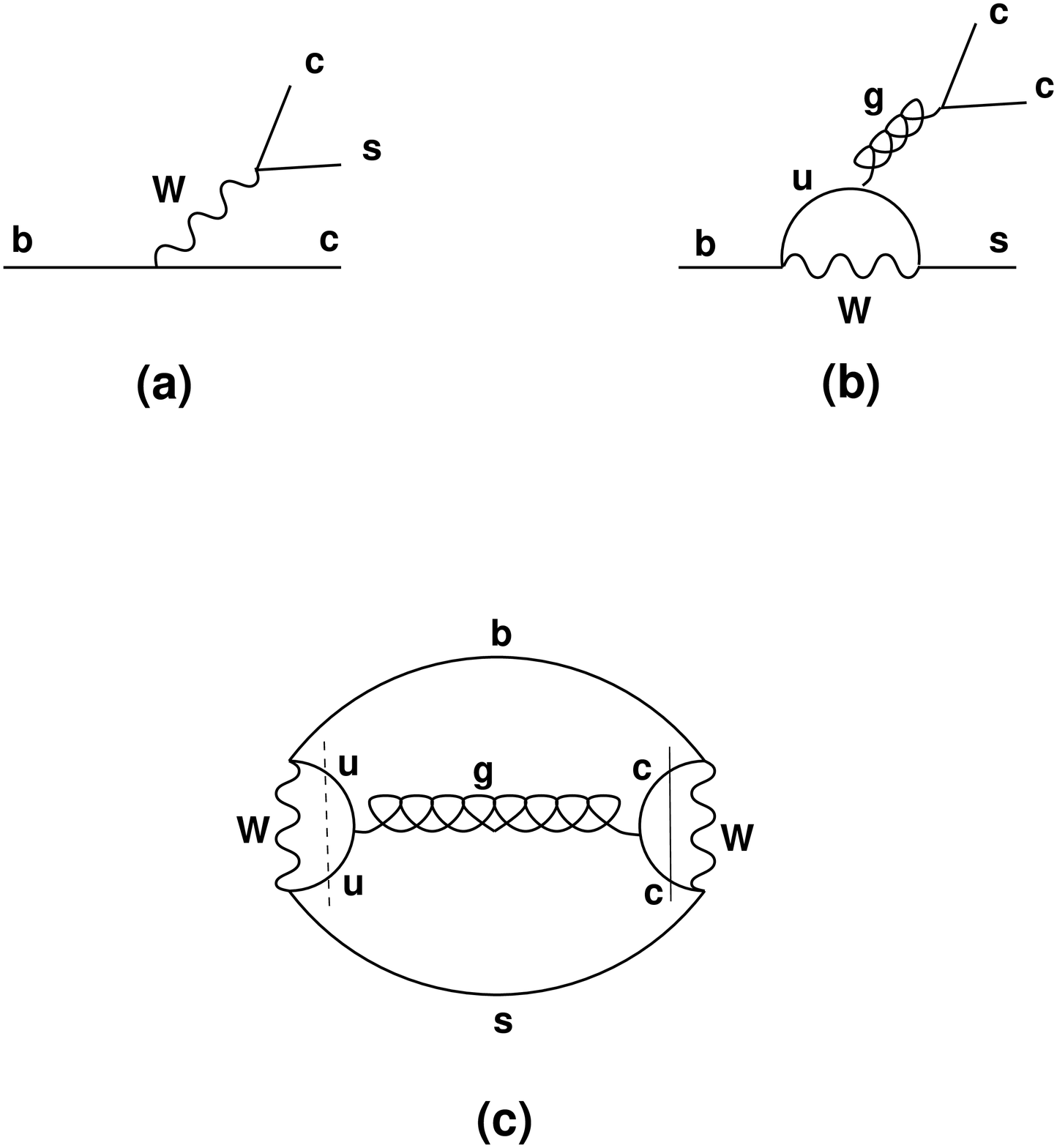,height=13cm}
\end{center}
\caption{\emph{
(a) Tree-level diagram for $b \to s c \bar c$,
(b) 1-loop (order $\alpha_s$) penguin diagram for $b \to s c \bar c$, and 
(c) a diagrammatic description for the compensating nature of the contribution 
from $b \to s c \bar c$
 and $b \to s u \bar u$ to the PRA's. For $b \to s c \bar c$:
the dashed line indicates an absorptive cut
 along inner particles ($u \bar u$) in the penguin diagram and the solid 
line refers to an external phase-space cut ($c \bar c$). 
For $b \to s u \bar u$ the role of the cuts are reversed: 
the dashed line is the external $u \bar u$ phase-space cut and the solid 
line indicates the $c \bar c$ absorptive cut in the penguin contribution.}}
\label{chapter2fig1}
\end{figure}

Let us first illustrate how these considerations apply to PRA's in $b$
decays. Consider, for example, the process $b\to sc\bar c$. The lowest order
non-vanishing contribution to the PRA arises here, at order $\alpha_s$,
from the interference of the tree graph in Fig.~\ref{chapter2fig1}(a) 
with the penguin
graph in Fig.~\ref{chapter2fig1}(b) 
which has an absorptive cut across the $u$ 
quark line.
The compensating process is then $b\to s u\bar u$ where, for this
process, the absorptive part, $\Im$m(loop), is driven by a cut on the
$c$ quark line in the loop. Thus

\beq
\Delta_{sc\bar c} (su\bar u) + \Delta_{su\bar u} (sc\bar c) =0
\label{deltascbarc} ~, \nonumber 
\eeq

\ni where the compensating nature of these two processes is illustrated
in Fig.~\ref{chapter2fig1}(c).

At this point it is instructive to discuss in some detail how the
cancellation follows from the Feynman diagrams combined in 
Fig.~\ref{chapter2fig1}(c). Let
us denote by $T_{sc\bar c}$, $T_{su\bar u}$ the tree-level contributions to
$b \to sc\bar c$, $b \to su\bar u$, respectively. Likewise, let us denote by
$P^q_{sc\bar c}$, $P^q_{su\bar u}$ the penguin contributions to 
$b \to sc\bar
c$, $b \to su\bar u$, respectively, 
with $q=u,c,t$ being the intermediate quark in the
penguin. We also denote the conjugate amplitudes as $\bar T_{sc\bar c}$,
$\bar T_{su\bar u}$, $\bar P^q_{sc\bar c}$ and $\bar P^q_{su\bar u}$. 

These amplitudes may be represented in terms of their magnitude and phase
as follows

\beq
\baa{ll}
T_{sc\bar c}  = e^{i\phi_c} |T_{sc\bar c}| & \bar T_{sc\bar c} =
e^{-i\phi_c}|T_{sc\bar c}| \\
T_{su\bar u} = e^{i\phi_u}|T_{su\bar u}| & \bar T_{su\bar u} =
e^{-i\phi_u}|T_{su\bar u}| \\
P^q_{sc\bar c} = e^{i\phi_q} e^{i\lambda_q^c}|P^q_{sc\bar c}| & \bar
P^q_{sc\bar c} = e^{-i\phi_q}e^{i\lambda_q^c}|P^q_{sc\bar c}| \\
P^q_{su\bar u} = e^{i\phi_q}e^{i\lambda_q^u}|P^q_{su\bar u}|\qquad & \bar
P^q_{su\bar u} = e^{-i\phi_q}e^{i\lambda_q^u}|P^q_{su\bar u}|
\eaa \label{cbarcsubarus}
\eeq

\ni Here $\phi_q$ is the CP-odd weak phase which has its origin in the
Lagrangian. In particular, the SM gives
$\phi_q=\arg(V_{qb}V^\ast_{qs})$, where $V_{qb}$ and $V_{qs}$ are 
the relevant CKM matrix elements. As can be seen in
Eq.~\ref{cbarcsubarus} this phase changes sign under CP\null.

The tree-level diagrams in perturbation theory have no additional phase,
however the penguin diagrams will, if the internal $q\bar q$ pair can be
on shell \cite{prl43p242}. 
These strong phases do not change sign under CP and 
are denoted here 
by $\lambda^u_q$ and $\lambda^c_q$. Note that in general
$\lambda^u_q,\lambda^c_q$ will depend also on the external momenta.

In this notation, the partial width differences that appear in
Eq.~\ref{deltascbarc} are given by the interference of the tree and penguin
amplitudes

\beqa
\Delta_{sc\bar c}(su\bar u) & = & \int 2 \left\{ \Re{\rm e} (T_{sc\bar
c}P^{u^\ast}_{sc\bar c}) - \Re{\rm e} (\bar T_{sc\bar c}\bar
P^{u^\ast}_{sc\bar c}) \right\} dPh(b\to sc\bar c) 
\nonumber\\
& = & 4\int |T_{sc\bar c}|\,|P^u_{sc\bar c}| \sin (\phi_c-\phi_u) \sin
(\lambda^c_u) dPh(b\to sc\bar c) 
\nonumber\\
\Delta_{su\bar u}(sc\bar c) & = & \int  2 \left\{ \Re{\rm e} (T_{su\bar
u}P^{c^\ast}_{su\bar u}) - \Re{\rm e}(\bar T_{su\bar u}\bar
P^{c^\ast}_{su\bar u}) \right\} dPh (b\to su\bar u) 
\nonumber\\
& = & -4\int|T_{su\bar u}|\,|P^c_{su\bar u}| \sin (\phi_c-\phi_u) \sin
(\lambda^u_c) dPh (b\to su\bar u) \nonumber\\
\eeqa

\ni where $dPh(\;)$ indicates an integral over the phase-space of the
decay. 

Referring to Fig.~\ref{chapter2fig1}(c) we see that the expression

\beq
\int |T_{sc\bar c}|\,|P^u_{sc\bar c}| \sin (\lambda^c_u) dPh (b\to s c\bar
c) ~,
\eeq

\ni is the interference of the absorptive part of $P^u_{sc\bar c}$
(indicated by the dashed line) with the tree-level graph 
(shown to the right of the solid line). It is identical to

\beq
\int |T_{su\bar u}| |P^c_{su\bar u}| \sin (\lambda^u_c) dPh (b\to u\bar us) ~,
\eeq

\ni since all the couplings are the same and we have just interchanged
the role of the internal phase-space of the $u \bar u$ 
cut (dashed line) with the
external phase-space $c \bar c$ cut (solid line). 
That is, for the PRA in the decay $b\to su\bar u$, the dashed line represents 
the external phase-space cut whereas the solid line indicates the absorptive 
$c \bar c$ cut.
We thus conclude that

\beq
\Delta_{sc\bar c} (su\bar u) = -\Delta_{su\bar u}(sc\bar c) ~,
\eeq

\ni as required.

The same kind of argument may be applied at higher orders in
perturbation theory as discussed in \cite{prd43p2909} in the case of $b$
decays, and is clearly an elaboration of the discussion above. 
For a further discussion of this example in the $S$-matrix
formalism see \cite{prd43p151}.

\begin{figure}
 \begin{center}
  \leavevmode
  \epsfig{file=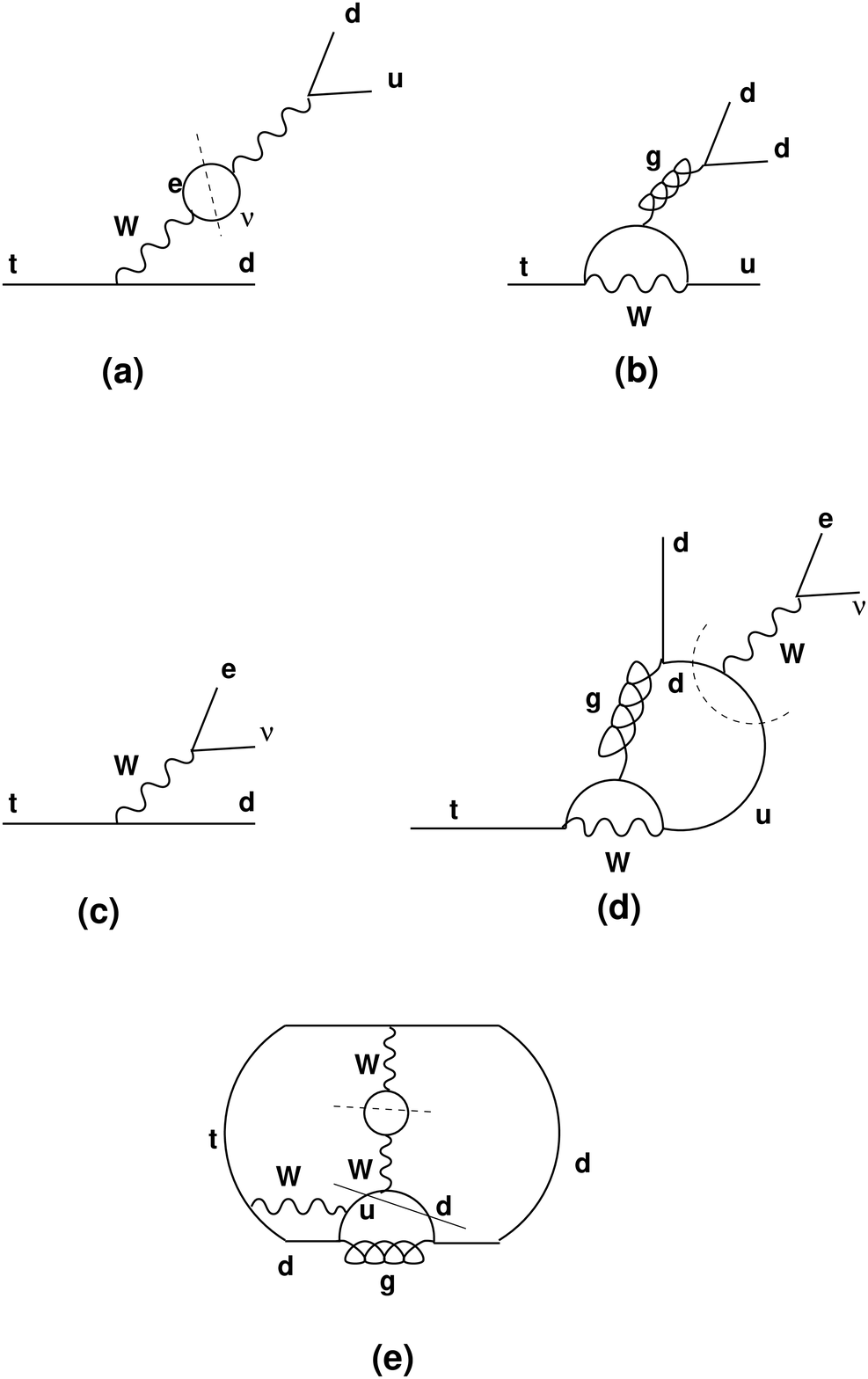,height=13cm}
\end{center}
\caption{\emph{Feynman diagrams contributing to the decays 
$t \to u d \bar d$ ($d=d,s$ or $b$ and $u=u$ or $c$) and 
to the decay $t \to d e \nu_e$:  
(a) ``Tree-level'' diagram for $t \to u d \bar d$ 
(with an imaginary part from the unstable on-shell W),
(b) 1-loop (order $\alpha_s$) penguin diagram for $t \to u d \bar d$, 
(c) The ``real'' tree-level diagram for $t \to d e \nu_e$,
(d) 2-loop contribution to $t \to d e \nu_e$ with an absorptive cut 
along the intermediate $du$ lines, and 
(e) A diagrammatic description for the compensating nature of the contribution 
of the 1-loop$\times$1-loop processes to a PRA in 
$t \to u d \bar d$ and the tree$\times$2-loop processes to a PRA in 
$t \to d e \nu_e$ (see text for discussion).}}
\label{chapter2fig2}
\end{figure}

Let us now consider the analogous example from top decays. PRA in
channels of 
the type $t\to u d \bar d$, $c d\bar d$, \dots ($d=d,s$ or $b$) 
arise through interference of
the ``tree'' graph in Fig.~\ref{chapter2fig2}(a) 
with the penguin in Fig.~\ref{chapter2fig2}(b). Since
$m_t>(m_W+m_b)$, the $W$ is on-shell and the $W$-propagator in
Fig.~\ref{chapter2fig2}(a) is complex, as the $W$ has 
an appreciable width. So the diagram shown in Fig.~\ref{chapter2fig2}(a)
has in fact an imaginary part (indicated by the dashed line) 
due to the $W$-width, which can dominate
over the imaginary part of the penguin diagram depicted 
in Fig.~\ref{chapter2fig2}(b). 
In fact, the 2-loop 
diagram corresponding to Fig.~\ref{chapter2fig2}(b) 
with a ``bubble'' on the internal $W$-propagator can dominate over
the 1-loop diagram.

The CP-CPT connection has two implications here, discussed by Soares
\cite{prl68p2102}. 
First\-ly, in calculating the PRA into a given channel such as $t\to
cb\bar b$, part of the total $W$-width that arises from $W\to c\bar b$
must be subtracted away as it represents rescattering of the final
state onto itself (see Eq.~\ref{pr3}). Numerically the most important
consequence here is for the channel $t\to ud\bar d$ as then the process
$W\to u\bar d$ is not Cabibbo suppressed and indeed contributes about
1/3 to the $W$-width.

A second interesting implication of the CP-CPT connection,
pointed by Soares \cite{prl68p2102},  
in particular Eq.~\ref{pr4}, is that
the PRA of 
leptonic decays of the top (e.g., $t\to de\nu_e$) can be deduced
from the PRA into its hadronic decays (e.g., $t\to dc\bar d$). In the
language of Eq.~\ref{pr4} these two reactions are the compensating
processes of each other, i.e., $I=dc\bar d$ and $J=de^+\nu_e$ in
Eq.~\ref{pr4} leading to

\beq
\Delta_{de\nu_e} (dc\bar d) = - \Delta_{dc\bar d} (de\nu_e)
\label{pr12} ~.
\eeq

\noindent The relevant Feynman diagrams that contribute to a PRA in 
the leptonic top decay $t\to de\nu_e$ are depicted in 
Fig.~\ref{chapter2fig2}(c) (tree-level diagram) and 
Fig.~\ref{chapter2fig2}(d) (2-loop diagram with 
an absorptive cut along the inner $du$ lines).  
A graphic illustration of how the compensating nature between the channels 
$t\to d u \bar d$ and $t\to de\nu_e$ comes about (i.e., Eq.~\ref{pr12}) 
is shown in Fig.~\ref{chapter2fig2}(e). 
In this figure the dashed line indicates the absorptive cut
that, due to the Cutkosky rule, is responsible for the FSI phase 
in the decay $t\to d u \bar d$,  
and
the solid line indicates the separation between the two contributing
diagrams to the PRA in $t\to d u \bar d$. 
Similarly, for the decay $t\to de\nu_e$ the roles of the dashed and 
solid lines are reversed such that the dashed line indicates the 
separation between the two contributing diagrams (shown in 
Figs.~\ref{chapter2fig2}(c) and \ref{chapter2fig2}(d)) and the solid line 
represents the necessary absorptive cut for the PRA in the decay channel
$t\to de\nu_e$.  
Notice that a direct calculation of the PRA in the
leptonic top decays would have required a calculation of a 2-loop
graph (Fig.~\ref{chapter2fig2}(d)). 
So this application of the CPT theorem is especially
noteworthy as through Eq.~\ref{pr4} the necessary calculation gets
simplified to dealing with two 1-loop graphs (namely 
Fig.~\ref{chapter2fig2}(a) and Fig.~\ref{chapter2fig2}(b)). 
Unfortunately all the PRA's resulting from such
interferences between the $W$-propagator and the penguin are much too
small to be of experimental interest, at least in the SM\null.

Before we finish this discussion on the CPT constraints on PRA's,
let us emphasize two important points.
First it should be very clear that these constraints are imposed only on
{\it a single\/} CP-violating observable, namely PRA\null. Indeed, in 
general,
many other ($T_N$-even) CP-violating observables can be constructed
(e.g., energy asymmetry, helicity asymmetry,\dots) which are similar to
PRA in that they  all require CP-conserving FSI
phase(s). CPT constraints are very specific; they {\it do not\/} affect
any of these other observables other than PRA\null. In particular the
stricture of CPT, that the rescattering of a final state onto itself
cannot give rise to an asymmetry, 
is specific to PRA only. Rescattering of a final
state onto itself {\it can\/} give rise to important nontrivial
experimental implications for all other CP-violating observables;
rescattering graphs can certainly cause observable  helicity or energy
asymmetry. Another way of viewing this situation, in light of the CPT
theorem, is to regard each portion of the phase-space  or each
polarization as a separate final state. Thus one portion of phase-space
or polarization state can compensate for another.

As a specific example of this, consider the Schmidt-Peskin effect namely
the helicity asymmetry \cite{prl69p410}:

\beq
\Delta N_{LR} 
= \frac{N(t_L\bar t_L) - N(t_R\bar t_R)}
{
{\rm all~}t\bar t
} 
\label{pr13}
\eeq

\noindent induced in the reaction $pp\to t\bar t+X$. Here $N(t_L\bar t_L)$
represents the number of pairs of left-handed $t$ and left-handed
$\bar t$ produced in the inclusive reaction. In a 2HDM 
with CP-violating phase(s) in the neutral Higgs sector, a
non-vanishing value of $\alpha_h$ arises from interference of 
tree-level diagrams with 1-loop diagrams involving neutral Higgs exchanges
in the loops (see Fig.~\ref{ppttfig2} in Chapter \ref{sec7}). 
In this case, the CP-conserving FSI phase required by
$\alpha_h$ is provided by an absorptive cut which explicitly arises 
from $t\bar t\to t\bar t$,
i.e., rescattering of a final state onto itself. So a very interesting
CP-violating effect, namely the helicity asymmetry in Eq.~\ref{pr13} -  
a CP-odd $T_N$-even
observable (similar to PRA with respect to this classification) - arises
from self-rescattering of a final state (we will return to a more detailed 
discussion of the Schmidt-Peskin 
effect in Chapter \ref{sec7}). Thus, in general, in
discussing CP violation phenomena or in calculations of CP violation
effects, self-rescattering graphs should not automatically be discarded, 
unless
of course one is specifically calculating PRA's. In particular, 
for the specific case discussed above, we may
regard $t_L\bar t_L$ and $t_R\bar t_R$ as compensating final states.

An important, although somewhat obvious consequence is that even when
PRA's are vanishingly small or exactly zero, other CP-odd observables
can be non-vanishing and can have interesting observational
consequences. 

Another interesting example arises from CP-violating phase originating
from a charged Higgs sector, as in the Weinberg model \cite{prl37p657} with
three 
doublets of Higgs fields. Important CP-violating effects arise,  
for example, in the leptonic decay $t\to b\tau\nu_\tau$
from the interference of the SM graph (with $W$ exchange) 
with the tree-level charged Higgs exchange graph 
(see Figs.~\ref{tdecaysfig3}(a) and \ref{tdecaysfig3}(c) 
in Chapter \ref{sec5}).

For this case, let us first consider the PRA

\beq
\alpha_\tau = \frac{\Gamma(t \to b\tau\nu) - \Gamma(\bar t\to \bar
b\bar\tau\bar\nu)}{\Gamma(t\to b\tau\nu) + \bar \Gamma(\bar t\to \bar
b\bar\tau\bar\nu)} \label{pr14} ~.
\eeq

\noindent For simplicity, let us assume that $m_{H^+}>m_t$; 
then a CP-conserving
absorptive part required for $\alpha_\tau$ arises from the 
$W$-boson ``bubble''
which contains all possible states other than $\tau\nu_{\tau}$ 
as required by
CPT.  However, because of the spin zero nature of the Higgs the 
$W^+$-$H^+$
interference 
has non-vanishing contributions only for the scalar part of the $W$.
This argument is most readily seen if one uses the Landau gauge for 
calculating
this interference (for more details see Chapter \ref{sec5}). 
In that case, the
scalar and vector components of the $W$ propagator are cleanly separated
according to their total angular momentum. Thus, graphs which pass through a
vector $W$ intermediate state will not interfere with graphs 
that pass through a
Higgs state. The Higgs must therefore interfere only with the Goldstone
propagator which corresponds to the decay of longitudinal $W$ 
into fermion pairs
$c\bar s$, $u\bar d$, $e\nu_e$\dots which are all suppressed by powers of the
fermion masses. Furthermore, the Goldstone propagator shows none of the
resonance enhancement associated with the vector component.  
Thus the $c\bar s$
is the most important contributor to the scalar component of the $W$-boson
``bubble''.  The $\tau\bar \nu$ and $c\bar s$ can be thought of as the
compensating processes. So, in fact, the PRA goes as

\beq 
\alpha_\tau \sim \frac{m^2_c}{m^2_t}\, \frac{m^2_\tau}{m^2_t}
\frac{\Gamma_W}{m_W} \label{pr15} ~, 
\eeq

\noindent and is extremely small \cite{prd51p3525}. 

However, as already mentioned before, just 
because PRA is vanishingly small does not, though, mean that there
are no CP violation effects. Indeed, very important and large
CP-violating asymmetries may arise in the decay $t \to b \tau \nu_\tau$. 
First of all, one can construct an
energy asymmetry:

\beq
\alpha_E = \frac{\langle E_\tau\rangle - \langle
E_{\bar\tau}\rangle}{\langle E_\tau\rangle + \langle E_{\bar\tau}
\rangle} \label{pr16} ~.
\eeq

\noindent That is, compare, e.g., the average energy of the $\tau$ in
$t\to b\tau\nu$ with that of the $\bar \tau$ in $\bar t\to \bar
b\bar\tau \bar\nu$. Then $\alpha_E$ will not suffer from the helicity
suppression or constraints of CPT on $\alpha_\tau$ and one expects

\beq
\frac{\alpha_E}{\alpha_\tau} \sim \frac{m^2_t}{m^2_c} \,
\frac{m^2_t}{m^2_\tau} \label{pr17} ~,
\eeq

\noindent as explicit calculations confirm.

Indeed, a CP-violating asymmetry even bigger than the energy asymmetry,
namely the transverse polarization asymmetry of the $\tau$, resides in
this $W^+$-$H^+$ interference. In fact, the transverse polarization
asymmetry is enhanced by another factor of $m_t/m_\tau$ compared to the
energy asymmetry as is shown in section \ref{sssec513}.

\subsection{Resonant $W$ 
effects and CP violation in top decays \label{ssec24}}

The large mass of the top $(m_t\simeq 174$ GeV) means that it decays
 to a three-body final state primarily through an on-shell $W$.
This fact is of particular interest in the study of CP violation in
such decays since there will be a large strong (i.e., CP-even) 
phase inherent in this
$W$-propagator. 

In particular, since the $W$-width is substantial ($\Gamma_W\sim2$
GeV), the transverse modes of the $W$ are controlled by the
Breit-Wigner propagator

\beq
G_T = \frac{1}{q^2_W-m_W^2+i m_W \Gamma_W} \label{rwe1} ~,
\eeq

\ni which will have a substantial strong phase.

The enhancement of the imaginary part of $G_T$ is evident by
considering that, at $q^2_W=m^2_W$, 

\beq
| \Im{\rm m} (G_T) | = (m_W\Gamma_W)^{-1} ~. 
\eeq

\ni The real part swings through 0 at this point but in the vicinity of
the resonance it will also be large. For instance, if
$|q^2_W-m^2_W|=m_W\Gamma_W$, then

\beq
|\Re{\rm e} (G_T)| = (2 m_W\Gamma_W)^{-1} ~.
\eeq

\noindent Since $\Gamma_W\sim {\cal O} (\alpha)$ 
this means that near $q^2\to m^2_W$
both the real and imaginary parts of the amplitudes 
for decays such as $t\to bu\bar d$,
$b c \bar b, \dots$ behave as if they are ${\cal O} (1)$ in the gauge coupling
constant. This phenomena is what we mean by ``resonance enhancement''.
The imaginary part here then provides the needed absorptive part
(i.e., FSI phase) to lead to enhancement of CP-odd $T_N$-even observables.
Likewise, near $q^2\sim m^2_W$, the real part can magnify the effect of
CP-odd $T_N$-odd observables.

The basic idea of FSI phase driven by particle widths in decays was
discussed in \cite{prl67p1979} (see also \cite{plb245p185}). 
Although
originally the 
discussion \cite{prl67p1979} 
took place in the context of the SM in conjunction with the
CKM phase, it should be completely clear that they can be equally well
used with non-standard sources of CP violation. Indeed,
as dealt in section \ref{sssec513}, the $W$-resonant effects provide a
significant enhancement of $T_N$-even and $T_N$-odd CP-violating
effects in top decays in the context of an extended Higgs sector.

\subsection{Effective Lagrangians and observables \label{ssec25}}

One tool that is often used to catalog the effects of new physics
at an energy scale, $\Lambda$, much higher than the electroweak 
scale, is the effective Lagrangian (${\cal L}_{eff}$) method.
If the underlying extended theory under consideration only becomes important 
at a scale $\Lambda$, then it makes sense to expand the Lagrangian 
in powers of $\Lambda^{-1}$ where the $\Lambda^0$ term is the 
SM Lagrangian and the other terms are the effective Lagrangian terms.

Simple dimensional arguments tell us that the operator which multiplies 
$\Lambda^{-n}$ must be of dimension $n+4$.
This restriction together with symmetry considerations implies that at 
each order in $\Lambda^{-1}$ there are only a finite number of possible
 terms. Conversely, this implies that experimantal tests for the existence 
of specific terms in ${\cal L}_{eff}$ is a relatively model independent 
\cite{bnlsum90} way to search for new physics. 

Here we are interested in top quark physics which violates 
CP and so, in the effective Lagrangian approach, the operators 
of interest are further restricted. For example, in Chapter \ref{sec4}
we discuss the top electric dipole moment (and related effects) 
which can arise from a dimension 5 term in the effective 
Lagrangian:

\begin{equation}
{\cal L}_{eff}^{[5]} \propto \bar t \sigma_{\mu \nu} \gamma_5 
t F^{\mu \nu} ~.
\end{equation}

\noindent At dimension 6, photons can interact with the top 
quark via a CP-violating operator such as

\begin{equation}
{\cal L}_{eff}^{[6]} \propto \left( \bar t \gamma_5 t \right) 
\left( F^{\mu \nu} F_{\mu \nu} \right) ~,
\end{equation}

\noindent which could arise, for instance, via a SUSY box diagram.

Let us now consider as a specific example effective Lagrangian terms 
which would contribute to the process $gg \to t \bar t$.
It is
useful to recall that in such an expansion, operators that are proportional
to the QCD equations of motion for the top or the gluon fields may be
eliminated by a field redefinition and are therefore redundant
\cite{bnlsum90}. 

There are a number of requirements that an operator has to satisfy for
it to be relevant to CP violation in top production 
in hadronic collisions. These are:

\begin{enumerate}
\item It must violate CP\null.
\item Its Feynman rules must include couplings to two or fewer gluons.
\item It must not be proportional to $q^2$ of one of the on-shell
gluons in the initial state.
\end{enumerate}

The need for the first requirement is clear. The second requirement
is present since the events in question have two gluons in the initial 
state and no gluons in the final states. If one wanted to consider 
experiments where additional gluon jets were detected in the final state, 
clearly one would have to generalize this requirement.

In constructing a basis of operators which satisfy the above conditions it
can be shown that \cite{cburgesscpwshp} one can eliminate, without loss of
generality, any operator which is equal to $0$ modulo the equations of
motion. Equivalently, if the difference of two operators is $0$
modulo equations of motion, then only one need to be included.

Here is a set of operators that we choose which satisfy the
requirements mentioned above and are of dimension six or less

\beqa
O_a & = & 
\bar t
\ \  
i [ {1\over 2} f_a (-\square)F^{\mu\nu}_i] \sigma_{\mu\nu} \gamma_5
T^i
\ \ 
t 
~, \nonumber \\
O_b & = & 
\bar t
\ \  
i [ f_b (-\square) F^{\mu\nu}_i F^i_{\mu\nu} ]\gamma_5
\ \ 
t 
~, \nonumber \\
O_c & = & 
\bar t 
\ \ 
\epsilon_{\alpha\beta\gamma\delta} 
[f_c (-\square)F^{\alpha\beta}_i F^{\gamma\delta}_i]
\ \ 
t 
~, \label{el1}\\
O_d & = & 
\bar t 
\ \ 
i [f_d (-\square) F^{\mu\nu}_j F^k_{\mu\nu} ] \gamma_5
d^{ijk} T^i
\ \ 
t 
~, \nonumber \\
O_e & = & 
\bar t 
\ \ 
\epsilon_{\alpha\beta\gamma\delta} 
[ f_c (-\square) F^{\alpha\beta}_j F^{\gamma\delta}_k d^{ijk}] 
T^i 
\ \ 
t
~, \nonumber 
\eeqa
\newcommand{\squareb}{\sqcap\mbox{\hglue-.65em}\sqcup}

\noindent where $F$'s are the gluon field strength tensor, 
$\squareb = D^\mu D_\mu$, the analytic functions
$f_a,\dots,f_e$ are form factors
and $T^i=\lambda^i/2$, $\lambda^i$ being the Gell-Mann
color matrices. 
Note, that in general, higher order terms in $\squareb$ will imply the 
existence of couplings to additional numbers of gluons.

As an illustration, let us consider further the operator $O_a$. This
operator essentially corresponds to 
the Chromo-Electric Dipole Moment (CEDM) form
factor. The experimental implications of the static analog of this
quantity were considered in \cite{prl69p2754}. We can expand the operator to
obtain the Feynman rule. The vertex for the one gluon interaction is

\beq
if_a (q^2) \bar t \sigma^{\mu\nu} \gamma_5 T^jt q_\mu \epsilon^j_\nu 
~, \label{el2}
\eeq

\noindent 
here $\epsilon_\mu$ is the polarization vector of the gluon. This 
is completely analogous to the EDM form factor. However
$O_a$ now also gives rise to a two gluon coupling given for on-shell
gluons by

\beqa
& & g_s \bar t\sigma^{\mu\nu} \gamma_5T^i tF^{ijk} \bigg[h_a(q^2) 
\epsilon^j_{1\mu}
\epsilon^k_{2\nu}  \nonumber \\
& & \quad + \left[ \frac{h_a(q^2)-h_a(0)}{q^2} \right] (q_2\cdot \epsilon^j_1
q_{2\mu} \epsilon^k_{2\nu} - 
q_1\cdot \epsilon^k_2 q_{1\mu} \epsilon^j_{1\nu}) \bigg]
~, \label{el3} 
\eeqa

\noindent 
where $q=q_1+q_2$. Note that the second 
term that appears is needed to maintain gauge invariance.

An important feature of $f_a$ (as well as of the other form factors) is
that the constant piece $f_a(0)$ must be real while, at $q^2\ne0$, $f_a$
may have an imaginary part due to the possibility of thresholds giving
rise to absorptive pieces. Indeed these type of effects have also been
considered in some particular extensions of the SM (see Chapter \ref{sec4}).

The phenomenology of the static CEDM (i.e., for $q^2 \simeq 0$) was considered
extensively in \cite{prl69p2754}. It was shown that in a hadron collider
with $\sim 10^{7} ~ t\bar t$ pairs, 
both of which decay leptonically, a precision
of about $5\times10^{-20}$ $g_s$-cm for $f_a(0)$ could be achieved. We
can extend this consideration with a simplifying assumption that $f_a$
is approximately constant above the $t\bar t$ threshold. We can then
introduce the quantity $f^\prime = f_a(4m^2_t)-f_a(0)$. With this
assumption and approximation we find that, under ideal conditions,
$\Re{\rm e}(f^\prime)$ and $\Im{\rm m}(f^\prime)$ 
can also be measured to a precision
of about (2--3)${}\times10^{-20}$ $g_s$-cm.

The Weinberg model with an extended Higgs sector provides a specific
example of non-standard physics where one can study this general feature
of the operator analysis above. We recall that the source of CP violation
now are the charged Higgs exchanges (see section \ref{sssec324}).  Since
$O_a$ is the only operator which gives a one gluon Feynman rule, the
electromagnetic form factor calculated in \cite{prl69p33} is the quantity
$f_a$ except for the replacement of $g_s$ with $e$, the electric charge.
The result of that reference is that thus far QCD yields a limit around
$5\times10^{-20}$ $g_s$-cm. In that work it is also explicitly
shown (see Chapter \ref{sec4}) that the $q^2$ dependence is rather mild.
Furthermore, above threshold the $\Im{\rm m}(f_a)$ was also shown to have
roughly the same ball park value. Thus, studies at a hadron collider could
exhibit CP-violating signals although admittedly the experimental
challenges are formidable. 

Another useful way to characterize the amplitude for $gg\to t\bar t$ is to
express it in terms of 
form factors.
There are three possible color structures which such an amplitude can 
have. If $A$, $B$ are the color indices of the gluon and $i$, $j$ the 
indices of the $t$ and $\bar t$, these color structures are

\begin{eqnarray}
\Delta&=&\delta_{AB}\delta_{ij} ~, \nonumber\\
D&=&d^{ABC}T^C_{ij} ~, \\
F&=&f^{ABC}T^C_{ij} ~. \nonumber
\end{eqnarray}

\noindent Let us define $P_{g1}$, $P_{g2}$ to be the momenta of the 
gluons and $P_{t1}$, $P_{t2}$ to be the momenta of the $t, \bar t$
quarks respectively. Let us further define the variables

\begin{eqnarray}
s = (P_{g1} + P_{g2})^2,~~ 
t = (P_{g1} - P_{t1})^2,~~ 
u = (P_{g1} - P_{t2})^2
\nonumber
\end{eqnarray}

\begin{eqnarray}
z = (t - u)/(s-2m_t^2)
\end{eqnarray}

\noindent Let $E_1$ and $E_2$ be the polarizations of the gluons in a
gauge where $E_1\cdot P_2 = E_2\cdot P_1 = 0$.

Here, we are interested in amplitudes which violate CP. These amplitudes 
must also be symmetric under the interchange of the two gluons. The 
helicity amplitudes which satisfy these conditions are

\begin{eqnarray}
a_1^n&=& 
f_1^n(s,z^2)
(E_1\cdot E_2)
(\bar t \gamma^5 t)
[D,\Delta,Fz]
\nonumber\\
a_2^n&=& 
f_2^n(s,z^2)
(\epsilon_{\mu\nu\sigma\rho}
E_{1}^\mu E_{2}^\nu P_{g1}^\sigma P_{g2}^\rho) 
(\bar t  t)
[D,\Delta,Fz]
\nonumber\\
a_3^n&=& 
zf_3^n(s,z^2)
(\epsilon_{\mu\nu\sigma\rho})  
E_{1}^\mu E_{2}^\nu P_{t1}^\sigma P_{t2}^\rho
(\bar t  t)
[D,\Delta,Fz]
\nonumber\\
a_4^n&=& 
zf_4^n(s,z^2)
E_1^\mu E_2^\nu
(\bar t \sigma_{\mu\nu}\gamma^5  t)
[D,\Delta,Fz]
\nonumber\\
a_5^n&=& 
f_5^n(s,z^2)
(\epsilon_{\mu\nu\sigma\rho}
E_{1}^\mu E_{2}^\nu (P_{g1}+ P_{g2})^\sigma) 
(\bar t \gamma^\rho t)
[D,\Delta,Fz]
\end{eqnarray}

\noindent
where $f_i$ is a function of $s$ and $z^2$ and
the notation  $[D,\Delta,Fz]$
means that the term may be multiplied by any of these color structures, 
the index $n=1,2,3$ respectively depending on which of these color 
structures apply. 

As an explicit example, at tree-level, the CEDM operator 
discussed above will contribute to the amplitude $a_4^3$.  

\subsection{Optimized observables \label{ssec26}}

\begin{figure}
 \begin{center}
  \leavevmode
  \epsfig{file=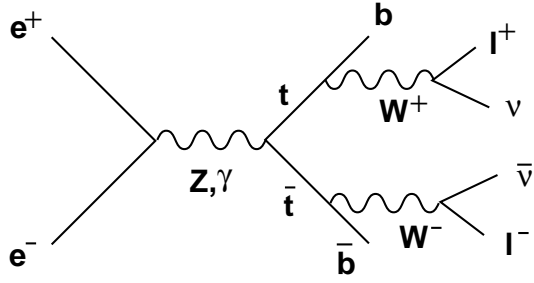,width=7 cm}
\end{center}
\caption{\emph{
Feynman diagram describing the process $e^+e^- \to t \bar t$ followed 
by the $ t \bar t$ decays $t \to b W^+ \to b \ell^+ \nu_\ell$ and 
 $\bar t \to \bar b W^- \to \bar b \ell^- \bar\nu_\ell$.}}   
\label{Feyn1}
\end{figure}

Due to the exceedingly short life-time of the top quark, measurement of
its couplings requires considering top production and decay
simultaneously. Consider, for example, the production and decay of $t\bar
t$ in $e^+e^-$ annihilation (see Fig.~\ref{Feyn1}),

\beq
e^+(p_+) + e^-(p_-) \to t(p_t) + \bar t (p_{\bar t}) \label{oo1} ~,
\eeq

\noindent with

\beq
t(p_t) \to b(p_b) + W^+(p_{W^+}) \label{oo2} ~,
\eeq

\noindent and

\beq 
\bar t(p_{\bar t}) \to \bar b ( p_{\bar b}) + W^-(p_{W^-}) \label{oo3} ~.
\eeq

\noindent Indeed each of the $W^\pm$ also decays leptonically or
hadronically (i.e., into jets). Thus

\beqa
W^+ (p_{W^+}) & \to & \ell^+(p_{\ell^+}) + \nu_\ell (p_\nu) ~,\nonumber \\
W^- (p_{W^-}) & \to & \ell^-(p_{\ell^-}) + \bar\nu_\ell (p_{\bar\nu})
\label{oo4} ~,\\
W^\pm & \to & j_1(p_{j_1}) + j_2 (p_{j_2}) ~. \nonumber 
\eeqa

\noindent This allows one to construct a multitude of observables,
involving momenta of the initial beam and various decay products, to
probe the presence of anomalous vertices in the top interactions. The case
of the CP 
violating dipole moment interactions is especially interesting to this
work. Such anomalous terms could occur at the $\bar tt\gamma$, $\bar ttZ$ or
$\bar ttg$ vertices corresponding to electric, weak or 
chromo-electric dipole
moment of the top quark. Of experimental interest are the values of the
corresponding form 
factors at $q^2=s$, the square of the c.m. energy rather than the moments
($q^2=0$) themselves. It is possible, therefore, that the form factor
is complex. The real and imaginary parts can thus lead to 
distinct experimental effects.

Two examples of simple (or ``naive'') observables that
can 
be used to probe the presence of imaginary part of the dipole moment
form factor are

\beqa
&& \frac{\langle E_{\ell^+} \rangle - \langle E_{\ell^-} \rangle}
{\langle E_{\ell^+} \rangle +
\langle E_{\ell^-} \rangle} ~, \\
&& \frac{\langle E_b \rangle - \langle E_{\bar b}\rangle}
{\langle E_b \rangle + \langle E_{\bar b} \rangle} ~.
\label{oo5} 
\eeqa

\noindent Examples of simple (or ``naive'') observables that
probe the presence of the real part of the dipole moment form factor are

\beq
\frac{\langle \epsilon_{\mu\nu\rho\sigma} p^\mu_{\ell^+} p^\nu_{\ell^-}
p^\rho_b p^\sigma_{\bar b} \rangle}{( p_{\ell^+} \cdot p_{\ell^-} p_b
\cdot p_{\bar b} )^{1/2}} \label{oo6} ~,
\eeq

\beq
\frac{(p^x_{\ell^+} p^y_{\ell^-} - p^y_{\ell^+} p^x_{\ell^-}) \mbox{
sgn } (p^z_{\ell^+} - p^z_{\ell^-} )}{(p_{\ell^+} \cdot
p_{\ell^-})^{1/2}} \label{oo7} ~.
\eeq

\noindent The ability to polarize the electron beams at a future 
$e^+e^-$ collider (e.g., the NLC) allows us to construct additional 
observables 
involving beam polarization.
Clearly, while many observables can be constructed to probe the dipole
moment, we may ask whether it is possible to construct an ``optimal
observable'' i.e., one which will be the most sensitive or will have
the largest ``resolving power''. A general procedure for constructing
an optimal observable was given in \cite{prd45p2405}. Here we will
briefly review the method.

Let us write the differential cross-section as a sum of two terms

\beq
\Sigma(\phi) = \Sigma_0(\phi) + \lambda\Sigma_1(\phi) \label{oo8} ~,
\eeq

\noindent where $\lambda$ is a parameter (e.g., dipole moment or
magnetic moment form factor) and $\phi$ is some phase-space variable
(including angular and polarization variables).

For an ideal detector that accurately records the value of $\phi$ for
each event that occurs, any method for determining the value of
$\lambda$ amounts to weighting the events with a phase-space-dependent
function $f(\phi)$ 
which we assume is CP-odd. Let us define

\beq
f^{(1)} (\lambda) = \int{f(\phi)} \Sigma{(\phi)} d\phi ~. \label{oo9}
\eeq

\noindent Thus the change due to the contribution from the presence of
$\lambda$ is

\beq
f^{(1)}(\lambda) = \lambda\int f(\phi)
\Sigma_1(\phi) d\phi \equiv \lambda f^{(1)}(\lambda=1) \label{oo11} ~.
\eeq

\noindent $f^{(1)}(\lambda)$ then has to be compared to the error in
its measurement. If $n$ events are recorded, the error is

\beq
\Delta f=  \left[ \frac{f^{(2)} \sigma - 
\left( f^{(1)}(\lambda) \right)^2 }{n \sigma^2} 
\right]^{1/2} 
\approx \left[ \frac{f^{(2)}}{n\sigma} \right]^{1/2} \label{oo12} ~,
\eeq

\noindent where

\beq
\sigma =\int\Sigma d\phi = \int\Sigma_0 d\phi \label{oo13} ~,
\eeq

\noindent is the total cross-section and 

\beq 
f^{(2)} = \int f^2\Sigma d\phi = \int f^2 \Sigma_0 d\phi ~.\label{oo14}
\eeq

\noindent Note that $f^{(1)}(\lambda) \propto \lambda$ but $f^{(2)}$
and $\sigma$ do not depend on $\lambda$ to first order;
we will assume that $\lambda$ is sufficiently small 
so that the above approximation is valid as well as 
$f^{(2)} \sigma >> \left( f^{(1)}(\lambda) \right)^2$.  
We now introduce the ``resolving power'', $R$, which measures the
effectiveness of an operator for determining $\lambda$. For
the function $f$, $R$ is defined as

\beqa
R &=& \frac{1}{n \sigma^2 \lambda^2}\; 
\frac{ \left( f^{(1)}(\lambda) \right)^2}{(\Delta f)^2}=
\frac{(f^{(1)}(1))^2 }{f^{(2)} \sigma} ~.\label{oo15}
\eeqa

\noindent The statistical significance $S$, with which the presence of a
nonzero value of $\lambda$ may be ascertained, is given by

\beq
S = \frac{ f^{(1)}(\lambda) }{\sigma \Delta f} = \lambda \sqrt{nR} ~. \label{oo16}
\eeq

\noindent Thus the larger $R$ is the more effective $f(\phi)$ is for
measuring $\lambda$. Clearly one would like to choose a function $f(\phi)$ 
to determine $\lambda$ for which $R$ is maximal.

A special case of such observables are asymmetries, which are
observables where $f(\phi)=\pm1$. For such
observables $f^{(2)}=\sigma$ so (see Eq.~\ref{oo15}),

\beq
R = (f^{(1)}(1))^2/\sigma^2 = \langle f\rangle^2 / \lambda^2 ~. 
\eeq

\ni In this case it is conventional to denote for an asymmetry $f$

\beq
\langle f\rangle = \alpha_f ~,
\eeq

\ni so that Eq.~\ref{oo16} becomes

\beq
S= |\alpha_f| \sqrt{n} ~.
\eeq

\ni Thus the quantity $\lambda\sqrt{R}$ is the natural generalization
of $|\alpha_f|$ for more general observables among which  
we would like to find the optimal one, i.e., that which maximizes the value of 
$R$ and therefore also the statistical significance with which it can 
be measured. 

In order to define such an optimal function (i.e., $f_{\rm opt}$) it is
useful to decompose an arbitrary function $f$ as \cite{prd45p2405}:

\beq
f = A \frac{\Sigma_1}{\Sigma_0} + \hat f \label{oo18} ~,
\eeq

\noindent where $A$ is defined as

\beq
A = \frac{\int f\Sigma_1 d\phi}{\int \frac{\Sigma^2_1}{\Sigma_0} d\phi} ~.
\label{oo19} 
\eeq

\noindent 
Since rescaling $f$ does not change the value of $R$ we can take $A=1$.
If we now expand the definition of $R$ to lowest order in $\lambda$ and
use the above decomposition we get

\beqa
\sigma R = \frac{|\int f\Sigma_1d\phi|^2}{\int f^2\Sigma_0 d\phi}
= \frac{|\int \frac{\Sigma^2_1}{\Sigma_0} d\phi|^2}{\int
\frac{\Sigma^2_1}{\Sigma_0} d\phi + \int \hat f^2 \Sigma_0 d\phi}
\label{oo21a} ~.
\eeqa

\noindent Since the two terms in the denominator are each positive it
is clear that $R$ is maximal when $\hat f=0$. Thus, from
Eq.~\ref{oo18},  

\beq
f=f_{\rm opt} = \frac{\Sigma_1}{\Sigma_0} \label{oo22} ~,
\eeq

\noindent maximizes $R$.

Another way to understand this derivation of $f_{\rm opt}$ is to
introduce the concept of a vector space as the set of all functions on
which we define the scalar product

\beq
g_1\cdot g_2 = \int g_1(\phi) g_2(\phi) \Sigma_0(\phi) d\phi
\label{oo23} ~,
\eeq

\noindent where $g_{1,2}$ are functions of $\phi$. If we denote
$f_0=\Sigma_1/\Sigma_0$ then in this notation Eq.~\ref{oo21a} may be
rewritten as

\beq
\sigma R = (f\cdot f_0)^2/f\cdot f \label{oo24} ~.
\eeq

\noindent Thus $R$ is maximized when $f=f_0=f_{\rm opt}$.

Let us first illustrate these consideration with a toy example.
Consider a $2 \to 2$ scattering process where the differential
cross-section takes the form

\beq 
\frac{d\sigma}{d\xi} = A+B\xi +C\xi^2 \label{oo25} ~,
\eeq

\noindent where $\xi =\cos\theta$ and $B,C\le A$. According to the
preceding discussion, the function most sensitive to $B$ is $f_B=\xi$.
Thus explicitly we find $\sigma=2A+\frac{2}{3}C$, $f^{(1)}_B =
\frac{2}{3}B$, $f^{(2)}_B = \frac{2}{3}A +\frac{2}{5}C$. Using
Eq.~\ref{oo15}, the resolving power for the optimum choice
($f_B=\xi$) is

\beqa
R(f_B) & = & \frac{1}{B^2} \, \frac{[\int^1_{-1}
B\xi^2d\xi]^2}{\int^1_{-1} Ad\xi \int^1_{-1} A\xi^2 d\xi} 
= \frac{1}{3} \, \frac{1}{A^2} \label{oo26} ~.
\eeqa

\noindent Suppose we consider measuring $B$ via another function, which
has the same symmetry properties as $f_B$ and is defined as:

\beq
g_B(\xi) = {+1 \atop -1} \qquad {{\rm if}\,\, \xi>0 \atop {\rm if}\,\,
\xi<0} \label{oo27} ~.
\eeq

\noindent Then one gets $g^{(1)}_B = B$, $g^{(2)}_B =2A + \frac{2}{3}
C$, so that $R(g_B)=\frac{1}{4}\,\frac{1}{A^2}$, which is not as good as
$R(f_B)$. 
\medskip

\noindent\underbar{Generalizations}
\medskip

This simple method outlined above for an optimized observable has been
widely used and a useful generalization has been proposed by Gunion,
Grzadkowski and He \cite{prl77p5172}.

In general, we will assume that the 
differential cross-section can be expressed as

\beq
\Sigma(\phi) \equiv \frac{d\sigma}{d\phi} = \sum_i c_if_i(\phi)
\label{oo28} ~,
\eeq

\noindent where $c_i$ are model-dependent coefficients and $f_i(\phi)$
are known functions of the final state phase-space variable(s), $\phi$.
Thus, $c_i$ may be couplings; some or all of which one is trying to
extract from experimental data. So, in analogy, with the discussion
above, the task is to find an optimal choice for the functions
$f_i(\phi)$ such that, with a fixed set of data, $c_i$ can be deduced
with maximal statistical precision.

The coefficients $c_i$ can be extracted by using the appropriate weighting
functions $w_i(\phi)$ such that $\int w_i(\phi) \Sigma(\phi) d\phi =
c_i$. As in the special example of the electric dipole moment discussed
above, in general, there are an infinite number of choices for 
$w_i(\phi)$ that satisfy this condition which is equivalent to

\beq
\int w_i(\phi) f_j(\phi) d\phi = \delta_{ij} \label{oo29} ~.
\eeq

\noindent It can be shown \cite{prl77p5172} that minimizing the statistical 
error is equivalent to finding the stationary point of 
$\int w_i w_j \Sigma (\phi) d\phi$. In other words, solving

\beq
\delta \int  w_i(\phi)w_j(\phi) \Sigma(\phi) d\phi =
0 \label{oo30} ~,
\eeq
 
\noindent subject to the condition in Eq.~\ref{oo29}. The solution 
to this condition is

\beq
w_i = \frac{\sum_j X_{ij} f_j(\phi)}{\Sigma(\phi)} \label{oo32} ~,
\eeq

\noindent where $X_{ij}=M^{-1}_{ij}$ and 

\beq
M_{ij} \equiv \int \frac{f_i(\phi) f_j(\phi)}{\Sigma(\phi)} d\phi
\label{oo33} ~. 
\eeq

\noindent The desired coefficients, $c_i$, are then given by

\beq
c_i = \sum_k X_{ik} I_k = \sum_k M^{-1}_{ik} I_k \label{oo34} ~,
\eeq

\noindent where

\beq
I_k \equiv \int f_k(\phi)d\phi \label{oo35} ~.
\eeq

\noindent It follows that with this method of 
determining $c_i$, the covariance matrix is

\beq V_{ij} \equiv \langle \Delta c_i \Delta c_j\rangle = M^{-1}_{ij}
\sigma_T/N \label{oo36} ~,
\eeq

\noindent where $\sigma_T$ is the total cross-section and $N=L_{eff}
\sigma_T$ is the total number of events, with $L_{eff}$ being the
effective luminosity, i.e., luminosity times the efficiencies.

It is important to note that since
the entire method has to be implemented numerically utilizing
experimental data, experimental cuts and efficiencies can be
incorporated. 
Indeed the method has the additional virtue that, due to its
generality, it can be applied to only a subset $\hat \phi$ of the
kinematic variables that can be determined if some (say $\bar \phi$)
among the total of $\phi$ cannot be determined. Situations such as this
can occur due to experimental limitations. For example, in the case of
the important reaction $e^+e^-\to t\bar tH^0$, $H^0={}$neutral Higgs, to fully
define a point in phase-space one must be able to identify the momenta
of the $t$ and the $\bar t$ and have no more than one invisible
particle. This requires the use of those final states in which one top
decays leptonically and the other hadronically so that the $t$ and
$\bar t$ are reconstructible. If this can be done, then this case
corresponds to all the kinematic variables $\phi$ being determined.

%
%
%
%

On the other hand, for those events in which the $t$, $\bar t$ both
decay purely hadronically or both decay leptonically then the situation
corresponds to only a part ($\hat\phi$) of the kinematic variables
being determined. The technique for optimization can then be applied by
using the subset of variables, $\hat\phi$, that can be observed. The
functions to be used can now be defined as 

\beq
\hat f_i(\hat\phi) \equiv \int f_i(\phi) d\bar\phi ~. \label{oo37}
\eeq

\noindent Then the entire procedure above can be used in conjunction
with these functions $\hat f$. In fact, the method should work even if
one or more of $\hat f$ are zero. For example, for the reaction 
$e^+e^-\to t\bar tH^0$, if experimentally one is unable to
distinguish $t$ and $\bar t$ then the $f_k(\phi)$ that is a CP-odd
function of the variables $\phi$ reduces to $\hat f_k=0$.

The simple method, first suggested in \cite{prd45p2405}, for extracting the
dipole moment of the top quark has been extensively applied with
appropriate modifications to deduce e.g., dipole moments of 
\cite{zpc24p403}
$\tau$ (from
$Z\to \tau^+\tau^-$), for determination of the three
gauge-boson couplings in $e^+e^- \to W^+W^-$ \cite{zpc62p397}, 
CP-violating phase(s) in $b$ decays 
\cite{prl74p220} as well as in
$t\bar tH^0$ production \cite{prd53p1162}. 
Many of these studies explicitly showed that optimized
observables can be very effective, often yielding improvements over the
``naive'' operators by as much as an order of magnitude.

The generalization discussed in \cite{prl77p5172} 
was originally used specifically to
study the reaction $e^+e^-\to t\bar tH^0$. 
It demonstrated extraction of the $t \bar t H^0$ and
$ZZH^0$ coupling $a$, $b$, and $c$ from the interaction terms:

\beqa
t\bar tH^0: & & \bar t(a+ib\gamma_5) t ~,\\
Z_{\mu} Z_{\nu} H^0: & & c g_{\mu\nu} \label{oo38} ~,
\eeqa

\noindent respectively. An important consequence in this application is
that the CP nature of a Higgs particle, i.e., whether it is CP-odd or
CP-even, may well be deducible from studies of some momenta correlations in
$e^+e^-\to t\bar tH^0$.

We will discuss in greater detail the applications of this method, 
i.e., optimization of observables, to the top dipole
moments (see sections 
\ref{sssec611}, \ref{sssec721} and \ref{sssec814}) and 
to $e^+e^-\to t\bar tH^0$ (see section \ref{ssec62}) in
this review.

\subsection{The naked top \label{ssec27}}

The large mass of the top quark makes it significantly different from
all the other quarks in many important ways. Since $m_t\sim 174$ GeV the
top readily undergoes two-body weak decays

\beq
t\to b+W \label{ttobw} ~,
\eeq

\noindent with $\Gamma\sim 1$ GeV\null.  The top life-time,
$\Gamma^{-1}\sim 1$ ${\rm GeV}^{-1}$, is therefore much shorter than the
``strong interaction time scale'' $\sim 1/\Lambda_{\rm QCD} \sim
10$ ${\rm GeV}^{-1}$. As a result, the top quark, in complete
contrast to all the other quarks, cannot appear in traditional bound
states \cite{plb181p157}. 
Studies of this ``naked top'' are therefore not masked by the
difficult non-perturbative effects of the so-called ``brown muck''.
Thus we should think of the top-quark as an elementary fermion; we can,
for instance, 
try to study its anomalous magnetic moment, i.e., $(g-2)_{\rm top}$
and even more importantly its electric dipole moment 
\cite{prd45p2405,plb279p389,prl69p33}. 
The magnetic moment receives
a large contribution from QCD and other SM interactions but the 
electric dipole moment
is CP-violating and cannot arise at least to 2-loop order in the SM 
\cite{prd18p1632} and so is
expected to be extremely small ($<< 10^{-27}$ e-cm).
Non-standard sources of CP violation can cause this to be significantly
bigger to the point that it may well be measurable.

Now, generically, the electric dipole moment and its generalization 
to other gauge fields (i.e., the corresponding weak and 
chromo-electric dipole moments) is an interaction of the type
$\sigma_t\cdot E$, i.e., the top spin with an external gauge field
(photon, $Z$ or gluon). Therefore, determination of the dipole moment
understandably involves tracking the top spin. Fortunately this is
possible for the top quark, just because it does not bind, even though it is
notoriously difficult to track the spin of $b$ or the other quarks. As
we shall see spin tracking provides an extremely important tool for 
studying all
aspects of CP violation in the top system, not just for probing its
electric dipole moment.

\subsection{Elements of top polarimetry \label{ssec28}}

Decays of the top quark are very effective analyzers of its spin. It is
indeed useful to introduce the notion of the ``analyzing power'' of a
decay. The analyzing power $(\epsilon^A_t)$ measures the degree to which
the momentum of a decay product $(A)$, in the reaction $t\to
A+{}$anything, is correlated with the top spin. For an angular distribution
given by 

\beq
\frac{d\Gamma}{d\cos\theta_A} = a \left( 1+\epsilon^A_t \cos\theta_A \right) 
\label{dgamdcos} ~,
\eeq

\noindent we can define $\epsilon^A_t$ in terms of the expectation 
value of $\cos\theta_A$

\beq 
\epsilon^A_t = 3 \langle \cos\theta_A\rangle \label{epsat} ~,
\eeq

\noindent where $\theta_A$ is the angle between $\vec p_A$ and the spin
of the top $(\vec S_t)$ in the top rest frame. 
Using the optimal observable introduced 
in section \ref{ssec26}, it should be clear 
that with a distribution such as in Eq.~\ref{dgamdcos}, the best way 
to determine the polarization of the top is to experimentally measure 
$\langle \cos\theta_A\rangle$.   

Using this method, it follows that 
the number of top decays that needs to be observed to determine the
top polarization to within 1-$\sigma$ is given by 

\beq
N^{(1\sigma)}_t =  \frac{3}{\left[ \epsilon^{A}_t \Pi_t \right]^2} 
\label{onesigma} ~,
\eeq

\noindent where $\Pi_t$ is the polarization of the top quark,

\beq
\Pi_t \equiv \frac{N(\uparrow) - N(\downarrow)}{N(\uparrow) + N(\downarrow)}
~,
\eeq

\noindent $N(\uparrow)$ being the number of top quarks with spins 
along the positive $z$ axis. Here
are three concrete handles on the top spin:

\begin{enumerate} 

\item Leptonic decays $t\to b\ell^+\nu_\ell$ are the best analyzers,
i.e., $\epsilon^\ell_t=1$ for the correlation between $\vec p_\ell$
and $\Pi_t$. 
Consider the decay chain $t\to bW^+\to be^+\nu_e$. In the limit that
the $b$ quark is massless, the V-A nature of the weak interaction forces
all the particles in the final state to be left-handed. Thus the amplitude
for the top-quark decay is proportional 
to factors of fermion spins times \cite{prl69p2754}:

\beq 
\left[ \bar u_b\gamma_\mu (1-\gamma_5) u_t \right] 
\left[ \bar u_\nu \gamma^\mu (1-\gamma_5) v_e \right] ~.
\label{tdeq2} 
\eeq 

\noindent On Fierz transformation this becomes 

\beq
\frac{1}{2} \left[ \bar u_b (1+\gamma_5) u^c_\nu \right] 
\left[ \bar v^c_e (1-\gamma_5) u_t \right] ~,
\label{tdeq3} 
\eeq 

\noindent where $u^c=C\bar u^T$, $C$ being the charge
conjugate matrix. This implies that, in the rest frame of the top quark,
the top spin is in the direction of the positron three-momentum.
This statement is true in the same sense that in the decay 
$\tau^- \to \pi^- \nu_\tau$ the $\tau$ is polarized in the direction 
of the $\pi$ momentum. Thus, for a top fully polarized in the $z$-direction, 
the angular distribution of the positron is $\propto (1 + \cos\theta_e)$, 
hence $\epsilon_t^e =1$.

\item For hadronic decays of the $W$, if the momentum of
the $W$ can be determined, then it is easy to see that 

\beq
\epsilon^W_t = \frac{m^2_t-2m^2_W}{ m_t^2+2m_W^2} \simeq 0.4 \label{epswt} ~.
\eeq

\noindent This reflects the fact that the $W$-momentum is an 
imprecise 
indicator of the $\bar d$ type jet in the decay $W^+ \to \bar d u$. Here
the $\bar d$ jet plays the same role as the $e^+$ in the previous case. 

\item From the above two points, it should be clear that 
we can increase $\epsilon$ from 0.4 to 1 if we can identify the 
$\bar d$-type jet, which plays the role of the $e^+$.
Since this jet is one of the two that reconstructs to the $W$
in method 2 we are approximating it with the $W$-momentum.
In the limit that $m_t$ is much greater than $m_W$, the two quark jets 
would coincide with the $W$-momentum and $\epsilon_t^W \to 1$; however, with 
the given top mass, the result is far from this limit.

In some cases one could determine which jet was the $\bar d$-type one by 
detailed examination of the decay products. This may be feasible  
for the case of a $b$-jet. However, $b$-jets are highly suppressed in 
$W$ decay so we will not consider this possibility further. Probably
such methods will not offer significant improvements over the above. 
 
On the other hand, we have not exploited the full information available in
the kinematics of the decay.  It was suggested in \cite{plb350p218} that
the energy distribution of the two jets could give additional information
concerning which is the desired $\bar d$-jet. In particular, 
in $W^+ \to \bar d u$, the
$\bar d$-jet is on the average less\footnote{In passing we wish to mention 
that while we find the $\bar d$-jet to be on average less energetic then 
the $u$-jet, Ref. \cite{plb350p218} found it to be the other way around.}  
energetic than the $u$-jet: and so, if we
take the  polarization to be in the direction of the less 
energetic of
these two jets, one would expect some improvement. Indeed, in this case
we obtain

\beq
\epsilon^{\rm low~energy~jet}_t \simeq 0.5 \label{epsbardt} ~.
\eeq

\noindent 
Further, if one uses the optimization methods discussed
in section \ref{ssec26} 
but restricts the experimentally available information to
quantities which are symmetric under the interchange of the two jets, then
one can improve this further

\beq
\epsilon^{\rm opt}_t \simeq 0.63 \label{epsbardt2} ~,
\eeq

\noindent
which represents the best result that can be obtained without knowing the 
identities of the two $W$-jets.

\end{enumerate}  
\pagebreak
\newcommand{\lra}[1]{\buildrel  \longrightarrow  \over {#1}}
\newcommand{\m}{\marginpar}

\section{Models of CP violation \label{sec3}}
\setcounter{equation}{0}

In this chapter we describe the key features of CP violation in 
the SM and in some popular models 
beyond the SM such as MHDM's (Multi-Higgs Doublet-Models) 
and SUSY models. In doing so, we 
emphasize the relevance of the new CP-violating phases that appear 
in these models, to top quark physics.

\subsection{{CP violation and the Standard Model \label{ssec31}}}

In the SM, CP violation emanates from a CP-odd phase in the 
CKM matrix \cite{ptp49652} which influences directly only 
the quark sector \cite{hepph9710551}. 
In this section we will briefly describe the 
properties of this flavor mixing CKM matrix.

\subsubsection{General remarks \label{sssec311}}

The ElectroWeak (EW) 
Lagrangian of the SM can be symbolically written in  the form
(for notation, see e.g., \cite{jarlskog}): 

\beq
{\cal L}={\cal L}(f,G)+{\cal L}(f,H)+{\cal L}(G,H)+{\cal L}(G) + V(H)~,
\label{smlag}
\eeq 
   
\halign{\ni #\hfil \quad& #\hfil \cr
where:&$f$=fermions (quarks, leptons) , \cr
&$G$=gauge-bosons ($\vec W$ and $B$) , \cr &$H$=the Higgs doublet. \cr }

The Lagrangian in Eq.~\ref{smlag} is constructed so that  
it is invariant under the local (space time dependent) symmetry group
SU(2)$_L\times U(1)_Y$. 

The purpose of this 
section in not to review the structure of this 
Lagrangian but rather to explore the salient features of its 
CP-violating part. We will therefore present a self-contained 
discussion only of its CP-nonconserving pieces. 

All CP violation in the SM originates from the term
${\cal L}(f,H)$. The hadronic part of this term is given by

\beq
{\cal L}^h(f,H)=\sum_{j,k=1}^N \left \{Y_{jk}^u\overline{(q,q')}_{jL}
     \pmatrix{\phi^{0*}\cr -\phi^{(-)}}q_{kR}+Y_{jk}^d \overline{(q,q')}_{jL}
  \pmatrix{\phi^{(+)}\cr \phi^0}q_{kR}'+\hbox{h.c.} \right \}, 
\eeq 

\ni where we introduced the multiplets of the quark weak eigenstates

\beq 
\pmatrix{q_{jL}\cr q_{jL}'},q_{jR},q_{jR}'~,\qquad j=1,2,\ldots,N~,
\eeq 

\ni and

\beq 
   q_R={1+\gamma_5\over 2}q~,\qquad q_L={1-\gamma_5\over 2}q~. 
\eeq

\ni Also, $j,k$ are family indices, $N$ denotes the number of families 
and
$Y_{jk}^u,Y_{jk}^d$ are the Yukawa couplings, which are arbitrary complex
numbers.
In our discussion we will consider $N=3$ corresponding to the SM 
with the three known families of fermions.
This Lagrangian has no fermion mass term; fermion masses must therefore 
be induced by Spontaneous Symmetry Breaking (SSB) 
of the $SU(2)\times U(1)$
symmetry of the scalar potential term $V(H)$. In the broken state, the 
scalar doublet assumes a Vacuum Expectation Value 
(VEV) and thus
one obtains the mass terms $M^u$ and $M^d$ of the charge 
2/3 and charge $-1/3$ quarks, respectively, for the weak 
eigenstates of the SU(2)$_L \times U(1)_Y$ gauge theory 
(i.e., ${\bar q}_{iR} M^u_{ij} q_{jL}$ and  ${\bar {q'}}_{iR} 
M^d_{ij} q'_{jL}$). These quark mass matrices are related to the 
Yukawa couplings via

\beq
M^u_{ij}=\frac{v}{\sqrt 2} Y_{ij}^u ~,~~~ M^d_{ij}=\frac{v}{\sqrt 2}
Y_{ij}^d ~,
\eeq 

\ni where $v$ is the VEV of the Higgs doublet. 
In general the mass matrices $M^{d,u}$ are 
not hermitian, and each one depends on 9 
complex unknown parameters. 
Since an arbitrary matrix $M$ can be written as $M=Hu$, with 
$H$ hermitian and $u$ unitary, there exists a field redefinition
such that $M^u$ and $M^d$
are hermitian, i.e.,
that $M^u=M^{u\dagger}$ and $M^d=M^{d\dagger}$ \cite{jarlprl55p1039}. 
A unitary transformation on the $u$ and $d$ quark fields gives the 
physical basis where the mass matrices are diagonal

\beqa
U_R^{\dagger} M^u U_L & = & {\rm diag}(m_u,m_c,m_t)~,\\
D_R^{\dagger} M^d D_L & = & {\rm diag}(m_d,m_s,m_b)~,
\eeqa

\ni where $U_R,~U_L,~D_R$ and $D_L$ are unitary matrices
that relate the weak eigenstates to the physical eigenstates.
It is worth mentioning already at this point that all 
CP violation in the SM emanates from the apparent mismatch between 
the gauge and mass (physical) eigenstates of the quark fields.

For the physical states thus defined, there is no longer an exact
$SU(2)$ identity between the left handed $d$ and $u$ quarks (since 
they are no longer the gauge eigenstates). To see this, write
the Lagrangian in terms
of the physical fields and drop 
the numerical factors and the coupling
constants. Thus 
one is left with the CP-violating charge current 
terms

\beq
X_C=W^+_{\mu}J_C^{\mu}+\hbox{h.c.}~,
\eeq

\ni where $W^+$ is the charged, spin 1, SU(2) vector-boson and

\beq
J_C^\mu=(\bar u,\bar c,\bar t)_L \gamma^\mu V \pmatrix{d\cr s\cr b}_L~.
\eeq 

\ni Here $u,c,t,d,s$ and $b$ are the quark mass eigenstates. 
The $3 \times 3$ unitary matrix $V$ 
will therefore be the product of the
unitary diagonalizing matrices since

\beq
V=U_L^{\dagger} D_L \equiv     \pmatrix{V_{ud}&V_{us}&V_{ub} \cr
     V_{cd}&V_{cs}&V_{cb}\cr V_{td}&V_{ts}&V_{tb}} \label{vgeneral}~. 
\eeq 

\ni Expressing the fermions in this basis, the
term $X_C$ is the one where all the CP violation 
in the SM resides. 
We will therefore consider the properties of $V$ in some detail.

CP conservation requires the matrix $V$ to be real
up to a trivial rephasing of the quark fields. 
In general, for 3 families of quarks 
it can be specified by 18 complex parameters of a general $3\times 3$ 
unitary matrix. However, 9 of these 18 parameters are eliminated by 
the unitarity constraints $V^{\dagger}_{ik}V_{kj}=\delta_{ij}$, 
therefore we are left with only 9 free parameters. Moreover, 
there is a freedom to absorb 5 phases into 5 out of the 6 left 
handed fields \cite{collider}: 

\beq
q_L \to e^{i\alpha(q_L)} q_L ~.
\eeq 

\ni Thus 5 more degrees of freedom can be removed. Therefore, 
one is left with only 4 physically independent parameters upon which 
$V$ depends. These four can then be parameterized in 
terms of three Euler-type rotation angles and one CP-violating phase.
Note that these four angles cannot be predicted within the SM and have to
be deduced from experiment (see below). 

In the parameterization that is called ``standard'' \cite{utriangle}:

\small
\beq 
V_{CKM}=\pmatrix{c_{12}c_{13}& s_{12}c_{13} &  s_{13}e^{-i\delta_{13}}  \cr
   -s_{12}c_{23} -c_{12}s_{23}s_{13} e^{i\delta_{13}}  & 
   c_{12}c_{23} - s_{12}s_{23}s_{13} e^{i\delta_{13}} & s_{23} c_{13} \cr
   s_{12}s_{23} - c_{12}c_{23} s_{13} e^{i\delta_{13}}  & 
   -c_{12} s_{23} - s_{12}c_{23} s_{13} e^{i\delta_{13}}& c_{23}c_{13} }  
\label{eq3.12}.
\eeq
\normalsize

\ni where, as usual, $c_{ij}=\cos\theta_{ij}$, $s_{ij}
=\sin\theta_{ij}$; the indices 1, 2, 3 are ``generation'' labels and
$\delta_{13}$ is the phase. 

With the advent of the $b$-quark lifetime measurements in 1983 
\cite{prl51p1022},
Wolfenstein \cite{wolfparam} made the important observation that the
magnitude of 
the CKM elements exhibit a specific 
hierarchical structure. The parameterization
proposed by Wolfenstein uses the Cabibbo angle, $s_{12}$, as an
expansion parameter making this hierarchy manifest by rewriting the
matrix in terms of the 4 parameters, $\lambda$, $A$, $\rho$
and $\eta$. These are defined as

\[
s_{12} = \lambda, \qquad s_{23}=A\lambda^2, \qquad
s_{13}e^{-i\delta_{13}} = A\lambda^3 (\rho-i\eta)~.
\]

\ni We then arrive at the Wolfenstein representation for the
$3\times 3$ matrix \cite{paganini,buras}:

\beq
\tiny
V_{\rm Wolf} = \pmatrix{ 1-\frac{\lambda^2}{2} - \frac{\lambda^4}{8} & \lambda
& A\lambda^3 (\rho-i\eta) \cr
-\lambda+ \frac{A^2\lambda^5}{2} (1-2\rho) -iA^2 \lambda^5\eta & 1
-\frac{\lambda^2}{2} - \lambda^4 \left(\frac{1}{8} + \frac{A^2}{2}
\right) & A\lambda^2 \cr
A\lambda^3 \left[1-\left(1-\frac{\lambda^2}{2}\right) (\rho+i\eta)\right] &
-A\lambda^2 \left(1-\frac{\lambda^2}{2}\right) [1+\lambda^2
(\rho+i\eta)] & 1-\frac{A^2\lambda^4}{2} }\normalsize + 
\displaystyle{{\cal O}}(\lambda^6) \label{wolf111}~.
\eeq

\ni For most purposes, a simpler form, with truncation to
order $\lambda^4$ suffices \cite{wolfparam}:

\beq
V_{\rm Wolf} = \pmatrix{1-\frac{\lambda^2}{2} & \lambda &
A\lambda^3(\rho -i\eta) \cr
-\lambda & 1-\frac{\lambda^2}{2} & A\lambda^2 \cr
A\lambda^3(1-\rho-i\eta) & -A\lambda^2 & 1 } + 
\displaystyle{{\cal O}}(\lambda^4) 
\label{eqvwolf}~.
\eeq

\ni Notice that the matrix is diagonal to a good approximation.
This is due to the fact that the couplings between quarks of the same
family are close to unity and the off-diagonal elements
 become smaller as the separation
between the families gets larger. 
Note that all CP violation in the CKM matrix is proportional to $\eta$
since this parameter gives a complex phase to the CKM matrix, 
in particular to $V_{ub}$ and $V_{td}$, in the above parameterization.

\subsubsection{The Jarlskog invariant \label{sssec312}}

There is a unique invariant way to parameterize CP nonconservation 
which emerges from the mixing matrix $V$ in Eq.~\ref{vgeneral}. 
That is, to introduce one 
invariant quantity as considered in
\cite{jarlprl55p1039,jarlplb208p268}, 
which will be independent of any phase convention of the quark fields 
and will enter in every CP-violating effect in the SM\null. 

It was shown in \cite{jarlplb208p268} that in order to obtain 
CP nonconservation in the SM, 14 conditions must be satisfied. 
These conditions can be 
expressed as the equation

\beq
{\rm det} C = -2 F^u F^d J \neq  0 \label{determinant}~,
\eeq

\ni where $C$ is the commutator of the mass matrices of 
the up and down quarks defined by

\beq
\left[ M^u,M^d \right] = iC ~.
\eeq

\ni Also, $F^u$ and $F^d$ are given by 

\beqa
F^u &\equiv& (m_t^2-m_c^2)(m_t^2-m_u^2)(m_c^2-m_u^2) ~,\\
F^d &\equiv& (m_b^2-m_s^2)(m_b^2-m_d^2)(m_s^2-m_d^2) ~,  
\eeqa

\ni and $J$, which is sometimes referred to as the 
{\it Jarlskog invariant}, is defined through 
\cite{jarlprl55p1039,jarlplb208p268}:

\beq
{\mIm}(V_{\alpha j} V_{\beta k} V_{\alpha k}^* 
V_{\beta j}^*) \equiv J 
\sum_{l,\gamma} 
\epsilon_{\alpha \beta \gamma} \epsilon_{jkl}~.
\eeq

\ni Here the indices $\gamma$ and $l$ are 
independently summed over while the other indices 
are not summed; Greek, Latin indices stand for up, down-quarks, respectively. 
The parameter $J$ 
is independent of the phase 
convention and, most importantly, it enters in any CP-violating 
effect within the SM\null. In the ``standard'' 
parameterization (i.e., Eq.~\ref{eq3.12}), $J$ is given by

\beq
J=s_{12}^2 s_{13} s_{23} c_{12} c_{13} c_{23} s_{\delta_{13}}
\label{jinvariant}~. 
\eeq
  
\ni From Eqs.~\ref{determinant}--\ref{jinvariant} the 
14 conditions for obtaining CP violation in the SM can be 
recovered. In particular, no two quarks with the same 
charge are degenerate (since that will make $F^u$ or $F^d$ go to zero). 
Also, none of the rotation angles (i.e., $\theta_{12}$, $\theta_{13}$,
$\theta_{23}$) 
and the phase $\delta_{13}$ are allowed to be 0 or $\pi$ (this 
condition assures that $J \neq 0$). 

One can use the invariant parameterization of the angles of 
the mixing matrix $V$ to further examine the structure of CP violation 
in the SM by introducing the ``unitarity triangles'' 
(sometimes called ``CP violation triangles'') 
\cite{prl53p1802}. 
To do so, define

\beqa
&&a_i \equiv V_{i1}V_{i2}^* ~,~ b_i \equiv V_{i2}V_{i3}^* ~,~  
c_i \equiv V_{1i}V_{2i}^* ~,\nonumber \\
&& d_i \equiv V_{2i}V_{3i}^* ~,~ e_i \equiv V_{i1}V_{i3}^* ~,~  
f_i \equiv V_{1i}V_{3i}^* \label{cpvectors} ~,
\eeqa

\ni where $i=1-3$. The unitarity of $V$ (i.e., 
$\sum X_i =0$ 
for $X_i=a_i,b_i,c_i,d_i,e_i$ or $f_i$) implies that each set of the three 
vectors $X_i$ in the complex plane defines a triangle. 
Note that often the name ``unitarity triangle'' refers only to the triangle 
resulting from $\sum e_i=0$ (see below).
All these 
6 triangles have an equal area $S_{\Delta}=J/2$. 
It is shown in 
\cite{jarlplb208p268} that the square of this area can be 
expressed in terms of only 4 independent moduli of the elements of $V$ as

\beq
S_{\Delta} = \lambda^{1/2}(x,y,z) \label{lamda}~,
\eeq

\ni where $\lambda(x^2,y^2,z^2) = [x^2-(y-z)^2] [x^2 - (y+z)^2]$.
Here $x,y,z$ can have two sets of values

\beqa
I)~~x=|V_{\alpha j}||V_{\alpha k}|~,~y=
|V_{\beta j}||V_{\beta k}|~,~z=|V_{\gamma j}||V_{\gamma k}|~,\nonumber \\
II)~~x=|V_{\alpha l}||V_{\beta l}|~,~y=
|V_{\alpha j}||V_{\beta j}|~,~z=|V_{\alpha k}||V_{\beta k}| \label{area} ~.    
\label{utriangs}
\eeqa

\ni For set I $\alpha \neq \beta \neq \gamma$ and $j \neq k$, 
while for set II $\alpha \neq \beta$ and $j \neq k \neq l$. 
Therefore, 
one can compute the angles of the unitarity triangles, up 
to an overall sign, in terms of only 4 moduli of the elements 
of $V$ (note that in Eq.~\ref{area} only 4 independent moduli 
enter $x,y,z$ for any of the allowed values of $\alpha,\beta,\gamma,i,j,l$).
In other words, the existence of CP violation or equivalently
$J^2\ne 0$, may be inferred in the SM with three generations if the 
three sides
of a unitarity triangle can form a triangle with non-zero area. The moduli
that enter into Eq.~\ref{utriangs} may be obtained from purely 
CP-conserving observations. However, an experimental measurement of 
CP violation is needed to fix the sign of $J$.

An example of such a unitary triangle is the one commonly used 
in $B$ physics studies \cite{paganini,buras}, which is constructed 
out of the three vectors $e_1,e_2$ and $e_3$ in Eq.~\ref{cpvectors} 
corresponding to $V_{ud}V_{ub}^*,V_{cd}V_{cb}^*$ and $V_{td}V_{tb}^*$, 
respectively, which we will discuss  below.

\subsubsection{Experimental constraints \label{sssec313}}

Let us now briefly summarize the experimental constraints on the elements
of the $3\times3$ CKM matrix. The best determined is
$|V_{us}|=\lambda\simeq s_{12}$, i.e., the sine of the 
Cabibbo angle. It has been determined 
from $K$ decays, e.g., 
$K^+\to \pi^0\ell^+\nu_\ell$ and from hyperon decays, e.g., $\Lambda\to
pe^-\bar\nu_e$. In addition, 
it is determined from a study
of charm production via neutrino beams as well as from decays of charm to
non-strange final states, albeit with much less precision. The average
of the two determinations gives \cite{paganini}

\begin{equation}
|V_{us}| = \lambda = 0.2205 \pm 0.0018 ~. 
\end{equation}

The next best determined element is $|V_{cb}|$ and, through it, 
the Wolfenstein parameter $A$ via

\begin{equation}
|V_{cb}| = A\lambda^2~.
\end{equation}

\ni $|V_{cb}|$ is deduced using semi-leptonic $B$ decays to
inclusive and exclusive final states, both at LEP and at CLEO\null. The
error in its determination is dominated by theory.
In recent years, heavy quark symmetry and heavy quark
effective theory \cite{prd42p2388}, have had a
significant impact in reducing the model dependence. The average of various
techniques now gives \cite{paganini}:

\beq
|V_{cb}| = 0.0397\pm0.0020~, 
\eeq

\ni thus

\beq 
A = 0.81 \pm 0.04~.
\eeq

\ni The other two parameters, $\rho$ and $\eta$, are poorly known.
Considerable theoretical and experimental effort is being directed to improve
their determinations. 
$B$ physics experiments at $e^+e^-$ based
$B$-factories as well as  
other facilities will surely improve 
our knowledge of $\rho$, and $\eta$. 
Indeed this
will be the focus of intense theoretical and experimental activity in
the near future in providing precision tests of the SM\null.

In the context of decays of the $b$ quark, it is very useful to 
consider the unitarity triangle which 
involves the $b\leftrightarrow
d$ elements 

\beq
V^\ast_{ud} V_{ub} + V^\ast_{cd}V_{cb} + V^\ast_{td} V_{tb} = 0~.
\eeq

\ni Using the parameterization in Eq.~\ref{eqvwolf}, and neglecting
contributions of ${\cal O}(\lambda^4)$ we can write

\beqa
V^\ast_{ud}V_{ub} & = & A\lambda^3 (\bar\rho -i\bar\eta) \label{elone} ~, \\
V^\ast_{cd}V_{cb} & = & -A\lambda^3 ~, \\
V^\ast_{td}V_{tb} & = & A\lambda^3 (1-\bar\rho +i\bar\eta) \label{eqvudcdtd}~.
\eeqa

\ni Here we have used the parameterization introduced in \cite{buras}:

\beq
\bar \rho = \rho \left(1-\frac{\lambda^2}{2}\right) ~~,~~ \bar\eta =
\eta \left(1-\frac{\lambda^2}{2}\right)~.
\eeq

\begin{figure}
\psfull
 \begin{center}
  \leavevmode
\epsfig{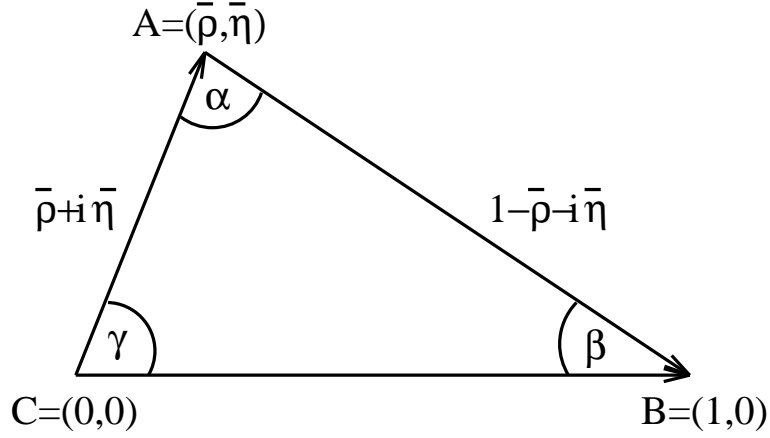}
 \end{center}
\caption{\emph{
Unitarity triangle in the complex $(\bar\rho,
\bar\eta)$ plane.}}
\label{modelsfig1}
\end{figure}

\ni Factoring out the common quantity $A\lambda^3$, the three elements
in Eqs.~\ref{elone} - \ref{eqvudcdtd} can be given a 
geometrical representation in the 
$(\bar\rho,
\bar\eta)$ plane of a triangle with apexes at $A(\bar\rho, \bar\eta)$,
$B(1,0)$ and $C(0,0)$ (see Fig.~\ref{modelsfig1}).

Referring to that figure we have

\beqa
|\lra{AC}| &=& \frac{1-\frac{\lambda^2}{2}}{\lambda} \,
\frac{|V_{ub}|}{| V_{cb}|} = \sqrt{\bar\rho^2 + \bar\eta^2} \nonumber ~,
\\
|\lra{AB}| &=& \frac{|V_{td}|}{\lambda|V_{cb}|} =
\sqrt{(1-\bar\rho)^2 + \bar\eta^2}~.
\eeqa

\ni The angles $\alpha$, $\beta$, $\gamma$ of this triangle 
(see Fig.~\ref{modelsfig1} for their definition) provide a 
basis for testing the SM especially with regard to its
description of CP violation phenomena and CKM unitarity.
In particular, unitarity implies that 
$\alpha+\beta+\gamma=180^\circ$.

Let us now briefly mention the key experimental and theoretical
ingredients that enter to provide the current bounds on $\bar\rho$,
$\bar\eta$ or alternatively on $\alpha$, $\beta$.

The evidence for CP violation from the $K^0-\bar K^0$
system through the indirect CP violation parameter, $\epsilon_K$,
plays a crucial role in constraining $\bar \eta$ and $\bar\rho$ via
\cite{hepph9704376}

\beqa
|\epsilon_K| & = & \frac{G^2_Ff^2_Km_Km^2_W}{6\sqrt{2}\pi^2\Delta m_K}
B_K A^2\lambda^6 \bar\eta \left[ -\eta_1 (1-\frac{\scriptstyle
\lambda^2}{\scriptstyle 2} ) S(x_c) \right. \nonumber \\
& & \left. + A^2\lambda^4 (1-\bar\rho -(\bar\rho^2 +\bar\eta^2)
\lambda^2)\eta_2 S(x_t) \right. \nonumber \\
& & \left. + \eta_3 S(x_c,x_t) \right] \label{eqepsilonk}~.
\eeqa

\ni The $\eta_i$ represent QCD corrections, evaluated to
Next-to-Leading Order (NLO), the $S$'s are 
functions of $x_q =
\frac{m^2_q}{m^2_t}$, $f_K$ is the Kaon decay constant  
and $B_K$ is the so called ``bag'' parameter. Perhaps the
best determination of $B_K$ comes from lattice 
calculations\footnote{The value given here is the renormalization - group
- invariant - $B_K$ often denoted as ${\hat B}_K$.} 
\cite{zzzz1}

\beq
B_K \simeq 0.9 \pm 0.1~.
\eeq

\begin{figure}
\psfull
 \begin{center}
  \leavevmode
\epsfig{file=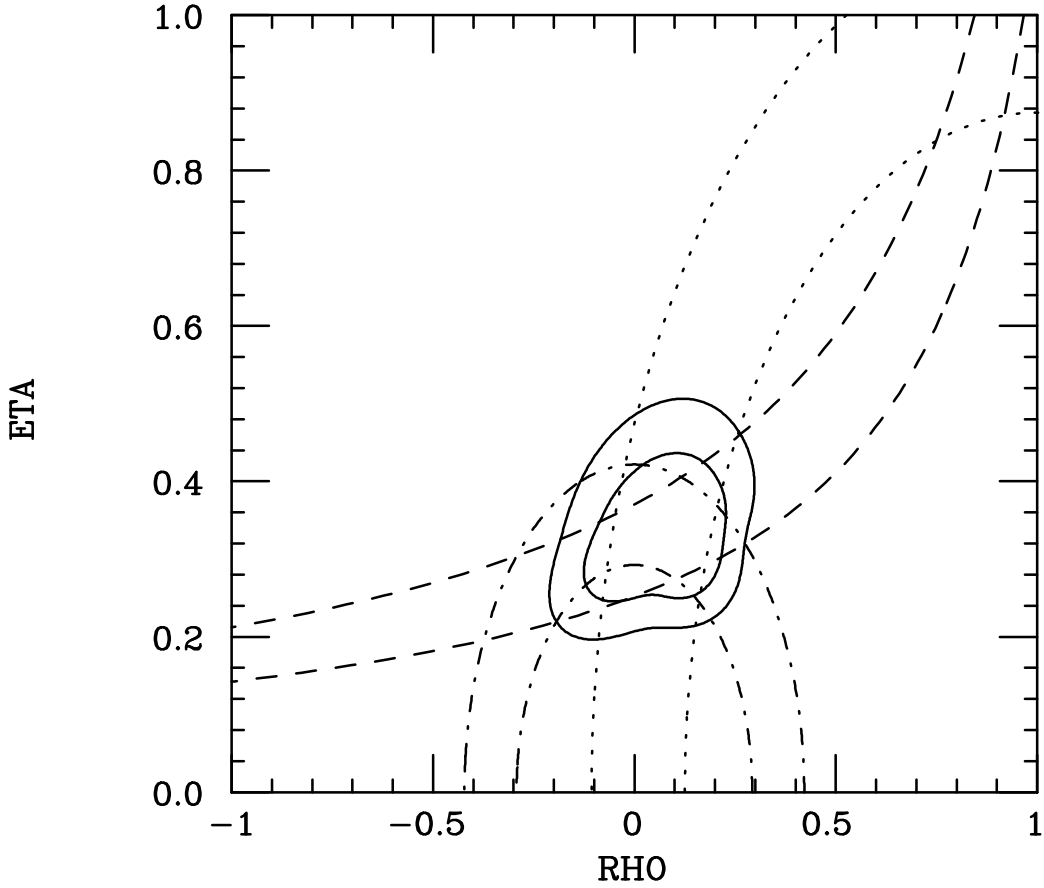,width=9cm}
 \end{center}
\caption{\emph{
The allowed regions for the SM parameters $\bar\rho$ and $\bar\eta$ are 
shown. The solid curves show the $68\%$ and $95\%$ confidence regions. The 
1-$\sigma$
bound originating from neutral $K_L$ decays is shown by the 
dashed curves.
The
1-$\sigma$
bound originating from the rate of $b\to c$ transitions is shown with the 
dot-dashed curve and
the
1-$\sigma$
bound originating from $B^0-\bar B^0$ oscillations is shown with the 
dotted curve.}}
\label{DAVID}
\end{figure}

\ni Using the experimental value of $|\epsilon_K|$
along with $G_F$, $m_W$, $m_t$, $B_K$ etc.\ in
Eq.~\ref{eqepsilonk}, 
leads to an allowed range for $\bar\rho$, $\bar\eta$ as shown in
Fig.~\ref{DAVID}.

$V_{td}$ and therefore $\bar\rho$, $\bar\eta$ also enter intimately in
controlling $\Delta m_d$, i.e., the mass difference between the two mass
eigenstates of the $B^0_d-\bar B^0_d$ system. Thus

\beq
\Delta m_d = \frac{G_F^2 m_W^2}{6\pi^2} A^2\lambda^6 [ (1-\bar\rho)^2
+\bar\eta^2] m_{B_d} f^2_{B_d} B_{B_d} \eta_B x_t F(x_t) \label{eqdeltamd}~. 
\eeq

\ni Here $F(x_t)$ is
calculated perturbatively and is
given to leading order in \cite{ptp65p297}.      
$\eta_B$, $f_{B_d}$ and $B_{B_d}$ 
have the same meaning as their values for the $K$ system, as in
Eq.~\ref{eqepsilonk}.
Once again
$f_{B_d}$, $B_{B_d}$ need to be calculated non-perturbatively and
lattice QCD provides perhaps their best determination. The results from
existing lattice calculations  are best summarized as 
\cite{paganini,zzzz1}: 

\beq
f_{B_d} = 165\pm 20 \pm 30 \; {\rm MeV} ~, 
\eeq

\ni and

\beq
B_{B_d} (2\; {\rm GeV}) = 1.0 \pm .10 \pm .15~.
\eeq

\ni Using the experimental value of $\Delta m_d$ in conjunction with $m_t$,
$G_F$, $f_{B_d}\sqrt{B_{B_d}}$ etc.\ in Eq.~\ref{eqdeltamd}, we can represent
the allowed range on the $\bar\rho$, $\bar\eta$ plane in the form of a
hyperbola (see Fig.~\ref{DAVID}). 

A non-trivial test of the CKM paradigm is obtained by examining
simultaneously the allowed range of $\bar\rho$,$\bar\eta$ through a
determination of $|V_{ub}|$ from semi-leptonic charmless $B$-decays.
In this regard various techniques are used to deduce
$\left|\frac{V_{ub}}{V_{cb}}\right|$ from exclusive and inclusive
semi-leptonic decays. For now the uncertainties in the theoretical
models of these transitions is quite substantial giving
\cite{utriangle,paganini}: 

\beq
\left|\frac{V_{ub}}{V_{cb}}\right| = .08 \pm .02 \Rightarrow
\sqrt{\bar\rho^2 + \bar\eta^2} = .35\pm .09 ~.
\eeq

\ni This compendium of experimental and theoretical information on
$\epsilon_K$, $\Delta m_d$ and $b\to
u\ell\nu$ is used to obtain the best fitted values \cite{paganini}:

\beq
\bar\rho = .10^{+.13}_{-.38}~,  \quad \bar \eta =
.33^{+.06}_{-.09} \label{eqbarrho}~.
\eeq

\ni The corresponding 68\% and 95\% CL contours are 
shown in Fig.~\ref{DAVID}.

In the future, the measurement of $B_s-\bar B_s$ oscillations
will
provide an extremely important test of the SM\null. The point is that the
ratio of the mass differences

\beq
\frac{\Delta m_d}{\Delta m_s} = \frac{m_{B_d}}{m_{B_s}} \, \frac{f^2_{B_d}
B_{B_d}}{f^2_{B_s} B_{B_s}} \, \frac{\eta_{B_d}}{\eta_{B_s}} \,
\frac{|V_{td} |^2}{|V_{ts}|^2}
\label{dmddmsratio}~,
\eeq

\ni will involve significantly less uncertainty due to hadronic
matrix elements (i.e., $f^2B$). 
Aside from the CKM elements, the most uncertain factor on the rhs of 
Eq.~\ref{dmddmsratio} is

\beq
r_{sd} \equiv \frac{f^2_{B_s} B_{B_s}}{f^2_{B_d} B_{B_d}}~.
\eeq

\ni However, theoretical uncertainties    
in 
extracting $|V_{td}|$ from 
$r_{sd}$ are expected to become smaller 
\cite{cbernard} in comparison to the errors in 
extracting $|V_{td}|$ from $\Delta m_d$ alone.

The experiments at LEP have already made significant progress in studying
$B_s-\bar B_s$ oscillations. A combined analysis of ALEPH, DELPHI and
OPAL leads to \cite{paganini,feindt}:

\beq
\Delta m_s > 8.0 ~ ps^{-1} \quad {\rm at~} 95\% ~ {\rm CL}~.
\eeq

\ni Incorporating this along with $\epsilon_K$, $\Delta m_d$ and $b\to
u\ell\nu$, into the $\bar\rho$, $\bar\eta$ constraints one finds
\cite{paganini}: 

\beq
\bar\rho = .11^{+.13}_{-.25}~,  \quad \bar\eta = .33 \pm.06~. 
\eeq

\ni Comparing this with Eq.~\ref{eqbarrho} we see that the LEP bound on $\Delta
m_s$ is already reducing the negative error on $\bar\rho$ appreciably
\cite{paganini}. Further slight changes (see the update of Buras 
in \cite{pl389749}) result from improvement, obtained by combining 
LEP/SLD/D0 data, of the limit on $\Delta m_s$ to read:

\beq
\Delta m_s > 12.4 ps^{-1} \quad {\rm at~} 95\% ~ {\rm CL}~,
\eeq  

\noindent and from the small increase of $|V_{ub}|/|V_{cb}|$ to 
\cite{vancouver98}:

\beq 
|V_{ub}|/|V_{cb}| = 0.091 \pm 0.016 ~.
\eeq

\noindent Translated to $\beta$ and $\gamma$ (see Fig.~\ref{modelsfig1}) 
yields:

\beq
\sin 2\beta = 0.71 \pm 0.13~, \qquad \sin\gamma=0.83 \pm 0.17 ~.
\eeq

\noindent The above value of $\sin 2\beta$ is consistent with the recent 
CDF result \cite{hepex9903002,www-cdf} $\sin 2\beta = 0.79^{+0.41}_{-0.42}$.

\subsection{Multi-Higgs doublet models \label{ssec32}}

In the SM the interaction of the only neutral Higgs-boson 
with fermions is automatically P
and C conserving as well as flavor conserving. This property is not 
valid
in general in models beyond the SM\null.
In this section we consider extensions to the SM involving the addition 
of extra Higgs doublets.

CP-violating effects in such models can originate in the scalar 
sector and be manifested in the physics of fermions, particularly the top 
quark.
For
such an effect to occur, two or more complex SU(2) doublets of Higgs
fields are required; this was first pointed out by Lee  \cite{prd8p1226}.
However, the mere presence of more than one doublet does not guarantee
CP violation in the Higgs sector.  For instance, a
CP-violating phase in the case of models with two Higgs doublets (2HDM) 
can be rotated out
of the Higgs sector entirely 
if one imposes various discrete symmetries as will be discussed 
below.
But if such phase(s) cannot be rotated away, this approach
leads to CP violation from neutral Higgs-boson interactions, from charged
Higgs-boson interactions, and perhaps in addition from the presence of a
non-vanishing phase in the CKM matrix.
In all cases, the various types of CP violation are presumably related at a
fundamental level
to CP violation in the Higgs potential,
but because of our ignorance of the Higgs sector, in
practice the parameters of each type of interaction are independent
and should be separately measured.

Another general feature of CP violation in an extended Higgs sector is
that larger effects are expected in heavier quarks systems (compared to the
usual SM approach), because Higgs-boson couplings to fermions are proportional 
to the fermions masses. 
This makes the top quark system an especially good testing ground 
for such phenomena.

CP violation in the Higgs sector can arise in models where the
Higgs potential may contain complex couplings. This might
lead directly to
a CP-violating interaction or to complex VEV's of the Higgs fields which can
induce CP-violating effects. In addition, as we shall see in some examples
below, it is also possible that a real potential can lead to a ground
state with a complex VEV, in which case CP is broken spontaneously. In any
case, there are generally a large number 
of parameters in these models so that
considerable experimental effort will eventually be required to 
determine them all. In particular, 
it is important to consider which predictions of
such models differ from the SM, so that might lead to early signs that extra
scalar fields are present. CP violation in top physics is especially
useful since the SM contributions to CP violation in top quark 
reactions are negligible
and the mass dependent coupling of 
the Higgs means that top quark physics is very sensitive to such effects.

It is convenient to classify CP symmetry-breaking in the scalar sector
into three different categories: hard (intrinsic), soft and spontaneous.
Hard or intrinsic CP violation refers to symmetry-breaking terms with
dimension four, for example, terms 
in the Lagrangian
with complex Yukawa coupling constants, 
or with self-coupling of scalar fields.
Soft breaking is associated with terms in the Lagrangian with
canonical dimension
less than four.
If the Lagrangian starting from the outset is CP invariant,
CP violation can still be achieved by introducing complex phases from the
VEV's of the scalar fields 
(i.e., spontaneous CP violation).

In the following we will consider simple versions of 2HDM
 and Three Higgs Doublets Model (3HDM) in which CP violation is manifested
in the interactions of neutral and charged Higgs particles with fermions.

\subsubsection{Two Higgs doublet models \label{sssec321}}

We start with a description of the most general 2HDM\null.
The Higgs potential for such a 2HDM is given by \cite{higgshunters}:

\beqa
V(\Phi) &=& -\mu_1^2 \Phi_1^{\dagger} \Phi_1 -\mu_2^2 \Phi_2^{\dagger}
\Phi_2 - (\mu_{12}^2 \Phi_1^{\dagger} \Phi_2 +  {\rm h.c.}) \nonumber \\
&+& \lambda_1 (\Phi_1^{\dagger} \Phi_1)^2 + 
\lambda_2 (\Phi_2^{\dagger} \Phi_2)^2 
+ \lambda_3 (\Phi_1^{\dagger} \Phi_1 \Phi_2^{\dagger} \Phi_2) + 
\lambda_4 (\Phi_1^{\dagger} \Phi_2)(\Phi_2^{\dagger} \Phi_1) \nonumber \\
&+& \frac{1}{2} [\lambda_5 (\Phi_1^{\dagger} \Phi_2)^2 + {\rm h.c.}] 
+[\lambda_6 \Phi_1^{\dagger} \Phi_1 \Phi_1^{\dagger} \Phi_2 +
\lambda_7 \Phi_2^{\dagger} \Phi_2 \Phi_1^{\dagger} \Phi_2 + {\rm h.c.}]
\label{mhdmeq1} ~, \nonumber \\
\eeqa 

\ni where $\Phi_i$ ($i=1,2$) are Higgs fields such that

\beqa
\Phi_i = \pmatrix{\phi_i^+ \cr \phi_i^0} \label{mhdmeq2} ~.
\eeqa

\ni The scalar spectrum of a 2HDM consists of three neutral 
and two charged Higgs-bosons which we will denote by ${\cal H}^k$ 
($k=1,2,3$) and $H^\pm$, respectively. 

The important parameters in $V(\Phi)$ that may drive CP violation in the
Higgs sector are: $\lambda_5,~\mu_{12}^2,~\lambda_6$ and $\lambda_7$. 
It is the different
choices of these parameters that will determine which 
type of mechanism is generating CP
violation in the 2HDM\null.

It should be noted that without the complex VEV's of the two 
doublets (or only one of them that can generate spontaneous CP violation), 
one needs at least two terms out of  $\lambda_5,~\mu_{12}^2,~\lambda_6$ and 
$\lambda_7$ to be non-zero 
in the Higgs potential (see Eq.~\ref{mhdmeq1})
in order to have CP nonconservation in the model. 
That is, if there is only one complex coefficient in the Higgs potential
prior to SSB, then SSB leads via the minimization condition,   
$\partial V/\partial \phi |_{\phi=0} = 0$, 
 to certain relations among the
parameters of the Higgs potential which, in turn, forces the single complex
coefficient to have a zero imaginary part if the VEV's are real. 
As an example, assuming that $\lambda_6=\lambda_7=0$ and only 
$\mu_{12}^2$ is complex 
in Eq.~\ref{mhdmeq1}, $\partial V/\partial \phi |_{\phi=0} = 0$ 
leads to $\Im{\rm m}(\mu_{12}^2) \propto \sin\theta$, where $\theta$ 
is a phase associated with a complex VEV (see Eq.~\ref{mhdmeq3}). 
Thus, if the VEV's are real (i.e., $\theta=0$), 
then $\Im{\rm m}(\mu_{12}^2) = 0$ and 
the imaginary part of the only complex coefficient $\mu_{12}^2$ in the 
Higgs potential (prior to SSB) gets rotated away.     
Therefore, in general, there could be several scenarios of generating CP
violation in a 2HDM emanating from $V(\Phi)$:
\begin{enumerate}
\item At least two terms in the potential should violate 
CP when the two VEV's 
are chosen
by a suitable field redefinition to be
real.

\item Only one term in the Higgs potential 
violates CP and the two VEV's have a non-zero
relative phase.

\item The Higgs potential is CP-conserving however the ground state
spontaneously violates CP by giving a non-zero phase between the two VEV's.
This scenario requires non-vanishing hard couplings of the type $\lambda_6$ 
and $\lambda_7$ (i.e., hard self-couplings of scalar fields).  

\item CP violation occurs everywhere it can, i.e., 
explicitly in the Higgs potential
and from a relative phase of the VEV's.
\end{enumerate}

Let us illustrate how the above scenarios work for the Higgs 
potential given in Eq.~\ref{mhdmeq1}.
We will focus on two out of the four 
possible scenarios above for breaking CP
symmetry that will result from two different
choices of the set $\lambda_5,~\mu_{12}^2,~\lambda_6$ and 
$\lambda_7$, i.e., scenarios 2 and 3 above.
In general, when there exists a relative phase $\theta$ 
between the two VEV's

\beqa
<\phi_1^0> = \frac{v}{\sqrt{2}} \cos\beta e^{i \theta} \ \ , \ \ 
<\phi_2^0> = 
\frac{v}{\sqrt{2}} \sin\beta 
\label{mhdmeq3} ~,
\eeqa  

\ni where  

\beqa
\tan\beta = \frac{v_2}{v_1}  \ \ {\rm and} \ \ v=\sqrt{v_1^2 + v_2^2} 
\label{mhdmeq4} ~,
\eeqa  

\ni then with $\lambda_5$ non-zero and real, 
CP violation can arise from non-zero complex entries 
of one or more of $\mu_{12}^2$, $\lambda_6$ and $\lambda_7$. If 
$\lambda_5$, $\mu_{12}^2$, $\lambda_6$ and $\lambda_7$ are all 
real then CP violation can
occur spontaneously (i.e., through the relative phase $\theta$). 
If one of $\mu_{12}^2$, $\lambda_6$ and $\lambda_7$ is 
complex then in addition to the complex VEV's above that appear after SSB,
there is an explicit CP nonconservation in $V(\Phi)$. However, whether 
CP violation 
is  spontaneous or explicit, the structure of the 
CP-violating sector of the model can be redefined to  depend only on the 
relative phase $\theta$.  

Let us consider, for example, the
particular choice $\lambda_6=\lambda_7=0$ 
which follows, for example, from 
type II 2HDM, that we will describe later in this 
section. Here CP violation can occur only 
if ${\mIm}[\lambda_5/(\mu_{12}^2)^2] \ne 0$ \cite{prd42p860}. 
 
If, on the other hand, $\arg(\lambda_5)=2 \arg(\mu^2_{12})=\psi$, 
then the field redefinitions 
$\Phi_1\to \exp(+i\psi/4)\Phi_1$ and 
$\Phi_2\to \exp(-i\psi/4)\Phi_2$ would eliminate this phase.
A phase difference between $\lambda_5$ and $\mu_{12}^2$
is therefore essential in order to get CP 
nonconservation in a 2HDM with no hard couplings 
of the type $\lambda_6$ and $\lambda_7$ 
and, in particular, $\lambda_5$ cannot vanish.
Let us 
remark at this point that in the Minimal  
Supersymmetric Standard Model (MSSM) with only two Higgs doublets,
$\lambda_5 = 0$ is required by the supersymmetric 
nature of the Lagrangian. Therefore, no CP violation
can arise from the pure Higgs sector in the MSSM. 
We will return to this
point when we will describe CP violation in SUSY models 
(see section \ref{ssec33}).

In models with extended Higgs sectors, an important experimental 
constraint is that the processes with
Flavor Changing Neutral Currents (FCNC) are severely suppressed. 
To understand the extent to which such models will give rise 
to FCNC, consider the
most general Yukawa interaction of quarks in a 
2HDM \cite{wu}:

\beqa
{\cal L}_Y = && \sum_{i,j} [ \ (\bar{u}_i,\bar{d}_i)_L (U^1_{ij} 
\tilde{\Phi}_1 +
U^2_{ij} \tilde{\Phi}_2) (u_j)_R + \nonumber \\ 
&&(\bar{u}_i,\bar{d}_i)_L 
(D^1_{ij} \Phi_1 +
D^2_{ij} \Phi_2) (d_j)_R \ +{\rm h.c.} \ ]~, 
\label{mhdmeq5}
\eeqa 

\ni where $i$ and $j$ are generation indices, 
$u$ and $d$ stand for up and down quarks, respectively, and
$\tilde{\Phi}=i \sigma_2 \Phi^*$ while 
$U^1$, $U^2$, $D^1$ and $D^2$ are
matrices in flavor space. In the general 
case presented above,         
where either both $U^1$ and $U^2$ or $D^1$ and $D^2$
are present in ${\cal L}_Y$, FCNC 
will appear in the model. To avoid
FCNC in a 2HDM, 
most models which have been considered
impose an {\it ad hoc} discrete symmetry on the
2HDM Lagrangian \cite{prd15p1958}; the idea being that such a symmetry 
may originate from
physics at a more fundamental level. 
We will return to this point later when we discuss CP violation in a
2HDM with no FCNC at tree-level.  

In the next two sections we examine two widely 
studied cases of the 
2HDM where CP violation arises from the Higgs sector. In 
section \ref{sssec322} we 
consider a model where FCNC are in fact present. In particular we take
$\lambda_5 > 0$ and real, and also $\mu_{12}^2,~\lambda_6,~\lambda_7$ 
are non-zero and real (corresponding to scenario 3 above).
Since FCNC  will be present, the parameters of the model will be 
constrained to keep them below the experimental limits.
In section \ref{sssec323} 
we consider a case which falls into scenario 2 and 
has flavor conservation built in at tree-level. In particular, 
$\lambda_6 = \lambda_7 = 0$ and a discrete symmetry (which is only 
softly broken) is imposed on the model \cite{plb279p389,npb386p63}.

\subsubsection{2HDM with CP nonconservation and FCNC \label{sssec322}}

In this model, no discrete symmetry on the Yukawa couplings of any kind is
imposed,  thus allowing for the presence of
all the couplings which appear in $V(\Phi)$ and ${\cal L}_Y$. 
However, in view of the low energy data on
FCNC, one has to require that the Flavor Changing (FC) parameters  meet 
those experimental constraints. One systematic
way that has been suggested to achieve this without fine 
tuning the parameters is
to impose an approximate global U(1) 
symmetry which acts only on fermions
(see \cite{wu,hepph9808453} and references therein). 
This symmetry will be responsible for the smallness 
of the  flavor changing couplings in this
model and it leads to the Cheng-Sher ansatz~\cite{prd35p3484,prd44p1461} 
which imposes a hierarchy on the terms of ${\cal L}_Y$ that can reasonably 
evade FCNC constraints. 
We choose not to concern ourselves with the technical 
aspects of imposing such a U(1) symmetry, but
rather concentrate on the CP-violating consequences of a 
2HDM with FCNC, the so called Model III.
A more detailed discussion
about the flavor changing parameters and their experimental 
constraints can be found, for example, in \cite{prd55p3156}.

We can rewrite ${\cal L}_Y$ for this model
in terms 
of the neutral and charged Higgs mass-eigenstates ${\cal H}^k$ and 
$H^{\pm}$, respectively. We divide ${\cal L}_Y$ into four terms
\cite{wu}:

\beq
{\cal L}_Y = (\sqrt{2} G_F)^{\frac{1}{2}} [{\cal L}_Y^{H^0} + 
{\cal L}_Y^{(H^0+FC)} +
\sqrt{2} ({\cal L}_Y^{H^{\pm}} +{\cal L}_Y^{(H^{\pm}+FC)})] 
\label{mhdmeq6}~,
\eeq

\ni where ${\cal L}_Y^{(H^0+FC)}$$({\cal L}_Y^{(H^{\pm}+FC)})$ 
contains the FC effects for the
neutral (charged) Higgs-bosons, and ${\cal L}_Y^{H^0}$$({\cal L}_Y^{H^{\pm}})$ 
has no flavor changing
effects other than the ones expected from the CKM matrix of the SM 
part of the theory which also factors into the fermion
charged Higgs coupling.

These terms are written 
as follows:

\beqa
 {\cal L}_Y^{H^0} & = & \sum_{i,k=1}^3 
( \bar{u}_L^i m_{u_i} \eta_{u_i}^{(k)} u_R^i  + 
\bar{d}_L^i m_{d_i} \eta_{d_i}^{(k)} d_R^i + {\rm h.c.} ) 
{\cal H}^k \label{mhdmeq7} ~,\\
 {\cal L}_Y^{H^{\pm}} & = & \sum_{i,j=1}^3 ( \bar{u}_L^i m_{d_j} \xi_{d_j} 
V_{ij} d_R^j H^+ + 
\bar{d}_L^i m_{u_j} \xi_{u_j} V_{ij}^{\dagger} u_R^j H^- \nonumber \\  
&& \quad ~~~ + {\rm h.c.} ) \label{mhdmeq8} ~,\\
 {\cal L}_Y^{(H^0+FC)} & = & \sum_{i \ne j,k=1}^3 ( \bar{u}_L^i 
\sqrt{m_{u_i}m_{u_j}} \mu_{ij}^{u(k)} u_R^j  + 
\bar{d}_L^i \sqrt{m_{d_i}m_{d_j}} \mu_{ij}^{d(k)} 
d_R^j \nonumber \\
& & \quad ~~~~ +{\rm h.c.} ) {\cal H}^k \label{mhdmeq9}~,\\
 {\cal L}_Y^{(H^{\pm}+FC)} & = & \sum_{i,j \ne j^{\prime}=1}^3 
( \bar{u}_L^i V_{ij^{\prime}} 
\sqrt{m_{d_{j^{\prime}}}m_{d_j}} \mu_{j^{\prime}j}^d d_R^j H^+ \nonumber
\\
& & \quad ~~~~+ \bar{d}_L^i V_{ij^{\prime}}^{\dagger} \sqrt{m_{u_{j^{\prime}}}
m_{u_j}} \mu_{j^{\prime}j}^u 
u_R^j H^-  + {\rm h.c.} ) \label{mhdmeq10}~.
\eeqa  

\ni Note that in Eqs.~\ref{mhdmeq9} and \ref{mhdmeq10} proportionality
of the FC couplings to the masses of the fermions participating in the
FC vertex is imposed. This idea was originally proposed in
\cite{prd35p3484,prd44p1461} and is often referred to in the
literature as the Cheng-Sher ansatz. With this proportionality
assumed, the severe experimental constraints on the FC couplings of
light quarks can be satisfied quite naturally, i.e., without fine
tuning.  

The three
neutral Higgs mass eigenstates, ${\cal H}^k$, are related to the real
and imaginary pieces of the two Higgs doublets by a $3\times 3$
orthogonal matrix $R$.  The parameters $\xi_{f_i}$ in Eq.~\ref{mhdmeq8} 
($f$ stands for
fermion) arise from the diagonal elements of the Yukawa couplings
$U^1,D^1$ or $U^2,D^2$ in Eq.~\ref{mhdmeq5}, whereas the
factors $\mu_{ij}^f$ in Eqs.~\ref{mhdmeq9} and \ref{mhdmeq10} 
arise from the off-diagonal elements. Using the
notation presented in \cite{wu}:

\beqa \xi_{f_i} &\simeq& \frac{\sin \delta_{f_i}}{\sin \beta \cos
\beta \sin \theta} e^{i \sigma_f (\theta - \delta_{f_i})} - \cot \beta
\label{mhdmeq11} ~,\\ \mu_{ij}^f &\simeq& \frac{1}
{\sin \beta} 
(e^{i \sigma_f (\theta - \delta_{f_j})} - 
\frac{\sin \delta_{f_j}}{\sin \theta} ) (S_1^F)_{ij} \label{mhdmeq12}~,
\eeqa  

\ni and

\beqa
\eta_{f_i}^{(k)} &=& f(R,\xi_{f_i}) \label{mhdmeq13}~,\\
\mu_{ij}^{f(k)} &=&  \mu_{ij}^f  f^{\prime}(R) \label{mhdmeq14}~. 
\eeqa     

\ni Here $f$ and $f^{\prime}$ are functions of $R$, $\xi_{f_i}$ and of $R$, 
respectively.
$S_1^F$ is an arbitrary off-diagonal real matrix.
$\theta$ is the relative phase between the two
VEV's  as defined in Eq.~\ref{mhdmeq3} 
and $\delta_{f_i}$ is the phase associated with 
the mass $m_{f_i}$ and is defined through

\beq
\sqrt{2} m_{f_i} e^{i \sigma_f \delta_{f_i}} \equiv 
(g_1^{F_i} \cos \beta e^{i \sigma_f \theta} + 
g_2^{F_i} \sin\beta ) v \label{mhdmeq15}~,
\eeq

\ni where $g_1^{F_i}$($g_2^{F_i}$) are diagonal 
elements of $U^1,D^1$($U^2,D^2$) defined in Eq.~\ref{mhdmeq5}, 
and $\sigma_f = + \ {\rm or} \ -$ 
for up or down fermion, respectively. 

Evidently, ${\cal L}_Y$ manifests four 
patterns of CP violation in
the 2HDM being considered, all of
which are being driven by the relative phase $\theta$ 
(which appears after SSB) between the two VEV's and 
by some definite choice of the parameters 
$\lambda_5$, $\lambda_6$, $\lambda_7$ and $\mu_{12}^2$ 
in $V(\Phi)$:
\begin{enumerate}
\item CP violation induced by the complex Yukawa couplings 
$\xi_{f_i}$ which appear both in the
neutral and the charged Higgs sectors.
\item Scalar-pseudoscalar mixing in the couplings of a neutral 
Higgs species with fermions. That is,
the mixing of the neutral imaginary and real parts of the two 
Higgs doublets which results from such
a model generates neutral Higgs mass eigenstates that do 
not have a definite CP-property. Thus the
couplings of neutral Higgs with fermions will have the generic form 

\beq
\Gamma_{{\cal H}^k f \bar{f}} = {\cal H}^k \bar{f} (a^k_f+ib^k_f\gamma_5)f
\label{mhdmeq16}~,
\eeq  

\ni where $a^k_f$ and $b^k_f$ are functions of $R$, $\tan \beta$
and $\xi_f$.
\item The phases in $\mu_{ij}^f$ which yield CP violation in FCNC 
interactions both in the neutral and the
charged Higgs sectors.
\item The usual CKM matrix $V$ which gives CP violation in the
charged Higgs interactions much like that of
the charged $W$-boson interactions in the SM\null.
\end{enumerate}

\subsubsection{2HDM with CP nonconservation and no FCNC \label{sssec323}}

There is a natural way suggested by 
Glashow and Weinberg \cite{prd15p1958} to have tree-level FCNC 
vanish. This idea that there may be a
discrete symmetry present in the 2HDM Lagrangian, also implies 
the vanishing of CP violation if this discrete symmetry is exact 
(see discussion below). 
Depending on the discrete symmetry 
imposed, we can then obtain different versions of a 2HDM. 
The two cases usually considered are 
the discrete symmetries $D_I$ and $D_{II}$
   
\beqa
D_{I}&:& \  \Phi_1;\Phi_2;(d_i)_R;(u_i)_R \to -\Phi_1;\Phi_2; - (d_i)_R; 
- (u_i)_R \label{mhdmeq17}~,\\
D_{II}&:& \  \Phi_1;\Phi_2;(d_i)_R;(u_i)_R \to -\Phi_1;\Phi_2; - (d_i)_R 
;(u_i)_R \label{mhdmeq18}~. 
\eeqa

\ni 
The models with these symmetries are referred to as type I and II,
respectively,
depending on whether the $-1/3$ and $2/3$ charge quarks are coupled 
to the same or to different scalar doublets.  

Let us describe now the 2HDM of type II (sometimes also referred to just
as Model II)   following the notation in \cite{npb386p63}.
With the discrete symmetry $D_{II}$, we can build the
Lagrangian for the 2HDM of type II as a special
case of 
the general 2HDM Lagrangian
described in the previous section, where:
\begin{description}
\item (a) In order that ${\cal L}_Y$ in Eq.~\ref{mhdmeq5} be 
invariant under $D_{II}$, $U^1_{ij}=D^2_{ij}=0$ is
required. Thus 

\beq
{\cal L}_Y = \sum_{i,j} [ \ (\bar{u}_i,\bar{d}_i)_L 
U^2_{ij} \tilde{\Phi}_2 (u_j)_R \ + \ (\bar{u}_i,\bar{d}_i)_L D^1_{ij} 
\Phi_1 (d_j)_R \ + {\rm h.c.} \ ]~, 
\label{mhdmeq19}
\eeq   

\ni and hence only $\Phi_2$ gives mass to charge $+2/3$ 
quarks and only $\Phi_1$ is responsible for the mass generation 
of the charge 
$-1/3$ quarks.
\item (b) If all the terms that are non-invariant under the operation of 
$D_{II}$ in
the  Higgs potential are zero and the symmetry is exact, then there is no CP
violation in the theory.  Therefore, allowing only for a one non-zero value of
the soft breaking term $\mu_{12}^2 \ne 0$, 
we have in the Higgs
potential of Eq.~\ref{mhdmeq1}, $\lambda_6=\lambda_7=0$. 
\item (c) Without loss of generality, $\lambda_5$ is chosen as real 
(this can be done by
``rephasing'' $\Phi_2$) and $\mu_{12}^2$ is chosen as complex, 
thus having explicit CP
violation already at the Lagrangian, in addition to the 
relative phase between the two VEV's which arises after SSB\null.
\end{description}

With the above three points 
and respecting the discrete symmetry (except for the soft
breaking terms which do not introduce FCNC at tree-level), 
the 2HDM of type II can be
extracted from the general 2HDM of the previous section 
by taking \cite{prl73p2809}:

\beq
\delta_d=\theta \ \ , \ \ \delta_u=0 \label{mhdmeq20}~,    
\eeq  

\ni thus

\beqa
\xi_d = \tan \beta \ \ , \ \ \xi_u = - \cot \beta \label{mhdmeq21}~.    
\label{xiud}
\eeqa  

\ni Also because of point (a) above, we have $\mu_{ij}^f=0$ in ${\cal L}_Y$,
thus 

\beq
{\cal L}_Y^{(H^0+FC)}={\cal L}_Y^{(H^{\pm}+FC)}=0 \label{mhdmeq22}~,
\eeq  

\ni and we are left with only two CP-violating 
mechanisms in this model:
\begin{enumerate}
\item The scalar-pseudoscalar mixing described in the 
previous section (i.e., Eq.~\ref{mhdmeq16}), where still 
$a^k_f$ and $b^k_f$ are functions of the $3\times 3$ Higgs 
mixing mass matrix $R$, $\tan \beta$ and $\xi_f$. Note, however, 
$\xi_f$ is now real and is given by Eq.~\ref{xiud} above.
\item The usual CKM matrix $V$ which gives CP violation 
in the charged Higgs interactions.
\end{enumerate}

For later use, let us introduce the following notation for
the ${\cal H}^k f \bar f$ and 
${\cal H}^k VV$ ($V=W$ or $Z$) parts of the Lagrangian in
a general 2HDM

\beqa
{\cal L}_{{\cal H}^k ff} &=& -\frac{g_W}{\sqrt 2} \frac{m_f}{m_W} {\cal H}^k 
\bar f \left( a_f^k + i b_f^k \gamma_5 \right) f \label{2hdmab}~,\\ 
{\cal L}_{{\cal H}^k VV} &=&  g_W m_W C_V c^k {\cal H}^k 
g_{\mu\nu} V^{\mu} V^{\nu} \label{2hdmc}~,
\eeqa

\ni where $C_{W;Z}\equiv1;m_Z^2/m_W^2$. 
Note that in the SM the couplings in Eqs.~\ref{2hdmab} and 
\ref{2hdmc} of
the only neutral Higgs present are $a_f=1/\sqrt {2},b_f=0$  and $c=1$, and
there is no phase in the ${\cal H}^k f \bar f$ coupling.
In Model II, for up quarks for example \cite{npb386p63}:

\beq
a_u^k=R_{1k}/\sin\beta~~,~~b_u^k=R_{3k}/\tan\beta~~,~~c^k=R_{1k}\sin\beta
+R_{2k}\cos\beta \label{mhdmeq23}~,
\eeq

\ni where $\tan\beta \equiv v_u / v_d$ and $v_u$($v_d$) is 
the VEV
responsible for giving mass to the up(down) quark. 
$R$ is the neutral Higgs
rotation matrix which can be parameterized by three Euler 
angles $\alpha_{1,2,3}$ as follows \cite{npb386p63}:

\beqa
R=\left(\matrix{ & c_1 & s_1c_3  & s_1s_3 \cr
&-s_1c_2 & c_1c_2c_3-s_2s_3 & c_1c_2s_3 + s_2c_3 \cr
&s_1s_2 & -c_1s_2c_3-c_2s_3 & -c_1s_2s_3 + c_2c_3 } \right) 
\label{2hdmrij} ~,
\eeqa

\ni where $s_i\equiv \sin \alpha_i$ and $c_i\equiv \cos \alpha_i$.  

A general feature of this class of 2HDM's, in which 
CP violation results from scalar-pseudoscalar mixing 
in the neutral Higgs interactions with fermions, 
is that only two out of the three neutral Higgs-bosons can 
simultaneously have a coupling to vector-bosons and a 
pseudoscalar-type of coupling to fermions. Denoting these two neutral 
Higgs-bosons by $h$ and $H$ with couplings $a_f^h,b_f^h,c^h$ and  
$a_f^H,b_f^H,c^H$  
(corresponding
to the light  and heavy neutral Higgs, respectively),  
an important aspect of these 2HDM's,  which
has crucial phenomenological implications for CP violation, is  
that these couplings are subject to the constraint $b_f^hc^h + b_f^Hc^H=0$.
This is a general feature of CP violation induced by mass mixing and 
is due to
the existence of a GIM-like cancellation, 
dictating that CP-odd effects must
vanish when  the two Higgs-bosons $h$ and $H$ are degenerate.

To end this section let us briefly comment on the existing 
experimental limits on the  neutral and charged Higgs 
masses and on the couplings $a_f^k,b_f^k,c^k$. 
There are very good reviews on this subject in the literature 
and we only wish to point out the highlights 
of those investigations.  
The existing limits are usually given on 
$\tan\beta$ and they depend on the mass of the charged 
Higgs-boson or the neutral Higgs-boson of the theory. They can be translated 
to bounds on $a_f^k,b_f^k,c^k$ using Eq.~\ref{mhdmeq23}. 
In particular, the limits are obtained from 
the experimental constraints using  
low energy data on $B -\bar B$, $D - \bar D$ and $K -\bar K$ mixing, 
$\epsilon_K$, 
$b \to u$, $b \to c$ and $b \to s \gamma$ transitions 
\cite{plb243p301,prd41p3421,hepph9406302,hepph9411436,mpla11p675}, 
on $B \to \tau \nu_\tau X$ decays 
\cite{plb406p337,prd56p5786}, on $(g-2)_\mu$ \cite{prd55p6968},  
or from high energy processes such 
as $e^+e^- \to Z h$ \cite{prl79p982}, the decay $t \to b H^+$
\cite{hepph9603391,prd59p055007,prl79p357} 
and from $Z$ decays \cite{mpla10p1057,hepph9607268}.
Typically they find that $\tan\beta < 1$ is allowed if the 
charged Higgs mass is above several hundred GeV (recall that a small 
$\tan\beta$ enhances the Higgs coupling to the top quark). In order to have 
$\tan\beta \lsim 0.5$ the charged Higgs mass is required to be 
typically $\gsim 500$ GeV\null. If $\tan\beta$ is sufficiently large, 
i.e., $\tan\beta \sim {\cal O}(m_t/m_b)$, then a light charged 
Higgs-boson is possible, of the order of several tens of GeV\null.  
Note also that there are theoretical approximate  lower and upper bounds
on $\tan\beta$ coming from perturbative considerations. That is, 
in order for a perturbative description to remain valid, $\tan\beta$ 
has to roughly satisfy $0.1 \lsim \tan\beta \lsim 100$ \cite{prd41p3421}. 
The lower (upper) 
limit corresponds to the a perturbative top (bottom) Yukawa coupling.

\begin{figure}
\psfull
 \begin{center}
  \leavevmode
  \epsfig{file=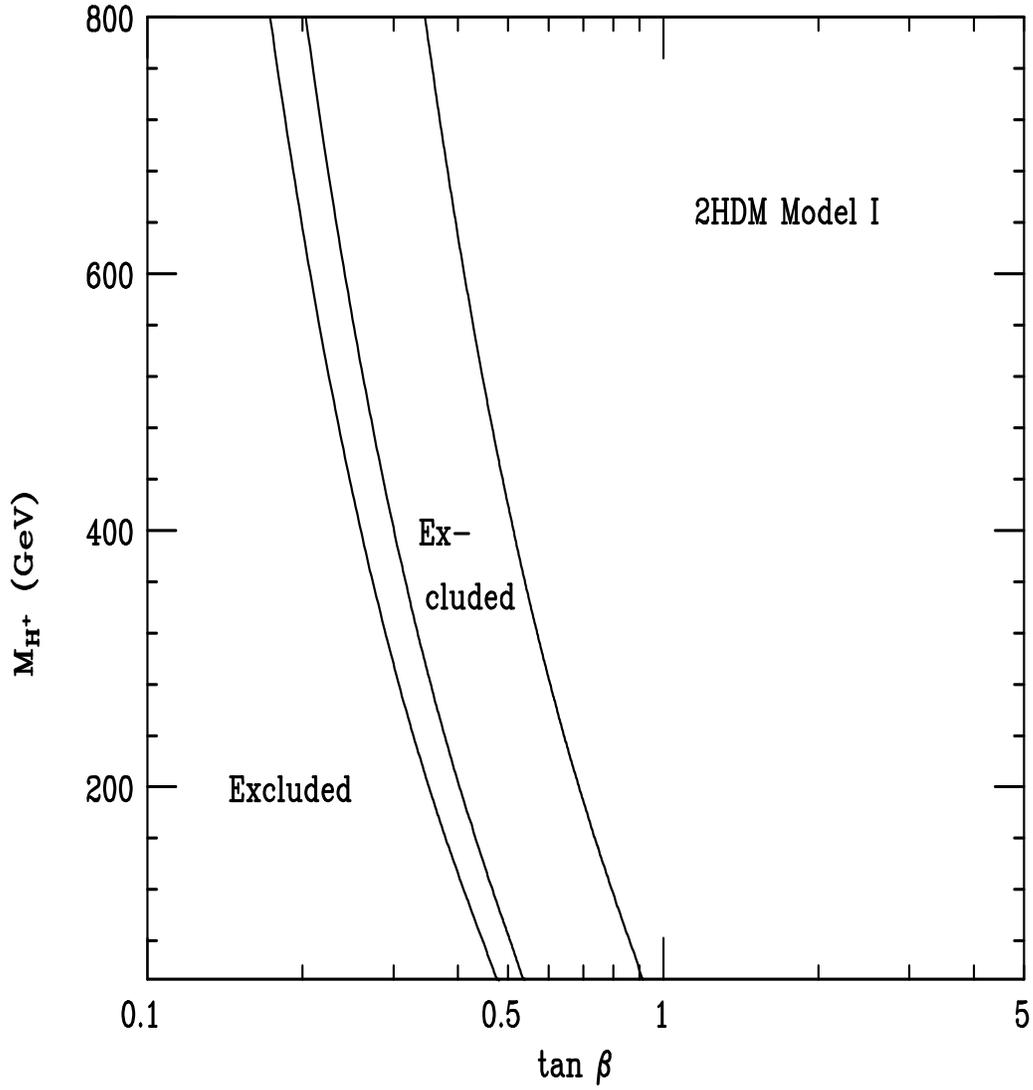,height=8cm,angle=270}
 \end{center}
\caption{\emph{The excluded regions in the $m_{H^\pm}-\tan\beta$ 
plane resulting
from the present CLEO bounds in Model I and for $m_t=175$ GeV. 
The excluded regions
are (from left to right) to the left of the first curve and 
between the second and third curves. Updated figure from 
\cite{hepph9406302} (see \cite{wethank}).}}
\label{modelsfig3}
\end{figure}

\begin{figure}
\psfull
 \begin{center}
  \leavevmode
  \epsfig{file=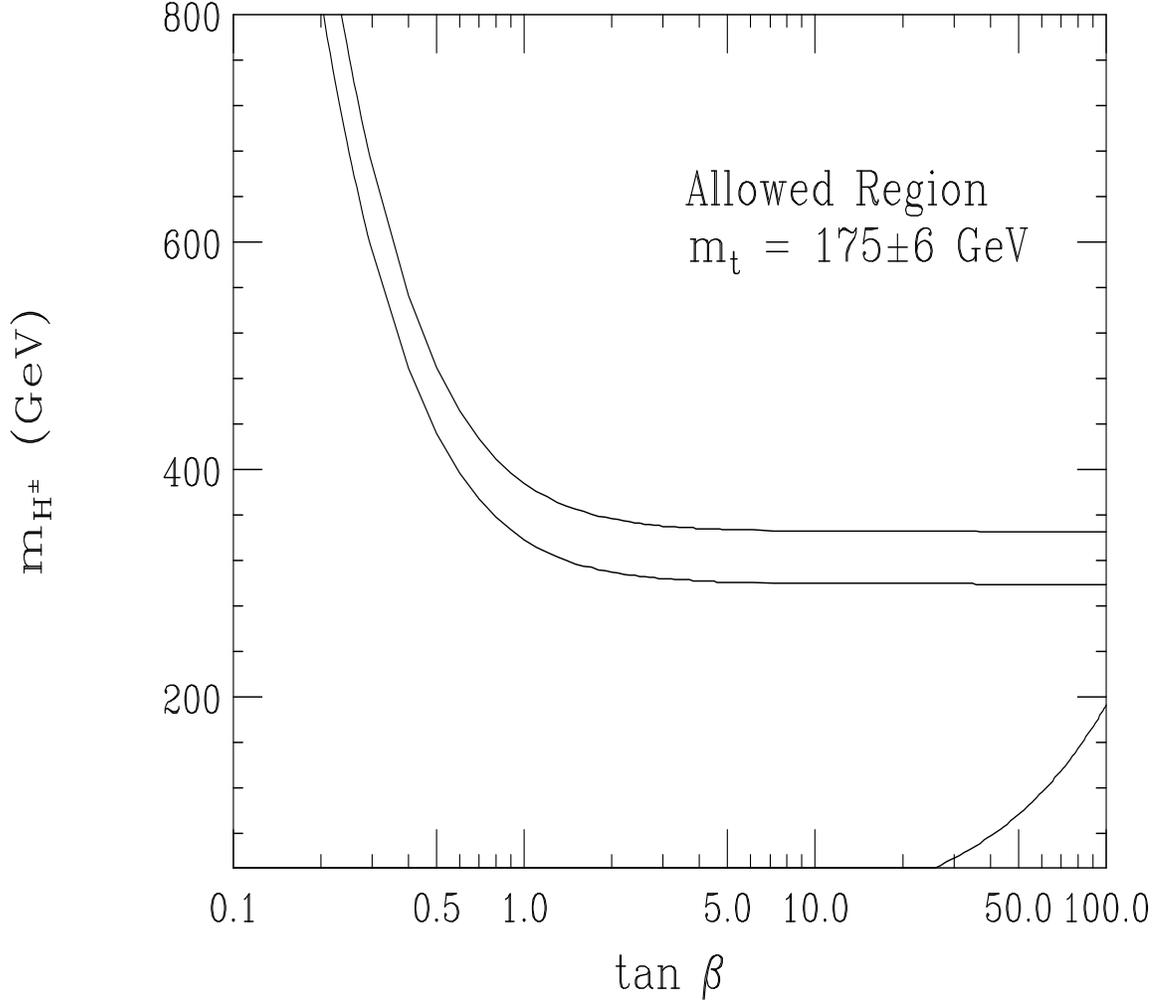,height=12cm}
 \end{center}
\caption{\emph{Constraints in the $m_{H^\pm}-\tan\beta$ 
plane in Model II 
from the CLEO bound on ${\rm Br}(B\to X_s\gamma)$. The excluded region
is to the left and below the curves. The upper line is for 
$m_t = 181$ GeV and the lower line is for $m_t =169$
GeV\null. We also display the restriction $\tan\beta/m_{H^\pm}>0.52$
GeV$^{-1}$ which arises from measurements of $B\to X\tau\nu$ as
discussed in \cite{prl60p182,plb357p630}. 
Figure taken from \cite{prd55p5549}.}}
\label{modelsfig4}
\end{figure}

The inclusive decay $B\to X_s\gamma$, which is equivalent at the quark
level to $b\to s\gamma$, plays a unique role in constraining the
parameter space of 2HDM's \cite{hepph9406302}. Recall first the CLEO
observation \cite{cleocollab} ${\rm Br}(B\to X_s\gamma) = (2.32 \pm .57 \pm
.35) \times 10^{-4}$ and the corresponding 95\% CL bounds of
$1\times10^{-4} < {\rm Br} 
(B \to X_s \gamma) < 4.2\times10^{-4}$. 
These were updated by CLEO to read 
${\rm Br} (B \to X_s \gamma) < (3.11 \pm 0.8 \pm 0.72) \times 10^{-4}$
where, at 95\% CL, $2.0 \times 10^{-4} < {\rm Br} (B \to X_s \gamma) < 
4.5 \times 10^{-4}$ \cite{ichep981011,hepex9901005}.
Furthermore, ALEPH presented the result \cite{plb429p169} 
${\rm Br} (B \to X_s \gamma) < (3.15 \pm 0.35_{\rm stat} \pm 0.32_{\rm syst} 
\pm 0.26_{\rm model} ) \times 10^{-4}$, consistent with CLEO.
In Model I, the two
contributions of the charged Higgs scale as $\cot^2\beta$ causing
significant enhancements to the decay rate for small values of
$\tan\beta$. An especially interesting aspect of Model I is that the
contribution of the two charged Higgs interfere destructively for some
values of the parameter space. Consequently, the lower and upper bounds
from CLEO enable us to place stringent constraints on the mass
$(m_{H^\pm}) - \tan\beta$ plane \cite{hepph9406302}. Most notably, for
small value of $\tan\beta$ (especially for $\tan\beta\lsim0.3$),
$m_{H^\pm}\lsim 1$ TeV gets excluded as can be seen from 
Fig.~\ref{modelsfig3} \cite{wethank}.

${\rm Br}(b\to s\gamma)$ also places important restriction on Model II
\cite{hepph9406302}, again, especially for small values of $\tan\beta$
(see Fig.~\ref{modelsfig4}) \cite{wethank,prl60p182,plb357p630,prd55p5549}. 
An
important difference with Model I is that ${\rm Br}(b\to s\gamma)$ is now
always larger than in the SM, independent of $\tan\beta$
\cite{hepph9406302}. Thus, $m_{H^\pm}\lsim300$ GeV is ruled out 
practically for
all values of $\tan\beta$. One can relax these bounds if 
MSSM contributions (chargino loops) 
are added to the charged Higgs in the loop \cite{hepph9406302}.

\subsubsection{Three Higgs doublet model \label{sssec324}}

As was mentioned above, if one chooses to adopt Natural Flavor 
Conservation (NFC),
enforced by some discrete symmetry (see e.g., Eqs.~\ref{mhdmeq17} and 
\ref{mhdmeq18}), 
then the minimum number of Higgs
doublets which allows 
CP violation in the gauge model with NFC is
three \cite{npb259p306,ijmpa7p1059} (the other possibility is 
two Higgs doublets plus one singlet but we choose not to discuss it).

 In the Weinberg 3HDM \cite{prl37p657,prd21p711}, 
CP violation in the
Higgs sector arises from the phase differences of the VEV's and from
the complex quartic terms in the Higgs potential (as will be shown later).
This model has many unknown parameters and it allows
the standard CKM mechanism as one of the sources of CP nonconservation.
Let us therefore present its
spontaneous version which has a very attractive
feature: spontaneous CP violation and NFC together with a real tree-level
CKM matrix (for three generations) 
as a starting point,
imply that CP violation is generated only after SSB
 via complex VEV's \cite{prl44p504,prd22p2901}. 
Thus, in such a scenario CP is a good symmetry of the 
Lagrangian before SSB and CP violation at tree-level 
comes solely from Higgs-boson exchanges once the VEV's are assigned 
a non-vanishing phase. 

The most general scalar
potential with three Higgs doublets consistent with NFC 
is given by\footnote{Unlike Eq.~\ref{mhdmeq1} which describes
the Higgs potential for a 2HDM, we use here the more generic form 
of the Higgs potential applicable also for arbitrary number of Higgs
doublets.}

\beqa 
 &V(\Phi)=\displaystyle{\sum_i} m_i^2\Phi_i^{\dag }\Phi_i+
       \sum_{i\ne j} a_{ij}(\Phi_i^{\dag}\Phi_i)(\Phi_j^{\dag}\Phi_j)+ \cr
      &\displaystyle{\sum_{i\ne j}} b_{ij}(\Phi_i^{\dag}\Phi_j)
    (\Phi_j^{\dag}\Phi_i)+
     \Bigl [\sum_{i\ne j} c_{ij}(\Phi_i^{\dag}\Phi_j)(\Phi_i^{\dag}\Phi_j)+
     \hbox{h.c.} \Bigr ] \label{mhdmeq24}~.
\eeqa   

\ni Hermiticity of the scalar  potential requires that the $m_i^2$ terms,
the $a_{ij}$ and the $b_{ij}$ be real while
$c_{ij}$ need only be hermitian. However, if CP is broken spontaneously,
all the parameters of $V(\Phi)$ can be chosen to be real before gauge
symmetry breaking and we choose to work in the latter scheme.
\par The Higgs doublets can be written as

\beq
 \Phi_i=\pmatrix{\phi_i^+\cr \phi_i^0} =
\pmatrix{\phi_i^+\cr {1\over \sqrt{2}}(v_i+\rho_i+i\eta_i)}
\label{mhdmeq25} ~,
\eeq  

\ni with

\beq
 v_i=|v_i|e^{i\theta_i} \label{mhdmeq26}~.
\eeq  

\ni Assuming that the third Higgs doublet does not couple to 
quarks and that
its VEV does participate in breaking SU(2)$\times$U(1), the general
Yukawa interactions consistent with NFC read

\beqa
 {\cal L}_Y&=&Y_{ij}^{(1)}\bar{D}_R^i (\phi_1^{+*}U_L^j+\phi_1^{0*}D_L^j)
     +Y_{ij}^{(2)}\bar{U}_R^i (\phi_2^0 U_L^j-\phi_2^+D_L^j)+ \nonumber \\ 
&   &\quad Y_{ij}^{(3)}
  \bar{E}_R^i (\phi_3^{+*}N_L^j+\phi_3^{0*}E_L^j)+\hbox{h.c.} 
\label{mhdmeq27}~,
\eeqa  

\ni where $U=(u,c,t)$, $D=(d,s,b)$, $E=(e,\mu,\tau)$ and
$N=(\nu_e,\nu_\mu,\nu_\tau)$.
The Yukawa couplings are chosen to be real as CP  is assumed to be
a good symmetry of ${\cal L}_Y$.

After SSB and after a phase redefinition of
the quark fields in order to obtain real quark mass matrices, we can rewrite
${\cal L}_Y$ as

\beq  
 {\cal L}^+_Y={\phi_1^{\prime +}\over v_1}\overline{U}_L KM_D D_R-
     {\phi_2^{\prime +}\over v_2}\overline{U}_R M_U KD_L+{\phi_3^{\prime +}
     \over v_3}\overline{N}_L M_E E_R+\hbox{h.c.}
\label{mhdmeq28}~,
\eeq  

\ni for charged Higgs-bosons, with $\phi_i^\prime =\phi_i 
e^{i\theta_i}$ and a real CKM matrix, denoted here by $K$. 
Also

\beqa 
 {\cal L}^0_Y&=&{\rho_1\over v_1}\overline{D}  M_D D+i{\eta_1\over v_1}
     \overline{D} M_D\gamma_5 D+{\rho_2\over v_2}\overline{U} M_U U-
     \nonumber \\
     & & \quad i{\eta_2\over v_2}\overline{U} M_U \gamma_5 U+{\rho_3\over v_3}
     \overline{E} M_E E+i{\eta_3\over v_3}\overline{E} M_E \gamma_5 E
\label{mhdmeq29}~,
\eeqa  

\ni for neutral Higgs-bosons, where $M_U$, $M_D$ and $M_E$ 
are diagonal mass
matrices. The Yukawa interactions of Eqs.~\ref{mhdmeq28} and \ref{mhdmeq29} 
are still CP invariant,
however, the scalar fields $\phi_i,\rho_i$ and $\eta_i$ are not the
physical Higgs states with definite masses.
\par The mass matrix in the basis of
$\phi_i^{\prime +}/v_i$ is

\beq
 m^2=\pmatrix{X_{12}+X_{13}&-X_{12}-iY&-X_{13}+iY \cr \cr
     -X_{12}+iY&X_{12}+X_{23}&-X_{23}-iY \cr  \cr
     -X_{13}+iY&-X_{23}+iY&X_{13}+X_{23}} \label{mhdmeq30}~,
\eeq  

\ni where

\beq 
 X_{ij}=\left [{1\over 2}b_{ij}+c_{ij}\cos (\theta_i-\theta_j)\right ]
     |v_i|^2 |v_j|^2 \label{mhdmeq31}~,
\eeq  

\ni and

\beqa 
 Y& = & -c_{12}v_1^2 v_2^2 \sin 2(\theta_1-\theta_2)
     =-c_{23}v_2^2 v_3^2 \sin 2(\theta_2-\theta_3) \nonumber \\
 & = & c_{13}v_1^2 v_3^2 \sin 2\theta_1 \label{mhdmeq32}~.
\eeqa  

\ni Since the parameter $Y$ is in general non-zero, it is evident that
CP violation in the charged Higgs-boson sector comes from the
imaginary part of the off-diagonal Higgs-boson mass matrix elements.

The unitary matrix which relates the weak eigenstates
$\phi_i^{\prime +}$ to the physical charged states $H_i^+$ is defined by

\beq
 \pmatrix{\phi^{\prime +}_1\cr \cr \phi^{\prime +}_2 \cr \cr
     \phi^{\prime +}_3}=U^+\pmatrix{G^+\cr \cr H_1^+ \cr \cr H_2^+}
\label{mhdmeq33}~,
\eeq

\ni where $G^+$ is the charged Goldstone-boson which 
is absorbed into the
$W^+$. $U^+$, which has three arbitrary phases, of which two 
can be
removed by a redefinition of $H_1^+$ and $H_2^+$, can be 
parameterized
exactly in the same way as the CKM matrix 
\cite{prd21p711}:

\beq
 U^+\equiv \pmatrix{\tilde c_1& \tilde s_1\tilde c_3
     & \tilde s_1\tilde s_3 \cr \cr -\tilde s_1 \tilde c_2
     &\tilde c_1\tilde c_2\tilde c_3+\tilde s_2\tilde s_3 e^{i\delta_H}
     &\tilde c_1\tilde c_2\tilde s_3-\tilde s_2\tilde c_3 e^{i\delta_H} \cr
     \cr -\tilde s_1 \tilde s_2&\tilde c_1\tilde s_2\tilde c_3-
      \tilde c_2\tilde s_3 e^{i\delta_H}&\tilde c_1\tilde s_2\tilde s_3+
      \tilde c_2\tilde c_3 e^{i\delta_H} }
\label{mhdmeq34}~,
\eeq 

\ni with $\tilde s_i \equiv \sin \tilde \theta_i$, 
$\tilde c_i \equiv \cos \tilde \theta_i$
and $\delta_H$ are the charged Higgs mixing angles and phase,
respectively.
\par From the gauge sector, it is straightforward to show that

\beq 
 (2\sqrt{2}G_F)^{-1}=|v_1|^2+|v_2|^2+|v_3|^2=v^2 \label{mhdmeq35}~,
\eeq  

\ni or that

\beq
 G^+={1\over v}(v_1\phi_1^{\prime +}+v_2\phi_2^{\prime +}
     +v_3\phi_3^{\prime +}) \label{mhdmeq36}~.
\eeq  

\ni It follows from Eqs.~\ref{mhdmeq33} and \ref{mhdmeq34} that

\beq
 v_1=\tilde c_1 v~,\qquad v_2=-\tilde s_1 \tilde c_2 v~,\qquad
     v_3=-\tilde s_1\tilde s_2 v \label{mhdmeq37}~.
\eeq  

\ni Therefore, the mixing angles 
$\tilde \theta_1,\tilde \theta_2$ are
determined by the VEV's $v_1,v_2$ and $v_3$ whereas $\tilde \theta_3$ and
$\delta_H$ depend on the parameters of the Higgs potential.
 
In terms of the Higgs mass eigenstates, the Yukawa 
interactions in Eq.~\ref{mhdmeq28} become

\beqa
 {\cal L}_Y^+ & = & 
(2\sqrt{2}G_F)^{1/2}\sum_{i=1}^2 (\alpha_i \overline{U}_LKM_D
 D_R+\beta_i \overline{U}_R M_U KD_L \nonumber \\
 & & \quad + \gamma_i \overline{N}_L M_E E_R)H_i^++\hbox{h.c.} 
\label{mhdmeq38}~,
\eeqa  

\ni with

\beqa
&\alpha_1={\tilde s_1\tilde c_3\over \tilde c_1}~,~
     \beta_1={\tilde c_1\tilde c_2\tilde c_3+
     \tilde s_2\tilde s_3 e^{i\delta_H}\over \tilde s_1\tilde c_2}~,~
     \gamma_1={\tilde c_1\tilde s_2\tilde c_3-
     \tilde c_2\tilde s_3 e^{i\delta_H}\over \tilde s_1\tilde s_2}~, \cr
     &\alpha_2={\tilde s_1\tilde s_3\over \tilde c_1}~,~
     \beta_2={\tilde c_1\tilde c_2\tilde s_3-
     \tilde s_2\tilde c_3 e^{i\delta_H}\over \tilde s_1\tilde c_2}~,~
     \gamma_2={\tilde c_1\tilde s_2\tilde s_3+\tilde c_2\tilde c_3
     e^{i\delta_H}\over \tilde s_1\tilde s_2}~,
\label{mhdmeq39}
\eeqa  

\ni and we see that

\beqa
 \hbox{\mIm }(\alpha_2\beta_2^*)& = & -\hbox{\mIm }(\alpha_1\beta_1^*)~,~
     \hbox{\mIm }(\alpha_2\gamma_2^*)=-\hbox{\mIm }(\alpha_1\gamma_1^*)~,
     \nonumber \\
      \hbox{\mIm }(\beta_2\gamma_2^*) & = & -\hbox{\mIm }(\beta_1\gamma_1^*)~.
\normalsize\label{mhdmeq40}
\eeqa  

\ni As in the charged Higgs case, we can write down an 
analogous $6\times 6$ real mass
matrix for the neutral scalar states. Then the
neutral Higgs-boson Yukawa interactions in Eq.~\ref{mhdmeq29} become 
\cite{ijmpa7p1059}:

\beqa 
{\cal L}_Y^0 & = & (2\sqrt{2}G_F)^{1/2}\sum_{i=1}^5
     (g_{1i}\overline{D} M_D D+g_{2i}\overline{D} M_D i\gamma_5 D+
      g_{3i}\overline{U} M_U U \nonumber \\
&  & \quad + g_{4i}\overline{U} M_U i\gamma_5 U+g_{5i}\overline{E} M_E E+
      g_{6i}\overline{E} M_E i\gamma_5 E)H_i^0 \label{mhdmeq41}~,
\eeqa  

\ni where the couplings $g_i$ are real. Since $\bar{\psi}\psi$ and
$\bar{\psi}i\gamma_5\psi$ have opposite 
$P$, $T$ and CP transformation properties, $P$ and CP can be 
violated through the exchange of neutral Higgs-bosons.

We will now briefly mention some of the more notable constraints on
this class of models \cite{prl70p1364,plb313p126,prl71p492,npb426p355}. One
class of restrictions that are of some 
importance follow (as in the case of the 2HDM) 
if we further assume that the Higgs sector of the theory is 
perturbative \cite{prd41p3421}. 
These lead to

\beq
|\alpha_i| \lsim 120 \quad , \quad |\beta_i| \lsim 6~. 
\eeq

\ni Since these complex coupling constants arise from the
diagonalization of the charged scalar mixing matrix, they obey the
relation

\beq
\sum_{i=1,2} \alpha_i \beta^\ast_i = 1~. 
\eeq

\ni Thus \cite{plb313p126}:

\beq
\mIm (\alpha_1\beta^\ast_1) = - \mIm (\alpha_2\beta^\ast_2) ~,
\eeq

\ni and 

\beq
\mIm (\alpha_i \beta^\ast_i) \le |\alpha_i\beta_i| \lsim 720~.       
\eeq

\ni Assuming for simplicity that one of the $H^+$, for instance $H^+_2$, 
is very
heavy, then $B$-$\bar B$ mixing imposes an important constraint on
$|\beta_1|$ \cite{prd41p3421,npb337p284}:

\beqa
|\beta_1| & \lsim & 2 \mbox{\rm~for~} m_{H_1} \sim \frac{1}{2} m_Z \nonumber
\\
& \lsim & 3 \mbox{~for~} m_{H_1} \sim 2 m_Z~. 
\eeqa

\ni Using $B\to \tau\nu X$, a constraint on $|\alpha_1|$ is deduced
\cite{prl60p182,plb298p409}: 

\beq
|\alpha_1| \lsim 2 m_{H_1} /\mbox{~GeV}~.
\eeq

\begin{figure}
\psfull
 \begin{center}
  \leavevmode
\epsfig{file=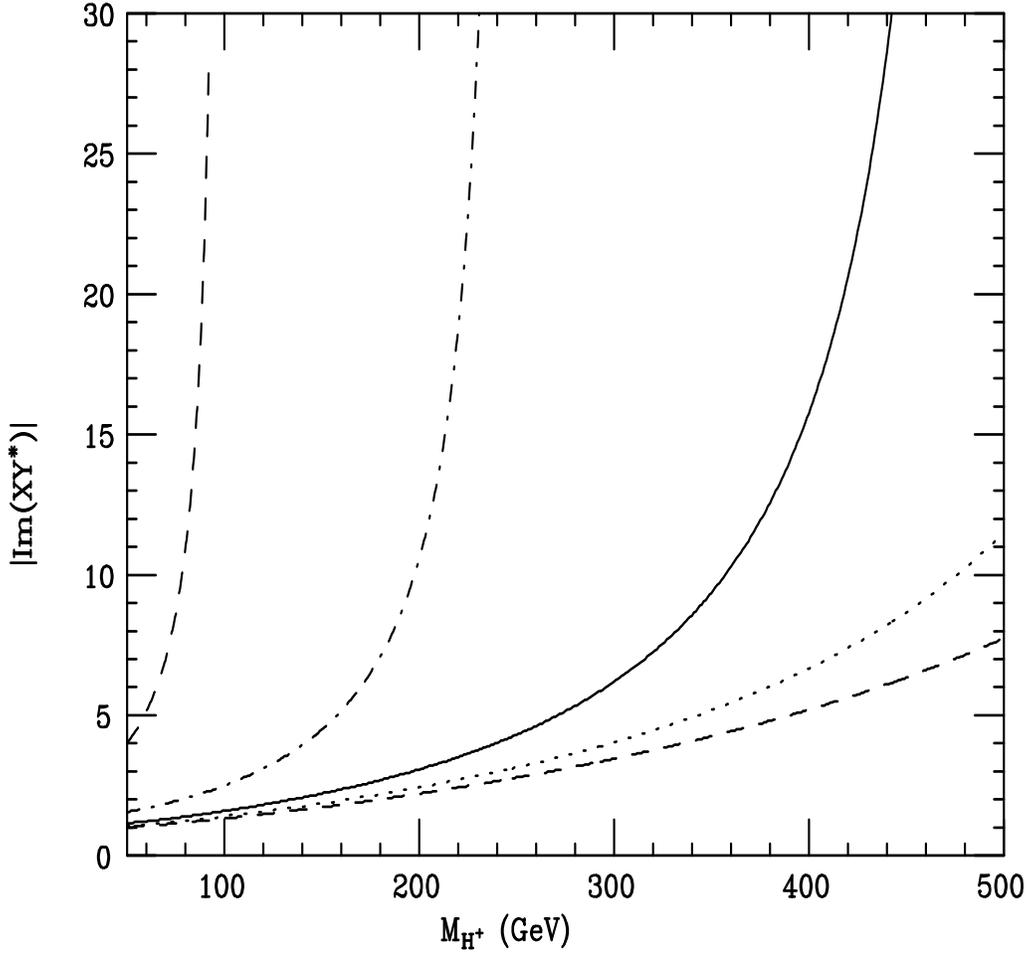,height=7cm,angle=270}
 \end{center}
\caption{\emph{Constraints on ${\mIm}(XY^*)|$ 
($\equiv |{\mIm}(\alpha \beta^*)|$ in our notation) as a function 
of the lightest charged Higgs-boson mass $M_{H^+} \equiv m_{H_1}$, with  
$m_{H_2}=100, ~250, ~500$ and $750$ GeV corresponding 
to (from left to 
right) the dashed, dashed-dotted, solid and dotted curves,
respectively. 
The bottom dashed curve represents the case where the $H_2^\pm$ 
contributions have been neglected. The allowed region lies 
to the right and below
the 
curves. $m_t=175$ GeV is used. Updated figure from \cite{hepph9406302} 
(see \cite{wethank}).}} 
\label{modelsfig5}
\end{figure}

A direct bound on $\mIm(\alpha\beta^*)$ comes from the 
electric dipole moment of the neutron ($d_n$)
\cite{npb364p10,ynir}:

\beqa
\mid \mIm (\alpha\beta^\ast) \mid & 
\lsim & 20 \mbox{~for~} m_{H_1} \sim \frac{1}{2} m_Z
\nonumber \\
& \lsim & 100 \mbox{~for~} m_{H_1} \sim 2m_Z~.
\eeqa

Interestingly enough, the strongest constraint so far actually comes 
from a CP-conserving process $b\to s\gamma$
\cite{plb313p126,npb426p355}. The amplitude for $b\to 
s\gamma$ receives contributions from terms 
proportional to 
$\mIm (\alpha_i\beta^\ast_i)$ 
that do not interfere with 
the other terms.

Thus these terms only enter quadratically in the expression for the
rate for $b\to s\gamma$. A conservative bound on $|\mIm
(\alpha\beta^\ast)|$ (where $\mIm(\alpha\beta^\ast) = \mIm
(\alpha_1\beta^\ast_1) = -\mIm (\alpha_2\beta^\ast_2)$) is obtained by
assuming that such a contribution saturates the measured rate for $b\to
s\gamma$. These constraints are displayed in Fig.~\ref{modelsfig5}
\cite{hepph9406302,wethank} as a
function of the light charged Higgs mass ($m_{H_1}$) for various values
of the heavier charged Higgs mass ($m_{H_2}$), subject to the
restriction $m_{H_1}\le m_{H_2}$. On the figure, the bottom solid curve
corresponds to the case when $m_{H_2}>\!\!> m_{H_1}$ so that the
contribution of the second charged Higgs is neglected. We note that the
constraints depend strongly on $m_{H_2}$ and they essentially disappear
when $m_{H_1} \simeq m_{H_2}$ due to a cancellation between the two
contributions \cite{hepph9406302,wethank} resulting from a GIM-like 
mechanism.

It is useful to note how stringent these constraints are. For example

\beqa
\mid \mIm(\alpha\beta^\ast) \mid & 
\lsim & 1.5 \mbox{~for~} m_{H_1}\sim \frac{1}{2}
m_Z \nonumber \\
& \lsim & 2.5 \mbox{~for~} m_{H_1}\sim 2m_Z~, 
\eeqa

\ni for $m_{H_2} \sim 500$ GeV.
A very important consequence of these tight bounds on $\mIm
(\alpha\beta^\ast)$ is that the charged scalar exchange can only make a
negligible contribution to the CP violation
parameters in $K\to 2\pi$, i.e.,
$\epsilon$ or $\epsilon^\prime$. Therefore, CP violation in the 3HDM
cannot be the sole source of the observed CP violation. We should note,
though, that in the original Weinberg model for three Higgs doublets,
CP is not assumed to be an {\it a priori\/} symmetry of the model. Thus
CP violation arises from complex quartic terms in the Higgs potential
as well as
from the phase differences of the VEV's. In addition, one has the
complex CKM phase as an independent source of CP violation which may be able
to accommodate the CP violation in the Kaon system.

\subsection{{Supersymmetric models \label{ssec33}}}

Needless to say,  the minimal SUSY 
extension of the SM is a very appealing theory \cite{susyreview}. 
Among its compelling features are: it allows for Radiative 
ElectroWeak Symmetry Breaking (REWSB), it unifies the gauge coupling 
constants, with masses of superpartners not much heavier then 
a TeV it gives a well-grounded explanation to the hierarchy 
problem and it provides a good dark matter 
candidate - the lightest SUSY particle. 

As far as CP violation
is concerned, new non-SM mechanisms are introduced in each 
version of such SUSY models 
\cite{hepph9701231,susyreview,npb255p413,prd59p016004}. 
It is again the top quark sector in these models that may exhibit 
large CP-violating effects due to its very large mass. In particular, the 
supersymmetric partners of the top quark 
(these two scalar particles are
often referred to as the stop and denoted by ${\tilde t}$),  
can be responsible for relatively large CP-violating phenomena.
Such effects are enhanced by the possibility of having large mass 
splittings between the two stops which is in turn due to the relatively 
large top mass.
These type of SUSY CP violation in top quark reactions has received 
considerable attention in the past few years. They are
all strongly dependent on the magnitude of the low energy phase of 
the soft trilinear breaking term $A_t$ in the SUSY Lagrangian and 
we will describe some of these works in the following chapters.

Another possible manifestation of the CP-violating phase 
$\arg(A_t)$ is Baryogenesis in the early universe.
It was shown that with  $\arg(\mu) \to 0$, $t$ squarks can mediate
the charge transport mechanism needed to generate 
the observed baryon asymmetry,
even with squark masses $\sim$ hundred GeV, 
provided that $\arg(A_t)$
is not much suppressed \cite{ptp98p1325}.
We will therefore emphasize here the phenomenological importance of 
possible CP-violating effects which may reside in 
${\tilde t}_L - {\tilde t}_R$ mixing  
and are therefore proportional to $\arg(A_t)$. Indeed, 
due to experimental constraints on the Neutron Electric Dipole 
Moment (NEDM), the possible 
phase in the Higgs 
mass term, i.e., $\arg(\mu)$, is expected to be small (see below). Thus 
$\arg(A_t)$ should be practically the only important SUSY CP-odd phase 
observable
in high-energy reactions.     
Of course, the most natural place to look for 
such
effects, 
driven by $\arg(A_t)$, is high energy processes involving 
the top quark.
Thus CP-violating effects of the top quark observable in the laboratory 
may have direct bearing on Baryogenesis 
in the early Universe.
 
It was stated in \cite{prd55p1611} that using the relations 
obtained from the Renormalization Group Equations (RGE) 
of the imaginary parts in the SUSY Lagrangian, 
combined with the severe constraint on the low energy phase of the 
Higgs mass parameter, $\mu$, from the present experimental 
limit \cite{prl82p904} on the NEDM,  
the phase in $A_t$ at low energy scales
is likely to be very small 
provided one imposes
some definite boundary conditions for the SUSY soft 
breaking terms. As a 
consequence, at high energies, any CP-nonconserving effect 
that is driven by $\arg(A_t)$ will then be suppressed leaving 
top quark reactions almost insensitive to CP-violating effects 
of a SUSY origin in models with these assumptions.

On the other hand we will describe below the 
key phenomenological 
features of 
a general MSSM 
and a GUT-scale N=1 minimal SUperGRAvity (SUGRA) model. 
We will demonstrate that the prediction made in \cite{prd55p1611} 
depends on the GUT-scale boundary conditions, 
and therefore may be significantly relaxed to yield a large 
CP-violating phase in the $A_t$ term compatible with the existing experimental 
limit on the NEDM\null. This should encourage SUSY CP violation 
studies in top quark systems as they may well be the 
only venue for constraining $\arg(A_t)$ in 
high energy experiments at colliders in the foreseeable future.

\subsubsection{General description and the SUSY Lagrangian \label{sssec331}}

The most general low energy softly broken minimal SUSY 
Lagrangian which is invariant under 
SU(3)$\times$SU(2)$\times$U(1) 
consists of three generations of quarks and leptons, 
two Higgs doublets and the
SU(3)$\times$SU(2)$\times$U(1) gauge fields, along with their SUSY 
partners, can be written as 
\cite{susyreview,npb221p495,prd41p4464,hepph9902340}:

\beq
{\cal L} = \ \mbox{kinetic terms} + \int d^2 \theta W + 
{\cal L}_{\rm soft} \ \ .
\eeq 

\ni Here $W$ is the superpotential and is given 
by\footnote{We do not 
include R-parity violating terms in the SUSY Lagrangian below, 
since we do not discuss in this review any CP-violating effect 
which may be driven by such terms. We only briefly mention in Chapter 
\ref{summary_chapter} 
the possible impact of R-parity violating SUSY interactions 
on CP violation studies in the top quark system.}

\beq
W = \epsilon_{ij}(g^{IJ}_U \hat{Q}^i_I \hat{H}^j_2 \hat{U}_J
+g^{IJ}_D \hat{Q}^i_I \hat{H}^j_1 \hat{D}^c_J + 
g^{IJ}_E \hat{L}^i_I \hat{H}^j_1 \hat{R}^c_J+ \mu\hat{H}^i_1\hat{H}^j_2)  \ \ .
\label{superw}
\eeq 

\ni $\epsilon_{ij}$ is the antisymmetric tensor with 
$\epsilon_{12}=1$ and
the usual convention was used for the superfields 
$\hat{Q},\hat{U},\hat{L},\hat{R}$ and $\hat{H}$ \cite{npb221p495}.
$I,J=1,2$ or 3 are generation indices and $i,~j$ are SU(2) indices.

${\cal L}_{\rm soft}$ consists of the soft breaking 
terms and can be divided into three pieces

\beq
{\cal L}_{\rm soft} \equiv {\cal L}_{\rm gaugino} + 
{\cal L}_{\rm scalar} + {\cal L}_{\rm trilinear} \ \ ,
\eeq

\ni which are the soft supersymmetry breaking gaugino, 
scalar mass terms and the trilinear coupling terms. 
These are given by \cite{prd41p4464}:

\beqa
{\cal L}_{\rm gaugino} &=& \frac{1}{2} (\tilde{m}_1 \lambda_B 
\lambda_B + \tilde{m}_2 \lambda_W^a \lambda_W^a +
\tilde{m}_3 \lambda_G^b \lambda_G^b) \label{softg}~,\\ 
{\cal L}_{\rm scalar}  &=& -m_{H_1}^2 |H^1_i|^2 -m_{H_2}^2 
|H^2_i|^2 - m_L^2 |L^i|^2 -  \nonumber \\
&&m_R^2 |R|^2 - m_Q^2 |Q^i|^2 - m_D^2 |D|^2 - m_U^2 |U|^2 
\label{softs}~,\\
{\cal L}_{\rm trilinear} &=& \epsilon_{ij} (g_U A_U Q^i H^j_2 U +
g_D A_D Q^i H^j_1 D +\nonumber \\
&&g_E A_E L^i H^j_1 R + \mu B H^i_1H^j_2) \label{tricouplings} ~,
\eeqa 

\ni where we have omitted  the generation 
indices $I$ and $J$ in the soft breaking terms.
The above scalar fields 
correspond to the superfields which were indicated in our 
notation by a ``hat''. $\lambda_B$, 
$\lambda_W^a$ (with $a=1, 2$ or 3) and $\lambda_G^b$ (with $b=1,\dots,8$) 
are the gauge superpartners of the 
U(1), SU(2) and SU(3) gauge-bosons, respectively. 
Also we remark that proportionality of the trilinear 
couplings to the Yukawa couplings (i.e., $g_U,g_D$ and $g_E$) 
is imposed in Eq.~\ref{tricouplings}.

\subsubsection{CP violation in a general MSSM \label{sssec332}}

We now turn to a discussion of the CP-odd phases in the theory.  
In general, when no further assumptions are imposed on the 
pieces of the Lagrangian in Eqs.~\ref{superw} and 
\ref{softg} - \ref{tricouplings}, 
there
are several possible new sources (apart from the usual 
SM CKM and strong $\theta$ phases) of CP nonconservation 
at the scale $\Lambda_S$ - where the soft breaking terms are generated.  
These are \cite{hepph9701231,susyreview,npb255p413}: the 
trilinear couplings $A_F$ (i.e., $F=U,D$ or $E$), the soft 
breaking Higgs coupling $\mu B$, the gauginos mass 
parameters $\tilde{m}_a$ ($a=1,2$ or 3) and the Higgs mass 
parameter $\mu$ in the superpotential. 
However, not all of them are physical and by a global phase 
change of one of the Higgs multiplets one can set $\arg(\mu B)=0$ 
ensuring real VEV's of the Higgs doublets and fixing the phase 
of $\mu$ to be $\arg(\mu)=-\arg(B)$. 
Moreover, in the absence 
of the soft breaking Lagrangian, the MSSM possesses an 
additional U(1) R-symmetry \cite{npb465p23}. 
Thus, with an 
R-transformation one can remove an additional 
phase from the theory, say from one of the soft 
gaugino masses $\tilde{m}_a$. 
The remaining physical 
phases are: one phase for each $\arg(A_f)$ (corresponding to 
a fermion $f$), $\arg(B)$ and $\arg(\tilde{m}_a)$, say for $a=1,2$.   
In the most general MSSM scenario, these remaining complex 
parameters at the $\Lambda_S$-scale cannot simultaneously 
be made real by redefining the phases of fields without 
introducing phases in the other couplings.

Of course, once the above phases are set to their 
$\Lambda_S$-scale values, they feed into the  
SUSY parameters of the theory at the EW-scale through the RGE. 
Instead of studying the RGE for the 
full theory, one needs only consider a complete subset 
of the RGE of the complex parameters in the effective 
theory. Taking only the top and bottom Yukawa couplings 
and neglecting small effects from the other Yukawa couplings, 
such a complete subset was given in \cite{prd55p1611}:

\beqa
\frac{d{\tilde{m}_a}}{dt}&=&2b_a \alpha_a \tilde{m}_a 
\label{firsteq}~,\\
\frac{d A_t}{dt}&=&2 c_a \alpha_a \tilde{m}_a +12 
\alpha_t A_t +2 \alpha_b A_b ~,\\
\frac{d A_b}{dt}&=&2 c^{'}_a \alpha_a \tilde{m}_a +
12 \alpha_b A_b +2 \alpha_t A_t ~,\\   
\frac{d A_{u,c}}{dt} &=& 2 c_a \alpha_a \tilde{m}_a +6 
\alpha_t A_t ~,\\
\frac{d A_{d,s}}{dt} &=& 2 c^{'}_a \alpha_a \tilde{m}_a +6 
\alpha_b A_b ~,\\
\frac{dB}{dt}&=&2 c^{''}_a \alpha_a \tilde{m}_a + 6 
\alpha_b A_b + 6 \alpha_t A_t ~,\\
\frac{d \alpha_t}{dt} &=& 2 \alpha_t (-c_a\alpha_a + 
6 \alpha_t +\alpha_b) ~,\\
\frac{d \alpha_b}{dt} &=& 2 \alpha_b (-c^{'}_a\alpha_a + 
6 \alpha_b +\alpha_t) ~,\\
\frac{d \alpha_a}{dt} &=& 2 b_a \alpha_a^2 \label{lasteq}~,
\eeqa 

\ni where $t \equiv {\rm ln} (Q/\Lambda_S) / 4 \pi$, $a$ is 
summed from 1 to 3 and $b_a=(33/5,1,-3),c_a=(13/15,3,16/3),
c_a^{'}=(7/15,3,16/3),c_a^{''}=(3/5,3,0)$. 
Also, $\alpha_{t}$ and $\alpha_{b}$ are related to the 
corresponding quark masses via

\beq
\alpha_{t(b)} = \frac{g_2^2}{8 \pi} \frac{m_{t(b)}^2}{m_W^2} 
\frac{1}{\sin^2\beta(\cos^2\beta)} ~.
\eeq
 
\ni We remark that in the 
general MSSM framework with 
arbitrary CP-violating phases at $\Lambda_S$, the above RGE are of 
less importance and with such boundary conditions almost any 
low energy CP-violating scenario can be generated. 
In particular, large CP-violating phases
at the EW-scale are not excluded in this unconstrained scenario.
However, in a more constrained SUSY version, one can 
reduce the number of the physical CP-odd phases in the 
theory. In this case the RGE given above are crucial 
for determining the SUSY CP-violating phases at the EW-scale 
\cite{prd55p1611}. We will return to this point later when 
we discuss the $N=1$ minimal low energy SUGRA - GUT model.

Let us consider now
the phenomenological consequences 
of CP nonconservation in such a general SUSY model. 
First we need
to describe briefly how these new CP-violating phases 
enter in reactions which are driven by supersymmetric particles. 
As it turns out, all CP violation in the low energy SUSY vertices 
is driven by diagonalization of the complex mass matrices of the 
sfermions, charginos and neutralinos. For more detailed investigations 
of the diagonalization procedure and extraction of the mass spectrum and 
CP-violating phases from these 
mixing matrices we refer the reader to the existing 
literature, see {\it e.g.}, 
\cite{susyreview,
prd41p4464,npb272p1,prd46p3025,prd49p4908,prl66p2565,prd45p4345,prd57p478}.
Here we only wish to briefly describe the key  features of the   
 formulation and the definitions.   

We denote by $M^2_{\tilde{f}}$ the mass squared matrix 
of the scalar partners of a fermion, and 
$M_{\tilde{\chi}}$ and $M_{\tilde{\chi}^0}$ are the 
mass matrices of the charginos
and neutralinos, respectively. Then, with the rotation 
matrices $Z_f$, $Z_N$, $Z^+$ and $Z^-$, we can define 

\beqa
Z^{\dagger}_f M^2_{\tilde{f}} Z_f
& = & {\rm diag}\left(m^2_{\tilde{f}_1}, m^2_{\tilde{f}_2}\right) \ \ ,\\
(Z^-)^{\dagger} M_{\tilde{\chi}} Z^+
& = & {\rm diag}\left(m_{\tilde{\chi}_1}, m_{\tilde{\chi}_2}\right) \ \ ,\\
Z_N^T M_{\tilde{\chi}^0} Z_N
& = & {\rm diag}\left(m_{\tilde{\chi}^0_1}, m_{\tilde{\chi}^0_2},
m_{\tilde{\chi}^0_3}, m_{\tilde{\chi}_4^0}\right) \ \ .  
\eeqa 

\ni $M^2_{\tilde{f}}$ is then given by 

\beq
%
%
%
%
M^2_{\tilde{f}} = 
\footnotesize
\left(\matrix{
m^2_f - \cos 2 \beta(T_{3f} - Q_f \sin^2 \theta_W)m_Z^2 + m^2_{\tilde{f}_L}
& - m_f(R_f\mu + A_f^*) \cr
- m_f(R_f\mu^* + A_f)
& m^2_f - \cos 2 \beta Q_f \sin^2 \theta_Wm_Z^2 + m^2_{\tilde{f}_R} 
}\right)  %
\normalsize
%
%
%
\label{mfgal} ~,
\eeq  \\

\ni where $m_f$ is the mass of the fermion $f$, 
$Q_f$ its charge and $T_{3f}$
the third component of the weak isospin of a left-handed 
fermion $f$. $m^2_{\tilde{f}_L}
(m^2_{\tilde{f}_R})$ is the low energy mass squared 
parameter for the left (right) sfermion
$\tilde{f}_L(\tilde{f}_R)$.  $R_f = \cot\beta(\tan\beta)$ for
$T_{3f}=\frac{1}{2}(-\frac{1}{2})$ where $\tan\beta = v_2/v_1$ 
is the ratio between the two VEV's of the two 
Higgs doublets in the model.

$M_{\tilde{\chi}}$ and $M_{\tilde{\chi}^0}$ are given 
by

\beq
M_{\tilde{\chi}} = \left(\matrix{
\tilde{m}_2 & \sqrt{2} m_W \sin\beta \cr
\sqrt{2} m_W \cos\beta & \mu } \right) , 
\eeq

\beq
M_{\tilde{\chi}^0} = 
\footnotesize
\left(\matrix{
\tilde{m}_1   & 0 & -m_Z\cos\beta\sin\theta_W & m_Z\sin\beta\sin\theta_W \cr
0 & \tilde{m}_2 & m_Z\cos\beta\cos\theta_W & -m_Z\sin\beta\cos\theta_W \cr
-m_Z\cos\beta\sin\theta_W & m_Z\cos\beta\cos\theta_W & 0 & -\mu\cr
m_Z\sin\beta\sin\theta_W & -m_Z\sin\beta\cos\theta_W & -\mu & 0 } \right) 
\normalsize ~,
\eeq

\ni where $\tilde{m}_1$($\tilde{m}_2$) is 
the mass parameter for the U(1)(SU(2)) gaugino.

Because of their relatively simple form, we will  
discuss below the way of parameterizing the CP-violating 
phases only of the sfermions 
and charginos diagonalizing matrices 
$Z_f$ and $Z^+, \ Z^-$, respectively. 
The diagonalization of the $4 \times 4$ neutralino mixing matrix with 
complex entries  
is more involved and may be estimated numerically, although 
in some limiting cases it may be approximated analytically (see e.g., 
\cite{prd55p1611}). If all elements of $Z_N$ are real then the diagonalization 
procedure can also be done analytically 
(see e.g., \cite{prd49p4908,prd45p4345}).

The mixing matrix of the sfermions 
is parameterized as

\beq
Z_f = \left( \matrix {\cos \theta_f & - e^{-i\beta_f}  \sin\theta_f \cr
 e^{i\beta_f}  \sin\theta_f & \cos\theta_f } \right) \label{zfmatrix} \ ,
\eeq

\ni where $\theta_f$ is the mixing angle and $\beta_f$ 
is the phase responsible for CP-violating phenomena 
in sfermions interactions with other particles in the theory, 
and is given by

\beq
\tan\beta_f = -\frac{|\mu| R_f \sin\alpha_{\mu}-|A_f|\sin\alpha_f}{|\mu| R_f \cos\alpha_{\mu} + |A_f|\cos\alpha_f} \ ,
\label{tanbeta}
\eeq 

\ni where we have used $A_f= |A_f|e^{i \alpha_f}$ and 
$\mu=|\mu|e^{i \alpha_{\mu}}$.
Recall that $R_f = \cot\beta(\tan\beta)$ for 
$T_{3f}=\frac{1}{2}(-\frac{1}{2})$ and $\tan\beta = v_2/v_1$.
Also, the mixing angle $\theta_f$ is given by

\beq
\tan\theta_f = \frac{- 2m_f\left| R_f\mu + A^*_f\right|}{\cos2\beta(2Q_f\sin^2\theta_W - T_{3f})
m^2_Z + m^2_{\tilde{f}_L} - m^2_{\tilde{f}_R}}
\label{tanteta} ~.
\eeq

\ni It is obvious from Eq.~\ref{tanbeta} and \ref{tanteta} that 
in the limit where all the quark masses are small except for $m_t$, only 
the phase of $A_t$ leads to CP-violating effects
(the limit $m_f \to 0$, 
$f \neq t$, is useful when considering high energy reactions). 
In particular, the other $A$-terms are multiplied by 
the light fermion masses (see also Eq.~\ref{mfgal}) and, therefore, 
they have negligible effect on any physical quantity 
evaluated at high enough energies. That is,  
the off-diagonal elements of $Z_f$ are zero in this limit 
(i.e., from Eq.~\ref{tanteta} we see that 
$\sin\theta_f \to 0$ when $m_f \to 0$)
and there is no mixing between 
the left and right
components of the superpartners of light quarks. Of course, 
this is not the case for the NEDM which is 
particularly sensitive to
the slight deviation from degeneracy of the supersymmetric 
partners of the $u$ and the $d$ quarks. We will return to a 
more detailed discussion on the ``SUSY-CP problem'' of the 
NEDM in the following section.

For a sfermion ${\tilde f}$, it is useful to  
adopt a parameterization for its 
${\tilde f}_L - {\tilde f}_R$ mixing such that the sfermions of 
different handedness are related to their mass eigenstates through 
the transformation

\beqa
\tilde f_L & = & Z_f^{11} \tilde f_1 +  Z_f^{12} 
\tilde f_2 ~, \nonumber \\
\tilde f_R & = &  Z_f^{21} \tilde f_1 +  Z_f^{22}
\tilde f_2 \label{recall}~,
\eeqa

\ni where $\tilde f_{1,2}$ are the two mass eigenstates. 
We note that, in the case where 
all CP violation arises from ${\tilde f}_L - {\tilde f}_R$ mixing, 
i.e., from the complex entries in the sfermion mixing matrix $Z_f$, 
it has to be proportional to

\beq
{\mIm} (\xi_f^i) \equiv {\mIm}(Z_f^{1i*}Z_f^{2i}) = \frac{(-1)^{i-1}}{2}
\sin2\theta_f \sin\beta_f \label{xit}~. 
\eeq      

\ni Clearly, $\xi_f^i \to 0$ if there is no mixing between 
the left and right sfermions such that 
they are nearly degenerate.
We will describe in the next chapters CP-nonconserving effects  
in top quark systems which are driven by the possibly large mass splitting 
between the two stop mass eigenstates and are therefore proportional
to ${\mIm} (\xi_t^i)$.   

The charginos mixing matrices are given by

\beq
Z^{\pm}=P^{\pm}O^{\pm} \label{zpo}~,
\eeq 

\ni where

\beqa
P^+ =  \left(\matrix{1 & 0 \cr
0 & -e^{i\alpha_2} } \right) ~~,~~
P^- =  \left(\matrix{e^{i\alpha_2} & 0 \cr
0 & -1 } \right) ~,
\eeqa

\ni and

\beqa
O^+ & = & \left(\matrix{\cos\theta_+ & -e^{-i\beta_+} \sin\theta_+  \cr
e^{i\beta_+} \sin\theta_+ & \cos\theta_+  } \right) \ , \\
O^- & = & \left(\matrix{e^{i\gamma_1}\cos\theta_- 
& -e^{-i(\beta_- - \gamma_2)} \sin\theta_-  \cr
e^{i(\beta_- + \gamma_1)} \sin\theta_- & e^{i\gamma_2} 
\cos\theta_-  } \right) ~.
\eeqa

\ni Here $\alpha_2 = \arg (\tilde{m}_2)$ 
and the CP-violating angles $\beta_{\pm}$ and $\gamma_{1,2}$ 
above are given by

\beqa 
\tan\beta_+ & \equiv & - \frac{\sin\theta}{\cos\theta + 
\frac{|\tilde{m}_2| \cot\beta}{|\mu|} }~, \\
\tan\beta_- & \equiv & \frac{\sin\theta}{\cos\theta + 
\frac{|\tilde{m}_2| \tan\beta}{|\mu|} }~, \\
\tan\gamma_1 & \equiv & - \frac{\sin\theta}{\cos\theta + 
\frac{2 |\tilde{m}_2| (m_{\tilde{\chi}_1}^2 - 
|\mu|^2)}{g^2 |\mu| v_1 v_2} }~, \\
\tan\gamma_2 & \equiv & \frac{\sin\theta}{\cos\theta + 
\frac{g^2 |\tilde{m}_2| v_1 v_2}{2 |\mu| 
(m_{\tilde{\chi}_2}^2 - |\tilde{m}_2|)} } \label{tgam2}~,
\eeqa 
 
\ni where $\theta \equiv \alpha_{\mu} + \alpha_2$ and 
$m_{\tilde{\chi}_i}$ ($i=1,2$)
 are the masses of the two charginos. 
$\theta_+$ and $\theta_-$ are also functions of 
$\tilde{m}_2,\mu,v_1$ and $v_2$ (see \cite{prd46p3025}).

\subsubsection{CP violation in a GUT-scale $N=1$ minimal SUGRA model 
\label{sssec333}}

Let us proceed by describing a more constrained supersymmetric 
model. 
In particular, 
we want to consider a spontaneously broken $N=1$ SUGRA, 
which apart from gravitational interactions, is essentially identical at
low  energies to a theory with  softly broken supersymmetry with GUT
motivated relations at the GUT mass scale.
One of the most appealing consequences of such 
a constrained SUSY scenario is that it allows 
REWSB of   
SU(2)$\times$U(1) with the fewest number of free parameters. 

According to conventional wisdom, complete 
universality of the soft supersymmetric parameters at 
the GUT-scale (or at the scale where the SUSY soft breaking 
terms are generated) is assumed.   
More explicitly, a common scalar mass $m_0$ and a common 
gaugino mass $M_{1/2}$ 
at the GUT-scale

\beqa
m^2_{H_1} & = & m^2_{H_2} = m^2_R = m^2_L = m^2_Q = 
m^2_U = m_D^2 = m_0^2 \ \ ,\nonumber \\
\tilde{m}_a & = & M_{1/2} \ \ , \ \ a=1,2 \ {\rm or} \ 3 ~, 
\eeqa 

\ni and also universal boundary conditions for 
the soft breaking trilinear terms are assumed

\beq
A_E = A_D = A_U \equiv A^G \label{universala}~.
\eeq

\ni Of course, the above relations do not survive 
after renormalization effects from the GUT-scale (which is usually taken
to be  $M_G \sim 2\times10^{16}$ GeV) to the
EW-scale are included.

It then follows that the universal parameters
of the minimal SUGRA model at the GUT-scale are:
$m_0, \ M_{1/2}, \ A^G, \ B^G, \ \mu^G$ and tan$\beta$.  
This is the most general set of independent parameters before REWSB.
However, a bonus of this economical framework is that 
REWSB occurs and the parameters $\mu^G$ and $B^G$ are no longer taken as
independent but are set by $m_0$, $M_{1/2}$ and 
$\tan\beta$ (the magnitude of $\mu$ is 
adjusted to give the appropriate $Z$-boson 
mass but the sign of $\mu$ remains as an independent parameter).
Thus, the number of independent 
parameters is reduced to five, namely
$m_0$, $M_{1/2}$, $A^G$, sign($\mu$) and $\tan\beta$. 

In this GUT motivated SUGRA theory there
are four possible new sources of CP nonconservation at the GUT-scale.
 These are \cite{npb255p413}: the universal  
trilinear coupling $A^G$, the Higgs mass parameter $\mu^G$, the gauge mass 
parameter $M_{1/2}$ and the parameter $B^G$. However, 
as was mentioned before, $M_{1/2}$ can be made real by an
$R$-transformation and by using one remaining phase freedom,  
a redefinition of the Higgs fields can set the
product $B^G \mu^G$ to be real so that the VEV's of the two 
Higgs fields in the theory are also made real.
We therefore have $\arg(\mu^G) = - \arg(B^G)$.
With this choice we are left with only 
two new SUSY CP-odd phases at the GUT-scale which 
are carried by $A^G$ and $\mu^G$, apart from 
the usual CKM phase that originates from the Yukawa 
couplings in the theory for three generations.

One can proceed by choosing a more constrained CP-violating 
sector by setting one of these two phases to zero 
(we will address to this possibility in the next section), 
or even a ``super'' constrained CP-violating sector by taking 
$\arg(\mu^G)=\arg(A^G)=0$, thus being left (at the GUT-scale) only with 
the usual CKM phase present as in the SM\null.
  
Note that having a universal phase for all trilinear couplings of  
sfermions at the GUT-scale, does not necessarily mean that all 
sfermions will
have the same phase 
(driven by their trilinear coupling $A_f$) at the EW-scale. 
That is, the GUT-scale phases $\arg(A^G)$ and $\arg(\mu^G)$ 
feed into the other parameters of the theory through renormalization 
effects from the GUT-scale to the EW-scale. In particular, 
these GUT-scale phases can produce different phases for 
different squarks and 
sleptons of different generations. However, irrespective of 
what those different phases are at the EW-scale, 
they can all be expressed in principle with the two 
new CP-odd phases of $A^G$ and $\mu^G$ through the 
RGE in Eqs.~\ref{firsteq}-\ref{lasteq}.
Thus, it is evident  
from the very simple structure of the 
evolution equations of the gaugino masses 
(see Eq.~\ref{firsteq}) that     
${\tilde m}_a$, $a=1-3$, are left with no phase at any 
scale. 
Moreover, the difference between the
three real low energy gaugino mass parameters comes from 
the fact that they undergo
a different renormalization as they evolve from the GUT-scale to the 
EW-scale, due to the different gauge structure of their interactions. 
In particular, they are related, at the EW-scale, by 
(see e.g., \cite{prd49p4908})

\beq
\frac{3\cos^2\theta_W}{5}\frac{{\tilde m}_1}{\alpha}=\sin^2\theta_W
\frac{{\tilde m}_2}{\alpha}=\frac{{\tilde m}_3}{\alpha_s} \label{mhalf}~,
\eeq

\ni where $\theta_W$ is the weak mixing angle and ${\tilde m}_3$ is
the low energy gluino mass. 

Notice now that with $\arg({\tilde m}_a)=0$, the RGE 
(Eqs.~\ref{firsteq}-\ref{lasteq}) simplify to a large extent, 
thus we have only to consider the evolution of the 
$A_f^G$'s and $B^G$ (or equivalently $\mu^G$) to the EW-scale. 
This constrained version of the MSSM has strong implications on 
the low energy phase in $A_t$ as was 
suggested in \cite{prd55p1611}. 
In particular, it was shown  
that with and without universal trilinear couplings at 
the GUT-scale and with some definite boundary 
conditions for them (for example, 
$\arg(\mu^G)=\arg(A_f^G)=0$ for fermion species $f$), 
the low energy CP-violating phase of $A_t$ induces 
potentially large phases in $A_u,A_d$ and $B$ at the 
EW-scale (through relations obtained from the above set of 
the RGE). This in turn gives rise to a large NEDM which 
is ruled out by the present experimental limit. 
Therefore, severe constraints on the low energy phases of 
$A_f$ and $\mu$ where obtained.
We will return to this point in the next section.          
 
\subsubsection{A plausible low energy MSSM framework 
and the ``SUSY-CP problem'' of the NEDM \label{sssec334}}

The EDM of the neutron, $d_n$, imposes
important phenomenological constraints on
SUSY models 
\cite{npb255p413,prd55p1611,prd46p3025,
prl66p2565,prd57p478,nedmreview,npb52ap78,constraints}.
In particular, with a low energy MSSM that 
originates from a GUT-scale SUGRA model 
(with complete universality of the soft 
breaking terms), keeping $d_n$ within its allowed 
experimental value (i.e., 
$|d_n| \lsim  10^{-25}$ e-cm \cite{prl82p904}) requires 
the ``fine tuning'' of the SUSY phases to be less than or of 
the order of $10^{-2}-10^{-3}$ for SUSY particle masses of 
the order of the EW symmetry breaking scale. We remark, though, that it has  
recently been claimed in \cite{prd57p478},  that 
in some regions of the SUSY parameter space, cancellations
among the different components of the neutron EDM may occur and 
such a severe fine tuning for either the SUSY masses or 
the SUSY CP-violating 
phases (i.e., at the order of $10^{-2}$) may not be necessary.
However, for such cancellations to occur SUSY parameters
have to be suitably arranged and, also, several large SUSY phases 
have to be present, which renders this scenario
less predictive and less attractive as well.

The NEDM can be written schematically in any low 
energy MSSM scenario in which the CP-phases originate from the 
$\mu$ and the $A_f$ terms as \cite{prd55p1611}:

\beq
\frac{d_n}{10^{-25}~{\rm e-cm}} = X^{\mu} 
\frac{|\mu|}{M_S}\sin\alpha_{\mu} + X^{A_u} 
\frac{|A_u|}{M_S}\sin\alpha_u + X^{A_d} \frac{|A_d|}{M_S}\sin\alpha_d ~,
\eeq

\ni where $M_S$ is the typical SUSY mass scale 
which may be used to describe the typical squark masses. 
It was found in \cite{prd55p1611} that typical values 
of the $X^i$'s in almost every such low energy SUSY realization 
are: $X^{A_u,A_d} > {\cal O}(1)$ and $X^{\mu} > {\cal O}(10^2)$ 
for $M_S \lsim 500$ GeV. Therefore, with\footnote{Since $M_S$ is the only SUSY 
mass scale associated with the squarks sector, it is 
natural to choose the soft breaking terms 
to be of ${\cal O}(M_S$).} 
$|A_f| \approx |\mu| \approx M_S$, 
only moderate bounds can be put on 
$\sin\alpha_u$ and $\sin\alpha_d$. In contrast, an 
unambiguous severe constraint is obtained on the low 
energy phase of the Higgs mass term $\mu$, namely 
$\sin\alpha_{\mu} < {\cal O}(10^{-2})$ for 
$M_S \lsim 500$ GeV 
(see also \cite{hepph9701231,prd46p3025,npb52ap78} and references therein). 

The important finding in \cite{prd55p1611} is that if 
a complete universality of the soft breaking terms 
is imposed at the GUT-scale and the two GUT phases 
are zero or very small, i.e., 
$\arg(A^G) \approx \arg(B^G) \lsim 0.1$,  
then this severe constraint on $\sin\alpha_{\mu}$ combined 
with the relations between \mIm($A_t$) and \mIm($\mu$), 
obtained from the RGE, leads to the comparable constraint 
$\sin\alpha_t< {\cal O}(10^{-2})$. 
With a universal $A$ term at the GUT-scale 
this will also imply (through relations obtained from the RGE) 
$\sin\alpha_{u,d}< {\cal O}(10^{-2})$. 
Moreover, the above strong constraints on the SUSY CP-violating 
phases hold even if the universality of the $A$ terms 
is relaxed at the GUT-scale as long as the GUT-scale 
CP-violating phases are kept very small. 
This strong constraint on the low energy phase of $A_t$ 
would also eliminate any possible SUSY CP-nonconserving 
effect in top quark systems.

While this scenario with very small CP-violating phases at the GUT-scale 
provides  an explanation of the smallness of the NEDM 
(in fact, the above constraints will drive the NEDM to a 
value of the order of $10^{-27}-10^{-28}$ e-cm), 
an unavoidable question then arises: 
why do the CP-violating phases happen to be so small 
wherever they appear? If so, then an underlying 
theory that screens the CP-violating phases is required. 

We therefore 
feel that a 
somewhat different phenomenological approach is needed, 
namely, the implication that
$\sin\alpha_t< {\cal O}(10^{-2})$ 
should be specially scrutinized in the top quark sector. 
The latter, being very sensitive to $\sin\alpha_t$ at high 
energies, may serve as a unique probe for searching for 
significant deviations from the above upper bound, 
$\sin\alpha_t< {\cal O}(10^{-2})$ (see e.g., \cite{prd57p1495}). 
Indeed, if at the GUT-scale the universality of the $A$ terms is relaxed 
and no assumption is made on the magnitude 
of the CP-violating phases, then  
one can construct a plausible low energy MSSM framework which 
incorporates an ${\cal O}(1)$ low energy phase for $A_t$, 
while leaving the NEDM within its experimental limit. 
The crucial difference in assuming non-universal boundary 
conditions for the soft breaking trilinear $A$ terms is that, 
in this case, there is no {\it a-priori} reason to believe that 
the low energy phases associated with the different $A_f$ terms 
are related at the EW-scale. In particular, the bounds on 
$\sin\alpha_{u,d}$ obtained from the experimental limit of the 
NEDM may not be used to constrain $\sin\alpha_t$.
In addition, when no further assumption is made on the magnitude 
of the CP-violating phases at the GUT-scale, 
a value of $\sin\alpha_{\mu} \to 0$ may be realized without 
contradicting any existing relation from the RGE. 

As in \cite{prd57p1495} we therefore take the following phenomenological
point of view  in constructing a plausible low energy set of the MSSM 
CP-violating phases and  mass spectrum:
\begin{enumerate}
\item $\sin\alpha_{\mu} \to 0$ as strongly implied from the analysis 
of the NEDM.
\item $\sin\alpha_u$, $\sin\alpha_d$ and $\sin\alpha_t$ 
are not correlated at the EW-scale, which implicitly 
assumes non-universal boundary conditions at the GUT-scale.
In particular, $\sin\alpha_u,~\sin\alpha_d$ may then be constrained 
only from the NEDM experimental limit, with no implications on the size 
of $\sin\alpha_t$. 
\item Motivated by the theoretical prediction of the unification
of the SU(3), SU(2) and U(1) gauge couplings when SUSY particles with a mass
scale around 1 TeV are folded into the RGE,
one may follow only the following 
traditional simplifying GUT assumption: there
is an underlying grand unification. As mentioned before, 
this leads us to have a common gaugino
mass parameter defined at the GUT-scale which can be 
made real by an R-rotation. Thus, 
${\tilde m}_a$, $a=1-3$ are left with no phase at any 
scale. 
Moreover, using the relation of Eq.~\ref{mhalf}, once the gluino mass 
is set at the 
EW-scale, the SU(2) and
U(1) gaugino masses ${\tilde m}_2$ and ${\tilde m}_1$, 
respectively, are determined.
\item The typical SUSY-scale is $M_S$ and all the squarks except 
for the light stop, are assumed to be
degenerate with mass $M_S$. 
\end{enumerate}

Note that in this low energy framework one is only left with 
the phases of the various $A_f$ terms at the 
EW-scale, out of which only $A_t$ plays a significant 
role in any high energy reaction. For $m_f \to 0$ 
the superpartners of the light quarks are practically 
degenerate and therefore the CP-violating effects from the phases 
of the other $A_f$ terms, corresponding to the light quarks, 
can be safely neglected. Note again that 
our approach, the EW$\to$GUT approach, assumes a set 
of phases at the EW-scale, subject to existing experimental 
data, which implicitly assumes arbitrary phases at the scale 
in which the soft breaking terms are generated.
As was mentioned above, this low energy 
CP violation scenario can naturally arise from a GUT-scale SUGRA model if 
only the universality of the $A$ terms is relaxed and no 
assumption is made on the magnitude of the GUT-scale phases. 

Also, the mass matrices of the neutralinos, 
$M_{{\tilde \chi}^0}$, and
charginos, 
$M_{{\tilde \chi}}$, depend on the low energy Higgs mass 
parameter $\mu$,
the two gaugino masses ${\tilde m}_2$ and ${\tilde m}_1$ (which are resolved
by the gluino mass) and  $\tan\beta$ (see previous sections) and are 
therefore real in this scenario.  
Thus, once
$\mu$, ${\tilde m}_3$ and $\tan\beta$ are set to their low energy values, 
the four physical
neutralino 
species $m_{{\tilde \chi}_n^0}$ ($n=1-4$) and the two physical 
chargino species
$m_{{\tilde \chi}_m}$ ($m=1,2$) are extracted by 
diagonalizing the real matrices $M_{{\tilde
\chi}^0}$ and $M_{{\tilde \chi}}$.
\begin{description} 
\item 5. \hspace*{0.45em} Finally, the resulting SUSY 
mass spectrum may now be subject 
to the existing experimental limits, i.e., limits on the masses of 
squarks, gluino, neutralinos, 
charginos, etc. (see e.g., \cite{susylimits}). 
\end{description}

With these assumptions, when $\arg(\mu) \to 0$, 
the leading contribution to a light quark EDM comes from gluino 
exchange, which with the approximation of degenerate $\tilde u$ 
and $\tilde d$ squark masses (which we will denote by 
$m_{\tilde q}$), can be written as \cite{prd46p3025}:

\beq
d_q(G)=\frac{2 \alpha_s}{3\pi} Q_qe m_q 
\frac{|A_q|\sin\alpha_q}{m_{\tilde q}^3} {\sqrt r} K(r) \label{qedm} ~,
\eeq  

\ni where $m_q$($m_{\tilde q}$) is the quark(squark) 
mass and $Q_q$ is its charge. Also, $r \equiv m_G^2/m_{\tilde q}^2$ 
(for the rest of this 
section we denote the gluino mass by $m_G$) and $K(r)$ is 
given by\footnote{Note that the function $K(r)$ in Eq.~\ref{kr} is slightly
different from that obtained in \cite{prd55p1611}. 
However, we find that numerically
the difference is insignificant and does not change our predictions
below.} 

\beq
K(r)= \frac{1}{(r-1)^3}\left( \frac{1}{2}+\frac{5}{2}r + 
\frac{r(2+r)}{1-r}{\rm ln}r \right) \label{kr} ~.
\eeq 

\ni Then, within the naive quark model, the NEDM can be obtained by
relating it to the $u$ and $d$ quarks EDM's 
(i.e., $d_u$ and $d_d$, respectively) as

\beq
d_n=(4d_d-d_u)/3 ~. 
\eeq

\begin{figure}
\psfull
 \begin{center}
  \leavevmode
\epsfig{file=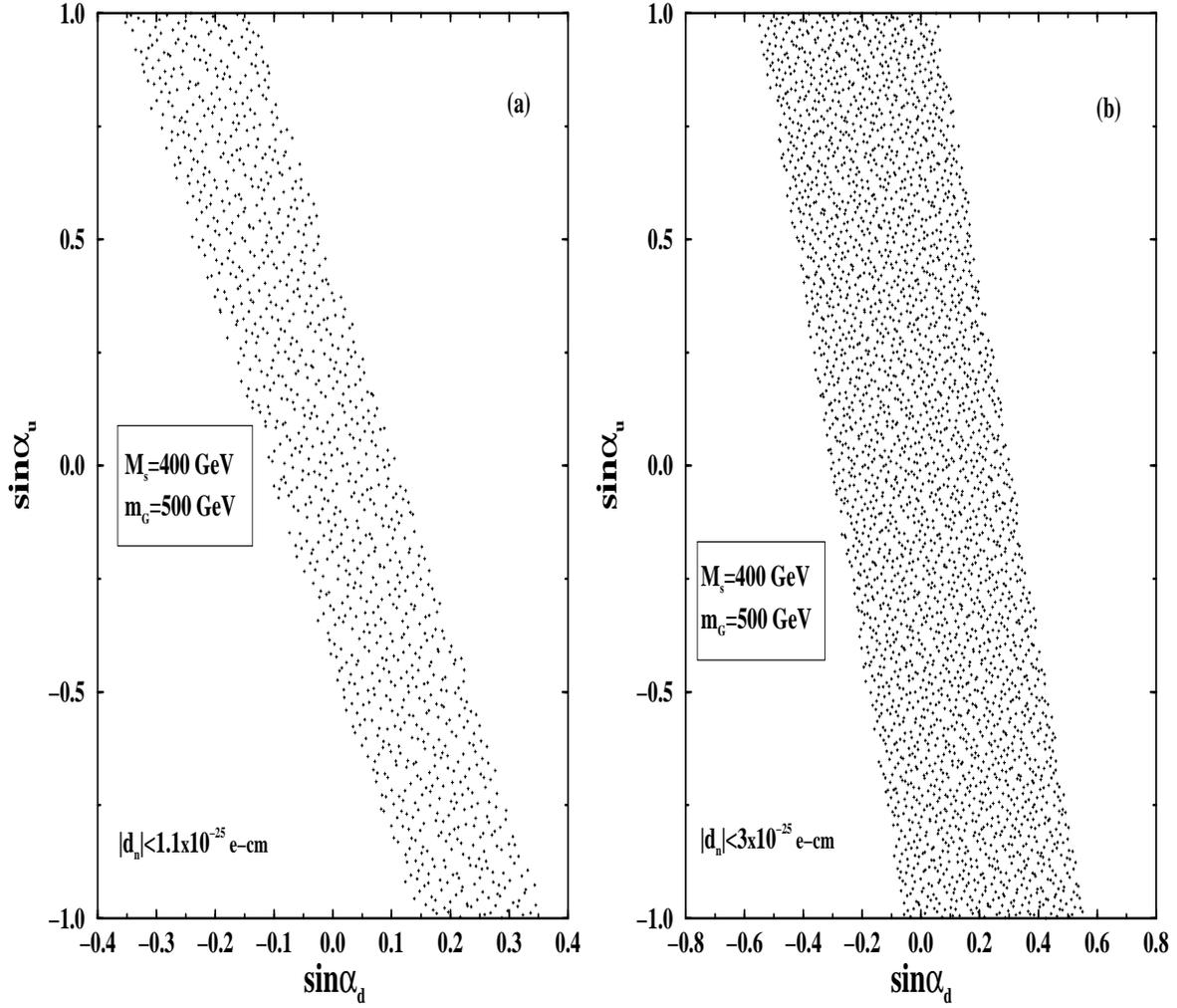
,height=13cm,bbllx=0cm,bblly=2cm,bburx=20cm,bbury=25cm,angle=0}
 \end{center}
\caption{\emph{The allowed regions in the $\sin\alpha_u - \sin\alpha_d$ plane
for the NEDM not to exceed (a) $1 \times 10^{-25}$ e-cm and (b) $3\times
10^{-25}$ e-cm. $M_S=400$ GeV and $m_G=500$ GeV is used. The shaded areas
indicate the allowed  regions. Figure taken from \cite{prd57p1495}.}} 
\label{modelsfig6}
\end{figure}

\ni We now consider $\arg(A_u)$ and $\arg(A_d)$ to be free
parameters 
of the model irrespective of $\arg(A_t)$.  
In Figs.~\ref{modelsfig6}(a) and \ref{modelsfig6}(b) we have plotted
the allowed regions in the $\sin\alpha_u - \sin\alpha_d$ plane for $|d_n|$
not to exceed $1 \times 10^{-25}$ e-cm (for the present experimental limit 
see \cite{prl82p904}) 
 and $3\times 10^{-25}$ e-cm. In calculating $d_n$
we assumed that the above naive quark model relation holds.
Although there is no doubt
that it can serve as a good approximation for an order of magnitude
estimate, 
it may still deviate from the true theoretical value which involves
uncertainties in the calculation  of the corresponding hadronic matrix 
elements. 
Note also that it was 
argued in \cite{plb377p83} 
that the naive quark model overestimates the NEDM, as the strange
quark may carry an appreciable fraction of the neutron spin which  can partly
screen the contributions to the NEDM coming from the $u$ and the $d$ quarks. 
To be on the safe side, we therefore slightly relax the
theoretical limit on $d_n$ in Fig.~\ref{modelsfig6}(b)
 to be $3\times 10^{-25}$ e-cm.   

We have used, for these plots, $m_{\tilde d}=m_{\tilde u}=M_S=400$ GeV,
$m_G=500$ 
GeV and for simplicity we also took $|A_u|=|A_d|=M_S$. As remarked before,
it is only natural to choose the mass scale of the
soft breaking terms according to our typical SUSY mass scale $M_S$. 
Also, we took the values for current quark
masses as $m_d=10$ MeV, $m_u=5$ MeV and $\alpha_s(m_Z)=0.118$.   

From Fig.~\ref{modelsfig6}(a) and in particular Fig.~\ref{modelsfig6}(b), 
it is evident that $M_S=400$ GeV
and $m_G=500$ GeV can be safely assumed, leaving ``enough room'' in the
$\sin\alpha_u 
- \sin\alpha_d$ plane for $|d_n|$ not to exceed $1 - 3 \times 10^{-25}$ e-cm.
We observe that while $\sin\alpha_u$ is basically not constrained,
$-0.35 \lsim \sin\alpha_d \lsim 0.35$ is needed for $|d_n|
< 1 \times 10^{-25}$ e-cm and $-0.55 \lsim \sin\alpha_d \lsim 0.55$ is
needed for $|d_n| < 3 \times 10^{-25}$  e-cm. Moreover, varying $m_G$ between
250 GeV to 650 GeV has almost no effect on the allowed areas in the
$\sin\alpha_u 
- \sin\alpha_d$ plane that are shown in Figs.\ref{modelsfig6}(a) 
and \ref{modelsfig6}(b).\footnote{We will take $M_S=400$ GeV and 
vary $m_G$ in this range in some 
of the CP-violating effects in collider experiments to be discussed 
in the following chapters.} 
That is, keeping
$M_S=400$ GeV and lowering $m_G$ down to 250 GeV, very slightly shrinks the
dark areas in Figs.\ref{modelsfig6}(a) and \ref{modelsfig6}(b), whereas, 
increasing $m_G$ up to 650 GeV slightly
widens them. Of course, $d_n$ strongly depends on the scalar mass $M_S$ -
increasing $M_S$ enlarges the allowed regions in 
Figs.\ref{modelsfig6}(a) and \ref{modelsfig6}(b) as expected
from Eq.~\ref{qedm}. It is also very interesting to note that,
in some instances, for a cancellation between the contributions of the $u$
and $d$ quarks to occur, $\sin\alpha_u,\sin\alpha_d > 0.1$ is essential rather
than being just possible. For example, with $|\sin\alpha_u| \gsim 0.75$,
$|\sin\alpha_d| 
\gsim 0.1$ is required in order to keep $d_n$ below its experimental limit.

We can therefore conclude that CP-odd phases in the $A_u$ and $A_d$ 
terms of the order
of ${\rm few} \times  10^{-1}$ can be accommodated without too much difficulty
with the existing experimental constraint on the NEDM even for typical SUSY
masses of $\lsim 500$ GeV\null. Therefore, we restate what is emphasized in 
\cite{prd57p1495}: somewhat in contrast to the commonly
held viewpoint we do not find that a ``fine-tuning''
at the level of $10^{-2}$ is necessarily required for the SUSY 
CP-violating phases
for squark masses of a few hundreds GeV or slightly heavier. 

\subsubsection{CP and the pure Higgs sector of the MSSM \label{sssec335}} 

Since the superpotential is required to be a 
function of only left (or only right) chiral superfields, 
it forbids the appearance of ${\hat H}_1^*$ and ${\hat H}_2^*$ in 
the superpotential $W$ in Eq.~\ref{superw}. 
Therefore, since 
a $\hat{Q}^i_I \hat{H}^j_1 
\hat{U}_J$ coupling in $W$ is prohibited
by gauge invariance,
only $H_2$ is responsible for giving mass to up 
quarks and $H_1$ to down quarks \cite{npb272p1}.
As a consequence, the requirement that there will be 
no ``hard'' breaking terms of the symmetry 
$\Phi_i \to - \Phi_i$ in the Higgs potential is 
automatically satisfied in a minimal supersymmetric 
model. That is, $\lambda_6=\lambda_7=0$ in Eq.~\ref{mhdmeq1}, 
for the Higgs potential 
in the MSSM\null. Moreover, no term of the form

\beq
( \Phi_1^{\dagger} \Phi_2 )^2 +{\rm h.c.} ~,
\eeq 

\ni appears in the Higgs potential of the 
MSSM \cite{npb272p1}, thus implying 
$\lambda_5=0$ in Eq.~\ref{mhdmeq1}. It 
is then straightforward to observe that with the above constraints on the 
pure Higgs sector of the MSSM, any phase which 
may appear in a complex soft breaking parameter (i.e., 
proportional to $\mu_{12}^2$ in Eq.~\ref{mhdmeq1}) 
can be removed by a redefinition 
of one of the Higgs doublet fields, thus also setting 
the relative phase between the two VEV's 
to zero. Therefore, the pure Higgs sector in the MSSM 
possesses no CP violation \cite{npb272p1}.
Of course, CP violation may emerge in interactions of 
the Higgs fields with the other fields in the theory 
due to the CP-violating phases carried by these latter fields.    
\pagebreak
\section{Top dipole moments \label{sec4}}
\setcounter{equation}{0}

\subsection{Theoretical expectations \label{ssec41}}

A non-vanishing value for the EDM of a fermion
is of special interest as it signifies the presence of CP-violating
interactions. We recall that the search for the EDM of the neutron and
that of the electron have intensified in recent years. 
Since the top is such an unusual fermion, in fact so
heavy that it is very unlikely to exist as a bound state with another
quark, it is clearly important to ask: What is its EDM? How can we
measure it, if at all? This topic has been of interest to many for the
past several years.

In the SM quarks cannot have an EDM at least to three loops 
\cite{nedmreview}. For the
electron the three loop contribution has been estimated to be
$d_e^{\gamma}(0)\sim 10^{-37}$ e-cm. Simple dimensional 
scaling then suggest for
the top the value $d_t^{\gamma}(0)\sim 10^{-31}$--$10^{-32}$ 
e-cm, much too
small to be observable.

In contrast, in extensions of the SM, e.g., MHDM's and SUSY models, 
this situation changes sharply and the Top 
Dipole Moment (TDM)\footnote{Strictly 
speaking, the term dipole moment refers to the static form factor (i.e., 
at $q^2=0$). Here we will mostly concern ourselves with the dipole moment 
form factor at $q^2 \neq 0$; for simplicity, we will still use the term
TDM throughout.}$^{,}$\footnote{We 
will use the abbreviation TDM in general for $d_t^{\gamma,Z,g}$ denoting  
the top quark EDM, weak-EDM (ZEDM) and Chromo-EDM (CEDM), 
unless we need to explicitly separate the type.}
can arise at the 1-loop level and as a result, the typical 
TDM is of the order of $10^{-18}-10^{-20}$ e-cm which is larger 
than the SM prediction by more than 10 orders of magnitude. 
The enhancement due to the large top mass is particularly evident
in some models with an extended Higgs sector for which the 
dipole moments often 
scale as $m^3_f$. Since at 1-loop light quarks (or neutron)
in these models can get dipole moment of order $10^{-26}$ e-cm, the
TDM could easily reach $10^{-20}$ e-cm or even more. 
It is at
that level that measurable consequences can arise.

Because of the unique importance of the top quark it should be
clear that measurements of the TDM will be extremely important. However,
it should also be clear that due to the extraordinary short life-time
of the top quark ($\lsim 10^{-24}$ sec) it will be extremely difficult
to actually measure the static (i.e., at $q^2=0$) TDM\null. Measurements
of some of the effects driven by the presence of a dipole moment form
factor may have a better chance. In fact, the TDM may be considered as a 
CP-odd form factor in the $\gamma t \bar t$, $Zt \bar t$ or $g t \bar t$ 
vertex that probes the interactions of a short-lived top quark with an 
off-shell $\gamma$,$Z$ or a gluon, respectively, and can be represented by

\begin{eqnarray}
\Gamma_{\mu}^{\gamma,Z,g} = id^{\gamma,Z,g}_t(s) \bar u_t(p_t) 
\sigma_{\mu \nu} \gamma_5 p^{\nu} v_{\bar t}(p_{\bar t}) \label{tdeq1}~,
\end{eqnarray}

\noindent where $s=q^2$, $q=p_t+p_{\bar t}$ 
and color indices for $\Gamma_{\mu}^{g}$ 
were suppressed. 
Therefore, depending on the masses in the loops, 
the TDM 
form-factor
can also develop an imaginary part (contrary 
to the static EDM of the electron or the quark) if 
the energy of the off-shell $\gamma$, $Z$ or gluon 
is sufficient for the particles in the loop to be 
on-shell, i.e., there is an absorptive cut.      

Since the dipole moment characterizes the effective coupling between the
spin of the fermion and the external gauge field, to extract the dipole
moment one needs information on the spin polarization of the top quark.
Fortunately, the left handed nature of the weak decays of the top allows
us to determine its polarization quite readily.  As discussed in section
\ref{ssec28}, top quark decays can analyze the initial polarization of the
top quark. For instance in the leptonic decay $t\to e^+ \nu_e b$, in
the rest frame of the top quark the top is 100\% polarized in the
direction of the $e^+$ momentum. This greatly simplifies calculation of
the top quark production followed by its subsequent decay. The problem is
then essentially reduced to calculating the production of a polarized top
quark. It is then straightforward to fold in the decay to the spin indices
of the top quark.  A serious limitation to be kept in mind about this
procedure is that it is only valid when the decays are governed by the SM,
since they assume the helicity structure of the SM. If non-standard
interactions make large contributions to the decay, the decay
distributions may be modified to the point that the polarimetry we have
discussed is only approximate. This point must be borne in mind when
considering the effects of new physics.

A detailed discussion of the feasibility of extracting the TDM in future
collider experiments such as $e^+e^- , pp \to t \bar t$, will be given in
subsequent sections.  In this section we consider the contribution to the
TDM which arises in extended Higgs sectors and SUSY models.


%
%

\subsection{Arbitrary number of Higgs doublets and a CP-violating 
neutral Higgs sector \label{ssec42}}

It is instructive to calculate the TDM in models with an arbitrary
number of Higgs doublets and singlets satisfying NFC (natural flavor
conservation) constraints. 
CP violation 
arises as a result of
scalar exchanges between quarks, being driven by the imaginary parts of
the complex quantities (e.g., $\wtZ_{1n}$) defined as \cite{prd42p860}:

\beqa
\frac{1}{(v_1)^2} \langle \phi^0_1\phi^0_1\rangle_q & \equiv &  \sum_n
\frac{\sqrt{2} G_F \wtZ_{1n}}{q^2 -m^2_{H_n}} \label{tdeq4} ~,\\
\frac{1}{(v_1v^\ast_2)^2} \langle \phi^+_2\phi^{{}^\ast+}_1\rangle_q 
& \equiv& \sum_n \frac{\sqrt{2} G_F Z_n}{q^2-m^2_{H^\prime_n}} 
\label{tdeq5}~,
\eeqa

\noindent where $v_1$ and $v_2$ are the VEV's of the neutral Higgs fields
$\phi^0_1$, $\phi^0_2$ and the summation runs over all the mass
eigenstates of neutral or charged scalars in the theory ($H_n$ or
$H_n^\prime$, respectively). Also, $\langle\chi\eta\rangle_q$ stands, for
any pair of scalar fields, $\chi$ and $\eta$, for the momentum dependent
quantity

\beq 
\langle \chi\eta\rangle_q = \int d^4x \langle
0|T[\chi(x)\eta(0)]|0 \rangle e^{-iqx} \label{tdeq6}~.
\eeq

\begin{figure}
\psfull
 \begin{center}
  \leavevmode
  \epsfig{file=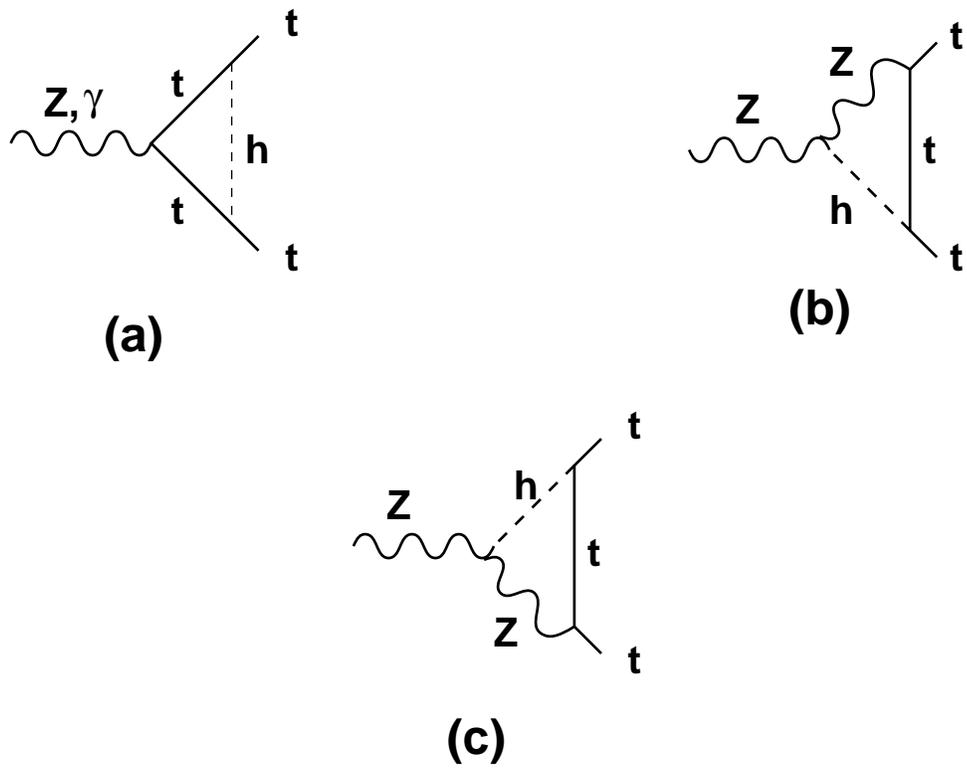,height=10cm}
 \end{center}
\caption{\emph{Feynman diagrams that contribute to the electric and 
weak top dipole moments in a two Higgs doublets model with CP-violating 
interaction of a neutral Higgs ($h$) with a top quark.}}  
\label{topdipolefig1}
\end{figure}

\noindent CP violation in the neutral Higgs sector then generates the
dominant 1-loop contribution to the EDM of the top, i.e., $d_t^\gamma$,  
through the 1-loop graph with the external photon line 
in Fig.~\ref{topdipolefig1}(a) (for this discussion, 
$h\equiv H_n$ in Fig.~\ref{topdipolefig1}(a)). 
This contribution is given by

\beqa
d_t^{\gamma}(0) & = & \frac{2\sqrt{2}}{3(4\pi)^2} G_F m_te\sum_n \Im{\rm m}
\wtZ_{1n} f \left[ \frac{m^2_{H_n}}{m^2_t}\right] \nonumber \\
& = & (1.4 \times 10^{-21} {\rm e-cm}) \left( \frac{m_t}{\rm GeV} 
\right) \sum_n \Im{\rm m}
\wtZ_{1n} f \left[\frac{m^2_{H_n}}{m^2_t} \right] \label{tdeq7}~,
\eeqa

\noindent where

\beq
f(r) = \left\{ 
\begin{array}{l}
1-\frac{r}{2} \ln r + \frac{r^2-2r}{\sqrt{r(4-r)}} \left[\arctan \left(
\frac{2-r}{\sqrt{r(4-r)}} \right) + \arctan \left( \frac{r}{\sqrt{r(4-r)}}
\right)\right] ~,~ {\rm if}\;r<4 \\
3-4 \ln 2 ~,~ {\rm if}\;r=4\\
1-\frac{r}{2} \ln r - \frac{r^2-2r}{\sqrt{r(r-4)}} \ln \left[
\frac{\sqrt{r} -\sqrt{r-4}}{2} \right] , {\rm if}\; r>4
\end{array}
\right. \label{tdeq8}
\eeq

\noindent and $r=m^2_{H_n}/m^2_t$ for any value of $n$. For
$r >> 4$, $f(r)$ approaches 
$\frac{1}{r} \left( \ln r - \frac{3}{2}\right)$
asymptotically. The $\wtZ_{1n}$'s satisfy some important sum
rules, for example \cite{prd42p860}:

\beq
\sum_n \Im{\rm m}\;\wtZ_{1n} = 0 \label{tdeq9}~,
\eeq

\noindent so that $d_t^\gamma$ will vanish if all the neutral Higgs-bosons
were degenerate; no such degeneracy is of course expected. For
illustrative purposes let us assume that the lightest neutral 
Higgs-boson, with
mass $m_h$, dominates the sum in Eq.~\ref{tdeq7}. Taking $m_h=100$ GeV and 
$m_h=2m_t$
($m_t=175$ GeV) and setting $\Im{\rm m}Z$'s to be of order unity, 
then $d_t^{\gamma}$ is about
$1.3 \times 10^{-19}$ e-cm and $5.6 \times 10^{-20}$ e-cm, respectively.
We note that for $m_h>m_t$, $d_t^{\gamma}(0)$ varies slowly with
$m_h$ \cite{prl69p33}.

For experimental purposes the top EDM at high $q^2$ may be more relevant.
This is given by \cite{prl69p33}:

\beq
d_t^{\gamma}(q^2) = \frac{2\sqrt{2}}{3(4\pi)^2} 
G_F m_te \sum_n \Im{\rm m} \wtZ_{1n}
f \left[ \frac{m^2_{H_n}}{m^2_t} , \frac{q^2}{m^2_t} \right]
\label{tdeq10} ~,
\eeq

\noindent where

\beq
f(r,s) = \int^1_0 dx \int^{1-x}_0 dy \frac{x+y}{(x+y)^2 + (1-x-y) r
-xys-i\epsilon} \label{tdeq11}~.
\eeq

\noindent For $q^2>4m^2_t$, $d_t^{\gamma}(q^2)$ develops an imaginary part. 
In the following section we will present explicit 
numerical results for the real and the imaginary parts 
of $d_t^{\gamma}(q^2)$ 
and $d_t^Z(q^2)$ in a 2HDM
with CP violation in neutral Higgs exchanges.
In this case, 
the top CEDM is immediately obtained by replacing 
the photon with a gluon in Fig.~\ref{topdipolefig1}(a) and, therefore, 
$\frac{2}{3}e$ with $g_s$ 
(the QCD coupling constant) 
in Eqs.~\ref{tdeq7} and \ref{tdeq10}. 
So, $d^g_t\sim1.5d^{\gamma}_t$ where $d^{\gamma}_t$ 
is in e-cm and $d^g_t$ in $g_s$-cm.
In theories with such a Higgs sector many CP violation effects are driven
by the top CEDM. The Schmidt-Peskin energy asymmetry in 
$pp\to t\bar t + X$
followed by top decay is one such interesting effect \cite{prl69p410}, 
and we will discuss it in detail in Chapter \ref{sec7}.\\


\newpage
~

\newpage


\subsection{Expectations from 2HDM's with CP violation 
in the neutral Higgs sector \label{ssec43}}

The general analysis given in the previous section holds, of course, 
for any number of Higgs doublets. However, 
let us now focus on the simplest extension of the Higgs sector.
That is, a 2HDM with CP violation in the neutral Higgs 
sector driven by a phase in the Higgs-fermion-fermion interaction 
\cite{plb279p389,npb386p63}. The example that we will explicitly 
consider here is the type II 2HDM; however,  
the analysis 
can also 
be applied to type I and III models with simple 
redefinitions of the couplings (see also 
section \ref{ssec32}).
In this model 
the dipole moment form factors for the top quark start 
to contribute at 1-loop order via the Feynman diagrams in 
Fig.~\ref{topdipolefig1}. 
The required CP-odd phase is provided by the ${\cal H}^k t \bar t$ 
Lagrangian piece in Eq.~\ref{2hdmab}, where ${\cal H}^k$, for $k=1,2,3$, 
stands for the three neutral Higgs particles in the model.  
The couplings $a_t^k,b_t^k$ in Eq.~\ref{2hdmab} 
depend on the three Euler angles, i.e., $\alpha_{1,2,3}$,  
which parameterize the neutral Higgs mixing matrix 
and on $\tan\beta$ which is the ratio between the two VEV's, 
$v_1$ and $v_2$, corresponding to the two Higgs doublets of the model 
(for more details see section \ref{ssec32}).  

The top EDM and ZEDM within this class of 2HDM's was considered 
in \cite{plb279p389,npb408p286}. 
They can be written as

\beqa
d^{\gamma}_t(s) & = & \sum^3_{k=1} d_k^\gamma(s) g_k \label{tdeq12} ~,\\
d^Z_t(s) & = & \frac{3g^t_V}{4\sin\theta_W\cos\theta_W} d^\gamma_t(s) +
\sum^3_{k=1} d'^{Z}_{t,k}(s) g'_k  \label{tdeq13}~,
\eeqa

\noindent where $g^t_V = 1/2 -4\sin^2\theta_W/3$ and in Model II

\beqa
g_k & = & a_t^k b_t^k = R_{1k}R_{3k} \cot\beta/\sin\beta  \label{tdeq14} ~,\\
g'_k & = & b_t^k c^k = R_{3k}
(R_{2k}\cos\beta + R_{1k}\sin\beta) \cot\beta  \label{tdeq15} ~.
\eeqa

\noindent The $3\times 3$ neutral Higgs mixing mass matrix $R$ 
is given in Eq.~\ref{2hdmrij}. 
$c^k$ in Eq.~\ref{tdeq15} is the coupling constant associated with the 
$ZZ{\cal H}^k$ vertex, see Eqs.~\ref{2hdmc} and \ref{mhdmeq23}.

From Fig.~\ref{topdipolefig1}(a),(b) and (c)  
one can extract the functions
$d_k^\gamma$ and $d'^{Z}_{t,k}$

\begin{eqnarray}
d_k^\gamma (s) &=& -\frac{eQ_t\sqrt{2} G_F m^3_t}{4\pi^2} \times 
\frac{C_{11}^a}{2}
 \label{tdeq16} ~,\\
d'^{Z}_{t,k}(s) &=& \frac{eg^t_V}{8\pi^2\sin\theta_W\cos\theta_W}
\sqrt{2} G_F m^2_Zm_t \times \frac{C_{12}^b +C_{11}^c -C_{12}^c}{2}
  \label{tdeq17}~,
\end{eqnarray}

\noindent where $Q_t=2/3$. The three-point loop form factors $C_x^i$, 
$x \in \left\{11,12 \right\}$ and $i=a,b,c$ 
corresponding to diagrams (a),(b),(c) in Fig.~\ref{topdipolefig1}, are given  
in our notation by

\begin{eqnarray}
C_x^a &=& C_x(m_{{\cal H}^k}^2,m_t^2,m_t^2,m_t^2,s,m_t^2) 
\label{tdeq0018} ~,\nonumber \\
C_x^b &=& C_x(m_t^2,m_{{\cal H}^k}^2,m_Z^2,m_t^2,s,m_t^2)  
\label{tdeq18} ~, \\
C_x^c &=& C_x(m_t^2,m_Z^2,m_{{\cal H}^k}^2,m_t^2,s,m_t^2) 
\label{tdeq018} ~,\nonumber
\end{eqnarray}
 
\noindent and $C_x(m_1^2,m_2^2,m_3^2,p_1^2,p_2^2,p_3^2)$ is defined 
in Appendix A. Analytical expressions for the imaginary parts of the 
three-point loop form factors, $C_x^i$, may be derived through the Cutkosky 
rule \cite{plb279p389}:

\begin{eqnarray}
\Im{\rm m}G_k(s) &\equiv& \Im{\rm m} \left( \frac{C_{11}^a}{2} \right)  = 
\pi \frac{\Theta(s-4m^2_t)}{m_{{\cal H}^k}\sqrt{s}} [
Z^{-3}_k \ln (1+ Z^2_k) - Z^{-1}_k ]  \label{tdeq19} ~, \\
\Im{\rm m}D_k(s) &\equiv& \Im{\rm m} \left(\frac{C_{12}^b +C_{11}^c 
-C_{12}^c}{2}\right)
  =  \pi\Theta [s-(m_Z+m_{{\cal H}^k})^2 ] \nonumber \\ 
& & \times \Biggl\{ A_k(s) + \left[ \frac{4m^2_t-m^2_Z -m_{{\cal H}^k}^2}{2(4m^2_t-s)} +
\frac{m_{{\cal H}^k}^2 - m^2_Z}{2s} \right] B_k(s) \Biggr\} ~, \nonumber \\
 \label{tdeq20} 
\end{eqnarray}

\noindent where

\begin{eqnarray}
&&Z_k = \sqrt{(s -4m^2_t)/ m_{{\cal H}^k}^2} ~,\\
&&A_k(s) = \frac{\sqrt{w_k}}{s (4m^2_t -s)}  \label{tdeq21}~,
\end{eqnarray}
and
\beq
\left. 
\begin{array}{lll}
B_k(s) & = -a^{-1/2} \ln \left| \frac{b_k+\sqrt{aw_k}}{b_k -
\sqrt{aw_k}} \right| & ~,~\mbox{for }a>0 \\
& = -2\sqrt{w_k}/b_k & ~,~\mbox{for }a=0 \\
& = -2(-a)^{-1/2} \arctan (\sqrt{-aw_k}/b_k) & ~,~\mbox{for }a<0
\end{array}
\right\}  \label{tdeq22} ~.
\eeq

\noindent also

\beqa
&&a  =  s(s-4m^2_t)  \label{tdeq23} ~,\\
&&b_k  =  s[s-(m^2_Z+m_{{\cal H}^k}^2)]  \label{tdeq24} ~,\\
&&w_k  =  [s-(m_Z+m_{{\cal H}^k})^2][s-(m_Z-m_{{\cal H}^k})^2] 
 \label{tdeq25}~.
\eeqa 

\noindent The
real parts are obtained from a dispersion relation

\begin{eqnarray}
\Re{\rm e} G_k(s) &=& \frac{1}{\pi} {\cal P} \int^{\infty}_{4m^2_t} ds^\prime
\frac{\Im{\rm m}G_k(s^\prime)}{s^\prime  - s}  \label{tdeq26}~.\\
\Re{\rm e} D_k(s) &=& \frac{1}{\pi} {\cal P} 
\int^{\infty}_{(m_Z+m_{{\cal H}^k})^2} ds^\prime
\frac{\Im{\rm m}D_k(s^\prime)}{s^\prime  - s}  \label{tdeq27}~.
\end{eqnarray}

\begin{figure}[htb]
\psfull
 \begin{center}
  \leavevmode
  \epsfig{file=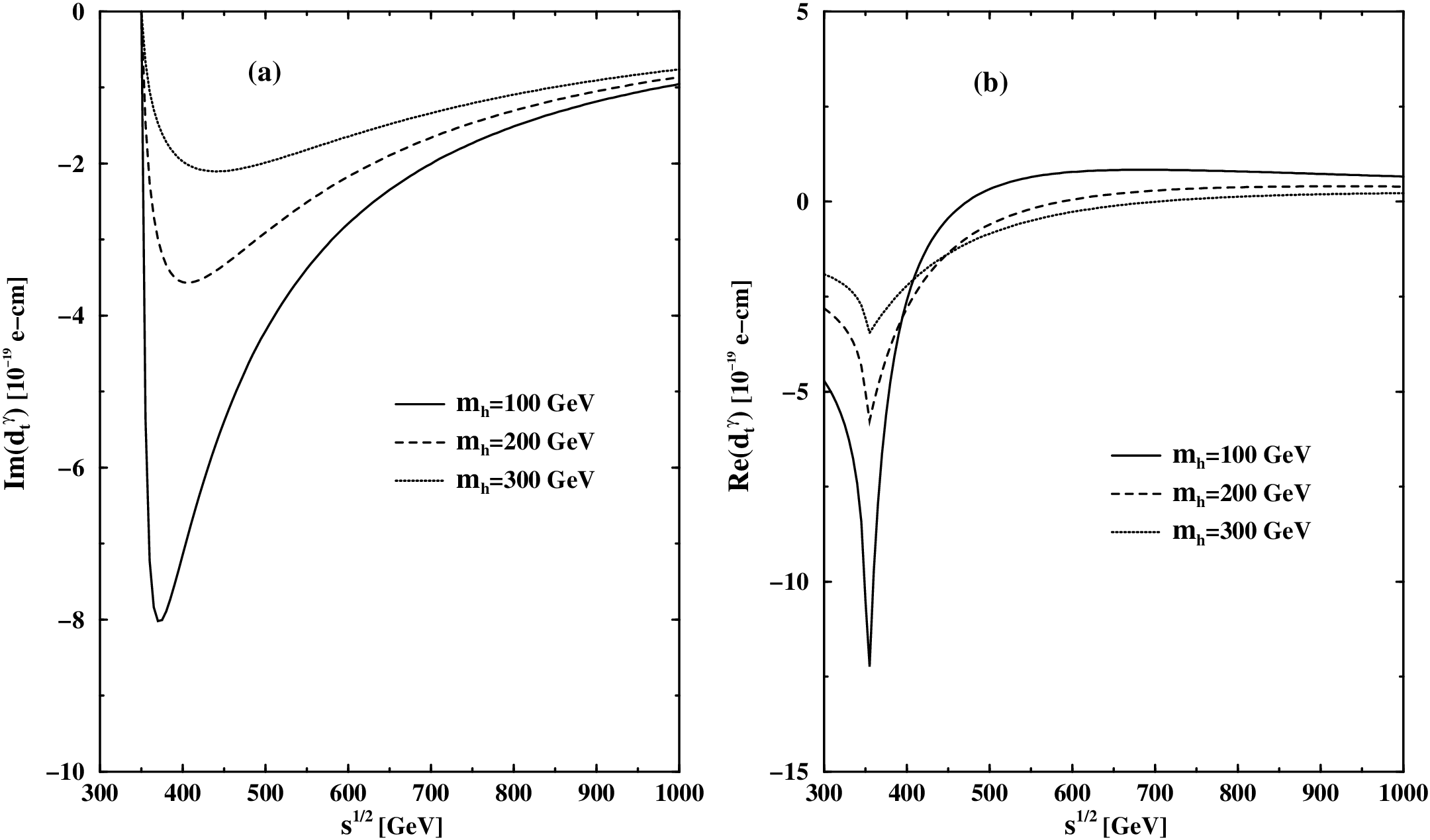,height=9.5cm
,width=9.5cm,bbllx=0cm,bblly=2cm,bburx=20cm,bbury=25cm,angle=0}
 \end{center}
\caption{\emph{ Imaginary (a) and Real (b) parts of $d_t^\gamma$ in 
units of $10^{-19}$ e-cm 
as a function of $\sqrt s$,  
for various  masses of the lightest neutral Higgs ($h$) 
and for $m_t=175$ GeV.}}
\label{topdipolefig2}
\end{figure}

\begin{figure}[htb]
\psfull
 \begin{center}
  \leavevmode
  \epsfig{file=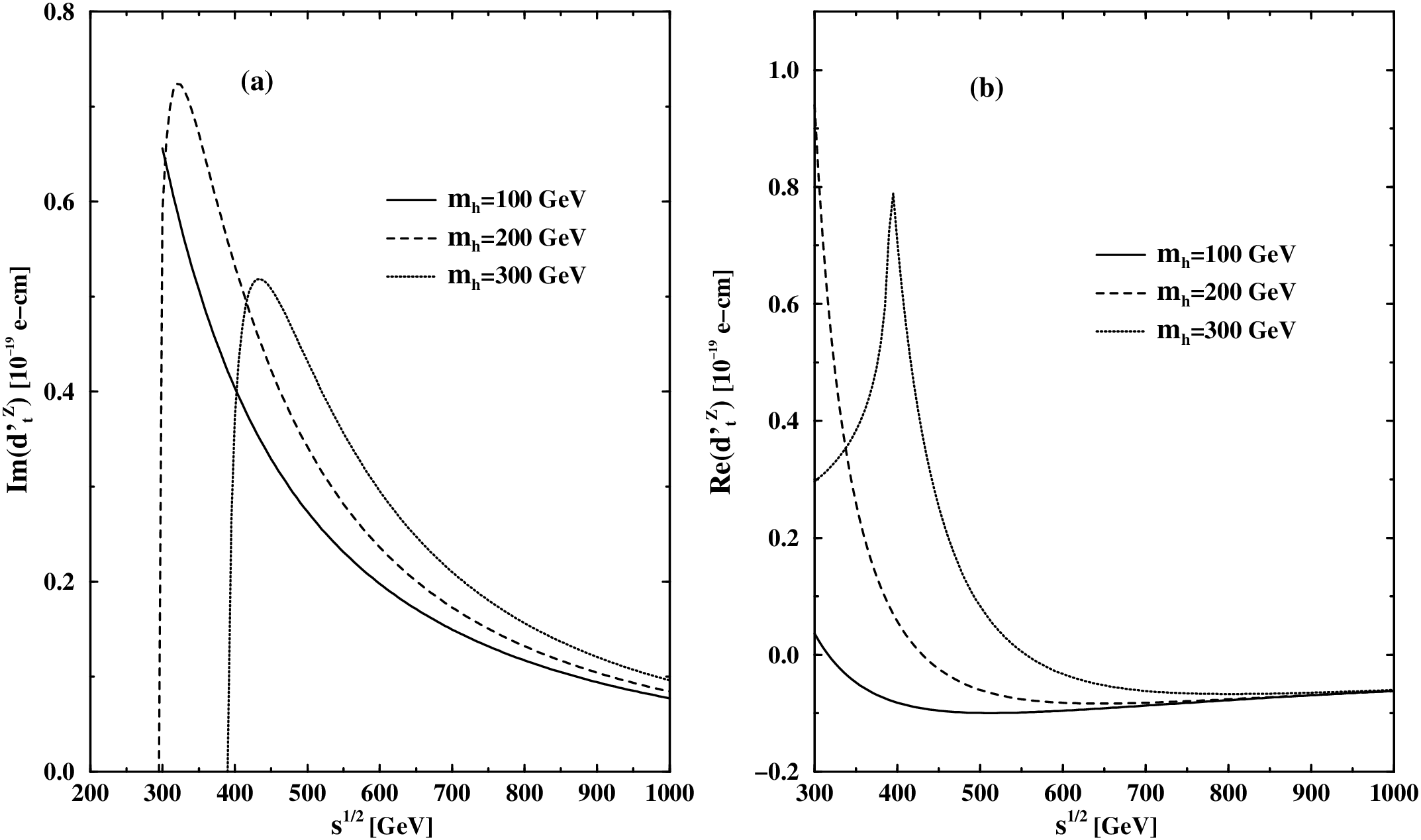
,height=9.5cm,width=9.5cm,bbllx=0cm,bblly=2cm,bburx=20cm,bbury=25cm,angle=0}
 \end{center}
\caption{\emph{ Imaginary (a) and Real (b) parts of $d'^Z_t$ in 
units of $10^{-19}$ e-cm
as a function of $\sqrt s$,  
for various masses of the lightest neutral Higgs ($h$) 
and for $m_t=175$ GeV.}}
\label{topdipolefig3}
\end{figure}

\ni Let us now assume again that the masses of the 
other two neutral Higgs
particles are considerably larger then the lightest one and, therefore, 
the lightest neutral Higgs dominates the 
sums in Eqs.~\ref{tdeq12} and \ref{tdeq13}. 
Recall that the CP-violating effects would vanish if the Higgs were
degenerate, i.e., $d^\gamma_t=d^Z_t=0$, due to the orthogonality properties
of the neutral Higgs mixing matrix.
With no loss of generality we denote 
the lightest neutral Higgs by $h$ with couplings $g_1,g'_1$,
corresponding to $k=1$ 
in Eqs.~\ref{tdeq14} and \ref{tdeq15}.   
Also, we scale out the couplings $g_1,g'_1$ and plot in  
Figs.~\ref{topdipolefig2}(a),(b) and \ref{topdipolefig3}(a),(b) 
 the real and imaginary parts of 
$d_t^\gamma$ and $d'^{Z}_t$ (recall that $d'^{Z}_t$ 
is the contribution to the ZEDM which 
arises through diagrams (b) and (c) in Fig.~\ref{topdipolefig1})   
for a variety
of Higgs masses, $m_h=100,~200,~300$ GeV and $m_t=175$ GeV. 

We can see from Figs.~\ref{topdipolefig2}(a),(b) that 
$\Re{\rm e} (d_t^\gamma)$ and $\Im{\rm  m}(d_t^\gamma)$
are typically $\sim 10^{-19}- 10^{-18}$ e-cm for $m_h=100-300$ GeV.
The peak in
the threshold region, shown in Fig.~\ref{topdipolefig2}(b), 
originates from a Coulomb-like singularity present in diagram (a)
of Fig.~\ref{topdipolefig1}. This is more pronounced, of course, 
for light Higgs masses. 
Note that the contribution from the diagram in 
Fig.~\ref{topdipolefig1}(a) to $\Re{\rm e}(d_t^Z)$ and $\Im{\rm  m}(d_t^Z)$
is smaller by a factor of 
$3g_V^t/4 \sin\theta_W \cos\theta_W \simeq 0.35$ than 
the contribution to 
$\Re{\rm e}(d_t^\gamma)$ and $\Im{\rm  m}(d_t^\gamma)$ 
as can be seen from Eq.~\ref{tdeq13}.

As is evident from Figs.~\ref{topdipolefig2} and \ref{topdipolefig3}, 
$\Re{\rm e}(d'^Z_t)$ and $\Im{\rm m}(d'^Z_t)$ are typically about one order 
of magnitude smaller than 
$\Re{\rm e}(d_t^\gamma)$ and $\Im{\rm m}(d_t^\gamma)$, respectively. 
Clearly, disregarding the 1-loop form factors, 
this difference is in part due to the different couplings in 
Eqs.~\ref{tdeq16} and \ref{tdeq17}. Thus one finds

\begin{eqnarray}
\frac{d_t^\gamma}{d'^Z_t} \approx 
\frac{2Q_t \sin\theta_W \cos\theta_W}{g_V^t} \times \frac{m_t^2}{m_Z^2} 
\simeq 10.7  \label{tdeq28}~.
\end{eqnarray}

\ni Moreover, the difference between the contributions from 
Fig.~\ref{topdipolefig1}(a) and   
Figs.~\ref{topdipolefig1}(b) and (c) 
becomes even more pronounced once the 2HDM couplings 
$g_1$ and $g'_1$ are included. In particular, 
from Eqs.~\ref{tdeq14} and \ref{tdeq15} one finds that, for $\tan\beta <1$,  
$g_1 \propto 1/\tan^2 \beta$ and $g'_1 \propto 1/\tan\beta$. Thus, 
as we will show below, the ratio $g_1/g'_1$ may even become as large as 
$\sim 10$ for $\tan\beta \lsim 0.5$.

\begin{figure}[htb]
\psfull
 \begin{center}
  \leavevmode
  \epsfig{file=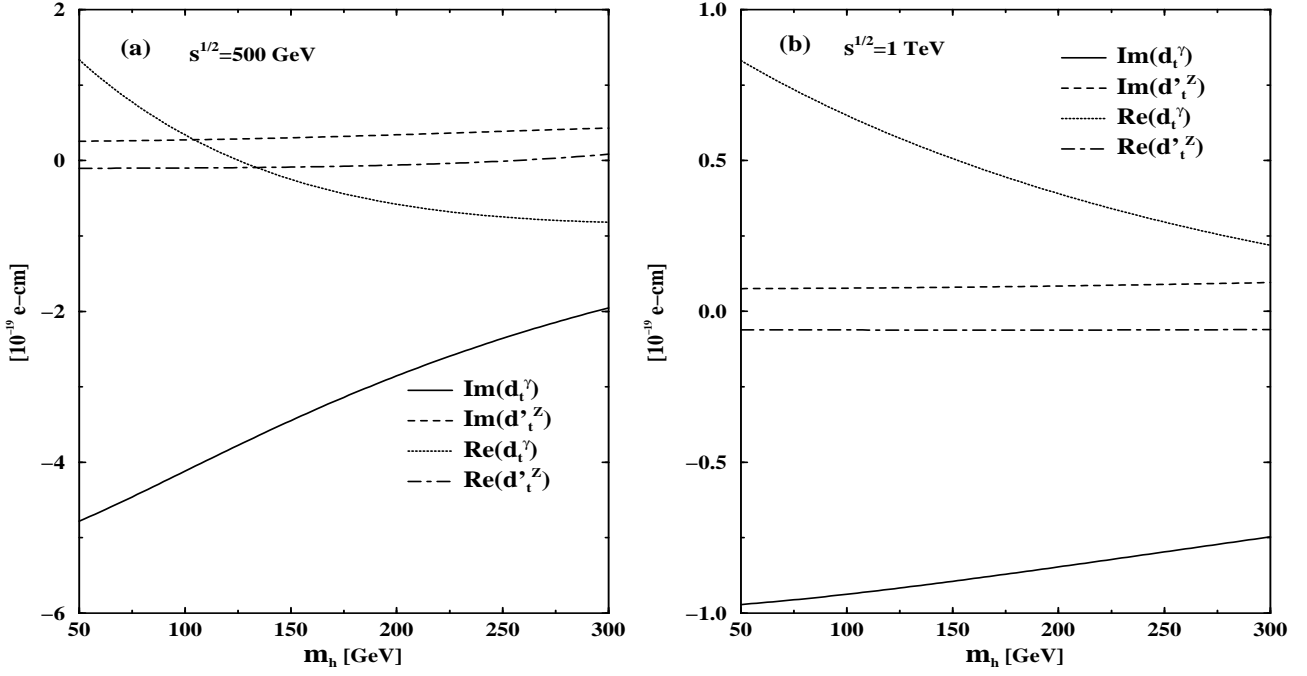,height=9.5cm
,width=9.5cm,bbllx=0cm,bblly=2cm,bburx=20cm,bbury=25cm,angle=0}
 \end{center}
\caption{\emph{Imaginary and Real parts of $d_t^\gamma$ and $d'^Z_t$ in 
units of $10^{-19}$ e-cm
as a function of $m_h$,  
for (a) $\sqrt s=500$ GeV and (b) $\sqrt s=1$ TeV 
and for $m_t=175$ GeV.}}
\label{topdipolefig4}
\end{figure}

In Figs.~\ref{topdipolefig4}(a) and (b) we show the dependence 
of the real and imaginary parts of $d_t^\gamma$ and $d'^Z_t$ on 
the mass of the lightest Higgs-boson $m_h$. 
Evidently, the dependence of $d'^Z_t$ on $m_h$ is rather 
insignificant,  
while $d_t^\gamma$ drops as $m_h$ increases.  

The
EDM's of the neutron and electron do not constrain the neutral 
Higgs mixing matrix in any significant way. 
Thus, as already mentioned above, 
its matrix elements, which enters $g_1$ and $g'_1$ 
in Eqs.~\ref{tdeq14} and \ref{tdeq15} could be of ${\cal O}(1)$.
Furthermore, for some versions of 2HDM, $\tan\beta<1$ is a viable
alternative. In this case $|g_1|, |g'_1|$ can even become larger
than one, further enhancing the dipole form factors. 
Let us now include the factors $g_1$ and $g'_1$ in calculating 
the top's EDM and ZEDM. For illustration, 
we choose three sets for the two Euler angles  
\footnote
{note that in the parameterization of Eq.~\ref{2hdmrij} in 
section \ref{sssec323}, $g_1$ and $g'_1$ are insensitive to $\alpha_3$ 
and it is sufficient to consider only 
$\alpha_1$ and $\alpha_2$}
$\left\{\alpha_1,\alpha_2\right\}$ 
and take $\tan\beta=0.3$. Set I: 
$\left\{\alpha_1,\alpha_2\right\}=\left\{\pi/4,\pi/2\right\}$
, Set II: 
$\left\{\alpha_1,\alpha_2\right\}=\left\{\pi/4,3\pi/4\right\}$ 
and Set III: 
$\left\{\alpha_1,\alpha_2\right\}=\left\{\pi/4,\pi/4\right\}$. 
Note that in Set I $g_1$ is maximized, in Set II $g'_1>1$ and in 
Set III $g_1$ and $g'_1$ have opposite relative signs such that 
the contributions of $d'^Z_t$ and $d_t^\gamma$
to $d_t^Z$ add. In particular, in terms of $g_1$ and $g'_1$ we have

\begin{eqnarray}
&&{\rm Set~I}:~~~g_1\simeq 5.8~,~g'_1\simeq 0.48  \label{tdeq29} ~,\\
&&{\rm Set~II}:~~~g_1\simeq 4.1~,~g'_1\simeq 1.14  \label{tdeq30}~,\\
&&{\rm Set~III}:~~~g_1\simeq 4.1~,~g'_1\simeq -0.46  \label{tdeq31} ~. 
\end{eqnarray}

\ni In Table~\ref{topdipoletab1} we give the real and imaginary 
parts of $d_t^\gamma$ and $d_t^Z$ (the total ZEDM including 
diagrams (a),(b) and (c) in Fig.~\ref{topdipolefig1}). 
For illustration, we present numbers for $m_h=100,~200,~300$ GeV 
and $\sqrt s=500,~1000$ GeV. We see from Table~\ref{topdipoletab1}
that $d_t^\gamma$ ranges from a few $\times 10^{-19}$ e-cm to 
a few $\times 10^{-18}$ e-cm. Also, as expected, 
$d_t^Z$ is typically smaller by 
about a factor of $\sim 3 - 4$.
The imaginary parts tend to be bigger by factors ranging from 2--10 for 
$\sqrt{s}=1000$, $500~GeV$ respectively.
In passing we also
note that the weak dipole moment form factors obtained here are about
an order of magnitude bigger than that found in \cite{prl69p33}.
Note that, here also, the top CEDM is immediately obtained by replacing 
the photon with a gluon and, therefore, $\frac{2}{3}e$ with $g_s$.

\begin{table}[htb]
\begin{center}
\caption[first entry]{\emph{Real and Imaginary parts of $d_t^\gamma$ and 
$d_t^Z$ in units of $10^{-19}$ e-cm, for $m_h=100,~200$ and 300 GeV and 
for $\sqrt s=500$ GeV and $\sqrt s=1$ TeV (in parenthesis). 
$\tan\beta=0.3$ and Set I,II,III means 
$\left\{\alpha_1,\alpha_2\right\}=\left\{\pi/4,\pi/2\right\},
\left\{\pi/4,3\pi/4\right\},\left\{\pi/4,\pi/4\right\}$, respectively.} 

\bigskip

\protect\label{topdipoletab1}}
\begin{tabular}{|r||r|r|r|r|} \cline{3-5}
\hline
Type of moment 
& $m_h$
& \multicolumn{3}{c|}{The different Sets of $\left\{\alpha_1,\alpha_2 \right\}$, $\tan\beta=0.3$}
\\ \cline{3-5}
$(10^{-19}~{\rm e-cm}) \Downarrow$ &$({\rm GeV}) \Downarrow$ & Set I & Set II &Set III\\ 
\hline
\hline
&100& $1.97(3.77)$ & $1.40(2.66)$ & 1.40(2.66) \\ \cline{2-5}
$\Re{\rm e}(d_t^\gamma)$& 200 & $-3.36(2.26)$ & $-2.38(1.60)$ & $-2.38(1.60)$\\ \cline{2-5}
& 300&  $-4.75(1.27)$ & $-3.36(0.90)$ & $-3.36(0.90)$ \\ \hline
\hline
& 100 &  $-23.89(-5.44)$ & $-16.88(-3.84)$ & $-16.88(-3.84)$ \\ \cline{2-5}
$\Im{\rm m}(d_t^\gamma)$ & 200 & $-16.56(-4.91)$ & $-11.70(-3.47)$ & $-11.70(-3.47)$ \\ \cline{2-5}
&300 &  $-11.34(-4.33)$ & $-8.02(-3.06)$ & $-8.02(-3.06)$\\ \hline \hline
& 100 &  $0.62(1.25)$ & $0.36(0.83)$ & $0.52(0.93)$ \\ \cline{2-5}
$\Re{\rm e}(d_t^Z)$ & 200 & $-1.17(0.74)$ & $-0.87(0.47)$ & $-0.78(0.57)$ \\ \cline{2-5}
&300 &  $-1.57(0.40)$ & $-1.04(0.24)$ & $-1.18(0.33)$\\ \hline \hline
& 100 &  $-7.96(-1.81)$ & $-5.41(-1.21)$ & $-5.85(-1.34)$ \\ \cline{2-5}
$\Im{\rm m}(d_t^Z)$ & 200 & $-5.45(-1.62)$ & $-3.58(-1.08)$ & $-4.12(-1.22)$ \\ \cline{2-5}
&300 &  $-3.64(-1.42)$ & $-2.22(-0.93)$ & $-2.91(-1.08)$\\ \hline
\end{tabular}
\end{center}
\end{table}


\newpage
~

\newpage
~

\newpage
~

\newpage

\subsection{Expectations from a CP-violating 
charged Higgs sector \label{ssec44}}

\begin{figure}[htb]
\psfull
 \begin{center}
  \leavevmode
  \epsfig{file=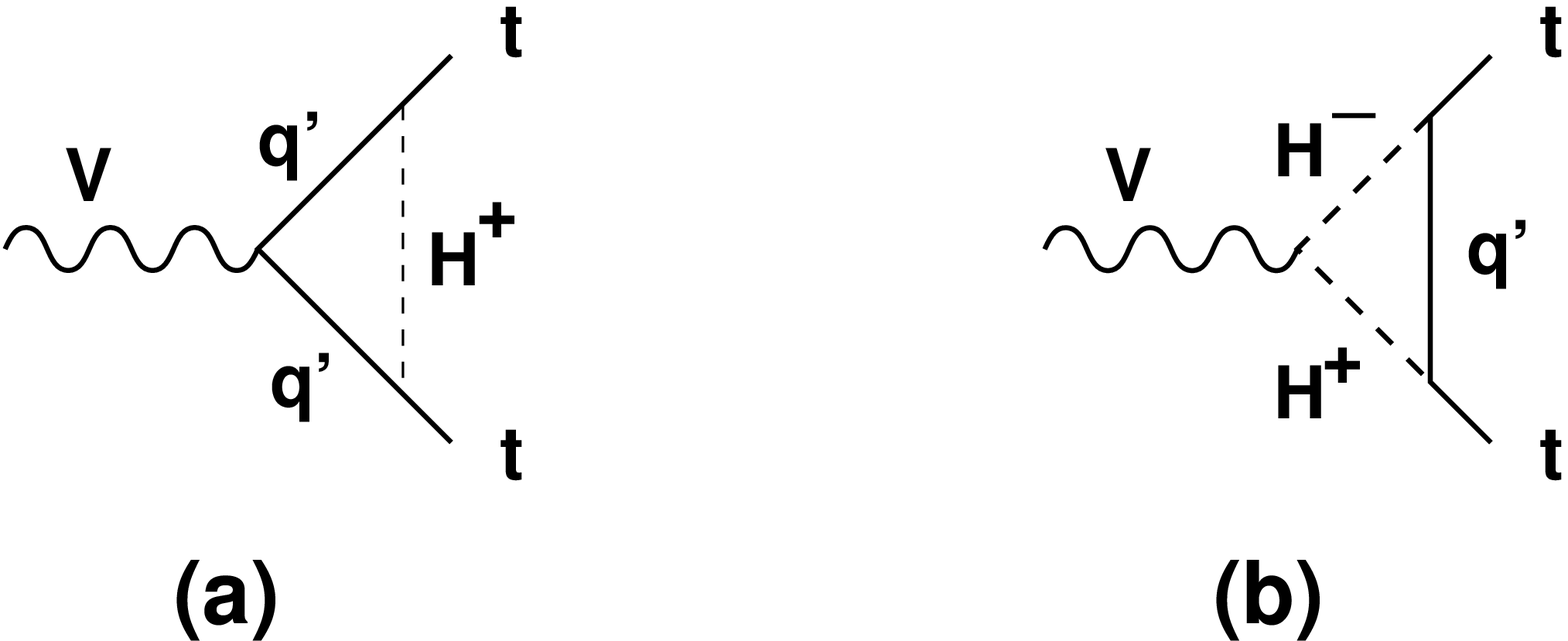
,height=5.5cm}
 \end{center}
\caption{\emph{Feynman diagrams that contribute to the electric, weak and 
chromo-electric top dipole moments due to CP-violating
interactions of a charged Higgs with a top quark. $V=\gamma,Z$ 
(also $V=$gluon in diagram (a)) 
and $q^{\prime}$ stands for $d$, $s$ or $b$-quark.
\label{fig1charge}}}
\end{figure}

In models with three or more Higgs doublets
\cite{prl37p657,prd21p711}, it is also
possible to have
CP violation in the charged Higgs sector. 
In this case a top EDM, ZEDM and CEDM
receives contributions from diagrams in Figs.~\ref{fig1charge}(a) and 
\ref{fig1charge}(b) 
with $q^\prime=\{d,s,b\}$.
In the case of the 3 Higgs doublet model as considered for example in
section \ref{sssec324} we may express the coupling of the lighter 
charged Higgs-bosons, say $H_2^+ \equiv H^+$,
to the third generation of quarks as (see also Eq.~\ref{mhdmeq38} for our 
notation):
  
\begin{equation}
{\cal L}_{tbH^+}= {g_W\over \sqrt{2} m_W} K_{tb}
\left( m_t \beta_2 \bar t_R b_L + m_b \alpha_2 \bar t_L b_R \right) + h.c ~,
\label{eqn_f} ~.
\end{equation}

\noindent where $K$ is the CKM matrix (we will assume that $K_{tb}\approx
1$); $\beta_2$ and $\alpha_2$ 
are complex parameters of the model (for a more complete 
description of these parameters see section \ref{sssec324}).
Also, we neglect contributions from the $H^+td$ and $H^+ts$ couplings 
as those will yield a TDM smaller by a factor of $\sim (m_d/m_b)^2$ and 
$\sim (m_s/m_b)^2$, respectively, compared to the TDM coming from the $H^+tb$ 
Lagrangian piece.  

The CP violation in this case is proportional to the quantity
$\Im{\rm m}(V)$ where $V \equiv (\beta_2 \alpha_2^*)$. 
We denote the coupling of a vector-boson to 
the $b$-quark by
  
\begin{equation}
-i\gamma^\mu(A^b_V+B^b_V\gamma^5) \label{eqn_g} ~,
\end{equation}
  
\noindent and the coupling of a vector-boson to the pair of charged Higgs by
  
\begin{equation}
iC_V^H (p_{H+}-p_{H-})^\mu \label{eqn_h} ~,
\end{equation}
  
\noindent where $p_{H\pm}$ are the in-going 
momenta of the charged Higgs-bosons.

In \cite{prd51p1034} we have calculated the contribution of the charged 
Higgs exchanges in Figs.~\ref{fig1charge}(a) and \ref{fig1charge}(b) 
to the $b$-quark dipole moment.
The top EDM, ZEDM or CEDM given by these diagrams 
may, therefore, 
be extracted from \cite{prd51p1034} by the replacement 
$m_b \leftrightarrow m_t$ and are thus given by
  
\begin{equation}
d_t^{\gamma,Z,g}=-{\alpha \over 4 \pi \sin^2 \theta_W}
\Im{\rm m}(V){m_b^2\over m_W^2}m_t
\left [ A^b_{\gamma,Z,g} C_{12}^a -C_{\gamma,Z,g}^H 
(C_{12}^b+{1\over 2} C_0^b ) \right]
\label{eqn_i} ~,
\end{equation}
  
\noindent where  
$C_x^i$, 
$x \in \left\{0,12 \right\}$ and $i=a,b$, are 
the three-point loop form factors  
corresponding to diagrams (a),(b) in Fig.~\ref{fig1charge} such that  

\begin{eqnarray}
C_x^a &=& C_x(m_{H^+}^2,m_b^2,m_b^2,m_t^2,s,m_t^2) \label{chargeloopform00} 
~,\nonumber \\
C_x^b &=& C_x(m_b^2,m_{H^+}^2,m_{H^+}^2,m_t^2,s,m_t^2)  
\label{chargeloopform} ~.
\end{eqnarray}
 
\noindent Here $s=(p_t+p_{\bar t})^2$ and 
$C_x$ is defined 
in Appendix A. Also

\begin{eqnarray}
&&A_\gamma^b=-\frac{1}{3}e~,~C_\gamma^H=e~,\\
&&A_Z^b=\frac{e}{2 \sin\theta_W \cos\theta_W} \left(-\frac{1}{2} 
+ \frac{2}{3} \sin^2\theta_W \right)~,~C_Z^H=e \cot 2\theta_W~,\\
&&A_g=g_s~,~C_g^H=0~,
\end{eqnarray}

\noindent where $e$ is the electric charge, 
$g_s$ is the strong coupling constant and $\theta_W$ is the weak mixing angle.
Note that in the case of the gluon, i.e., the CEDM,
 only Fig.~\ref{fig1charge}(a) enters 
as the $gH^+H^-$ coupling is absent.  

Using $m_{H^+}=200$ GeV 
(also $m_t \simeq 175$ GeV) and denoting
$d_t^{\gamma,Z,g}\equiv \Im{\rm m}(V)\delta_t^{\gamma,Z,g}(s)$ 
we find that for $ 500~{\rm GeV} < \sqrt s < 1000~{\rm GeV}$

\begin{eqnarray}
\Im{\rm m} \delta_t^{\gamma,Z,g}~, \Re{\rm e} \delta_t^{\gamma,Z,g} 
\lsim {\rm few}\times 10^{-22} ~e-{\rm cm},g_s-{\rm cm} ~,
\end{eqnarray}

\noindent where, in fact, for $\sqrt s=1000$ GeV,   
$\Re{\rm e} \delta_t^{\gamma,Z} \lsim 10^{-23} ~e-{\rm cm}$. 
Moreover, as $m_{H^+}$ is increased the TDM drops rapidly and, for example, 
for $m_{H^+}=500$ GeV, we find that, typically,  
$\Im{\rm m} \delta_t^{\gamma,Z,g}$ and $\Re{\rm e} \delta_t^{\gamma,Z,g}$ 
are smaller than $\sim 10^{-22}$ e-cm or $g_s$-cm.
Such small TDM, residing in charged Higgs exchanges, is expected simply 
by comparing $\delta_t^{\gamma,Z,g}$ above to the corresponding terms
for neutral Higgs exchanges which were given in the previous sections. 
In particular, one can immediately observe that the 
contribution to the TDM from charged Higgs exchanges is 
naively suppressed by a factor of $\sim (m_b/m_t)^2$ with respect to 
the neutral Higgs exchanges wherein the TDM was found to be typically 
at the order of $\sim 10^{-19}$ e-cm.  

Let us briefly consider what constraints can be placed on the
parameter, $\Im{\rm m}(V)$, by use of the experimental results on the 
EDM of the neutron (NEDM) and on the decay $b \to s \gamma$.
For our discussion we will simply use the
bound given in \cite{prl82p904} on the NEDM
  
\begin{equation}
d^e_n \lsim 10^{-25} ~ \mbox{e-cm} \label{eqn_new1} ~.
\end{equation}
  
\noindent Using our
previous results
given in Eq.~\ref{eqn_i}
one can deduce an expression for the electric dipole
moment of the light $(u,d)$-quarks in the charged Higgs model 
(see \cite{prd51p1034}). 
%
%
Making the simplifying assumption that the
NEDM equals that of its valence quarks one finds (as has
been noted before \cite{prl37p657}) that there is significant
uncertainty from the
numerical value of the mass of the light quark to be used in the above
formula. For the purpose of obtaining an upper bound on $\Im{\rm m}(V)$ we can
take the 
current mass to be 
$\sim10$ MeV\null. Then for
$m_t=180$ GeV, $m_{H^+}$ in the range 200--500 GeV, we find
from Eq.~\ref{eqn_new1} and using Eq.~\ref{eqn_i} for the light $u,d$-quarks 
that 
$\Im{\rm m}(V) \lsim 10$ \cite{prd51p1034}. 
The experimental data on the decay 
$b \to s\gamma$ can also constrain this parameter. 
In particular, it was shown in \cite{hepph9406302,plb313p126} that, 
for $m_{H^+}=200-500$ GeV
and neglecting the effects of the second charged Higgs of the model,
a conservative upper bound of

\begin{eqnarray}
\Im{\rm m}(V) \lsim 3-9~,
\end{eqnarray}

\noindent can be placed from the CLEO 
measurement of 
the decay rate of $b \to s \gamma$ (see also discussion 
in section \ref{sssec324}).\footnote{We note, however, that the 
limit on $\Im{\rm m}(V)$ 
gets weaker as the masses of the two charged Higgs approach the same value,  
due to a GIM-like cancellation of the CP-violating effect mediated by the two.
For example, if $m_{H^+}=350$ GeV and 
the mass of the second charged Higgs is 
$\sim 500$ GeV, then $\Im{\rm m}(V)\gsim 10$ does not 
contradict the upper bounds on $b \to s \gamma$, and,  
of course, no such limit exist at all if the two charged 
Higgs are degenerate \cite{hepph9406302}.}
It turns out, however, that 
even when using $\Im{\rm m}(V)= 10$, in conjunction with numerical values for
$\delta_t^{\gamma,Z,g}(q^2)$ 
as given before, the TDM
is expected to be $\lsim 10^{-21}$ e-cm in this class of charged 
Higgs mediated CP violation. Thus,
it is 
typically smaller by at least an order of magnitude than what one would expect 
from CP-violating neutral Higgs 
exchanges and from the MSSM (see the following section). 

Finally, it should be noted that in any given MHDM
 with new 
mechanisms of 
CP violation in the charged Higgs sector, i.e., three and more Higgs doublets, 
the neutral Higgs sector will also acquire new CP-violating phases which in 
general cannot be screened (see section \ref{sssec324}). 
Therefore, as was already mentioned before, 
in the top quark case, CP-violating neutral Higgs 
exchanges dominate the charged Higgs by a typical factor of 
$\sim (m_t/m_b)^2$. One power of $(m_t/m_b)$ 
originates
from the ratio between the neutral and charged Higgs couplings to 
fermions and another power of $(m_t/m_b)$ comes in from the necessary mass 
insertion in the propagator of the fermion in the loop. Thus,
the charged Higgs contribution is negligible 
compared to one from a neutral Higgs.
Note, however, that in 
light quark systems charged Higgs exchanges are expected 
to yield the dominant CP-violating effects since the above argument is 
basically reversed \cite{prd51p1034}.


\newpage
~

\newpage

\subsection{Expectations from the MSSM \label{ssec45}}

Within the MSSM the TDM can also arise already at 1-loop 
even without generation mixing. As was mentioned in section \ref{ssec33}, 
in general, the required CP-violating phases 
are provided by the chargino and neutralino mixing matrices as 
well as the squarks ${\tilde q}_L - {\tilde q}_R$ mixing matrices. 
If one assumes GUT-scale unification which leads to a common gaugino mass 
at the GUT-scale, then the phase in the gaugino mass term can be 
rotated away, leaving the gaugino masses phaseless at any scale. 
Thus one is left with only three phases 
(neglecting generation mixing between the squarks) 
relevant for the TDM at the EW-scale; $\alpha_{\mu}$, $\alpha_{t}$ 
and $\alpha_{b}$ which arise from the Higgs mass term $\mu$ 
and the stop and sbottom mixing matrices, respectively (for the 
definitions of $\alpha_{\mu}$, $\alpha_{t}$ 
and $\alpha_{b}$ see Eq.~\ref{tanbeta} in section \ref{sssec332}). 
We recall (see Eqs.~\ref{zfmatrix} and 
\ref{recall} in section \ref{sssec332}) that, 
for a sfermion ${\tilde f}$, it is useful to  
adopt a parameterization for its 
${\tilde f}_L - {\tilde f}_R$ mixing such that the sfermions of 
different handedness are related to their mass eigenstates 
through 
the transformation

\beqa
\tilde f_L & = & \cos \theta_f \tilde f_1 - e^{-i\beta_f} \sin\theta_f
\tilde f_2 ~, \nonumber \\
\tilde f_R & = & e^{i\beta_f} \sin\theta_f \tilde f_1 + \cos \theta_f
\tilde f_2 \label{tdeq32}~,
\eeqa

\noindent where $\tilde f_{1,2}$ are the two mass eigenstates 
(i.e. physical states)
and the phase 
$\beta_f$ is related to the phase $\alpha_f$ by Eq.~\ref{tanbeta}. 
The contribution to the CP-violating TDM 
which arise 
from the above ${\tilde f}_L - {\tilde f}_R$ mixing matrix
will always be proportional 
to the quantity (see also Eq.~\ref{xit})

\begin{eqnarray}
\xi_{CP}^f \equiv  2 |\xi_f^i| = 
\sin2 \theta_f \sin\beta_f \label{tdeq33}~.
\end{eqnarray}

\noindent Clearly, from Eq.~\ref{tdeq32} we see 
that $\xi_{CP}^f \to 0$ if 
the two sfermions are nearly degenerate.

\begin{figure}[htb]
\psfull
 \begin{center}
  \leavevmode
  \epsfig{file=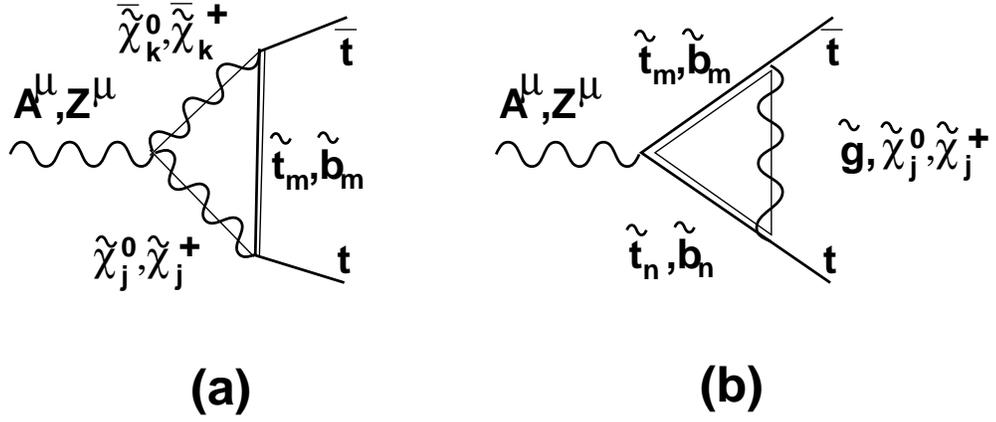,height=5.5cm}
 \end{center}
\caption{\emph{
Feynman diagrams contributing to $d_t^\gamma$ and 
$d_t^Z$: 
(a) with two fermions and one scalar in the loop, 
(b) with two scalars and one fermion in the loop.
}}
\label{topdipolefig5}
\end{figure}

In the MSSM the TDM can therefore acquire a non-vanishing value 
through the Feynman diagrams depicted in Fig.~\ref{topdipolefig5}. 
 One can then distinguish between the 
following three contributions:
\begin{enumerate}
\item Gluino contribution, $d_{t(\tilde g)}^{\gamma,Z}$, with 
$\tilde t {\tilde t}^* \tilde g$ in the loop 
(see Fig.~\ref{topdipolefig5}(b)).
\item Chargino contribution, $d_{t({\tilde\chi}^+)}^{\gamma,Z}$, 
with ${\tilde\chi}^+ {\tilde\chi}^- \tilde b$ 
(see Fig.~\ref{topdipolefig5}(a)) and 
$\tilde b {\tilde b}^* {\tilde\chi}^+$ in the loop 
(see Fig.~\ref{topdipolefig5}(b)).
\item Neutralino contribution, $d_{t({\tilde\chi}^0)}^{\gamma,Z}$, 
with ${\tilde\chi}^0 {\tilde\chi}^0 \tilde t$ 
(see Fig.~\ref{topdipolefig5}(a))
and $\tilde t {\tilde t}^* {\tilde\chi}^0$ in the loop
(see Fig.~\ref{topdipolefig5}(b)).       
\end{enumerate}
The gluino contribution was considered in 
\cite{plb315p338,zpc61p599,plb305p384,npb460p235}. It is 

\begin{eqnarray}
d_{t(\tilde g)}^{\gamma}&=& \frac{\alpha_s}{3\pi} Q_t  
\xi_{CP}^t m_{\tilde g} \times 
\left( C_0^{11}+C_{11}^{11}-C_0^{22}-C_{11}^{22} \right) \label{tdeq34}~,\\
d_{t(\tilde g)}^{Z} &=& \frac{\alpha_s}{6\pi \sin\theta_W \cos\theta_W}  
 \xi_{CP}^t m_{\tilde g} \times \nonumber \\ 
&& \left\{ \left( \cos^2 \theta_t - \frac{4}{3}\sin^2\theta_W \right) 
\left( C_0^{11}+C_{11}^{11}-C_0^{21}-C_{11}^{21} \right) + \right. \nonumber \\
&& \left. \left( \sin^2 \theta_t - \frac{4}{3}\sin^2\theta_W \right) 
\left( C_0^{12}+C_{11}^{12}-C_0^{22}-C_{11}^{22} \right) \right\} 
\label{tdeq35}~, 
\end{eqnarray}

\noindent where the three-point 1-loop form factors, $C_x^{ij}$, 
$x \in \left\{0,11\right\}$ and $i,j=1,2$,  
are given by:  

\begin{eqnarray}
C_x^{ij} &=& C_x(m_{\tilde g}^2,m_{{\tilde t}_i}^2,
m_{{\tilde t}_j}^2,m_t^2,s,m_t^2) \label{tdeq36}~,
\end{eqnarray}
 
\noindent and $C_x$ is defined 
in Appendix A. Also, $\theta_t$ and 
$\xi_{CP}^t$ are defined in Eqs.~\ref{tdeq32} and \ref{tdeq33},
respectively. 

The chargino contribution was given in \cite{npb507p35}:  
\begin{eqnarray} 
d^{\gamma}_{t(\tilde{\chi}^{+})}
&=& - \frac{\alpha \, Y_{t}}{4\pi\sin^2\theta_{W}} 
\times \nonumber \\
&& \sum_{m,j = 1}^{2} \hspace{-2mm} m_{\tilde{\chi}^{+}_{j}}
\left( C_{11}^{m,jj} - C_{12}^{m,jj}
- \frac{1}{6} ( C_{0}^{j,mm} + C_{11}^{j,mm}) \right)
\times \nonumber\\ 
&& 
~~~\left\{
    ( 1 - (-)^{m} \cos 2\theta_b )
    \, \Im{\rm m} [ U_{j1} V_{j2} ] +  \right. \nonumber\\
&&~~~  \left. Y_{b} (-)^{m} \sin 2\theta_b
    \, \Im{\rm m} [ U_{j2} V_{j2} e^{i\beta_{b}} ] \right\}
\label{tdeq37} ~, \\
d^{Z}_{t(\tilde{\chi}^{+})}
&=& \frac{\alpha \,}{16\pi\sin^3\theta_{W} \cos\theta_W} 
\times 
    \left( f_{1}^+ + f_{2}^+ + s_{1}^+ + s_{2}^+ \right)~,
\end{eqnarray}
\noindent where
\begin{eqnarray}
f_{1}^+
&=& - Y_{t} \sum_{m,j,k = 1}^{2} \hspace{-1mm} 
m_{\tilde{\chi}^{+}_{j}} \left(C_{11}^{m,jk}-C_{12}^{m,jk}\right)
\times \nonumber \\
&&~~~\left\{ 
  \left( \delta_{jk} ( 1 + 2 \cos 2\theta_{W} ) 
  + |U_{k1}|^{2} - |V_{k2}|^{2} \right) \times \right.
\nonumber \\ & & \left. ~~~
  \left( ( 1 - (-)^{m} \cos 2\theta_{b} )
    \, \Im{\rm m} [ U_{j1} V_{j2} ]
  + Y_{b} (-)^{m} \sin 2\theta_{b}
    \, \Im{\rm m} [ U_{j2} V_{j2} e^{i\beta_{b}} ]
  \right) \right.
\nonumber\\ && \left. ~~~
- Y_{b} ( 1 - \delta_{jk} ) (-)^{m} \sin 2\theta_{b}
    \, \Im{\rm m} [ U_{j2} V_{j2} e^{i\beta_{b}} ] \right\} ~, \\
f_{2}^+
&=& 
- m_{t} Y_{b} \sin 2\theta_{b}
  \Im{\rm m} [ U_{11}^{*} U_{12}^{\,} e^{i\beta_{b}} ] \times \nonumber \\
&& ~~~\sum_{m=1}^{2} (-)^{m}
  \left( C_{11}^{m,12} + C_{21}^{m,12} - 2 C_{12}^{m,12} - 
2 C_{23}^{m,12} \right) ~,
\\
s_{1}^+
&=& - Y_{t}
\sum_{k,m,n=1}^{2} \hspace{-1mm}
m_{\tilde{\chi}^{+}_{k}}
  \left( C_{0}^{k,mn} + C_{11}^{k,mn} \right) \times \nonumber \\
& & ~~~
  \left( \frac{2}{3} \sin^{2} \theta_{W} \delta_{mn}
  + \frac{1}{2} ( 1 - (-)^{n} \cos 2\theta_{b} )
  \right) \times  
\nonumber\\ && ~~~
\left\{
  \left( 1 - (-)^{m} \cos 2\theta_{b} \right)
    \, \Im{\rm m} [ U_{k1} V_{k2} ] + \right. \nonumber \\ 
  && \left. ~~~Y_{b} (-)^{m} \sin 2\theta_{b}
    \, \Im{\rm m} [ U_{k2} V_{k2} e^{i\beta_{b}} ] \right\} ~,\\
s_{2}^+
&=& 
- m_{t} Y_{b} \sin 2\theta_{b}
\sum_{k=1}^{2} 
  \left( C_{11}^{k,12} + C_{21}^{k,12} - 2 C_{12}^{k,12} - 
2 C_{23}^{k,12} \right)
\times \nonumber \\
  &&~~~\Im{\rm m} [ U_{k1}^{*} U_{k2}^{\,} e^{i\beta_{b}} ] 
\label{tdeq38}~,
\end{eqnarray}

\noindent where $m,n$ are sbottom indices and 
$j,k$ are the chargino indices. Thus, 
the above three-point 1-loop form factors $C_x^{\ell,mn}$ and 
$C_x^{m,jk}$ for $m,n=1,2$ and $\ell=j~or~k=1,2$,  
$x \in \left\{0,11,12,21,23\right\}$,  
are given by  

\begin{eqnarray}
C_x^{\ell,mn} &=& C_x(m_{{\tilde\chi}_\ell^+}^2,m_{{\tilde b}_m}^2,
m_{{\tilde b}_n}^2,m_t^2,s,m_t^2) \label{tdeq39}~,\\
C_x^{m,jk} &=& C_x(m_{{\tilde b}_m}^2,m_{{\tilde\chi}_j^+}^2,
m_{{\tilde\chi}_k^+}^2,m_t^2,s,m_t^2) \label{tdeq40}~,
\end{eqnarray}
 
\noindent and $C_x$ is defined 
in Appendix A. Also, $Y_t$ and $Y_b$ are the top and bottom Yukawa 
couplings

\begin{eqnarray}
Y_t = \frac{m_t}{\sqrt 2 m_W \sin\beta}\label{tdeq41}~~,~~
Y_b = \frac{m_b}{\sqrt 2 m_W \cos\beta}\label{tdeq42}~,
\end{eqnarray}

\noindent and, as usual, $\tan\beta$ is the ratio between the  
VEV's of the two Higgs doublets in the theory.
Furthermore, note that the phase $\alpha_\mu$, although not explicitly 
appearing in the above, is contained in $U$ and $V$ which 
are the $2\times 2$ matrices that diagonalize 
the chargino mass matrix and, in the notation used in 
section \ref{sssec332}, we have $U^* \equiv (Z^-)^\dagger$ and 
$V^* \equiv Z^+$. The definitions of $Z^+$ and $Z^-$ are given 
in section \ref{sssec332} by Eqs.~\ref{zpo} - \ref{tgam2}. 

\noindent The neutralino contribution was also given in \cite{npb507p35}:  
\begin{eqnarray} 
d^{\gamma}_{t(\tilde{\chi}^{0})} 
&=& 
  \frac{\alpha}{12\pi\sin^2\theta_{W}} 
 \sum_{k=1}^{4} \sum_{m=1}^{2} 
 m_{\tilde{\chi}^{0}_{k}}
  \left( C_{0}^{k,mm} + C_{11}^{k,mm} \right) \times \nonumber \\
&&
\left\{ 
  (-)^{m} \sin 2\theta_{t} 
  \Im{\rm m} [ ( h_{Lk}^{2} - f_{Lk} f_{Rk}^{*} )
         e^{-i \beta_{t}} ] \right.
\nonumber \\ && \left.
- ( 1 - (-)^{m} \cos 2\theta_{t} ) \Im{\rm m} [ h_{Lk} f_{Lk}^{*} ] 
- \right. \nonumber \\
&&\left. ~~~( 1 + (-)^{m} \cos 2\theta_{t} ) \Im{\rm m} [ h_{Lk} f_{Rk} ] 
\right\} ~, \label{tdeq43} \\
d^{Z}_{t(\tilde{\chi}^{0})} 
&=& \frac{\alpha}{16\pi\sin^3\theta_{W} \cos\theta_W} 
\times
    \left( 2 f_{1}^0 + 2 f_{2}^0 + s_{1}^0 + s_{2}^0 \right)
\label{tdeq44}~,
\end{eqnarray}
\noindent where
\begin{eqnarray}
f_{1}^0 &=& 
\frac{1}{2} \sum_{j,k=1}^{4} \sum_{m=1}^{2}
 m_{\tilde{\chi}^{0}_{j}} \left(C_{11}^{m,jk} - C_{11}^{m,jk}\right)
 \times \nonumber \\
&& \left\{
  (-)^{m} \sin 2\theta_{t} \,
  \Im{\rm m} [ O_{jk}^{\prime\prime}
         ( f_{Lj} f_{Rk}^{*} - f_{Lk} f_{Rj}^{*} )
         e^{-i \beta_{t}} ] \right.
\nonumber \\ & & \left.
- ( 1 + (-)^{m} \cos 2\theta_{t} )
  \Im{\rm m} [ O_{jk}^{\prime\prime}
         ( h_{Lj}^{*} f_{Rk}^{*} - h_{Lk}^{*} f_{Rj}^{*} ) ] \right.
\nonumber \\ & & \left.
- ( 1 - (-)^{m} \cos 2\theta_{t} )
  \Im{\rm m} [ O_{jk}^{\prime\prime}
         ( h_{Lj} f_{Lk}^{*} - h_{Lk} f_{Lj}^{*} ) ] \right\} ~,
\label{tdeq45}\\
f_{2}^0 &=& 
\frac{1}{2} m_{t} \sum_{j<k}^{4} \sum_{m=1}^{2} 
  \left( C_{11}^{m,jk} + C_{21}^{m,jk} - 2 C_{12}^{m,jk} 
- 2 C_{23}^{m,jk} \right) \times
\nonumber \\ & &
\left\{
  2 (-)^{m} \cos 2\theta_{t} \,
  \Im{\rm m} [ h_{Lj} O_{jk}^{\prime\prime} h_{Lk}^{*} ] \right.
\nonumber \\ & & \left.
- 2 (-)^{m} \sin 2\theta_{t} \,
  \Im{\rm m} [ ( f_{Lj}^{*} - f_{Rj} ) O_{jk}^{\prime\prime} h_{Lk}^{*}
         e^{i \beta_{t}} ] \right.
\nonumber \\ & & \left.
+ ( 1 - (-)^{m} \cos 2\theta_{t} )
  \Im{\rm m} [ f_{Lj}^{*} O_{jk}^{\prime\prime} f_{Lk} ] \right.
\nonumber \\ & & \left.
+ ( 1 + (-)^{m} \cos 2\theta_{t} )
  \Im{\rm m} [ f_{Rj} O_{jk}^{\prime\prime} f_{Rk}^{*} ]
\right\} ~, \\
s_{1}^0
&=& 
\frac{1}{2} \sum_{k=1}^{4} \sum_{m,n=1}^{2} 
m_{\tilde{\chi}^{0}_{k}}
  \left( C_{0}^{k,mn} + C_{11}^{k,mn} \right) \times \nonumber \\
& & 
\left\{ (-)^{m} 
  \Bigl(
    \frac{8}{3} \sin^2\theta_{W} \delta_{mn}
  - ( 1 - (-)^{n} \cos 2\theta_{t} )
  \Bigr)
  \Im{\rm m} [ ( h_{Lk}^{2} - f_{Lk} f_{Rk}^{*} )
         e^{-i \beta_{t}} ] \right. 
\nonumber \\ & &  \left.
- \Bigl(
    \frac{8}{3} \sin^2\theta_{W} \delta_{mn}
    ( 1 + (-)^{m} \cos 2\theta_{t} )
  - (-)^{m+n} \sin^{2} 2\theta_{t}
  \Bigr)
  \Im{\rm m} [ h_{Lk} f_{Rk} ] \right. 
\nonumber \\ & & \left.
- \Bigl(
    \frac{8}{3} \sin^2\theta_{W} \delta_{mn}
  - ( 1 - (-)^{n} \cos 2\theta_{t} )
  \Bigr) 
  ( 1 - (-)^{m} \cos 2\theta_{t} ) \times \right. \nonumber \\
&&  \left. \Im{\rm m} [ h_{Lk} f_{Lk}^{*} ] 
\right\} ~, \\
s_{2}^0
&=& 
m_{t} \sin 2\theta_{t}
\sum_{k=1}^{4}
  \left( C_{11}^{k,12} + C_{21}^{k,12} - 2C_{12}^{k,12} 
- 2 C_{23}^{k,12} \right) \times \nonumber \\
  &&~~~\Im{\rm m} [ h_{Lk}^{*} ( f_{Lk}^{*} - f_{Rk} )
         e^{i\beta_{t}} ] \label{tdeq46}~.
\end{eqnarray} 

\ni For the neutralino contribution above, 
$m,n$ are stop indices and 
$j,k$ are the neutralino indices. Thus, 
the above three-point 1-loop form factors, $C_x^{\ell,mn}$ and 
$C_x^{m,jk}$ for $m,n=1,2$ and $\ell=j~or~k=1,2$,  
$x \in \left\{0,11,12,21,23\right\}$,  
are given by  

\begin{eqnarray}
C_x^{\ell,mn} &=& C_x(m_{{\tilde\chi}_\ell^0}^2,m_{{\tilde t}_m}^2,
m_{{\tilde t}_n}^2,m_t^2,s,m_t^2) \label{tdeq47}~,\\
C_x^{m,jk} &=& C_x(m_{{\tilde t}_m}^2,m_{{\tilde\chi}_j^0}^2,
m_{{\tilde\chi}_k^0}^2,,m_t^2,s,m_t^2) \label{tdeq48}~,
\end{eqnarray}
 
\noindent and $C_x$ is again defined 
in Appendix A. Also, 
$f_{Lj},f_{Rk}$ are gaugino 
couplings and $h_{Lj}$ are higgsino couplings that contain the large 
Yukawa coupling $Y_{t}$. $O_{jk}^{\prime\prime}$ contains elements 
of the neutralino mixing matrices. 
The factors $f_{Lj},f_{Rk},h_{Lj},O_{jk}^{\prime\prime}$ 
are all given in 
\cite{npb507p35}.  

In order to be able to estimate the size of the TDM one has 
to choose a plausible set of the SUSY masses 
and parameters involved, i.e., 
${\tilde m}_2$, $\mu$, $\tan\beta$, $m_{{\tilde t}_k}$, $m_{{\tilde b}_k}$,
$\cos\alpha_{t}$, $\cos\alpha_{b}$ 
and the phases $\alpha_{\mu}$, $\beta_{t}$ 
and $\beta_{b}$. A reference set of 
parameters was chosen in \cite{npb507p35}:

\begin{center}
\begin{tabular}{|rcrl|rcrl|rcrl|}
  \hline
  ${\tilde m}_2$ &=& 230, 360 & \hspace{-4mm} GeV &
    $m_{\tilde{t}_{1}}$ &=& 150 & \hspace{-4mm} GeV &
      $m_{\tilde{b}_{1}}$ &=& 270 & \hspace{-4mm} GeV \\
  $|\mu|$ &=&      250 & \hspace{-4mm} GeV &
    $m_{\tilde{t}_{2}}$ &=& 400 & \hspace{-4mm} GeV &
      $m_{\tilde{b}_{2}}$ &=& 280 & \hspace{-4mm} GeV \\
  $\tan\beta$ &=& 2 & &
    $\theta_{t}$ &=& $\frac{\pi}{9}$ & &
      $\theta_{b}$ &=& $\frac{\pi}{36}$ & \\
  $\alpha_{\mu}$ &=& $\frac{4 \pi}{3}$ & &
    $\beta_{t}$ &=& $\frac{\pi}{6}$ & &
      $\beta_{b}$ &=& $\frac{\pi}{3}$ & \\
  \hline
\end{tabular}
\end{center}  

Note that with a common gaugino mass at the GUT-scale all 
the low-energy gaugino mass parameters are related and are proportional 
to ${\tilde m}_2$ - the SU(2) gauginos mass term at the GUT-scale. 
For example, $m_{\tilde g} \approx 3 {\tilde m}_2$ 
(for more details see section \ref{ssec33}).



The real and imaginary parts of the top EDM and ZEDM, 
including all contributions in diagrams (a) and (b) in 
Fig.~\ref{topdipolefig5} 
(i.e., gluino, chargino and neutralino contributions), 
with the above set of the relevant SUSY parameters and 
as a function of the c.m. energy of the collider, 
$\sqrt s$ , are given in Fig.~\ref{topdipolefig6}. 
We see that, typically:

\begin{eqnarray}
\Re{\rm e}d^{\gamma,Z}_t(s)~,~\Im{\rm m}d^{\gamma,Z}_t(s) 
\sim 10^{-20} - 
10^{-19}~{\rm e-cm} \label{tdeq49}~.
\end{eqnarray}

\noindent Note that this results is about one order of 
magnitudes smaller 
than what is expected in the 2HDM discussed in section \ref{ssec43}.

Consider now the low energy MSSM scenario described in section \ref{sssec334}.
There we have taken $\alpha_{\mu} \to 0$, 
motivated by the experimental bound on the NEDM which strongly 
implies that $\alpha_{\mu} < 10^{-2} - 10^{-3}$. Moreover, all squarks except 
from the light stop were assumed to be degenerate with 
a mass $M_S$. 
It is instructive to evaluate the TDM in this limit in which the only 
relevant CP-odd phase resides in ${\tilde t}_L - {\tilde t}_R$ mixing and 
is proportional to $\sin\beta_{t}$ (or equivalently to $\sin\alpha_{t}$ - the 
phase in the top trilinear soft breaking term $A_t$). 
In this framework 
$d_{t({\tilde \chi}^+)}^{\gamma,Z}=0$ since 
the two sbottom particles are
degenerate.
Moreover,  
$d_{t({\tilde \chi}^0)}^{\gamma,Z}$ gets its contribution only from terms 
proportional to $\sin\beta_{t}$ in 
Eqs.~\ref{tdeq43}-\ref{tdeq46} where,
in general, one finds that 
$d_{t({\tilde \chi}^0)}^{\gamma,Z} < d_{t(\tilde g)}^{\gamma,Z}$. 
Thus, in this scenario the TDM can be approximated by considering only  
the gluino exchange diagram in Fig.~\ref{topdipolefig5}(b) for which only
the masses 
$m_{{\tilde t}_1},m_{{\tilde t}_2}$ and $m_{\tilde g}$ are relevant.



In Figs.~\ref{topdipolefig7}, \ref{topdipolefig8} and \ref{topdipolefig9} 
we have plotted the imaginary and real parts of 
$d_{t(\tilde g)}^{\gamma,Z}$ in the above 
MSSM scenario with $\xi_{CP}^t=1$ and with the approximation 
$d_t^{\gamma,Z} \simeq d_{t(\tilde g)}^{\gamma,Z}$,  
as a function of $\sqrt s,m_{\tilde g}$ and 
$m_{{\tilde t}_1}$ (the light stop mass), respectively. 
Our reference set of masses for these figures are
$m_{{\tilde t}_1}=50$ GeV and $m_{{\tilde t}_2}=m_{\tilde g}= 400$ GeV. 
We again see that, typically,    
$\Re{\rm e},\Im{\rm m} (d_t^{\gamma,Z}) \sim 10^{-20}-10^{-19}$ e-cm.
From Fig.~\ref{topdipolefig8}(a) we see that there is a small enhancement 
in the imaginary part of the TDM as the gluino mass gets smaller and, for 
example, we find 
$|\Im{\rm m} (d_t^{\gamma})| \simeq 3.25 \times 10^{-19}$ e-cm for 
$m_{\tilde g} =200$ GeV. 
Fig.~\ref{topdipolefig9}(b) illustrates how the TDM vanishes 
when the two stop mass eigenstates are degenerate, i.e., 
$m_{{\tilde t}_1}=m_{{\tilde t}_2}=400$ GeV.

It is important to note that within the MSSM, unlike in MHDM cases, 
the top CEDM cannot be calculated simply by replacing the 
off-shell photon with an off-shell gluon. 
The reason is that SUSY models give rise to an additional 
$g \tilde g \tilde g$ coupling (i.e., gluon-gluino-gluino coupling). 
Thus, in addition to 
replacing the photon with the gluon in 
Fig.~\ref{topdipolefig5}(b), a full calculation of 
$d_t^{g}$ has to include the additional diagram 
with $\tilde g \tilde g \tilde t$ in the loop. This effect was 
considered in the context of CP violation in $pp \to t \bar t + X$ by 
Schmidt \cite{plb293p111}, and we will return to it in 
Chapter \ref{sec7}.


\newpage
~

\newpage
~

%
%
\begin{center}
\setlength{\unitlength}{1mm}
%
%
\begin{picture}(133,120)(3,0)
\put( 7,15){\mbox{\epsfig{file=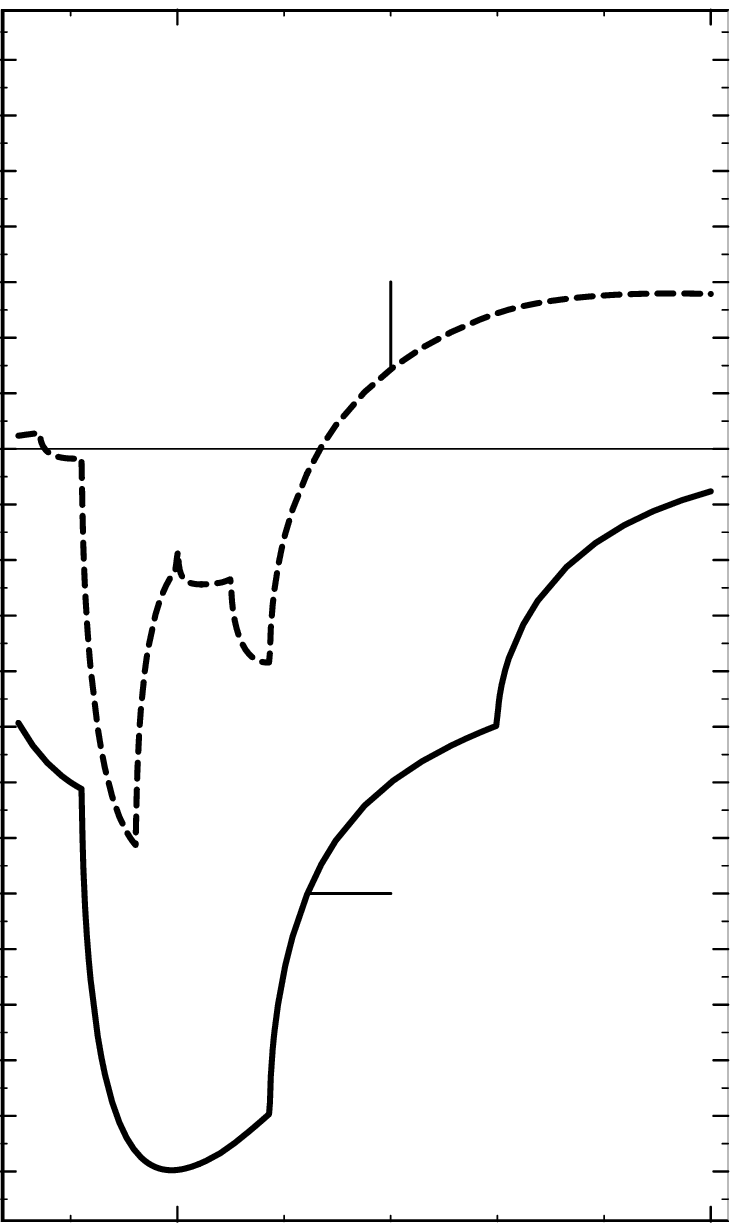,width=63mm}}}
\put( 41,100 ){\makebox(0,0)[b ]{$\Im{\rm m}{\tilde d}_t^Z$}}
\put( 42,47.5){\makebox(0,0)[ l]{$\Im{\rm m}d_t^\gamma$}}
\put(  6,123  ){\makebox(0,0)[ r]{\small  9}}
\put(  6,104.5){\makebox(0,0)[ r]{\small  }}
\put(  6, 85.5){\makebox(0,0)[ r]{\small  0}}
\put(  6, 66.5){\makebox(0,0)[ r]{\small }}
\put(  6, 47.5){\makebox(0,0)[ r]{\small -9}}
\put(  6, 28.5){\makebox(0,0)[ r]{\small }}
\put( 13  ,14){\makebox(0,0)[t]{\small 400}}
\put( 22.2,14){\makebox(0,0)[t]{\small 500}}
\put( 31.4,14){\makebox(0,0)[t]{\small 600}}
\put( 40.6,14){\makebox(0,0)[t]{\small 700}}
\put( 49.8,14){\makebox(0,0)[t]{\small 800}}
\put( 59  ,14){\makebox(0,0)[t]{\small 900}}
\put( 70, 3){\makebox(0,0)[b ]{$\sqrt{s}/$GeV}}
\put( 38, 2){\makebox(0,0)[b ]{(a)}}
%
\put(70,15){\mbox{\epsfig{file=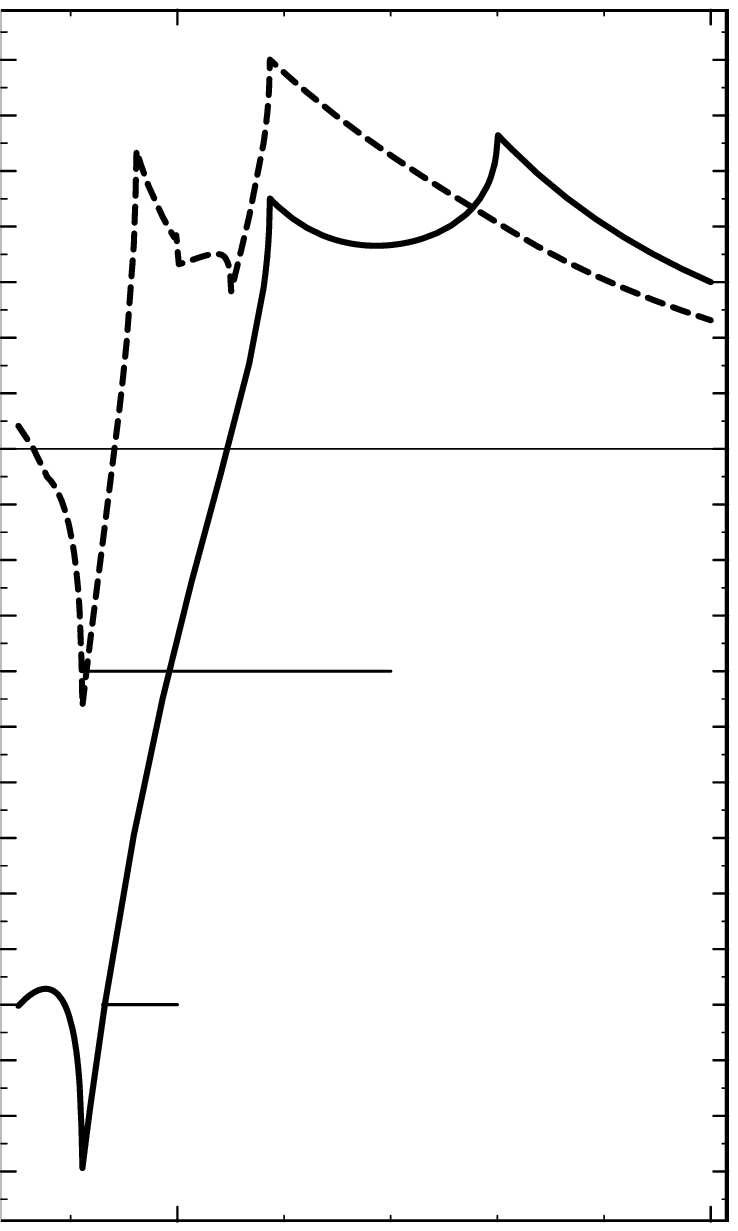,width=63mm}}}
%
\put(104,66.5){\makebox(0,0)[ l]{$\Re{\rm e}{\tilde d}_t^Z$}}
\put( 86,37){\makebox(0,0)[ l]{$\Re{\rm e}d_t^\gamma$}}
\put(134,123  ){\makebox(0,0)[ l]{\small  9}}
\put(134,104.5){\makebox(0,0)[ l]{\small  }}
\put(134, 85.5){\makebox(0,0)[ l]{\small  0}}
\put(134, 66.5){\makebox(0,0)[ l]{\small }}
\put(134, 47.5){\makebox(0,0)[ l]{\small -9}}
\put(134, 28.5){\makebox(0,0)[ l]{\small }}
\put( 76  ,14){\makebox(0,0)[t]{\small 400}}
\put( 86.2,14){\makebox(0,0)[t]{\small 500}}
\put( 94.4,14){\makebox(0,0)[t]{\small 600}}
\put(103.6,14){\makebox(0,0)[t]{\small 700}}
\put(112.8,14){\makebox(0,0)[t]{\small 800}}
\put(122  ,14){\makebox(0,0)[t]{\small 900}}
\put(101, 2){\makebox(0,0)[b ]{(b)}}
\end{picture}\\
\setlength{\unitlength}{1pt}
\end{center}
\begin{figure}[htb]
\psfull
 \begin{center}
  \leavevmode
 \end{center}
\caption{\emph{$d_t^\gamma(s)$ and ${\tilde d}_t^Z(s)$ 
(note that $d_t^Z(s) \equiv 
\frac{{\tilde d}_t^Z(s)}{2 \cos\theta_W \sin\theta_W}$)
in units of $10^{-20}$ e-cm, for the reference parameter set with 
\mbox{${\tilde m}_2 = 230$~GeV}. Note that    
in (a) \mbox{$\Im{\rm m} d_t^\gamma (s)$} (full line), 
\mbox{$\Im{\rm m} {\tilde d}_t^Z (s)$} (dashed line), 
and in (b) \mbox{$\Re{\rm e} d_t^\gamma (s)$} (full line), 
\mbox{$\Re{\rm e} {\tilde d}_t^Z (s)$} (dashed line). 
Figure taken from \cite{npb507p35}.}}
\label{topdipolefig6}
\end{figure}

\newpage
~

\begin{figure}[htb]
\psfull
 \begin{center}
  \leavevmode
  \epsfig{file=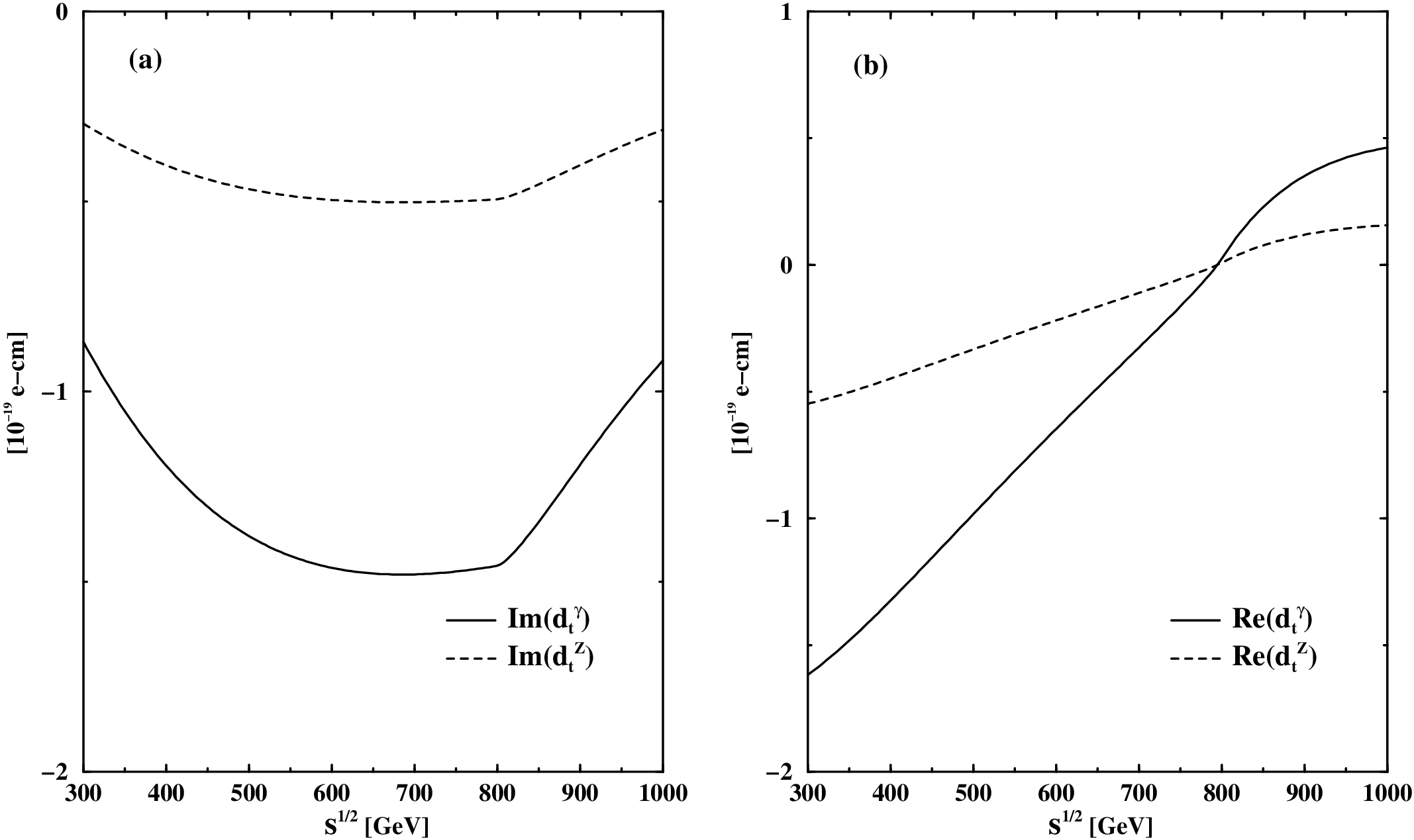
,height=9.5cm,width=9.5cm,bbllx=0cm,bblly=2cm,bburx=20cm,bbury=25cm,angle=0}
 \end{center}
\caption{\emph{Imaginary (a) and Real (b) parts of $d^\gamma_t$ and 
$d^Z_t$, due to only the gluino exchange diagram depicted 
in Fig.~\ref{topdipolefig5}(b), in 
units of $10^{-19}$ e-cm and  
as a function of $\sqrt s$. We use:  
$m_{{\tilde t}_1}=50$ GeV, $m_{{\tilde t}_2}=m_{\tilde g}=400$ GeV and  
$m_t=175$ GeV.}}
\label{topdipolefig7}
\end{figure}

\newpage
~

\begin{figure}[htb]
\psfull
 \begin{center}
  \leavevmode
  \epsfig{file=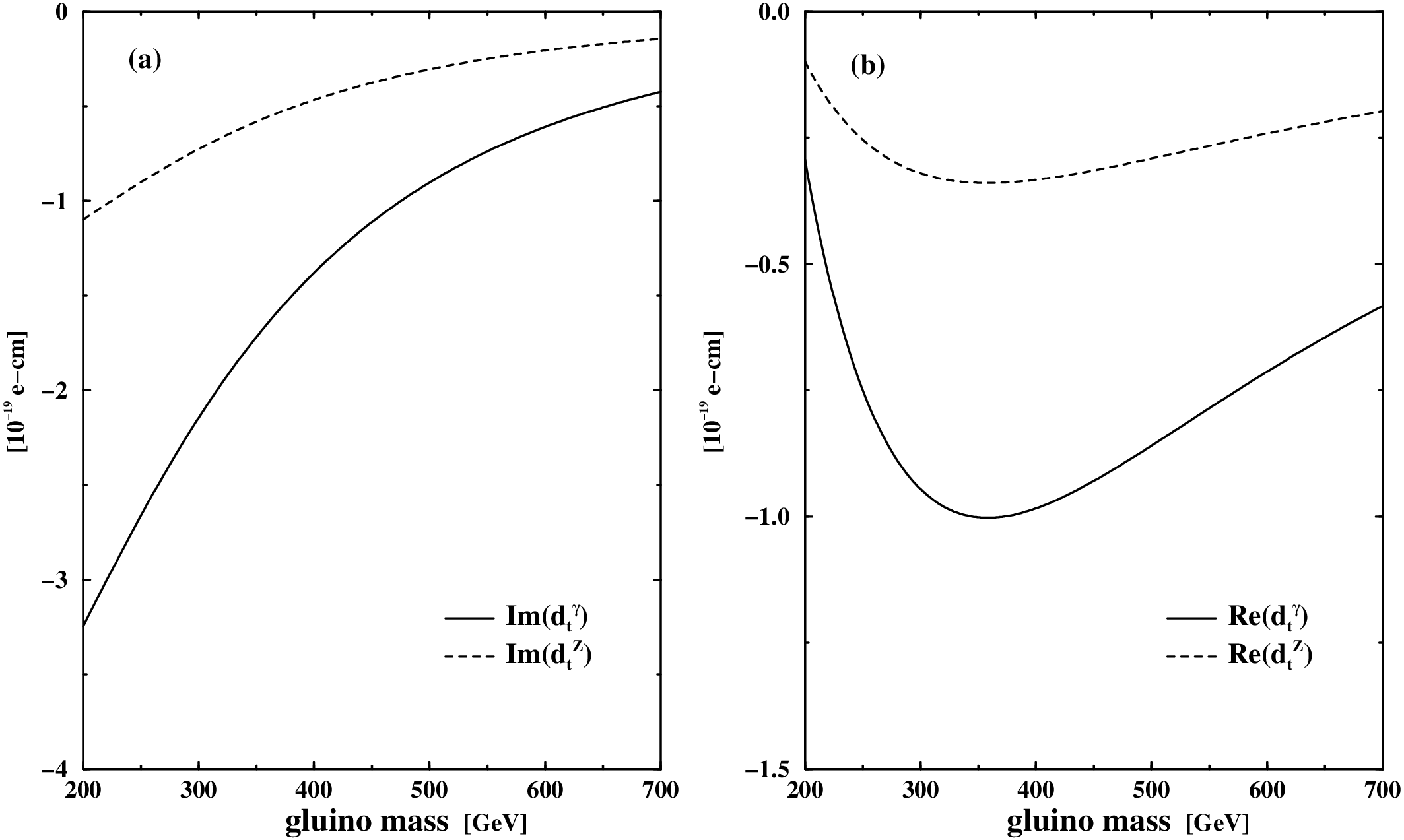
,height=9.5cm,width=9.5cm,bbllx=0cm,bblly=2cm,bburx=20cm,bbury=25cm,angle=0}
 \end{center}
\caption{\emph{Imaginary (a) and Real (b) parts of $d^\gamma_t$ and 
$d^Z_t$, emanating from only the gluino exchange diagram depicted 
in Fig.~\ref{topdipolefig5}(b), in 
units of $10^{-19}$ e-cm and 
as a function of the gluino mass $m_{\tilde g}$. We use: $\sqrt s=500$ GeV,  
$m_{{\tilde t}_1}=50$ GeV, $m_{{\tilde t}_2}=400$ GeV and  
$m_t=175$ GeV.}}
\label{topdipolefig8}
\end{figure}

\newpage
~

\begin{figure}[htb]
\psfull
 \begin{center}
  \leavevmode
  \epsfig{file=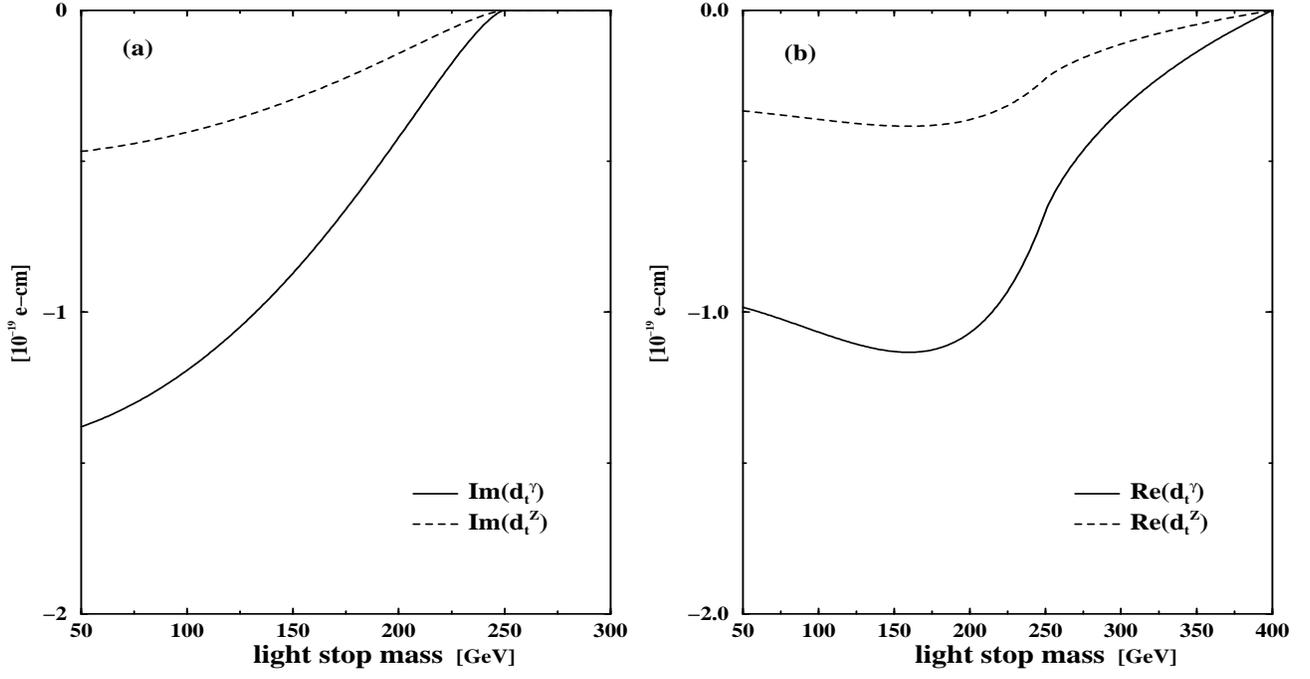
,height=9.5cm,width=9.5cm,bbllx=0cm,bblly=2cm,bburx=20cm,bbury=25cm,angle=0}
 \end{center}
\caption{\emph{Imaginary (a) and Real (b) parts of $d^\gamma_t$ and 
$d^Z_t$, emanating from only the gluino exchange diagram depicted 
in Fig.~\ref{topdipolefig5}(b), in 
units of $10^{-19}$ e-cm and 
as a function of the light stop mass $m_{{\tilde t}_1}$. 
We use: $\sqrt s=500$ GeV,  
$m_{{\tilde t}_2}=m_{\tilde g}=400$ GeV and  
$m_t=175$ GeV.}}
\label{topdipolefig9}
\end{figure}

\newpage


\subsection{Top dipole moments - summary \label{ssec46}}

In this section we have performed a detailed investigation of the 
top quark dipole moments in models beyond the SM in which 
new CP-violating phases appear rather naturally.
The models that we have considered are MHDM's and the MSSM.
In Table~\ref{dipolesumtable} we summarize our 
numerical results for the expected TDM 
within these class of models and for comparison we also write the 
expected size of the TDM in the SM.
For illustration, for each model the TDM is evaluated by setting the 
corresponding CP-violating quantities to 
plausible representative values, compatible with   
existing experimental limits if any.  

Let us also summarize below 
the salient features of the CP-violating 
mechanisms of 
these models which give rise to a non-vanishing TDM, and specify 
our choice of values used in Table~\ref{dipolesumtable}
 for the CP-violating quantities of each model.  

\begin{itemize}

\item {\bf MHDM's:}\\
In general one can distinguish between two types of Higgs mediated 
CP-violating contributions to the TDM in MHDM's:

\underline{1. TDM from neutral Higgs exchanges:}\\

In any MHDM with or without NFC and with new CP-violating phases in the 
neutral Higgs sector, the neutral Higgs-fermion-fermion 
interaction Lagrangian may be generically expressed as (say for $h$ - 
the lightest neutral Higgs)

\begin{eqnarray}
{\cal L}_{h ff}=-\frac{g_W}{\sqrt 2} \frac{m_f}{m_W} h \bar f \left( 
a_f^h + i b_f^h \gamma_5 \right) f \label{sumeq1}~.
\end{eqnarray}

\noindent The $h Z Z$ vertex which 
is also needed for calculating the TDM in the case of 1-loop neutral 
Higgs exchanges is given by (see Eq.~\ref{2hdmc})

\begin{eqnarray}
{\cal L}_{hZZ}=g_W \frac{m_Z^2}{m_W^2} c^h h g_{\mu \nu} Z^\mu Z^\nu 
\label{sumeq1b}~.
\end{eqnarray}

\noindent The CP-violating TDM then arises 
from the interference between the scalar, $a_t^h$, and the 
pseudoscalar, $b_t^h$, couplings in Eq.~\ref{sumeq1} 
and the interference between 
the scalar, $a_t^h$, and the $h Z Z$ coupling $\propto c^h$ 
in Eq.~\ref{sumeq1b}. 

The numbers in the fourth column in 
Table~\ref{dipolesumtable} are given for masses of a neutral Higgs 
in the range 100--300 GeV 
(assuming that the masses of the 
other neutral Higgs particles in these models are much heavier) 
for
\begin{eqnarray}
a_t^h=b_t^h=c^h=1~,
\end{eqnarray}
\noindent and they hold for any MHDM, i.e., 
a 2HDM of type I and II with NFC, a 2HDM of type III with FCNC in the 
neutral Higgs sector or  
for three or more Higgs doublets which have the generic 
$h t t$ interaction Lagrangian 
in Eqs.~\ref{sumeq1} and \ref{sumeq1b}.

\underline{2. TDM from charged Higgs exchanges:}\\
 
In models with three or more doublets the charged Higgs sector can acquire
new CP-odd phases which can give rise to the CP-violating TDM. 
Again, one can parameterize a generic (assumed lightest) 
charged Higgs-up quark-down quark 
CP-violating 
interaction Lagrangian, which will appear in such models, as

\begin{equation}
{\cal L}_{H^+ud}= {g_W\over \sqrt{2} m_W} K_{ud}
 \left(m_u \beta_2 \bar u_R d_L + m_d \alpha_2 \bar u_L d_R \right) +h.c.
\label{sumeq2} ~,
\end{equation}    

\noindent where $K$ is the SM CKM matrix and $u$ and $d$ denote charge
$+2/3$ and $-1/3$ quarks, respectively. The CP-violating TDM will then be
proportional to $\Im{\rm m} (V) \equiv \Im{\rm m}(\beta_2 \alpha_2^*)$. 
As mentioned before, the interaction Lagrangian in Eq~\ref{sumeq2} is not
the only source of CP violation in this class of models and one also has
to take into account the CP-violating neutral Higgs contributions arising
from the $hff$ coupling in Eq.~\ref{sumeq1}.  In fact, for the TDM we find
that the CP-odd effect from a $H^+tb$ coupling in Eq.~\ref{sumeq2} is much
smaller, i.e., typically by a factor of $\sim (m_b/m_t)^2$, than the one
from the $htt$ coupling in Eq.~\ref{sumeq1}. 

The numbers in the fifth column in 
Table~\ref{dipolesumtable} are given for masses of the charged 
Higgs in the range 200--500 GeV (again assuming that the masses of the 
other charged Higgs particles of these models are much heavier and therefore 
their contribution is negligible) and for
\begin{eqnarray}
\Im{\rm m}(V)=5~,
\end{eqnarray} 
\noindent and they represent only the charged Higgs 
contribution to the TDM in any MHDM with three or more doublets, 
which have the generic 
$H^+ tb$ interaction vertex 
in Eqs.~\ref{sumeq2}.

\item {\bf MSSM:}\\
In the MSSM, as was shown in the previous section, 
if one neglects the phase in the Higgs mass parameter $\mu$ 
(as strongly implied from the existing limit on the NEDM) 
and the small mass splitting between the left and right superpartners 
of the light quarks,  
then the dominant contribution to the TDM arises from 1-loop 
gluino exchange. In that case the TDM emanates from 
${\tilde t}_L-{\tilde t}_R$ mixing and 
is proportional to the CP-violating 
quantity

\begin{equation}
\xi_{CP}^t \equiv \sin(2 \theta_t) \sin(\beta_t)~,
\end{equation}

\noindent where the angle $\beta_t$ represents the CP-phase in the 
soft trilinear breaking term associated with the top, i.e. $A_t$.  

The numbers in the sixth column in 
Table~\ref{dipolesumtable} are given for gluino masses in the range 
200--500 GeV, for
\begin{eqnarray}
\xi_{CP}^t=1~,
\end{eqnarray} 
and for
\begin{eqnarray}
m_{{\tilde t}_1}=50~{\rm GeV}~,\\
m_{{\tilde t}_2}=400~{\rm GeV}~, 
\end{eqnarray}
\noindent where $m_{{\tilde t}_{1,2}}$ are the masses of the two 
stop mass-eigenstates.
They therefore represent only the dominant gluino   
contribution to the TDM in any low energy supersymmetric framework  
in which all squarks except from the stop are degenerate and 
the phase in the Higgs mass parameter, $\mu$, is neglected. 

\end{itemize}
\footnotesize
\begin{table}
\begin{center}
\caption[first entry]{\emph{The contribution to the 
top quark EDM ($d_t^\gamma(s)$) and ZEDM ($d_t^Z(s)$) 
form factors,  
in units of 
e-cm,  
at $s=(p_t+p_{\bar t})^2=500^2,~1000^2$ GeV$^2$, 
for the SM (where it is a purely guess-estimate) 
and for some of its extensions.
4th column shows results for  
neutral Higgs exchanges in any MHDM's 
with a CP-violating $htt$ coupling of the form 
$(g_W/\sqrt 2)(m_t/m_W)( 
a_t^h + i b_t^h \gamma_5 )$, an $hZZ$ coupling  
$g_W (m_Z^2/m_W^2) c^h g_{\mu \nu}$ 
and with $a_t^h=b_t^h=c^h=1$.
5th column is for charged Higgs exchanges 
in any MHDM's of three or more doublets
with a CP-violating $H^+tb$ coupling of the form  
$(g_W/\sqrt{2} m_W)
\left[ m_t U_t (1 + \gamma_5)/2 + m_b U_b (1 - \gamma_5)/2 \right]$ 
and with $\Im{\rm m}(U_tU_b^*) \equiv \Im{\rm m} (V)=5$.
Only the contribution from the lightest neutral or charged
Higgs is retained.
6th column shows the results for the MSSM where only the dominant 
1-loop gluino exchange diagram with gluino masses 
$m_{\tilde g}=200-500$ GeV is considered, 
in which CP violation arises from 
${\tilde t}_L-{\tilde t}_R$ mixing and is proportional 
to $\xi_{CP}^t \equiv \sin(2 \theta_t) \sin(\beta_t)$, 
where $\theta_t$ and $\beta_t$ are the angle and phase that parameterize
the ${\tilde t}_L-{\tilde t}_R$ mixing matrix. The numbers 
are given for $\xi_{CP}^t=1$ and for stop masses of 50 GeV (light stop) 
and 400 GeV (heavy stop).} 

\bigskip
\protect\label{dipolesumtable}}
\begin{tabular}{|r||r||r|r|r|r|} \cline{2-6}
\hline \hline
type of moment & $\sqrt s$ & Standard& neutral Higgs 
& charged Higgs & Supersymmetry \\
$(e-cm)~~ \Downarrow$ &$({\rm GeV}) \Downarrow$ & Model 
& $m_{h}=100 -300$ & $m_{H^+}=200 -500$ & $m_{\tilde g}=200-500$ \\ 
\hline
\hline
&500& & $(4.1-2.0)\times 10^{-19}$ & $(29.1-2.1)\times 10^{-22}$ 
& $(3.3-0.9)\times 10^{-19}$ \\
$|\Im{\rm m}(d_t^\gamma)|$& & $< 10^{-30}$ & & & \\ 
& 1000& & $(0.9-0.8)\times 10^{-19}$ &$(15.7-1.0)\times 10^{-22}$ 
& $(1.2-0.8)\times 10^{-19}$ \\ \hline
\hline
&500& & $(0.3-0.8)\times 10^{-19}$ 
& $(33.4-1.5)\times 10^{-22}$& $(0.3-0.9)\times 10^{-19}$ \\ 
$|\Re{\rm e}(d_t^\gamma)|$& & $< 10^{-30}$ & & & \\ 
& 1000 & & $(0.7-0.2)\times 10^{-19}$  
& $(0.3-2.7)\times 10^{-22}$ & $(1.1-0.3)\times 10^{-19}$ \\ \hline
\hline
&500& &$(1.1-0.2)\times 10^{-19}$ &$(15.8-2.5)\times 10^{-22}$ 
& $(1.1-0.3)\times 10^{-19}$  \\ 
$|\Im{\rm m}(d_t^Z)|$& & $< 10^{-30}$ & & & \\ 
& 1000& & $(0.2-0.2)\times 10^{-19}$& $(9.2-1.2)\times 10^{-22}$ 
&$(0.4-0.3)\times 10^{-19}$   \\ \hline
\hline
&500& & $(1.6-0.2)\times 10^{-19}$ & $(22.9-0.8)\times 10^{-22}$ 
& $(0.1-0.3)\times 10^{-19}$ \\ 
$|\Re{\rm e}(d_t^Z)|$& & $< 10^{-30}$ & & & \\ 
& 1000& &$(0.2-1.4)\times 10^{-19}$ 
&$(0.6-1.9)\times 10^{-22}$ & $(0.4-0.1)\times 10^{-19}$ \\ \hline
\hline
\end{tabular}
\end{center}
\end{table}
\normalsize
\pagebreak
\noindent We observe from Table~\ref{dipolesumtable} that the expected 
magnitudes of the  
real and imaginary parts of the top EDM and ZEDM in these models 
for masses in the loops of, typically, several hundreds GeV and 
energy scales of 500 - 1000 GeV are

\begin{eqnarray}
{\rm Neutral~Higgs:}~~&&\Im{\rm m}d_t^{\gamma,Z} \sim {\rm few} 
\times 10^{-20}-10^{-19}~{\rm e-cm}~,\\
&&\Re{\rm e}d_t^{\gamma,Z} \lsim 10^{-19}~{\rm e-cm}~,
\end{eqnarray}
\begin{eqnarray}
{\rm Charged~Higgs:}~~&&\Im{\rm m}d_t^{\gamma,Z} \sim {\rm few} 
\times 10^{-22}-10^{-21}~{\rm e-cm}~,\\
&&\Re{\rm e}d_t^{\gamma,Z} \sim {\rm few} \times 10^{-23}-10^{-21}~{\rm e-cm}~,
\end{eqnarray}
\begin{eqnarray}
{\rm Supersymmetry:}~~&&\Im{\rm m}d_t^{\gamma,Z} \sim {\rm few} 
\times 10^{-20}-10^{-19}~{\rm e-cm}~,\\
&&\Re{\rm e}d_t^{\gamma,Z} \lsim 10^{-19}~{\rm e-cm}~.
\end{eqnarray}

\noindent The top CEDM $d_t^g$, given in units of $g_s$-cm,
 can be estimated within these models 
as follows: (i) In the neutral Higgs exchange case it is simply given 
by multiplying $d_t^\gamma$ by $1/Q_t$, where $Q_t=2/3$ is the top quark 
charge. (ii) In the charged Higgs case one cannot simply replace the 
off-shell photon with an off-shell gluon  
and $d_t^g$ has to be explicitly calculated from Eq.~\ref{eqn_i}. 
Nonetheless, we find 
that  $d_t^g/(g_s$-cm) $\approx d_t^\gamma/$(e-cm).
(iii) In the MSSM case it is also not possible to extract the top CEDM from
the top EDM by the simple exchange of a photon with a gluon since 
there is an additional graph with two gluino 
propagators in the loop coming from a new $g \tilde g \tilde g$ coupling. 
We have not estimated this additional contribution here. 

To conclude this section, we have shown that in MHDM's 
with 
CP violation in the neutral or the charged Higgs sector
and in 
the MSSM,  
the TDM is always bounded to be smaller then about 
$\sim 10^{-18}$ e-cm. This is rather discouraging since, as we 
will see in the next few sections, the attainable limits on the TDM 
that can be obtained in future $e^+e^-$ and hadronic colliders seem to fall 
short by about one order of magnitude compared to the above model dependent 
expectations for these quantities.    

Basically, the strategy that we will describe in the following chapters 
for such investigations of the various TDM 
in future colliders is to incorporate      
an effective Lagrangian approach which elaborates 
new effective interactions
of a dipole moment type at the $tt \gamma,Z,g$ vertices, 
with dimensions greater than 4 which can provide 
for a model independent investigation of new physics beyond the SM. 
The effects of such phenomenological vertices can then be studied in future 
$e^+e^-$ and hadron colliders.  
Of course, a hadron collider is appropriate for the study of the 
top CEDM and is not a very good environment for studying
the top EDM and ZEDM couplings to
a photon and a $Z$-boson; 
the EDM and ZEDM will obviously be masked by the gluon
dynamics which will govern top quarks production in a hadron\-ic collider.
Therefore, a more natural place for such studies will be an $e^+e^-$ collider.
We will discuss later the feasibility of extracting 
information on the various TDM's in both hadron and $e^+e^-$ 
colliders through an investigation of CP-odd and even 
CP-even observables, e.g., 
cross-sections.   
It should be noted that the information that 
can be obtained on
a CP-odd quantity by studying its effect on a CP-even observable is much
less than what might be learned about the various EDM's 
of the top by measuring
a non-vanishing CP-odd observable driven by 
these CP-odd effective couplings.
In particular, folding into a given amplitude the various 
CP-violating EDM interaction terms,
the corresponding differential cross-section will acquire a 
CP-odd piece driven by the interference of the tree-level
process with the EDM's interactions 
(to leading order only one EDM effective
coupling has to enter in each diagram). 
Then with an appropriate CP-odd observable, 
which linearly depends on the EDM of the top, one can, in principle,
analyze directly and separately the possible CP-violating effects that can
arise from each EDM interaction in collider experiments 
such as $e^+e^-~{\rm or}~pp \to t \bar t$.
We will discuss CP-violating effects in these 
reactions in the following chapters.
\pagebreak

\newcommand{\mImag}{\,\mbox{\small $\Im$m\,}} 
\newcommand{\eReal}{\,\mbox{\small $\Re$e\,}} 

\def\gesim{\lower0.5ex\hbox{$\:\buildrel >\over\sim\:$}}
\def\lesim{\lower0.5ex\hbox{$\:\buildrel <\over\sim\:$}}
\def\emph{\it}

\def \be {\begin{equation}}
\def \ee {\end{equation}}
\def \bea{\begin{eqnarray}}
\def \eea{\end{eqnarray}}

\def \n{\noindent}


\section{CP violation in top decays \label{sec5}}
\setcounter{equation}{0}

In  this chapter we discuss CP violation in various top decays. In
particular, we will consider two-body decays, i.e., $t \to d_k W$, 
with $k=1,2,3$, the generation
index, three-body decays as well as radiative decays.  The following 
CP-violating asymmetries will be reviewed (not for all decay modes) :

\begin{itemize}
\item PRA (Partial Rate Asymmetry).
\item PIRA (Partially Integrated Rate Asymmetry).
\item Energy  asymmetry.
\item $\tau$  polarization asymmetry.
\end{itemize}

\n Although the $\tau$ polarization asymmetry tends to be the largest effect
in models with CP violation phase(s) in the charged Higgs exchanges, for the
sake of generality and completeness, we will first discuss the other effects.
Asymmetries such as top polarization asymmetry, although intimately related
to the top decays, but for which  most, if not
all discussions  in   the literature  are specific to the 
production process, are discussed in Chapters \ref{sec6}, \ref{sec7} and 
\ref{sec8}. Models included are:

\begin{itemize} 
\item SM (Standard Model).
\item  2HDM (Two Higgs Doublet Model).
\item 3HDM (Three Higgs Doublet Model).
\item MSSM (Minimal Supersymmetric Standard Model).
\end{itemize}

\n Only PRA's are considered within the  framework of all the models, while
only the predictions of the 3HDM for $t\to b\tau\nu_\tau$
are presented for all the above asymmetries.
      
In addition, for the PRA, the Form Factor (FF) approach to the $tbW$ vertex
will be presented. In this approach, the FF's can assume complex 
values, thus emulating physical cuts in higher order Feynman diagrams, leading
to non-vanishing 
CP-odd,
$T_N$-even observables such as PRA\null. This can be
contrasted~\cite{zpc56p97,prd45p124,prd56p5928}, with  the effective
Lagrangian approach, assuming that all new particles lie above $m_t$\mbox{,}
where  the coefficients are real. There, only 
$T_N$-odd CP-violating asymmetries  can emerge~\cite{prd56p5907}.

All CP-violating top decay asymmetries, within the SM, that have been
studied  so far are found to be too small to be measured. The same 
conclusion holds for CP-violating top production asymmetries. This results
from severe GIM~\cite{gim}, or even double-GIM cancellations due
to the fact that the masses of $d$, $s$, $b$ are too small compared to
the top quark mass. The obvious conclusion is that an observation of CP
violation in top quark decays, will serve as a very strong indication for 
the existence of  
new physics beyond the SM.

\subsection{Partial rate asymmetries \label{ssec51}}

In most models, the PRA, defined in Eq.~\ref{pra11} in section \ref{ssec22}
is found to be small. 
This can be readily understood, in a model-independent
way, from the CP-CPT connection discussed in section~\ref{ssec23}. Let us
consider, for example, what seems to be the main decay  of the top quark
$t \to b W^+$. Due to CPT, to have a non-vanishing PRA, at least one
additional decay channel should be available for the $t$. In other words,
in the limit that $t\to bW$ becomes the only decay channel possible, then
PRA has to vanish due to the fact that PRA then tends to become equal to
the asymmetry in the total widths of $t$ and $\bar t$ which is constrained
by CPT to vanish. In the SM, by 
virtue of $V_{tb} \simeq 1$, there is very  little competition to $t \to
b W^+$, and the PRA   turns out indeed to be tiny. Larger asymmetries are 
obtained for other decay channels, but their rates are too small to result
in an experimentally interesting signal. In models beyond the SM,
the situation is slightly better since there is a possibility for new
particles to be produced in top decays, leading to the absorptive part
necessary for PRA\null. The 
largest credibly   possible  predicted PRA, is $\sim  0.3 \%$
for $t \to b W^+$ in the MSSM with low $tan \beta$ 
which arises mainly since the top can have an appreciable decay rate into 
a 
$\tilde t \tilde \chi^0$
(i.e. the stop and neutralino final state)
in this scenario
(see below). 
However,  
once the window which allows SUSY final states in $t$
decays, such as $\tilde t \tilde{\chi}^0$ is closed, then
the PRA in top quark decays become vanishingly small in this model too, i.e., 
the MSSM.  
In  the following, we
elaborate on some of the issues mentioned above, and more.

\subsubsection{PRA in the SM \label{sssec511}}
 
\n \underline{$t \to d_k W^+$} \\

\begin{figure}[htb]
 \begin{center}
  \leavevmode
  \epsfig{file=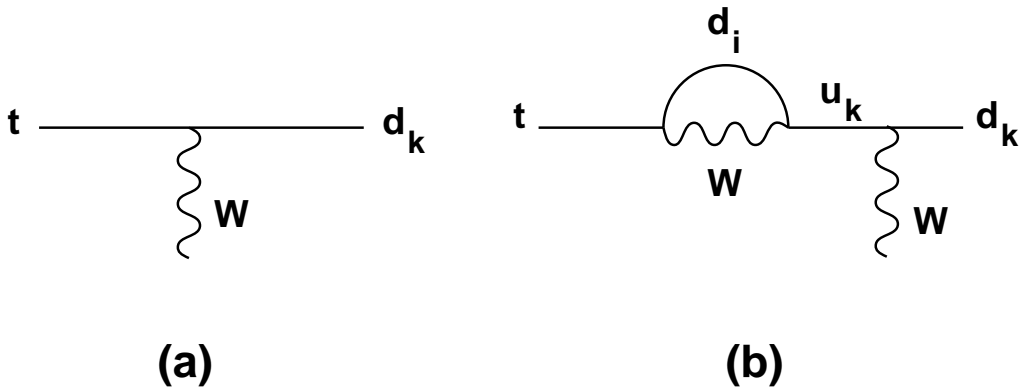
,height=5cm}
\end{center}
\caption{\emph{
(a) Tree-level diagram for $t \to d_k W$ ($k=1,2,3$ for $d,s,b$).
(b) Example of a 1-loop 
diagram for $t \to d_k W$.}}
\label{tdecaysfig1}
\end{figure}

CP violation via PRA in the process $t \to d_k W^+$,  where
$k=1,2,3$ is the generation index, was discussed
in~\cite{zpc56p97,plb319p526,prd58p095008}. The PRA results from 
interference of 
the two diagrams in Fig.~\ref{tdecaysfig1}, i.e.\ from tree$\times$loop 
interference, where
the loop is the non-diagonal $t-u_j$ self-energy. By ``interference'' we
actually mean ``the difference between interferences for the process and
its CP conjugated process''. The loop contributes the necessary imaginary
part $\mImag \Sigma(m_{i}^{2})$ through  the cut on $d_i W$. One has to sum
over $i,j$, and obviously $d_i \neq d_k$ and $u_j  \neq t$. Before
continuing to discuss the above contribution of the absorptive part to the
PRA, let us digress to show that, as stated in~\cite{plb319p526}, the
CP-CPT connection \cite{npb352p367,prd43p2909,prd43p151} 
forbids the self-energy of the $W$ from contributing to the PRA\null. This
will be shown within the SM; the proof holds  for other models too, as
can be easily generalized.

\n Consider the case $d_k=b$ (the generalization to $k=1,2$ is trivial).
The result of~\cite{plb319p526} is that the PRA in $t \to b W^+$  comes from
the interference of diagram (a) with diagram (b) in
Fig.~\ref{tdecaysfig1}, where the absorptive part is provided by the  $sW$
cut in diagram (b) (the $dW$ cut is negligible here). Thus we can symbolically
write that the result of~\cite{plb319p526} corresponds to

\be   
\Gamma(t \to b W^+)-\Gamma(\bar{t} \to \bar{b} W^-)=\eReal\: {\rm a}(b W)
\times \mImag\: {\rm b}(b W; sW\;{\rm cut})~, \label{GK1}
\ee   

\n where a, b denote the contribution of diagrams (a), (b), respectively,
and the arguments of their real, $\eReal$, and imaginary, $\mImag$, 
parts denote
the final state,  with the additional information about the relevant cut
in $\mImag$.  Note that the CKM as well as numerical factors are suppressed in
Eq.~\ref{GK1}.
Now, the rate difference in Eq.~\ref{GK1} can be written as

\bea
\Gamma(t \to b W^+)-\Gamma(\bar{t} \to \bar{b} W^-)&
=&\eReal\:1 (b u_k \bar{d}_k )\times
\mImag\:3 (b u_k \bar{d}_k ; sW\;{\rm cut}) \nonumber \\
&& +\eReal\:1 (b \ell \bar{\nu}_\ell)\times
\mImag\:3 (b \ell \bar{\nu}_\ell; sW\;{\rm cut}) ~, \nonumber \\
\label{GK2}
\eea

\begin{figure}[htb]
 \begin{center}
  \leavevmode
  \epsfig{file=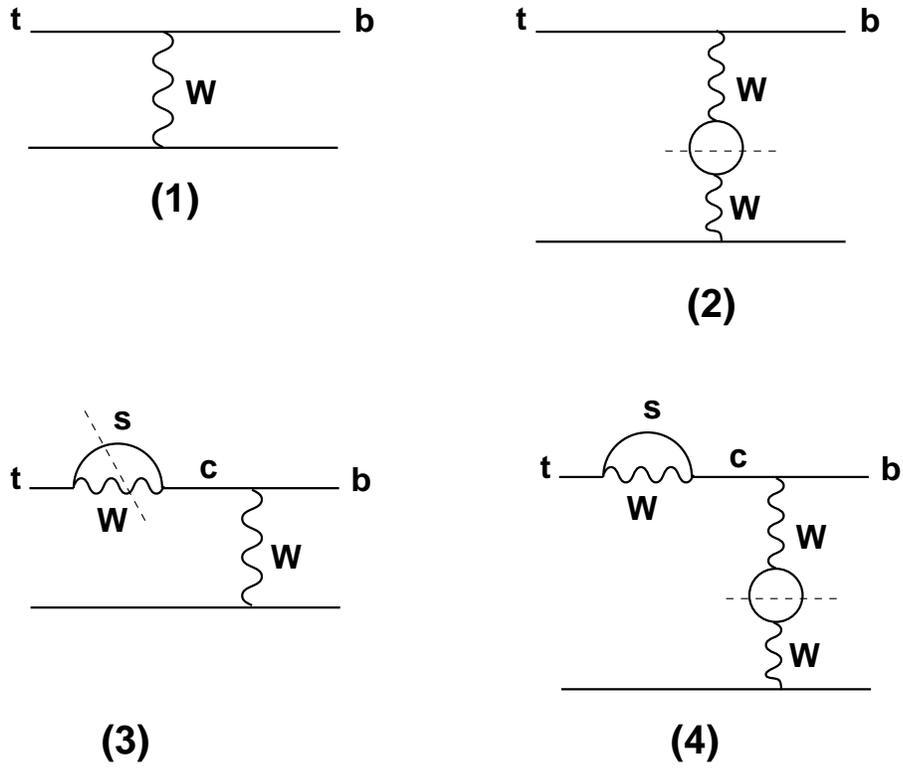
,height=10cm}
\end{center}
\caption{\emph{
The four diagrams considered in the proof that
the $W$ self-energy diagram does not  contribute to CP violation 
for  $t \to d_k W$,  within the  SM\null. Dashed lines indicate cuts. 
Similar considerations hold for other models.}}
\label{tdecaysfig2}
\end{figure}

\n where 1, 3 stand for diagrams (1), (3), respectively, in 
Fig.~\ref{tdecaysfig2}.
Also $u_k \bar{d}_k$ denote summation over all quark pair states that the $W$
can decay into, and    $\ell \bar{\nu}_\ell$ means summation over lepton
generations. Other $qW$ cuts give negligible contributions to the PRA
for $t \to b W^+$. To prove that the PRA contains no term  from the
interference of  diagram (2) with (3), and of diagram (1) with (4), we have to 
show that, after summing over all final 
states\footnote{Note that near resonance
diagrams (2) and (4) of Fig.~\ref{tdecaysfig2}
 become ${\cal O}(g^2)$ and ${\cal O}(g^4)$, respectively.}

\be
 \eReal\:3 \times \mImag\:2+ \mImag\:4 \times \eReal\:1=0  \label{GK3}~.
\ee

\n But for a specific final state and cut (say, the $b c \bar{s}$
final state, with a $\tau \bar{\nu}_\tau$ cut) in the $\mImag$'s,
there is a compensating contribution (say, the $b \tau \bar{\nu}_\tau$
final state, with a $c \bar{s}$ cut). Thus, summing, for each  
final state, over all possible cuts on the $W$ line (rescattering excluded),
then summing over final states, it is easy to show that Eq.~\ref{GK3}
holds for  $t \to b W$. We are thus left with $\eReal\:{\rm a} (bW) \times
\mImag\:{\rm b}(bW; sW\;{\rm cut})$, which is, by the way, compensated by
$\eReal\:{\rm a} (sW) \times \mImag\:{\rm b} (sW; bW\;{\rm cut})$.


Let us now recapitulate the calculation of \cite{plb319p526}, for the  
CP-violating PRA for $t \to b W$, defined as

\bea
{\cal A}_k \equiv \frac{\Gamma(t \to d_k W^+)-\Gamma(\bar{t} \to \bar{d}_k
 W^-)}  {\Gamma(t \to d_k W^+)+\Gamma(\bar{t} \to \bar{d}_k
 W^-)}  \label{GK4}~,
\eea
 
\n in the SM, with $k=3$. The  two interfering amplitudes,
Fig.~\ref{tdecaysfig1}(a) and \ref{tdecaysfig1}(b), have a relative CP-odd  
phase and Fig.~\ref{tdecaysfig1}(b) has the required  absorptive phase. 
They are given by

\bea
&&A_k^a=V_{tk}^* \hat {A}_k ~,\\
&&A_k^b=- \hat {A}_k \sum_{i,j} V_{ti}^* V_{ji} V_{jk}^* 
\frac{m_t^2}{m_t^2 - m_j^2} \Sigma(m_i^2) ~,
\eea

\n  where

\be
\hat {A}_k = -\frac{ig_W}{\sqrt 2} \bar u (p_k) \gamma_{\mu} 
L u(p_t) \epsilon^{\mu} ~.
\ee

\noindent $\epsilon^{\mu}$ is the $W$-boson polarization vector and 
$L=(1- \gamma_5)/2$. The PRA  was then found to be

\be
{\cal A}_k = -2 \sum_{i,j} 
\frac{\mImag\; ( V_{tk}^* V_{ti} V_{ji}^* V_{jk} )}{|V_{tk}|^2} 
\frac{m_t^2}{m_t^2 - m_j^2} \mImag\; \left[ \Sigma(m_i^2) \right] 
\label{aprak}~.
\ee  
     
\n Thus the effect is proportional to the Jarlskog invariant $J$ 
(recall that 
$J=|\Im{\rm m}\left(
V_{\beta j}
V_{\beta k}^*  
V_{\gamma k}
V_{\gamma j}^* \right)|$, for any 
$\beta,\gamma,j$ 
and $k$)
as
expected,   
being the only CP-violating parameter in the SM (see section~\ref{sssec312}),
and to the absorptive phase in the self-energy diagram of   
Fig.~\ref{tdecaysfig1}(b). Note that the asymmetry is doubly-GIM suppressed
being summed  over both $i$ and $j$. Therefore, with the Wolfenstein
parameterization for the CKM elements 
(see Eqs.~\ref{wolf111} and \ref{eqvwolf} in section \ref{sssec311}), 
$\mImag ( V_{tk}^* V_{ti} V_{ji}^* V_{jk} ) \sim \lambda^6$, 
one expects the leading contribution to be of the form \cite{plb319p526}:

\be 
\Gamma_k - {\bar {\Gamma}}_k \sim \lambda^6 
\frac{ \Delta m_u^2 \Delta m_d^2 }{m_W^4} ~,
\ee

\noindent where $\Delta m_u^2 \equiv m_t^2 - m_u^2$ and 
$\Delta m_d^2 \equiv m_b^2 - m_d^2$.
It was  found in~\cite{plb319p526} that

\be
{\cal A}_k= \frac{\alpha}{8 \sin^2\theta_W} 
\frac{m_b^2 m_c^2}{m_W^4}  f_W(y) \; \eta \; {\cal C}_k \label{GK15}~,
\ee

\n independent of the Wolfenstein parameter $A$. $f_W$ 
originates from the imaginary part of the self-energy and is given by

\bea
&&f_W(y)=-3 \frac{(1+y^4)}{y^6} ~,\nonumber\\
&&y \equiv m_t/m_W \label{GK14} ~,
\eea

\n and

\bea
&&{\cal C}_d = -\left[(1-\rho)^2 +\eta^2 \right]^{-1}~,\\
&&{\cal C}_s = \lambda^2~,\\
&&{\cal C}_b = -\lambda^6 \frac{m_s^2}{m_b^2} ~.
\eea

\n The largest asymmetry is obtained for $d_k=d$ , ${\cal A}_d\simeq
10^{-9}$, requiring at least $10^{23}$ top quarks! Even smaller
asymmetries are predicted for $t \to sW$ and $t \to bW$. Thus the discussion
of PRA's for  $t \to d_k W$   in the SM is of purely an academic value.
Let us note, in passing, that in the SM with four
generations  (a currently unfashionable approach) ${\cal A}_k$
may be substantially enhanced \cite{plb319p526}.\\

\n \underline{$t \to cV$}\\

The branching ratios for the rare flavor changing
processes $t \to c \gamma$ ~\cite{prd44p1473,prl65p827,prd43p268,
prd43p3591}, $t \to c g$ ~\cite{prd44p1473,prd45p178} and
 $t \to c Z$ ~\cite{prd44p1473,plb286p106}
in the SM are: $4.9 \cdot 10^{-13}, ~ 1.4 \cdot 10^{-13}$ and 
$4.4 \cdot 10^{-11}$, respectively. The CP-violating PRA's,
resulting from the interference of two penguin diagrams, 
are largest for the even rarer (by more than an order of magnitude)
decays $t \to u \gamma$ ~\cite{prd43p3591} and  $t \to u g$ ~\cite{prd45p178} 
 where they are $\simeq 0.2\%$. A slightly larger
asymmetry is obtained for  $t \to u g^{\ast} \to u \bar{u} u $
 ~\cite{prd45p178}. \\

\n \underline{Three-body $t$ decays} \\

In the above we have discussed interferences of the
type $\mImag A_{\rm loop} \times \eReal A_{\rm tree}$~,
where the loop is the off-diagonal $t-u_j$ self-energy
(in contrast, in the MSSM, see below, ``loop'' stands for ``vertex
corrections''), and 
of the type  $\mImag A_{\rm penguin} \times \eReal A^{'}_{\rm penguin}$~.
  The prime  indicates that the imaginary and real
parts of the penguins must have a different weak phase. This is possible
since three different quark amplitudes are in the penguin graph.
PRA in the SM for the six  three-body top decays of the type

\be
t \to c q \bar{q}\,\,\,\,\,{\rm and}\,\,\,\,t \to u q \bar{q}
\,\,\,,\,\,\,q=d,s,b
\label{3body}
\ee
    
\n was considered in Ref.~\cite{prl67p1979,prl68p2102}.
There,  PRA from the interference $\mImag A_{\rm tree} \times \eReal A_{\rm
penguin}$ for each of the final states in Eq.~\ref{3body} was calculated.
This interference is present in top decays due to the fact that
$m_t>m_W+m_b$~, thus endowing the tree diagram with a well-defined
CP-even phase from the $W$-width~\cite{plb245p185}. 
Though the $W$-width turns a 
``tree''
diagram into a ``loop'' diagram, we will keep loosely calling it as a
``tree''. The above interference is there in addition
to terms of the type: $\eReal A_{\rm tree} \times \mImag A_{\rm penguin}$, 
which
are analogous to  the interference \cite{prl43p242} that leads to the CP
violation in $b$ decays, such as $b \to s u \bar{u}$ (see the discussion
in section \ref{ssec23}). 
These  interferences arise since for each specific channel
$t \to q^{'} q \bar{q}$, where $q^{'}=c\,\,\, {\rm or}\,\,\, u$, 
there are two classes of possible paths:
\begin{enumerate}
\item The tree process $t\to qW$ followed by $W\to q^{'}\bar q$.
\item The 1-loop penguin process $t\to q{'} g^{\ast}$ 
followed by $g^{\ast} \to q \bar{q}$. 
\end{enumerate}

\n  The two Feynman amplitudes, tree and penguin, have different weak and
different strong  phases. Consequently, the interference of these two
channels, provides the necessary condition mentioned in section \ref{ssec22} 
for the 
observability of CP violation. Note that penguin$\times$penguin
terms are relatively small, except in the absence of tree terms.

\n The PRA for the three-body decays 
$t \to q^{'} q \bar{q}$ where $q^{'}=c\,\,\, {\rm or}\,\,\, u$
and $q=d,s,b$ is~\cite{prl67p1979,prl68p2102}:

\bea
{\cal A}&=&\frac{8g_sm_W\Gamma_W}{3\pi g_W^2m^6_t Z_V} \sum_{j=b,s}
\mImag (V^\ast_{tq} V_{q^\prime q} V_{tj} V^\ast_{q^\prime j})
 \nonumber\\
& &\times \int^{(m_t-m_q)^2}_{(m_{q^\prime}+m_{\bar q})^2} dm^2_{q^\prime \bar
q} \int^{(m^2_{q^\prime q})\max}_{(m^2_{q^\prime q})\min} dm^2_{q^\prime q}
 \nonumber\\
& &\cdot \frac{[\hat{\Gamma}_Wm_W \eReal(T_j-T_d) - K \mImag(T_j-T_d)]}{(\bar
q+q)^2 (K^2+\Gamma^2_Wm^2_W)}~, \label{pra3body}
\eea

\n where $Z_V = |V^\ast_{tq} V_{q^\prime q}|^2 (1-x) (1+x-2x^2)$
with $x=m^2_W/m^2_t$, $K=m^2_q+m^2_{q^\prime} + 2\bar q\cdot q^\prime -
m^2_W$; $T_j$ is a lengthy expression for the resulting trace, where
$j$ indicates the virtual quark exchange in the loop (for which we have 
used CKM unitarity).
Furthermore, $\hat{\Gamma}=\Gamma_W-\Gamma(W \to\ q^\prime q)$,
thus excluding rescattering~, i.e.,  $q^\prime q \not \to q^\prime q$ 
\cite{prl68p2102}. The largest   
contribution to ($T_j-T_d$) is from $j=b$, i.e., $(T_b-T_d)>> (T_s-T_d)$.
As a result, the asymmetry is completely negligible for the most abundant 
 final state, $t\to b c \bar b$ as $\mImag(V^\ast_{tb}
V_{q^\prime b} V_{tb} V^\ast_{q^\prime b})=0$. If one resurrects
$j=s$ for  $t\to b c \bar b$ , an absurdly small PRA, of the order
of $10^{-11}$~, emerges.
Indeed, despite the resonance enhancement, the asymmetries in all the final
states are very small. The largest effect is for $t\to dc\bar d$.
Even for this mode the asymmetry is too small to be of any experimental
relevance,  ${\cal A} \sim 10^{-5}$ for $m_t=180$ GeV\null.

We briefly note~\cite{prl68p2102}
that interferences of the type $\eReal$(tree)$\times\mImag$(2-loops)~,
 where the higher order loops have both a virtual $g$ and
a  virtual $W$~, for processes such as $t \to d e \nu_e$, are required
\cite{npb352p367,prd43p2909,prd43p151} to compensate the partial
width  differences for the reactions  in Eq.~\ref{3body} 
(see Fig~\ref{chapter2fig2} and the discussion in section \ref{ssec23}).
Consequently, from such mechanisms, one gets asymmetries of the order of
$10^{-7}$ for  $t \to dc \bar s$~, and there is even a tiny asymmetry    
in the SM for the leptonic decay $t \to d e^+ \nu_e$.

\subsubsection{PRA in a 2HDM \label{sssec512}}

In a 2HDM, there are 5 physical Higgs particles: 3 neutral,
${\cal H}^k$, $k=1,2,3$ and 2 charged ones, $H^\pm$. 
If one imposes discrete symmetry~\cite{prd15p1958} 
to avoid FCNC at tree-level, then there is no CP violation in the
charged Higgs sector,  (unlike the case in the 3HDM, discussed below)
and CP violation then exists only in the neutral Higgs sector 
(see section~\ref{ssec32}).
This class of 2HDM is further subdivided into Model I and Model II (see
section~\ref{sssec323}). However, there is a version of 2HDM, 
the so-called Model III (see section \ref{sssec322}),
where no discrete symmetry is imposed; the model then admits 
tree-level FCNC\null.
In Model III large tree-level FCNC may then be avoided for the light quarks
by  assuming \cite{prd35p3484} that the couplings $hq_iq_j$ are 
proportional to $(m_im_j)^{1/2}$. Note that in the class of 
2HDM's with NFC (i.e., Model I
or II), CP violation resides in flavor-diagonal ${\cal H}^k$ exchanges;
consequently CPT forbids PRA in $t \to d_k W$ to arise. 
In model III PRA in $t \to d_k W$ need not vanish, though it is expected to 
be very small due to the small 
${\cal H}^k d_i d_j \propto \sqrt {m_{d_i} m_{d_j}} 
/m_W$ coupling ($i \neq j$). \\

\n \underline{$t \to d_k W^+$} \\

The effect of a charged Higgs with the CKM phase was 
considered in \cite{plb319p526} for $t \to d_k W^+$.
Assuming  $m_{H^+}+m_i < m_t$, then, since a charged 
Higgs exchange is added to the $W$ exchange in Fig.~\ref{tdecaysfig1}(b),
one has to make the following substitution in Eq.~\ref{GK15} (where a
large $1/ \tan \beta \equiv v_d/v_u$ is assumed)

\be
f_W(y) \longrightarrow f_W(y)+\frac{f_H(x,y)}{\tan^2 \beta}~.
\ee

\n $f_W(y)$ is defined in Eq.~\ref{GK14} and 

\bea
&&f_H(x,y)=\frac{(1-x^2)^2}{x^4 y^2} ~, \nonumber\\
&&x \equiv m_t/m_H ~.
\eea

\n Taking into account the experimental bounds on $\tan \beta$, there is no
increase in the tiny PRA, ${\cal A}_k$, over what was obtained in the
SM\null. This conclusion also holds for  other observables governed  by  a
charged Higgs exchange unless the CP-violating couplings to fermions are 
different from the SM.       

\subsubsection{PRA in a 3HDM \label{sssec513}}
 
The 3HDM (Weinberg Model for CP violation 
\cite{prl37p657}), as described   in section \ref{sssec324}, can  cause 
CP violation effects, through its CP-violating couplings of a charged 
Higgs-boson to fermion pairs.  
There are no tree-level FCNC within this model.  The physical charged Higgs 
states are  $H_1^\pm$ and  $H_2^\pm$,   where, for simplicity, 
it is usually assumed that
$m_{H_1} >> m_{H_2}$,  thus decoupling $H_1$ from all predictions.
It is also easy to show that any CP violation asymmetry vanishes for
$m_{H_1}=m_{H_2}$ through a GIM-like mechanism.

In the 3HDM,
the interesting new
phenomenon  is the existence of a CP-violating  coupling $\gamma_2$ in the
leptonic term  in the Yukawa part of the Lagrangian 
(see Eq.~\ref{mhdmeq38})

\be
{\cal L}_{H^+_2}=\frac{g_W}{\sqrt{2}m_W}\left( m_t \beta_2 \bar{t}_R b_L +  
m_\tau \gamma_2 \bar{\nu}_L \tau_R \right) H_2^+ + h.c.~,  
\label{3HDMlag}
\ee

\n where $V_{tb}=1$ is assumed. Note the  proportionality to the mass
of the charged lepton, which prompted studies of CP violation in the
reaction $t \to b \tau \bar{\nu}_\tau$. The reaction 
$t \to b c \bar{s}$ is not that useful in view of the difficulty in
identifying  $c$, and especially $s$
jets. In addition, the $\tau$ mode will enable, by following the $\tau$
decay products, measurements of spin related CP-violating observables.

It is convenient~\cite{prd21p711} to parameterize CP violation in the
Yukawa couplings  with a CKM-like matrix (the SM CKM matrix itself
is assumed to be real), then $\beta_2$ and $\gamma_2$ in Eq. \ref{3HDMlag}
are given by (see Eq.~\ref{mhdmeq39})

\bea
\beta_2 & = & \frac{\tilde{c}_1 \tilde{c}_2  \tilde{s}_3-\tilde{s}_2
\tilde{c}_3 e^{i \delta_H}} {\tilde{s}_1 \tilde{c}_2}~,\label{Uu} \\
\gamma_2 & = & \frac{\tilde{c}_1 \tilde{s}_2  \tilde{s}_3+\tilde{c}_2
\tilde{c}_3 e^{i \delta_H}} {\tilde{s}_1 \tilde{s}_2}~,\label{Ul}
\eea

\n where $\tilde{s}_i\equiv\sin(\tilde{\theta}_i)$ and 
$\tilde{c}_i\equiv\cos(\tilde{\theta}_i)$,
and $\tilde{\theta}_i$, $\delta_H$ are parameters of the
model.\footnote{Note a typographical error  
in $\beta_2$ in~\protect\cite{prd21p711}.}
CP violation will be proportional to combinations such as 

\be   
\mImag (U)=\mImag (\beta_2^\ast \gamma_2)~. \label{ImU}
\ee \\

\n \underline{$t \to b \tau^+ \nu_\tau$} \\

Let us define the PRA

\be
{\cal A}_\tau=\frac{\Gamma(t \to b \tau^+ \nu_\tau)-\Gamma(\bar{t} \to \bar{b}
\tau^-\bar{\nu}_\tau)}{\Gamma(t \to b \tau^+ \nu_\tau)+\Gamma(
\bar{t} \to \bar{b}\tau^-\bar{\nu}_\tau)} \label{atbtn}~.
\ee

\n The lowest order contribution to ${\cal A}_\tau$ arises 
due  to the interference 
of the SM $W$ mediated tree diagram (see diagram (a) in 
Fig.~\ref{tdecaysfig3})
with the 3HDM $H^+$ mediated tree diagram 
(diagram (c) in Fig.~\ref{tdecaysfig3}). We assess the contribution 
of the second graph making two simplifying assumptions: 
(i) $m_{H^+} \equiv m_{H_2} << m_{H_1^+}$, this allows us to 
neglect the effect of the heavier charged
Higgs, $H_1$. Furthermore, we also assume that 
(ii) $m_{H^+} > m_t$, thus the $H^+$ width becomes irrelevant. 
${\cal A}_\tau$ will then be proportional to
$\mImag(W-{\rm tree}) \times \eReal(H-{\rm tree})$, where, in analogy to 
our previous discussion in section~\ref{sssec511}, the 
CP-violating CKM-like angular function was factored out. 
It is easy to see that,
because of the chirality miss match, only the longitudinal part of the 
$W$-propagator contributes to ${\cal A}_\tau$. 
In other words, decomposing the
$W$-propagator in the unitary gauge as

\be
D^{W}_{\mu\nu}=i \left(-g_{\mu\nu}+\frac{q_\mu q_\nu}{q^2}\right)G_T+  
i\frac{q_\mu q_\nu}{q^2}G_L ~, \label{Wprop}
\ee

\n only $\mImag G_L$ will appear in ${\cal A}_\tau$ 
\cite{prd49p289,9504386}.  
In fact ${\cal A}_\tau$, obtained from the interference between
diagrams (a) and (c) in Fig.~\ref{tdecaysfig3},  is proportional to

\be
{\cal A}_\tau \propto \mImag (U) \int_{m_{\tau}^2}^{(m_t-m_b)^2}\!f(q^2)\mImag
\hat{G}_L ~, \label{calA} 
\ee

\n where $\hat{G}_L$ indicates that $\tau\nu$ is missing from
$G_L$ to respect CPT invariance and  
$f(q^2)$ is a phase-space function.

Now, while the transverse part of the W-propagator in Eq.~\ref
{Wprop} resonates, i.e.,   

\be
G_T \simeq \frac{1}{q^2-m_W^2+i\Gamma_W m_W}~,
\label{GT}
\ee

\n for $q^2 \simeq m_W^2$, there is no such enhancement for $G_L$.
This is  one of the  reasons  for  an extremely small asymmetry,
$10^{-8}$ ~\cite{9504386} or smaller~\cite{prd51p3525},
from tree$\times$tree 
interference. The other reasons are the proportionality   
of both the Higgs coupling and $\mImag \hat{G}_L$ to small
fermion masses.\footnote{While there is agreement in the literature as
to  the above facts, there is controversy - into  which we do not enter
here (due in part to the fact that it has no observational
consequences) - regarding the form of 
$\mImag G_L$ ~\cite{prd51p3525,prd49p289,9504386,prd47p1741,ijmpa11p563} 
to  be inserted  in Eq.~\ref{calA}.}

The next logical step is to capitalize on the resonance behavior
of $\mImag G_T$  and the fact that, unlike $\mImag G_L$, it is not proportional
to small masses, by  considering interferences that are higher  order in  the 
weak interaction coupling constant~\cite{prl70p1364}. Thus, PRA  from
interferences of the type $\mImag {\rm (a)} \times \eReal {\rm (b)}$ and 
$\eReal {\rm (c)} \times \mImag {\rm (d)}$,  is calculated; (a)-(d) 
denote the diagrams
in Fig.~\ref{tdecaysfig3}, where  (b) and (d) 
represent  all box diagrams. Bremsstrahlung is included, but 
diagrams yielding ${\cal A}_\tau \to 0$ for $G_L\to 0$, see
below, were neglected. The asymmetry is then given by

\be
{\cal A}_\tau \simeq \frac{\left[\mImag (U) / 512\pi^3m^3_t \right]
\int dq^2 du \mImag
\hat G_T \eReal ({\rm tree} \times {\rm box})}{\Gamma (t\to bW\to
b\tau\nu)}~,
\ee

\n where $\mImag (U)$ is defined in Eq.~\ref{ImU} and  

\bea
u & = & (p_\tau+p_b)^2 \quad , \quad q^2=(p_\nu
+p_\tau)^2~,\label{uq^2} \\
\mImag \hat G_T & = & - \frac{m_W\hat\Gamma_W}{(q^2-m^2_W)^2+ (\Gamma_W
m_W)^2} ~, 
\eea

\n with

\be
\hat\Gamma_W = \Gamma_W -\Gamma_{W \to \tau\nu}~.
\ee
 
\n The maximal value of ${\cal A}_\tau$ turns out to be negative, and  of 
order $10^{-5}$. Subsequently, the following contributions to 
${\cal A}_\tau$,  
explicitly neglected in ~\cite{prl70p1364}, were calculated 
in~\cite{9504386}:

\begin{itemize}

\item Since the integration in Eq.~\ref{calA} reaches up to 
$(m_t-m_b)^2$, it includes a region with $q^2 > m_W^2$,
for which a $W-\gamma$ loop in $G_L$ has to be 
taken into account.             

\item Imaginary parts can also appear in box and vertex
diagrams corresponding to $t \to b W$.  

\end{itemize}

\n Both new terms are non-resonant and do not suffer
small mass suppression from fermion loops in $G_L$.   They turn out to give 
a large correction, of about 50\% and of the same sign, as
compared to the value of  ${\cal A}_\tau$ calculated in \cite{prl70p1364}
using only $G_T$. The minimal number of $t$-quarks required to  observe CP
violation in  PRA within the 3HDM is - although 
many orders of magnitudes larger than the best leptonic
SM result (i.e., $t \to d \ell \nu$ \cite{prl67p1979}) - of the
order  of  $10^{10} - 10^{11}$ and thus not very promising.    
As we will see later, one can do much better in the 3HDM by
considering PIRA rather than PRA\null.

\subsubsection{PRA in the MSSM \label{sssec514}}

An extremely interesting possibility, investigated in 
~\cite{prd57p1495,plb319p526,plb320p299,hepph9904228}, is that the 
CP-violating PRA in
two-body modes  (that was found to be extremely small in the SM, 2HDM and
in the 3HDM)  
may receive appreciable contribution from new SUSY CP-odd phases. 
For example, consider the PRA ${\cal A}_3$ in Eq.~\ref{GK4} for the 
main top decay $t \to bW$; the Feynman diagrams that can potentially
contribute to ${\cal A}_3$ in the MSSM are depicted 
in Fig.~\ref{tdecaysfig4}. 

Recall that since ${\cal A}_3$ is $T_N$-even it requires an absorptive
phase in the Feynman amplitude. This necessitates radiative corrections to
the $t\to bW$ to at least 1-loop order and, in particular,
 the SUSY particles 
exchanged in the loops have to be light enough such that absorptive cuts 
will arise. Of course, in addition to the strong phase from 
FSI (Final State Interactions), 
a CP-odd phase is needed. We recall that, 
in the MSSM, with the most general boundary 
conditions for the soft breaking parameters at the scale where they 
are generated and ignoring generation mixing,
only three places remain in the SUSY
Lagrangian that can give rise to CP phases that cannot be rotated away:
The superpotential contains a complex coefficient $\mu$ in the term
bilinear in the Higgs superfields and the soft-supersymmetry breaking
operators introduce two further complex terms, the gaugino masses $\tilde m$
and the left and right-handed squark mixing terms. The latter, being 
proportional to the trilinear soft breaking terms (i.e., the $A_q$ terms)
and to $\mu$, may be complex in general (for more details see 
section \ref{ssec33}). It is clear then that,
in general, 
there are many sources of CP-violating phases. Therefore, reliable
predictions cannot be made unless we make some simplifying assumptions.

Let us first describe a convenient way to derive the PRA ${\cal A}_3$.  
Following \cite{prd57p1495}, 
the $t \to bW^+ $ and $\bar t \to \bar bW^-$ decay vertices 
can be parameterized as follows

\bea
J^{\mu(t)}_ k & \equiv & i \frac{g_W}{\sqrt 2} \sum_{P=L,R} 
\bar{u}_b \left( \frac{{\cal D}_{1(k)}^P p_t^{\mu}}{m_t} + 
{\cal D}_{2(k)}^P \gamma^{\mu} \right)
Pu_t \label{jt}~, \\
J^{\mu(\bar t)}_ k & \equiv & i \frac{g_W}{\sqrt 2} \sum_{P=L,R} 
\bar{v}_t \left( \frac{{\bar {\cal D}}_{1(k)}^P p_t^{\mu}}{m_t} + 
\bar{{\cal D}}_{2(k)}^P \gamma^{\mu} \right)
Pv_b \label{jbart}~, 
\eea

\n where ${\cal D}_{1(k)}^{L,R}$ and ${\cal D}_{2(k)}^{L,R}$ 
defined in Eqs.~\ref{jt} and \ref{jbart},  contain the CP-violating
phases as well as the  absorptive phases of the decay diagram $(k)$
($k=a,b,c$ or $d$ corresponding to diagrams (a),(b),(c) or (d) 
in Fig.~\ref{tdecaysfig4}). 
The important contributions to the
${\cal D}$'s above are likely to come from those diagrams in which one of
the two on-shell superparticle is the Lightest Supersymmetric Particle (LSP),
e.g., the neutralino in our case. Such is the case for diagrams (b) and (d) in 
Fig.~\ref{tdecaysfig4}. Also, with a very light stop (i.e., $\sim 50$ GeV)
an absorptive cut  can arise from diagram (a) in Fig.~\ref{tdecaysfig4} 
if the gluino mass is below $\lesim 130$~GeV\null. The current
experimental bounds  on the superparticles involved in the loop of diagram
(c) in  Fig.~\ref{tdecaysfig4} 
are already stringent enough that they are unlikely to have absorptive
parts for $m_t\sim175$ GeV~\cite{susylimits}.

Let us now write ${\cal A}_3$ in the most general case with 
no assumptions on the masses of the SUSY particles and taking 
into account all four diagrams in Fig.~\ref{tdecaysfig4} 
\footnote{In pages to follow
we will evaluate ${\cal A}_3$ within a plausible set of the low energy 
SUSY parameter space.}. In terms of the 
scalar $({\cal D}_{1(k)})$ and vector $({\cal D}_{2(k)})$ form factors, the 
PRA ${\cal A}_3$ is given by

\be
{\cal A}_3 = \sum_k \left[ \frac{(x-1)}{2(x+2)} 
\eReal({\cal D}_{1(k)}^R + {\bar {\cal D}}_{1(k)}^L) + 
\eReal({\cal D}_{2(k)}^L - {\bar {\cal D}}_{2(k)}^L) 
\right] \label{adecay}~,
\ee

\n where $x \equiv m_t^2/m_W^2$ and the sum is carried out over all decay
diagrams in Fig.~\ref{tdecaysfig4}  (i.e., $k={\rm a,b,c}$ and d).
It is easy to show that if one defines

\bea
{\cal D}_{1(k)}^R &\sim& e^{i \delta_s^{1(k)}} 
\times e^{i \delta_w^{1(k)}} \label{relation1}~,\\
{\cal D}_{2(k)}^L &\sim& e^{i \delta_s^{2(k)}} 
\times e^{i \delta_w^{2(k)}} \label{relation2}~,
\eea

\n where $\delta_s^{1(k)},\delta_s^{2(k)}$ are the CP-even 
absorptive phases (i.e., FSI phases) 
and $\delta_w^{1(k)},\delta_w^{2(k)}$ are the CP-odd 
phases associated with diagrams (a)--(d) in Fig.~\ref{tdecaysfig4}, then

\bea
{\bar {\cal D}}_{1(k)}^L &\sim& - e^{i \delta_s^{1(k)}} \times 
e^{-i \delta_w^{1(k)}} \label{relation3}~,\\
{\bar {\cal D}}_{2(k)}^L &\sim& 
e^{i \delta_s^{2(k)}} \times e^{-i \delta_w^{2(k)}} 
\label{relation4}~.
\eea 

\n We then get for the scalar form factors in Eq.~\ref{adecay}

\bea   
\eReal({\cal D}_{1(a)}^R + {\bar {\cal D}}_{1(a)}^L) &=& 
- \frac{8}{3} \frac{\alpha_s}{\pi} m_t m_{\tilde{g}}  
{\cal O}_a^1 \Im{\rm m}C^a_{12} \label{aalrd}~, \\
\eReal({\cal D}_{1(b)}^R + {\bar {\cal D}}_{1(b)}^L) &=& 
- \frac{\alpha}{\pi \sin^2\theta_W} m_t \left[   
m_t {\cal O}_b^1 \Im{\rm m}\left( C^b_{22} - C^b_{23} \right) 
\right. \nonumber \\ 
&& + \left. m_{\tilde{\chi}_n^0} {\cal O}_b^2 \Im{\rm m}C^b_{12} \right] ~,\\
\eReal({\cal D}_{1(c)}^R + {\bar {\cal D}}_{1(c)}^L) &=& 
\frac{\alpha}{\pi \sin^2\theta_W} m_t \left[ 
m_t {\cal O}_c^1 \Im{\rm m}\left( C^c_{23} - C^c_{22} \right) 
\right. \nonumber \\
&& \left. - 
m_{\tilde{\chi}_m} {\cal O}_c^2 \Im{\rm m}\left( C^c_{11} - C^c_{12} 
\right) \right. \nonumber \\ 
&& \left. + m_{\tilde{\chi}_n^0} {\cal O}_c^3 
\Im{\rm m}\left( C^c_{0} + C^c_{11} \right) \right]~, \\
\eReal({\cal D}_{1(d)}^R + {\bar {\cal D}}_{1(d)}^L) &=&
\eReal({\cal D}_{1(c)}^R + {\bar {\cal D}}_{1(c)}^L) \left( 
m_{\tilde{\chi}_n^0} \to - m_{\tilde{\chi}_m},
 m_{\tilde{\chi}_m} \to m_{\tilde{\chi}_n^0},\right. \nonumber\\ 
&&\left. {\cal O}_c^i \to {\cal O}_d^i, 
\Im{\rm m}C^c_{ij} \to \Im{\rm m}C^d_{ij} \right) ~,
\eea   

\n while the vector form factors in Eq.~\ref{adecay} are given
by\footnote{We note that in \cite{prd57p1495} there is a misprint in one of
the terms   proportional to $m_W^2$ in the form factor 
$\eReal({\cal D}_{2(c)}^L - {\bar {\cal D}}_{2(c)}^L)$.  
The correct form of this term is given in Eq.~\ref{mistake}.}

\bea
\eReal({\cal D}_{2(a)}^L - {\bar {\cal D}}_{2(a)}^L) &=& 0 ~, \\
\eReal({\cal D}_{2(b)}^L - {\bar {\cal D}}_{2(b)}^L) &=& 
- \frac{\alpha}{\pi \sin^2\theta_W} {\cal O}_b^1 \Im{\rm m}C^b_{24} ~, \\ 
\eReal({\cal D}_{2(c)}^L - {\bar {\cal D}}_{2(c)}^L) &=& 
\frac{1}{2} \frac{\alpha}{\pi \sin^2\theta_W} \left[  
{\cal O}_c^1 \left( m_t^2 \Im{\rm m}\left( C^c_{22} - C^c_{23} \right) 
\right. \right. \nonumber \\
&& \left. \left. + 
m_W^2 \Im{\rm m}\left( C^c_{11} + C^c_{21}-
C^c_{12} - C^c_{23} \right) + 2 \Im{\rm m}C^c_{24} \right)
\right. \nonumber\\ 
&&\left.  
+ m_t m_{\tilde{\chi}_m} {\cal O}_c^2  
\Im{\rm m}\left( C^c_{11} - C^c_{12} \right) \right. \nonumber\\ 
&&\left. - m_t m_{\tilde{\chi}_n^0} {\cal O}_c^3
\Im{\rm m}\left( C^c_0 + C^c_{11} - C^c_{12} \right) \right. \nonumber\\   
&& \left. -  m_{\tilde{\chi}_m} m_{\tilde{\chi}_n^0} 
{\cal O}_c^4 \Im{\rm m}C^c_0 \right]  \label{mistake}~, \\
\eReal({\cal D}_{2(d)}^L - {\bar {\cal D}}_{2(d)}^L) &=& 
\eReal({\cal D}_{2(c)}^L - {\bar {\cal D}}_{2(c)}^L) \left( 
m_{\tilde{\chi}_n^0} \to - m_{\tilde{\chi}_m}, 
m_{\tilde{\chi}_m} \to m_{\tilde{\chi}_n^0}\right.,\nonumber\\ 
&& \left.{\cal O}_c^i \to 
{\cal O}_d^i, \Im{\rm m}C^c_{ij} \to \Im{\rm m}C^d_{ij} \right) \label{dalrd}~.
\eea

\n Here $\Im{\rm m}C^k_x$, $x \in \left\{0,11,12,21,22,23,24 \right\}$ and 
$k=a-d$,
are the 
imaginary parts, i.e., absorptive parts,  
of the three-point form factors associated with the
1-loop integrals in diagrams (a)-(d) in Fig.~\ref{tdecaysfig4}.
The $C^k_x$ are given by \cite{prd57p1495}:

\bea
C^a_x &=& C_x(m_{{\tilde b}_j}^2,m_{{\tilde t}_i}^2,m_G^2,m_W^2,m_t^2,m_b^2)~, \\
C^b_x &=& C_x(m_{{\tilde b}_j}^2,m_{{\tilde t}_i}^2,m_{{\tilde\chi}_n^0}^2,m_W^2,m_t^2,m_b^2)~, \\
C^c_x &=& C_x(m_{{\tilde\chi}_n^0}^2,m_{{\tilde\chi}_m}^2,m_{{\tilde b}_j}^2,m_W^2,m_t^2,m_b^2)~, \\
C^d_x &=& C_x(m_{{\tilde\chi}_m}^2,m_{{\tilde\chi}_n^0}^2,m_{{\tilde t}_i}^2,m_W^2,m_t^2,m_b^2)~,
\eea

\ni and $C_x(m_1^2,m_2^2,m_3^2,p_1^2,p_2^2,p_3^2)$ is defined in appendix A.
The indices $i,j=1,2$ stand for the two stop, sbottom mass eigenstates, 
respectively, and $m=1,2$ and $n=1-4$ correspond to 
the two charginos and four neutralinos mass eigenstates, respectively;
also, $m_G$ is the gluino mass.   

The ${\cal O}_k^i$'s in Eqs.~\ref{aalrd}-\ref{dalrd} contain the SUSY 
CP-odd phases for the decay
diagrams and they were given  in \cite{prd57p1495}. 
There, also the required
Feynman  rules for calculating the above PRA were given.
For example, ${\cal O}_d^i$, for $i=1-4$, containing the SUSY CP-odd phases
which appear in diagram (d)\footnote{As will be shown below, in our case 
this diagram will give rise to the leading contribution to ${\cal A}_3$.} 
are

\bea
{\cal O}_d^1 &=& -\Im{\rm m}\left(K^- M^1 \right) \label{od1chap5}~, \\
{\cal O}_d^2 &=& \Im{\rm m}\left(K^- M^2 \right)~, \\
{\cal O}_d^3 &=& \Im{\rm m}\left(K^+ M^2 \right)~, \\
{\cal O}_d^4 &=& -\Im{\rm m}\left(K^+ M^1 \right) \label{od4chap5} ~.
\eea

\ni Here we have defined

\bea
K^+ &\equiv& Z_N^{2n*} Z_{1m}^- + \frac{1}{\sqrt 2} Z_N^{3n*} Z_{2m}^-
\label{kp} ~, \\ 
K^- &\equiv& Z_N^{2n} Z_{1m}^{+*} - \frac{1}{\sqrt 2} Z_N^{4n} Z_{2m}^{+*}
\label{km} ~, \\
M^1 &\equiv& \frac{1}{\sqrt 2} \frac{m_t}{M_W \sin\beta} \left( Z_{2m}^{+} 
L^{+*} \xi_t^{i*} - \sqrt 2 Z_{1m}^{+} Z_N^{4n*} \xi_t^i \right) \nonumber \\
&& + \frac{1}{\sqrt 2} \left( \left(  \frac{m_t}{M_W \sin\beta} \right)^2
|Z_t^{2i}|^2 Z_{2m}^{+} Z_N^{4n*} -  
\sqrt 2 |Z_t^{1i}|^2 Z_{1m}^{+} L^{+*} \right) ~, \nonumber \\
\\
M^2 &\equiv& \frac{1}{\sqrt 2} \frac{m_t}{M_W \sin\beta} \left( \frac{4}{3}
\tan\theta_W |Z_t^{2i}|^2 Z_{2m}^{+}  
Z_N^{1n} + \sqrt 2 |Z_t^{1i}|^2 Z_{1m}^+ Z_N^{4n} \right) \nonumber \\
&& - \left( \frac{1}{\sqrt 2} \left(  \frac{m_t}{M_W \sin\beta} \right)^2
Z_{2m}^{+} Z_N^{4n} \xi_t^{i*} +  
\frac{4}{3} \tan\theta_W  Z_{1m}^{+} Z_N^{1n} \xi_t^i \right) ~, 
\nonumber \\
\eea

\ni and 

\bea
\xi_t^i &\equiv& Z_t^{1i*} Z_t^{2i} \label{chap5xit}~, \\
L^{\pm} &\equiv& \frac{1}{3} \tan\theta_W Z_N^{1n} \pm Z_N^{2n} \label{lpm}
~.
\eea

\ni In Eqs.~\ref{kp}-\ref{lpm}, $Z_t,~Z_N$ and $Z^-,~Z^+$ are the mixing 
matrices of the stops, neutralinos and charginos, respectively 
(i.e., with indices $i,n$ and $m$), which are 
defined in section \ref{sssec332}.

Obviously, to obtain an estimate of the numerical value of the asymmetry,
one needs to know the definite form for the mixing matrices and various
other parameters. Not knowing these makes it very difficult to give a
reliable quantitative prediction for the asymmetry. 
Therefore, one has to choose a reference set of the SUSY spectrum 
subject to theoretically motivated assumptions as well as 
experimental data. Such a reference set which constructs a plausible 
low energy MSSM framework was described in \cite{prd57p1495} 
(and is also described in section \ref{sssec334}). 
The key assumptions made there are:

\begin{itemize} 

\item There is an underlying grand unification which leads to the 
relation in Eq.~\ref{mhalf} between $U(1)$ and $SU(2)$ gaugino masses and 
the gluino mass. 

\item All squarks except  the lighter stop 
(with a mass denoted hereafter by $m_l$) are degenerate with a mass 
$M_S$; in the analysis below we set $M_S=400~GeV$. 

\item The gluino mass is varied subject to $m_G>250~GeV$~\cite{susylimits}.

\item The parameters are chosen subject to the upper limit on the NEDM, 
$d_n < 1.1 \times 10^{-25}$ e-cm~\cite{prd541}. 
In particular, the Higgs parameter $\mu$ is chosen to be real 
as strongly implied from this upper bound on the NEDM when the 
squark masses are below $\sim 1$ TeV\null. 

\end{itemize}

\n With the above criteria one is left with only one CP-odd phase arising from 
${\tilde t}_L - {\tilde t}_R$ mixing. That is, 
when $\mu$ is real
all the elements in ${\cal O}_k^i$ above except from $\xi_t^i$, defined in 
Eq.~\ref{chap5xit}, are real. 
Recall that the stop squarks of different
handedness are related to their mass eigenstates $\tilde t_+$, $\tilde
t_-$ through the following transformations (see section \ref{sssec332})

\bea
\tilde t_L & = & \cos \theta_t \tilde t_- - e^{-i\beta_t} \sin\theta_t
\tilde t_+ ~, \nonumber \\
\tilde t_R & = & e^{i\beta_t} \sin\theta_t \tilde t_- + \cos \theta_t
\tilde t_+ \label{pra510}~.
\eea

\n The asymmetry is thus proportional to the quantity (see also 
Eqs.~\ref{xit} and \ref{tdeq33})

\be
\xi_{CP}^t \equiv 2 |\xi_t^i| =  \sin(2 \theta_t) \sin(\beta_t) \label{xicptchap5}~,
\ee

\ni where $\xi_t^i$ is defined in Eq.~\ref{chap5xit}.

Although the CP-odd phases in the squarks sector generate the 
NEDM, the resulting restrictions on the 
CP-phases in the $\tilde t_L-\tilde t_R$ mixing are rather weak.  
As we
have demonstrated in section \ref{sssec334},  the main contribution to the 
NEDM (when $\mu$ is real) comes from the mixing of the
superpartners of the  lighter squarks.  Therefore, if the trilinear soft
breaking terms $A_u,A_d$ and $A_t$ are not  correlated at the EW-scale,
as is the case in our low energy MSSM framework 
described in section \ref{sssec334}, then  
it is not unreasonable to study the effects of maximal CP
violation in the stop  sector, i.e., $\xi_{CP}^t = 1$ without contradicting
the current limit  on the NEDM\null. 

With no further assumptions, the reference parameter set consists 
of $M_S,m_l,m_G,\mu,\tan\beta$ and $\xi_{CP}^t$. 
The neutralinos and charginos masses are extracted by diagonalizing 
the corresponding mass matrices which are functions of 
$\mu,m_G$ and $\tan\beta$ (see section \ref{sssec332}).   
Note that the consequences of such a low energy
MSSM scenarios  on the various diagrams in Fig.~\ref{tdecaysfig4}
 that can potentially contribute to the PRA, ${\cal A}_3$, are: 

\begin{itemize}
\item For $m_G \gesim 250$ GeV diagram (a) does not have the needed 
absorptive cut and thus does not contribute to the PRA\null.
\item Diagram (c) does not have a CP-violating phase 
when ${\rm arg}(\mu)=0$.
\end{itemize}

\n This simplifies our discussion to a great extent and we are therefore
left with only two diagrams that can contribute  to ${\cal A}_3$. These are
diagrams (b) and (d),  where in fact we find that, by far, 
the leading contribution comes from diagram (d). In particular, we have
calculated the PRA effect, ${\cal A}_3$, arising from diagrams (b) and
(d), 
for
$\arg(\mu)=0$, $m_{\tilde q}=M_S=400$ GeV 
and subject to
$m_l  > 50$ GeV,  $m_G > 250$ 
GeV, the LSP (in our case the neutralino) mass to be above 20 GeV and
the mass of the lighter chargino to be above 65 GeV\null.

In Figs.~\ref{tdecaysfig5} and \ref{tdecaysfig6}  we plot ${\cal A}_3$ for
two values of $\tan\beta$ which  correspond to a low ($\tan\beta=1.5$) and
high ($\tan\beta=35$) $\tan\beta$  scenarios, where the SUSY mass parameters
are varied subject to all the above  constraints and maximal CP violation is
taken in the sense  that $\xi_{CP}^t = 1$, thus presenting ${\cal A}_3$ in
units  of $\sin2\theta_t \sin\beta_t$. In particular, in
Figs.~\ref{tdecaysfig5}(a) and \ref{tdecaysfig5}(b)  we plot the asymmetry as
a function  of $\mu$ for several values of $m_G$ and for $\tan\beta=1.5$
and $\tan\beta=35$, respectively. In Figs.~\ref{tdecaysfig6}(a) and 
\ref{tdecaysfig6}(b) the asymmetry is plotted as a function of the gluino
mass $m_G$ for several values of $\mu$ and for $\tan\beta=1.5$ and
$\tan\beta=35$, respectively.  In both figures we set $M_S=400$ GeV and
$m_l=50$ GeV\null.   Evidently, from Figs.~\ref{tdecaysfig5} and 
\ref{tdecaysfig6} we see that a PRA in $t \to b W$ 
is very small over the whole range
of our SUSY parameter space. In particular, we  always find

\be
|{\cal A}_3| < 0.3\% ~.
\ee

\n Of course the asymmetry further drops as the mass of the lighter 
stop, $m_l$, is increased and vanishes when $m_l \gesim 130$ GeV since in 
that  case there is no absorptive cut in the relevant contributing 
diagrams. Also, we find that the PRA is almost insensitive to $\tan\beta$
in the range  $\tan\beta \gesim 10$ and that  ${\cal A}_3 \sim 0.3\%$ become
possible only for  $\tan\beta \sim {\cal O}(1)$.

The asymmetry we find is therefore somewhat small compared to the estimates
of Grzadkowski and Keung (GK) \cite{plb319p526} and  of Christova and
Fabbrichesi (CF) \cite{plb320p299}.  In the GK limit only
the gluino exchange of diagram (a) was considered. They utilized the 
CP-violating, quark-squark-gluino interaction, occurring with coupling
strength $g_s$ (the QCD coupling) and the $W {\tilde t} {\tilde b}$ 
interaction

\bea
{\cal L}_{\tilde qq\lambda} & = & i\sqrt{2} g_s [ \tilde t^\ast_L T^a
(\bar\lambda^a t_L) + \tilde t^\ast_R T^a (\bar\lambda^a t_R)] 
+ (t\leftrightarrow b) \nonumber \\
& & - \frac{ig_W}{\sqrt{2}} V_{tb} \tilde b^+_L 
\stackrel{\leftrightarrow}{\partial}_\mu \tilde t_L
W^{-\mu} + \hbox{h.c.} \label{pra9}~.
\eea

\n As in our case, the most important source of CP violation is then the
phase in the $\tilde t_L-\tilde t_R$ mixing and, therefore, their effect
is also proportional to $\xi_{CP}^t$ defined above. However, the GK limit
is applicable only if  $m_t > m_G + m_{\tilde t}$, so that an absorptive cut
can  occur in diagram (a). In the best case, GK found a $\sim 1\%$
asymmetry for $m_G=m_{\tilde b}=100$ GeV\null.  

On the other hand, in the CF limit, numerical results were
given only for the neutralino exchange diagram (i.e., diagram (b)) wherein
the CP-odd phase was chosen to be proportional to $\arg(\mu)$ and maximal
CP violation with regard to $\arg(\mu)$ was taken. This can be
parameterized by introducing a single CP-violating phase~\cite{plb320p299}:

\be
f^b_n N^\ast_{n4} \simeq \frac{1}{2} \sin\delta_{CP} \label{pra6}
\ee

\n where $f^b_k$ and $N_{k4}$ appear in the $\tilde q
q\tilde{\chi}^0_n$ Lagrangian

\bea
L_{\tilde qq\tilde{\chi}^0} & = & \frac{1}{2} g_W \sum_{n,f} \bar q_f \left[
f^f_n (1+\gamma_5) - \frac{\sqrt{2} m_f}{2m_WB_f} N^\ast_{n,5-f}
(1-\gamma_5) \right] \tilde{\chi}^0_n\tilde q_{fL} \nonumber \\
& & + \frac{1}{2} g_W \sum_{n,f} \bar q_f \left[ g^f_n (1-\gamma_5) -
\frac{\sqrt{2}m_f}{2m_WB_f} N^\ast_{n,5-f} (1+\gamma_5) \right]
\tilde{\chi}^0_n \tilde q_{fR} \nonumber \\
& & + \hbox{h.c.} \label{pra7} ~,
\eea

\n and are defined, together with $g_n^f$ and $B_f$, in~\cite{plb320p299}.
For maximal CP violation, i.e., $\sin\delta_{CP} = 1$, and with
$m_{\tilde q}=m_{\tilde{\chi}^+}=100$ GeV and $m_{\tilde{\chi}^0}=18$ GeV, 
CF find ${\cal A}_3 \simeq 2\%$.
So an asymmetry in the main two-body mode, $t \to b W$, of a
few percent can occur in their limit.

However, these relatively large PRA's, reported by GK and CF in
~\cite{plb319p526,plb320p299} suffer from the following drawbacks: 

\begin{itemize}
\item For the GK limit, 
$m_G+m_{\tilde t}<m_t$
is now essentially  disallowed by the current experimental
bounds. 
\item For the CF limit, $\arg(\mu) \gesim 10^{-2}$ is an unnatural
choice in view of the stringent constraints on this phase coming from the
experimental 
limits on the NEDM as discussed in section \ref{sssec334}. 
\item For both the GK and
CF limits, the large asymmetry arises once the masses of the superpartners
of the light quarks are set to 100 GeV\null. Again, this is a rather
unnatural choice  as it is theoretically very hard, if at all possible, to
meet the NEDM experimental limits when the masses of the squarks (except
for  the lighter stop) are of the order of 100 GeV\null. Besides, the current 
experimental limits disfavor down squarks 
lighter than about 200 GeV\null. 
\end{itemize}

We also remark that PRA in $t \to b W$ within 
the more constrained N=1 SUGRA model was investigated in \cite{hepph9801294}, 
where similar numerical results 
(i.e. ${\cal A}_3<0.3\%$)
for ${\cal A}_3$ were obtained. 
  
To conclude this section, although PRA's in the range of 
$\sim 10^{-3} - 10^{-2}$ 
in the main two-body
decays of $t$, $\bar t$ are appreciable, their measurements is likely to
be very hard. Presumably an $e^+e^-$ collider (NLC) could be suitable
due to its cleanliness. However, there may be about 10,000 to 50,000 $t\bar t$
events a year. Therefore, bearing experimental efficiency factors, 
under the best of circumstances only an asymmetry of the order a few percent
could be measured in the NLC\null. The LHC, being able to produce 
 $10^7 -10^8$ $t \bar t$ pairs, might seem more
 appropriate for a measurement of 
such a small PRA. However, for a measurement of a $\sim 0.3\%$ asymmetry, 
experimental systematics can pose serious limitations.

\subsubsection{PRA within the form factor approach \label{sssec515}}

\n \underline{$t \to bW$} \\
 
The basic idea in the form factor approach is to write a
model independent  coupling, then investigate the dependence of
various asymmetries on the form factors involved 
\cite{zpc56p97,prd45p124,prd56p5928,npb388p53,plb287p237,plb316p137}. 
Thus one can write  the amplitude for $t \to b W^+$ as the sum

\be
{\cal M}_{tbW} \equiv {\cal M}^0_{tbW} + {\cal M}^1_{tbW}~,
\ee

\n where ${\cal M}^0_{tbW}$ is the amplitude 
at the lowest order in the SM which is given by

\be
{\cal M}^0_{tbW} = -\frac{g_W}{\sqrt 2} V_{tb} 
\epsilon_{\mu}^* (p_{W^+}) \bar u_t(p_b) 
\gamma^{\mu} L u_t(p_t)~.
\ee

\n In this equation 
$\epsilon_{\mu}(p_{W^+})$ is the polarization vector of $W^+$ with
four momentum  $p_{W^+}$ and $p_b,p_t$ are the four momenta of $b,t$, 
respectively. 
${\cal M}^1_{tbW}$ contains the new 
CP-violating interactions and can be written in general 
(for on-shell $W^+$ and in the limit $m_e=0$) as follows

\beq
{\cal M}^1_{tbW}= -\frac{g_W}{\sqrt 2} V_{tb} \epsilon^{*}_\mu (p_{W^+})
{\bar u}_b(p_b) \left\{\sum_{P=L,R} \left( 
f_1^P \gamma^{\mu} P + i \frac{f_2^P}{m_W}
\sigma^{\mu}_{\nu} p_{W^+}^{\nu} P \right) \right\}  u_t(p_t)  
\label{ff1a}~,
\eeq

\noindent where $P=L~{\rm or}~R$, $L(R)=(1-(+)\gamma_5)/2$
and the form factors  $f_1^P$ and $f_2^P$ are complex in general; they can 
both have an absorptive phase and a CP-violating phase. 
Note also that, in the SM, $f_1^L=1$; $f_1^R=f_2^L=f_2^R=0$, 
at tree level.

Similarly, the non-standard part of the 
amplitude for $\bar t \to \bar b W^-$ is defined as

\beq
\bar{{\cal M}}^1_{tbW}= -\frac{g_W}{\sqrt 2} V_{tb}^* \epsilon_\mu (p_{W^-})
{\bar v}_t(p_{\bar t}) \left\{\sum_{P=L,R} \left( 
{\bar f}_1^P \gamma^{\mu} P + i \frac{{\bar f}_2^P}{m_W}
\sigma^{\mu}_{\nu} p_{W^-}^{\nu} P \right) \right\}  v_b(p_{\bar b})  
\label{ff1b}~. 
\eeq

\n In general the form factors $f_i^P$ and ${\bar f}_i^P$ can be 
further simplified to the form

\begin{eqnarray}
f_i^P &\equiv& f_{i,CPC}^P \times f_{i,CPV}^P \label{fip}~,\\
\bar f_i^P &\equiv& \bar f_{i,CPC}^P \times \bar f_{i,CPV}^P \label{barfip}~,  
\end{eqnarray}

\n where the indices $CPC$ and $CPV$ stand for the CP-conserving and 
CP-violating parts in the above form factors. In particular, 
$f_{i,CPC}^P$, $\bar f_{i,CPC}^P$ can be complex due to an absorptive 
phase (FSI phase), and $f_{i,CPV}^P$, $\bar f_{i,CPV}^P$ are complex 
in the presence of a non-zero CP-violating phase. 

In terms of the CP-conserving and 
CP-violating parts of these form factors in Eqs.~\ref{fip} and \ref{barfip}, 
it is 
useful to note that the following relations exist (between the form factors 
associated with  $t \to b W^+$ and those related to $\bar t \to \bar b W^-$)

\begin{eqnarray} 
f_{1,CPC}^L = \bar f_{1,CPC}^L 
~~&{\rm and}&~~f_{1,CPC}^R = \bar f_{1,CPC}^R \label{rel1}~,\\
f_{1,CPV}^L = \left( \bar f_{1,CPV}^L \right)^* 
~~&{\rm and}&~~f_{1,CPV}^R = \left( \bar f_{1,CPV}^R \right)^* ~,\\
f_{2,CPC}^L = \bar f_{2,CPC}^R 
~~&{\rm and}&~~f_{2,CPC}^R = \bar f_{2,CPC}^L~,\\ 
f_{2,CPV}^L = \left( \bar f_{2,CPV}^R \right)^* 
~~&{\rm and}&~~f_{2,CPV}^R = \left( \bar f_{2,CPV}^L \right)^* \label{rel4}~.
\end{eqnarray} 
 
\n Using the relations above it is then easy to show that 
any CP-violating observable must always be proportional to any one of 
the combinations: $(f_1^L - \bar f_1^L)$, $(f_1^R - \bar f_1^R)$, 
$(f_2^L - \bar f_2^R)$ or $(f_2^R - \bar f_2^L)$. In particular, 
a CP-odd, $T_N$-even quantity (like the PRA) will be proportional 
to the real parts of these combinations, e.g., 
$\Re{\rm e}(f_2^L - \bar f_2^R)$, but a CP-odd, $T_N$-odd quantity will be 
proportional to their imaginary parts, e.g., $\Im{\rm m}(f_2^L - \bar f_2^R)$.

Assuming these form factors to be  purely CP-violating, i.e., 
$\Re{\rm e}(f_{i,CPV}^P) = 0$ and $\Re{\rm e}(\bar f_{i,CPV}^P) = 0$,
the PRA  defined in  Eq.~\ref{GK4} for $t \to b W$ can be expressed as

\be
{\cal A}_3 = \sum_{i=1}^2 \sum_{P=L,R} 
a_i^P \eReal (f_i^P) ~.
\ee

\n In this context 
it was found that\footnote{Note the slight difference between 
our definition of the form factors $f_i^P$, $\bar f_i^P$ and 
the definition presented in \cite{zpc56p97}.} \cite{zpc56p97} 
$a_1^L\simeq 0.7, 
a_1^R\simeq -0.04,
a_2^L\simeq 0.04$ and $a_2^R\simeq -0.7$~. We thus see that the PRA is more
sensitive to $f_1^L$ and  $f_2^R$ than to $f_1^R$ and $f_2^L$.


\newpage
~

\newpage

\newpage
~

\begin{figure}[htb]
 \begin{center}
  \leavevmode
  \epsfig{file=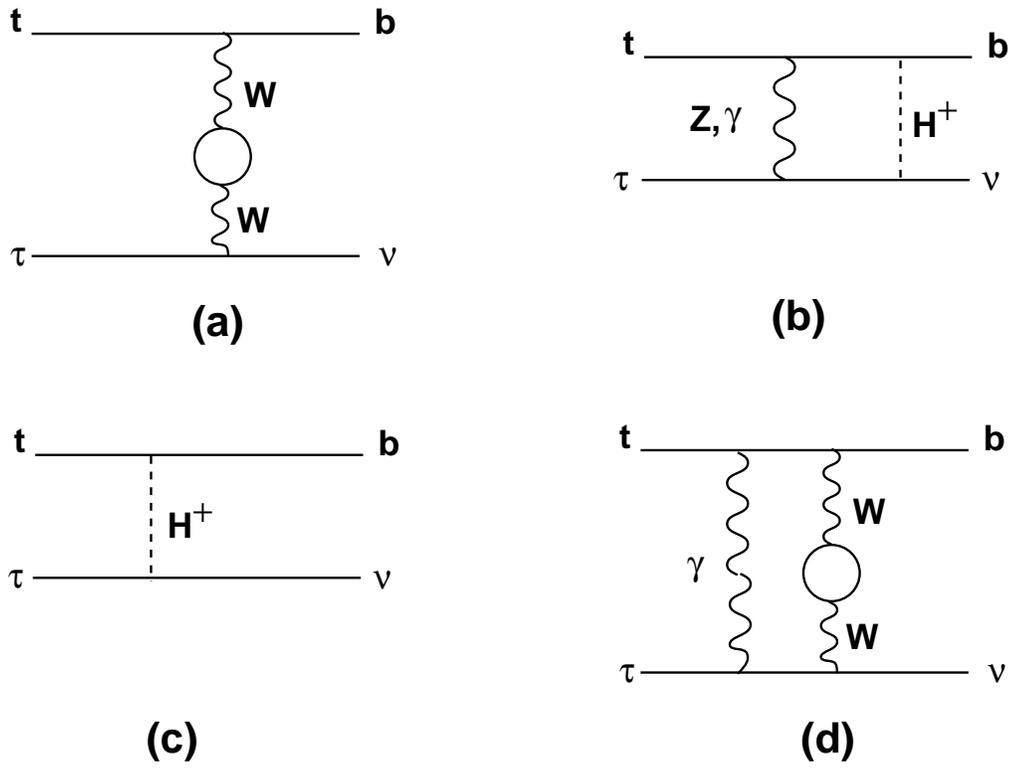
,height=10cm}
\end{center}
\caption{\emph{
Tree-level diagrams and a representative set 
of box diagrams considered for PRA in $t \to b \tau\nu$, within the 3HDM.}}
\label{tdecaysfig3}
\end{figure}

\newpage

\newpage
~

\begin{figure}[htb]
\begin{center}
  \leavevmode
  \epsfig{file=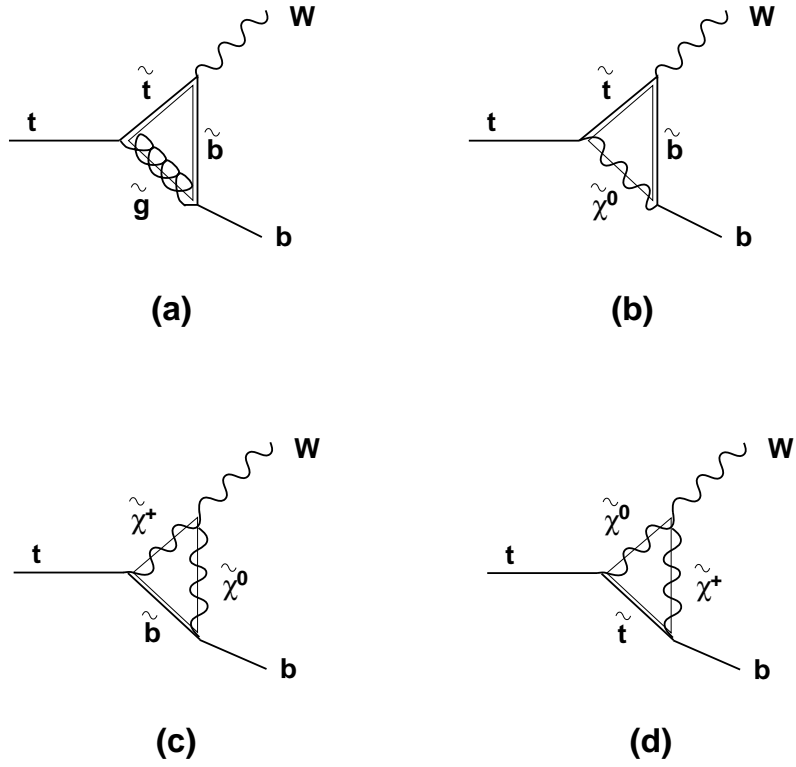
,height=10cm}
\end{center}
\caption{\emph{The SUSY induced 1-loop Feynman diagrams that contribute 
to CP violation in the main top decay $t \to bW$. $\tilde\chi$ is 
the chargino, $\tilde\chi^0$ is the neutralino, $\tilde g$ is the 
gluino and $\tilde t,\tilde b$ are the stop and sbottom particles, 
respectively.}}
\label{tdecaysfig4}
\end{figure}

\newpage

\newpage
~

\begin{figure}[htb]
 \begin{center}
  \leavevmode
  \epsfig{file=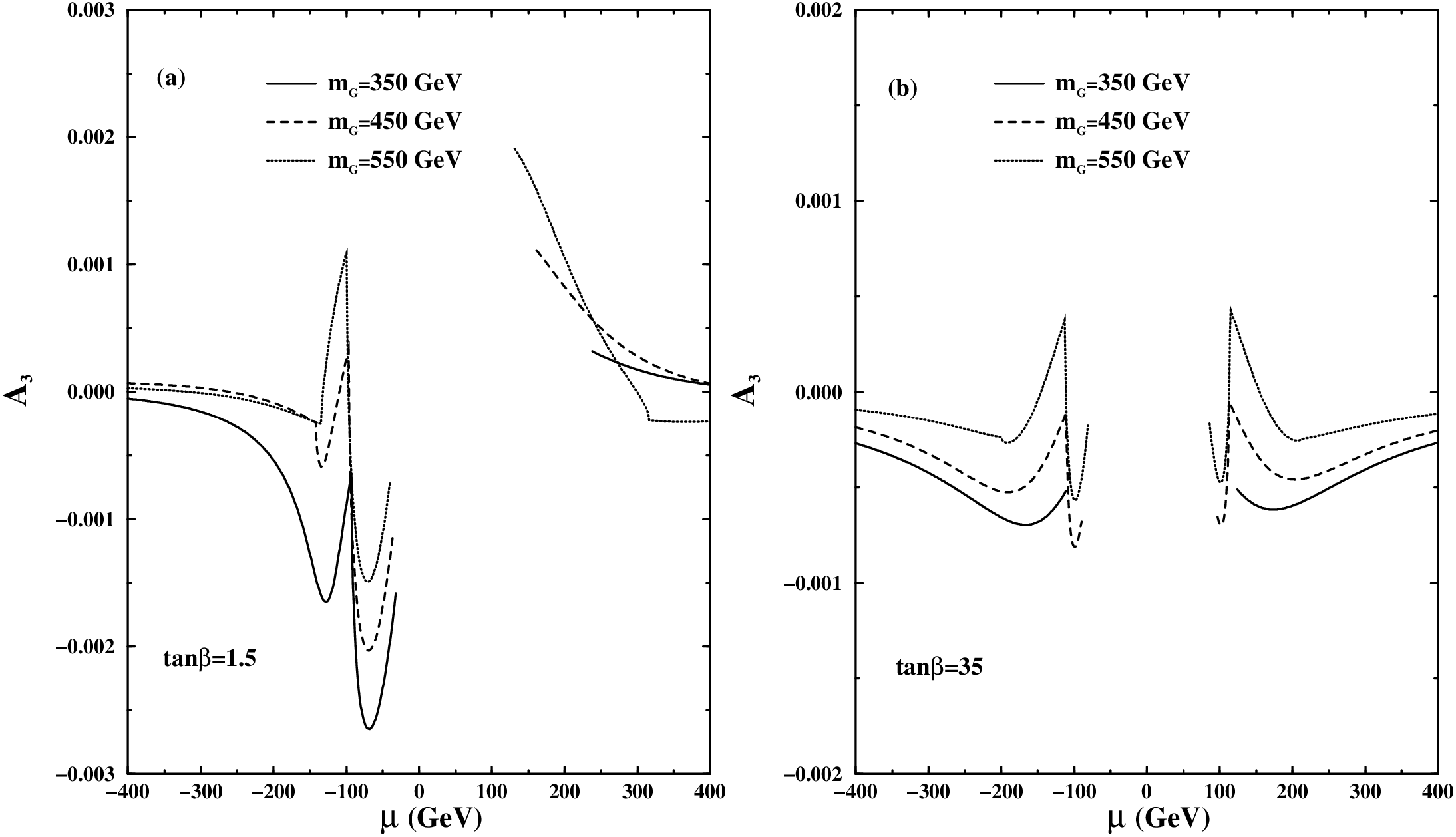
,height=8cm,width=9cm,bbllx=0cm,bblly=2cm,bburx=20cm,bbury=25cm,angle=0}
 \end{center}
\caption{\emph{The SUSY induced PRA ${\cal A}_3$
in the main top decay $t \to b W$, as a function of $\mu$, for several values 
of $m_G$ and for (a) $\tan\beta=1.5$
and (b) $\tan\beta=35$. $M_S=400$ GeV, $m_l=50$ GeV is used.
Figure taken from \cite{prd57p1495}.}}
\label{tdecaysfig5}
\end{figure}

\newpage 
~

\begin{figure}[htb]
 \begin{center}
  \leavevmode
  \epsfig{file=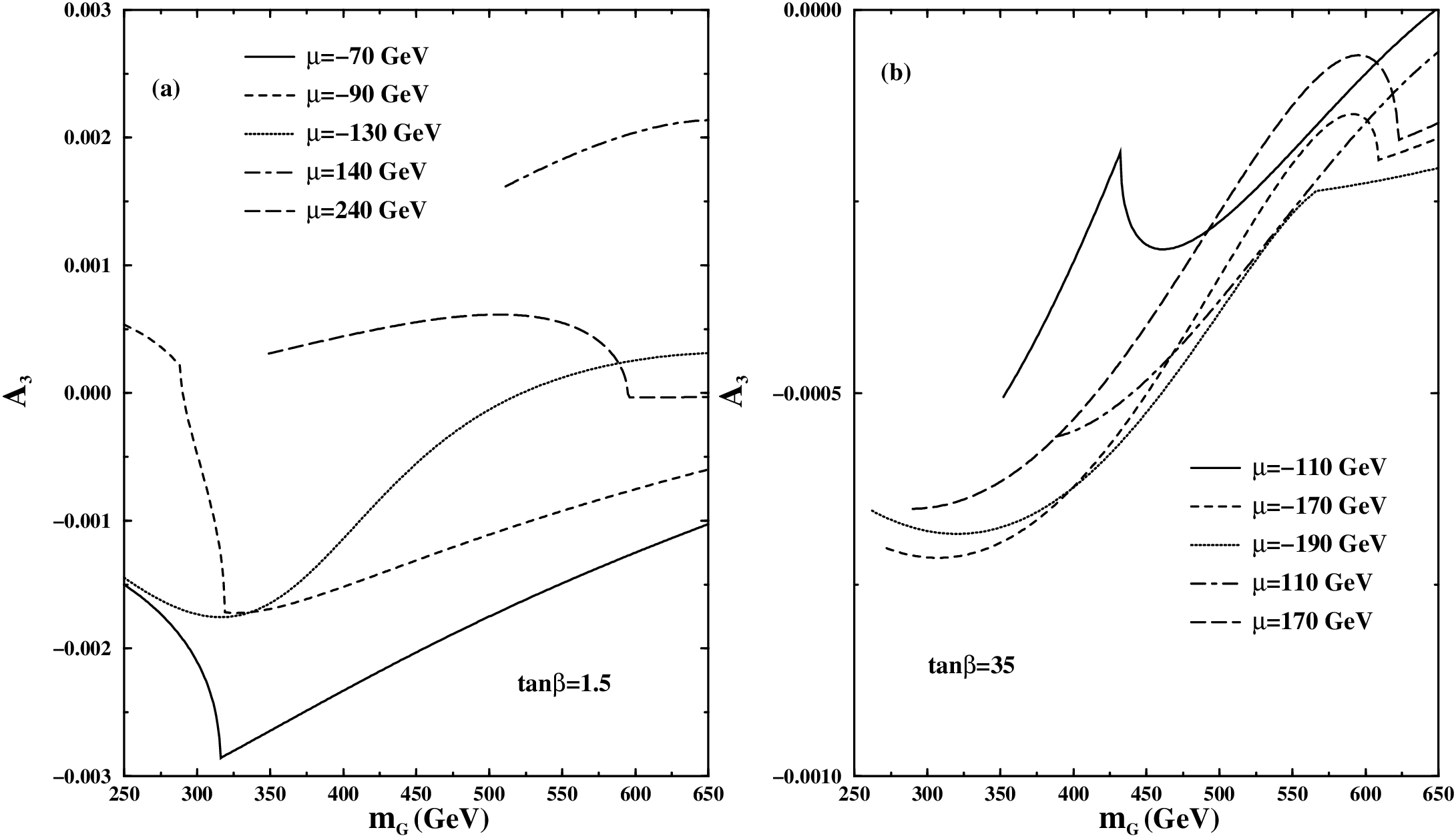
,height=8cm,width=9cm,bbllx=0cm,bblly=2cm,bburx=20cm,bbury=25cm,angle=0}
 \end{center}
\caption{\emph{The SUSY induced PRA ${\cal A}_3$
in the main top decay $t \to b W$, as a function of $m_G$, for several values 
of $\mu$ and for (a) $\tan\beta=1.5$
and (b) $\tan\beta=35$. $M_S=400$ GeV, $m_l=50$ GeV is used.
Figure taken from \cite{prd57p1495}.}}
\label{tdecaysfig6}
\end{figure}

\newpage


\subsection{Partially integrated rate asymmetries \label{ssec52}}

In the rest of the chapter we will discuss CP asymmetries for  the $t\to
b\tau\nu_\tau$ decay within the 3HDM,   starting with PIRA\null. This 
asymmetry is defined as follows

\be
{\cal A}_{PIRA} \equiv \frac{\Gamma_{PI} (t\to b\tau^+\nu_\tau) - 
\Gamma_{PI}
(\bar t \to \bar b\tau^-\bar \nu_\tau)}{\Gamma_{PI} (t\to
b\tau^+\nu_\tau) + \Gamma_{PI}(\bar t\to \bar b\tau^-\bar\nu_\tau)},
\ee

\n where $\Gamma_{PI}$ stands for the partially integrated width,
i.e., the width obtained by integrating over only a part of  phase-space,
rather than over the full kinematic range available to $u$ and $q^2$, defined
in Eq.~\ref{uq^2}. It is easy to see that,
unlike the PRA, 
the PIRA is non-vanishing even
for $m_f\to 0\;(f \neq \tau)$.  The reason is that in
the calculation of the PRA, when the integration over the full range of $u$
is performed, $p^\mu_\tau$ from the $\tau \nu_\tau$ loop sandwiched between
the $W$ and the $H^+$, necessarily gets replaced by $q^\mu$. But then, the
contribution of the transverse part of the $W$-propagator (i.e., $G_T$)
vanishes since $q^\mu\rho^{\mu\nu}_T=0$ ($\rho_T^{\mu \nu} \equiv -g^{\mu
\nu} + q^{\mu}q^{\nu}/q^2$).  On the other hand,
when we calculate the PIRA, then the relevant 
integration is over only a part of the full
kinematic range of $u$ which allows $G_T$ to contribute even to the 
$W^+$-tree${}\times H^+$-tree interference.

Let us consider the integration over $u$ for a fixed $q^2$ in the rest
frame of the $W$-boson, i.e., $\;\vec q=0$. The integration over $u$ is
now equivalent to that over the angle $\theta$ between ($-\vec
p_\tau$) and $\vec p_b$. Define the PIRA over positive values of
$\cos\theta$ to be ${\cal A}_+$. Then, explicit calculation of the
$W^+$-tree${}\times H^+$-tree interference 
(i.e., diagrams (a) and (c) in Fig.~\ref{tdecaysfig3}) 
yields~\cite{prl70p1364}:

\be
{\cal A}_+=\frac{\sqrt{2}}{4 \pi} \frac{G_F m^{2}_{\tau} r_{WH} \mImag (U)}
{(2+r_{Wt})(1-r_{WH}){\cal B}(W \to \tau \nu_\tau)}~, \label{Aplus} 
\ee

\n where $\mImag (U)$ is defined in Eq.~\ref{ImU}, 
$r_{Wt}=m_W^2/m_t^2$ and $r_{WH}=m_W^2/m_H^2$. For $200\lesim
m_H\lesim 300$ GeV, $\;{\cal A}_+\sim{}$ a few $\times 10^{-3}$ 
(see Table~\ref{tablePIRA}) so it is enhanced by two orders of magnitude over
the  PRA, ${\cal A}_\tau$, in Eq.~\ref{atbtn}. 
Even larger asymmetries are likely for $m_H<m_t$.

\begin{table}
\begin{center}
\begin{tabular}{cccccc}  \hline \hline \\ \vspace{1mm}
$m_{H^+}$ & $\tilde{s}_1$ & $\tilde{s}_2$  & $\tilde{s}_3$ & 
${\cal A}_+$
& $({\cal A}^2_+ {\cal B}_+)^{-1}$ \\ \hline
200 & 0.252 & 8.29 $\times 10^{-3}$ & 0.707 & 2.9 $\times 10^{-3}$ &
3.0 $\times 10^6$ \\  
300 & 0.210 & 9.99 $\times 10^{-3}$ & 0.707 & 1.5 $\times 10^{-3}$ &
1.2 $\times 10^7$ \\ \hline \hline 
\end{tabular} 
\end{center}
\caption[dummy]{\emph{
Results for the PIRA ${\cal A}_+$, see Eq.~\ref{Aplus},
and $({\cal A}_+^2 {\cal B}_+)^{-1}$,
in $t \to b \tau \nu_\tau$ within the Weinberg model for CP violation. 
$\tilde{s}_i \equiv \sin(\tilde{\theta}_i)$, where $\tilde{\theta}_i$~, 
$\delta_H=\pi/2$, are CKM-like angles chosen to maximize  
the charged Higgs coupling $\mImag (U)$ (see Eq.~\ref{ImU}).
Note $({\cal A}_+^2{\cal B}_+)^{-1}$ is the number of $t\bar t$ pairs 
required to observe the asymmetry to $1-\sigma$.
${\cal B}_+ \simeq 0.04$ is the appropriate branching fraction.
Table taken from~\cite{prl70p1364}, updated to $m_t=180$ GeV\null.
}}\label{tablePIRA}
\end{table}

A related PIRA was investigated in~\cite{plb289p440}. 
There, a PRA for $t \to b \tau \nu_{\tau}$ 
was defined, specifically for $gg \to \bar t t$
in a hadron collider. The imposition
of experimental cuts (for details see~\cite{plb289p440}), turns    
the asymmetry into PIRA\null. Looking for the maximal CP-violating
effect by choosing the most favorable values for the three CKM-like angles
and $\delta_H=\pi/2$, subject to experimental constraints, 
asymmetries of the order of a few$\times 10^{-4}$ were
obtained from tree$\times$tree interference with a resonant $W$.

\subsection{Energy asymmetry \label{ssec53}}

Another explicit example of how an interesting CP-violating asymmetry
can be sizable even when the PRA is vanishingly small is the energy
asymmetry. Specifically, let us define~\cite{prl70p1364}:

\be
{\cal A}_E = \frac{\langle E_{\tau^+} \rangle - \langle
E_{\tau^-}\rangle}{\langle E_{\tau^+}\rangle + \langle E_{\tau^-}
\rangle} 
\ee

\n where $\langle E_{\tau^+} \rangle$ is the average of the $\tau^+$
energy in $t\to b\tau^+\nu$, etc. In the calculation of $\langle
E_{\tau^+}\rangle$ the integrand is of course equal to that
for the PRA with just an additional factor of $E_{\tau^+}$. Now $G_T$ does
contribute even when the integration is over the full range of $u$.
Explicitly one finds

\be
{\cal A}_E = \frac{\sqrt{2}}{12\pi} \; \frac{G_F m^2_\tau r_{WH}
(1-r_{Wt}) \Im{\rm m}(U)}{(1 +3r^2_{Wt} +2r_{Wt}) (1-r_{WH}) [B(W\to
\tau\nu_\tau)] } ~.
\ee

\n The energy asymmetry is closely related to the PIRA and in
fact numerically~\cite{prl70p1364}:

\be
{\cal A}_E \sim {\cal A}_+/3~.
\ee

\n 
Indeed, both are weighted CP-odd observables constructed from the outgoing 
momenta. Observables constructed in this way
have the drawback that they
are proportional to $m_\tau$. This factor of $m_\tau$ is in addition to
a factor of $m_\tau$ in the Yukawa coupling. The latter cannot be
dispensed with, as long as we are dealing with $t\to b\tau\nu$. However,
the additional power of $m_\tau$ entering these asymmetries can be
overcome by examining the transverse polarization of the $\tau$ as we
will discuss next.


\subsection{$\tau$-polarization asymmetry \label{ssec54}}

The advantage of using a
polarization asymmetry over an energy or
a rate asymmetry is that
the latter asymmetries go as
$m^2_\tau/m^2_H$, where one power of $m_\tau$ comes from the Yukawa
coupling at the $H\tau\nu_\tau$ vertex. The second power of $m_\tau$
comes from the trace over the lepton loop in $W^+$-$H^+$ interference,
i.e.,

\be
Tr[\gamma^\mu (\not\! p_\tau + m_\tau) (1-\gamma_5)
\not\! p_\nu ] = 4m_\tau p^\mu_\nu . 
\ee 

\n The only way to avoid this power
of $m_\tau$ is to avoid summing over the spin ($s_\tau$) of $\tau$ in
the preceding trace. Then the trace will take the form

\be
Tr[\gamma^\mu (\not\! p_\tau + m_\tau)(1+ \gamma_5 \not\! s_\tau)
(1-\gamma_5) \not\! p_\nu] \stackrel{m_{\tau} \to 0}{\longrightarrow} 
4i\epsilon (\mu, s_\tau, p_\tau,p_\nu) \dots
\ee

\n Thus 
the $W^+$-tree$\times$$H^+$-tree
interference will make a 
contribution to the transverse
polarization of the $\tau$, i.e., to $\vec s_\tau\cdot (\vec p_\tau
\times\vec p_\nu)$
without suffering a suppression by an additional power of $m_{\tau}$
(i.e. in addition to the Yukawa coupling)
so this asymmetry will be enhanced over the PIRA and
energy asymmetries  by a factor of about $m_t/m_\tau\sim100$!

We will consider the following CP asymmetries that involve the 
$\tau$ polarization~\cite{prl71p492}:

\bea
{\cal A}_y & \equiv & \frac{\tau^+(\uparrow) - \tau^+(\downarrow) + 
\tau^-(\uparrow) - \tau^-(\downarrow)}
{\tau^+(\uparrow) + \tau^+(\downarrow) + \tau^-(\uparrow) + 
\tau^-(\downarrow)}~, \\
{\cal A}_z & \equiv & \frac{\tau^+(\uparrow) - \tau^+(\downarrow) - 
\tau^-(\uparrow) + \tau^-(\downarrow)}
{\tau^+(\uparrow) + \tau^+(\downarrow) + \tau^-(\uparrow) + 
\tau^-(\downarrow)}~ \label{eq585},  
\eea    

\n where for ${\cal A}_y$ (${\cal A}_z$) the arrows indicate the spin up
or down in the direction $y(z)$. The reference frame is defined to be the
$\tau$ rest frame, such that the $t$ momentum is in the $-x$ direction
(i.e. the $x$ axis is the boost axis from the top to the $\tau$ frame), the
$y$ axis is defined to be in the decay plane with a positive $y$ component
for the $b$ momentum. The $z$ axis is defined by the right-hand rule.
Also, ${\cal A}_y$ (${\cal A}_z$) is CP-odd, $T_N$-even (CP-odd, $T_N$-odd),
and is therefore proportional to the absorptive (dispersive) part of the
bubble in the $W$-propagator. When integrated over the entire phase-space,
${\cal A}_y$ and ${\cal A}_z$ give

\be
{\cal A}_y = - \frac{9}{64}  \frac{g_W^2 \mImag (U) \sqrt{x_{\tau}x_W}}
{(1+2x_W)(x_H-x_W)},
\ee    

\be
{\cal A}_z = \frac{9}{64 \pi} \frac{g_W^2 \mImag (U) \sqrt{x_l}}{(1-x_W)^2 
(1+2x_W) x_H} f(x_W,y_W,x_H) \label{aztaupol}~.
\ee     

\n Recall that 
$\Im{\rm m} (U)$ is given by Eqs.~\ref{Uu}, \ref{Ul} and \ref{ImU}, 
$x_j \equiv m_j^2/m_t^2$ and $y_W \equiv 
\Gamma_W^2 /m_t^2$. Also, $f$ is defined as the integral 

\be
f(x_W,y_W,x_H) \equiv \int_0^1 \frac{(\lambda -x_W)
x_H(1-\lambda)\sqrt{\lambda}}{[(\lambda -x_W)^2 + x_W y_W](x_H-\lambda)}
d \lambda \label{fxwywxh} ~,
\ee 

\n and $\lambda \equiv (p_e+p_\nu)^2/m_t^2$.

A fully integrated polarization asymmetry as in Eq.~\ref{aztaupol},
but weighting events differently for different ranges of the
$\tau\nu_{\tau}$ invariant mass,  was also given in~\cite{prl71p492}:

\be
{\cal A}_z^{\prime} = \frac{9}{64 \pi} \frac{g_W^2 \mImag (U) 
\sqrt{x_l}}{(1-x_W)^2 (1+2x_W) x_H} f^{\prime}(x_W,y_W,x_H) ~,
\ee       

\n where $f^{\prime}$ is the integral of Eq.~\ref{fxwywxh}
 except that $\lambda -x_W$ is replaced by $|\lambda -x_W|$.

The results for the above asymmetries, where experimental 
constraints were imposed on the relevant 3DHM parameters (for details see 
~\cite{prl71p492}), are a few percents for ${\cal A}_y$ and ${\cal A}_z$,
and a few tens of percents for ${\cal A}_z^{\prime}$. As expected, the
$\tau$-polarization asymmetries are much larger and, therefore, perhaps better
suited to look for CP
violation within the 3HDM, than  any other asymmetry.   

\subsection{CP violation in top decays - summary \label{ssec55}}

As discussed in Chapter \ref{sec2},
CP violation can manifest in decays of particles.
Such CP-violating signals may be driven by new physics 
containing new heavy, particles. 
Thus, the large mass of the top may cause 
enhancments of CP violation in top decays as compared to 
the situation in
light quarks decays.
CP-odd signals in top decays, therefore, 
are attractive venues for such studies.
    
In this chapter we have discussed several types of 
CP-violating asymmetries in two and three-body top decays. 
In particular, PRA, PIRA, energy asymmetry in the top decay products
and $\tau$-polarization asymmetries in the three-body decay $t \to b \tau \nu$.
In the SM, the CP violation in the top decays is found to be  
vanishingly small. 
This fact makes CP violation in top decays 
an extremly useful place for searching for new 
physics.  
We have, of course,  also considered
CP violation in top decays in extensions of the SM such as MHDM's and SUSY.

We found that a sizable CP-violating PRA can arise in the main top decay, 
$t \to b W$, in SUSY models. In particular, a stop-neutralino-chagrino loop 
in the $tbW$ vertex can give rise to a PRA of the order of $0.1\%$ if the 
SUSY parameter space turns out to be favorable. Such a PRA is, in principle, 
within the reach of the LHC provided that detector systematics can  
be kept sufficiently under control.

A much bigger CP-violating signal is expected in the three-body decay, $t
\to b \tau \nu$, in a MHDM's with new CP-odd phases in the charged Higgs
sector, e.g., a 3HDM.  Indeed such CP violation may arise already at
tree-level and is best observed through a CP-violating transverse
$\tau$-polarization asymmetry. In a favorable scenario this asymmetry may
be as large as a few tens of percents, requiring $\sim 1000$ top quarks
for its detection. This is a particularly gratifying result since over
$10^4$ top quark pairs are expected to be produced in the future
colliders.  The decay $t \to b \tau \nu$ is therefore a very promising
place to look for new signals of CP violation in top decays. 

\pagebreak

\newcommand{\verl}{\rule{0.05em}{1.5ex}\,}
\newcommand{\verr}{\,\rule{0.05em}{1.5ex}}

\section{CP violation in 
\protect\boldmath ${e^+e^-}$ collider experiments \label{sec6}}
\setcounter{equation}{0}

The energy of circular $e^+e^-$ machines cannot be increased beyond the
energy of LEP-II, due to  heavy losses to the synchrotron
radiation. Therefore, the next step in $e^+e^-$ physics will involve 
linear colliders only, the ``existence proof'' 
of which has been demonstrated at the 100 GeV scale by SLC\null.
For recent reviews on linear colliders see \cite{eereview1}. 
The luminosity of future $e^+e^-$ colliders is projected to be
${\cal L} \approx 10^{34}$ cm$^{-2}$sec$^{-1}$, corresponding 
to a yearly integrated luminosity of ${\cal L} \approx 100$ fb$^{-1}$. 
The working assumption usually is to take it as tens of fb$^{-1}$ 
at the lower end of the scale of c.m. energies, and as hundreds of 
fb$^{-1}$ at its upper scale, corresponding to higher c.m. energies, 
to compensate for the decreasing 
cross-sections. In the first stage, the c.m. energy will cover the range 
approximately between LEP-II and 500 GeV, eventually reaching perhaps 1.6 
TeV and 
hopefully even 2 TeV\null.  Furthermore, beam polarization - which can
help in clarifying many of the physics issues - is an interesting option.
In this context recall that the SLC achieved polarization as
high as $70 - 80\%$.

\subsection{$e^+e^- \to t \bar t$ \label{ssec61}}

In an high energy $e^+e^-$ collider running with c.m.  energies of
500--2000 GeV and an integrated luminosity of ${\cal L} \sim {\cal
O}(100)$ fb$^{-1}$, $10^4-10^5$ pairs of $t \bar t$ will be produced
mainly through the simple reaction $e^+e^- \to \gamma,Z \to t \bar t$. 
This facility, especially due to its relatively clean environment, may
therefore be thought of as a very efficient ``top factory''and it is
expected that many of the rare phenomena associated with top quark systems
will be intensely studied there. Here we will focus on CP violation in the
overall reaction

\beq
e^+ e^- \to \ura{t}{b W^+}\!\!\!\!\!\!\!\!\!\!\!\!\!\!\!\!\!
+ \dra{\bar t}{\bar bW^-} \label{eetteq1} ~. 
\eeq

\noindent Decays of the $W$ also need to be included; the leptonic
channels ($W\to\ell\nu_\ell$, $\ell=e$, $\mu$) are perhaps the cleanest
although experimental simulations suggest that $W$'s could be detected
through jet topologies as well \cite{hepph9606201}.  In what follows we
will not entertain the theoretical possibility that there is additional CP
violation in $W^+,W^-$ decays and will focus only on effects directly
related to the top quark. 

In general, in the limit $m_e=0$, the $\gamma$(or $Z$)$t\bar t$ vertex can
be modified to include the top magnetic and electric(or weak) dipole moments

\beq
-i [\gamma^\mu (A^V_t + B^V_t\gamma_5) + \sigma^{\mu\nu} q_\nu (ic^V_t
+ d^V_t \gamma_5)] \label{eetteq2}~,
\eeq

\noindent where $c_t^V$ and $d_t^V$, for $V=\gamma$ or $Z$, 
are the magnetic and electric dipole moment
form factors of the top quark at $q^2=s$, assuming that they are deduced
by the use of the reaction in Eq.~\ref{eetteq1}.
The tree-level SM values for these parameters are 

\beqa
A^\gamma_t & = & \frac{2}{3}e ~,\nonumber \\
B^\gamma_t & = & 0 ~,\nonumber \\
A^Z_t & = & \frac{e}{\sin\theta_W\cos\theta_W} \left(\frac{1}{4} -
\frac{2}{3}\sin^2\theta_W\right) ~,\label{eetteq3} \\
B^Z_t & = & -\frac{e}{4\sin\theta_W\cos\theta_W} ~,\nonumber \\
c^V_t & = & d^V_t = 0 ~.\nonumber
\eeqa

\noindent Note also that, in the SM, the $Ve^+e^-$ couplings 
in the notation of Eq.~\ref{eetteq2} are

\beqa
A^\gamma_e & = & -e ~,\nonumber \\
B^\gamma_e & = & 0 ~,\nonumber \\
A^Z_e & = & \frac{e}{\sin\theta_W\cos\theta_W} \left(-\frac{1}{4} +
\sin^2\theta_W\right) \label{eetteq4}~,\\
B^Z_e & = & \frac{e}{4\sin\theta_W\cos\theta_W} ~.\nonumber
\eeqa 

\noindent The magnetic form factor, which is 
CP-conserving, 
has a significant SM contribution at 1-loop due to QCD corrections
and therefore is of lesser interest. Since in $e^+e^- \to t \bar t$ we have
$q^2=s>4m^2_t$, these form factors are in general complex. In particular,
with regard to the EDM form factors: 
$\Re{\rm e}d_t^{\gamma,Z}(q^2)$ is $T_N$-odd, and 
$\Im{\rm m}d_t^{\gamma,Z}(q^2)$ 
is $T_N$-even and, of course, all of these four quantities are CP-odd.

Similar to the production vertex, the $tbW^+$ and $\bar t \bar b W^-$ decay 
amplitudes may have CP-violating pieces. In order to take into account this
possibility, for on-shell $W^+$ and in the limit $m_e=0$, 
the decay amplitudes
for $t \to b W^+$ may be decomposed with the most general form factors as 
(see also section \ref{sssec515}) 

\beq
{\cal M}_{tbW}= -\frac{g_W}{\sqrt 2} V_{tb} \epsilon^{*}_\mu (p_{W^+})
{\bar u}_b(p_b) \left\{ \gamma_{\mu}L + \sum_{P=L,R} \left( 
f_1^P \gamma^{\mu} P + i \frac{f_2^P}{m_W}
\sigma^{\mu}_{\nu} p_{W^+}^{\nu} P \right) \right\}  u_t(p_t)  
\label{eetteq5a}~,
\eeq

\noindent where $\epsilon^{*}_{\mu}(p_{W^+})$ is the polarization vector of
$W^+$ with four  momentum $p_{W^+}$ and $p_t,p_b$ are the four momenta of
the $t,b$ respectively. $P=L~{\rm or}~R$  where $L(R)=(1-(+)\gamma_5)/2$
and the form factors  $f_1^P$ and $f_2^P$ are complex in general.

Similarly the amplitude for $\bar t \to \bar b W^-$ is defined as

\beq
\bar{{\cal M}}_{tbW}= -\frac{g_W}{\sqrt 2} V_{tb}^* \epsilon_\mu (p_{W^-})
{\bar v}_t(p_{\bar t}) \left\{ \gamma_{\mu}L + \sum_{P=L,R} \left( 
{\bar f}_1^P \gamma^{\mu} P + i \frac{{\bar f}_2^P}{m_W}
\sigma^{\mu}_{\nu} p_{W^-}^{\nu} P \right) \right\}  v_b(p_{\bar b})  
\label{eetteq5b}~. 
\eeq

\n Furthermore, some useful relations exist 
(see Eqs.~\ref{fip}--\ref{rel4}) between pairs of 
$(f_i^P,\bar f_i^P)$ in terms of their CP-conserving and CP-violating parts.
In particular,
CP-violating observables associated with top decays 
must always be proportional to any one of 
the combinations: $(f_1^L - \bar f_1^L)$, $(f_1^R - \bar f_1^R)$, 
$(f_2^L - \bar f_2^R)$ or $(f_2^R - \bar f_2^L)$, such that 
a CP-odd, $T_N$-even quantity will be proportional 
to the real parts of these combinations, 
but a CP-odd, $T_N$-odd quantity will be 
proportional to their imaginary parts (for details see section \ref{sssec515}).

\begin{figure}
 \psfull
 \begin{center}
  \leavevmode
\epsfig{file=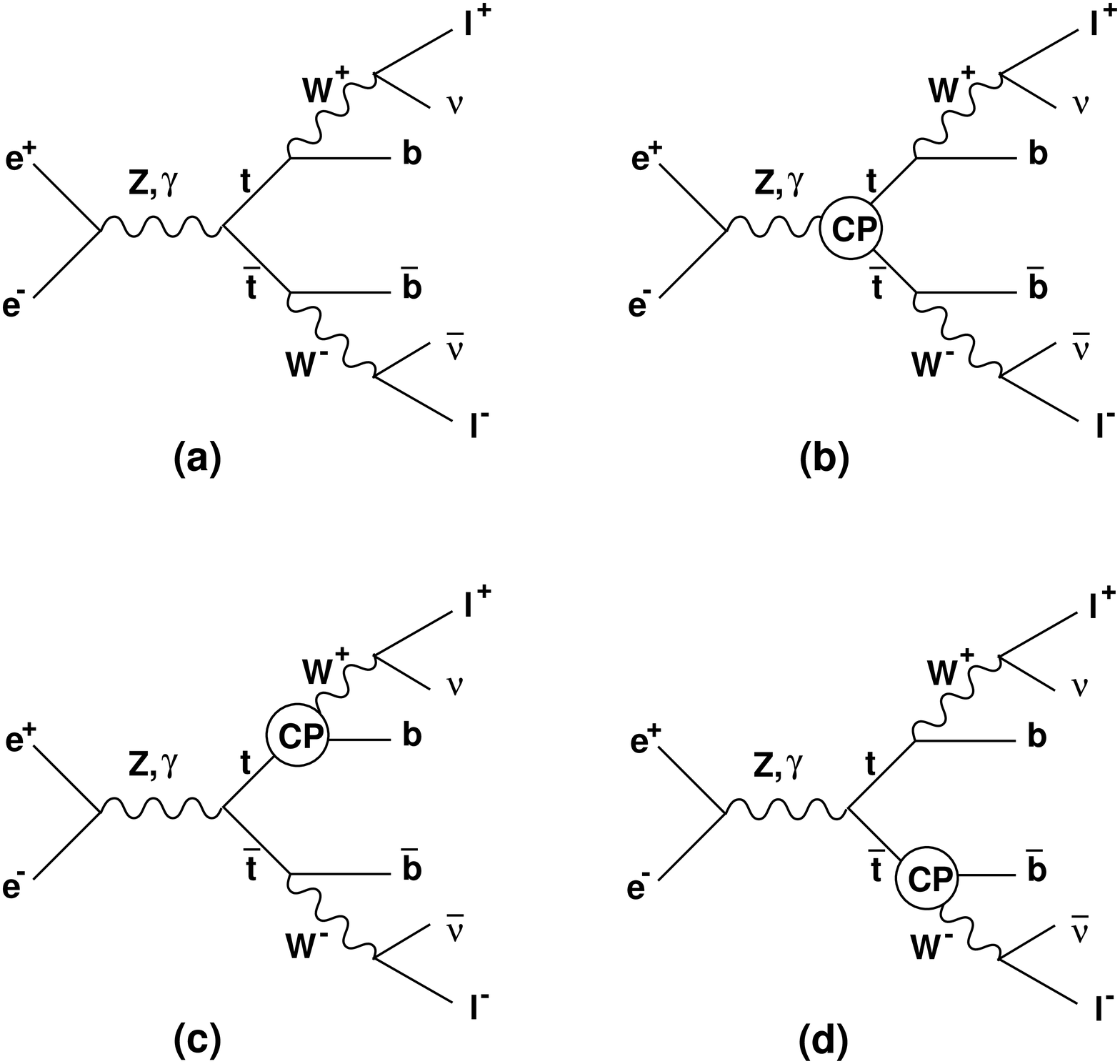
,height=10cm}
 \end{center}
\caption{\emph{Tree-level (a) and CP-violating 
amplitudes (b),(c),(d) to leading 
order in the SM couplings and in CP-violating form factors.}}
\label{eettfig1}
\end{figure}

CP violation effects in $e^+e^- \to t \bar t \to bW^+ \bar b W^-$ 
may thus enter in both the production and the decay vertices of the top  
and the anti-top\footnote{For a comprehensive treatment of the helicity
amplitudes  for $e^+e^- \to t \bar t$ and for the subsequent top decays $t
\to b \ell \nu_l$ in the presence  of the CP-violating couplings in
Eqs.~\ref{eetteq2}, \ref{eetteq5a} and \ref{eetteq5b}, 
see e.g., \cite{prd45p124,hepph9704312}.}. 
To leading order in the CP-violating form factors 
present in the production or the decay of the top, one has to include
interferences of diagrams (b)--(d)  with the SM diagram (a) in 
Fig.~\ref{eettfig1}.

In principle, in order to experimentally separate CP-nonconserving effects
in the production  vertex from the decay vertex, one has to construct 
appropriate observables with sensitivity to only one CP-violating vertex,
i.e., either production or decay (see e.g., \cite{prd54p5412}), or 
alternatively some simplifying assumptions have to be made.

It is important to note that the description of the $Vt\bar t$ vertex for
$e^+e^-\to t\bar t$ in terms of Eq.~\ref{eetteq2} is not necessarily
sufficient. For example, in the MSSM, 1-loop box diagrams with exchanges 
of SUSY particles may contribute to CP violation in $e^+e^- \to t \bar t$.
In these type of diagrams a  $V t \bar t$ vertex is obviously absent 
(such effects were investigated in \cite{plb315p338,hepph9812298}). 
However, note that in 2HDM's with CP violation in
the neutral Higgs sector,  the general parameterization in Eq.~\ref{eetteq2}
holds as  there are no 1-loop box diagrams contributing to 
$e^+e^- \to t \bar t$, in the limit $m_e=0$.     

\subsubsection{Optimized observables \label{sssec611}}

As mentioned before, in the reaction $e^+e^- \to t \bar t \to bW^+ \bar b
W^-$, CP violation can arise from both the production (see Eq.~\ref{eetteq2})
and the decay (see Eqs.~\ref{eetteq5a} and \ref{eetteq5b}) of the top. 
In this problem many momenta
are available, so several $T_N$-odd  triple correlations can be constructed
all of which, in principle, can have non-vanishing expectation values that
are proportional  to $\Re{\rm e} d_t^{\gamma,Z}(q^2)$. 
Similarly several $T_N$-even
(CP-odd) observables can be constructed to measure 
$\Im{\rm m} d_t^{\gamma,Z}(q^2)$. 
The sensitivity to $\Re{\rm e}d_t^{\gamma,Z}(q^2)$ or 
$\Im{\rm m}d_t^{\gamma,Z}(q^2)$ can
vary considerably amongst the observables. It is, therefore, useful to
devise a general procedure that represents a rough measure of the
sensitivity of the  
observables in such situations. Thereby, one is lead to consider the
possibility of constructing ``optimized observables'', i.e., observables that
have the maximum statistical sensitivity. Recall that 
for a given number of $t\bar
t$ events, the optimized observables will yield the smallest attainable limit
on the real  and the imaginary parts of $d_t^{\gamma,Z}(q^2)$.
The basic idea of the optimized observables was first outlined in
\cite{prd45p2405}; the general  recipe for construction of such observables
is given in section \ref{ssec26}. 

Optimal observables have by now been used extensively  
in \cite{prd45p2405,zpc72p461,hepph9602273,npb484p17,plb391p172}.
In \cite{prd45p2405,zpc72p461,hepph9602273} CP violation 
in the top decays was ignored and the CP-odd effect was attributed solely 
to the EDM ($d_t^\gamma$) and ZEDM ($d_t^Z$) 
of the top in the $\gamma t \bar t$ and $Zt \bar t$ production 
vertex. Indeed, in \cite{zpc61p599} it was shown that, in model calculations 
such as 2HDM and MSSM, the dipole moment in the $t \bar t$ 
production leads to larger CP-nonconserving effects than what 
might be expected in the top decays. 

In \cite{npb484p17,plb391p172} the optimization technique was employed to
the overall reaction $e^+e^- \to t \bar t \to bW^+ \bar b W^-$, where
CP violation from both the production and the decay vertices of the top were
investigated. 

Using the general $Vtt$ vertex in Eq.~\ref{eetteq2}, 
the differential cross-section for $e^+e^- \to t \bar t$ may be expressed as

\beq
\Sigma(\phi) d\phi =  \Sigma_0(\phi) d\phi + \sum_{V=\gamma,Z} \left( \Re{\rm
e}  d_t^V(s) \Sigma_{\Re{\rm e}(d_t^V)} (\phi) + \Im{\rm m} d_t^V(s) 
\Sigma_{\Im{\rm m}(d_t^V)}(\phi) \right) d\phi \label{eetteq6} ~.
\eeq

\noindent  As was explained in \cite{prd45p2405}, the simplest optimized
observables for the real and imaginary  parts of $d_t^{\gamma,Z}(q^2)$ are

\beq
{\cal O}_R^{\gamma,Z} =  \Sigma_{\Re{\rm e}(d_t^{\gamma,Z})}/\Sigma_0 ~~,~~
{\cal O}_I^{\gamma,Z}  = \Sigma_{\Im{\rm m}(d_t^{\gamma,Z})}/\Sigma_0 
\label{eetteq7}~. 
\eeq

\noindent These optimal observables are constructed simply from the available
four momenta in $e^+e^- \to t \bar t$
and the subsequent decays.  It was found in \cite{prd45p2405}
that, for example,  with $10^4$ $t \bar t$ events in an NLC running
at c.m.\ energies of $\sqrt s =500$ GeV,  $\Re{\rm e}(d_t^{\gamma,Z})$ 
and $\Im{\rm m}(d_t^{\gamma,Z})$  
of about $\sim {\rm few} \times 10^{-17}$ e-cm become
accessible at the  $1$-$\sigma$ significance level. Recall that in model
calculations, such as MHDM's and SUSY,  the size of top EDM and ZEDM are
typically  at the level of $\lsim 10^{-18}$ e-cm 
(see Chapter \ref{sec4})
if one pushes the CP-violating phases of these models to their largest allowed
values.  Thus, the $1$-$\sigma$ limit obtained in \cite{prd45p2405} is 
at least one order of magnitude above the theoretical expectation for these
dipole  form factors within extensions of the SM\null. 

Consider now the reaction $e^+e^- \to t \bar t \to bW^+ \bar b W^-$ where
the $W^+,W^-$ further decay leptonically via $W \to \ell \nu_l$ or
hadronically, i.e., to up and down quark jets. Of course, 
as is well known, 
the top quark  decay occurs in an extremely short time  and one
measures directly only  the momenta of the decay products of the top. 
The optimization technique may be therefore improved to include all available
4-momenta in a given  decay scenario of the $t \bar t$  
as, for example, 
was done in
\cite{zpc72p461,hepph9602273}.  The basic idea there was to translate the
CP-odd  top spin correlations generated by the dipole moments in $e^+e^-
\to t \bar t$  to correlations among momenta of the decaying products 
of the $t$ and $\bar t$. For this purpose, the most promising decay scenario
is the single-leptonic decay channels, i.e., when one, say the $t$, decays
leptonically  and the other, i.e., $\bar t$, 
decays hadronically or vice versa

\beqa
t  \enskip  \bar{t}  \quad & \to & \quad \ell^+(q_+) + \nu_\ell + b
   + \overline{ X}_{\rm had}(q_{\bar X})\label{eetteq8}~, \\
  t  \enskip  \bar{t} \quad & \to & \quad  X_{\rm had}(q_X) +
\ell^-(q_-) +  \bar \nu_\ell     + \bar{b} \  \ .
\label{eetteq9}
\eeqa

\noindent The decay scenarios in Eqs.~\ref{eetteq8} and \ref{eetteq9} (each
of which has a branching ratio, $B_{\ell} \sim 0.15$, if $\ell=e,\mu$) 
allow for the reconstruction of the $\bar t$ and $t$ momentum, respectively,
which, in turn, gives the rest frames of these quarks. The fact that the rest 
frames of $t$ and $\bar t$ may be accessible in these decay modes allows one
to use CP-odd observables in terms of lepton unit momenta 
${\hat q}_{\pm}^*$ in the corresponding top rest frames,  instead of ${\hat
q}_{\pm}$ - defined in the $e^+e^-$ c.m.\ frame.  

The optimal observables that they used are, again, simply the ratio 
between the CP-odd and CP-even differential cross-sections and are 
given in \cite{zpc72p461}. However, in their optimal observables,   
the differential cross-section corresponds to the overall production 
and decay of the $t \bar t$ and the leptonic momenta are taken 
in the corresponding $t$, $\bar t$ rest frames. With the optimal observables
they \cite{hepph9602273} calculated 
the best 1-$\sigma$  sensitivity to the CP-violating dipole moments
form factors $\Re{\rm e}(d_t^{\gamma}),\Re{\rm e}(d_t^{Z}),\Im{\rm
m}(d_t^{\gamma}), \Im{\rm m}(d_t^{Z})$ assuming 100\% tagging efficiency 
of the single-leptonic decay modes of $t \bar t$ in Eqs.~\ref{eetteq8} and 
\ref{eetteq9}; 
these are given in Table~\ref{eetttab1}. We thus see 
that beam polarization (of the incoming electrons), denoted here by 
$P_e$, may
increase the sensitivity  to $\Re{\rm e}(d_t^{\gamma})$ 
and $\Im{\rm m}(d_t^{Z})$
by almost an order of magnitude. Evidently, the results in
Table~\ref{eetttab1} imply that a NLC running with a c.m. energy 
of $\sqrt s =500$
GeV  and an integrated luminosity of 20 fb$^{-1}$, or $\sqrt s =800$ GeV
and an integrated luminosity of 50 fb$^{-1}$,  will be able to probe, at
1-$\sigma$ and in  the best cases, real and imaginary parts of the TDM, 
typically  of the order of  a few${}\times 10^{-19}-10^{-18}$. 
As in their analysis, this may be achieved
by investigating the single-leptonic decay mode of $t \bar t$.   
This improves the limits obtained in \cite{prd45p2405} by about one order
of magnitude. However,  at the 3-$\sigma$ significance level, the 
corresponding sensitivities are typically ${\rm few} \times 10^{-18}-10^{-17}$
and so are still about an order of magnitude above the
expectations from the models such as MHDM's and SUSY.

It should be noted again that a CP-violating dipole moment at the $t \bar
t \gamma$ and $t \bar t Z$ vertices may not, in general, account for the
entire CP violation effect in $t \bar t$ production. As mentioned before,
this will, for example, be the case in the MSSM where 1-loop CP-violating
box diagrams can cause an additional CP-odd effect \cite{hepph9812298}. We
note, however, that specifically in the MSSM, these box contributions to
CP violation in $t \bar t$ cannot significantly enhance the CP-violating
signal. In fact, in some ranges of the relevant SUSY parameter space the
contribution of the box graphs comes with an opposite sign relative to the
top dipole moments, such that cancellations may occur, thus decreasing the
net CP violation effect in $t \bar t$ production \cite{hepph9812298}.

\begin{table}
\begin{center}
\caption[first entry]{\emph{Attainable 1-$\sigma$ sensitivities 
to the CP-violating
dipole moment form factors in units of $10^{-18}$ e-cm, with ($P_e=\pm 1$) 
and without ($P_e=0$)
beam polarization. $m_t=180$ GeV\null. Table 
taken from \cite{hepph9602273}. \protect\label{eetttab1}}}
\bigskip

\begin{tabular}{|c|c|c|c||c|c|c|}
\hline
  & \multicolumn{3}{c||}{$20 {\mbox{\ fb}}^{-1},\sqrt{s}=500{\mbox{\ GeV}}$} &
\multicolumn{3}{c|} {$50 {\mbox{\ fb}}^{-1},\sqrt{s}=800{\mbox{\ GeV}}$} \\
  & $P_e=0$ & $P_e=+1$ & $P_e=-1$ & $P_e=0$ & $P_e=+1$ &  $P_e=-1$ \\
\hline
$\delta(\Re{\rm e}d_{t}^{\gamma})$ & 4.6 & 0.86 & 0.55 & 1.7 & 0.35 & 0.23 \\
\hline
$\delta(\Re{\rm e}d_{t}^{Z})$ & 1.6 & 1.6 & 1.0  & 0.91 & 0.85 & 0.55 \\
\hline 
$\delta(\Im{\rm m}d_{t}^{\gamma})$ & 1.3 & 1.0 & 0.65 & 0.57 & 0.49 & 0.32 \\
\hline
$\delta(\Im{\rm m}d_{t}^{Z})$ & 7.3 & 2.0 & 1.3  & 4.0 & 0.89 & 0.58 \\
\hline
\end{tabular} 
\end{center}
\end{table}

A more complete investigation was carried out in
\cite{npb484p17,plb391p172}. There the CP-violating form factors from both
the production and the decay amplitudes of the $t$ and $\bar t$ were
included. For this purpose, they used the single-leptonic energy spectrum

\beq
\frac{1}{\sigma^\pm}{\frac{d\sigma}{dx}}^{\!\pm}
=\sum_{i=1}^{3}c_i^\pm f_i(x) \label{eetteq10} ~,
\eeq

\noindent where 

\beq
c_1^\pm =1,\ \ c_2^\pm =\mp \xi,\ \
c_3^+=-\Re{\rm e}(f_2^R),\ \ c_3^-=-\Re{\rm e}(\bar{f}_2^L) \label{eetteq11}~.
\eeq

\noindent The functions $f_i(x)$ are all given in \cite{npb484p17} and 
$f_2^R$, $\bar f_2^L$ are defined in Eqs.~\ref{eetteq5a} and \ref{eetteq5b}.
Also, $\pm$ indicates the charge of the lepton and

\beq
x(\bar x)=2\frac{E^{\ell^+}(E^{\ell^-})}{m_t} 
\sqrt {\frac{(1-\beta)}{(1+\beta)}} ~, \label{eetteq12}
\eeq

\noindent $E^{\ell^+}(E^{\ell^-})$ being the energy of $\ell^+(\ell^-)$ 
in the $e^+e^-$ c.m.\ frame and $\beta=\sqrt{1-4m^2_t/s}$.
Specifically, for $m_t=180$ GeV and $\sqrt s =500$ GeV\footnote{Notice the 
difference in our notation (Eq.~\ref{eetteq2}) and the one used 
in \cite{npb484p17} for the top dipole moments. The translation is 
$D_{\gamma,Z} = (4m_t \sin\theta_W/e) \times i d_t^{\gamma,Z}$.} 
%
%
%

\beq 
\xi \simeq -\frac{1.76 \times 10^{16}}{[{\rm e-cm}]} \times \left( 
1.06 \Im{\rm m}(d_t^\gamma) + 0.18  \Im{\rm m}(d_t^Z) \right) 
\label{eetteq15}~, 
\eeq

\noindent In \cite{npb484p17}, optimal observables that may be used to
separately measure the CP-violating form factors in the production or the
decay vertex were given.  Their optimal observables utilize the
single-leptonic energy spectrum in Eq.~\ref{eetteq10}. With the
optimization technique they showed that $\xi$ and $\Re{\rm e}(f_2^R -
{\bar f}_2^L)$ may be extracted individually from the difference in the
$\ell^+$ versus $\ell^-$ energy spectra by convoluting the differential
energy spectrum with approximately chosen kernel functions

\beqa
&&\xi = \frac{1}{2} \int dx \left[ \frac{1}{\sigma^+}\frac{d\sigma^+}{dx} - 
 \frac{1}{\sigma^-}\frac{d\sigma^-}{dx} 
\right] \Omega_{\xi} \label{eetteq16}~,\\
&&\Re{\rm e}(f_2^R - {\bar f}_2^L) = 
\int dx \left[ \frac{1}{\sigma^+}\frac{d\sigma^+}{dx} - 
 \frac{1}{\sigma^-}\frac{d\sigma^-}{dx} 
\right] \Omega_{f} \label{eetteq17}~, 
\eeqa

\noindent where $\Omega_f,\Omega_{\xi}$ are functions of $f_i(x)$ defined
in Eq.~\ref{eetteq10} and are given in  \cite{plb391p172}. The minimal values
of  $|\xi|$ and $|\Re{\rm e}(f_2^R - {\bar f}_2^L)|$ that can be obtained with
a statistical significance  $N^{ttV}_{SD}$ and $N^{tbW}_{SD}$, respectively, 
at the NLC with $\sqrt s = 500$ GeV and $m_t=180$ GeV, can then 
be computed \cite{npb484p17}:

\beqa
&&|\xi|^{\rm min} = 11.3 \frac{N_{SD}^{ttV}}
{\epsilon_L \times {\rm pb}^{1/2}} \label{eetteq18} ~,\\
&&|\Re{\rm e}(f_2^R - {\bar f}_2^L)| = 13.1 \frac{N_{SD}^{tbW}} 
{\epsilon_L \times {\rm pb}^{1/2}} \label{eetteq19} ~.  
\eeqa
 
\noindent Here $\epsilon_L$ represents the square root of the effective
luminosity for the single-leptonic $t \bar t$ pairs at the NLC\null. Thus,
$\epsilon_L \equiv \sqrt {\epsilon_{tt} {\cal L}}$, where ${\cal L}$ 
is the integrated luminosity at the NLC and $\epsilon_{tt}$ is the tagging 
efficiency for the single-leptonic mode. Note that, in the best case, 
$\epsilon_{tt} =B_\ell \approx 15\%$, if one assumes 100\% efficiency in
measuring the single leptons from $t \bar t$. 

We see from Eqs.~\ref{eetteq18} and \ref{eetteq19} that, for example,
 with ${\cal L}=100$ fb$^{-1}$ and $\epsilon_{tt}= 0.15$ 
we have $\epsilon_L \approx 122.5$ pb$^{-1/2}$ (note that with these values 
one has $\sim 9000$ single-leptonic $t \bar t$ events). Therefore, with this
number,  a 3-$\sigma$ detection of  $|\xi|$ will be possible for  
$|\xi| \approx 0.28$. Using the relation between $\xi$ and 
$\Im{\rm m}(d_t^{\gamma,Z})$  in Eq.~\ref{eetteq15} we then get the following  
3-$\sigma$ equality

\beq
1.06 \Im{\rm m}(d_t^{\gamma}) + 0.18 \Im{\rm m}(d_t^{Z}) \simeq 
1.6 \times 10^{-17}~{\rm e- cm} \label{eetteq20}~.
\eeq
  
\noindent Eq.~\ref{eetteq20} implies that $\Im{\rm m}(d_t^{\gamma,Z})$ of 
the order of $\sim 10^{-17}$ e-cm may be detected at the 3-$\sigma$ level
at the NLC, running  with c.m.\ energy of $\sqrt s =500$ GeV and with an
integrated  luminosity of ${\cal L}=100$ fb$^{-1}$, using the optimal
observables suggested in \cite{npb484p17}. This result is again about
one order of magnitude better compared  to the results obtained in 
\cite{prd45p2405} and it is comparable to the results shown in 
Table~\ref{eetttab1} which  were obtained in \cite{hepph9602273}. 

As for the CP-violating form factors in the decay amplitude, 
we can use Eq.~\ref{eetteq19}
 to get the 3-$\sigma$ limit on $|\Re{\rm e}(f_2^R - {\bar
f}_2^L)|$. For $\epsilon_L \approx 122.5$ pb$^{-1/2}$

\beq
|\Re{\rm e}(f_2^R
- {\bar f}_2^L)| \approx  0.32 \label{eetteq21}~.
\eeq

\noindent In Chapter \ref{sec5} we have discussed the theoretical 
expectations
for couplings  such as $f_2^R$ and $\bar f_2^L$ in the SM and its extensions. 
The SM prediction for such form factors, induced by the CKM matrix, is much
too small to be observed. Moreover,  even within MHDM's and the MSSM the
resulting 3-$\sigma$ limit in  Eq.~\ref{eetteq21} falls short by at least
one order of magnitude.  

Finally, in \cite{plb391p172} the single-leptonic channel was compared to the 
double-leptonic mode, i.e., when both $t$ and $\bar t$ decay leptonically, 
using the optimization technique. 
It was found there that the single-leptonic mode comes out favorable 
by about a factor of 2.

\subsubsection{Naive observables constructed from 
momenta of the top decay products \label{sssec612}}

Various types of ``naive'' observables to deduce the real and imaginary parts
of the non-standard form factors in top production and decay 
were considered  in 
\cite{plb279p389,prd45p2405,zpc61p599,npb388p53,zpc72p461,hepph9602273,hepph9802352,hepph9803426,hepph9809290}.
These observables are constructed simply from correlations between momenta 
of the decaying products of the top quark. The basic idea again utilizes the
fact that weak decays of the top quark act as very efficient analyzer
of the top spin. So the momenta of the decay products (via $t\to b\ell
\nu$) can be used to construct the observables with the right
transformation properties. In general, as expected, the ``naive'' operators 
are less effective than the optimal ones, sometimes by as much as an order
of magnitude. However, it is important to bear in mind that this
advantage pertains only with respect to statistical errors; in actual
experimental considerations, systematic errors will also need to be
taken into account and that could offset some of the advantage of the
optimized observables.

In \cite{prd45p2405}, amongst the various possible correlations, 
the best simple operators that they found are 

\beqa
\epsilon_{\mu\nu\sigma\rho} P^\mu_b Q^\nu_Z H^{+\sigma}H^{-\rho} & {\rm
for} & \Re{\rm e} d^\gamma_t ~,\nonumber \\
\epsilon_{\mu\nu\sigma\rho} P^\mu_e Q^\nu_Z H^{+\sigma}H^{-\rho} & {\rm
for} & \Re{\rm e} d^Z_t \label{eetteq22}~, \\
H^- \cdot Q_Z & {\rm for} & \Im{\rm m} d^\gamma_t ~{\rm and}~ 
\Im{\rm m} d^Z_t ~,
\nonumber 
\eeqa

\noindent where the momenta above are given by

\beqa
P_b & = & p_b - p_{\bar b} ~,\nonumber \\
P_e & = & p^+_e - p^-_e ~,\nonumber \\
Q_Z & = & p^+_e + p^-_e \label{eetteq23} ~,\\
H^\pm & = & 2(E^+_W \cdot p_t) E^+_W \pm 2(E^-_W \cdot p_t) E^-_W ~,\nonumber
\eeqa

\noindent and $E_W$ is the $W$-boson polarization for the reaction $W(p_{W})
\to \ell(p_\ell) \nu_{\ell}(p_{\nu})$. Because of the
left-handed nature of the coupling of $W$ to leptons its polarization
can be constructed from the momenta of the decay products as 

\beq
E_W^\mu = \frac{Tr [ \not p_\nu \not\omega_0 \not p_\ell \gamma^\mu
(1-\gamma_5)]}{4\sqrt{p_\nu \cdot \omega_0 p_\ell\cdot \omega_0}}
\label{eetteq24} 
\eeq

\noindent Here $\omega_0$ is an arbitrary light-like vector that
determines the phase convention for the polarization. The expression above
requires the $\nu$ momenta. Recall that the final state consists of six
particles $b\ell^+\nu_\ell\bar b\ell^-\bar\nu_\ell$. Of these only four
are directly observable as the neutrinos escape detection. However, by
imposing the following conditions, the momenta of the missing neutrinos
may in fact be inferred. The conditions that need to be imposed are:  (1)
conservation of four-momentum together with the conditions that (2) the
lepton and a neutrino reconstruct to the $W^\pm$ mass, (3) the $b$-quark
together with a lepton and neutrino reconstruct the $t$, $\bar t$ mass and
(4) the neutrinos are massless.

It was then found in \cite{prd45p2405} that the use of the $W$ 
polarization in the operators of Eq.~\ref{eetteq22} can easily improve the 
sensitivity to the dipole moments by factors of 10--50 when compared 
to simple correlations which do not involve the $W$ polarization vector.
As compared to the optimal observables discussed in their work, 
the observables in Eq.~\ref{eetteq22} are less effective, typically,
by about a factor of 2--10. 

In \cite{plb279p389,npb388p53}, the following CP-odd and $T_N$-odd
correlations were considered 

\beqa
\hat T_{ij} &=& (\hat q_--\hat q_+)_i \frac{(\hat q_-\times\hat q_+)_j}{|
\hat q_-\times \hat q_+|} + (i\leftrightarrow j) \label{eetteq25}~,\\
\hat A_1 &=& \hat p_+ \cdot \frac{(\hat q_-\times\hat q_+)}{|
\hat q_-\times \hat q_+|} \label{eetteq26}~,
\eeqa

\noindent where $\hat p_+$ is the unit momentum of the incoming 
positron and $\hat q^{\pm}$ are the unit momenta of a charged decay
product from $t\to A, \bar t\to \bar B$ in the overall c.m.\ system.
Thus, $\hat q_{\pm}$ are the directions of a charged lepton or a $b$
jet. An interesting property of these correlations is that 
they are not sensitive to CP-violating effects in the $t$ and $\bar t$ 
decays and, therefore, they can be expressed in terms of only the real part 
of the EDM and ZEDM form factors, $\Re{\rm e}(d_t^{\gamma,Z})$. The mean 
values of ${\hat T}_{ij}$ and $\hat A_1$ plus the 
conjugate ones are given by \cite{npb388p53}:

\beqa
\langle \hat T_{ij}\rangle_{A \bar B} + \langle\hat T_{ij}\rangle_{B \bar
A} &=& 2  \frac{\sqrt s}{e} (c_{\gamma,AB} 
\Re{\rm e}d^\gamma_t + c_{Z,AB} \Re{\rm e}d_t^Z) s_{ij} ~, \\
\langle \hat A_1 \rangle_{A \bar B} +  \langle\hat A_1\rangle_{B \bar A}
&=& 2 \frac{\sqrt s}{e} (r_{\gamma,AB} 
\Re{\rm e}d^\gamma_t + r_{Z,AB} \Re{\rm e}d_t^Z) ~,
\label{eetteq27} 
\eeqa

\noindent where, identifying the $z$-axis with the $e^+$ beam axis

\beq
s_{ij}= \frac{1}{2} \left( \hat p_{+i} \hat p_{+j} - 
\frac{1}{3} \right) = {\rm diag} \left(-\frac{1}{6},
-\frac{1}{6},\frac{1}{3} \right) \label{eetteq28}~. 
\eeq

\noindent The coefficients $c_{\gamma,Z}$ and 
$r_{\gamma,Z}$ depend on the specific
decay channel and were calculated as a function of $s$ in \cite{npb388p53} 
for correlations among $AB=\ell^+\ell^-$, $b\bar b$ and $\ell^+ b +
\ell^- \bar b$.

Possible CP-odd, $T_N$-even correlations that use the momenta of the decay
products of $t$ and $\bar t$ were also examined  in \cite{npb388p53}:

\beqa
\hat Q_{ij} &=& (\hat q_+ + \hat q_-)_i (\hat q_--\hat q_+)_j + 
(i\leftrightarrow j) \label{eetteq29}~,\\
\hat A_2 &=& \hat p_+ \cdot (\hat q_+ + \hat q_-) \label{eetteq30} ~.
\eeqa

\noindent In contrast to the $T_N$-odd observables in Eqs.~\ref{eetteq25} 
and \ref{eetteq26}, the $T_N$-even observables $\hat Q_{ij}$ and 
$\hat A_2$, which
acquire absorptive phases,  are sensitive also to the combinations 
$\Re{\rm e}(f_i^P-\bar f_i^P)$ (recall that $P=L$ or $R$), 
e.g., $\Re{\rm e}(f_2^R - \bar f_2^L)$,  
of the form factors in the decay amplitudes of Eqs.~\ref{eetteq5a} and 
\ref{eetteq5b}. In 
\cite{npb388p53}, CP-odd effects in the decay process  were neglected when
evaluating the two CP-odd, $T_N$-even  observables in Eqs.~\ref{eetteq29}
and \ref{eetteq30}. In terms of the imaginary parts of the EDM and 
ZEDM of the top they obtained

\beqa
\langle \hat Q_{ij}\rangle_{A \bar B} +  \langle\hat Q_{ij}\rangle_{B \bar
A} &=& 2 \frac{\sqrt s}{e} (q_{\gamma,AB} 
\Im{\rm m}d^\gamma_t + q_{Z,AB} \Im{\rm m}d_t^Z) s_{ij} ~, \\
\langle \hat A_2 \rangle_{A \bar B} + \langle\hat A_2\rangle_{B \bar A}
&=& 2 \frac{\sqrt s}{e} (p_{\gamma,AB} 
\Im{\rm m}d^\gamma_t + p_{Z,AB} \Im{\rm m}d_t^Z) ~.
\label{eetteq31} 
\end{eqnarray}

\noindent Here, again, the coefficients $q_{\gamma,Z}$ and $p_{\gamma,Z}$
depend on the specific  channel and were given as a function of 
$s$ in \cite{npb388p53} for correlations among 
$AB= \ell^+\ell^-$, $b\bar b$ and $\ell^+ b + \ell^- \bar b$.

From the simultaneous measurement of the pairs $\langle \hat T_{ij} \rangle,
\langle\hat A_1 \rangle$  and $\langle \hat Q_{ij} \rangle,\langle \hat A_2
\rangle$ and for a given c.m.\ energy, $\Re{\rm e}(d_t^{\gamma,Z})$ and 
$\Im{\rm m}(d_t^{\gamma,Z})$ can be disentangled, respectively. Assuming 
$10^4$
available $t \bar t$ events at  each c.m.\ energy, the 1-$\sigma$ statistical
sensitivity  to $\Re{\rm e}d_t^\gamma$ and $\Re{\rm e}d_t^Z$, using 
only the 33 component of $\hat T_{ij}$, are

\beqa
\delta \left( \Re{\rm e}d_t^\gamma \right) 
&=& \frac{e}{\sqrt {s N_{\rm eff}}} \frac{\sqrt
{\langle \left(3r_Z \hat T_{33} -  c_Z \hat A_1 \right)^2 \rangle}}{| c_\gamma
\cdot  r_Z - c_Z \cdot r_\gamma |} \label{eetteq32}~,\\
\delta \left( \Re{\rm e}d_t^Z \right) 
&=& \frac{e}{\sqrt {s N_{\rm eff}}} \frac{\sqrt {\langle
\left(3r_\gamma \hat T_{33} -  c_\gamma \hat A_1 \right)^2 \rangle}}{| c_Z
\cdot  r_\gamma - c_\gamma \cdot r_Z |} \label{eetteq33}~,
\eeqa

\noindent where $N_{\rm eff}=10^4 \times {\rm Br}(t \to A) \times {\rm
Br}(\bar t \to \bar B)$.  Similar relations can be obtained for 
$\Im{\rm m}d_t^\gamma$ and $\Im{\rm m}d_t^Z$  
using $\hat Q_{33}$ and $\hat A_2$. 

Using this formalism, Ref.~\cite{npb388p53} 
presents the one standard deviation accuracies in measuring the
real and imaginary parts of  $d_t^{\gamma,Z}$, again assuming $10^4$ $t
\bar t$ events  at c.m.\ energy $\sqrt s=500$ GeV and with $m_t=175$ GeV  

\beqa
&&\Re{\rm e}d_t^\gamma \approx 1 \times 10^{-17} ~~,~~ 
\Im{\rm m}d_t^\gamma \approx 1.2 \times 10^{-17} \label{eetteq34}~,\\
&&\Re{\rm e}d_t^Z \approx 5 \times 10^{-18} ~~,~~ 
\Im{\rm m}d_t^Z \approx 7.5 \times 10^{-17} \label{eetteq35}~, 
\eeqa

\noindent where 
the best sensitivity was obtained for correlations between the
$\ell^+ \ell^-$ momenta.  
We note that the sensitivity to $\Re{\rm e}d_t^{\gamma,Z}$
is slightly better than to $\Im{\rm m}d_t^{\gamma,Z}$. 
  
Cuypers and Rindani \cite{plb343p333} have investigated the effect of polarized
incoming electron beams.  They found that the sensitivity of observables
of the type $A_1$ and $A_2$  in Eqs.~\ref{eetteq26} and \ref{eetteq30} 
to the real and imaginary parts of $d_t^{\gamma,Z}$, respectively, can be
enhanced if the incoming  electron beam is longitudinally polarized.
Note also \cite{zpc72p461}, the type of observables in 
Eqs.~\ref{eetteq25}, \ref{eetteq26}, \ref{eetteq29} and \ref{eetteq30} may
be improved (with respect to their sensitivity  to the dipole moments) 
if one replaces the momentum of the charged lepton in the $e^+e^-$ c.m.\ 
frame by its momentum in the top (or anti-top) rest frame, in the 
single-leptonic $t\bar t$ channel. 

As was shown in the previous section, the CP-violating effects 
in the top decay $t \to b W^+$ and its conjugate may be isolated using 
the optimization procedure. This may also be achieved in some limiting cases
 using naive observables. 
In \cite{zpc61p599,npb388p53} a naive 
observable that projects onto CP violation in the top decay vertex 
was suggested

\beq
{\cal O}_1 = \hat p_+ \cdot \left[ \frac{(\hat q_{\bar\ell} \times \hat
q_b)} {|\hat q_{\bar\ell} \times \hat q_b|} - \frac{(\hat q_{\ell} \times
\hat q_{\bar b})} {|\hat q_{\ell} \times \hat q_{\bar b}|} \right]
\label{eetteq36}~. 
\eeq

\noindent Although, in general, the expectation value of ${\cal O}_1$ above
receives contributions  both from the CP violation in $t \bar t$ production and
in the decay amplitude, it was shown in \cite{npb388p53} that close to
threshold, i.e., $\sqrt s \simeq 2 m_t$, the contribution from the $t \bar
t$ production vertex vanishes.  For example, with the correlation ${\cal
O}_1$, for an appropriate $e^+e^-$ collider with 
$m_t=150~GeV$ and 
c.m.\ energy 
$\sqrt s \simeq 2m_t$, they find

\beq
\langle {\cal O}_1 \rangle \approx 0.15 \times 
\Im{\rm m}(f_2^R - \bar f_2^L)
\label{eetteq37}~,
\eeq

\noindent where $f_2^R,\bar f_2^L$ are form factors in the decay 
amplitudes defined
in Eqs.~\ref{eetteq5a} and \ref{eetteq5b}.  Assuming $3 \times 10^3$ 
$t\bar t$ events in which $t \to b \ell^+
\nu_{\ell}$ and  $\bar t \to \bar b \ell^- {\bar\nu}_{\ell}$, 
they found that, to 1-$\sigma$,

\beq
\delta \left( \Im{\rm m}(f_2^R - \bar f_2^L) \right) 
\approx 0.1
\label{eetteq38}~,
\eeq

\noindent can be determined from a measurement of ${\cal O}_1$. Again, the
limit in Eq.~\ref{eetteq38} falls short from model predictions for these
form factors (see Chapter \ref{sec5}).  This is easily understood from
Eq.~\ref{eetteq37} which implies an asymmetry of the order of $\sim
10^{-3}$ for $\Im{\rm m}(f_2^R - \bar f_2^L) \sim 0.1$.  Recall that the
typical asymmetries in extensions of the SM that we have described in
Chapter \ref{sec5} are $\lsim$ a few times $10^{-3}$ and in the SM they
are a lot smaller than that.

A related CP-odd and $T_N$-odd asymmetry that projects only onto CP-violating
effects in the decay processes  of the $t$ and $\bar t$ was suggested by 
Grzadkowski and Keung, \cite{plb316p137}.  
This asymmetry was defined by partially integrating 
over the azimuthal angle of $\ell^+(\ell^-)$ in the $W^+(W^-)$ rest frame
and subtracting the integration  in the range $\left\{-\pi,0 \right\}$ from
the  integration in the range $\left\{0,\pi \right\}$ and it is essentially
proportional to the triple  product $p_\ell \cdot (p_b \times p_W)$. When
evaluated within the MSSM,  the asymmetry was found to be $\sim 10^{-4}-
10^{-3}$  which again falls short by at least one order of magnitude from
the limits that are anticipated to be obtainable, 
through the study of $e^+e^- \to t \bar t \to \ell^+ \ell^- \nu_\ell
{\bar \nu}_\ell b \bar b$
in a future $e^+e^-$ high energy
collider.


\subsubsection{Improved sensitivity using energy and angular 
distributions of top decay products and polarized electron 
beams \label{sssec613}}

Several differential leptonic asymmetries with respect to the charged lepton
in $t \to b \ell \nu_{\ell}$ and  for longitudinal electron (positron)
beam polarizations, $P_e$ ($P_{\bar{e}}$), have been  suggested in 
\cite{prd45p2405,npb484p17,plb391p172,plb349p379,prd50p4372}.

Consider the CP-violating $t-\bar t$ spin correlation ${\hat
p}_t \cdot (s_t-s_{\bar t})$. This spin correlation simply translates to
the asymmetry 

\beq
\Delta N_{LR} = \frac{[N(t_L\bar t_L) - N(t_R \bar t_R)]}{\hbox{all\ }
t\bar t} \label{eetteq39}~,
\eeq 

\noindent suggested first by Schmidt and Peskin (SP) in \cite{prl69p410}  
in the context of $t \bar t$ production in hadron colliders
(the SP effect will be discussed in more detail in 
Chapter \ref{sec7}).

Now, $\Delta N_{LR}$ is related to the asymmetry in the energy spectrum 
defined as \cite{npb408p286,plb349p379,prd50p4372}: 

\beq
 A_E(x)\;=\;\f{1}{\sigma}\,\left[\f{d\sigma}{dx\,(l^+)}\:-\:\f{d\sigma}
{dx\,(l^-)}\right] \label{eetteq40}~,
\eeq

\noindent through a simple multiplication by kinematic functions 
present in the lepton energy distribution functions (see \cite{npb484p17}).
The energy asymmetry in Eq.~\ref{eetteq40} is between distributions of $l^+$
and $l^-$ at the same value of $x\:=\:x(l^+)\:=\:x(l^-)\:=\:4\,
E(l^{\pm})\!/\!\sqrt{s}$. Note that when CP violation is present in the top
production and not in the top decay, then $A_E(x) \propto \Delta N_{LR} \propto
\xi$, where $\xi$ is  defined in Eq.~\ref{eetteq15}.

An up-down asymmetry, $A_{ud}$, was also studied in 
\cite{npb408p286,plb349p379}:

\beq
A_{ud}= \int_{-1}^{+1} A_{ud}(\theta) d\,\cos\theta \label{eetteq41} ~,
\eeq

\noindent where  

\beq
A_{ud}(\theta)=\f{1}{2\,\sigma} \left[\f{d\,\sigma(l^+,{\rm
up})} {d\,\cos\theta}-\f{d\,\sigma(l^+,{\rm
down})}{d\,\cos\theta}+\f{d\,\sigma(l^-,{\rm up})}{d\,\cos\theta}-
\f{d\,\sigma(l^-,{\rm down})}{d\,\cos\theta}\right] \label{eetteq42}~,
\eeq

\noindent and up/down refers to
$(p_{l^{\pm}})_y\;\raisebox{-1.0ex}{$\stackrel{\textstyle>}{<}$}\;0,\; 
\:(p_{l^{\pm}})_y$ being the $y$ component of $\vec{p}_{l^{\pm}}$ with respect
to a coordinate system chosen in the $e^+\,e^-$ c.m.\ frame so that the
$z$-axis is along $\vec{p}_t$, and the $y$-axis is along
$\vec{p}_e\,\times\,\vec{p}_t$.  The $t\bar{t}$ production plane is thus
the $xz$ plane. $\theta$ refers to the angle between $\vec{p}_t$ and
$\vec{p}_e$ in the c.m.\ frame.  Note that the asymmetry $A_{ud}$ 
is related to spin components of  the top transverse to the production
plain and, therefore, it is a $T_N$-odd quantity.  

Three additional asymmetries were considered in \cite{plb349p379}.
The combined up-down and forward-backward asymmetry

\beq
A_{ud}^{\rm fb}\;=\;\int_{0}^{1}\!A_{ud}(\theta)\,d\cos\theta\:-\:
\int_{-1}^{0}\!A_{ud}(\theta)\,d\cos\theta \label{eetteq43}~, 
\eeq

\noindent with $A_{ud}(\theta)$ given in Eq.~\ref{eetteq42}.
The left-right asymmetry

\beq
A_{lr}\;=\;\int_{-1}^{+1}\!A_{lr}(\theta)\:d\cos \theta \label{eetteq44}~,
\eeq

\noindent where

\beq
A_{lr}(\theta)=\f{1}{2\,\sigma}\left[\f{d\,\sigma(l^+,{\rm
left})}{d\,\cos \theta}-\f{d\,\sigma(l^+,{\rm right})}
{d\,\cos \theta}+\f{d\,\sigma(l^-,{\rm left})}{d\,\cos
\theta}-\f{d\,\sigma(l^-,{\rm right})}{d\,\cos \theta}\right] 
\label{eetteq45}~, 
\eeq

\noindent and left/right refers to $(p_{l^{\pm}})_x\:\raisebox{-1.0ex} 
{$\stackrel{\textstyle>}{<}$} \:0$. The combined left-right and
forward-backward asymmetry 

\beq
A_{lr}^{\rm fb}\;=\;\int_{0}^{1}\!A_{lr}(\theta)\,d\cos \theta\:-\:
\int_{-1}^{0}\!A_{lr}(\theta)\,d\cos \theta \label{eetteq46}~, 
\eeq

\noindent with $A_{lr}(\theta)$ given in Eq.~\ref{eetteq45}.

To leading order in the dipole form factors and on ignoring CP violation
in the $t$ and $\bar t$ decays, all the above five CP-odd asymmetries are
linear functions of $d_t^\gamma$ and $d_t^Z$. The asymmetries
$A_E,A_{lr},A_{lr}^{\rm fb}$ are $T_N$-even and are therefore proportional
to $\Im{\rm m}d_t^{\gamma,Z}$, while $A_{ud},A_{ud}^{\rm fb}$ are $T_N$-odd
and are proportional to $\Re{\rm e}d_t^{\gamma,Z}$. The $T_N$-even asymmetries
can be symbolically written as 

\beq
{\cal A}_i \equiv a^\gamma_i(P_e,P_{\bar e}) 
\Im{\rm m}d_t^\gamma + a^Z_i(P_e,P_{\bar e}) \Im{\rm m}d_t^Z \label{eetteq47}~.
\eeq 

\noindent Similarly, for the $T_N$-odd observables, one obtains

\beq
{\cal B}_i \equiv b^\gamma_i(P_e,P_{\bar e}) 
\Re{\rm e}d_t^\gamma + b^Z_i(P_e,P_{\bar e}) 
\Re{\rm e}d_t^Z \label{eetteq48}~.
\eeq   

\noindent where ${\cal A}_i=A_E,A_{lr}$ or $A_{lr}^{\rm fb}$ and ${\cal
B}_i=A_{ud}$ or $A_{ud}^{\rm fb}$. The functions $a^{\gamma,Z},b^{\gamma,Z}$
depend,  among other parameters, on the polarizations of the incoming electron
and positron beams, $P_e$  and $P_{\bar e}$, respectively, 
and are explicitly given in \cite{plb349p379}.  
 
\begin{table}[htb]
\begin{center}
\caption[first entry]{\emph{$90\%$ confidence limits on the real and imaginary
parts of the top dipole form factors $d_t^\gamma$ and $d_t^Z$, 
in  units of $10^{-17}$ e-cm, from
different asymmetries. In the unpolarized case , the asymmetries $A_{ud}$,
$A_{ud}^{\rm fb}$ are  together used to get limits on $\Re{\rm e}
d_t^{\gamma,Z}$ and $A_{lr}$, $A_{lr}^{\rm fb}$ to obtain the 
limits on $\Im{\rm m}d_t^{\gamma,Z}$. 
In the polarized case, the limits obtained
from  $A_{ud}$ and  $A_{lr}$  are denoted by (a) and the ones from 
$A_{ud}^{\rm fb}$ and $A_{lr}^{\rm fb}$ are denoted by (b). The numbers are
for the single-leptonic $t \bar t$ mode, for $m_t=174$ GeV, $\sqrt s=500$ GeV
and an integrated  luminosity of ${\cal L}=10$ fb$^{-1}$.
\protect\label{eetttab2}}}
\bigskip
\begin{tabular}{|ll|c|c|c|c|}
\hline
\multicolumn{2}{|l|}{Case}&$|\Re{\rm e}d_t^{\gamma}|$&$|\Re{\rm e}
d_t^Z|$&$|\Im{\rm m}d_t^{\gamma}|$&$|\Im{\rm m}d_t^Z|$\\[2mm]
\hline
unpolarized&&54.4\hskip 5mm&15.9&  7.9&62.4\\[3mm]
(a)  polarized($P_e= \pm$0.5)&&2.3&2.3&2.3&9.1\\[3mm]
(b)  polarized($P_e=\pm$0.5)&&12.5&9.1&2.3&7.9 \\[3mm]
\hline
\multicolumn{6}{c}{}\\
\multicolumn{6}{c}{}\\
\multicolumn{6}{c}{}
\end{tabular}
\end{center}
\end{table}

It is evident from Eqs.~\ref{eetteq47} and \ref{eetteq48} that, without beam
polarization, by measuring only  one asymmetry of the type ${\cal A}_i$ and/or
${\cal B}_i$,  one can extract information only on one combination of 
$\{ \Im{\rm m}d_t^\gamma$, $\Im{\rm m}d_t^Z\}$ and/or 
$\{ \Re{\rm e}d_t^\gamma$,
$\Re{\rm e}d_t^Z\}$, respectively.  However, any two asymmetries with the same
$T_N$  property can be used to determine two independent combinations of
the corresponding real or imaginary  parts of $d_t^\gamma$ and $d_t^Z$,
thus, giving  $\Im{\rm m}d_t^\gamma$ and $\Im{\rm m}d_t^Z$ 
or $\Re{\rm e}d_t^\gamma$
and $\Re{\rm e}d_t^Z$ independently.  Such an analysis was described in the
previous section  where the observable pairs $\hat T_{33},\hat A_1$ and 
$\hat Q_{33},\hat A_2$ were used to find the sensitivity of a high energy
$e^+e^-$ collider to  $\Re{\rm e}d_t^\gamma,\Re{\rm e}d_t^Z$ and $\Im{\rm m}
d_t^\gamma,\Im{\rm m}d_t^Z$, respectively.  However, it was suggested in 
\cite{plb349p379} that if, in addition, beam polarization is included, 
then one $T_N$-even($T_N$-odd) asymmetry is sufficient to determine 
$\Im{\rm m}d_t^\gamma$  and 
$\Im{\rm m}d_t^Z$($\Re{\rm e}d_t^\gamma$ and $\Re{\rm e}d_t^Z$)
independently by measuring  this asymmetry for different polarizations. 
Both approaches were adopted in \cite{plb349p379}. Their best results are
summarized in  Table \ref{eetttab2} where $90\%$ confidence level limits
are given for $\sqrt s=500$ GeV  and $m_t=174$ GeV\null. We can see from 
Table \ref{eetttab2} that the second approach, of incorporating beam 
polarization, increases the sensitivity to both the real and imaginary parts
of the  EDM and ZEDM of the top, in some cases, by about one order 
of magnitude. With $50\%$ beam polarization,  the 90\% confidence level
limits for  $\Im{\rm m}d_t^\gamma,\Im{\rm m}d_t^Z,\Re{\rm e}d_t^\gamma$ 
and $\Re{\rm e}d_t^Z$ are again at best around 
$\sim {\rm few} \times 10^{-17}$ e-cm. 

An interesting differential asymmetry that combines information from both
the production and decay vertices of the top was suggested in
\cite{plb391p172}: 

\beq
A_{\ell\ell}\equiv \frac{\left(\int\!\!\!{\int_{}}_{x<\bar{x}}\!\!\!\!\!
~~~dxd\bar{x}\frac{d^2\sigma}{dxd\bar{x}}
-\int\!\!\!{\int_{}}_{x>\bar{x}}\!\!\!\!\! 
~~~dxd\bar{x}\frac{d^2\sigma}{dxd\bar{x}}\right)}
{\int\!\!\!\int dxd\bar{x}\frac{d^2\sigma}{dxd\bar{x}}}\:.
\label{eetteq49}
\eeq

\noindent This asymmetry utilizes the double-leptonic energy distribution 
in $e^+e^- \to t \bar t \to \ell^+ \ell^- \nu_\ell {\bar \nu}_\ell b \bar b$

\beq
\frac{1}{\sigma} \frac{d^2 \sigma}{dx~d \bar x} =
\sum_{i=1}^{3} c_i f_i(x,\bar x) \label{eetteq50}~,
\eeq

\noindent where $x,\bar x$ are defined in Eq.~\ref{eetteq12} and

\beq
c_1=1~~,~~c_2=\xi~~,~~c_3=-\frac{1}{2} 
\Re{\rm e}(f_2^R-{\bar f}_2^L) \label{eetteq51}~.
\eeq

\noindent $f_2^R,{\bar f}_2^L$ and $\xi$ are defined in 
Eqs.~\ref{eetteq5a}, \ref{eetteq5b} and \ref{eetteq15}, respectively.  
In terms of $\xi$ 
(or equivalently of $\Im{\rm m}d_t^{\gamma,Z}$)  
and $\Re{\rm e}(f_2^R-{\bar f}_2^L)$ 
and with $\sqrt s=500$ GeV, $m_t=180$ GeV and the SM parameters 
they obtained the simple relation

\beq
A_{\ell\ell} = -0.34 \xi - 0.31 \Re{\rm e}(f_2^R-{\bar f}_2^L) 
\label{eetteq52}~.
\eeq

\begin{figure}[htb]
\begin{center}
\leavevmode
\epsfig{file=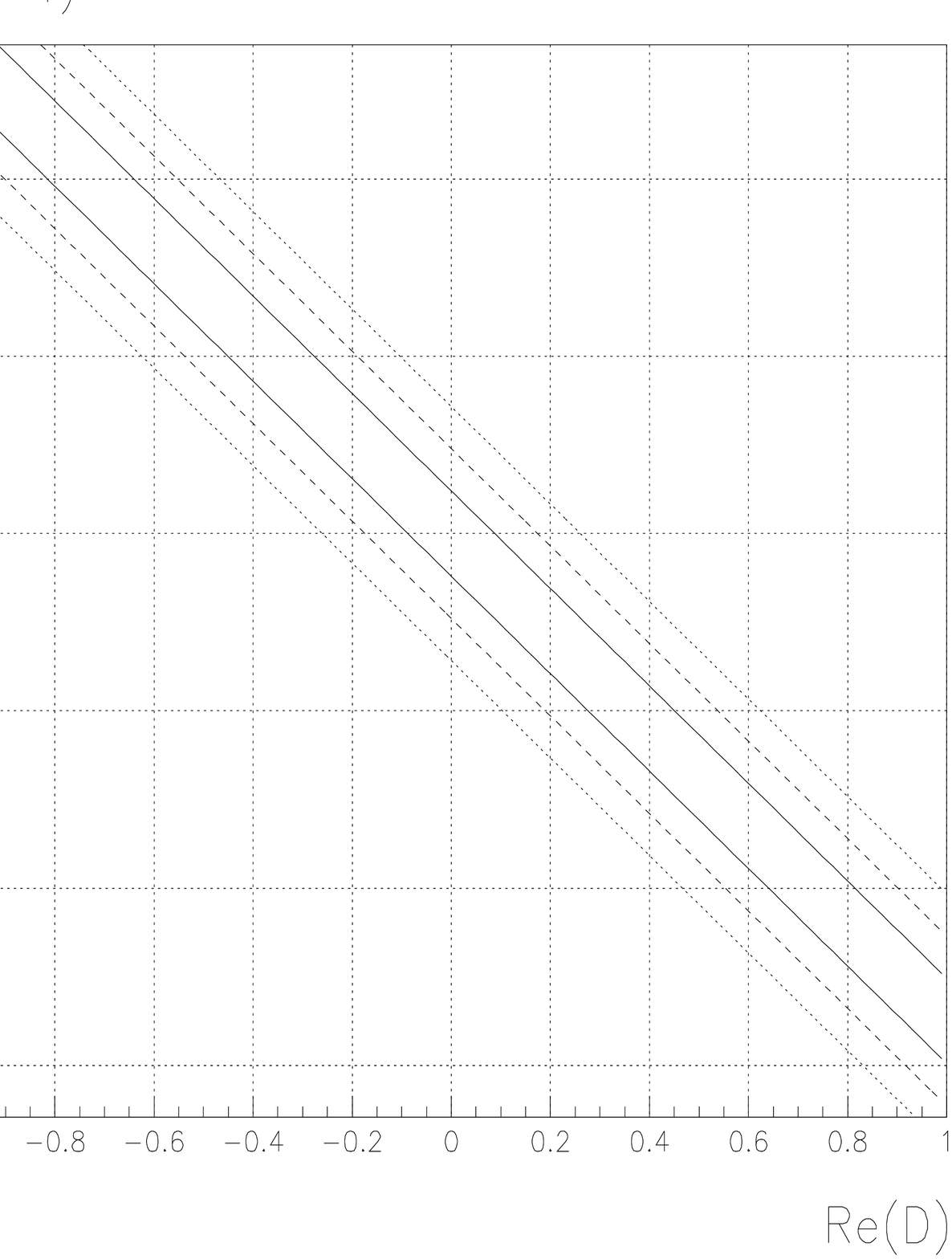
,height=9cm}
\end{center}
\caption{\emph{One can verify the asymmetry $A_{\ell\ell}$ in 
Eq.~\ref{eetteq49} to be non-zero
at 1-$\sigma$, 2-$\sigma$ and 3-$\sigma$ level when the parameters 
$\Re{\rm e}(D_{\gamma,Z})$ (horizontal axis) and 
$\Re{\rm e}(f^R_2-\bar{f}^L_2)$ (vertical axis) are outside the
two solid lines, dashed lines and dotted lines, respectively,
given $\sim 700$ double-leptonic $t \bar t$ events in a 500 GeV $e^+e^-$ 
collider. Recall that 
$D_{\gamma,Z}$ is related to $d_t^{\gamma,Z}$ via: 
$D_{\gamma,Z} = (4m_t \sin\theta_W/e) \times i d_t^{\gamma,Z}$, i.e., 
$\Re{\rm e}(D_{\gamma,Z})$ corresponds to $\Im{\rm m}d_t^{\gamma,Z}$.
Figure taken from \cite{plb391p172}.}}
\label{eettfig2}
\end{figure}

\noindent 
Given a number of available double-leptonic $t \bar t$ events in the NLC and
using the  relation in Eq.~\ref{eetteq52}, one can now plot the 1-$\sigma$, 
2-$\sigma$
and 3-$\sigma$ detectable   regions in the 
$\Im{\rm m}d_t^{\gamma,Z} - \Re{\rm
e}(f_2^R-{\bar f}_2^L)$  plane. These regions are shown in 
Fig.~\ref{eettfig2}
for $\sim 700$ double-leptonic $t \bar t$ events in a 500 GeV collider.  
We see from Fig.~\ref{eettfig2} that a 3-$\sigma$ detection of CP violation
through $A_{\ell \ell}$ is possible  in a wide range of the parameters 
$\xi$ and $f_2^R,{\bar f}_2^L$. However, if one parameter is very small, then
it requires the other to be relatively large. Thus, for example, if $\Re{\rm
e}(f_2^R-{\bar f}_2^L)=0$ then, at 3-$\sigma$, 
$|\Im{\rm m}(d_t^{\gamma,Z})| \gsim 1.7 \times 10^{-17}$ e-cm (or, with the 
notation used in \cite{plb391p172}, $|\Re{\rm e}(D_{\gamma,Z})|
\gsim 0.3$, see also Fig.~\ref{eettfig2}). 
This is again comparable to obtainable limits from other observables
described in this section and, thus, falls short by about one order of
magnitude when compared to model dependent predictions for the top dipole
moment.

Finally, interesting CP-violating asymmetries which involve correlations
among $b$-quarks from $t \bar t\to bW^+ \bar b W^-$ were suggested by
Bartl {\it et al.} in \cite{hepph9802352,hepph9803426,hepph9809290}.  They
utilized the angular \cite{hepph9802352} and energy \cite{hepph9803426}
distributions of $b$ and $\bar b$ with initial beam polarization, in
$e^+e^- \to t \bar t$ followed by $t \to bW^+$ and $\bar t \to \bar b
W^-$, to construct CP-violating asymmetries that can disentangle CP
violation in the $t \bar t$ production mechanism from CP violation in the
top decay.  Unfortunately, their asymmetries, when evaluated within the
MSSM, range from $10^{-4}$ to $10^{-3}$ at best, and therefore also seem
to be too small to be detectable at a future NLC.



\newpage
~

\newpage
~

\newpage



\subsection{$e^+e^-\to t\bar t h, t \bar t Z$, examples of tree-level CP
violation \label{ssec62}}

Let us now turn our attention to the reactions

\beqa
&& e^+(p_+)+e^-(p_-) \to t (p_t) + \bar t (p_{\bar t}) + h(p_h) 
\label{eetthzeq1} ~, \\
&& e^+(p_+)+e^-(p_-) \to t (p_t) + \bar t (p_{\bar t}) + Z(p_Z) 
\label{eetthzeq2} ~,
\eeqa

\noindent which may exhibit large CP violation asymmetries in a
2HDM \cite{prd53p1162,plb419p340,hepph9710355,prd60p115018}. 
Note that for reasons 
discussed below, for  the CP-violating effects in $e^+e^- \to
t \bar t h$ as well as in $e^+e^- \to t \bar t Z$, only two out of the three
neutral Higgs  of the 2HDM are relevant. We denote these two neutral Higgs
particles by $h$ and $H$  corresponding to the lighter and heavier Higgs-boson,
respectively. In some instances we denote a neutral Higgs 
by ${\cal H}$, then ${\cal H}=h$ or $H$ is to be understood.

A novel 
feature of these reactions is that the effect arises at the tree-graph level.
As a consequence, one can construct new type of 
asymmetries which are $\propto$ (tree$\times$tree)/(tree$\times$tree) and 
are therefore a-priori of ${\cal O}(1)$. This stands in contrast to loop 
induced CP-violating effects in $t \bar t$ production 
for which the CP asymmetries, in general, are 
$\propto$ (tree$\times$loop)/(tree$\times$tree) and are therefore 
suppressed by additional small couplings to begin with, i.e., at 1-loop, 
typically, by $\alpha$ - the fine structure constant.  

\begin{figure}
 \psfull
 \begin{center}
  \leavevmode
  \epsfig{file=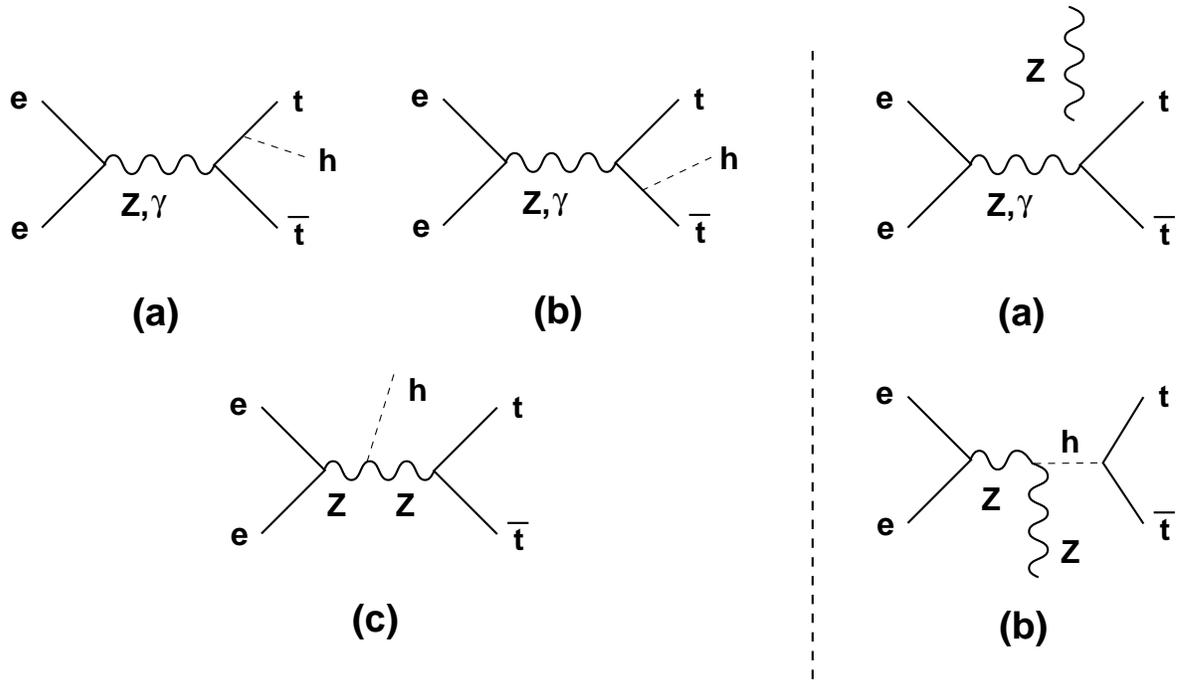
,height=9cm}
 \end{center}
\caption{\emph{
Tree-level Feynman diagrams contributing to $e^+e^-\to t\bar{t} h$ 
(left hand side) and $e^+e^-\to t\bar{t} Z$ (right hand side) in the unitary
gauge, in a 2HDM.  For $e^+e^-\to t\bar{t} Z$, diagram
(a) on the right hand side represents 8 diagrams in which either 
$Z$ or $\gamma$
are exchanged in the s-channel  and the outgoing $Z$ is emitted from 
$e^+,e^-,t$ or $\bar t$.
}}
\label{eetthzfig1}
\end{figure}

Indeed, we will show below that CP violation at the level of tens of a percent
is possible in the reactions in Eqs.~\ref{eetthzeq1} and \ref{eetthzeq2}. 
Basically, for the $t\bar th (t\bar tZ)$ final states,  
Higgs($Z$) emission off the
$t$ and $\bar t$ interferes with the Higgs($Z$)  emission off the $s$-channel
$Z$-boson (see Fig.~\ref{eetthzfig1})  
\cite{prd53p1162,plb419p340,hepph9710355}. 
We find that the processes $e^+e^- \to t \bar t h$ and $e^+e^- \to t \bar
t Z$ provide two independent, but analogous, promising probes to search for
the signatures of the  same CP-odd phase, residing in the $t \bar t$-neutral 
Higgs 
coupling, if the value of $\tan\beta$ (the ratio between the two VEV's in
a 2HDM) is in the vicinity of 1. In particular, they serve as good examples
for large CP-violating effects that could  emanate from $t$ systems due to
the large mass of the top quark  and, thus, they might illuminate the role
of a neutral Higgs particle in CP violation. 

Although these reactions are not meant (necessarily) to lead to the discovery
of a neutral Higgs,  they will, no doubt, be intensely studied at the 
NLC since they stand out as very important channels for a variety of 
reasons. In
particular, they could perhaps provide a unique opportunity  to observe the
top-Higgs Yukawa couplings directly  
\cite{prl77p5172,mpla7p1765,prd50p7042,hepph9609453,npb367p257,prd61p015006}. 
In \cite{prl77p5172,hepph9609453}, 
using a generalization of the optimal observables 
technique outlined below (see also section \ref{ssec26}), Gunion \etal\ have
extended the initial work \cite{prd53p1162} on CP violation  in $e^+ e^-
\to t \bar t h$ to include  a detailed cross-section analysis such that
all Higgs  Yukawa couplings combinations are extracted (see below). 
A similar analysis which also uses the optimized observable technique 
for $e^+e^- \to t \bar t Z$
is 
given in~\cite{prd60p115018}.
A detailed cross-section analysis of the reaction $e^+e^- \to t \bar t Z$
in the SM was performed by Hagiwara \etal\ \cite{npb367p257}.
There, it was found that the Higgs exchange contribution of diagram (b) on 
the right hand side of Fig.~\ref{eetthzfig1} will be almost invisible in a 
TeV $e^+e^-$ collider for neutral Higgs masses in the range $m_h < 2m_t$.
Interestingly, we will show here that, if the scalar sector is doubled, then
the lightest neutral Higgs ($h$) 
may reveal itself through CP-violating interactions
with the top quark even if  $m_h <2m_t$. Obviously, a non-SM (e.g., larger) 
top-Higgs Yukawa coupling can cause an enhancement in the rates for both 
the $t \bar t h$ and $t \bar t Z$ final states. Thus, a ``simple'' 
cross-section study for these reactions may also come in handy for searching 
for new physics. However, one should keep in mind that, from the 
experimental point of view,
asymmetries, i.e., 
ratios of cross-section, are easier to handle and, in particular, 
CP-violating signals are very distinctive evidence for new physics.

%
%
%
%
%

This section will be divided to three parts. In the first part we 
present a detailed analysis of  the tree-level CP violation in the reactions 
$e^+ e^- \to t \bar t h$ and $e^+ e^- \to t \bar t Z$ which manifests
itself as a $T_N$-odd correlation of momenta. In the second 
part we will consider the generalized optimization technique 
developed by Gunion \etal\ and its application to 
the reaction $e^+ e^- \to t \bar t h$. In the last part we will 
discuss CP violation in the Higgs decay $h \to t \bar t$, where we take the
Higgs to be produced  through the Bjorken mechanism $e^+ e^- \to Z h$.

\newpage
~

\newpage

\subsubsection{Tree-level CP violation \label{sssec621}}

In the unitary gauge the reactions in Eqs.~\ref{eetthzeq1} and \ref{eetthzeq2} 
can proceed via the Feynman diagrams depicted in Fig.~\ref{eetthzfig1}. 
We see that for $e^+ e^- \to t \bar t Z$, diagram (b) on the right hand
side of Fig.~\ref{eetthzfig1},  in which $Z$ and ${\cal H}$ are produced 
(${\cal H}=h~or~H$ is either a real or a virtual particle, i.e.\ $m_{\cal
H} >2m_t$ or $m_{\cal H}<2m_t$, respectively)  followed by ${\cal H} \to
t \bar t$,  is the only place where new CP-nonconserving dynamics from
the Higgs  sector can arise, being proportional to the  CP-odd phase in
the $ t \bar t {\cal H}$ vertex. As mentioned above, in both the $t\bar th$ and
the $t\bar tZ$ final state  cases, CP violation arises due to interference of
the diagrams  where the neutral 
Higgs is coupled to a $Z$-boson with the diagrams where
it is radiated off the $t$ or $\bar t$. We note that 
in the $t\bar tZ$ case there is no CP-violating contribution coming 
from the interference between 
the diagrams with the $ZZ{\cal H}$ coupling and the diagrams where 
the $Z$-boson is emitted from the incoming electron or positron lines (not
shown in Fig.~\ref{eetthzfig1}).  
 
The relevant pieces of the interaction Lagrangian involve the $t\bar t {\cal
H}^k$ Yukawa and the  $ZZ{\cal H}^k$ couplings and are given in 
Eqs.~\ref{2hdmab} and  \ref{2hdmc}. There, ${\cal H}^k$ ($k=1,2$ or 3) are the
three neutral Higgs scalars in the theory. As usual the three couplings
$a_t^k$, $b_t^k$ and $c^k$  in Eqs.~\ref{2hdmab} and  \ref{2hdmc} are
functions of $\tan\beta \equiv v_2/v_1$ (the ratio of the two VEV's) and of
the three mixing angles $\alpha_1$, $\alpha_2$, $\alpha_3$ which characterize
the Higgs mass matrix in Eq.~\ref{2hdmrij} (for details 
see section \ref{sssec323}). 

As was also mentioned in section \ref{sssec323}, 
only two out of the three neutral
Higgs can simultaneously  have a coupling to vector-bosons and a pseudoscalar
coupling to fermions.   Therefore, only those two neutral 
Higgs particles are relevant
for the present discussion and,  without loss of generality we denote them
as ${\cal H}^1=h$ and  ${\cal H}^2=H$ with couplings $a_t^h,b_t^h,c^h$ and
$a_t^H,b_t^H,c^H$, corresponding to the light, $h$, and heavy, $H$, neutral
Higgs, respectively. This implies  the existence of a ``GIM-like''
cancellation, namely,  when both $h$ and $H$ contribute to CP violation, then 
all CP-nonconserving effects, being proportional to $b_t^hc^h + b_t^Hc^H$,
must vanish when the two Higgs states $h$ and $H$ are degenerate. In the
following we set the mass of the heavy  Higgs, $H$, to be $m_H=750$ GeV or
1 TeV\null.   

In the process $e^+e^- \to t \bar t h$, a Higgs particle is produced 
in the final state, therefore, the heavy Higgs-boson, $H$, is not important
and  this ``GIM-like'' mechanism is irrelevant. Note that there is
an additional diagram contributing to $e^+ e^- \to t \bar t h$,  
which involves the $ZhH$ coupling and is not shown in Fig.~\ref{eetthzfig1}.
This diagram is, however,  negligible compared to the others 
for the large $m_H$ values used here.
In contrast, in the process $e^+ e^- \to t \bar t Z$, the Higgs is exchanged
as a virtual or a real particle and the effect of $H$ is, although small
compared  to $h$, important in order to restore the ``GIM-like'' cancellation
discussed above.  
  
For both the $t\bar th$ and $t\bar tZ$ final states processes, we denote the tree-level
polarized  Differential Cross-Section (DCS) 
by  $\Sigma_{(j)f}$, where
$f=t\bar th$
or $f=t\bar tZ$ corresponding  to the $t\bar th$ or $t\bar tZ$ final states, respectively,
and $j=1(-1)$ for  the left(right) polarized incoming electron beam. 
$\Sigma_{(j)f}$ can be subdivided into its CP-even ($\Sigma_{+(j)f}$) and
CP-odd  ($\Sigma_{-(j)f}$) parts

\beq
\Sigma_{(j)f} = \Sigma_{+(j)f} + \Sigma_{-(j)f} \label{eetthzeq4} ~.
\eeq

\noindent The CP-even and CP-odd DCS's can be further subdivided into different
terms which correspond to the various Higgs coupling combinations and which
transform as even or odd (denoted by the letter $n$) under $T_N$. For both
final states, $f=t\bar th$ and $f=t\bar tZ$,  we have

\beqa
\Sigma_{+(j)f} &=& \sum_i g_{+f}^{i(n)}
 F_{+(j)f}^{i(n)} \label{eetthzeq5} ~~,~~ {\rm CP-even} ~, 
\nonumber\\
\Sigma_{-(j)f} &=& \sum_i g_{-f}^{i(n)}
 F_{-(j)f}^{i(n)} \label{eetthzeq6} ~~,~~{\rm CP-odd}~,
\eeqa

\noindent where $g_{+f}^{i(n)}, g_{-f}^{i(n)}$, $n=+ ~or~ -$, are different
combinations   of the Higgs couplings $a_t^{\cal H},b_t^{\cal H},c^{\cal H}$ 
and $F_{+(j)f}^{i(n)},F_{-(j)f}^{i(n)}$, again with $n=+~or~-$, are
kinematical functions of phase space which transform like $n$ under $T_N$.  

Let us first write the Higgs coupling combinations for the CP-even part. In 
the case of $e^+ e^- \to t \bar t h$, neglecting the imaginary part 
in the s-channel $Z$-propagator, we have 
four relevant coupling combinations \cite{prl77p5172,prd53p1162,hepph9710355}:

\beq
g_{+t\bar th}^{1(+)}=(a_t^h)^2~,~g_{+t\bar th}^{2(+)}=(b_t^h)^2~,~
g_{+t\bar th}^{3(+)}=(c^h)^2~,~g_{+t\bar th}^{4(+)}=a_t^h c^h 
\label{eetthzeq7}~.
\eeq

\noindent In the case of $e^+ e^- \to t \bar t Z$, apart from the SM 
contribution,
which corresponds to interference terms  among the four SM diagrams
represented by diagram (a) on  the right hand side of Fig.~\ref{eetthzfig1},
and keeping terms proportional  to both the real and imaginary parts of the
Higgs propagator, $\Pi_{\cal H}$, we get \cite{plb419p340,hepph9710355}:

\beqa
g_{+t\bar tZ}^{1(+)}&=&(a_t^{\cal H}c^{\cal H})   \Re{\rm e}(\Pi_{\cal H})~,~
g_{+t\bar tZ}^{2(-)}=(a_t^{\cal H}c^{\cal H})     \Im{\rm m}(\Pi_{\cal H}) 
\label{eetthzeq8}~, \nonumber\\
g_{+t\bar tZ}^{3(+)}&=&(a_t^{\cal H}c^{\cal H})^2 \Re{\rm e}(\Pi_{\cal H})~,~
g_{+t\bar tZ}^{4(+)}=(a_t^{\cal H}c^{\cal H})^2   \Im{\rm m}(\Pi_{\cal H})
\label{eetthzeq9}~, \nonumber\\
g_{+t\bar tZ}^{5(+)}&=&(b_t^{\cal H}c^{\cal H})^2 \Re{\rm e}(\Pi_{\cal H})~,~
g_{+t\bar tZ}^{6(+)}=(b_t^{\cal H}c^{\cal H})^2   \Im{\rm m}(\Pi_{\cal H})
\label{eetthzeq10} ~,
\eeqa

\noindent where

\beq
\Pi_{\cal H} \equiv \left(s+m_Z^2-m_{\cal H}^2-
2p \cdot p_Z +i m_{\cal H} \Gamma_{\cal H} \right)^{-1} 
\label{eetthzeq11}~,
\eeq 

\noindent $p \equiv p_-+p_+$ and $\Gamma_{\cal H}$  is the width of ${\cal
H} = h ~{\rm or}~ H$.
 
The CP-odd coupling combinations are 

\beqa
g_{-t\bar th}^{1(-)}=b_t^h c^h \label{eetthzeq12}~,  
\eeqa

\noindent for the $t \bar t h$ final state and 

\beqa
g_{-t\bar tZ}^{1(-)}=b_t^{\cal H} c^{\cal H} \Re{\rm e}(\Pi_{\cal H}) ~,~
g_{-t\bar tZ}^{2(+)}=b_t^{\cal H} c^{\cal H} \Im{\rm m}(\Pi_{\cal H}) 
\label{eetthzeq13}~,
\eeqa

\noindent for $t \bar t Z$ final state.

\begin{figure}[htb]
 \psfull
\begin{center}
  \leavevmode
  \epsfig{file=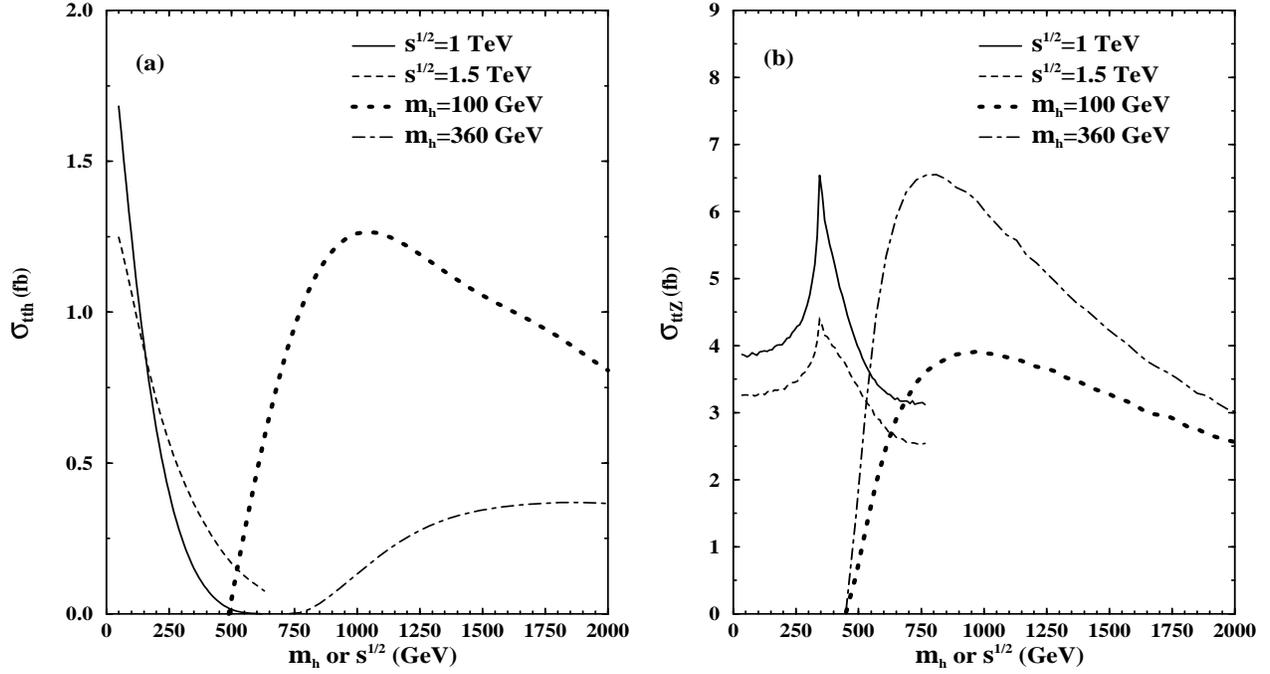,height=9.5cm,width=9cm,bbllx=0cm,
 bblly=2cm,bburx=20cm,bbury=25cm,angle=0}
\end{center}
\caption[dummy]{\emph{The cross-sections (in fb) for: (a) 
the reaction $e^+e^-\to t\bar{t}h$
with  $\tan\beta=0.5$ and (b) the reaction $e^+e^-\to t\bar{t}Z$ with 
$\tan\beta=0.3$, assuming unpolarized  electron and positron beams, 
for Model~II with set~II and as a function of $m_h$ (solid and dashed lines)
and $\sqrt s$ (dotted and dotted-dashed lines). Set~II means
$\left\{\alpha_1,\alpha_2,\alpha_3 \right\} \equiv \left\{\pi/2,\pi/4,0 \right
\}$. Figure taken from \cite{hepph9710355}.}}
\label{eetthzfig2}
\end{figure}

The CP-even pieces, $\Sigma_{+(j)f}$, yield the corresponding 
cross-sections (recall that $f=t\bar th$ or $t\bar tZ$)

\beq
\sigma_{(j)f} = \int \Sigma_{+(j)f} (\Phi) d \Phi \label{eetthzeq14}~,
\eeq

\noindent where $\Phi$ stands for the phase-space variables. In
Fig.~\ref{eetthzfig2}(a) and \ref{eetthzfig2}(b) we plot the unpolarized
cross-sections, $\sigma_{t\bar th}$ and $\sigma_{t\bar tZ}$ as a function
of $m_h$ and $\sqrt s$, for Model~II (i.e., 2HDM of type 2 as defined in
section \ref{sssec323}), with $m_H=750$ GeV and the set of values
$\left\{\alpha_1,\alpha_2,\alpha_3 \right\}= \left\{\pi/2,\pi/4,0
\right\}$ which we denote as set~II\null. Set~II is also adopted later
when discussing the CP-violating effect. Furthermore, for the $t\bar th$
final state we choose $\tan\beta=0.5$ while for $t\bar tZ$ we choose
$\tan\beta=0.3$.  Afterwards, we will discuss the dependence of the
CP-violating effect on $\tan\beta$ in the $t\bar th$ and $t\bar tZ$ cases.
One can observe the dissimilarities in the two cross-sections
$\sigma_{t\bar th}$ and $\sigma_{t\bar tZ}$: while $\sigma_{t\bar th}$ is
at most $\sim 1.5$ fb, $\sigma_{t\bar tZ}$ can reach $\sim 7$ fb at around
$\sqrt s \approx 750$ GeV and $m_h \gsim 2 m_t$. $\sigma_{t\bar th}$ drops
with $m_h$ while $\sigma_{t\bar tZ}$ grows in the range $m_h \lsim 2m_t$. 
$\sigma_{t\bar tZ}$ peaks at around $m_h \gsim 2m_t$ and drops as $m_h$
grows further.  Moreover, $\sigma_{t\bar th}$ peaks at around $\sqrt s
\approx 1(1.5)$ TeV for $m_h=100(360)$ GeV, while $\sigma_{t\bar tZ}$
peaks at around $\sqrt s \approx 750$ GeV for both $m_h=100$ and 360
GeV\null. As we will see later, these different features of the two
cross-sections are, in part, the cause for the different behavior of the
CP asymmetries discussed below.

Let us now concentrate on the CP-odd, $T_N$-odd effects in $e^+ e^- \to t
\bar t h,~ t \bar t Z$, emanating from the $T_N$-odd pieces in
$\Sigma_{-(j)t\bar th},~\Sigma_{-(j)t\bar tZ}$, respectively.  From
Eqs.~\ref{eetthzeq12} and \ref{eetthzeq13} it is clear that the
CP-violating pieces $\Sigma_{-(j)t\bar th},~\Sigma_{-(j)t\bar tZ}$ have to
be proportional to $b_t^h c^h$ (in the $t\bar tZ$ case there is an
additional similar piece corresponding to the heavy Higgs $H$).  The
corresponding CP-odd kinematic functions, $F_{-(j)t\bar
th}^{1(-)},~F_{-(j)t\bar tZ}^{1(-)}$, being $T_N$-odd, are pure tree-level
quantities and are proportional to the only non-vanishing Levi-Civita
tensor present, $\epsilon(p_-,p_+,p_t,p_{\bar t})$, when the spins of the
top are disregarded.  The explicit expressions for $F_{-(j)f}^{1(-)}$ are
(recall that $j=1(-1)$ for left(right) polarized incoming electron beam)

\beqa
F_{-(j)t\bar th}^{1(-)}& = & - \frac{1}{\sqrt 2}  \left( \frac{g_W^3}{c_W^3}
\right)^2 \frac{m_t^2}{m_Z^2} \Pi_{Zh} \Pi_Z T_t^3 c_j^Z 
\epsilon(p_-,p_+,p_t,p_{\bar t}) \times \nonumber\\ 
&& \left\{ j(\Pi_t^h + \Pi_{\bar t}^h) \left[ (s-s_t-m^2_h) (3w_j^- -w_j^+)
+ m_Z^2 (w_j^- -w_j^+)  \right] + \right. \nonumber \\
&& \left. T_t^3 c_j^Z  \Pi_Z (\Pi_t^h-\Pi_{\bar t}^h) f \right\}
\label{eetthzeq15} ~,\\
F_{-(j)t\bar tZ}^{1(-)}& = &  - \sqrt 2 \left( \frac{2g_W^3}{c_W^3} \right)^2 
\frac{m_t^2}{m_Z^2} \Pi_Z T_t^3 c_j^Z \epsilon(p_-,p_+,p_t,p_{\bar t}) \times
\nonumber\\ 
&& \left\{ j (\Pi_t^Z+\Pi_{\bar t}^Z) 
\left[ m_Z^2 w_j^- + (s_t-s) w_j^+ \right] + \right. \nonumber\\
&& \left. T_t^3 c_j^Z  \Pi_Z (\Pi_t^Z-\Pi_{\bar t}^Z) f \right\} 
\label{eetthzeq16}~,
\eeqa

\noindent where $s\equiv 2p_-\cdot p_+$ is the c.m.\ energy of the colliding
electrons, $s_t\equiv (p_t+p_{\bar t})^2$ and  $f\equiv (p_--p_+) \cdot
(p_t+p_{\bar t})$. Also 

\beqa
\Pi_{t(\bar t)}^h & \equiv & \left(2p_{t(\bar t)} 
\cdot p_h +m^2_h \right)^{-1} ~,~ 
\Pi_{t(\bar t)}^Z \equiv \left(2p_{t(\bar t)}\cdot p_Z + m^2_Z \right)^{-1}
\label{eetthzeq17} ~, \\
\Pi_Z &\equiv& \left(s-m_Z^2\right)^{-1}~,~
\Pi_{\gamma} \equiv s^{-1}~,~
\Pi_{Zh} \equiv \left((p - p_h)^2 - m^2_Z \right)^{-1} 
\label{eetthzeq18}~.  
\eeqa

\noindent and 

\beq
w_j^{\pm} \equiv \left( s_W^2Q_t - \frac{1}{2}T_t^3 \right)
c_j^Z  \Pi_Z \pm  Q_t s_W^2c_W^2 \Pi_{\gamma} \label{eetthzeq20} ~,
\eeq

\noindent where $s_W(c_W)$ is the $\sin(\cos)$ of the weak mixing angle
$\theta_W$, $Q_f$ and $T_f^3$ are the charge and the $z$-component of the
weak isospin of  a fermion, respectively, and $c_{-1}^Z=1/2-s_W^2,~
c_{1}^Z=-s_W^2$.

Since at tree-level there cannot be any absorptive phases, CP-violating
asymmetries only of the $T_N$-odd type are expected to occur in
$\Sigma_{-(j)f}$. Note that in the $t\bar tZ$ case there is also a CP-odd,
$T_N$-even piece, $ b_t^{\cal H} c^{\cal H} {\rm Im}(\Pi_{\cal H}) \times
F_{-(j)t\bar tZ}^{2(+)}$ (see Eq.~\ref{eetthzeq13}), in the DCS\null.
However, being proportional to the absorptive part coming from the Higgs
propagator, it is not a pure tree-level quantity.

Simple examples of observables that can trace the tree-level CP-odd effect in
$e^+ e^- \to t \bar t h;  t \bar t Z$ are \cite{prd53p1162}:

\beq
O = 
\frac{\vec{p}_-\cdot(\vec{p}_t\times\vec{p}_{\bar{t}})}{s^{3/2}} 
 ~~,~~ 
O_{\rm opt} (t\bar th) = 
\frac{\Sigma_{-(j)t\bar th}}  {\Sigma_{+(j)t\bar th}}
~;~~~
O_{\rm opt} (t\bar tZ) = 
\frac {\Sigma_{-(j)t\bar tZ}} {\Sigma_{+(j)t\bar tZ}}
\label{eetthzeq21}~.
\eeq

\noindent Here $O_{\rm opt} (t\bar th;t\bar tZ)$ are optimal observables
in the sense that the statistical error in the measured asymmetry is
minimized \cite{prd45p2405}. Note also that only the $T_N$-odd part of
$\Sigma_{-(j)t\bar tZ}$ is involved. As mentioned before, since the final
state consists of three particles, using only the available momenta, there
is a unique antisymmetric combination of momenta that can be formed.
Thus, both observables are proportional to $\epsilon(p_-,p_+,p_t,p_{\bar
t})$. Furthermore, $O_{\rm opt} (t\bar th;t\bar tZ)$ are related to $O$
through a multiplication by a CP-even function. In the following we focus
only on the CP-odd effects coming from the optimal observables. We remark,
however, that the results for the simple observable $O$ exhibit the same
behavior, though slightly smaller then those for $O_{\rm opt}$. 

The theoretical statistical significance, $N_{SD}$, in which an asymmetry
can be measured in an ideal experiment is $N_{SD}= A \sqrt L \sqrt
{\sigma}$ ($\sigma = \sigma_{t\bar th},\sigma_{t\bar tZ}$ for the $t\bar
th,t\bar tZ$ final states, respectively), where for the observables $O$
and $O_{\rm opt}$, the CP-odd quantity $A$, defined above, is

\beq
A_O \approx \langle O \rangle/\sqrt{\langle O^2\rangle} ~~,~~
A_{\rm opt} \approx \sqrt{\langle O_{\rm opt} \rangle} 
\label{eetthzeq22} ~.
\eeq

\noindent Also, $L$ is the effective luminosity for fully reconstructed
$t \bar t h$ or $t \bar t Z$ events. In particular,  we take $L=\epsilon
{\cal L}$, where ${\cal L}$ is the total yearly integrated  
luminosity and $\epsilon$ is the overall efficiency for   
reconstruction of the $t \bar t h$ or $t \bar t Z$ final states.

\begin{figure}[htb]
 \psfull
 \begin{center}
  \leavevmode
  \epsfig{file=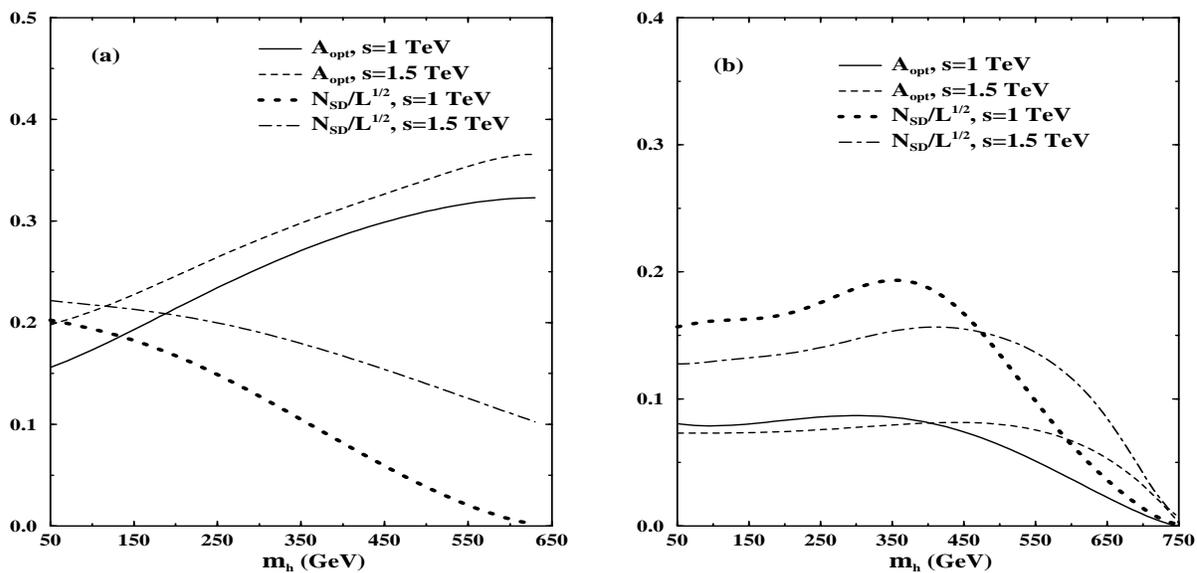,height=8cm,width=9cm,bbllx=0cm,
 bblly=2cm,bburx=20cm,bbury=25cm,angle=0}
 \end{center}
\caption[dummy]{\emph{The asymmetry, $A_{\rm opt}$, and scaled statistical
significance, $N_{SD}/\sqrt L$, for the optimal observable  $O_{\rm opt}$
for: (a) the reaction $e^+e^-\to t\bar{t}h$ with  $\tan\beta=0.5$ and (b)
the reaction $e^+e^-\to t\bar{t}Z$ with $\tan\beta=0.3$, as a function of
the light Higgs mass $m_h$, for $\sqrt s=1$ TeV and 1.5 TeV\null. All graphs
are with set~II of the parameters, as in Fig.~\protect\ref{eetthzfig2}, 
figure taken from \cite{hepph9710355}.}}
\label{eetthzfig3}
\end{figure}

In the following numerical analysis we have used set~II defined before for
the angles $\alpha_{1,2,3}$, i.e., $\left\{\alpha_1,\alpha_2,\alpha_3
\right\}= \left\{\pi/2,\pi/4,0 \right\}$ \footnote{Recall that for the
$t\bar th$ final state we choose $\tan\beta=0.5$ while for the $t\bar tZ$
final state we take $\tan\beta=0.3$}.  Figs.~\ref{eetthzfig3}(a) and
\ref{eetthzfig3}(b) show the expected asymmetry and statistical
significance in the unpolarized case, corresponding to $O_{\rm opt}$ in
Model~II for the $t\bar th$ and $t\bar tZ$ final states, respectively. The
asymmetry is plotted as a function of the mass of the light Higgs ($m_h$)
where again, $m_H=750$ GeV in the $t\bar tZ$ case.  We plot $N_{SD}/\sqrt
L$, thus scaling out the luminosity factor from the theoretical prediction
\footnote{as a reference value, we note that for $L = 100$ fb$^{-1}$,
$N_{SD}/\sqrt L =0.1$ will correspond to a 1-$\sigma$ effect}.

We remark that set~II corresponds to the largest CP effect, though not
unique since we are dealing with angles, i.e., $\alpha_{1,2,3}$, 
which may be rotated by $\pi$ or $\pi/2$ leaving the relevant combinations of 
angles with the same value (e.g., $b_t^h \propto \sin\alpha_1 \sin\alpha_2$). 
In the $t\bar th$ case $\tan\beta=0.5$ is favored, however, 
the effect mildly depends on $\tan\beta$ in the range 
$ 0.3 \lsim \tan\beta \lsim 1$ \cite{prd53p1162,hepph9710355}.  
In the $t\bar tZ$ case, the effect is practically insensitive to $\alpha_3$ and
is roughly proportional to $1/\tan\beta$,  it therefore drops as $\tan\beta$
is increased.  Nonetheless, we find that $N_{SD}/\sqrt L>0.1$, even in the
unpolarized  case for $\tan\beta \lsim 0.6$ \cite{plb419p340,hepph9710355};
note that $N_{SD}/\sqrt L$ here is dimensionless if $L$ is in fb$^{-1}$.

From Fig.~\ref{eetthzfig3}(a) we see that, 
in the $t\bar th$ case, as $m_h$ grows
the asymmetry increases while the statistical  significance drops, in part
because of the decrease in the cross-section. Evidently, the asymmetry
can become quite large; it  ranges from $\sim 15\%$, for 
$m_h \lsim 100$ GeV, 
to $\sim 35\%$ for $m_h \sim 600$ GeV. Indeed, the
CP-effect is more significant for smaller masses  of $h$, wherein $A_{\rm
opt}$ is smaller.  In contrast, Fig.~\ref{eetthzfig3}(b) shows that, 
in the $t\bar tZ$ case, $A_{\rm opt}$ stays roughly fixed at around $7-8 \%$
for $m_h \lsim 2m_t$,  and then drops till it totally vanishes at
$m_h=m_H=750$ GeV, due to  the ``GIM-like''  mechanism discussed above. The
scaled statistical significance  $N_{SD}/\sqrt L$ behaves roughly as $A_{\rm
opt}$. That is,     $N_{SD}/\sqrt L \approx 0.1$--0.2 in the mass range 
$50 ~{\rm GeV} \lsim m_h \lsim 350 ~{\rm GeV}$,   
for both $\sqrt s=1$ and 1.5 TeV\null.   

\begin{figure}[htb]
 \psfull
 \begin{center}
  \leavevmode
  \epsfig{file=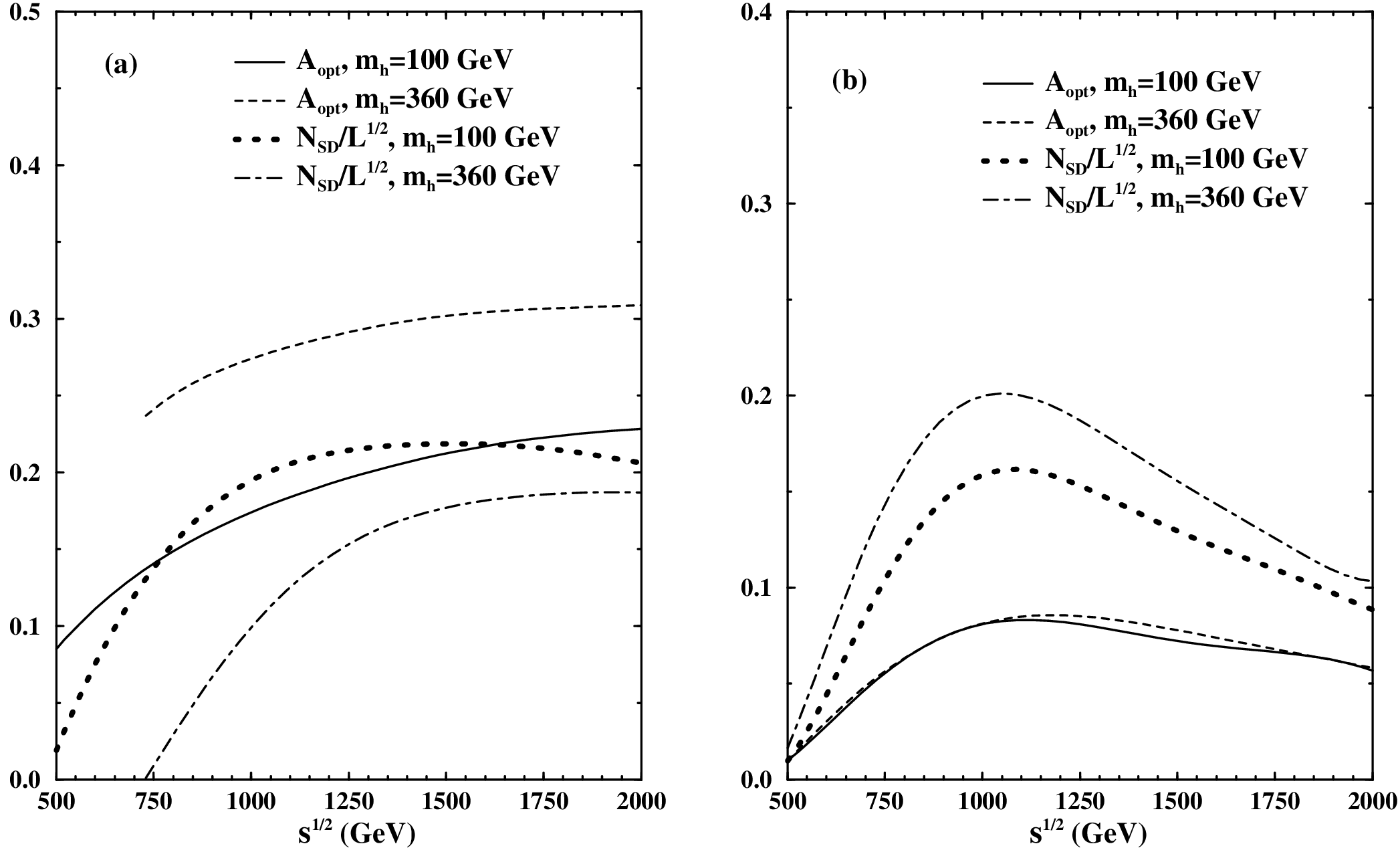,height=8cm,width=9cm,bbllx=0cm,
 bblly=2cm,bburx=20cm,bbury=25cm,angle=0}
 \end{center}
\caption[dummy]{\emph{The asymmetry, $A_{\rm opt}$, and scaled 
statistical significance,
$N_{SD}/\sqrt L$, for the optimal observable  $O_{\rm opt}$ for: (a) the
reaction $e^+e^-\to t\bar{t}h$ with  $\tan\beta=0.5$ and (b) the reaction
$e^+e^-\to t\bar{t}Z$ with $\tan\beta=0.3$, as a function of the c.m.\ energy
$\sqrt s$, for  $m_h=100$ GeV and $m_h=360$ GeV\null. All graphs are with
set~II  of the parameters, as in Fig.~\ref{eetthzfig2}.
Figure taken from \cite{hepph9710355}.}}
\label{eetthzfig4}
\end{figure}

Figs.~\ref{eetthzfig4}(a) and \ref{eetthzfig4}(b) 
show the dependence of $A_{\rm
opt}$ and $N_{SD}/\sqrt L$ on the  c.m.\ energy, $\sqrt s$, for the
$t\bar th$
and $t\bar tZ$ cases, respectively. We see that, in the case of $t\bar th$, 
the CP-effect 
peaks at $\sqrt s \approx 1.1(1.5)$ TeV for $m_h=100(360)$ GeV and 
stays roughly the same as $\sqrt s$ is further increased to 2 TeV\null. 
In the case of $t\bar tZ$, the statistical significance is 
maximal at around $\sqrt
s \approx 1$ TeV and then  decreases slowly as $\sqrt s$ grows for
both $m_h=100$ and 360 GeV\null. Contrary to the $t\bar th$ case, where a
light $h$ is favored, 
in the  $t\bar tZ$ case, the effect is best for $m_h \gsim 2m_t$. 
In that range,
on-shell $Z$ and $h$ are produced followed by the $h$ decay $h \to t \bar
t$, thus, the  Higgs exchange diagram becomes more dominant.           

In Tables~\ref{eetthztab1} and \ref{eetthztab2} we present $N_{SD}$ for
$O_{\rm opt}$, for the $t\bar th$ and $t\bar tZ$ cases, respectively, in Model~II with
set~II, and we also compare the effect of beam polarization with the unpolarized
case. As before, we take $\tan\beta=0.5$ and $\tan\beta=0.3$ for the
$t\bar th$
and $t\bar tZ$ cases, respectively,  where for the $t\bar tZ$ case we 
present
numbers for both  $m_H=750$ GeV (shown in the parentheses) and $m_H=1$ TeV,
to demonstrate the sensitivity of the CP-effect to the mass of the heavy
Higgs. For illustrative purposes, we choose $m_h=100,160$ and 360 GeV and
show the  numbers for $\sqrt s =1$ TeV  with ${\cal L}=200$ [fb]$^{-1}$ and
for  $\sqrt s =1.5$ TeV with ${\cal L}=500$ [fb]$^{-1}$ (see \cite{nlc}).
In both cases we take $\epsilon=0.5$ assuming that there is no loss  of
luminosity when the electrons are polarized.\footnote{Clearly if 
the efficiency for $t \bar
th$ and/or $t\bar tZ$  reconstruction is $\epsilon=0.25$, then our numbers 
would correspondingly require 2 years of running.} 

Evidently, for both reactions, left polarized incoming electrons can probe
CP violation  slightly better than unpolarized ones. We  see that in the
$t\bar th$ case the CP-violating effect 
drops as the mass of the light Higgs ($h$) grows,
while  in the $t\bar tZ$ case it grows with $m_h$. In particular, we find 
that with $\sqrt s =1.5$ TeV and for $m_h \gsim 2m_t$ the effect is
comparable for both  the $t\bar th$ and the $t\bar tZ$ cases where 
it reaches above
3-$\sigma$  for negatively polarized electrons. With a light Higgs mass 
in the range $100 ~{\rm GeV} \lsim m_h \lsim 160 ~{\rm GeV}$, the
$t\bar th$ case
is more sensitive to $O_{\rm opt}$ and the  CP-violating effect can reach
$\sim 4$-$\sigma$ for left polarized electrons.  In that light Higgs mass range,
the CP-violating effect reaches  slightly below 2.5-$\sigma$ for the
$t\bar tZ$
case.  For a c.m.\ energy of $\sqrt s =1$ TeV and $m_h =360$ GeV, the
$t\bar tZ$
case is much more sensitive to $O_{\rm opt}$ and  the effect can reach 
2.2-$\sigma$ for left polarized electron beam.  However, with that c.m.\ energy,
the $t\bar th$ mode gives a larger CP-odd effect in the range 
$m_h \sim 100 -160$ GeV.

\begin{table}[htb]
\begin{center}
\caption[first entry]
{\emph{
The statistical significance, $N_{SD}$, in which 
the CP-nonconserving effects in $e^+e^- \to t \bar t h$ can be detected in
one year of running of a future high energy  collider with either unpolarized
or polarized incoming electron beam.  We have used $\tan\beta=0.5$, a yearly
integrated luminosity of ${\cal L}=200$ and 500 [fb]$^{-1}$ for 
$\sqrt s=1$ and 1.5 TeV, respectively, and an efficiency reconstruction
factor of  $\epsilon=0.5$  for both energies. 
Recall that $j=1(-1)$ stands  for right(left) polarized electrons. 
Set~II means 
$\left\{\alpha_1,\alpha_2,\alpha_3 \right\} \equiv 
\left\{\pi/2,\pi/4,0 \right\}$. 
Table taken from \cite{hepph9710355}. \protect\label{eetthztab1} 
}}

\bigskip

\begin{tabular}{|r||r|r|r|r|} \cline{3-5}
\multicolumn{2}{c||}{~~} & 
\multicolumn{3}{c|}{$e^+e^- \to t \bar t h$ (Model~II with Set~II)}\\ \hline
$\sqrt{s}$ & j
& \multicolumn{3}{c|}{$O_{\rm opt}$}
\\ \cline{3-5}
$({\rm TeV}) \Downarrow$ &$({\rm GeV}) \Rightarrow$ & $m_h=100$ 
& $m_h=160$ &$m_h=360$\\ 
\hline
\hline
&-1& $2.2$ & $2.0$ & 1.1 \\ \cline{2-5}
1 & unpol & $2.0$ & $1.9$ & 1.0\\ \cline{2-5}
& 1 &  $1.8$ & $1.7$ & 0.9 \\ \hline
\hline
& -1 &  $4.0$ & $3.9$ & 3.2 \\ \cline{2-5}
1.5 & unpol & $3.6$ & $3.5$ & 2.9 \\ \cline{2-5}
&1 &  $3.2$ & $3.1$ & 2.6\\ \hline 
\end{tabular}
\end{center}
\end{table}
\begin{table}[htb]
\begin{center}
\caption[first entry]{\emph{The same as Table~\ref{eetthztab1} but for 
$e^+e^- \to t \bar t Z$, with $\tan\beta=0.3$. In this reaction, effects
of the heavy  Higgs, $H$, are included and  $N_{SD}$ is given for both
$m_H=750$ GeV (in parentheses)  and $m_H=1$ TeV\null.
Table taken from \cite{hepph9710355}.
\protect\label{eetthztab2}
}}

\bigskip
\begin{tabular}{|r||r|r|r|r|} \cline{3-5}
\multicolumn{2}{c||}{~~} & \multicolumn{3}{c|}{$e^+e^- \to t \bar t Z$ (Model~II with Set~II)}\\ \hline
$\sqrt{s}$ & j
& \multicolumn{3}{c|}{$O_{\rm opt}$}
\\ \cline{3-5}
$({\rm TeV}) \Downarrow$ &$({\rm GeV}) \Rightarrow$ & $m_h=100$ & $m_h=160$ &$m_h=360$\\ 
\hline
\hline
&-1& $(1.8)~1.7$ & $(1.8)~1.8$ & (2.2)~2.2 \\ \cline{2-5}
1 & unpol & $(1.6)~1.6$ & $(1.7)~1.6$ & (2.0)~2.0 \\ \cline{2-5}
& 1 &  $(1.5)~1.5$ & $(1.5)~1.5$ & (1.8)~1.8 \\ \hline
\hline
& -1 &  $(2.3)~2.9$ & $(2.4)~3.0$ & (2.8)~3.3 \\ \cline{2-5}
1.5 & unpol & $(2.1)~2.6$ & $(2.1)~2.7$ & (2.5)~3.0 \\ \cline{2-5}
&1 &  $(1.8)~2.3$ & $(1.8)~2.3$ & (2.1)~2.6\\ \hline 
\end{tabular}
\end{center}
\end{table}

Let us now summarize the above results and add some concluding remarks. We
have shown that an extremely interesting CP-odd signal may arise at
tree-level in the reactions $e^+ e^- \to t \bar t h$ and $e^+ e^- \to t
\bar t Z$. The asymmetries that were found are $\sim{}$15\%--35\% in the
$t\bar th$ case and $\sim{}$5\%--10\% for the $t\bar tZ$ final state. 
These asymmetries may give rise in the best cases, i.e., for a favorable
set of the relevant 2HDM parameters, to $\sim 3 - 4$-$\sigma$, CP-odd,
signals in a future $e^+e^-$ collider running with c.m.\ energies in the
range 1 TeV${}\lsim \sqrt s \lsim 2$ TeV\null.

Note, however, that the simple observable, $O$, as well as the optimal
one, $O_{\rm opt}$, require the identification of the $t$ and $\bar t$ and
the knowledge of the transverse components of their momenta in each $t\bar
th$ or $t \bar t Z$ event.  Thus, for the main top decay, $t \to b W$,
 the most suitable scenario is when either the $t$ or the $\bar
t$ decays  leptonically and the other decays hadronically.
Distinguishing between $t$ and $\bar t$ in the double-hadronic decay case
will require  more effort and still remains an experimental challenge.
If, for example, the identification of the charge of the $b$-jets
coming from  the $t$ and the $\bar t$ is possible, then the difficulty 
in reconstructing the transverse components of the $t$ and $\bar t$ 
momenta may be surmountable by using the momenta of the decay products
in the processes $e^+e^-\to t\bar th\to bW^+\bar bW^- h$ and $e^+e^-\to t\bar
tZ\to bW^+\bar bW^- Z$. For example, the observable  

\beq 
O_b = \frac{\epsilon (p_-, p_+, p_b, p_{\bar b}) } {s^2} \label{eetthzeq23} ~,
\eeq 

\noindent may then be used. We have considered this observable for the
reaction   $e^+e^-\to t\bar th\to bW^+\bar bW^- h$ in \cite{prd53p1162}.
We found there that, close to threshold, this observable is not very
effective. However, at higher energies,   $O_b$ is about as sensitive as the
simple triple product  correlation $O$ defined in Eq.~\ref{eetthzeq21} and,
therefore, only slightly less sensitive then $O_{\rm opt}$.

Note also that for the light Higgs mass, $m_h=100$ GeV, the most suitable
way to detect the Higgs  in $e^+e^-\to t\bar th \to bW^+\bar bW^- h $  is
via $h \to b \bar b$  with branching ratio $\sim 1$.  For $m_h \gsim 2m_t$,
and specifically with set~II used above,  there are two competing Higgs
decays, $h \to t \bar t$ and  $h \to W^+ W^-$, depending on the value of 
$\tan\beta$. For example, for $\tan\beta=0.5$, as was chosen above, one has
${\rm Br} (h \to t \bar t) \approx 0.77$ and  ${\rm Br} (h \to W^+ W^-)
\approx 0.17$, thus, the  $h \to t \bar t$ mode is more 
suitable. Of course,
$h \to t \bar t$ will dominate more for smaller values of $\tan\beta$ and
less if $\tan\beta > 0.5$. In particular,  for $\tan\beta=0.3(1)$ one has 
${\rm Br} (h \to t \bar t) \approx 0.89(0.57)$ and ${\rm Br} (h \to W^+ W^-)
\approx 0.08(0.32)$.

Finally, as emphasized before, the final states $t\bar th$ and $t\bar tZ$,
in particular the $t\bar th$, are expected to be the center of
considerable attention at a linear collider.  Extensive studies of these
reactions are expected to teach us about the details of the couplings of
the neutral Higgs to the top quark~\cite{prd61p013002}. Thus, it is
gratifying that the same final states promise to exhibit interesting
effects of CP violation. It would be very instructive to examine the
effects in other extended models. Numbers emerging from the 2HDM that was
used, especially with the specific value of the parameters, should be
viewed as illustrative examples. The important point is that the reactions
$e^+e^-\to t\bar th \to bW^+\bar bW^- h $ and $e^+e^-\to t\bar tZ \to
bW^+\bar bW^- Z$ appear to be very powerful and very clean tools for
extracting valuable information on the parameters of the underlying model
for CP violation.

\subsubsection{Generalized optimization technique and extraction 
of various Higgs couplings \label{sssec622}}

An optimization technique was employed by Gunion \etal\  in
\cite{prl77p5172} for the process $e^+e^-\to t\bar th$. It was shown that
this  reaction may provide a powerful tool for extracting the $t\bar th$ Yukawa
couplings and the $ZZh$ couplings.  
A similar analysis for the reaction $e^+e^-\to t\bar t Z$ was likewise 
considered 
in~\cite{prd60p115018}.
This technique is outlined in section
\ref{ssec26}.  
The basic idea is that the nature of the Higgs particle, i.e., whether
it is CP-odd or  CP-even, may well be distinguishable through studies of 
momentum correlations in $e^+e^-\to t\bar th$.  In particular, greater 
information on the detailed dependence of $\Sigma_{(j)t\bar th}(\Phi)$ on the 
 variable $\Phi$ is extracted to deduce limits that can be obtained on the
different Higgs couplings combinations in  
Eqs.~\ref{eetthzeq7} and \ref{eetthzeq12}.  
As described before, the differential $e^+e^-\to t\bar
th$ cross-section contains  five distinct terms which are explicitly written
in Eqs.~\ref{eetthzeq7}  and \ref{eetthzeq12}. The only CP-violating component
is  $b_t^h c^h$, while the others enter into the total cross-section as in 
Eq.~\ref{eetthzeq14}.   

Gunion \etal\ have investigated two issues. For a given c.m.\ energy
and integrated luminosity at the NLC,  they have examined:
\begin{enumerate}
\item The 1-$\sigma$ error in the determination 
of the couplings $a_t^h,b_t^h,c^h$, by fixing $\chi^2(t\bar th)=1$ 
for a given input model with couplings $g_{\pm t\bar th}^{0,i(n)}$.
\item To what degree of statistical significance can a model be ruled 
out, given a certain input model.
\end{enumerate}
Let us now elaborate more on how these two points were studied in
\cite{prl77p5172}. With the optimal technique outlined in \cite{prl77p5172},
Gunion \etal\ used unique weighting functions such that the statistical
error in the determination of the various  $g_{\pm t\bar th}^{i(n)}$ in 
Eqs.~\ref{eetthzeq7} and \ref{eetthzeq12} is minimized. They write

\beq
g_{\pm t\bar th}^{i(n)} = \sum_k M_{ik}^{-1} I_k \label{eetthzeq24}~,
\eeq

\noindent where $I_k$, $M_{ik}$ and the appropriate weighting 
functions are given in \cite{prl77p5172} (see also section \ref{ssec26}).
Then, given an input model, for which the couplings are denoted with the
superscript 0 as $g_{\pm t\bar th}^{0,i(n)}$, one can compute the confidence
level  at which parameters of choice, different from the input model, can be
ruled out

\beq
\chi^2(t\bar th) = \sum_{i,j=1}^5 \left( g_{\pm t\bar th}^{i(n)} -
g_{\pm t\bar th}^{0,i(n)}
\right)  \left( g_{\pm t\bar th}^{j(n)} -
g_{\pm t\bar th}^{0,j(n)} \right) V_{ij}^{-1} \label{eetthzeq25}~,
\eeq

\noindent where 

\beq
V_{ij} \equiv  < \Delta g_{\pm t\bar th}^{i(n)}   
\Delta g_{\pm t\bar th}^{j(n)} > = \frac{M_{ij}^{-1} 
\sigma_{t\bar th}}{N_{t\bar th}} \label{eetthzeq26}~,
\eeq

\noindent is the covariance matrix. $N_{t\bar th} = L \sigma_{t\bar th}$ 
is the total
number of $t \bar t h$ events, with  $L$ the effective luminosity defined
previously  (after the efficiency factor,  $\epsilon$, is included).
Therefore, the sensitivity of $\chi^2(t\bar th)$ to the couplings 
$a_t^h,b_t^h,c^h$
is determined  by the covariance matrix directly. 

Three input models were considered in \cite{prl77p5172}. These correspond
to different choices  of the set of parameters $a_t^h,b_t^h$ and $c^h$ 
(see Eqs.~\ref{2hdmab} and \ref{2hdmc} in section \ref{sssec323}), 
as follows:
\begin{description}
\item (I) A SM neutral Higgs, with $a_t^h=1/\sqrt 2,b_t^h=0,c^h=1$.
\item (II) A pure CP-odd neutral Higgs, 
with $a_t^h=0,b_t^h=1/\sqrt 2,c^h=0$.
\item (III) A CP-mixed neutral Higgs, with $a_t^h=b_t^h=1/2,c^h=1/\sqrt 2$. 
\end{description}

Given the couplings $g_{\pm t\bar th}^{0,j(n)}$ in the above input models, they
calculated $\chi^2(t\bar th)$ as a function  of the 
location in $a_t^h,b_t^h,c^h$
parameter space, from  which the 1-$\sigma$ error in any one of these 
parameters was determined. Their results are shown in Table~\ref{eetthztab3}.

%
%
%

\def\rtw{${1\over \sqrt 2}$}
\begin{table}[hbt]
\caption[fake]{\emph{
The 1-$\sigma$ errors in $a_t^h$, $b_t^h$, 
and $c^h$ are given,
for the three Higgs coupling cases I, II and III. 
Results are given for unpolarized beams and for 
100\% negative $e^-$ polarization ($P_e=-1$)
also 
$\sqrt s=1$ TeV, $m_h=100$ GeV, $m_t=176$ GeV and $L=50$ fb$^{-1}$.
Table taken from 
\cite{prl77p5172}.}}
\begin{center}
\begin{tabular}{|c|c|c|c|c|c|c|}
\hline
\ & \multicolumn{3}{c|}{Unpolarized $e^-$} & \multicolumn{3}{c|}{$P_e=-1$} \\
Case & $a_t^h\pm\Delta a_t^h$ & $b_t^h \pm \Delta b_t^h$ & $c^h\pm\Delta c^h$ 
     & $a_t^h\pm\Delta a_t^h$ & $b_t^h\pm\Delta b_t^h$ & $c^h\pm\Delta c^h$ \\
\hline
 & & & & & & \\
I & \rtw$+0.030\atop -0.047$ & 0$+0.76\atop -0.76$ & 1$+0.51 \atop -0.82$
  & \rtw$+0.026\atop -0.040$ & 0$+0.70\atop -0.70$ & 1$+0.51\atop -0.77$ \\
 & & & & & & \\
\hline
 & & & & & & \\
II & 0$+0.19\atop-0.19$ & \rtw$+0.066\atop-0.099$ & 0$+0.58\atop-0.58$ 
   & 0$+0.18\atop-0.18$ & \rtw$+0.056\atop-0.085$ & 0$+0.55\atop-0.55$ \\
 & & & & & & \\
\hline
 & & & & & & \\
III & $\frac{1}{2}$$+0.053\atop -0.062$ & $\frac{1}{2}$$+0.22\atop-0.44$ & 
 \rtw $+0.57\atop-0.80$ 
    & $\frac{1}{2}$$+0.047\atop -0.052$ & $\frac{1}{2}$$+0.19\atop-0.33$ & 
 \rtw $+0.56\atop-0.73$ \\
 & & & & & & \\
\hline
\end{tabular}
\end{center}
\label{eetthztab3}
\end{table}

We see that $a_t^h$ is well determined in all input models; in input models
I and III, where  $a_t^h \neq 0$, the 1-$\sigma$ error is at the few percent 
level. In the same manner, $b_t^h$ is best determined in input model II, where
$\Delta b \approx 10$--15\%, at 1-$\sigma$.  
The error in $c^h$ is above
the 50\% level in  all three input models. However, this can be improved by 
considering the reaction at hand, i.e., $e^+e^- \to t \bar t h$, combined
with  information extracted from the $e^+e^-\to Zh$ cross-section as was
done in \cite{hepph9609453}.  Assuming real Higgs production and  disregarding
the subsequent $h$ decay in the reaction  $e^+e^-\to Zh$, the cross-section
$\sigma (e^+e^-\to Zh)$  contains one useful Higgs coupling combination,
$g_{+t\bar th}^{3(+)} = (c^h)^2$. Therefore, the sensitivity to $(c^h)^2$ is
increased when the above technique  is also applied to $\sigma (e^+e^-\to Zh)$. 

\begin{table}[hbt]
\caption[fake]{\emph{The number of standard deviations, $\sqrt{\chi^2}$,
at which a given input model (I, II or III) can be distinguished from
the other two models, are tabulated, for $\sqrt s,m_h,m_t$ and $L$ as
in Table~\ref{eetthztab3}. Table taken from 
\cite{prl77p5172}.}}
\begin{center}
\begin{tabular}{|c|c|c|c|c|c|c|}
\hline
\ & \multicolumn{3}{c|}{Unpolarized $e^-$} & \multicolumn{3}{c|}{$P_e=-1$} \\
\ & \multicolumn{3}{c|}{Trial Model} & \multicolumn{3}{c|}{Trial Model} \\
Input Model &  I & II & III & I & II & III \\
\hline
I   &  -  & 9.5 & 4.8  &  -  & 11  & 5.5 \\
II  & 34  &  -  &  17  & 40  & -   & 20  \\
III & 6.3 & 6.3 &  -   & 7.3 & 7.3 &  -  \\
\hline
\end{tabular}
\end{center}
\label{eetthztab4}
\end{table}

Let us continue with a discussion of the work of 
Ref.~\cite{prl77p5172} in which only the
$t\bar th$ final state was considered. The ability to distinguish
between different models was  
furthermore investigated by  using the optimization technique.
This is shown in Table~\ref{eetthztab4}. We see 
from this table,
for  example, that if the Higgs is the SM one, then the pure CP-odd 
case (input model II) and the equal CP-mixture case (input model III) 
are ruled out at the 9.5-$\sigma$ and 4.8-$\sigma$ level, respectively. Note
also  that negative beam polarization slightly improves the results. 

Finally, the ability for determining a non-zero CP-violating component, $b_t^h
c^h$, was also investigated in  \cite{prl77p5172}. They found that with
$m_h=100$ GeV,  $L=100$ fb$^{-1}$ 
\footnote{this is the value considered by us in
the previous section \ref{sssec621} 
for which the results in Tables~\ref{eetthztab1} and 
\ref{eetthztab2} are given}, a non-zero $b_t^h c^h$ coupling can be
established at a level better then 1-$\sigma$ in a 1 TeV $e^+e^-$ collider.


\subsubsection{CP asymmetries in $e^+ e^- \to Zh$ and in 
the subsequent Higgs decay $h \to t \bar t$ \label{sssec623}}

We now consider the process $e^+e^-\to Zh$ followed by $h\to t\bar
t$. As we have discussed above, in  general, one cannot ignore the
SM-like diagrams of class (a) on the right hand side of  Fig.~\ref{eetthzfig1}
when analyzing CP violation in the reaction  $e^+ e^- \to t \bar t Z$. 
Moreover,
inclusion of those diagrams  and interfering them with diagram (b) on the right
hand 
side of  Fig.~\ref{eetthzfig1}, gives a bonus in the appearance of tree-level
CP violation  in $e^+ e^- \to t \bar t Z$. However, let us assume  that the
Higgs has already been  discovered, with a mass of $m_h > 2m_t$, by the time
a high energy $e^+e^-$ collider starts its first run. In such a scenario,
one should in principle be able to separate the contribution of
the Higgs 
exchange graph  in $e^+ e^- \to t \bar t Z$ from the rest of the SM-like 
diagrams which lead to the same final state, by imposing  a suitable 
cut on the
invariant $t \bar t$ mass.   Taking this viewpoint, we will consider 
Higgs production via $e^+ e^- \to h Z$ and CP violation in the subsequent
Higgs decay $h \to t \bar t$. 

A general method for tracing CP-odd and CP-even $t,\bar t$ 
spin-correlations, 
in $h \to t \bar t$, was introduced in \cite{prd56p90}. There,
it was assumed that  an on-shell Higgs, with $m_h > 2m_t$, is
produced  through, for example, $e^+ e^- \to Zh, \ell^+ \ell ^- h$ or  
even $\mu^+ \mu^- \to h$, and that its rest system can be reconstructed.
In \cite{plb338p71}, a helicity asymmetry  in $h \to t \bar t$ was suggested,
where $e^+ e^- \to Zh$ was explicitly assumed as the Higgs production 
mechanism. We will describe below these two works. Other related works can
be found in  \cite{plb350p218,plb314p104,prd48p3225,prd49p4481,prd49p4548,mpla9p205,zpc64p21,plb339p127}.     

In the method suggested in \cite{prd56p90}, the decay $h \to t \bar t$ 
stands out as an independent decay process, in which top spin-asymmetries 
can be formed. Consider, for example, the observables

\beqa
{\cal S}_{1} &=& {\bf \hat {k_t}}\cdot({\bf s}_{t} - {\bf s}_{\bar t}) 
\label{eetthzeq27} ~,\\
{\cal S}_{2} &=& {\bf \hat {k_t}}\cdot({\bf s}_{t}\times{\bf s}_{\bar t}) 
\label{eetthzeq28}~,\\
{\cal S}_{3} &=& {\bf s}_{t}\cdot{\bf s}_{\bar t} \label{eetthzeq29}~,
\eeqa  

\noindent where ${\bf s}_{t}({\bf s}_{\bar t})$ is the spin operator of
$t(\bar t)$ and ${\bf k_t}$ is the top quark 3-momentum  in the $t \bar t$
c.m.\ frame.  ${\cal S}_{1}$ and ${\cal S}_{2}$ are CP-odd, where ${\cal
S}_{1}$  is $T_N$-even and ${\cal S}_{2}$ is $T_N$-odd. Therefore, 
a non-zero expectation value of ${\cal S}_{1}$ will also require 
absorptive parts, while $<{\cal S}_{2}> \neq 0$ can be generated 
already at the tree-level. ${\cal S}_{3}$ is CP-even and is also 
generated at the lowest order (i.e., tree-level). For a general $t \bar th$
Yukawa interaction  
Lagrangian as in Eq.~\ref{2hdmab}, it was found in \cite{prd56p90} that 
the spin-asymmetries, ${\cal S}_{1,2,3}$ in
Eqs.~\ref{eetthzeq27}--\ref{eetthzeq29}, depend only on  
one combination of the couplings $a_t^h$ and $b_t^h$

\beq
r_t = \frac{b_t^h}{a_t^h + b_t^h} \label{eetthzeq30}~,
\eeq

\noindent which takes values between 0 to 1, i.e., $0 \leq r_t \leq 1$, where
the lower limit corresponds to $b_t^h=0$ and the upper one to $a_t^h=0$.

These observables can be translated to
correlations between momenta of  the $t$ and $\bar t$ decay products. 
To do so, one can define decay scenarios of the $t$ and $\bar t$, through
which both the $t$ and $\bar t$ momenta  and spins can be reconstructed in the
most efficient way \cite{prd56p90}:   

\beq
{\cal A}:\,\left\{\,\,
\begin{array}{l}
t\to W^+ +b \to \ell^++\nu_{\ell}+b \\ 
\bar{t}\to W^-+\bar{b}\to q+\bar{q}'+\bar{b}\,\,
\end{array}\right. ~.
\label{eetthzeq31}
\eeq

\noindent The sample ${\bar{\cal A}}$ is defined by the charge conjugate
decay  channels of the $t\bar{t}$ pairs

\beq
{\bar{\cal A}}:\,\left\{\,\,
\begin{array}{l}
t\to  W^++b\to \bar{q}+q'+b \\ 
\bar{t}\to W^- +\bar b \to \ell^-+\bar{\nu}_{\ell}+\bar{b} \,\,
\end{array}\right. ~.
\label{eetthzeq32}
\eeq 

\noindent In these decay samples, either the $t$ or $\bar t$ decays 
leptonically while the other decays hadronically. Each of these
samples has a branching fraction of about 2/9 of all $t \bar t$ pairs. 

With these decay scenarios Ref.~\cite{prd56p90} found 
the  momentum correlations ${\cal
O}_{1,2,3}$ which trace  the spin correlations ${\cal S}_{1,2,3}$,
respectively, 

\beqa
\!\!\!\!\!\!\!\!\<{\cal O}_1\>&=&\<\hat{\bf k} _t\cdot \hat{\bf p}_{\ell^+}^*\>_{\cal A} 
+\<{\hat {\bf k}}_t\cdot {\hat {\bf p}}_{\ell^-}^*
\>_{\bar {\cal A}} = \frac{2}{3} \<{\cal S}_1\> \label{eetthzeq33}~,\\ 
\!\!\!\!\!\!\!\!\<{\cal O}_2\>&=& \<{\hat {\bf k}}_{t}\cdot ({\hat {\bf p}}_{\ell^{+}}^*
\!\times {\hat {\bf p}}_{\bar b}^*)\>_{\cal A}-
\<{\hat {\bf k}}_{t}\cdot ({\hat {\bf p}}_{\ell^-}^*\times 
{\hat {\bf p}}_{b}^*)\>_{\bar {\cal A}} = 
\frac{8}{9} \left( \frac{1-2x}{1+2x} \right)
\<{\cal S}_2\> ~,\nonumber \\
\label{eetthzeq34} \\
\!\!\!\!\!\!\!\!\<{\cal O}_3\>&=& \<\hat{\bf p}_{\ell^+}^*\cdot
\hat{\bf p}_{\bar{b}}^*\>_{\cal A} 
+\<{\hat {\bf p}}_{\ell^{-}}^{*}\cdot{\hat {\bf p}}_{b}^*
\>_{\bar {\cal A}} = \frac{8}{9} \left( \frac{1-2x}{1+2x} \right)
\<{\cal S}_3\> \label{eetthzeq35}~,
\eeqa 

\noindent where ${\hat{\bf p}}_{\ell^+}^*$(${\hat{\bf p}}_{\ell^-}^*$)
is the flight direction of ${\ell^+}$(${\ell^-}$) in the $t$($\bar t$) quark
rest system. Similarly,  ${\hat {\bf p}}_{b}^*$(${\hat {\bf p}}_{\bar b}^*$)
is the unit momentum  of the $b$(${\bar b}$) in the $t$($\bar{t}$) rest
system. Also, $x \equiv m_W^2/m_t^2$ and the factor $(1-2x)/(1+2x)\approx
0.41$ measures the spin analyzing quality of the $b(\bar{b})$ (see section 
\ref{ssec28}).
 
\begin{figure}[htb]
 \psfull
\begin{center}
\leavevmode
 \epsfig{file=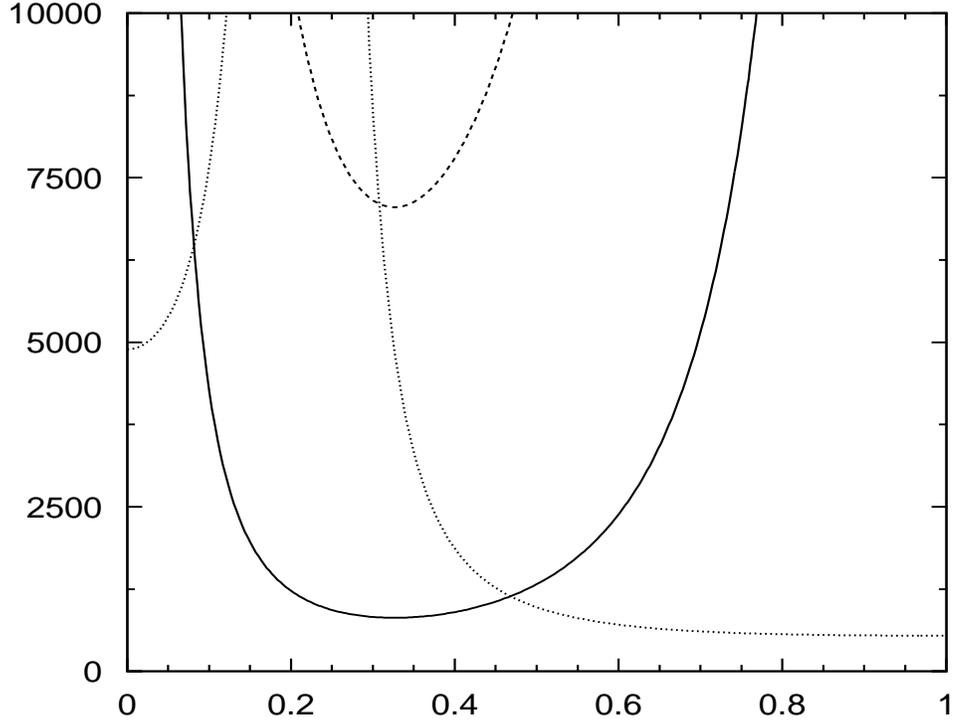,height=10cm,width=12cm,
 bbllx=0cm,bblly=2cm,bburx=20cm,bbury=25cm,angle=0}
 \end{center}
\caption{\emph{Number of events $h \to t\bar{t}$ to establish a non-zero 
correlation $\<{\cal S}_{1,2,3}\>$ with 3 standard deviation significance,
as a  function of $r_{t}$ (see Eq.~\ref{eetthzeq30}) and 
for a fixed Higgs mass of $m_h=400$ GeV.  
The dashed line represents the result for $N^{1}_{t\bar{t}}$,
the solid line is the result for   $N^{2}_{t\bar{t}}$ and the dotted line
is the result for   $N^{3}_{t\bar{t}}$. $m_t=175$
GeV, figure taken from \cite{prd56p90}.}}
\label{eetthzfig5}
\end{figure}

The number $N_{t\bar{t}}^{1,2,3}$ of $h \to t\bar{t}$ events that is required  
to establish a nonzero correlation  $\<{\cal O}_{1,2,3}\>$ 
at the $N_{SD}$ - standard deviations - significance level are given by

\beq 
N^{1,2,3}_{t\bar{t}}= \frac{N_{SD}^2}
{{\rm Br}({\cal A}) \times A^2_{1,2,3}} \label{eetthzeq36} ~,
\eeq 

\noindent where

\beq
 A_{1,2,3} = \frac{\<{\cal O}_{1,2,3}\>}
{\sqrt {\<{\cal O}_{1,2,3}^2\>}} \label{eetthzeq37} ~,
\eeq 

\noindent and ${\rm Br}({\cal A})={\rm Br}(\bar {\cal A})$ is the branching 
ratio of the decay samples ${\cal A}$ or $\bar {\cal A}$. In particular,
disregarding the $\tau$ leptons we have  ${\rm Br}({\cal A})=4/27$.   The
number of events needed for a 3-$\sigma$ (i.e., $N_{SD}=3$)  observation
of the spin-correlations, ${\cal S}_{1,2,3}$, of  
Eqs.~\ref{eetthzeq27}--\ref{eetthzeq29} are given in  Fig.~\ref{eetthzfig5},
for $m_h=400$ GeV,  ${\rm Br}({\cal A})=4/27$ and
 as a function of the parameter
$r_t$ defined in  Eq.~\ref{eetthzeq30}. These can be simply obtained from
$N^{1,2,3}_{t\bar{t}}$ in  Eq.~\ref{eetthzeq36}, by using the relations between
${\cal O}_{1,2,3}$ and  ${\cal S}_{1,2,3}$ given in 
Eqs.~\ref{eetthzeq33}--\ref{eetthzeq35}.

It should be noted again that, while ${\cal O}_{2,3}$ are non-zero
already at the tree-level, ${\cal O}_{1}$, being $T_N$-even, requires
an absorptive phase and, therefore, its non-zero contribution first 
arises only at the 1-loop level in perturbation theory. Thus, 
${\cal O}_{1}$ is expected to be less effective (as can be seen 
from Fig.~\ref{eetthzfig5}). In \cite{prd56p90} the 
1-loop QCD corrections to ${\cal O}_{1,2,3}$ were computed. 
For the operators, ${\cal O}_{2,3}$, the QCD corrections 
were found to be of the order of a few percent compared to the 
leading tree-level contribution, and were therefore neglected. 

We see from Fig.~\ref{eetthzfig5} that, for example, values of $r_t$ in the
range, $0.18 \lsim r_t \lsim 0.52$,  would give rise to a 3-$\sigma$ CP-odd
effect in   ${\cal O}_{2}$ with a data sample of $N^{2}_{t\bar{t}} \approx
1500$, i.e., $\sim 1500$  
($h \to t \bar t$) events. Note also that a simultaneous
measurement of  ${\cal O}_{2}$ and ${\cal O}_{3}$, with these
 $\sim 1500$ 
events, would have a 3-$\sigma$  sensitivity to $r_t$ from 0.18
to its maximal  value of 1. Furthermore,  production of a few thousands of
$h \to t \bar t$ events  through the Bjorken mechanism,
$e^+ e^- \to hZ$, is indeed feasible. For example, if a CP-mixed neutral
Higgs (with both scalar and pseudoscalar  couplings $a_t^h$ and $b_t^h$)
with a mass  $m_h=400$ GeV, has a $ZZh$ coupling $c^h$ of a SM strength, 
then the cross-section for  
$e^+ e^- \to hZ$ is of the order of a few fb's for 
$e^+ e^-$ c.m.\ energies in the range
 500  GeV${}\lsim \sqrt s \lsim 1000$ GeV\null.\footnote{The cross-section 
for $e^+ e^- \to hZ$ is given, for example, in \cite{collider}. 
With $m_h=400$ GeV and for $c^h=1$, i.e., the SM $ZZh$ coupling,
$\sigma(e^+ e^- \to hZ) \sim 3.5(8.5)$ 
fb for $\sqrt s = 500(1000)$ GeV.}   
Therefore, with a yearly integrated luminosity  of ${\cal L} \sim 10^{2}$
fb$^{-1}$, hundreds of $hZ$ pairs can be produced and, thus, 
a $\sim 2$-$\sigma$ limit on $r_t$  may be achievable.

An interesting CP-violating helicity asymmetry in $h \to t \bar t$ was 
suggested in \cite{plb338p71}:

\beq
{\cal O}_{tt}^h=\frac{\Gamma(++)-\Gamma(--)}{\Gamma(++)+\Gamma(++)}
\label{eetthzeq38} ~,
\eeq 

\noindent where $\Gamma(++)$ and $\Gamma(--)$ are the decay widths 
of the lightest
Higgs-boson $h$, into a pair of $t \bar t$ with the indicated helicities.
Since under CP: $(++)\leftrightarrow(--)$, non-zero ${\cal O}_{tt}^h$ 
would be a signal of CP violation.

\begin{figure}[htb]
 \psfull
\begin{center}
\leavevmode
 \epsfig{file=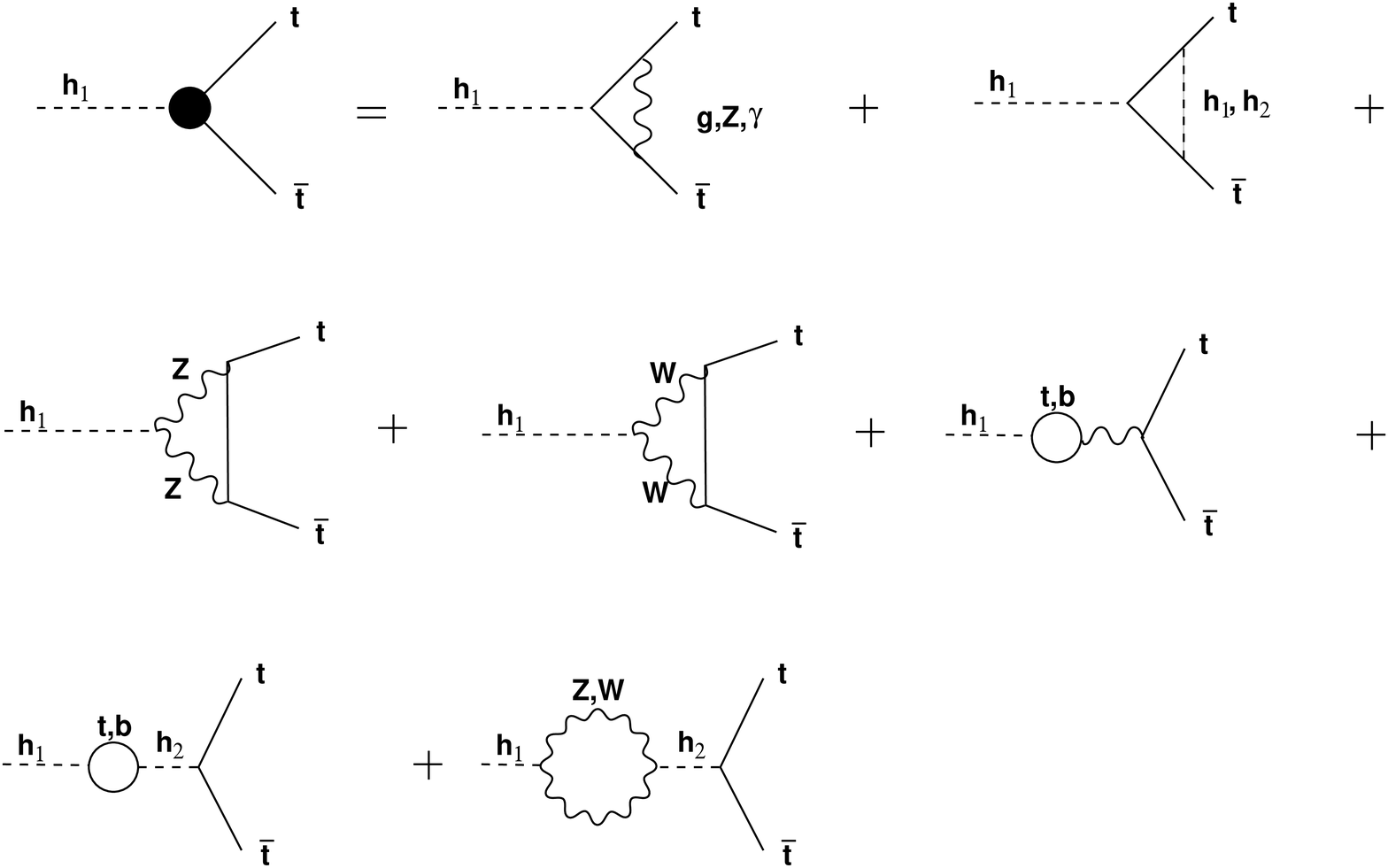
,height=9cm}
\end{center}
\caption{\emph{The 1-loop diagrams contributing to the 
asymmetry ${\cal O}_{tt}^h$
in a 2HDM\null.}} \label{eetthzfig6}
\end{figure}

${\cal O}_{tt}^h$ is CP-odd but $T_N$-even, therefore, it requires  a
CP-odd  as well as a CP-even  absorptive phase (i.e., FSI phase). 
As mentioned several times
before, in a 2HDM with a CP-mixed neutral Higgs, the CP-odd phase is provided
by the simultaneous presence of the scalar and  pseudoscalar $t\bar th$
couplings  in the $t\bar th$ interaction Lagrangian.  The FSI
absorptive phase is generated 
at the 1-loop level  from the diagrams in Fig.~\ref{eetthzfig6}. The 
expressions for the different contributions to ${\cal O}_{tt}^h$ 
corresponding to the different diagrams in Fig.~\ref{eetthzfig6} are given 
in \cite{plb338p71}. There, it was found that with $m_t=180$ GeV and 
$m_h \gsim 2m_t$, ${\cal O}_{tt}^h$ of the order of $50\%$ is possible. 
Explicitly, assuming the Higgs to be produced through the Bjorken mechanism, 
$e^+e^- \to hZ$, the statistical significance, $N_{SD}$, with which this
asymmetry can  potentially be detected is

\beq
N_{SD}= \sqrt L \sqrt {\sigma(e^+e^- \to hZ)} \times 
{\cal O}_{tt}^h \label{eetthzeq39} ~.
\eeq

\noindent In \cite{plb338p71} an effective 
integrated luminosity of $L = 85$ fb$^{-1}$
was assumed and a scan for maximal  $N_{SD}$ was performed as a function of
the 2HDM parameters  $\tan\beta$ and $\alpha_{1,2,3}$. It was found that
with  $\tan\beta=0.5$, for example, and with $m_t=180$ GeV,  $m_h \gsim 2m_t$,
up to a 7-$\sigma$ detection of  a CP-violating signal from ${\cal O}_{tt}^h$
is feasible,  if the Higgs is produced via $e^+e^- \to hZ$. It should be
emphasized, however, that there is a potential  background to the $t \bar t$
pairs (coming from the subsequent  Higgs decay in $e^+e^- \to hZ$) from the
SM-like diagrams included in Fig.~\ref{eetthzfig1}. Therefore, 
as mentioned before, in order for
such a study to be practical one has to know  the mass of the Higgs prior
to the actual experiment and, with a sufficient mass resolution, 
demand that the invariant $t \bar t$ mass reconstructs the Higgs.

\newpage
~

\newpage
~

\newpage
~

\newpage

\newpage
~

\newpage

\newpage
~

\newpage

\subsection{$e^+e^- \to t \bar t g$ \label{ssec63}}

Given the importance of the top pair production at the NLC, it should be
clear that the associated gluon emission will also receive considerable
attention. Of course, the gluons will be radiated off top quarks quite
readily 
once the  threshold for top pair production will be reached. One important
advantage of the reaction $e^+e^- \to t \bar t g$  is that it is rich in
exhibiting several different types of CP  asymmetries which can be driven
by 1-loop effects induced by  extensions of the SM\null. For example,
exchanges of neutral  Higgs from MHDM's with 
CP violation in the scalar sector, or exchanges of SUSY particles which 
carry a CP-odd phase in their interaction vertices, could  give rise to both
$T_N$-odd and $T_N$-even type CP-violating  dynamics. Therefore, both CP-odd,
$T_N$-odd and CP-odd, $T_N$-even type observables can be used to extract
information on the real and imaginary parts of the amplitude. With three
particles in the final state there are enough linearly independent
momenta available so that 
the construction of CP-odd, $T_N$-odd observables is straightforward;
there is no need to involve the spins of the top.       

\begin{figure}[ht]
 \psfull
 \begin{center}
  \leavevmode
  \epsfig{file=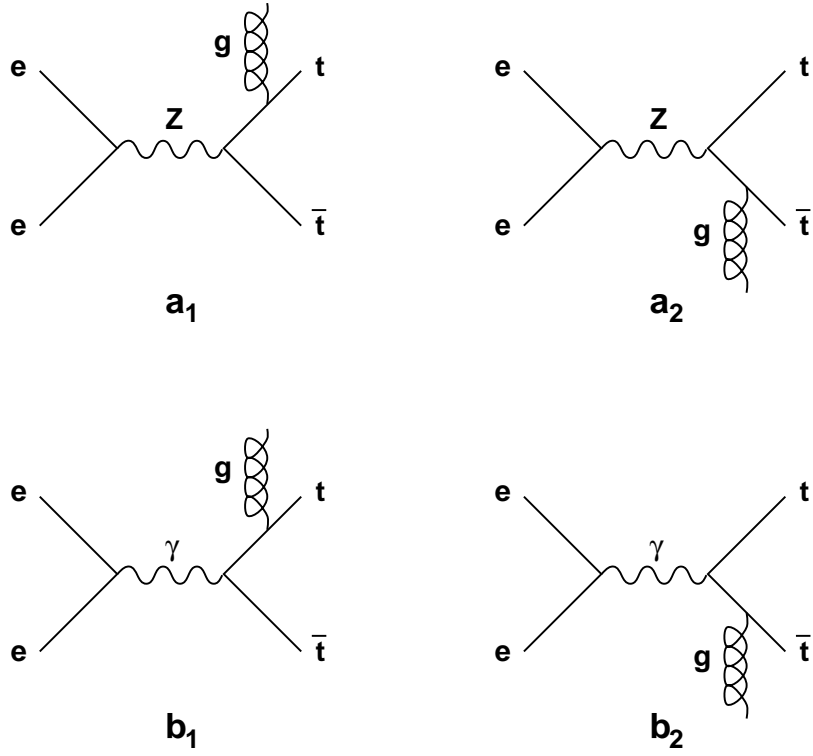
,height=10cm}
 \end{center}
\caption{\emph{Tree-level Feynman diagrams contributing to $e^+e^- \to t \bar
t g$ (for $q=t$).}} \label{eettgfig1}
\end{figure}

Following our work in \cite{zpc72p79}, we give below the full
analytical formulae of the tree-level DCS (Differential Cross-Section) as
well as a description for extracting  the 1-loop CP-violating DCS that can
be used for any given model.  The SM tree-level diagrams, are depicted in
Fig.~\ref{eettgfig1}.

The incoming polarized electron-positron current can be written as

\beq
J_e^{\mu(j)} = \bar{v}_e(p_+)\gamma^\mu P_j u_e(p_-) \label{eettgeq1}~,
\eeq

\noindent where $P_j=\frac{1}{2}(1+j\times\gamma_5)$ and
 $j=-1(1)$ for left
(right) handed incoming electrons.  $p_+(p_-)$ are the 4-momentum of the
positron (electron) and $p=(p_++p_-)$ is the 4-momentum of the
$s$-channel gauge-boson (the contribution from an $s$-channel
Higgs vanishes for $m_e \to 0$).

We also define the following constants

\beq
{\cal C}_Z = \frac{(4\pi\alpha)}{2c^2_Ws^2_W}T^ag_s\Pi_ZC^{e}_j \ \ , 
\ \    
{\cal C}_\gamma = (4\pi\alpha)T^ag_s\Pi_{\gamma}Q_q \label{eettgeq2}~,
\eeq

\noindent where $T^a$ is the appropriate SU(3) generator, $g_s$ is the strong
coupling constant, $Q_q$ is the charge of the quark and $c_W(s_W)$ stands
for $\cos \theta_W(\sin\theta_W)$, respectively. Also
$C^{e}_j = C^e_L (C_R^e)$ for $j=-1(1)$ with $C^f_L = - 2
I^f_3+2Q_fs^2_W$ and $C^f_R = 2Q_fs^2_W$. $\Pi_Z$ and $\Pi_\gamma$
are the gauge-boson propagators given by

\beq
\Pi_\gamma = \frac{1}{p^2} \ \ , \ \ \Pi_Z = \frac{1}{p^2-m^2_Z} 
\label{eettgeq3} ~. 
\eeq

\noindent Then the tree-level matrix element is given by

\beq
{\cal M}^0 \equiv {\cal M}_{a1} + {\cal M}_{a2} + {\cal M}_{b1}
+ {\cal M}_{b2} \label{eettgeq4} ~,
\eeq

\noindent where 
${\cal M}_{a1}, {\cal M}_{a2}, {\cal M}_{b1}$ and ${\cal M}_{b2}$
are obtained from diagrams $(a_1),(a_2),(b_1)$ and $(b_2)$ 
in Fig.~\ref{eettgfig1}, respectively,
all emanating from the SM\null.  We thus get

\beqa
{\cal M}^0 &=& \frac{1}{2}J_e^{\mu(j)} \bar{u}(p_q)
\left\{ {\cal C}_Z\left[\Pi_qT_{(a1)\mu} - \Pi_{\bar{q}}T_{(a2)\mu} \right] +
\right.       \nonumber \\
& &\left. 2{\cal C}_\gamma\left[\Pi_{\bar{q}}T_{(b1)\mu} -
\Pi_qT_{(b2)\mu} \right]\right\}v(p_{\bar{q}}) \label{eettgeq5} ~.
\eeqa

\noindent Here $p_q(p_{\bar{q}})$ is the 4-momentum of the outgoing quark
(anti-quark),  $p_g$ is the gluon's 4-momentum and the quark and anti-quark 
propagators are given by

\beq
\Pi_q = \frac{1}{2p_q\cdot p_g} \ \ , \ \
\Pi_{\bar{q}} = \frac{1}{2p_{\bar{q}} \cdot p_g} \label{eettgeq6} ~. 
\eeq

\noindent Furthermore, the hadronic vector elements in Eq.~\ref{eettgeq5} are

\beqa
T_{(a1)_\mu} &=& \not\eta(p\hspace{-0.20cm}/_q + p\hspace{-0.20cm}/_g +
m_q)\gamma_\mu C_{LR}^{+} \label{eettgeq7} ~, \\
T_{(a2)_\mu} &=& \gamma_\mu C_{LR}^{+}
(p\hspace{-0.20cm}/_{\bar{q}} + p\hspace{-0.20cm}/_g -m_q)\not\eta 
\label{eettgeq8} ~, \\
T_{(b1)\mu} &=& T_{(a1)_\mu} (C_{LR}^{+} \to 1) \label{eettgeq9} ~, \\
T_{(b2)\mu} &=& T_{(a2)_\mu} (C_{LR}^{+} \to 1) \label{eettgeq10} ~,
\eeqa

\noindent $\eta_\alpha$ being the polarization vector of the gluon and
$C_{LR}^{+}= C_{L}^{q}L + C_{R}^{q}R$, where $L(R) = P_{j=-1}(P_{j=1})$.

\begin{figure}[htb]
 \psfull
 \begin{center}
  \leavevmode
 \epsfig{file=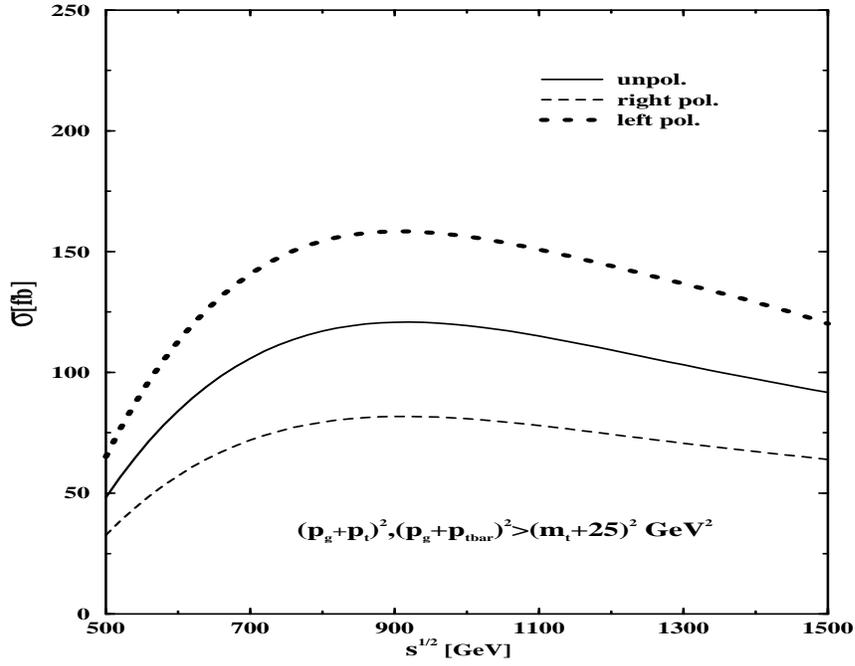
,height=8cm,width=8cm,bbllx=0cm,bblly=2cm,bburx=20cm,bbury=25cm,angle=0}
 \end{center}
\caption[dummy]{\emph{The cross-section for the reaction  $e^+ e^- \to t \bar t
g$ (in fb) as a function  of the c.m.\ energy $\sqrt s$, for unpolarized
(solid line),  negatively polarized (dotted line) and positively polarized
(dashed line) incoming electron beam. The cuts 
$(p_g+p_t)^2 \geq (m_t+m_0)^2$ and 
$(p_g+p_{\bar t})^2 \geq (m_t+m_0)^2$, $m_0=25$ GeV, 
are imposed. 
Figure taken from \cite{zpc72p79}.}}
\label{eettgfig2}
\end{figure}

In Fig.~\ref{eettgfig2} we have plotted the tree-level cross-section as a
function of the c.m. energy in an $e^+e^-$ collider for polarized and
unpolarized incoming electron beam.  To facilitate experimental identification
as well as to avoid infrared singularities we have imposed a cut on the
invariant mass of the jet pairs so that $(p_g+p_t)^2$ and $(p_g+p_{\bar t})^2
\geq (m_t+m_0)^2$ where we have taken $m_0=25$ GeV\null. This cut, which
effectively cuts the gluon energy, also removes soft gluon emission from the
secondary $b$-quarks of the top decays. We see from Fig.~\ref{eettgfig2}
 that with an
integrated luminosity of ${\cal L} \sim 200$ fb$^{-1}$, a 1 TeV (500 GeV)
$e^+e^-$ collider will be able to produce about $ \sim 3\times 10^4$ ($ \sim
1\times 10^4$) $t\bar tg$ events.

In a given model, the CP-violating corrections for the
reaction

\begin{equation}
e^-(p_-)e^+(p_+) \to q(p_q)\bar{q}(p_{\bar{q}})g(p_g) ~,
\end{equation}

\noindent requires the
calculation of the corresponding 1-loop diagrams. Let us write the general
form of the 1-loop matrix elements that violate CP as ${\cal M}^\rho_\sigma$.
For a given underlying model, the subscript indicates the diagram and
the superscript indicates which gauge particle is exchanged in the
$s$-channel. Thus

\beq
{\cal M}^\rho_\sigma = J_e^{\mu(j)}\bar{u}(p_q)H^\rho_{\sigma\mu}
v(p_{\bar{q}}) \label{eettgeq11} ~,
\eeq

\noindent where $H^{\rho}_{{\sigma}\mu}$ is the ``hadronic vector''
corresponding to each diagram and exchanged quanta. Denoting the complete 
CP-violating 1-loop contribution by 

\beq
{\cal M}^V = \sum_\rho\sum_\sigma {\cal M}^\rho_\sigma \label{eettgeq12}~,
\eeq

\noindent the ${\cal M}^V$ can be calculated within a given model, and  
the polarized CP-nonconserving DCS to 1-loop is then obtained from the
interference terms between the 1-loop and the Born amplitudes

\beq
\sum({\cal M}^V{\cal M}^{0*} + {\cal M}^0 {\cal M}^{V*}) \label{eettgeq13}~.
\eeq

\noindent Here the sum is carried over the polarizations of
$e^+$, $t$, $\bar{t}$ and $g$.

\newpage
~

\newpage
~

\newpage

\subsubsection{2HDM and CP violation in 
$e^+e^- \to t \bar t g$ \label{sssec631}} 

\begin{figure}
 \psfull
 \begin{center}
  \leavevmode
  \epsfig{file=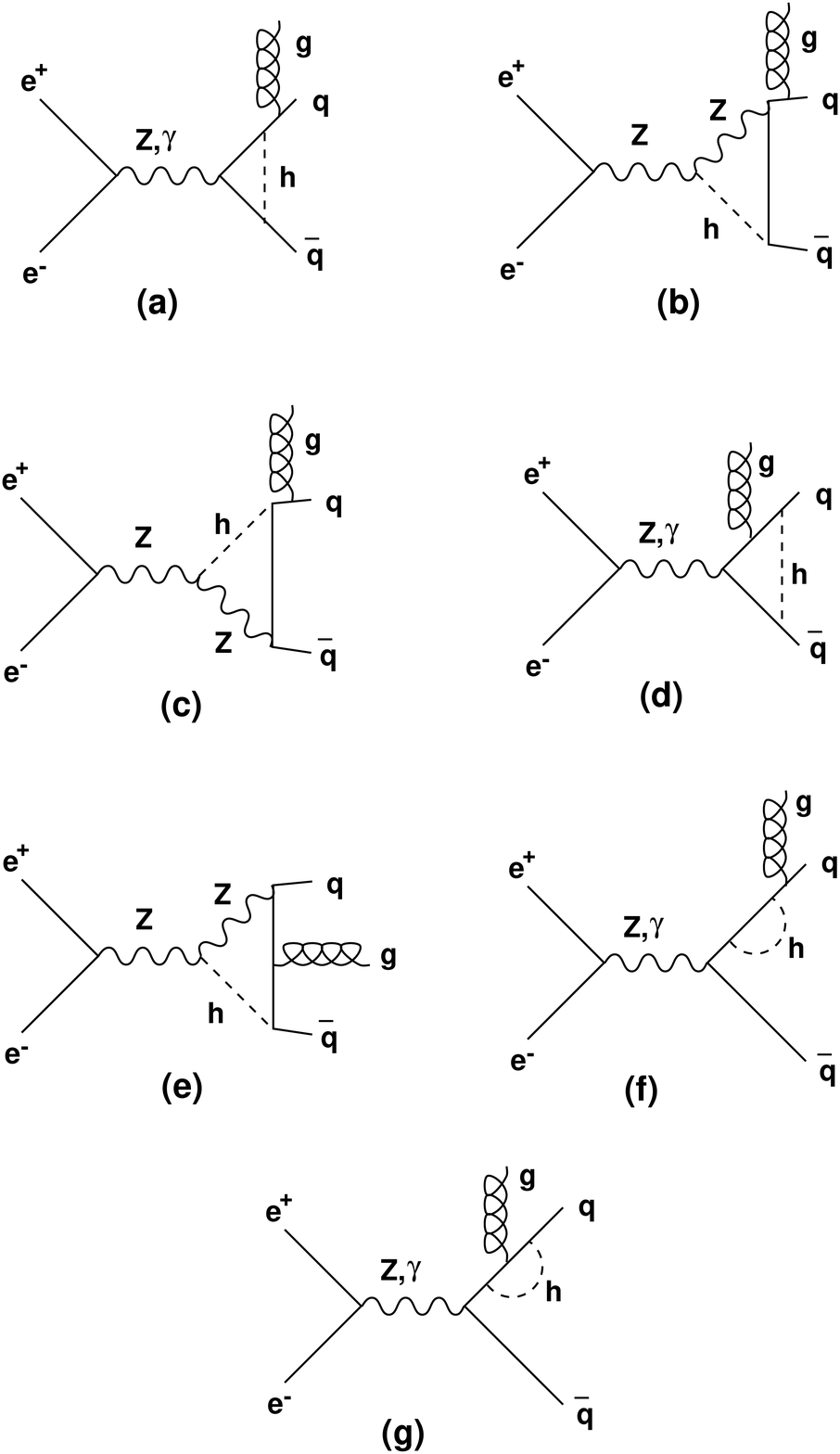
,height=14cm}
\end{center}
\caption{\emph{CP violating Feynman diagrams contributing 
to  $e^+e^- \to q \bar
q g$ to 1-loop order in a 2HDM ($h$ is a neutral Higgs-boson). 
Diagrams with permuted vertices
(i.e., $q \to \bar q$) are  not shown.}}
\label{eettgfig3}
\end{figure}

In a 2HDM, CP-violating neutral Higgs exchanges, at 1-loop order, can give
rise to the Feynman diagrams depicted in Fig.~\ref{eettgfig3} \cite{zpc72p79}.
We take the limit $m_e\to 0$, thus neglecting all the diagrams that are
proportional to the electron mass.  This includes any diagram that involves
electron coupling to the Goldstone modes, hence proportional to $m_e$.

The relevant Feynman rules for the diagrams in Fig.~\ref{eettgfig3}
can be extracted from parts of the Lagrangian involving the 
$f \bar f {\cal H}^k$ and $ZZ {\cal H}^k$ couplings 
$a_f^k,~b_f^k$ and $c^k$ defined in Eqs.~\ref{2hdmab} and 
\ref{2hdmc}, respectively.\footnote{Recall 
that ${\cal H}^k$ stand for any one of 
the three, i.e., $k=1,2$ or 3, neutral Higgs in a 2HDM.}
Again, for simplicity, we will consider only one light 
neutral Higgs, $h$, assuming
that the remaining two are considerably heavier.
Furthermore, for simplicity, whenever necessary we set 
the masses ($m_H$) of the remaining two neutral Higgs particles 
to be 1 TeV, i.e., assume that they are degenerate.

All the CP-violating terms in the 1-loop amplitudes corresponding to
 the diagrams in Fig.~\ref{eettgfig3}
emerge through interference of the scalar coupling $a_q^h$ (for a quark $q$) 
with the
pseudo-scalar coupling $b_q^h$ in any exchange of a neutral Higgs. 
In the
diagrams where the Higgs exchange is generated at the $ZZh$ 
vertex, the CP-violating
terms will be proportional to $b_q^h \times c^h$. 
The CP-violating 1-loop
amplitude can then be  calculated (for details see \cite{zpc72p79}) and  
the DCS can be schematically written as

\beq
\Sigma(\phi) = \Sigma_0(\phi) + \Sigma^{\Re{\rm e}}_1(\phi) + \Sigma^{\Im{\rm
m}}_1 (\phi) \label{eettgeq16}~.
\eeq
 
\noindent Here $\Sigma_0(\phi)$ is a CP-even piece and $\Sigma_1$ is
a CP-odd piece which  is further subdivided into two parts that depend on
the real and imaginary components of the amplitude.
Thus, CP-odd, $T_N$-odd and CP-odd, $T_N$-even effects will emanate from
$\Sigma^{\Re{\rm e}}_1(\phi)$ and $\Sigma^{\Im{\rm m}}_1 (\phi)$, respectively.

To estimate the CP-violating effects of both the $T_N$-even 
and $T_N$-odd types, the following two  CP-odd 
observables were considered in \cite{zpc72p79} 
for the reaction $e^+e^- \to t \bar t g$

\beqa
{\rm {T_N-even}}&:&~~~O_{i1} \equiv
\frac{\vec{p}_-\cdot(\vec{p}_t+\vec{p}_{\bar{t}})}{s} \label{eettgeq17}~, \\ 
{\rm {T_N-odd}}&:&~~~O_{r1}  \equiv  \frac{\vec p_- \cdot (\vec p_t\times
\vec p_{\bar t})}{s^{3/2}} \label{eettgeq18} ~. 
\eeqa

\noindent  A non-vanishing expectation value of any one of these would signal
CP violation so that experimental searches for them can be performed without
recourse to any model.  However, as was mentioned in section \ref{ssec26}, 
within
the context of any given model one can also construct optimal observables
i.e., those observables which will be the most sensitive to 
CP violation effects in that model \cite{prd45p2405},
 
\beq
O_{i\rm opt} \equiv \Sigma^{\Im{\rm m}}_1/\Sigma_0 \quad , \quad O_{r\rm
opt} \equiv \Sigma^{\Re{\rm e}}_1 /\Sigma_0 \label{eettgeq19}~,
\eeq

\begin{figure}[htb]
 \psfull
 \begin{center}
  \leavevmode
  \epsfig{file=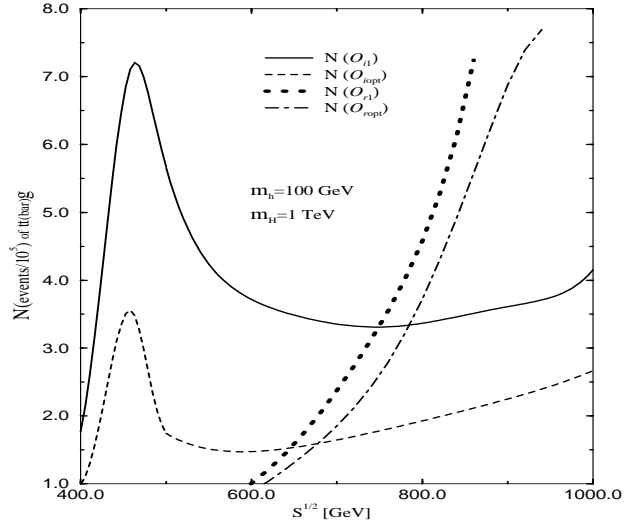,height=6cm,width=6cm,bbllx=0cm,bblly=2cm,bburx=20cm,bbury=25cm,angle=0}
 \end{center}
\caption{\emph{Number of events (in units of $10^5$) needed to 
detect CP violation
via $\langle O_{i1}\rangle$, $\langle  O_{r1}\rangle$, $\langle O_{i{\rm
opt}}\rangle$ and $\langle O_{r{\rm opt}}\rangle$ to 1-$\sigma$ level, as a
function of the total beam energy, $\sqrt s$, for left-handed polarized 
incoming electron beam. 
$m_h=100$ GeV, $m_H=1$ TeV and $a_t^h=b_t^h=c^h=1$ are used. Also, the cuts 
$(p_g+p_t)^2 \geq (m_t+m_0)^2$ and 
$(p_g+p_{\bar t})^2 \geq (m_t+m_0)^2$, $m_0=25$ GeV, are imposed. 
Figure taken from \cite{zpc72p79}.}}
\label{eettgfig4}
\end{figure}

\begin{figure}
 \psfull
 \begin{center}
  \leavevmode
  \epsfig{file=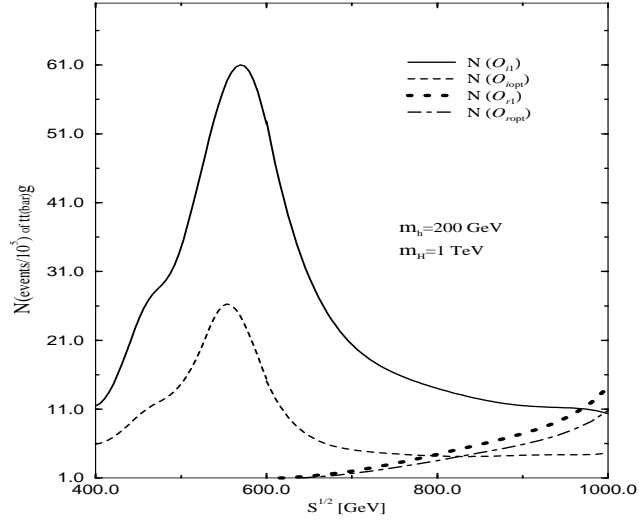
,height=6cm,width=6cm,bbllx=0cm,bblly=2cm,bburx=20cm,bbury=25cm,angle=0}
 \end{center}
\caption[dummy]{\emph{Same as Fig.~\protect\ref{eettgfig4} 
except $m_h=200$ GeV\null. Figure taken from \cite{zpc72p79}.}}
\label{eettgfig5}
\end{figure}

\begin{figure}
 \psfull
 \begin{center}
  \leavevmode
 \epsfig{file=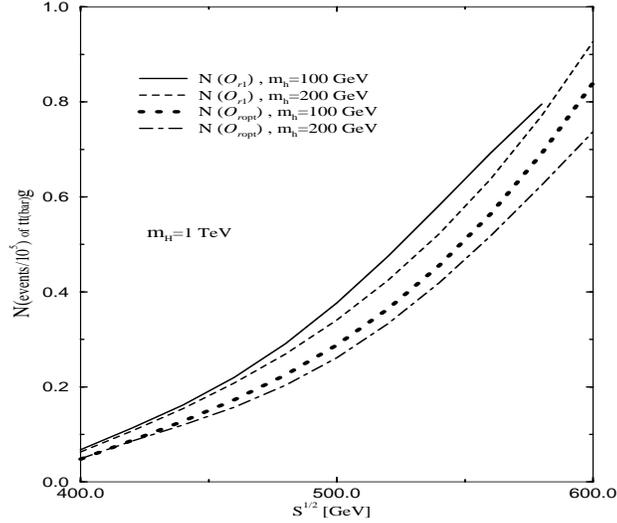,height=6cm,width=6cm,bbllx=0cm,bblly=2cm,bburx=20cm,bbury=25cm,angle=0}
 \end{center}
\caption[dummy]{\emph{Number of events (in units of $10^5$) needed to detect CP
violation via $\langle O_{r1}\rangle$ and $\langle O_{r{\rm opt}} \rangle$
to 1-$\sigma$ level as a function of total beam energy in the range $\sqrt{s}
= 400-600$ GeV for $m_H=1$ TeV, $m_h=100$ and 200 GeV\null.
The rest of the parameters are as in Figs.~\protect\ref{eettgfig4} and 
\protect\ref{eettgfig5}. Figure taken from \cite{zpc72p79}.}}  
\label{eettgfig6}
\end{figure}

The number of events needed in order to detect a CP-odd signal at 
the 1-$\sigma$ level via each of the above four CP-violating observables,
is shown in  Figs.~\ref{eettgfig4} and \ref{eettgfig5} for $m_h=100$ and 200
GeV, respectively.  
In Fig.~\ref{eettgfig6} we have magnified the range $\sqrt s =400
-600$  GeV using the same Higgs masses.  In these figures we have focused
on the case of left polarized  incoming electrons while in 
Table~\ref{eettgtab1} we give a brief comparison of the left, right and
unpolarized electron beam cases. Also, the following assumptions are made:
\begin{description}
\item 1) As mentioned before a cut on the invariant mass of the jet pairs was
imposed, so that $(p_g+p_t)^2$ and $(p_g+p_{\bar t})^2 \geq (m_t+m_0)^2$, 
where we have taken $m_0=25$ GeV\null.
\item 2) The $T_N$-odd observables, being proportional to the dispersive 
parts of the loop integrals, are sensitive also to the mass of the 
two heavier neutral Higgs particles, $m_H$. For simplicity, we have chosen
these two Higgs particles to be degenerate  with a mass of 1 TeV\null.
\item 3) We set the relevant Higgs couplings to unity. 
That is, $a_t^h=b_t^h=c^h=1$ (for $q=t$)
which serves our purpose of finding the order of magnitude of the CP-odd
signal that can arise in this reaction. In fact, CP violation in 
$e^+ e^- \to t \bar t g$ is found to be dominated by the terms proportional 
to $a_t^h \times b_t^h$. With regard to that, we note that, for low values 
of $\tan\beta$, i.e., $\tan\beta \lsim 0.5$, the product 
$a_t^h \times b_t^h$ can reach values above $\sim 5$ and, 
therefore, our choice 
above of $a_t^h \times b_t^h =1$ is rather conservative.
\end{description}       

\begin{table}[htb]
\begin{center}
\begin{tabular}{|p{0.7cm}|p{1cm}|p{1cm}|p{1cm}|p{1cm}|p{1cm}|p{1cm}|p{1cm}|p{1cm}|p{1cm}|}
\hline
& & \multicolumn{2}{c|}{$O_{i1}$} & \multicolumn{2}{c|}{$O_{r1}$} &
\multicolumn{2}{c|}{$O_{i{\rm opt}}$} & \multicolumn{2}{c|}{$O_{r{\rm
opt}}$} \\
\cline{3-10}
\protect\footnotesize$\sqrt{s}$ & \protect\footnotesize $j$ &
\protect\footnotesize$m_h=100$ &
\protect\footnotesize$m_h=200$ & \protect\footnotesize$m_h=100$ &
\protect\footnotesize$m_h=200$ &  \protect\footnotesize$m_h=100$ &
\protect\footnotesize$m_h=200$ & \protect\footnotesize$m_h=100$ &
\protect\footnotesize$m_h=200$ \\
\hline  & -1 & 1.8  & 11.5  & 0.07 & 0.05 & 1.0  & 6.0  & 0.05 & 0.05 \\
\cline{2-10}
400 & unpol.& 22.5  & 134.8  & 0.05 & 0.05 & 6.5  & 37.0  & 0.05 & 0.04 \\
\cline{2-10}
& 1 & 2.3  & 17.1  & 0.05 & 0.04& 1.3 & 8.4 & 0.04 & 0.03 \\
\hline
\multicolumn{10}{|c|}{} \\
\hline  & -1 & 3.4  & 20.0  & 2.2   & 2.1   & 1.7  & 5.1  & 1.9  & 1.7   \\
\cline{2-10}
700 & unpol.& 48.6  & 263.9  & 2.2    & 1.9   & 12.2  & 38.6  & 1.8   & 1.4   \\
\cline{2-10}
& 1 & 4.5 & 30.8  & 1.9   & 1.8   & 2.0  & 5.9 & 1.5   & 1.2   \\
\hline
\multicolumn{10}{|c|}{} \\
\hline  & -1 & 4.0  & 10.5  & 14.4   & 14.1   & 2.6  & 4.5  & 11.6   & 10.8   \\
\cline{2-10}
1000 & unpol.& 63.4  & 158.2  & 14.3   & 10.3   & 20.5  & 35.3  & 10.8   & 8.4    \\
\cline{2-10} & 1 & 5.0  & 14.0  & 14.1   & 10.3   & 3.1  & 5.4  & 10.7
& 8.3    \\
\hline
\end{tabular}
\end{center}
\caption[dummy]{\emph{The unpolarized case is compared with left 
polarization ($j=-1$)
and right polarization ($j=+1$) of the $e^-$. The number of events
in units of $10^5$ needed for detection of asymmetries, to 1-$\sigma$ level
are given.  The values of $\sqrt{s}$ and $m_h$ are given in GeV\null. The
results for  the  $T_N$-odd observables are given for $m_H=1$ TeV, where
$m_H$ is the  mass of  the  two heavy Higgs (see also text). 
Table taken from  \cite{zpc72p79}.
\protect\label{eettgtab1}}}
\end{table}

Summarizing briefly the numerical results presented in 
Figs.~\ref{eettgfig4}--\ref{eettgfig6} and in Table~\ref{eettgtab1},  
we can see that for the optimal observable $O_{i {\rm opt}}$, for both a
500 GeV and a 1 TeV $e^+e^-$ collider,  the number of needed $t\bar tg$ events
in order to detect a 1-$\sigma$  CP-odd signal is comparable and is around 
 few${}\times 10^5$ with neutral Higgs masses in the range
 100 GeV${}< m_h < 200$
 GeV\null. With $O_{r {\rm opt}}$ the number of needed $t\bar tg$ events 
at c.m. energies around 1 TeV is few${} \times 10^6$. However, we see from
Fig.~\ref{eettgfig6} that, at a c.m.\ energy of 500 GeV and for $100~ {\rm
GeV} < m_h < 200~ {\rm GeV}$,  a 1-$\sigma$ measurement of $O_{r {\rm opt}}$
will require ${\rm few} \times 10^4$ $t\bar tg$ events.  
From Table~\ref{eettgtab1}
we see that for the $T_N$-even (i.e., $O_{i1}$ and $O_{i{\rm opt}}$) cases
the polarization makes a significant difference and improves their
effectiveness by about an order of magnitude 
or even more. For these it seems that
the left-polarized case is marginally better than the right one. 

Bearing in mind that with an integrated luminosity of ${\cal L} \sim 200$
fb$^{-1}$, about $\sim 10^4 ~ t\bar tg$  will be produced in a 500 GeV NLC, 
and
few${} \times 10^4$ in a  1 TeV NLC (see Fig.~\ref{eettgfig2}),   
the observability of a non-vanishing value for $O_{r {\rm opt}}$, to the
1-$\sigma$ level,  in a NLC with c.m. energies of 500 GeV is marginal, 
while $O_{i {\rm opt}}$ falls short by about an order of magnitude. 
Also, with a 1 TeV NLC that can produce up to $3 \times 10^4$ $t \bar t g$'s 
a year, 
the CP-odd signal from $O_{r {\rm opt}}$ falls short by almost two orders
of magnitude, while the number of events needed to  detect a CP-odd effect
through $O_{i {\rm opt}}$  is one order of magnitude away from the expected
number of  available events in such a collider.

Clearly, although the CP-violating effects driven by neutral Higgs exchanges
that were found in \cite{zpc72p79}  
fall short
by at least one order of  magnitude for a 3-$\sigma$ detection, 
this does not rule out the possibility of larger effects in other
extensions of the SM (e.g., SUSY). Therefore, theoretical and
experimental studies of CP violation in the process $e^+e^-\to t\bar t
g$ may still be worthwhile.

\subsubsection{Model independent 
constraints on top dipole moments \label{sssec632}}

The effects of anomalous EDM ($d_t^\gamma$), ZEDM ($d_t^Z$) and CEDM ($d_t^g$)
couplings of the top  quark to a photon, $Z$-boson and a gluon, respectively,
in $e^+ e^- \to  t \bar t g$ were considered in 
\cite{hepph9610373,plb424p125}.

Let us write an effective top quark interaction with a neutral gauge-bosons
$V=\gamma,Z$ or $g$, which involves the top magnetic and electric 
dipole moments (see also section \ref{ssec25})

\beq
{\cal L}_V = \frac{i g_V}{2m_t}\bar t
\sigma_{\mu \nu}q^{\nu} \left( \kappa_V - i {\tilde
\kappa}_V \gamma_5 \right)t F_V^{\mu} \label{eettgeq20}~.
\eeq

\noindent Here $g_{\gamma}=g_W s_W =e$, $g_Z=g_W/2c_W$ and 
$g_g=g_s$, where 
$g_W(g_s)$ is the
weak(strong) coupling constant, $c_W\equiv \cos\theta_W$, $q$ is the
gauge-boson 4-momentum, and
$F_V$, $V =A,Z$ or $G$, 
is the appropriate gauge field (color index is suppressed).
Moreover, in Eq.~\ref{eettgeq20}
we have introduced the CP-conserving(violating) dimensionless
effective anomalous couplings $\kappa_V$($\tilde\kappa_V$) 
of the top quark to a gauge-boson $V=\gamma,Z$ or $g$. 
Note that 
$\kappa_V$ and $\tilde\kappa_V$ are related to $c_t^V$ 
(the magnetic-like dipole moments) 
and to $d_t^V$ (the electric-like dipole moments), defined in 
Eq.~\ref{eetteq2}, via

\begin{eqnarray}
\kappa_V &\equiv& \frac{2m_t}{g_V} \times c_t^V \label{dimct}~,\\
\tilde\kappa_V  &\equiv& \frac{2m_t}{g_V} \times d_t^V \label{dimdt}~,
\end{eqnarray} 
 
\noindent and that $\kappa_V$ and $\tilde\kappa_V$ are, in general,
complex. In particular, for the convenience of the reader we note that
${\tilde \kappa}_{V} \sim 0.1(1)$ corresponds to a top electric-like
dipole moment coupling of $d_t^V \sim 0.55(5.5) \times 10^{-17}$ $g_V$-cm.

The implications of the effective top couplings in Eq.~\ref{eettgeq20} can
be studied by either considering CP-even or CP-odd observables in the
reaction $e^+e^- \to t \bar t g$. Of course, it should be clear that an
analysis of CP-even quantities like cross-sections and the shape of the
gluon energy spectrum \cite{hepph9610373}, or CP-even combination of
polarization asymmetry \cite{plb424p125}, can place rather mild
constraints on the absolute values of various EDM's as only ${\tilde
\kappa}_V^2$ enters into such quantities.  Let us summarize below the
limits that can be obtained on the various EDM couplings of
Eq.~\ref{eettgeq20} by analyzing the effects of CP-even and CP-odd
quantities on the reaction $e^+e^- \to t \bar t g$. \\

\noindent{\bf 6.3.2a CP-even observables}\\

In \cite{hepph9610373} it was suggested that the process $e^+e^- \to t \bar
t g$ can be used to obtain limits on the anomalous dipole-like couplings
of the top to $\gamma,g$ and $Z$ through the analysis of the associated gluon
energy spectrum.  

If the couplings of the top to the neutral gauge-bosons $\gamma,Z$ and $g$
are altered by the effective magnetic and electric-like interactions  
in Eq.~\ref{eettgeq20}, then by allowing one or more of the different
$\kappa_V$'s and ${\tilde \kappa}_V$'s to be non-zero, the shape of the gluon
energy  spectrum in the process $e^+e^- \to t \bar t g$ can be modified. 
In \cite{hepph9610373}, Monte Carlo data samples (assuming that the SM is
correct) were generated and then a fit to the general expressions for the
$\kappa_V - |{\tilde \kappa}_V|$ dependent spectrum were performed with which
a 95\% CL allowed region in the $\kappa_V - |{\tilde \kappa}_V|$ was obtained.
This procedure is done for each gauge-boson separately. That is, in analyzing
the limits that can be placed on the various dipole moment couplings, only
one pair of $\kappa_V,{\tilde \kappa}_V$ 
corresponding to one neutral gauge-boson, $V=\gamma,Z$ or $g$,
was allowed to have a non-zero value. As we have mentioned before the
usefulness of the bound on the magnetic moment is rather limited as it
receives significant contribution from QCD\null.

\begin{figure}[htb]
 \psfull
 \begin{center}
  \leavevmode
  \epsfig{file=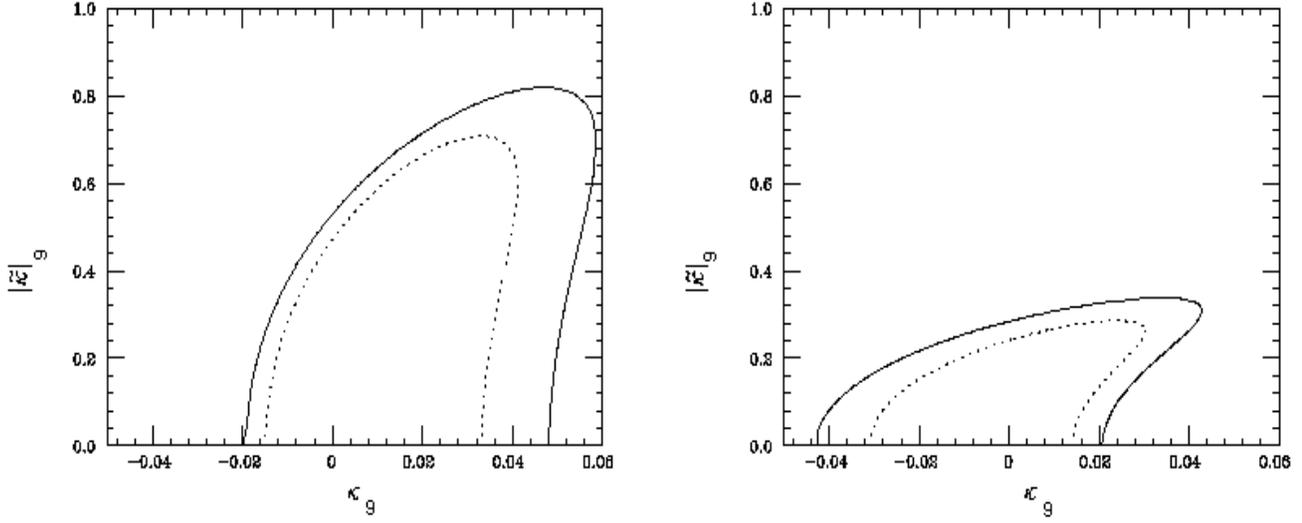,
height=7.5cm,width=7.5cm,bbllx=0cm,bblly=2cm,bburx=20cm,bbury=25cm,angle=270}
 \end{center}
\caption{\emph{$95\%$ CL allowed region in the $\kappa_g-|{\tilde \kappa}_g|$ 
(recall that ${\tilde \kappa}_g \equiv (2m_t/g_s) \times d_t^g$) plane 
obtained from fitting the gluon spectrum. On the left above: $E_g^{min}$=25 
GeV at a 500 GeV NLC assuming an integrated luminosity of 50(solid) or 
100(dotted) fb$^{-1}$. On the right above: 
for a 1 TeV collider with $E_g^{min}=
50$ GeV and luminosities of 100(solid) and 200(dotted) fb$^{-1}$.
Figure taken from \cite{hepph9610373}.}}
\label{eettgfig7}
\end{figure}

Summarizing now the limits obtained in \cite{hepph9610373},
Fig.~\ref{eettgfig7} shows the $95\%$ CL allowed region in the
$\kappa_g-|{\tilde \kappa}_g|$ plane for both a 500 GeV and a 1 TeV
$e^+e^-$ NLC. We see that the CEDM coupling, ${\tilde \kappa}_g$, can be
bounded in a 500 GeV NLC to $|{\tilde \kappa}_g| \lsim 0.6 - 0.8$. For the
CMDM coupling, for whatever it is worth, the allowed values are $\kappa_g
\sim \pm {\rm few} \times 10^{-2}$, with integrated luminosities of
$50-100$ fb$^{-1}$ and with a cut on the gluon energy of $E_g>25$
GeV\null. In a 1 TeV NLC, the limit on the CEDM coupling is approximately
twice as strong as what can be achieved in a 500 GeV NLC\null.

It should be noted that in \cite{plb424p125} two other CP-even quantities
were considered: the cross-section itself and a CP-even  combination of the
top and anti-top polarizations. The limits obtained  there for the CEDM of
the top, ${\tilde \kappa}_g$,  are somewhat weaker then those shown in
Fig.~\ref{eettgfig7}.

In analyzing the anomalous EDM's of the top to a photon and a $Z$-boson, the
``normalized'' gluon energy distribution was used 
in \cite{hepph9610373}: 

\beq
\frac{d{\cal R}}{dz} = \frac{1}{\sigma(e^+e^- \to t \bar t)} \frac{d \sigma(e^+e^-
\to t \bar t g)}{dz} \label{eettgeq21}~, 
\eeq
 
\begin{figure}
 \psfull
 \begin{center}
  \leavevmode
  \epsfig{file=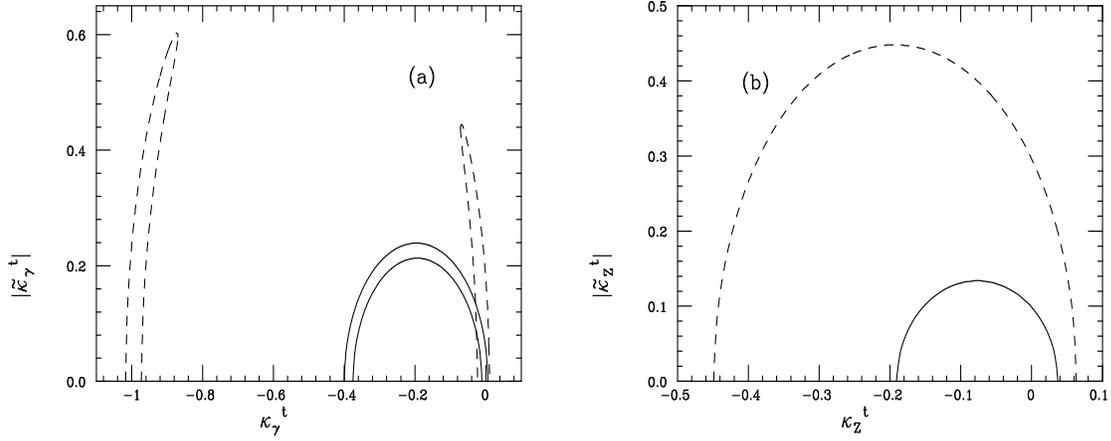,height=6cm,width=6cm
,bbllx=0cm,bblly=2cm,bburx=20cm,bbury=25cm,angle=270}
 \end{center}
\caption[dummy]{\emph{$95\%$ CL allowed regions obtained for the 
anomalous couplings
(a) $\kappa_\gamma,{\tilde \kappa}_\gamma$ and  (b) $\kappa_Z,{\tilde
\kappa}_Z$ 
at a 500(1000) GeV NLC, assuming a luminosity of 50(100) fb$^{-1}$, lie within
the dashed(solid) curves  (recall that 
${\tilde \kappa}_{\gamma} \equiv (2m_t/e)
\times d_t^{\gamma}$ and ${\tilde \kappa}_{Z} \equiv (2m_t/g_W/2c_W)
\times d_t^{Z}$). The gluon energy range $z \equiv 2E_g/\sqrt s \geq
0.1$ was  used in the fit. Only two anomalous couplings are allowed to be
non-zero at a time. Figure taken from \cite{hepph9610373}.}} 
\label{eettgfig8}
\end{figure}

It was then found that the EDM coupling of the top quark to a photon, ${\tilde
\kappa}_{\gamma}$, can be constrained at the NLC by studying 
the reaction $e^+e^-
\to t \bar t g$.  From Fig.~\ref{eettgfig8}(a) we  see that at a 500 GeV NLC
with integrated luminosity of 50 fb$^{-1}$,  only long narrow bands
around $\kappa_{\gamma} \sim -1$ or $0$ are allowed which then gives 
$|{\tilde \kappa}_{\gamma}| \lsim 0.4$--0.6. In a 1 TeV NLC with integrated 
luminosity of 100 fb$^{-1}$, one circular narrow band between 
$-0.4 \lsim \kappa_{\gamma}
\lsim 0$ is allowed giving $ 0 \lsim |{\tilde \kappa}_{\gamma}| \lsim 0.2$.

The anomalous EDM coupling of the top quark to the $Z$, 
${\tilde \kappa}_Z$, is much
less constrained. In particular, from Fig.~\ref{eettgfig8}(b) we see that with
a 500 GeV NLC and integrated luminosity of 50 fb$^{-1}$, $ 0 \lsim |{\tilde
\kappa}_Z| \lsim 0.5$ is allowed if $ -0.5 \lsim
\kappa_Z \lsim 0.1$.  However, with a 1 TeV NLC and integrated luminosity
of 100 fb$^{-1}$,  the allowed region in the $\kappa_Z - {\tilde \kappa}_Z$
plane is considerably reduced. Namely, $ 0 \lsim |{\tilde \kappa}_Z| \lsim
0.1$ can  be achieved if $ -0.2 \lsim \kappa_Z \lsim
0$.  Also, as was shown in \cite{hepph9610373}, doubling the integrated
luminosity does not  increase the sensitivity of the NLC to these anomalous
couplings of the top quark.                      
 
To conclude, note that while the process $e^+e^- \to t \bar t$ will
presumably be more appropriate for the exploration of CP-odd effects driven
by the top dipole moment couplings to 
$\gamma$ and $Z$ (see section \ref{ssec61}), the reaction
$e^+e^- \to t \bar t g$ might be  the only place for searching for the CEDM
of the top quark at the NLC\null. In that sense, an investigation of the
effects of ${\tilde \kappa}_g$ on  CP-odd quantities in the $t \bar tg$ 
final state 
at an $e^+e^-$ collider, is worthwhile. This was done in \cite{plb424p125} 
by constructing a genuinely CP-odd observable out of the top and anti-top
polarizations and is described below. \\

\noindent{\bf 6.3.2b CP-odd observables}\\

An interesting CP-odd observable was suggested in \cite{plb424p125}.
This observable involves the top polarization and is defined as

\beq
\Delta\sigma^{(-)} = \Frac{1}{2} \Big[\, \sigma(\uparrow) - \sigma(\downarrow)
+ \overline{\sigma}(\uparrow) - \overline{\sigma}(\downarrow) \,\Big]
\label{eettgeq22}~, 
\eeq

\noindent where $\sigma (\uparrow)$, $\overline{\sigma} (\uparrow)$ refer
respectively to the cross-sections for top and anti-top with a 
positive spin component in its direction of flight,
and $\sigma (\downarrow)$, $\overline{\sigma} (\downarrow)$
are the same quantities with a corresponding negative spin component. 

In \cite{plb424p125} the sensitivity of $\Delta\sigma^{(-)}$ to the CEDM
of the top,  ${\tilde \kappa}_g$, was studied. Note that $\Delta\sigma^{(-)}$ 
is CP-odd and $T_N$-even and therefore depends on the imaginary parts 
of the combinations of couplings 
$\Im{\rm m}(\kappa_g^*{\tilde \kappa}_g)$ and 
$\Im{\rm m}({\tilde \kappa}_g)$ 
in Eq.~\ref{eettgeq20}. The 90\% CL limits on the
values of $\Im{\rm m}(\kappa_g^*{\tilde \kappa}_g)$ and 
$\Im{\rm m}({\tilde \kappa}_g)$
were obtained from~\cite{plb424p125}:

\beq
\epsilon\, {\cal L}\,\left\vert\,\Dsigmam(\kappa_g,{\tilde \kappa}_g)- 
\Dsigmam_{\scriptscriptstyle
SM} \,\right\vert = 2.15 \,\sqrt{{\cal L}\,\left\vert\,
\sigma_{\scriptscriptstyle SM}
(\uparrow)+\overline{\sigma}_{\scriptscriptstyle SM}
(\uparrow)\,\right\vert\:} \label{eettgeq23} ~.
\eeq

\noindent where in the above expressions, the subscript ``SM" denotes the
value expected in the standard model, with $\kappa_g={\tilde \kappa}_g=0$;
$\epsilon$ is the top detection efficiency and ${\cal L}$ is the integrated 
luminosity.  

\begin{figure}
\vskip 65mm
 \includegraphics{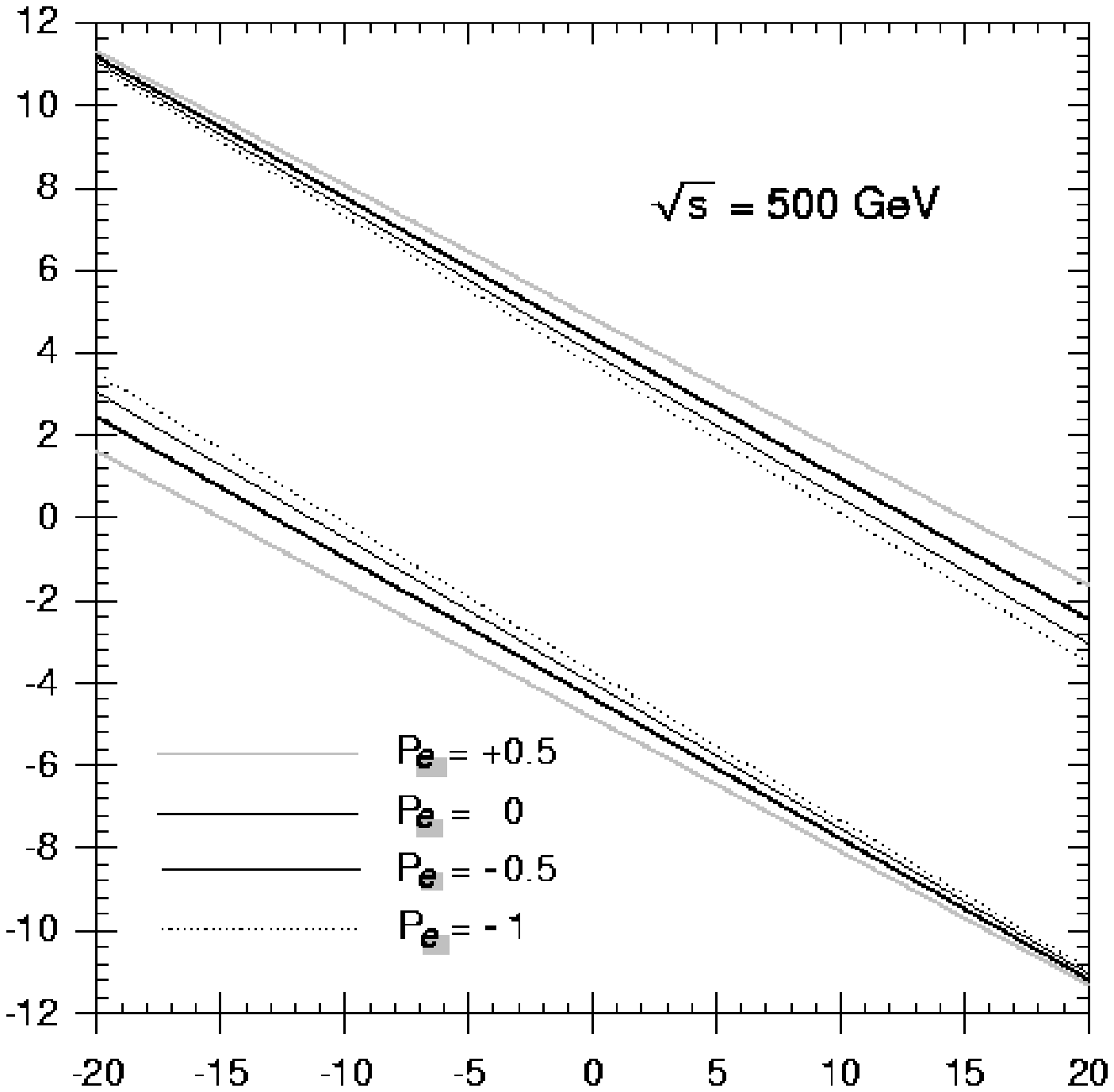}
 \includegraphics{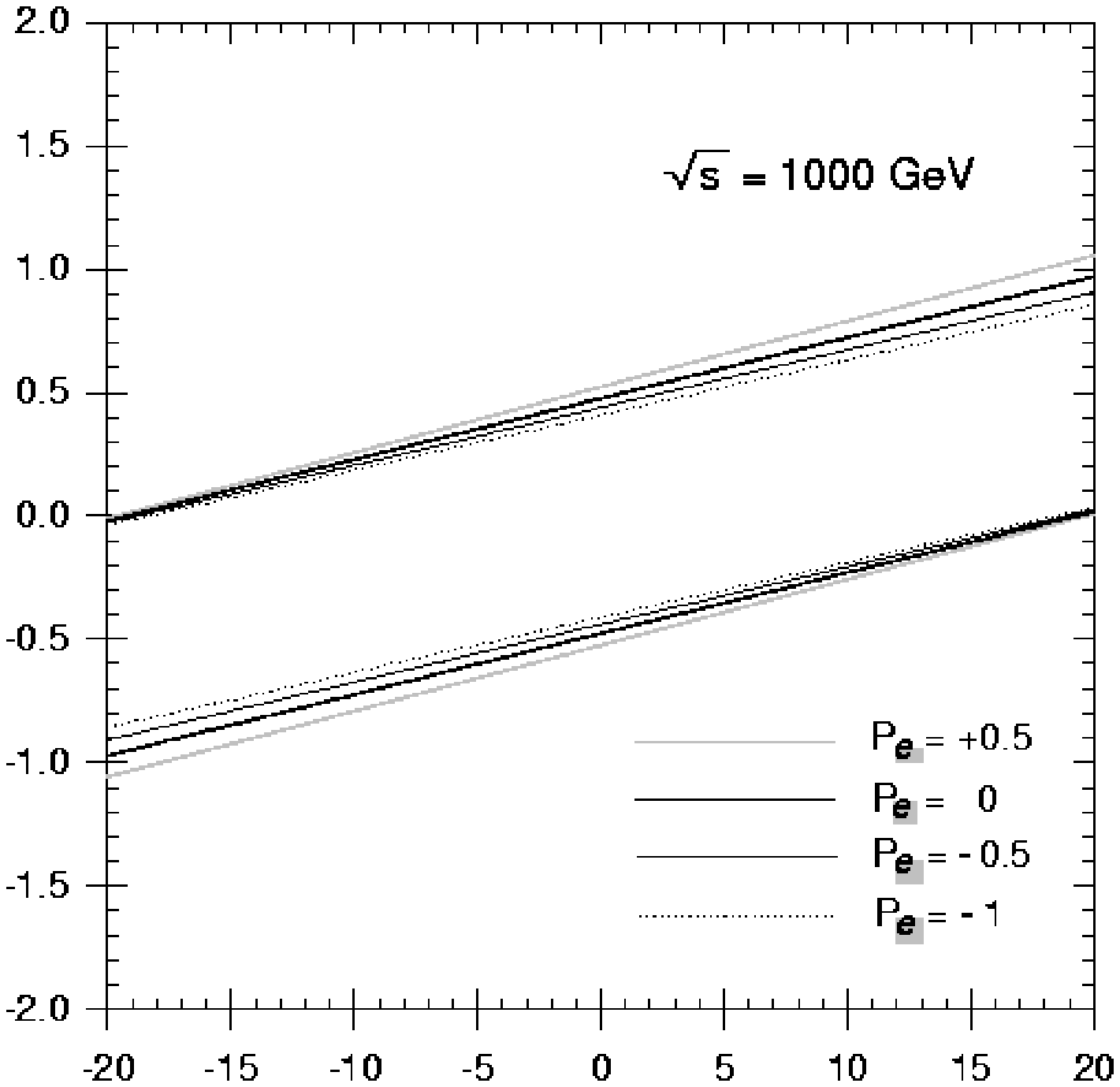}
\caption[dummy]{\emph{$\Dsigmam$ contour plots 
in the 
 $\Im{\rm m}(\kappa_g^* 
{\tilde \kappa}_g)$ 
(vertical axis)--$\Im{\rm m}({\tilde \kappa}_g)$ (horizontal axis) plane,
with 90\% confidence 
level at c.m.\
energies $\sqrt{s}=500$~GeV (left side) and $\sqrt{s}=1000$~GeV (right side),
and for different values of the electron beam polarization: $P_e=+0.5$, 0, 
$-0.5$, $-1$. Figure taken from \cite{plb424p125}.}}
\label{eettgfig9}
\end{figure}

Eq.~\ref{eettgeq23} gives contours in the 
$\Im{\rm m}(\kappa_g^*{\tilde \kappa}_g)$ - 
$\Im{\rm m}({\tilde \kappa}_g)$ plane which are shown 
in Figs.~\ref{eettgfig9}(a) and \ref{eettgfig9}(b)
for c.m.\ energies of $\sqrt s=0.5$ and 1 TeV, respectively, 
for an integrated
luminosity of ${\cal L}=50$ fb$^{-1}$ and for $\epsilon=0.1$.  Also different
polarizations of the incoming electron beam, $P_e$,  are considered in 
Figs.~\ref{eettgfig9}(a), (b). The allowed regions in 
Figs.~\ref{eettgfig9}(a), (b)
are the  bands lying between the upper and lower straight lines. 
We see that the dependence of $\Delta\sigma^{(-)}$ 
on the electron beam polarization is rather mild.

\begin{figure}[htb]
 \psfull
  \begin{center}
  \leavevmode
  \epsfig{file=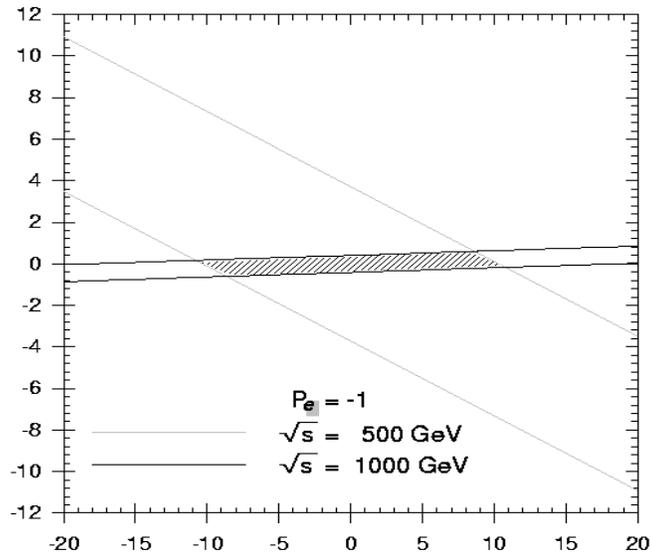,height=6cm,width=6cm,bbllx=0cm,bblly=2cm,bburx=20cm,bbury=25cm,angle=0}
 \end{center}
\caption[dummy]{\emph{Intersecting area in the 
 $\Im{\rm m}(\kappa_g^* {\tilde \kappa}_g)$--$\Im{\rm m}({\tilde \kappa}_g)$ 
plane
resulting from two independent 
$\Dsigmam$
measurements at $\sqrt{s}=500$~GeV and $\sqrt{s}=1000$~GeV. $P_e=-1$ is used. 
See also caption to
Fig.~\ref{eettgfig9}. Figure taken from \cite{plb424p125}.}} \label{eettgfig10}
\end{figure}

It was also suggested in \cite{plb424p125} that a measurement of
$\Delta\sigma^{(-)}$ at two different c.m.\ energy  may improve the limits
on $\Im{\rm m}(\kappa_g^*{\tilde \kappa}_g)$ 
and $\Im{\rm m}({\tilde \kappa}_g)$. This is
demonstrated in Fig.~\ref{eettgfig10}  where it is shown that from
measurements 
of $\Delta\sigma^{(-)}$ at $\sqrt s=0.5$ and 1 TeV, the possible limits are

\beq
-0.8 < \Im{\rm m} (\kappa_g^*{\tilde \kappa}_g) < 0.8 ~~,~~
-11 < \Im{\rm m} ({\tilde \kappa}_g) < 11 \label{eettgeq24}~.
\eeq

\noindent Although dealing with a genuine CP-odd observable, the limits
in Eq.~\ref{eettgeq24}  are still weaker by about an order of magnitude 
than those obtained through the study of the 
gluon jet energy distribution. Note however, that those limits are
placed on the imaginary part of ${\tilde \kappa}_g$ while the limits 
obtained through the study of the gluon jet energy distribution 
are set on the absolute value of the top CEDM\null.



\newpage
~

\newpage
~

\newpage
~

\newpage
~

\newpage



\subsection{CP violation via $WW$ 
fusion in $e^+ e^- \to t \bar t {\nu}_e {\bar {\nu}}_e$ \label{ssec64}}

At the NLC with a very high c.m.\ energy, above 1 TeV, the 
t-channel $W^+W^-$ fusion subprocesses $W^+ W^- \to t \bar t$, where the two 
$W$-bosons are emitted from the initial $e^+e^-$ beams, starts to dominate 
over the simple $s$-channel production mechanism of a pair of $t \bar t$, 
i.e., $e^+e^- \to \gamma,Z \to t \bar t$. As it turns out \cite{hepph9607481}, 
the reaction
  
\beq
e^+ e^- \to W^+W^- \nu_e {\bar {\nu}}_e \to 
t \bar t {\nu}_e {\bar {\nu}}_e ,~\label{ttnunu} 
\eeq

\noindent can potentially exhibit large CP-violating phenomena, driven by 
CP-odd phases in the neutral Higgs sector in MHDM's. 

\begin{figure}[htb]
\begin{center}
\leavevmode
 \epsfig{file=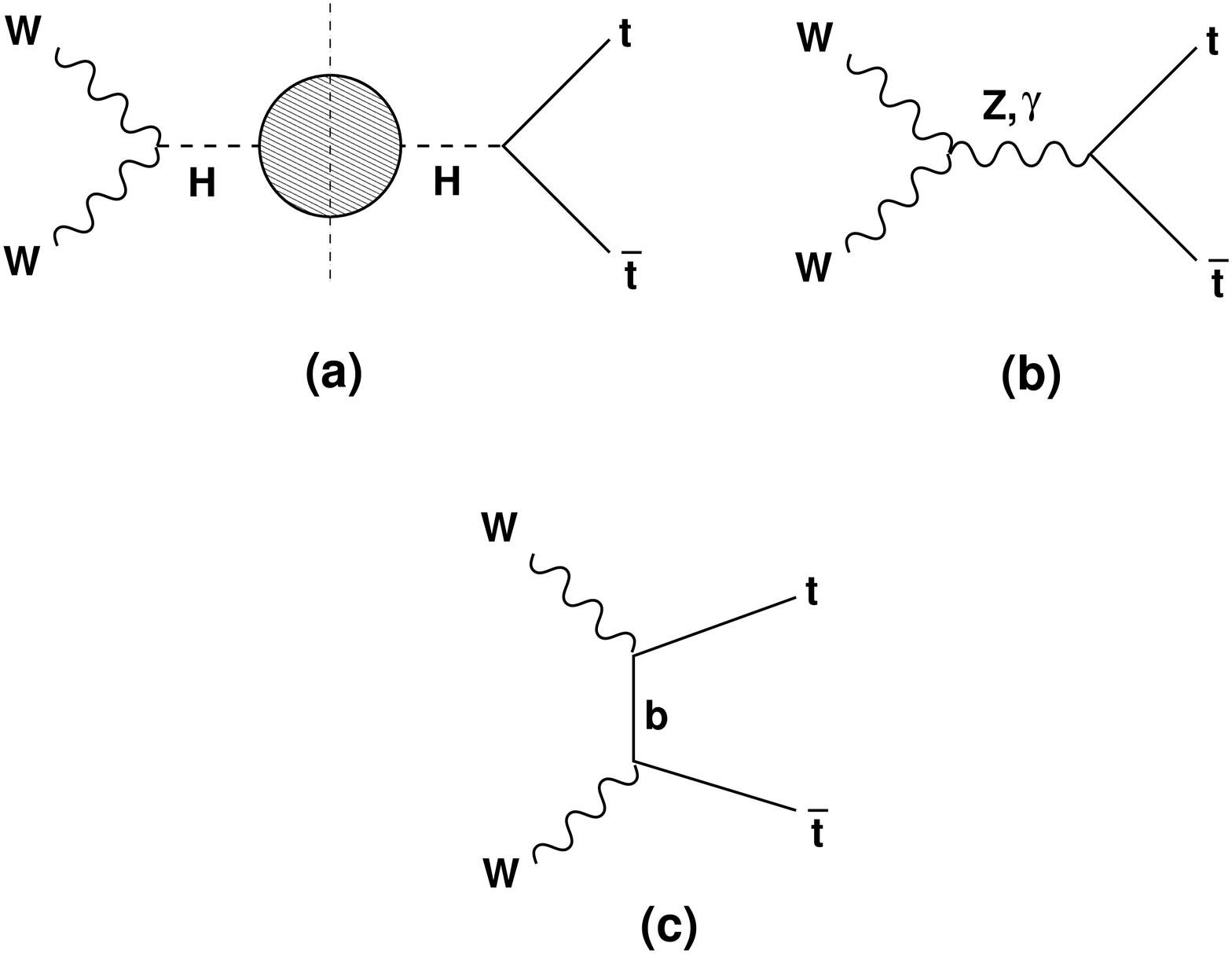
,height=8cm}
 \end{center}
\caption{\emph{The Feynman diagrams that participate in the subprocess 
$W^+W^-\to t\bar t$. The blob in diagram (a) represents the width of the
Higgs resonance and the cut across the blob is to indicate the
imaginary part.}}
\label{wwttfig1}
\end{figure}

To lowest order there are four Feynman graphs, shown in Fig.~\ref{wwttfig1}, 
relevant to the reaction in Eq.~\ref{ttnunu}. Indeed, at 
large c.m.\ energies, i.e., as $s/m_W^2$ becomes very large, 
the cross-section for the reaction in Eq.~\ref{ttnunu} is dominated 
by collisions of longitudinally polarized $W$'s and 
the subprocess $W^+W^- \to t \bar t$ shown in Fig.~\ref{wwttfig1}, 
when calculated in the Effective $W$-boson Approximation \cite{wwapprox},  
serves as a good approximation to the reaction 
$e^+ e^-  \to t \bar t {\nu}_e {\bar {\nu}}_e$.


The key point here, as suggested in \cite{hepph9607481}, is again 
to construct CP-odd observables utilizing the top polarization, which in 
turn can be traced through the top decays. Following \cite{hepph9607481}, 
in the rest frame of the $t$ one defines the basis vectors: 
$-e_z \propto ({\vec p}_{W^+} + {\vec p}_{W^-})$,  
$e_y \propto {\vec p}_{W^+} \times {\vec p}_{W^-}$ and $e_x=e_y \times e_z$. 
For the anti-top one uses a similar set of the definitions in the $\bar t$ 
rest frame related by charge conjugation: 
$- \bar e_z \propto ({\vec p}_{W^-} + {\vec p}_{W^+})$,  
$\bar e_y \propto {\vec p}_{W^-} \times {\vec p}_{W^+}$ and 
$\bar e_x=\bar e_y \times \bar e_z$. 
Now let $P_j$ (for $j=x,y$ or $z$) be the polarization of $t$ along 
$e_x,e_y,e_z$ and similarly, $\bar P_j$ the polarization of $\bar t$ along 
$\bar e_x,\bar e_y,\bar e_z$. One can then combine information from 
the $t$ and $\bar t$ systems and define the following asymmetries

\beqa
A_x=\frac{1}{2}(P_x + \bar P_x)~~&,&~~B_x=\frac{1}{2}(P_x - \bar P_x) ~,
\nonumber \\
A_y=\frac{1}{2}(P_y - \bar P_y)~~&,&~~B_y=\frac{1}{2}(P_y + \bar P_y) 
\label{wwtteq1}~,\\
A_z=\frac{1}{2}(P_z + \bar P_z)~~&,&~~B_z=\frac{1}{2}(P_z - \bar P_z)~, 
\nonumber
\eeqa

\noindent where it is easy to verify that, within the above 
coordinate systems, the $A$'s are CP-odd and the $B$'s are CP-even. Moreover,
$\left\{ A_x,B_y,A_z \right\}$ are CP$T_N$-odd whereas 
$\left\{ B_x,A_y,B_z \right\}$ are CP$T_N$-even. 

We note here that the CP-even spin observable $B_y$, 
being proportional to the imaginary part of the Higgs propagator 
in Fig.~\ref{wwttfig1}(a), is also useful for experimentally measuring 
the Higgs width \cite{hepph9607481}. 
However, since here we are only interested in CP non-conservation effects in 
the reaction $e^+ e^- \to t \bar t {\nu}_e {\bar {\nu}}_e$, 
we will focus below on results obtained for the CP-odd observables, i.e., 
the $A$'s in Eq.~\ref{wwtteq1}.        

Let us consider a 2HDM with the $t \bar t{\cal H}^k$ and $W^+W^- {\cal H}^k$ 
Lagrangian pieces of Eqs.~\ref{2hdmab} and \ref{2hdmc}, respectively. 
Here also, the   
simultaneous presence of the scalar, $a_t^k$, and pseudoscalar couplings, 
$b_t^k$, in Eq.~\ref{2hdmab} is required 
for a non-zero expectation value of the CP-violating 
asymmetries $A_x,A_y,A_z$. Therefore, since only two 
out of the three neutral Higgs particles, i.e., $k=1,2$ or 3,  
(say, $h \equiv {\cal H}^1$ and $H \equiv {\cal H}^2$ for 
the lighter and heavier ones, respectively) 
have  
a simultaneous scalar and pseudoscalar couplings to $t \bar t$ 
(see section \ref{sssec323}), 
the third neutral Higgs need not be considered.
$A_z$ is expected to receive significant contributions from loop 
corrections. Therefore, we focus below on $A_x$ and $A_y$ only 
(see discussion in \cite{hepph9607481}).   

\begin{figure}[htb]
\begin{center}
\leavevmode
 \epsfig{file=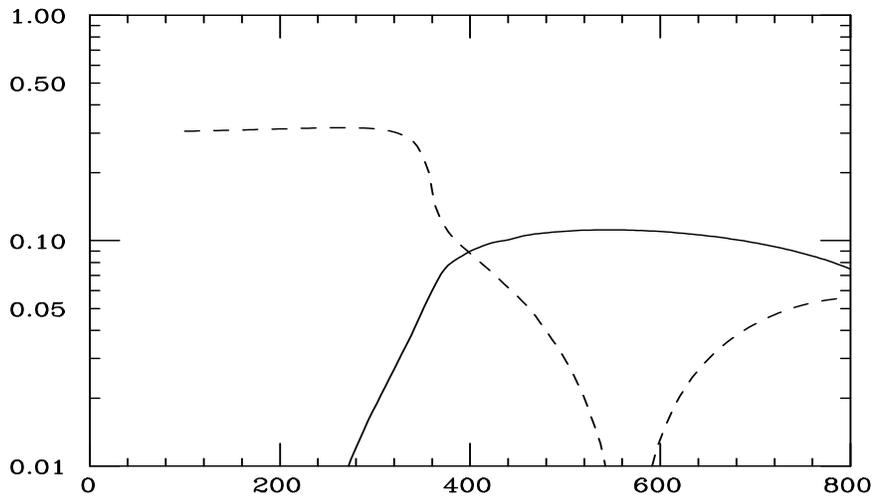
,height=9cm,width=6cm,bbllx=0cm,bblly=2cm,bburx=20cm,bbury=25cm,angle=90}
 \end{center}
\caption{\emph{The asymmetries $A_x$ (solid) and $A_y$ (dashes) 
integrated over $\hat s$ as a function of 
$m_h$ for $\sqrt{s}=1.5$ TeV and $m_H=1$ TeV.
The coupling parameters are 
for $\tan\beta=0.5$ and $\left\{\alpha_1,\alpha_2,\alpha_3 \right\}=
\left\{-\pi/2,\beta,-\pi/2 \right\}$ as described in the text. 
Figure taken from 
\cite{hepph9607481}.}}
\label{wwttfig2}
\end{figure}

The two asymmetries $A_x,A_y$ are shown in Fig.~\ref{wwttfig2}, for 
a NLC with a c.m.\ energy of $\sqrt s = 1.5$ TeV, 
as a function of the lighter Higgs mass 
$m_h$. The heavier Higgs mass is fixed to $m_H=1$ TeV\null. Also, 
for illustration, we use $\tan\beta=0.5$ and choose 
$\left\{\alpha_1,\alpha_2,\alpha_3 \right\}=
\left\{-\pi/2,\beta,-\pi/2 \right\}$, where $\alpha_1,\alpha_2$ 
and $\alpha_3$ are the three Euler angles that specify the $3 \times 3$ 
orthogonal mixing matrix of the three neutral Higgs-bosons 
(see Eq.~\ref{2hdmrij}). 
With this set of 
parameters the $t \bar t h,~t\bar tH,~W^+W^-h$ and $W^+W^-H$ 
couplings are fixed according 
to Eqs.~\ref{2hdmab}, \ref{2hdmc}, \ref{mhdmeq23} and \ref{2hdmrij} in 
section \ref{sssec323}.

We  observe from Fig.~\ref{wwttfig2} that for a wide range of the 
lighter Higgs mass the asymmetries are appreciable. In particular, 
$A_x$ is about $10\%$ for $m_h \sim 400$--800 GeV whereas $A_y$ is 
around $30\%$ for $m_h \sim 100 - 300$ GeV\null. Although not shown in 
Fig.~\ref{wwttfig2}, the asymmetries vanish when $m_h=m_H$ due to 
a GIM-like cancellation as explained in section \ref{sssec323}. 

As mentioned above, in order to measure those top polarization asymmetries, 
one needs the momentum of the $t$ and $\bar t$ 
decay products in a given decay scenario. In \cite{hepph9607481} two 
such decay scenarios useful for top polarimetry were considered:

\begin{enumerate}

\item The decay $t \to W^+ b$ followed by $W^+ \to \ell^+ \nu$, where 
$\ell=e,\mu$; in this case only the hadronic decays of $\bar t$ are included.
This case occurs with a branching ratio of $B_1 \approx (2/9)(2/3)=4/27$.

\item The decay $t \to W^+ b$ followed by $W^+ \to {\rm hadrons}$. Now 
the decay of $\bar t$ to a $\tau^-$ is excluded. 
This case occurs with a branching ratio of $B_2 \approx (2/3)(8/9)=16/27$.

\end{enumerate}

In the case of the leptonic decay of $t$ 
(or equivalently $\bar t$), the angular distribution of the lepton is 
$\propto (1+R_1 P \cos\eta_\ell )$, where $P$ is its polarization,
$\eta_\ell$ is the angle between the polarization axis and the momentum
of the lepton in the top rest frame and 
$R_1=1$ in the SM\null.  
Thus, the optimal method to obtain the value of $P$ is to use 
$P=3 \langle \cos\eta_\ell \rangle /R_1$ (see section \ref{ssec28}). 
Similarly, in the case of 
the hadronic $t$ decay (or equivalently $\bar t$), one uses the distribution 
of the $W$ momentum in the top frame which is 
$\propto (1+R_2 P \cos\eta_W )$ to extract the top polarization, 
where $R_2=(m_t^2-2m_W^2)/(m_t^2+2m_W^2)$. 
Therefore, in this case $P=3 \langle \cos\eta_W \rangle /R_2$.

Hence, bearing in mind that the leptonic decay of the top is self
polarizing, the number of events needed to obtain a 3-$\sigma$ signal in
the $t, \bar t$ decays of case 1 above is \cite{hepph9607481}:

\beq
N_{t \bar t}^{3\sigma}=(27/2) (R_1^2 B_1 +R_2^2 B_2)^{-1} \times a^{-2} ~,
\eeq

\noindent where $a$ is the asymmetry in question (either $A_x$ or $A_y$). 
Therefore, given the above numbers for $R_1,R_2,B_1$ and $B_2$, numerically
$N_{t \bar t}^{3\sigma} \approx 52 a^{-2}$, thus requiring some 5200 events
for an asymmetry of 10\%. In fact, this can be further improved in the case
of the hadronic decays of $t$ (case 2 above) by observing that
\cite{plb350p218} the less energetic of the two jets from the decay of the
$W$ is more likely to be the $\bar d$-type quark as noted in section
\ref{ssec28}.  In particular, in case 2 one obtains $N_{t \bar t}^{3\sigma}
\approx 32 a^{-2}$, thus reducing the requirement to 3200 events for an
asymmetry of 10\% \cite{hepph9607481}.  For an asymmetry of about 30\%,
which was found possible in the case of $A_y$ (see Fig.~\ref{wwttfig2}),
only a few hundred events will be needed. 

Indeed, the cross-section for the reaction in Eq.~\ref{ttnunu} was
calculated in \cite{hepph9607481} for the case of a 2HDM with the couplings
described above and it was found to be at the level of a few fb, reaching
$\gsim 10$ fb for $350~ {\rm GeV} \lsim m_h \lsim 550~ {\rm GeV}$.  Thus,
given that at $\sqrt s=1.5$ TeV the projected luminosity could be about $5
\times 10^{34}$ cm$^{-2}$ s$^{-1}$ \cite{nlc}, a cross-section of 10 fb
would yield about 5000 events rendering it feasible to detect asymmetries
$\gsim 10\%$. The conclusion is therefore that the top polarization
asymmetries for the reaction $e^+ e^- \to t \bar t \nu_e {\bar {\nu}}_e$
are accessible to the NLC and can serve as a powerful probe of CP violation
driven by the neutral Higgs sector of a 2HDM\null. However, this last
statement must be taken with some caution since, the two neutrinos in the
final state, carry a substantial amount of missing energy and may therefore
pose a problem in reconstructing the $t$ and $\bar t$ rest frames, as
required for measuring the polarization asymmetries in question when the
$t$ or $\bar t$ decays leptonically.  No such problem arises if both
the $t$ and the $\bar t$ decay purely hadronically but, in that case, it
remains to be seen if it will be possible to distinguish $t$ from $\bar t$
which is also required for measuring $A_x$ and $A_y$. 

An interesting generalization of this work \cite{hepph9607481} is to
consider instead the reaction $e^+e^-\to t\bar t e^+e^-$. Now the fusion
takes place via neutral gauge-bosons ($\gamma,Z$). Although there may be
some loss of the cross-section, to compensate that, there is also the
advantage that the difficulties in reconstructing the rest frames of $t$,
$\bar t$ may be far less formidable.

\newpage
~

\newpage

\newpage
~

\newpage

\pagebreak

\section{CP violation in 
\boldmath ${p p}$ collider experiments \label{sec7}}
\setcounter{equation}{0}

The LHC is a $pp$ collider at CERN, with c.m. energy
of 14 TeV, scheduled to start running around 2005 (For a recent review
on machine parameters see \cite{ppreview1}.)
Its design luminosity is ${\cal L} = 10^{34}$ cm$^{-2}$sec$^{-1}$,
corresponding to a yearly integrated luminosity of 100 fb$^{-1}$.
A low luminosity first stage of 10 fb$^{-1}$ is usually
assumed in articles discussing physics at the LHC\null.
The issues discussed in the following section, will be
relevant for the future CMS and ATLAS experiments (for a review
see \cite{ppreview2}); heavy ions and LHC-B will not be discussed in
the present work. For recent reviews on the physics at
LHC, see \cite{ppreview2} and \cite{ppreview3}.

\subsection{$pp\to t \bar t +X$: 
general comments \label{ssec71}}

In hadronic collisions $t \bar t$ pairs are produced through the
parton level subprocesses $q \bar q \to t \bar t$ and $gg \to t \bar
t$. The latter, $gg$ fusion process, dominates over the quark--anti-quark
annihilation in a multi-TeV $pp$ collider. 
For example, at the LHC, $\sigma(gg \to t \bar t ) \sim 90\%$ and 
$\sigma(q \bar q \to t \bar t ) \sim 10\%$ are expected. 
It is therefore 
important to investigate the expected CP violation effects in $pp\to t
\bar t +X$ that can arise from CP non-conservation in the subprocess
$gg \to t \bar t$.

Note that the simple $q \bar q$ fusion process is the analog of the 
$e^+e^- \to t \bar t$ production mechanism where an
s-channel gauge-boson is exchanged.
In the case of $q \bar q \to g \to t \bar t$, the
CP-odd effect can therefore be attributed to the CEDM ($d_t^g$) of the
top present at the $gt \bar t$ vertex. In contrast,
the $gg$ production process gives rise to a much richer possibility
of CP-violating interactions and the resulting asymmetries
in $gg \to t \bar t$ need not be related merely to the CEDM
of the top quark. This fact can be readily seen in model
calculations (such as 2HDM and MSSM to be discussed below),
where additional CP-violating 1-loop box diagrams as well 
as $\hat s$-channel resonant neutral Higgs exchange 
become relevant. 

We will first discuss an effective Lagrangian approach in which all
CP-violating effects are assumed to originate only from the
CEDM of the top. We will then present model dependent analysis of
CP non-conservation in $gg \to t \bar t$ where all possible CP-violating
operators are taken into account.

As will be shown, the typical size of the CP-violating asymmetries in
$pp \to t \bar t +X$ is $\sim 10^{-3}$. 
Although, naively one may expect 
such asymmetries to be within the experimental reach
of the LHC, which is expected to produce $\sim 10^7 - 10^8$
$t \bar t$ pairs, there are at least two types of hurdles that make
this objective very difficult to attain. First there is the detector
dependent systematics which are expected to present serious limitations
for asymmetries at the $\sim10^{-3}$ level. Another serious difficulty is that
the initial state ($pp$) is not an eigenstate of CP\null. Therefore one
expects fake asymmetries to arise at some level even though the
underlying interactions do not violate CP\null. These backgrounds are
process dependent and the fake asymmetries that they produce needs to
be much smaller in comparison to the CP-violating signal 
that
is of interest.
In some cases, e.g., an $\hat s$-channel resonant Higgs exchange within
a 2HDM, as will be described in section \ref{sssec732}, 
by employing clever cuts on the $t \bar t$ invariant mass one can 
obtain asymmetries at the percent level. 
In these cases, the CP signal is more robust and may be within 
the reach of the LHC if the 2HDM parameter space turns out favorable.

\subsection{$pp\to t \bar t +X$: 
general form factor approach and
the CEDM of the top \label{ssec72}}

\noindent As already mentioned in previous sections, in close analogy to the 
EDM and the weak($Z$)-EDM of the top, one
can generalize the top quark-gluon effective
Lagrangian to include terms of dimension 5 which can give rise to 
a CEDM for the top quark (see Eq.~\ref{eettgeq20} in section \ref{sssec632}).
In general, the CEDM coupling, $d_t^g$, may be considered
as a form factor. Its momentum dependence is generated by effective
Lagrangian operators of dimension greater than 5. In model dependent
calculations, this from factor may acquire momentum dependent
imaginary parts as well as real parts.
In momentum space, similar to
the EDM and weak-EDM cases, the CEDM modifies the $ttg$ interaction to read
(we will not concern ourselves here with the CP-conserving Chromo-magnetic 
dipole moment of the
top)

\be
-i  T_a \left(  g_s \gamma^\mu + d_t^g \sigma^{\mu \nu} \gamma_5 k_{\nu}
\right) \label{ttgpptt}~,
\ee

\ni where $k=p_t+p_{\bar t}$ is the gluon four-momentum and
$p_t(p_{\bar t})$ is the $t(\bar t)$ four momentum.

\begin{figure}[ht]
\psfull
 \begin{center}
  \leavevmode
  \epsfig{file=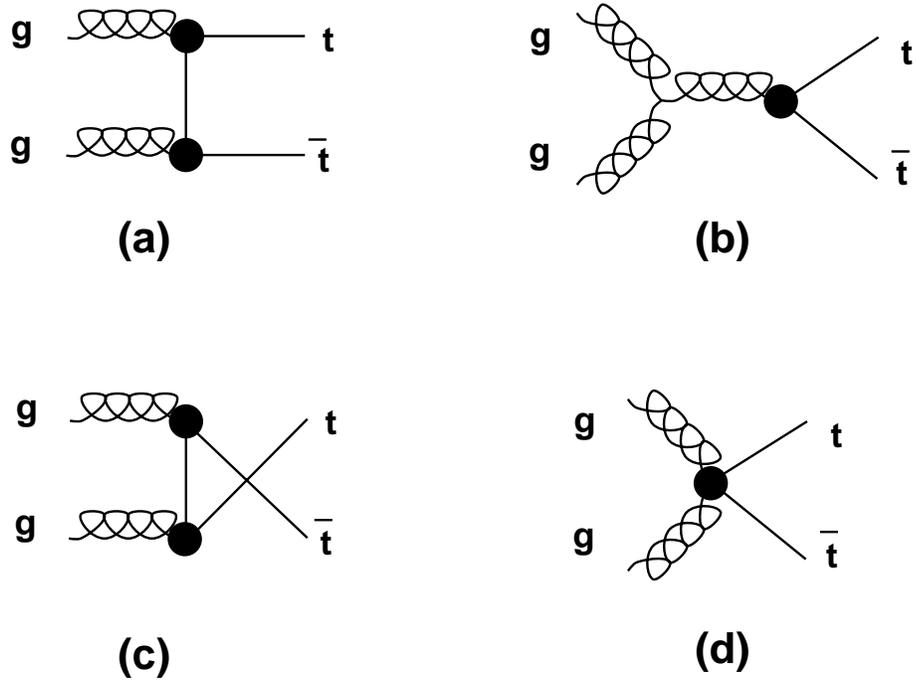
,height=9cm}
 \end{center}
\caption{{\emph Feynman diagrams 
contributing to $gg \to t \bar t$ in the presence
of a top CEDM in the $t \bar t g$ vertex which is denoted by the heavy dot.}}
\label{ppttfig1}
\end{figure}

The subprocess $gg \to t \bar t$ then proceeds through diagrams (a)--(d)
in Fig.~\ref{ppttfig1} where the heavy dots indicate the vertices
modified by the CEDM of the top defined in Eq.~\ref{ttgpptt}.
Diagram (d) involves an additional dimension 5 $ttgg$ contact term and
is needed to preserve gauge invariance (see section \ref{ssec25}). 
Assuming
that $d_t^g$ is small enough such that one can expand  the matrix
element squared to first order in $d_t^g$, the differential
cross-section for the subprocess $gg \to t \bar t$ can be written,
similar to the $e^+e^- \to t \bar t$ case, as \cite{prl69p2754}:

\beq
\Sigma(\phi) d\phi =  \Sigma_0(\phi) d\phi +
\left[ {\eRe} d_t^g(\hat s) \Sigma_{{\eRe}}
(\phi) + {\mIm} d_t^g(\hat s) \Sigma_{{\mIm}}(\phi) \right]
d\phi \label{pptt2} ~,
\eeq

\ni where ${\hat s}=x_1 x_2 s$  and $x_1,x_2$ are the gluons momentum
fractions. In Eq.~\ref{pptt2} above, the gluon structure functions are
included and thus $\phi$ represents the final state phase space
including the gluon momentum fraction variables. Also, with no
summation over the $t$ and $\bar t$ spins $s_t$ and $s_{\bar t}$,
respectively, the CP-odd differential cross-sections
$\Sigma_{{\eRe}}(\phi)$ and $\Sigma_{{\mIm}}(\phi)$ are
functions of $s_t,s_{\bar t}$. Therefore, because of the correlation  
between the top spin and the momentum of the charged 
lepton from the top decay $t \to b W^+ \to b \ell^+ \nu_{\ell}$,
Eq.~\ref{pptt2} with the $s_t$ and $s_{\bar t}$
dependence, gives in fact the differential cross-section
for the complete process $gg \to t \bar t$ including
the subsequent leptonic decay chains of the tops.


\newpage
~

\newpage

\subsubsection{Optimal observables \label{sssec721}}

With the effective Lagrangian in Eq.~\ref{eettgeq20} in 
section \ref{sssec632} and by ignoring
operators of dimension greater than 5, only the effect of a constant
real $d_t^g$ was investigated in \cite{prl69p2754}. Indeed, in model
calculations to be described below, the real part of the CEDM form
factor is a constant to a good approximation. Similar to the $e^+e^-
\to t \bar t$ case, an optimal $T_N$-odd, CP-violating observable for
$gg \to t \bar t$ was defined in \cite{prl69p2754} as

\beq
{\cal O} =  \frac{\Sigma_{{\eRe}}}{\Sigma_0}
\label{opt1pptt}~.
\eeq

\noindent In a realistic hadronic collider however, not all momenta which
enter into the problem are immediately observable. For example, with
leptonic decays of both $t$ and $\bar t$, the momenta of the neutrinos and
the longitudinal momenta of the initial gluons are not observed. As was
shown in \cite{prl69p2754}, this leads to a twofold or fourfold ambiguity
(depending on the number of solutions to the kinematics which results in a
quartic equation) in determining the neutrinos momenta. To bypass this
difficulty an ``improved'' optimal observable, that averages over the
reconstruction ambiguity, was suggested in \cite{prl69p2754}: 

\beq
{\cal O}\,^\prime \equiv \frac{\sum_i \Sigma_{{\eRe}}(\phi_i)}
{\sum_i \Sigma_0(\phi_i)} \label{opt2pptt}~,
\eeq

\ni where the sum is over the different possible reconstructions of the
neutrino and anti-neutrino momenta from the leptonic $t$ and $\bar t$
decays, respectively. 

Using the optimal observables ${\cal O}$ and ${\cal O}\,^\prime$,
the attainable 1-$\sigma$ limits on ${\eRe}d_t^g$, assuming $10^7$
dilepton $t \bar t$ decays, were given in \cite{prl69p2754}. Note that
one can consider also leptonic-hadronic and purely hadronic decays of 
the $t \bar t$
pairs. Due to the branching ratios of the $W$-boson, $10^7$ leptonic
$t \bar t$ pairs implies a sample of $\sim 6\times 10^7$ leptonic-hadronic
$t \bar t$ pairs and $\sim 9\times 10^7$ hadronic $t \bar t$ pairs. With
$m_t=175$ GeV, for the ``simple'' optimal observable ${\cal O}$, with
dilepton $t \bar t$ pairs, it was found that the 1-$\sigma$ limit is 
${\eRe}d_t^g \sim 2.8 \times 10^{-20}$ $g_s$-cm. For the ``improved''
optimal observable ${\cal O}\,^\prime$, ${\eRe}d_t^g \sim 3.0 \times
10^{-20}$ $g_s$-cm with dilepton or hadronic $t \bar t$ pairs and
${\eRe}d_t^g \sim 2.0 \times 10^{-20}$ $g_s$-cm with leptonic-hadronic
$t \bar t$ pairs. Comparing the 1-$\sigma$ limits on ${\eRe}d_t^g$
attainable with dileptonic $t \bar t$ pairs and through the use of the
optimal observables ${\cal O}$ and ${\cal O}\,^\prime$ in
Eqs.~\ref{opt1pptt} and \ref{opt2pptt}, respectively, we see that the
reconstruction ambiguity does not cause any significant changes.
Evidently, with these optimal observables, ${\eRe}d_t^g$ may be
measured to a precision of $ \sim 10^{-20}$ $g_s$-cm. ${\cal
O}\,^\prime$ with the leptonic-hadronic $t\bar t$ channel seems to be
the most sensitive to ${\eRe}d_t^g$. 
Comparable results for ${\eRe}d_t^g$ were found in \cite{hepph9805358} 
by using the same type of optimal observables. 
Ref. 
\cite{hepph9805358} has also extended the analysis of 
\cite{prl69p2754} by including effects of the imaginary 
part of $d_t^g$. They found that the attainable limit at the LHC for 
$\Im{\rm m}d_t^g$ is of the same order, i.e., 
$\Im{\rm m}d_t^g \sim 10^{-20}$ $g_s$-cm, although slightly 
better than the one for ${\eRe}d_t^g$.     

This result is encouraging since, as we have discussed in Chapter
\ref{sec4} the CEDM of the top may be $\gsim 10^{-20}$ $g_s$-cm in some
extensions of the SM, e.g., MHDM's and the MSSM.


\subsubsection{Observable correlations between
momenta of the top decay products \label{sssec722}}

It is also instructive to consider simple observables constructed
exclusively out of momenta which are directly observed. With the decays
$t \to b \ell \nu_{\ell}$ and $\bar t \to \bar b \bar\ell
{\bar\nu}_{\ell}$, the momenta $p_\ell,~p_{\bar\ell},~p_b$ and $p_{\bar
b}$ will be directly observed and observables which involve
correlations between those momenta are the most appropriate. Two such
CP-odd, $T_N$-odd correlations were considered in \cite{prl69p2754}:

\bea
f_1 &=& \frac{\epsilon_{\mu \nu \sigma \rho}
p_e^{\mu} p_{\bar e}^{\nu} p_b^{\sigma}
p_{\bar b}^{\rho}}{(p_e \cdot p_{\bar e}
p_b \cdot p_{\bar b})^{1/2}} ~,\\
f_2 &=& (p_e^x p_{\bar e}^y - p_e^y
p_{\bar e}^x) 
\cdot
{\rm sgn}(p_e^z - p_{\bar e}^z)
(p_e \cdot p_{\bar e})^{1/2} ~,
\eea

\ni where ${\rm sgn}(X) = +1$ for $X \ge 0$ and $-1$ for $X < 0$.

The attainable 1-$\sigma$ limits on ${\eRe}d_t^g$ for the observables
$f_1$ and $f_2$, with $m_t=175$ GeV and assuming $10^7$ dileptonic $t
\bar t$ pairs, were also given in \cite{prl69p2754}. The findings were
for $f_1$: ${\eRe}d_t^g \sim 5.3 \times 10^{-20}$ $g_s$-cm and
for $f_2$: ${\eRe}d_t^g \sim 3.0 \times 10^{-19}$ $g_s$-cm.
We see that the limit that might be achieved with $f_1$ is about an
order of magnitude smaller than that from $f_2$. However, $f_2$ depends
only on the lepton momenta and is, therefore, easiest to determine
experimentally. Also, the limit from $f_1$ is about 2 times weaker then
the one obtained from the ``improved'' optimal observable
discussed previously.

In \cite{plb415p67} CP-odd $T_N$-even observables which might be used
to probe the imaginary part of the CEDM, i.e., ${\mIm}d_t^g$, were
considered

\bea
A_E &=& E_{\bar\ell} - E_\ell ~,\\
Q_{33} &=& 2(p_{\bar\ell}^z + p_\ell^z)
(p_{\bar\ell}^z - p_\ell^z) - \frac{2}{3}
({\vec p}_{\bar\ell}^{~2} -{\vec p}_{\ell}^{~2})~.
\eea

\ni $A_E$ is the energy asymmetry between $\ell$ and $\bar\ell$ and
$Q_{33}$ is an asymmetry originally suggested by Bernreuther {\it et al.}
in \cite{npb388p53} (see section \ref{sssec612}). In a $pp$ collider with
$\sqrt s=14$ TeV, an integrated luminosity of 10 fb$^{-1}$ and an
acceptance efficiency of $\epsilon=10\%$, taking only leptonic ($\ell =
e,\mu$) $t \bar t$ pairs and assuming $m_t=175$~GeV the following
1-$\sigma$ limits on ${\mIm}d_t^g$ were obtained through the observables
$A_E$ and $Q_{33}$

\bea
A_E ~&:&~~~ |{\mIm}d_t^g|=8.58 \times
10^{-19}~~g_s{\rm -cm} \label{aepptt}~,\\
Q_{33} ~&:&~~~ |{\mIm}d_t^g|=2.05
\times 10^{-18}~~g_s{\rm -cm} \label{q33pptt}~.
\eea

\ni Thus the limits on the imaginary part of the top CEDM
are weaker by about an order of magnitude than those that might be obtained
on the real part of the top CEDM, using the optimal
observables discussed before.


\subsubsection{Polarized proton beams \label{sssec723}}

A very interesting CP-violating polarization rate  asymmetry was
originally suggested by Gunion {\it et al.} in \cite{prl71p488}, for
Higgs production through $gg$ fusion in a $pp$ collider. This asymmetry
was applied to $pp \to t \bar t +X$ in \cite{plb415p67}. The basic idea
is that, if the gluons in a polarized proton are polarized, then the
initial CP-odd gluon-gluon configuration allows to probe CP-violating
effects without 
requiring
full reconstruction of the $t \bar t$ final state. The
polarization rate asymmetry is defined as

\be
A_{pr} \equiv \frac{\sigma_+ - \sigma_-}{\sigma_+ + \sigma_-} ~,
\ee

\ni where $\sigma_{\pm}$, in the subprocess $gg \to t \bar t$, is the
cross-section for $t \bar t$ production in collisions of an unpolarized
proton with a proton of helicity $\pm$. Clearly, $A_{pr}$ is CP-odd and
$T_N$-even and therefore can only probe the imaginary part of the top
CEDM. Of course, a crucial point for such an analysis is the degree of
polarization that can be achieved for gluons in the $pp$ collider. The
amount of gluon polarization in a positively polarized proton beam is
defined by the structure functions difference $\Delta g(x) = g_+(x)
-g_-(x)$. The structure functions of polarized gluons, $g_{\pm}$, are
not well known and depend on the amount of the proton's spin carried by
the gluons. In \cite{plb415p67} the following parameterization was
adopted ($g$ is the unpolarized gluon distribution)

\be
\Delta g(x)=\left\{\begin{array}{ll}
             g(x)       & (x > x_c)\\
            (x/x_c)g(x) & (x < x_c)
                   \end{array}\right.  ,
\ee

\ni where $x_c \sim 0.2$ yields a value of $\Delta g \sim 2.5$ at
$Q^2=10~{\rm GeV}^2$. The above distribution was actually evaluated at
$Q^2=100~{\rm GeV}^2$ disregarding any scale evolution.

\begin{table}[htb]
\centering
\caption{{\emph The number of $t \bar t$ events $N$, the ratio
$\Delta\hat{N}/N$ (see text), 
and the attainable 1-$\sigma$
limits on $|{\mIm}d_t^g|$, for various $p_{_T}$-cuts
with $\sqrt{s}=14$ ${\rm TeV}$, $m_t = 175$
${\rm GeV}$ and ${\cal L}=10$ ${\rm fb}^{-1}$.
Table taken from \cite{plb415p67}.
\label{tab1pptt}}}

\vskip 0.5cm

\begin{tabular}{c|ccc}
\hline\hline
 $p_{T}$-cuts(GeV) & $N(\times 10^6)$ & $\Delta\hat{N}/N$
          & $|\Im{\rm m}d_t^g| (\times 10^{-20}g_s{\rm -cm})$\\ \hline
        0   &  2.62 & 1.44 &  0.766 \\
        20  &  2.55 & 1.42 &  0.788 \\
        40  &  2.36 & 1.37 &  0.847 \\
        60  &  2.08 & 1.30 &  0.951 \\
        80  &  1.74 & 1.22 &  1.107 \\
        100 &  1.41 & 1.14 &  1.313 \\ \hline\hline
\end{tabular}
\end{table}

The 1-$\sigma$ attainable limits on $\left|{\mIm}d_t^g\right|$
were calculated in \cite{plb415p67} and are given
in Table \ref{tab1pptt}, for various transverse-momentum
cuts and for $\sqrt{s}=14$ TeV,
$m_t = 175$ GeV, ${\cal L}=10$ ${\rm fb}^{-1}$
and an efficiency acceptance of 10\%.
Also, in Table \ref{tab1pptt} $N=N_+ + N_-$ is the total
number of $t \bar t$ events, $\Delta\hat{N}=N_+ - N_-$ and $N_+(N_-)$
is the number of $t \bar t$ events predicted for
positively(negatively) polarized proton. They have
included all possible $t$ decay modes so that the net
branching ratio was taken as unity. We see that, even
with high $p_T$-cuts, it is possible to put a 1-$\sigma$
limit on $|{\mIm}(d_t^g)|$ up to the order of $10^{-20}$
$g_s$-cm in the LHC with polarized incoming protons.
This limit is more stringent than the ones obtained
in Eqs.~\ref{aepptt} and \ref{q33pptt} through the leptonic
correlations $A_E$ and $Q_{33}$, respectively, and is of
the same order as that obtained on the real part of the top CEDM
with the optimal observables discussed before. \\


\begin{figure}
\psfull
 \begin{center}
  \leavevmode
\epsfig{file=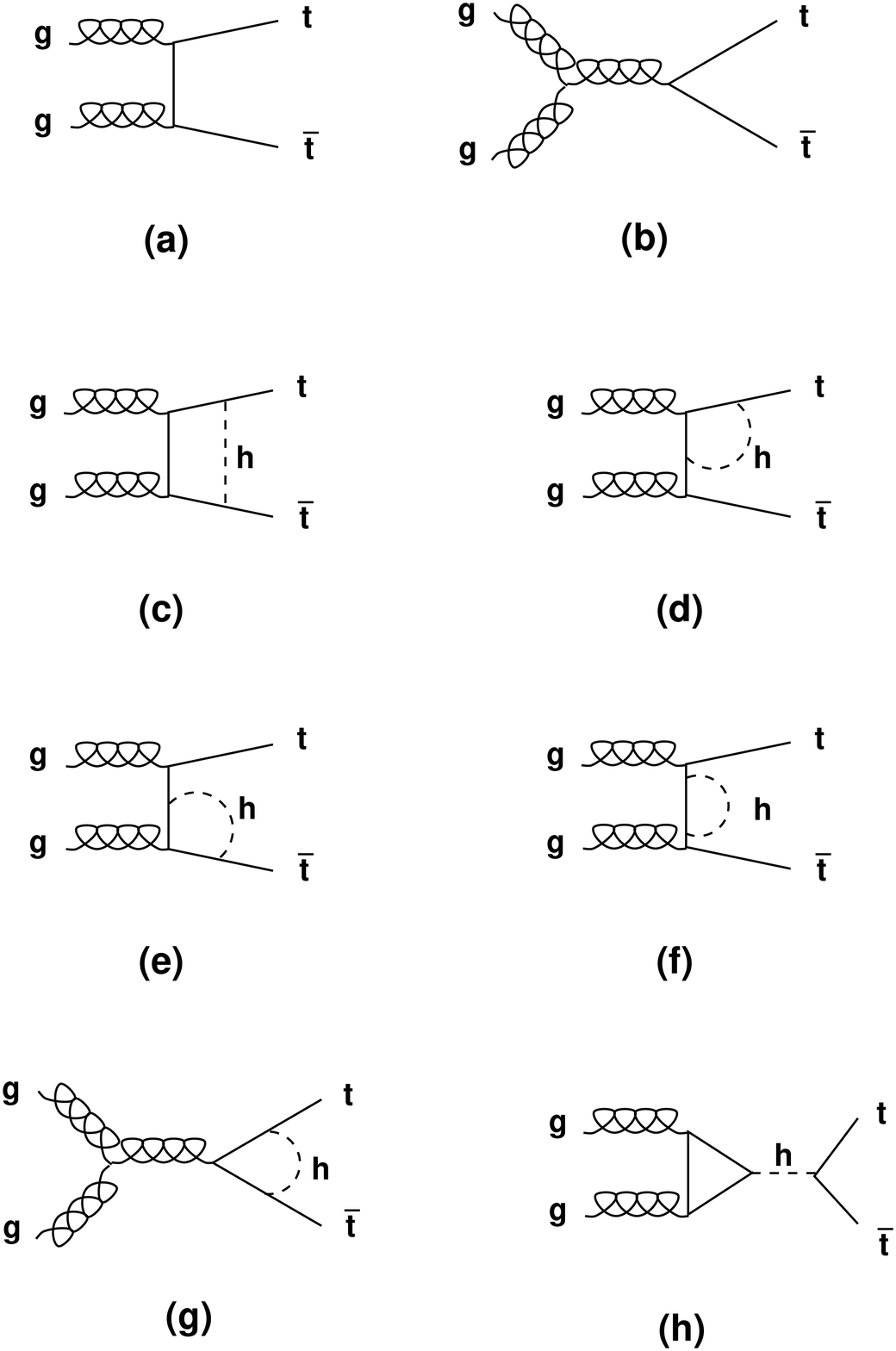
,height=12cm}
 \end{center}
\caption{
{\emph 
Feynman diagrams
for the tree
level QCD and neutral Higgs exchanges
(denoted by the dashed lines) 
which contribute to the production density
matrix for $gg \to t \bar t$. Diagrams with
crossed gluons are not shown.}}
\label{ppttfig2}
\end{figure}

\newpage


\subsection{2HDM and CP violation in $pp \to t \bar t +X$ \label{ssec73}}

\begin{figure}
\psfull
 \begin{center}
  \leavevmode
  \epsfig{file=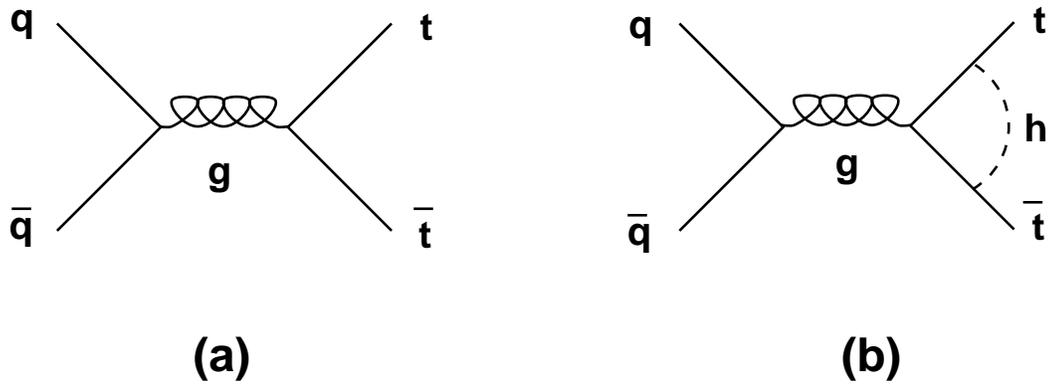
,height=5cm}
 \end{center}
\caption{{\emph Born level QCD and relevant neutral Higgs exchange
(denoted by the dashed line) Feynman diagrams
for $q \bar q \to t \bar t$.}}
\label{ppttfig3}
\end{figure}

\ni In a 2HDM with the CP-violating $t \bar t {\cal H}$ couplings in
Eq.~\ref{2hdmab}, neutral Higgs exchanges can give rise to CP violation in
$gg \to t \bar t$ and $q \bar q \to t \bar t$ at the 1-loop order in
perturbation theory.  In Fig.~\ref{ppttfig2}(c)-(h) all possible 1-loop
CP-violating Higgs exchanges in $gg \to t \bar t$ are drawn and in
Fig.~\ref{ppttfig3}(b) the only CP-violating 1-loop diagram for $q \bar q
\to t \bar t$ is shown. Interference of diagrams (c)-(h) with the SM
tree-level diagrams (a) and (b) in Fig.~\ref{ppttfig2} and interference of
diagram (b) with diagram (a) in Fig.~\ref{ppttfig3} can then give rise to
CP non-conservation effects in $gg$ and $q \bar q$ fusion, respectively.
One can then identify various CP-violating correlations to trace the
resulting CP-odd quantities which appear in the corresponding differential
cross-sections.

Here also we assume that two out of the three neutral Higgs particles in
the 2HDM model are very heavy or have very small CP-violating couplings,
such that either way their effects decouple. Thus, only the couplings of
the lightest neutral Higgs (denoted by $h$) are important and there will
be only one dimensionless CP-odd quantity relevant for the study of CP
violation in $q \bar q, gg \to t \bar t$. Using the notation in
\cite{plb314p104,prd49p4481,hepph9812387} in conjunction with our
parameterization in Eq.~\ref{2hdmab}, this quantity is

\begin{equation}
\gamma_{CP} \equiv - 2 a_t^h b_t^h \label{gcp} ~,
\end{equation}

\noindent where $a_t^h$ and $b_t^h$ are
defined by the $t\bar t h $ (say $h={\cal H}^1$) coupling in 
Eq.~\ref{2hdmab} and are functions of
$\tan\beta$ - the ratio between the two VEV's in the Higgs potential and
of the three Euler angles which parameterize the Higgs mixing matrix
(for details see section \ref{sssec323}).

Below we present two very interesting approaches of probing CP violation in
$pp \to t \bar t +X$. The first is the Schmidt and Peskin (SP) approach
\cite{prl69p410}, which utilizes the distribution of the
leptonic decay products of the top. 
The second is the Bernreuther and
Brandenburg (BB) approach \cite{plb314p104,prd49p4481,hepph9812387}, 
which studies
the CP-violating effect in the resonant $\hat s$-channel Higgs shown in
Fig.~\ref{ppttfig2}(h).

\newpage
~

\newpage


\subsubsection{Schmidt-Peskin signal \label{sssec731}}

Schmidt-Peskin (SP) proposed  \cite{prl69p410} a signature for CP
violation in production and decays of $t\bar t$ pairs for hadron
colliders, namely via the reaction

\be
pp(\bar p) \to \dra{t}{\dra{W^+}{\ell^+\nu_\ell}\hbox{\hspace{-.45in}}b}
\hbox{\hspace{-.48in}}
\ura{\bar t}{\ulra{W^-}{\ell^- \bar\nu_\ell}\hbox{\hspace{-.5in}}\bar b}
\hbox{\hspace{-.5in}}+X  \label{sp1} ~.
\ee

\ni Despite the complexity of the reaction and the hadronic
environment, the signal for CP violation that they suggest,
i.e., the lepton energy asymmetry

\be
\alpha_E = \frac{\langle E_{\ell^+} \rangle -
\langle E_{\ell^-} \rangle}{\langle
E_{\ell^+} \rangle + \langle E_{\ell^-} \rangle} \label{sp2} ~.
\ee

\ni is very simple and robust.
Such an asymmetry can only arise from non-SM sources such as
an extended Higgs sector or supersymmetry. The size of the asymmetry is
unfortunately rather small $\sim10^{-3}$.

Since this asymmetry is CP-odd and $T_N$-even, it requires an
absorptive part to the Feynman amplitude. Such an absorptive part
is already present
(see Figs.~\ref{ppttfig2}(c),(g),(h) and \ref{ppttfig3}(b))
as $t\bar t$ pair production requires the kinematic threshold

\be
\hat s > 4m^2_t \label{sp3}~,
\ee

\ni where $\hat s$ is the square of the energy in the subprocess
$q\bar q$ or $gg$ c.m.\ frame. In particular,  when a neutral Higgs
exchange leads to the CP-violating phase (as in their study), then the
absorptive part due to the threshold condition in Eq.~\ref{sp3}
arises even if $m_{h}^2 > \hat s$.

Note again that for the subprocess $q\bar q\to t\bar t$,
the underlying cause of CP violation in extended Higgs models 
is the CEDM of the top quark.
Of course given the extremely short life time of the top
quark ($\sim10^{-24}$ sec) the CEDM as such (i.e., at $q^2=0$) is
extremely difficult to be seen. Consider, however,
the asymmetry \cite{prl69p410}:

\be
\Delta N_{LR} = \frac{[N(t_L\bar t_L) - N(t_R \bar t_R)]}{\hbox{all\ }
t\bar t} \label{sp6} ~,
\ee

\ni where $N(t_L\bar t_L)$ is the number 
of $t_L\bar t_L$ pairs
produced via $q\bar q (gg) \to t\bar t$, etc. \footnote{Note that in this 
notation,
$\bar t_L$ means an anti-top quark with momentum, for instance, along
$+z$ and spin along $-z$}. 
Clearly $\Delta N_{LR}$ is 
CP-odd and $T_N$-even. The $q\bar q$ contribution to $\Delta N_{LR}$
arises from interference of Fig.~\ref{ppttfig3}(b)
with the lowest order graph for $q\bar q\to t\bar t$ depicted in
Fig.~\ref{ppttfig3}(a). They found

\be
\Delta N_{LR} = \frac{2\beta}{3-\beta^2} {\eRe}(F_{2A})
\label{sp7} ~,
\ee

\ni where $\beta= (1-4m_t^2/ \hat s)^{1/2}$ and, in their notation, 
$F_{2A}(\hat s)$ is the
CEDM form factor and ${\eRe}(F_{2A})$
involves the absorptive part of the Feynman integral

\beqa
{\eRe}(F_{2A}) & = & \frac{1}{8\pi} \left( \frac{m_t}{v} \right)^2
 \frac{4m^2_t}{\hat s \beta} \gamma_{CP} \nonumber \\
& & \times \left[ 1-\frac{m^2_h}{\hat s \beta^2} \ell n \left( 1+
\frac{\hat s \beta^2}{m^2_h} \right) \right] \label{sp8} ~.
\eeqa

\ni Here $m_h$ is the mass of the lightest neutral Higgs and $\gamma_{CP}$
is defined in Eq.~\ref{gcp}. 
It is easy to understand  \cite{prl69p410}
the effect intuitively:  for ${\hat s} >>  m_t^2$, the gluon
will predominantly  couple to  $t_L{\bar t}_R$ or  $t_R{\bar t}_L$.
However, when $\beta \to 0$,  $t_L{\bar t}_L$ and  $t_R{\bar t}_R$,
which are related to each  other via CP, are dominantly
produced,  which may thus lead to $\Delta N_{LR} \ne 0$.
The resulting asymmetry at the parton level, $\Delta N_{LR}$,
for the subprocess $q \bar q \to t \bar t$ for $m_t=175$ GeV,
$\gamma_{CP}=1 / \sqrt 2$ and for different values of $m_h$ and
$\sqrt {\hat s}$, is found to be of order $10^{-3}$.

For $gg\to t\bar t$ the calculation is more involved.  In particular, in
addition to the $\hat t$-channel $h$ exchange, now an $\hat s$-channel
Higgs exchange graph is also present (see Fig.~\ref{ppttfig2}(h)). There
is in fact constructive interference between these two channels for
$m_h<2m_t$. The result for the asymmetry in the $gg$ fusion case, but
without the $\hat s$-channel Higgs exchange (see Fig.~\ref{ppttfig3}(b)),
was also given in \cite{prl69p410}.  Near threshold, i.e., $\sqrt {\hat s}
\gsim 2m_t$, the asymmetry in the $gg$ fusion case is about twice as big
as that of the $q \bar q$ fusion case. However, although larger than the
$q \bar q$ fusion subprocess, it is again at the level of $10^{-3}$.
Adding the $q{\bar q}$ and $gg$ subprocesses, then $\Delta N_{LR}$ can
reach optimistically $\sim 10^{-2}$, for low values of $m_h$ and
$\tan\beta$. In any case, the $gg$ initial state gives rise to a much
richer possibility of CP-violating operators and, as was mentioned before,
the resulting asymmetry cannot be attributed merely to the CEDM of the top
quark. Indeed, as noted in \cite{plb314p104,prd49p4481,hepph9812387}, the
$\hat s$-channel neutral Higgs exchange that was ignored in
\cite{prl69p410}, can give rise to larger asymmetries in $gg \to t \bar t$
and may be attainable at the LHC. We will return to this effect in the
next section.

As has been emphasized at several places in this review, the fact that
top decays are a powerful spin analyzer comes in extremely 
handy here too in leading to a detectable signature. The CP violation in the
production process causes the polarization asymmetry above, which leads
to an asymmetry in the energies of the charged leptons emerging from
$t$ and $\bar t$ decays.

The distribution of the charged lepton in the $t$-rest frame is given
by

\be
\frac{d^2\Gamma}{dE_\ell d \cos\psi} = \frac{d\Gamma}{dE_\ell}\; \frac{1
+\cos\psi}{2} \label{sp10} ~,
\ee

\ni where $\psi$ is the angle between the top spin and the lepton
momentum. When the top quark is boosted to the $q\bar q (gg)$ c.m.\ frame,
Eq.~\ref{sp10} provides the correlation between the helicity of the
top and the energy of the decay lepton. The resulting energy spectrum
for $t_L(\bar t_R)$ and $t_R(\bar t_L)$ is
significantly different from each other as was shown in \cite{prl69p410}.
Clearly, their findings indicate that the
energy spectrum of the leptons, serves as a useful spin analyzer.

The asymmetry in $pp$ collision is calculated by folding in as usual
the parton distributions. For this purpose SP used the
parton density functions proposed in \cite{zpc39p21}.
The effects of the
longitudinal boost of the parton-parton collision are eliminated by
considering the transverse energy ($E_T$) of the leptons. The resulting
asymmetry is \cite{prl69p410}:

\be
\Delta N(E_T) = \frac{d\sigma/dE_{T,\ell^+} -
d\sigma/dE_{T,\ell^-}}{d\sigma /dE_{T,\ell^+} + d\sigma/dE_{T,\ell^-}}
\label{sp11}~,
\ee

\ni and it was calculated in \cite{prl69p410} for
$m_h=100$ GeV, $m_t=150$ GeV and $\gamma_{CP}= 1/ \sqrt 2$.
Unfortunately, numerically it is again only of order $10^{-3}$.

Let us briefly discuss the background for these type of 
CP-violating asymmetries
in $pp \to t \bar t+X$. As was mentioned at the beginning of this chapter,
the initial state ($pp$) at hadron Supercolliders, such as the LHC, 
 is not an eigenstate of CP. Consequently energy asymmetry in the
decay lepton spectrum are not necessarily due to CP violation. The
point is that the protons in the initial state produce more energetic
quarks than anti-quarks. Also the reaction $q\bar q\to t\bar t$ has a
small forward-backward asymmetry induced by $\alpha_s$ corrections.
Thus the top quarks  produced by this reaction tend to have a
slightly higher
energy than $\bar t$, leading to an asymmetry in the energy of the decay
lepton. Such an effect, originating from higher order QCD corrections,
causes an irreducible background.

Fortunately this background is very small. First of all, $q\bar q$
annihilation is subdominant at such $pp$ collider energies 
and the leading reaction
$gg\to t\bar t$ is free from such a forward-backward  
asymmetry. Also, 
as mentioned before,
the background
to the asymmetry arises from higher order (QCD) radiative corrections.
Furthermore, since the forward-backward asymmetry mainly affects longitudinal
variables, its effect on the transverse energy asymmetry
in Eq.~\ref{sp11} would cancel if there were no lepton acceptance cuts.

This background can be crudely estimated from the electromagnetic
analog of the forward-backward asymmetry for $e^+e^-\to\mu^+\mu^-$. The
analogous asymmetry is crudely estimated by the replacement $\alpha \to
[(d^{abc})^2 / 32]\alpha_s=(5/12)\alpha_s$. SP in \cite{prl69p410} 
used the approximate
formula in \cite{plb43p403} which allows them to get an estimate for
massless $t\bar t$ pairs. This approximation tends to overestimate this
background. For numerical estimates SP also impose a cutoff on the gluon
energy of $\Delta E/E=0.3$. The resulting background was found to be
of the order of $10^{-4}$. Therefore, it is much smaller than the
desired CP-violating effect and also it is essentially independent of
the lepton energy.

\subsubsection{${\hat s}$-channel resonance Higgs effects -
Bernreuther-Brandenberg approach
\label{sssec732}}

For $m_{h}>2m_t$, as noted before, there is an interesting $\hat
s$-channel Higgs contribution to $gg\to t\bar t$ shown by diagram (h) in
Fig.~\ref{ppttfig2}. This was explored in some detail in
\cite{plb314p104,prd49p4481} who recently improved their analysis in
\cite{hepph9812387}.  For the simple on-shell decay $h\to t\bar t$, a
large $t\bar t$ spin-spin correlation can be induced already at the
tree-level if $h$ is not a CP eigenstate, as happens in a class of 
2HDM's.\footnote{A more detailed analysis of the possible spin-spin
correlations in $h \to t \bar t$ is given in section \ref{sssec623}.} In a
2HDM, with the $t \bar t h$ coupling of Eq.~\ref{2hdmab} 
\footnote{recall that the
lightest neutral Higgs is assumed to be $h={\cal H}^1$, i.e., $k=1$ in
Eq.~\ref{2hdmab}}, this spin-spin correlation is given by
\cite{plb314p104}: 

\be
\langle \hat k_t \cdot (\vec s_t \times \vec s_{\bar t} ) \rangle =
\frac{\gamma_{CP} \beta_t}{(b_t^h)^2 + (a_t^h)^2\beta^2_t} \label{bb1} ~,
\ee

\ni where $\beta_t=(1-4m^2_t/m^2_h)^{1/2}$, $s_t,s_{\bar t}$ are the spin
operators of $t$ and $\bar t$, respectively, $\hat k_t$ is the unit vector
of the momentum of the top quark and $\gamma_{CP}$ is defined in
Eq.~\ref{gcp}. It is remarkable that this CP-violating spin-spin
correlation can, in principle, be as large as 0.5. In practice, though,
this decay has to be coupled to some particular production process and the
final asymmetry can vary significantly between different processes.
Moreover, for $pp$ collisions, there is an interference between the
continuum and the resonant $t\bar t$ production which tends to diminish
the spin-spin correlation. For the gluon-gluon fusion, the CP-violating
expectation value of $\langle \hat k\cdot (\vec s_t\times \vec s_{\bar t}
)\rangle$ was calculated in \cite{prd49p4481}.  The resulting asymmetry
was found to be at best only a few percent and falls significantly short
compared to the value of 0.5 mentioned above. In fact, when this is
translated to an asymmetry that utilizes the $t$ and $\bar t$ decay
products, as was done in \cite{plb314p104,prd49p4481}, the signal to noise
ratio for such an asymmetry was found to be at best $\sim 10^{-3}$.

The same non-vanishing spin-spin correlation of Eq.~\ref{bb1} can arise in
$q \bar q \to t \bar t$. The asymmetry for the $q \bar q$ fusion
subprocess was also calculated in \cite{prd49p4481} for the same set of
parameters as in the $gg$ fusion case. As expected, in the case of $q \bar
q$ fusion, the asymmetry is about one order of magnitude smaller than $gg$
fusion, since in this channel the resonant Higgs graph is absent.
Furthermore, the asymmetry gets smaller with growing Higgs-boson masses as
opposed to the $gg$ fusion case which we now discuss in some detail.

Let us now focus on an improved analysis of the results mentioned above. 
This was recently suggested by Bernreuther, Brandenburg and Flesch (BBF)
in \cite{hepph9812387}.  In their analysis the basic idea was to include
new cuts on the $t \bar t$ invariant mass which significantly improved
their previous results in \cite{prd49p4481}. 

For the case when both $t$ and $\bar t$ 
decay leptonically, consider the CP-violating 
observables \cite{hepph9812387}:

\bea
Q_{1} & = & \hat k_t \cdot \hat q_+ - \hat k_{\bar t} \cdot \hat q_-
\label{qabs} ~,
\eea
and
\bea
Q_{2} & = & (\hat k_t -  \hat k_{\bar t} )\cdot 
(\hat q_- \times \hat q_+)/2
\label{qdisp} ~,
\eea

\noindent where $\hat k_t,~\hat k_{\bar t}$ are the $t,~\bar{t}$ 
momentum 
directions in the parton c.m. system
and $\hat q_-$, $\hat q_+$ are the $\ell^+$, $\ell^-$ momentum directions
 (from $t \to b \ell^+ \nu_\ell$ and $\bar t \to \bar b \ell^- \bar\nu_\ell$) 
in the $t$ and  $\bar{t}$ rest frames, respectively.
Note that the decay channels to $\ell^+,~ \ell^-$ (disregarding the 
$\tau$ lepton) may include all different combinations of $e$ and $\mu$ and, 
in \cite{hepph9812387}, all possible combinations 
were summed over.
It is useful to note that $Q_2$ in Eq.~\ref{qdisp} is in fact a 
transcription of the spin-spin correlation $\hat k_t \cdot (\vec s_t \times 
\vec s_{\bar t})$ defined in Eq.~\ref{bb1} above, and is a $T_N$-odd 
quantity. Also, 
$Q_1$ traces the spin-spin correlation $\hat k_t \cdot (\vec s_t - 
\vec s_{\bar t})$ which corresponds to the CP asymmetry 
$\Delta N_{LR}$ defined in Eq.~\ref{sp6} 
and was suggested originally in \cite{prl69p410}. Clearly 
it is $T_N$-even, thus requiring an absorptive phase.

For the leptonic-hadronic decay channel of the $t \bar t$ pairs, 
it is useful to consider the two possible decay scenarios: sample ${\cal A}$ 
in which the $t$ decays leptonically and $\bar t$ decays hadronically 
, and 
sample $\bar {{\cal A}}$ 
in which the $t$ decays hadronically and $\bar t$ decays leptonically 
(see Eqs.~\ref{eetthzeq31} and \ref{eetthzeq32}, respectively,
in section \ref{sssec623}). 
One can then define the following CP-violating quantities  
with respect to samples ${\cal A}$ and $\bar {{\cal A}}$ \cite{hepph9812387}:

\bea
{\cal E}_{1} & = & \< {\cal O}_1 \>_{\cal A} - 
\< \bar{\cal O}_1 \>_{\bar {\cal A}}
\label{cale1} ~,
\eea 
\bea
{\cal E}_{2} & = & \< {\cal O}_2 \>_{\cal A} + 
\< \bar{\cal O}_2 \>_{\bar {\cal A}}
\label{cale2} ~,
\eea  

\noindent where

\bea
{\cal O}_1 \equiv \hat k_t \cdot \hat q_+ ~~&,&~~ 
\bar{\cal O}_1 \equiv \hat k_{\bar t} \cdot \hat q_- ~, 
\nonumber\\
{\cal O}_2 \equiv \hat k_t \cdot (\hat q_+ \times \hat q_{\bar b}) ~~&,&~~ 
\bar{\cal O}_2 \equiv \hat k_{\bar t} \cdot (\hat q_- \times \hat q_{b}) ~.
\eea 

\noindent Here $\hat q_b$ and $\hat q_{\bar b}$ denote the momentum 
direction of the $b$ and $\bar b$ jets in the $t$ and $\bar t$ rest frames,
respectively. Again, ${\cal E}_{1}$ effectively 
corresponds to the spin-spin correlation $\hat k_t \cdot (\vec s_t - 
\vec s_{\bar t})$ and is a $T_N$-even observable, and 
${\cal E}_{2}$ traces the spin-spin correlation 
$\hat k_t \cdot (\vec s_t \times \vec s_{\bar t})$ and is therefore a 
$T_N$-odd quantity. 
\begin{figure}
\unitlength1.0cm
\begin{center}
\begin{picture}(13.6,13.)
\put(2,5){\psfig{figure=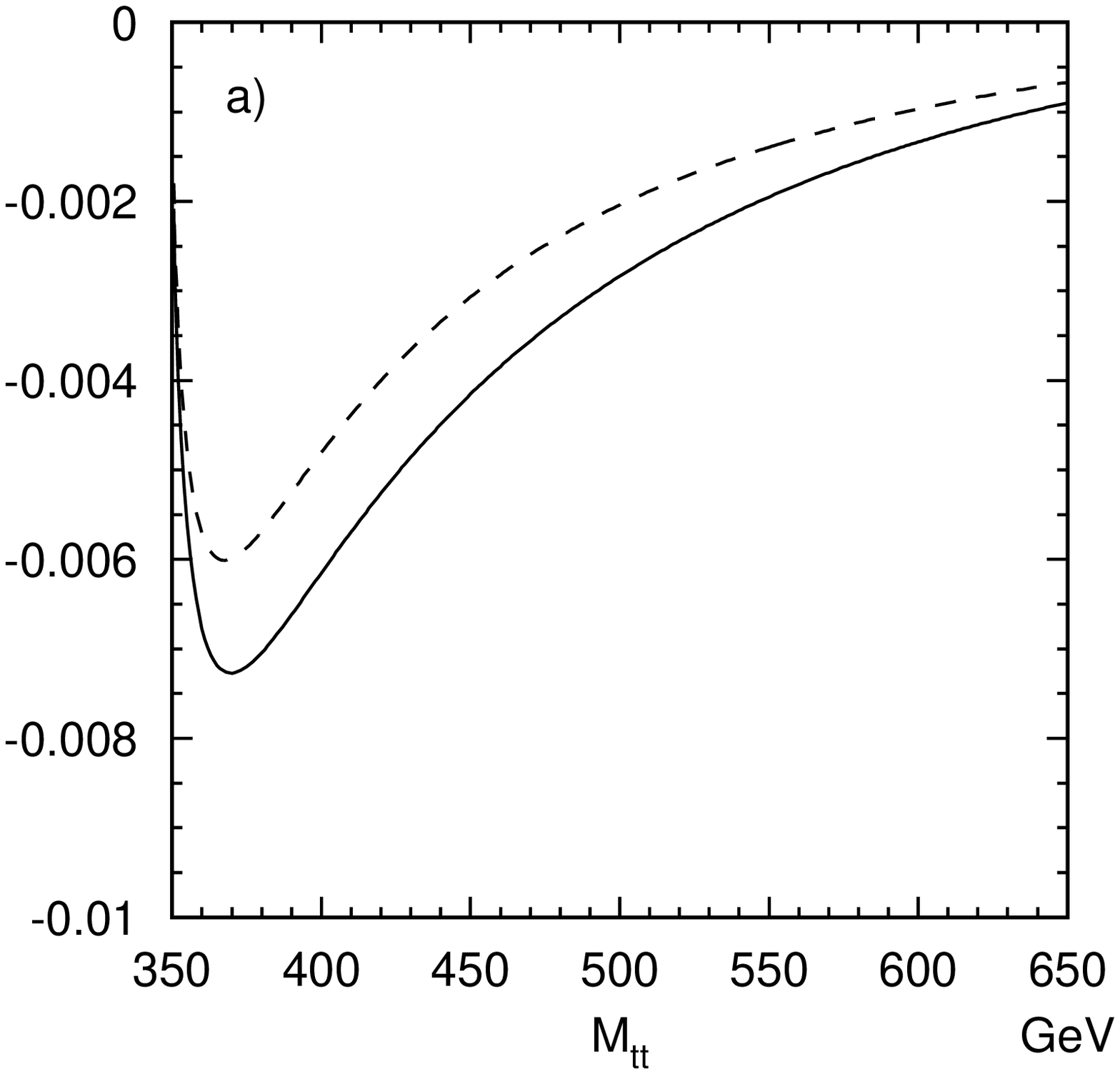,height=5cm,width=5cm}}
\put(7,5){\psfig{figure=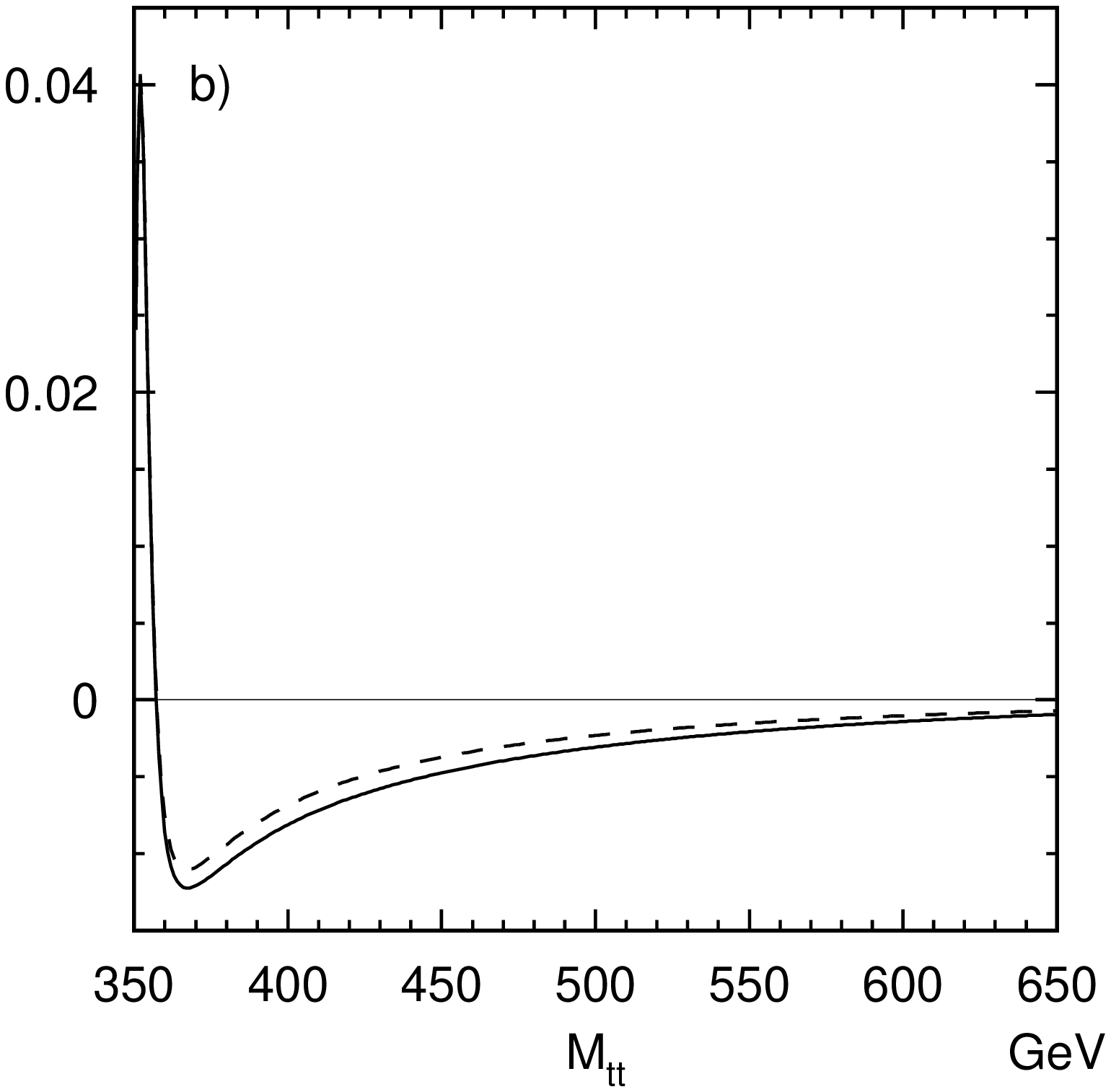,height=5cm,width=5cm}}
\put(2,0){\psfig{figure=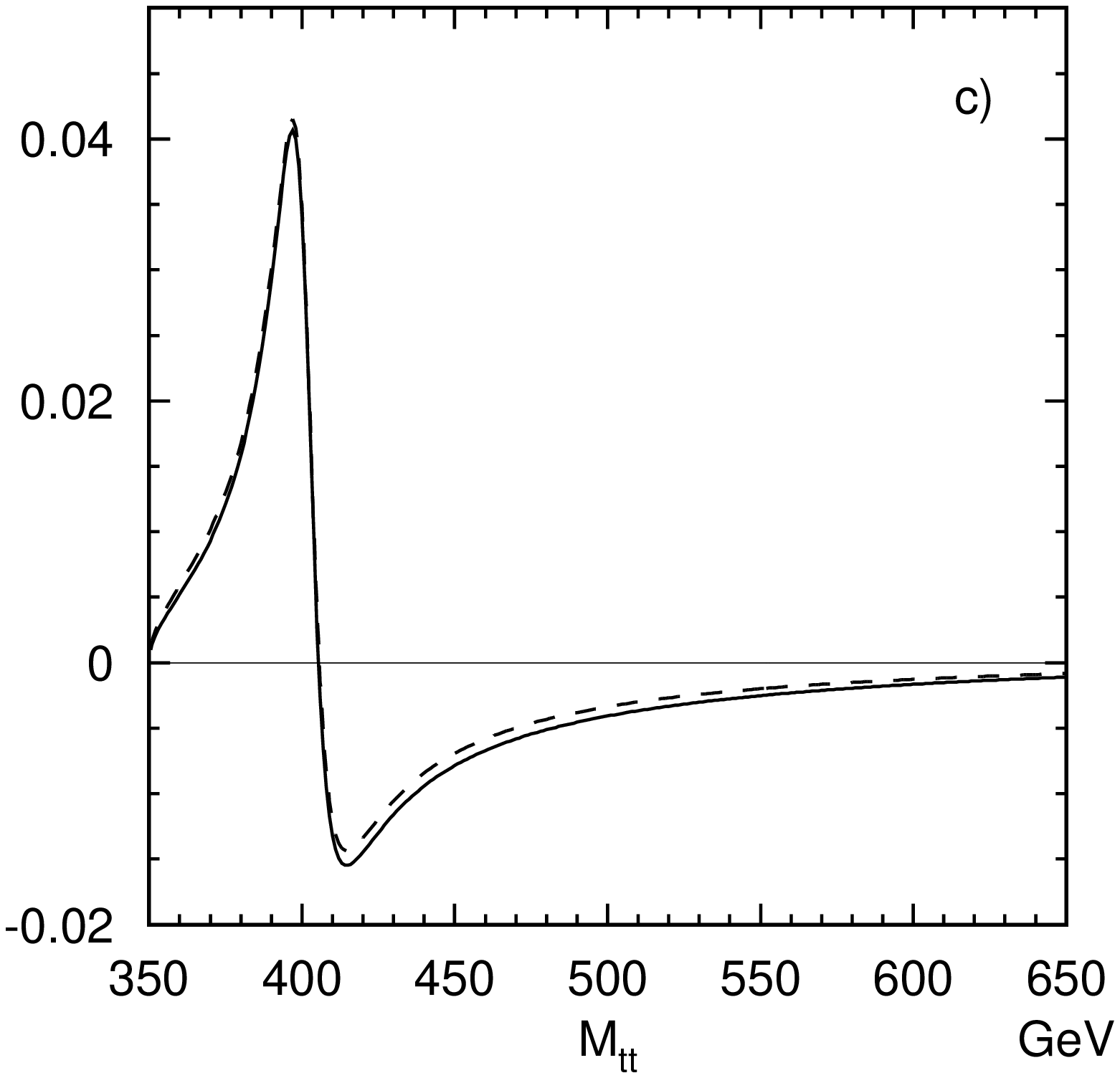,height=5cm,width=5cm}}
\put(7,0){\psfig{figure=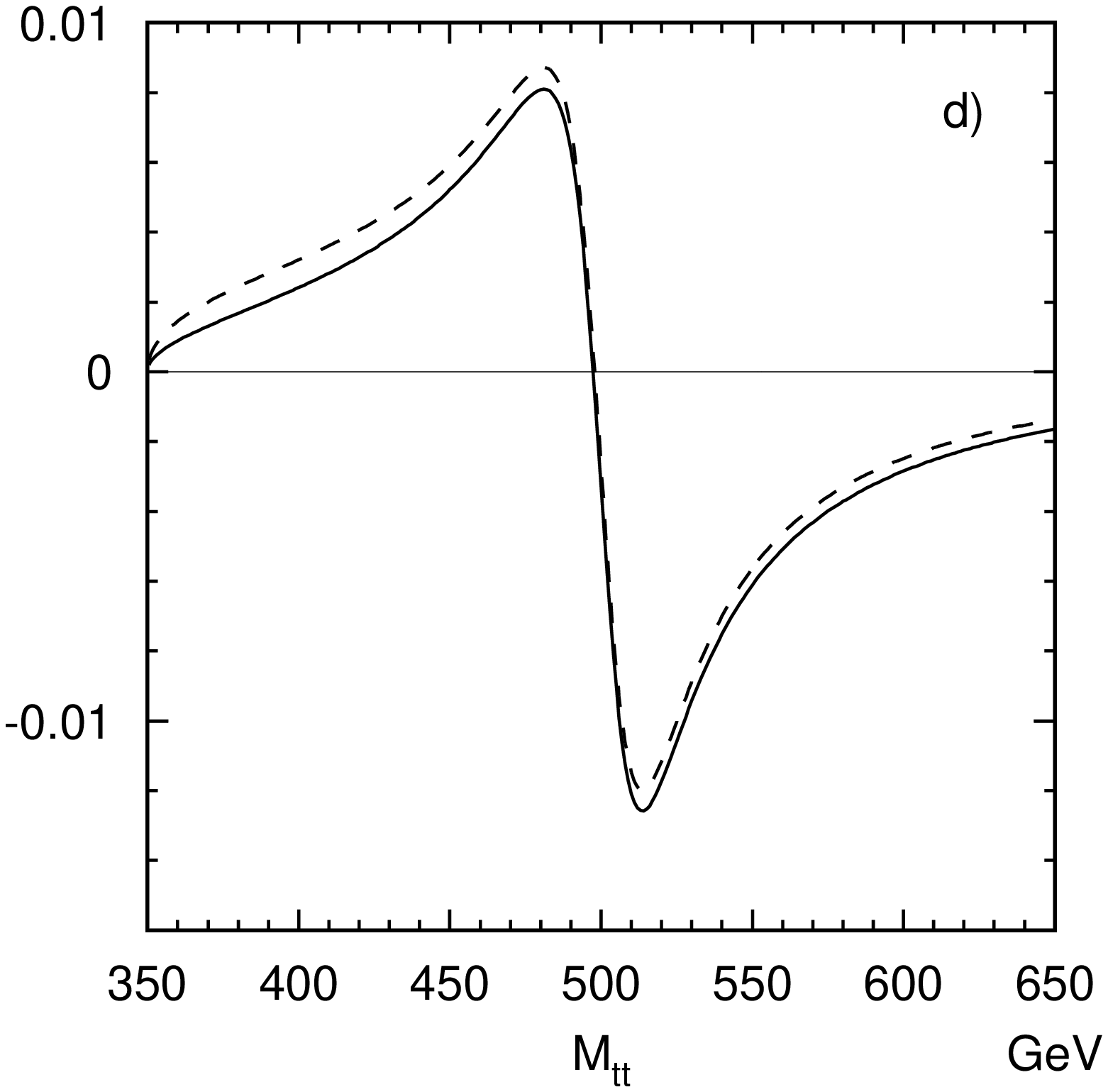,height=5cm,width=5cm}}
\end{picture}
\\[6pt]
\end{center}
\caption{{\emph 
$d<Q_2>/d M_{tt}$ as a function of $M_{tt}$
($Q_2$ is defined in  Eq.~\ref{qdisp}) 
at $\sqrt{s}=14$ TeV, for
reduced Yukawa couplings $c^h=0$, $a^h_t=1/\sqrt 2$, $b_t^h=-1\sqrt 2$, 
i.e., $\gamma_{CP}=1$ (see text), 
and Higgs-boson masses: (a) $m_h=320$ GeV, (b) $350$ GeV,
(c) $400$ GeV, and (d) $500$ GeV, in the dilepton channel. The dashed 
line represents the resonant and the solid line the sum of the
resonant and non-resonant $h$ contributions. Figure taken from 
\cite{hepph9812387}.}}
\label{ppttbbfig2}
\end{figure}

In \cite{hepph9812387} it was shown that, 
in the region $m_h > 2m_t$,
the magnitude of the asymmetry increases significantly and the dominant
contribution comes from the interference of the CP-violating 
terms in the amplitude of the neutral Higgs 
resonant diagram in Fig.~\ref{ppttfig2}(h)
with the Born amplitude. They have evaluated the dependence of 
the differential expectation values of 
$Q_1$ and $Q_2$ on the $t \bar t$ invariant mass, $M_{t \bar t}$.   
An example of such a dependence for the $T_N$-odd observable 
$Q_2$ and in the dilepton decay channels of the $t \bar t$ pairs 
is shown in Fig.~\ref{ppttbbfig2}. In this figure 
the resonant contribution in $gg \to h \to t \bar t$ is compared 
with the resonant $+$ the remaining $h$ contribution 
(i.e., all the non-resonant graphs), for different values of $m_h$, for 
$\sqrt s =14$ TeV and setting the CP-violating quantity, $\gamma_{CP}$, 
from the 
$t \bar t h$ vertex to be equal to 1. 
Also, since the CP-violating effect is sensitive 
to the neutral Higgs total width, it is therefore sensitive to 
the $ZZh,WWh$ ``reduced'' coupling $c^h$ 
(defined 
in Eq.~\ref{2hdmc} for $h={\cal H}^1$, i.e., $k=1$), 
which determines the decay rates 
$\Gamma(h \to ZZ)$ and $\Gamma(h \to WW)$ \cite{hepph9812387}.
In Fig.~\ref{ppttbbfig2}, $c^h=0$ was chosen, in which case the 
above decay channels of a neutral Higgs to the massive gauge-bosons are 
forbidden at tree-level.

Clearly, looking at 
Fig.~\ref{ppttbbfig2}, the non-resonant contributions are negligible 
with respect to the $\hat s$-channel $h$ contribution which in turn 
gives rise to a CP asymmetry at the level of a percent when 
$M_{t \bar t}$ is in the vicinity of $m_h$. 

\newpage
~

\begin{table}[ht]
\begin{center}
\caption{{\emph 
The expectation value of $Q_2$ and its sensitivity at the LHC 
with $\sqrt{s}=14$ TeV and an integrated luminosity of 100 fb$^{-1}$,
for the dilepton $t \bar t$ decay channels. The $M_{t\bar{t}}$ intervals are
chosen below $m_h$ such that: for $m_h=370,~400,~500$ GeV, $\< Q_2 \>$
was integrated over $\Delta M_{t \bar t} = 15,~40,~80$ GeV, in the 
$M_{t\bar{t}}$ ranges
355-370, 360-400 and 420-500~GeV,
respectively. 
For each pair 
($m_h$, $c^h$) the first column is $\<Q_{2}\>$ in percent and the 
second column is the sensitivity in standard deviations.
The rows correspond, in descending 
order, to ($a_t^h,b_t^h$) = $(1,-1)/\sqrt 2$, $(1,-0.3)/\sqrt 2$ 
and $(0.3,-0.3)/\sqrt 2$, i.e., to $\gamma_{CP}=1,~0.3$ and $0.09$, 
respectively.
Numbers for $m_h$ are in GeV. The non-resonant $h$ contributions have been
neglected for these values of $m_h$. Table taken from 
\cite{hepph9812387}.
\label{ppttbbtab4}}}
\vspace{5mm}
\begin{tabular}{|r|c|c|c|c|c|c|}
\hline
 & \multicolumn{6}{c|}{$c^h$}  \\
 $m_h$ & \multicolumn{2}{c|}{0.0} &  \multicolumn{2}{c|}{0.2}&\multicolumn
{2}{c|}{0.4}\\
\hline
\raisebox{-4.6ex}[0pt][0pt]{370} 
  & $ 4.4    $ & $ 29.8   $ & $ 4.1 $ & $ 27.4   $ & $ 3.3    $ & $ 20.9 $  \\ 
  & $ 3.9    $ & $ 23.4   $ & $ 2.9 $ & $ 16.7   $ & $ 1.6    $ & $ 9.0  $  \\ 
  & $ 1.2    $ & $ 6.6    $ & $ 0.75  $& $ 4.1    $ & $ 0.39   $ & $ 2.1  $  \\ 
\hline
\raisebox{-4.6ex}[0pt][0pt]{400} 
  & $ 2.3    $ & $ 24.4   $ & $ 2.1  $ & $ 22.8   $ & $ 1.8    $ & $ 18.7 $  \\ 
  & $ 1.3    $ & $ 13.4   $ & $ 1.1  $ & $ 11.1   $ & $ 0.75   $ & $ 7.5  $  \\ 
  & $ 0.49   $ & $ 4.9    $ & $ 0.35 $ & $ 3.5    $ & $ 0.21   $ & $ 2.1  $  \\ 
\hline
\raisebox{-4.6ex}[0pt][0pt]{500} 
  & $ 0.65   $ & $ 8.6   $ & $ 0.59 $ & $ 7.9    $ & $ 0.46   $ & $ 6.0  $  \\ 
  & $ 0.31   $ & $ 4.1   $ & $ 0.26$ & $ 3.5    $ & $ 0.18   $ & $ 2.4  $  \\ 
  & $ 0.14   $ & $ 1.9   $ & $ 0.10 $ & $ 1.4    $ & $ 0.06   $ & $ 0.77 $  \\ 
\hline
\end{tabular}
\\[6pt]
\end{center}
\end{table}

\newpage

The sharp peaks observed in the range $M_{t \bar t} \sim m_h$ give 
an extra handle in an attempt to enhance the CP signal. 
Indeed, Figs.~\ref{ppttbbfig2}(b), (c) and (d) 
show that, in the case of $m_h > 2m_t$, $\< Q_2 \>$ changes sign 
as one goes from $M_{t \bar t} \lsim m_h$ to $M_{t \bar t} \gsim m_h$, 
such that integrating over $M_{t \bar t}$ will diminish the CP-violating
 effect.
Therefore, choosing 
appropriate $M_{t \bar t}$ mass bins below or above $m_h$, allows for 
a significant enhancement of  
the CP-odd signal. This is demonstrated in Table \ref{ppttbbtab4}, where
the three values $c^h=0,~0.2$ and 0.4 were 
considered (see discussion above).    
Also, in Table \ref{ppttbbtab4},  
the left column gives the expectation value of $Q_2$ in percent and 
the right column shows the statistical significance in which this CP 
effect can be measured at the LHC with an integrated luminosity 
of 100 fb$^{-1}$. 
The $M_{t \bar t}$ intervals (i.e., mass bins) in Table \ref{ppttbbtab4} 
where chosen below $m_h$ (see caption of Table \ref{ppttbbtab4}). Also 
the rows in Table \ref{ppttbbtab4} 
correspond in descending order to $\gamma_{CP}=1,~0.3,~0.09$.

The numbers for the statistical significance of the CP-violating signal 
that were found in \cite{hepph9812387} and are shown in Table \ref{ppttbbtab4} 
 are quite remarkable. In most cases the CP-violating signal is well above 
the 3-$\sigma$ level, perhaps even a lot better than 5-$\sigma$
in the best case. 
As an example, note that with $m_h=370$ GeV and $\gamma_{CP}$ as low as 
$0.09$, 
the observable $Q_2$ can yield a 7-$\sigma$ effect with an appropriately 
chosen interval for  $M_{t \bar t}$. Recall that values as large as 
$\gamma_{CP} \sim 4$, corresponding to $\tan\beta \lsim 0.5$, are still 
allowed by present experimental data (see section \ref{sssec323}).    
 
A few additional remarks are in order regarding the 
analysis in \cite{hepph9812387}: 

\begin{enumerate}
\item The expected statistical significance for a CP-odd signal from  
the observable ${\cal E}_2$ in Eq.~\ref{cale2} 
corresponding to the leptonic-hadronic channel 
(i.e., lepton $+$ jet from the $t \bar t$ pairs) was found to be comparable 
to that of $Q_2$ discussed above. Moreover, for the $T_N$-even observables 
$Q_1$ (see Eq.~\ref{qabs}) and 
the corresponding one for the leptonic-hadronic channel ${\cal E}_1$ 
(see Eq.~\ref{cale1}) 
 the CP-violating signal, although somewhat smaller, may yield more than 
a 3-$\sigma$ effect as long as $\gamma_{CP} \gsim 0.3$.     

\item Apart from the cuts on the $M_{t \bar t}$ invariant mass, i.e., 
the chosen intervals/mass bins, Ref.~\cite{hepph9812387} 
employed additional cuts on the rapidities of the $t$ and $\bar t$ 
and on the transverse momenta of the final state charged leptons and quarks 
in both the dilepton channel and leptonic-hadronic channel samples.   

\item It is important to note that the $T_N$-odd observables $Q_2$ and 
${\cal E}_2$ are insensitive to CP violation from the subsequent $t, \bar t$ 
decays to leading order in the CP-violating couplings. This is ensured by 
$CPT$ invariance \cite{hepph9812387}. Moreover, 
the $T_N$-even observables $Q_1$ and ${\cal E}_1$, although may acquire 
contributions from CP-violating absorptive parts in the $t, \bar t$ decays, 
at least in the 2HDM case, these absorptive parts are absent in the limit 
of vanishing $b$-quark mass (see also related discussion in section 
\ref{sssec512}). Therefore, for the 2HDM case, both the $T_N$-even and the
$T_N$-odd quantities in Eqs.~\ref{qabs}-\ref{cale2} are ``clean'' probes 
of CP violation in the production mechanism of $t \bar t$ at the LHC.

\end{enumerate}

Refs.~\cite{plb314p104,prd49p4481} considered possible contaminations to
an asymmetry of the type $Q_2$ (or equivalently the spin-spin correlation
in Eq.~\ref{bb1}). Again, the key point is that the dominant $gg$ fusion
subprocess is free from undesired CP-conserving background to $Q_2$.
Therefore, background considerations are relevant only for the case of $q
\bar q \to t \bar t$.  Refs.~\cite{plb314p104,prd49p4481} estimated the
CP-conserving background to $Q_2$ to be of order $10^{-6}$ which is about
3 orders of magnitude smaller than the actual asymmetry. The reason for
that is that $T_N$-odd observables such as $Q_2$ do not receive
contributions from CP-invariant interactions at the Born level but only
from absorptive parts. Thus, in the case of $q \bar q \to t \bar t$, the
main background comes from order $\alpha_s^3$ and $\alpha_s^2 \alpha_{\rm
weak}$ absorptive parts \cite{plb314p104,prd49p4481}. 

Finally, let us note that the optimization technique (with no additional
cuts), i.e., the use of optimal observables, employed in
\cite{hepph9805358} for CP violation in $gg \to t \bar t$, yield roughly
the same results as those obtained in \cite{hepph9812387}. That is,
optimal observables can be sensitive to values down to $\gamma_{CP} \sim
0.1$ with no cuts on the $t \bar t$ invariant mass.


\subsection{SUSY and CP violation in $pp \to t \bar t+X$ \label{ssec74}}

\begin{figure}[ht]
\psfull
 \begin{center}
  \leavevmode
  \epsfig{file=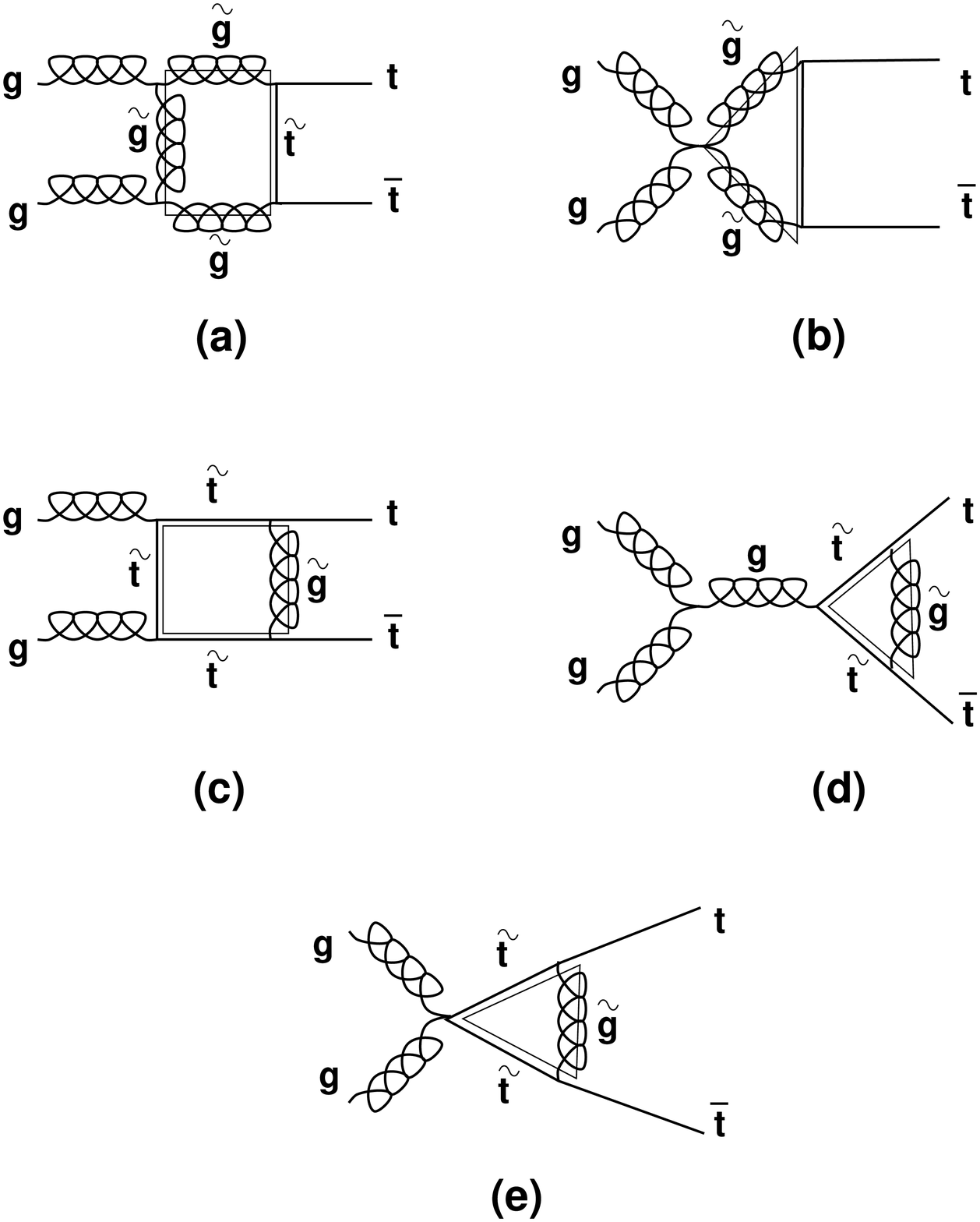
,height=12cm}
 \end{center}
\caption{{\emph 1-loop Feynman diagrams
contributing to CP violation in $gg \to t \bar t$ in supersymmetry.
Diagrams with crossed gluons are not shown.}}
\label{ppttfig4}
\end{figure}

\ni As we have discussed in previous sections,
1-loop exchanges of SUSY particles may give rise to
CP-violating phenomena in top systems which are driven
by SUSY CP-odd phases in the supersymmetric vertices.
Such SUSY CP-violating 1-loop effects in
$gg \to t \bar t$ were investigated by Schmidt \cite{plb293p111}.
In the case of $gg \to t \bar t$, only exchanges of
gluinos and stops are relevant and are shown in
Fig.~\ref{ppttfig4}. The only CP-odd phase arises
then from the off diagonal elements of the
${\tilde t}_L - {\tilde t}_R$ mixing matrix.
Writing again (see also sections \ref{sssec332} and \ref{ssec45}) 
the top squarks of different
handedness in terms of their mass eigenstates,
$\tilde t_+$, $\tilde
t_-$, as

\beqa
\tilde t_L & = & \cos \theta_t \tilde t_- - e^{-i\beta_t} \sin\theta_t
\tilde t_+ ~, \nonumber \\
\tilde t_R & = & e^{i\beta_t} \sin\theta_t \tilde t_- + \cos \theta_t
\tilde t_+ \label{pra10chap7}~,
\eeqa

\ni the asymmetry will then be proportional to
the quantity

\bea
\xi_{CP}^t = \sin(2 \theta_t) \sin(\beta_t) ~.
\eea


\ni Schmidt neglected possible CP-nonconserving effects in the subdominant
process $q \bar q \to t \bar t$ and considered the asymmetry $\Delta
N_{\rm LR}$ defined in Eq.~\ref{sp6} only for $gg \to t \bar t$. As
mentioned before, $\Delta N_{\rm LR}$ being CP-odd and $T_N$-even,
requires both a CP-odd phase and an absorptive phase. Such absorptive
phases are present in diagrams (a), (b) of Fig.~\ref{ppttfig4} if the
c.m.\ energy is large enough to produce on-shell gluino ($\tilde g$) pairs
and in diagrams (c)--(e) in Fig.~\ref{ppttfig4} if the c.m.\ energy of the
colliding gluons is sufficient to produce on-shell top squark ($\tilde t$)
pairs.  Obviously, if SUSY particles have masses of $O(1~TeV)$, then this
condition is satisfied at the LHC in which the c.m. energy of the
colliding protons is 14 TeV. Thus, $\Delta N_{\rm LR}$ is a sum of the two
contributions

\be
\Delta N_{\rm LR} \equiv \frac{\int d\,\cos\theta \left[ A^{\tilde g}
(\cos\theta) + A^{\tilde t}(\cos\theta) \right]}
{({\rm all} ~ t \bar t)}~,
\ee

\ni where $\theta$ is the  production angle
of the top quark in the $gg$ c.m.\ frame and $A^{\tilde g}(A^{\tilde
t})$ is the contribution from on-shell gluino (stop) pairs.
Schmidt found

\bea
A^{\tilde g}(\cos\theta) &=& \frac{9 \pi \alpha_s^2 \beta_t \delta}
{64 \hat s} \sum_{\sigma=1}^2 (-1)^\sigma
\Theta(\sqrt {\hat s} - 2 m_{\tilde g}) \nonumber \\
&\times& \left\{ \frac{1}{1- \beta_t \cos\theta}
\left[ \frac{\beta_t}{\beta_{\tilde g}} (1- \beta_{\tilde g}^2)
K_0^- - (1- \beta_t^2) K_1^- + \beta_t \beta_{\tilde g}
\sin^2 \theta (K_2^- - K_3^-) \right] \right. \nonumber \\
&+& \frac{1}{1+ \beta_t \cos\theta}
\left[ \frac{\beta_t}{\beta_{\tilde g}}
(1- \beta_{\tilde g}^2) K_0^+ - (1- \beta_t^2) K_1^+ \right. \nonumber \\
&+& \left.\left.\beta_t \beta_{\tilde g} \sin^2 \theta (K_2^+
- K_3^+) \right] \right\}~,
\eea

\ni and

\bea
A^{\tilde t}(\cos\theta) &=& \frac{9 \pi \alpha_s^2
\beta_t \delta}{64 \hat s} \sum_{\sigma=1}^2 (-1)^\sigma
\Theta(\sqrt {\hat s} - 2 m_{\sigma}) \nonumber \\
&\times& \left\{ \left( \frac{1/81}{1 - \beta_t \cos\theta}
+ \frac{10/81}{1 + \beta_t \cos\theta} \right)
\left[ \frac{\beta_t}{\beta_\sigma} (1- \beta_\sigma^2)
{\bar K}_0^- - \beta_t \beta_\sigma \sin^2
\theta {\bar K}_3^- \right] \right. \nonumber \\
&+& \left( \frac{1/81}{1 + \beta_t \cos\theta} +
\frac{10/81}{1 - \beta_t \cos\theta} \right)
\left[ \frac{\beta_t}{\beta_\sigma} (1- \beta_\sigma^2)
{\bar K}_0^+ \right. \nonumber \\
&-& \left.\left.\beta_t \beta_\sigma \sin^2
\theta {\bar K}_3^+ \right] \right\} ~,
\eea

\ni where $\beta_i=(1-4m_i^2/\hat s)^{1/2}$ and
the index $\sigma$ refers to the two mass
eigenstates of the stop. Also, here

\be
\delta \equiv \alpha_s \frac{m_{\tilde g} m_t}{\hat s \beta_t} \xi_{CP}^t ~,
\ee

\ni and the form factors $K_i^{\pm}$
($i=0-3$) are given in \cite{plb293p111}.

The asymmetry $\Delta N_{\rm LR}$ was calculated for several values of
$m_{{\tilde t}_-},m_{{\tilde t}_+}$ (the masses of the two stop
eigenstates) and $m_{\tilde g}$, and for $m_t=150$ GeV, $\xi_{CP}^t=-1$
(i.e., its maximal negative value). In general, the asymmetry was found to
be dominated by the amplitudes which contain the intermediate gluinos
(Fig.~\ref{ppttfig4}(a) and \ref{ppttfig4}(b)) even if the intermediate
stops in Fig.~\ref{ppttfig4}(c)-(e) can go on their mass shell. For
example, with $m_{\tilde g}=210$ GeV, $m_{{\tilde t}_-}=100$ GeV and
$m_{{\tilde t}_+}=500$ GeV, Schmidt found that the asymmetry at the parton
({\it i.e.} gluons) level can reach $\sim 2\%$ if the c.m.\ energy of the
colliding gluons is around 450 GeV. Note that, in this c.m.\ energy, and
with the above stops masses, $\Delta N_{\rm LR}$ receives contributions
only from diagrams (a) and (b) in Fig.~\ref{ppttfig4} since there is no
absorptive cut along the stops lines in diagrams (c)--(e) in
Fig.~\ref{ppttfig4}. These results for $\Delta N_{\rm LR}$ are about an
order of magnitude larger than what was found in the 2HDM case
\cite{prl69p410} and it is roughly comparable to the $\hat s$-channel
Higgs resonant effect \cite{plb314p104,prd49p4481,hepph9812387}.

However, in a more realistic study, one will have to integrate over the
structure functions of the incoming gluons and present asymmetries in the
overall reaction $pp \to t \bar t +X$. Unfortunately, folding in the
gluons structure functions, the CP-violating asymmetry drops again to the
level of $10^{-3}$. For this purpose Schmidt again considered the
asymmetry in the transverse energy of the leptons defined in
Eq.~\ref{sp11}. He \cite{plb293p111} found that, similar to the 2HDM case
when no additional cuts are made (like the ones suggested in
\cite{hepph9812387} and were described in the previous 
section), at the
LHC with $\sqrt s =14$ TeV and with typical SUSY masses of a few hundreds
GeV, the asymmetry $\Delta N(E_T)$ of Eq.~\ref{sp11} is again typically of
the order of $\sim {\rm few} \times 10^{-3}$. Schmidt also examined the
non-CP-violating background, and again found it to be negligible compared
to the CP-violating effect.

Clearly, the Schmidt SUSY CP-violating effect in $gg \to t \bar t$ as it
stands, is smaller than that of Ref.~\cite{hepph9812387}, i.e. the signal
caused by the resonant Higgs contribution.  This is mainly due to the
appreciable improvement that can be achieved with appropriate cuts on the
$t \bar t$ invariant mass, as was discussed in the previous section. Note
however, that by using optimal observables for extracting information on
CP violation in $gg \to t \bar t$, it was shown in \cite{hepph9805358}
that the signal to noise ratio for the CP-violating effect in $gg \to t
\bar t$ (driven by the diagrams in Fig.~\ref{ppttfig4} that were
considered by Schmidt) can reach the percent level after folding in the
gluons structure functions.  This allows a 1-$\sigma$ detection of the
CP-odd effect even for smaller values of $\xi_{CP}^t \sim 0.1$.

\newpage
~

\newpage

\pagebreak

\section{CP violation in 
\boldmath ${p \bar p}$ collider experiments \label{sec8}}
\setcounter{equation}{0}

Shortly after the demise of the SSC, it was suggested to upgrade the
energy of the Tevatron. More recently, substantial, two stage, upgrades in
luminosity without a factor of two or so increase in energy remain as
viable options. For a review see \cite{ppbarreview1}. In the previous run
at the Tevatron, at c.m.\ energy of 1.8 TeV, the D$\emptyset$ and CDF
experiments  accumulate  more than 0.2 fb$^{-1}$ of integrated
luminosity. In the first upgrade, called Run II, ${\cal L}$ will be
increased from its current peak value of $2 \times 10^{31}$
cm$^{-2}$sec$^{-1}$ to $10^{32}$ cm$^{-2}$sec$^{-1}$ (or even to twice
this value). In the second stage, so called Run III (or TeV-33), the
luminosity will be further increased to ${\cal L} \approx 10^{33}$
cm$^{-2}$sec$^{-1}$. The working hypotheses are that in Runs II and III
the integrated luminosity will be 2 fb$^{-1}$ and 30 fb$^{-1}$,
respectively, with a modest increase of c.m.\ energy to 2 TeV\null. In
addition, the D$\emptyset$ and CDF detectors 
are also being upgraded.

\subsection{$p \bar p \to t \bar t +X$ \label{ssec81}}

Contrary to the LHC  $pp$ collider, where $t \bar t$ pairs are produced
predominantly through the $gg$ fusion process $gg \to g \to t \bar t$,
in the Tevatron $p \bar p$ collider with c.m.\ energy of $\sqrt s \simeq
2$ TeV the main production mechanism of $t \bar t$ pairs is the $q \bar
q$ fusion, $q \bar q \to g \to t \bar t$. In particular, the $q \bar q$ fusion
process is responsible for about $90\%$ of the cross-section $p \bar p
\to t \bar t +X$. Being so, the processes $p\bar p \to t \bar t, t \bar t
g +X$, where $g$ stands for an extra gluon jet in $t\bar t$ production,
will presumably be sensitive to the CEDM and CMDM of the top quark, 
which can be incorporated as 
effective interactions at the $tt g$ vertex. As already mentioned in
previous chapters, being a CP-odd quantity, a non-vanishing CEDM coupling
might give rise to observable CP violation in top systems in such a hadron
collider. If so, this will be a  clue for new physics, as in the
SM the CP-nonconserving effects in the reactions $p\bar p \to t \bar t,
t \bar t g +X$ are extremely small.

In principle, CP-nonconserving effects due to the CEDM can be searched
for through a study of either CP-even or CP-odd correlations  in the
reactions $p\bar p \to t \bar t, t \bar t g +X$. Of course, CP-odd
correlations are expected to be more sensitive to the CEDM than CP-even
observables as the former are linear in CEDM whereas the latter are
proportional to the square of the CEDM form factor.

We will first present a study of the sensitivity of
some CP-even observables to the CEDM, and then describe some
interesting CP-odd correlations that may be applied
to  $p\bar p \to t \bar t+X$.

\subsubsection{CP-even observables in $p\bar p \to t \bar t +X$ and $p\bar
p \to t \bar t + jet +X$ \label{sssec811}}

The effective Lagrangian for the interaction between the top quark and a
gluon, that includes the CEDM and CMDM form factors of the top, 
is given in Eq.~\ref{eettgeq20} (for $V=g$). Recall that 
the effective interaction at the $ttgg$ vertex, absent in the SM, is also 
required to ensure gauge invariance (see discussion in section \ref{ssec25}).

The CMDM and CEDM dimensionless
form factors $\kappa_g$ and ${\tilde \kappa}_g$, which are 
defined in Eq.~\ref{eettgeq20},  
can develop an imaginary part.\footnote{we will loosely refer here to 
$\kappa_g$ and ${\tilde \kappa}_g$ as the CMDM and CEDM 
form factors, respectively. They are related to the dimensionful 
CMDM and CEDM via Eqs.~\ref{dimct} and \ref{dimdt}.} However, 
these imaginary parts vanish at
zero momentum transfer, 
and are only present if an on-shell intermediate state exists.
Using form factors as a probe for new physics is most useful when the novel
states can only be produced virtually and so here we consider
the case where $\kappa_g$ and ${\tilde \kappa}_g$  
are purely real.

\begin{figure}[htb]
\psfull
\begin{center}
\leavevmode
\epsfig{file=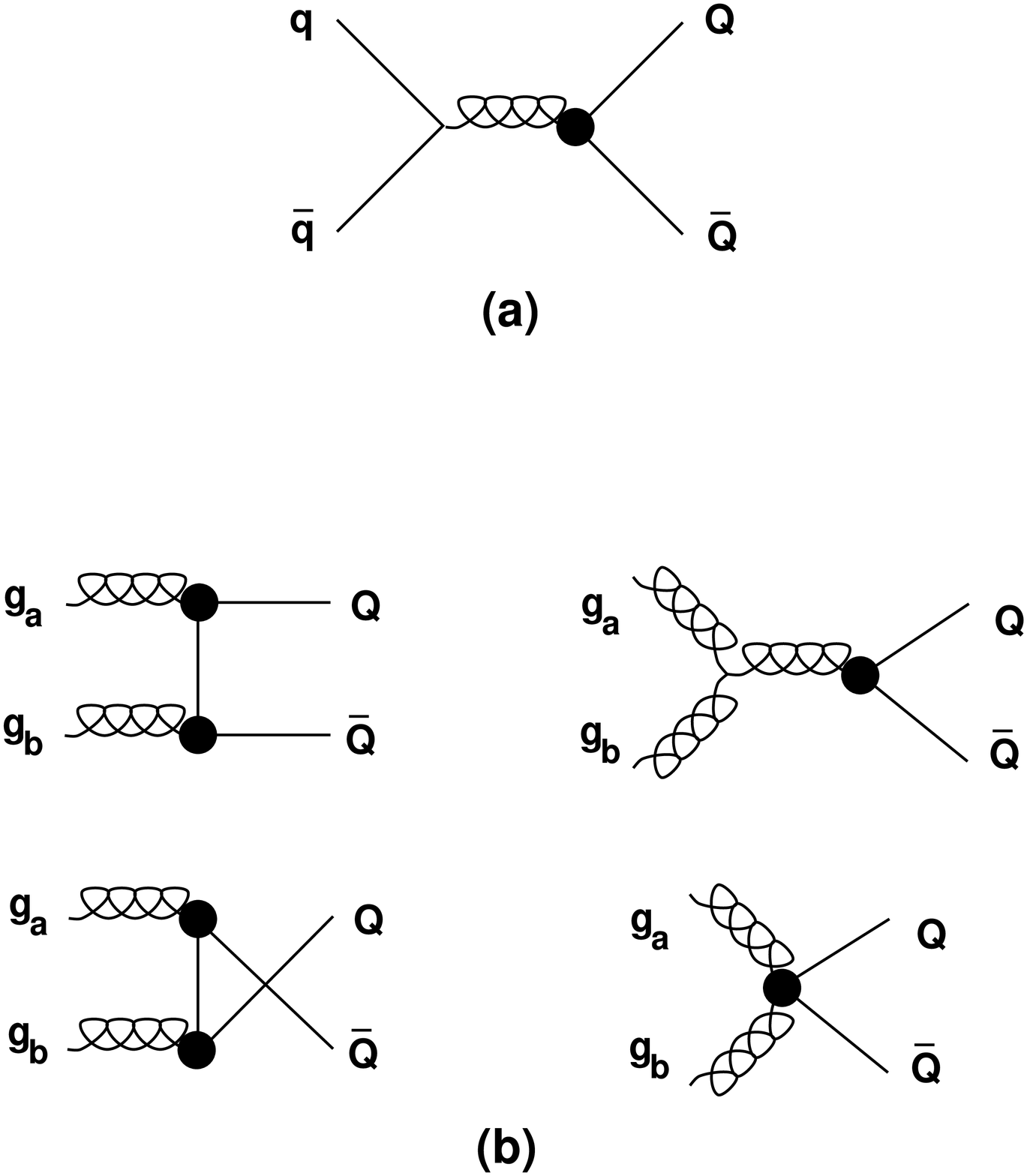,
height=10cm}
 \end{center}
\caption{\emph {Feynman diagrams for the process (a) $q\bar q\to Q \bar Q$,
and (b) $gg\to Q\bar Q$ with $Q=t$. The heavy dots represent the effective 
vertices involving $\kappa_g$ and ${\tilde \kappa}_g$ (see also text).}}
\label{ppbarttfig1}
\end{figure}

As mentioned above, the process $p \bar p \to t \bar t +X$ can proceed at
the parton level via: (a) $q \bar q \to t \bar t$, or (b) $g g \to t \bar
t$. With the effective interactions in Eq.~\ref{eettgeq20} and the
additional $ggtt$ effective vertex, the Feynman diagrams contributing to
these two processes are shown in Fig.~\ref{ppbarttfig1}. The parton level
cross-sections for $p \bar p \to t \bar t+X$ are then given by
\cite{prl69p2754,prd53p3604}: 

\beqa
\label{tt1}
\frac{d\hat\sigma_{q\bar q\to t\bar t}}{d\hat t} &=&
\frac{8\pi\alpha_s^2}{9\hat s^2} \left[
\frac{1}{2} -v +z -\kappa_g + \frac{1}{4}(\kappa_g^2 - {\tilde \kappa}_g^2)
+ \frac{v}{4z}(\kappa_g^2 + {\tilde \kappa}_g^2) \right ] ~, \\
\label{tt2}
\frac{d\hat\sigma_{gg\to t\bar t}}{d\hat t} &=&
\frac{\pi \alpha_s^2}{12\hat s^2} \biggr [ \left(\frac{4}{v}-9 \right)
\left(\frac{1}{2} -v +2z(1-\frac{z}{v}) - \kappa_g (1- \frac{\kappa_g}{2})
 \right)\nonumber \\
&+&  \frac{1}{4} (\kappa_g^2 + {\tilde \kappa}_g^2) \left( \frac{7}{z}
(1-\kappa_g)
+\frac{1}{2v}( 1+\frac{5\kappa_g}{2}) \right ) \nonumber \\
&+& \frac{1}{16} (\kappa_g^2 + {\tilde \kappa}_g^2)^2 \left( -\frac{1}{z} +
\frac{1}{v} + \frac{4v}{z^2} \right )
\biggr ] ~,
\eeqa

\ni where

\beq
z= \frac{m_t^2}{\hat s} ~~,~~ v=\frac{(\hat t -m_t^2)(\hat u -m_t^2)}{{\hat
s}^2} ~.
\eeq

\ni The process $p \bar p \to t \bar t j +X$, where $j$  stands for
a jet, can proceed via the following parton level subprocesses: (a) $q
\bar q \to t \bar t g$, (b) $gg\to t \bar t g$, or (c) $q(\bar q) g \to
t \bar t q  (\bar q) $ where an extra light quark jet is radiated.
The number of diagrams for these subprocesses is large. A detailed
description of the calculation of the cross-section for the reaction $p
\bar p \to t \bar t j+X$ is given in \cite{prd53p3604}.

\begin{figure}[htb]
\psfull
\begin{center}
\leavevmode
\epsfig{file=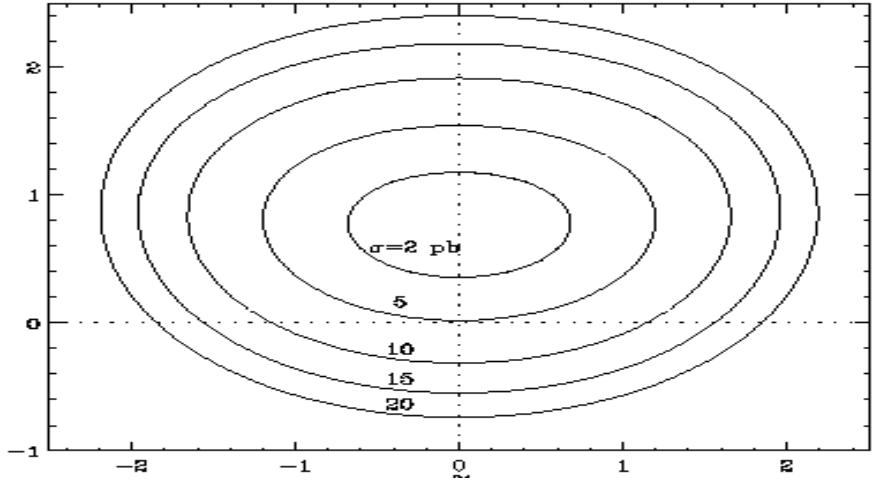,height=9cm,width=12cm,bbllx=0cm,bblly=2cm,
bburx=20cm,bbury=25cm,angle=0}
 \end{center}
\caption{\emph{Contours of the $p \bar p \to t\bar t+X$ cross-sections,
in pb, in the
${\tilde \kappa}_g$(horizontal axis)--$\kappa_g$(vertical axis) 
plane. Figure taken from \cite{prd53p3604}.}}
\label{ppbarttfig2}
\end{figure}

A plot of the dependence of the total cross-section for $t \bar t$
production on $\kappa_g$ and ${\tilde \kappa}_g$ was given in
\cite{prd53p3604} and is shown in Fig.~\ref{ppbarttfig2}. Of course, as
expected, the cross-section is  symmetric about ${\tilde \kappa}_g=0$ as
only ${\tilde \kappa}_g^2$ appears in the cross-section, being a CP-even
observable. The SM point is given by $\kappa_g={\tilde \kappa}_g=0$ and the
SM cross-section at this point is  $\sigma(t \bar t) \sim 5~{\rm pb}$.
Therefore, a measurement of $\sigma(t \bar t)> 5~{\rm pb}$  
will indicate the existence
of a non-zero $\kappa_g$ or ${\tilde \kappa}_g$. 

In \cite{prd53p4875} an ``anomalous cross-section'' was defined as

\beq
\Delta \sigma \equiv \sigma(\kappa_g,{\tilde \kappa}_g)-\sigma(0,0) ~,
\eeq

\ni and a plot of $\Delta \sigma$ in the $(\kappa_g,{\tilde \kappa}_g)$
plane was given. From the current experimentally allowed region of
$\Delta \sigma$, not surprisingly, they found that a rather large CEDM
coupling is permitted. For example, for certain values of $\kappa_g$,
they found that 
$0.5 \lsim |{\tilde \kappa}_g| \lsim 1.5$ is allowed, within 1$\sigma$.
This corresponds 
to $2.8 \times 10^{-17}~g_s{\rm -cm} \lsim |\Re{\rm e}d_t^g| \lsim 
8.3 \times 10^{-17}~g_s{\rm -cm}$, at 1-$\sigma$.

\begin{figure}[htb]
\psfull
\begin{center}
\leavevmode
\epsfig{file=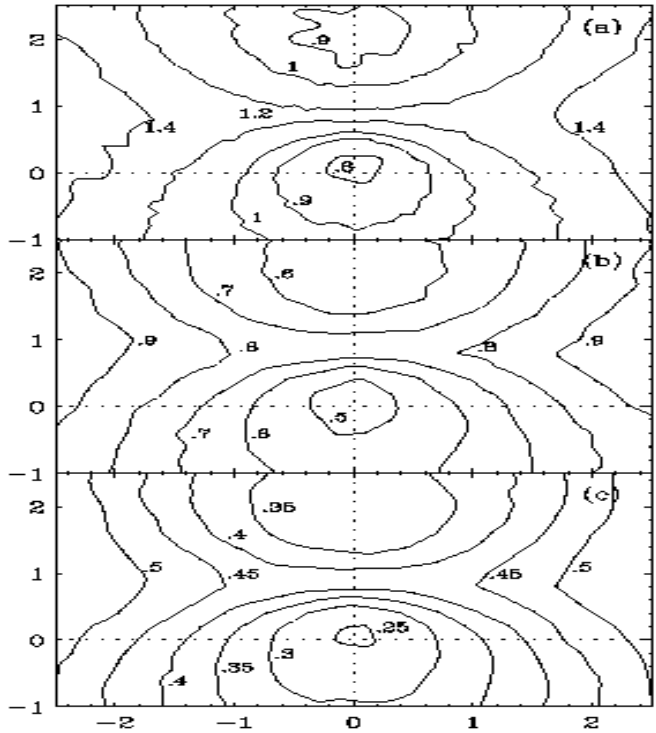,height=9cm,width=12cm,bbllx=0cm,bblly=2cm,
bburx=20cm,bbury=25cm,angle=0}
 \end{center}
\caption{\emph{Contours of the ratio
${\cal R}=\sigma(t\bar tj)/\sigma(t\bar t)$  in the
${\tilde \kappa}_g$(horizontal axis)--$\kappa_g$(vertical axis) plane,
with $p_T(j) > $5, 10 and 20 GeV in (a), (b),
and (c), respectively. Figure taken from \cite{prd53p3604}.}}
\label{ppbarttfig3}
\end{figure}

Considering again an associated extra jet,  $j$, in top pair production
at the 2 TeV Tevatron, it was shown in \cite{prd53p3604} that not much
will be gained from a simple cross-section analysis of the $t\bar t j$
final state compared to the $t\bar t$ final state.  However, a
more interesting CP-conserving quantity was further suggested in
\cite{prd53p3604}; it was shown that the ratio of the two
cross-sections, $ {\cal R}= \sigma(t\bar t j) / \sigma(t\bar t)$, can be
used to further constrain the CEDM and CMDM couplings.
Fig.~\ref{ppbarttfig3}(a), (b) and (c) show the main behavior of the ratio
$\cal R$ in the $(\kappa_g,{\tilde \kappa}_g)$ plane for $p_T(j)>5,~10,~20$
GeV, respectively, where $p_T(j)$ is the transverse momentum of the jet
$j$. The most interesting feature is the one depicted
in Fig.~\ref{ppbarttfig3}(a); we see that for the cut $p_T(j)>5$
GeV, at some regions in the $(\kappa_g,{\tilde \kappa}_g)$ plane, ${\cal
R}>1$. This happens as a consequence of the new  $ttg$ and $ttgg$
vertices (absent in the SM) in $tt$ + jet production. 

A discussion on the possible limits that can be put on the CEDM and CMDM
of the top by analyzing CP-even observables in single top production at
the Tevatron was presented in \cite{prd53p6218}. There, single top
production can occur either via the $q \bar q \to t \bar b$ or the
$W$-gluon fusion subprocess $Wg\to t \bar b$.  An analysis of the
cross-section for the reaction $Wg\to t \bar b$ was performed
\cite{prd53p6218}, trying to further constrain the CEDM and CMDM of the
top quark which may enter the $ttg$ vertex present in this reaction. 
However, it was found that, for both the Tevatron and the LHC, the
sensitivity of the $W$-gluon fusion subprocess leading to a single top
quark for the CEDM and CMDM couplings is more than an order of magnitude
smaller than the usual top pair production channel.

\subsubsection{CP-odd observables in $p\bar p \to t \bar t +X$ 
\label{sssec812}}

As already mentioned before, if all non-standard effects reside in the $t
\bar t g$ coupling,\footnote{Recall that this is not necessarily true in 
model calculations, e.g., in the MSSM -  
see discussion in section \ref{ssec74}.} 
then they can be parameterized by the effective $ttg$
and $ttgg$ interaction Lagrangian.
However, upon neglecting the $gg$ fusion process (which is a good approximation under the conditions of the Tevatron upgrade with c.m. 
energy of $\sqrt s=2$ TeV), the $ttgg$ contact
term plays no role and the only relevant effective interaction for the
reaction $q \bar q \to t \bar t$ is the effective $ttg$ CEDM interaction in
Eq.~\ref{eettgeq20}. Similarly, the $tbW$ vertex (with an on-shell $W$-boson) 
may give rise to additional non-standard couplings 
as in Eqs.~\ref{eetteq5a} and \ref{eetteq5b}, which 
may cause CP violation in the top and anti-top decays. 
Therefore, in the overall $t \bar t$ production and decays, 
CP violation is parameterized by non-zero values of the CEDM, $d_t^g$, in 
the production vertex, and/or by the quantities $f_2^{L,R} - \bar f_2^{R,L}$ 
(defined in Eqs.~\ref{eetteq5a} and \ref{eetteq5b}) 
emanating from the $t, \bar t$ decays.     

In order to detect
such CP-violating couplings one has to construct appropriate
CP-odd observables.
Following \cite{plb415p193}, let us again consider two decay
scenarios for $t$  and $\bar t$. In the first one, both
$t$ and $\bar t$ decay leptonically

\beq
p \bar p \rightarrow t \bar t \rightarrow \ell^+ \ell^- X
\, , \label{ppbartteq6}
\eeq

\ni while in the second scenario only one of the
$t$ or $\bar t$ decays leptonically (see also Eqs.~\ref{eetthzeq31} 
and \ref{eetthzeq32} in section \ref{sssec623})

\beq
p \bar p \rightarrow t \bar t \rightarrow \ell^+ X ~~,~~
p \bar p \rightarrow t \bar t \rightarrow \ell^- X ~.
\label{ppbartteq7}
\eeq

\noindent The processes in Eq.~\ref{ppbartteq7} have better statistics than
the one in Eq.~\ref{ppbartteq6} and give the best signature for the top quark
identification. Within these decay scenarios
two possible CP-odd observables were considered in \cite{plb415p193}
which we will describe below.

\subsubsection{Transverse energy asymmetry of charged leptons \label{sssec813}}

The transverse energy asymmetry of the charged leptons was
originally suggested by Peskin and Schmidt in \cite{prl69p410}
for $t \bar t$  production in a $pp$ collider (see Eq.~\ref{sp11}) 
and was discussed
in detail in sections \ref{ssec73} and \ref{ssec74}. 
Recall that in the case where
both $t$ and $\bar t$ decay leptonically it can be defined as 
(using here the notation in \cite{plb415p193})

\beq
A_T= {\sigma (E_T^- > E_T^+)-\sigma (E_T^+ > E_T^-) \over
\sigma (E_T^- > E_T^+)+\sigma (E_T^+ > E_T^-)}~. \label{ppbartteq8}
\eeq

\ni The result for the expected CP-violating 
asymmetry in the transverse energy of the muons ($A_T^\mu$)
was given in \cite{plb415p193}.  
They  considered $A_T^\mu$ as a function  of the imaginary part of 
the top CEDM, $\Im{\rm m}d_t^g$ 
(recall that $A_T$ is $T_N$-even thus requiring 
an absorptive phase), and of $\Re{\rm e}(f_2^R - \bar f_2^L)$. Also, 
the usual CDF cuts were applied. 
They found that CP violation from the production mechanism,
{\it i.e.}, $\propto \Im{\rm m}d_t^g$, is larger then that arising 
from the decay process,
{\it i.e.}, $\propto \Re{\rm e}(f_2^R - \bar f_2^L)$. 
For example, they found that with $\Im{\rm m}d_t^g \sim 10^{-17}$ $g_s$-cm 
and $\Re{\rm e}(f_2^R - \bar f_2^L) \sim 0.2$ one can
obtain an asymmetry around the $\sim 10 \%$ level.

In order to understand the feasibility of extracting such values for 
the CP-violating couplings in production and decays of the $t \bar t$, it 
is useful to decompose $A_T^\mu$ as follows \cite{plb415p193}:

\beq
A_T^\mu \equiv c_P A_P + c_D A_D \label{ppbartteq9} ~,
\eeq

\ni where the dimensionless couplings $c_P$ and $c_D$ are

\beq
c_P \equiv \frac{m_t}{g_s} \Im{\rm m}d_t^g ~~,~~ 
c_D \equiv \frac{1}{2} \Re{\rm e}(f_2^R - \bar f_2^L) \label{cpcd}~. 
\eeq

\ni Then, in terms of $c_P$ and $c_D$, the 
statistical significance for $A_T^\mu$ determination is given by

\beq
N_{SD}^T \equiv |c_P A_P +c_D A_D|\sqrt{N_{\ell \ell}} \label{ppbartteq10}~,
\eeq

\ni where $N_{\ell\ell}$ is the number of dilepton events expected to
be $\sim 80$ and $\sim 1200$ at an integrated luminosity $L=2$ and 30 
fb$^{-1}$, respectively \cite{hepph9704243}. $A_P$ and $A_D$ can be
calculated and, thus, it was found in \cite{plb415p193} that a 3-$\sigma$
effect will require the following relations to be satisfied

\bea
|2.5 c_P + 0.9 c_D|\geq 1 & {\rm for} & L=2\; \fmbarn^{-1}
\label{ppbartteq11}~,\\
|9.8 c_P + 3.3 c_D|\geq 1 & {\rm for} & L=30\; \fmbarn^{-1}
\label{ppbartteq12}~.
\eea

\ni Clearly,  
$A_T^\mu$ is more sensitive to CP violation in the production mechanism 
than in the $t \bar t$ decays. 
So, for example, $L=30\; \fmbarn^{-1}$ allows for an observation of
$c_P=c_D=0.08$ at $\sqrt{s}=2 \tev$. Note that $c_P=0.08$ 
corresponds to $\Im{\rm m}d_t^g = 8.8 \times 10^{-18}$ which, again, 
is more than an order of magnitude larger then what is expected for the CEDM 
in beyond the SM scenarios such as MHDM's and SUSY (see Chapter \ref{sec4}). 
Similarly, the resulting 3-$\sigma$ limit 
$\Re{\rm e}(f_2^R - \bar f_2^L) = 0.16$ (corresponding 
to $c_D=0.08$) falls short by about one order of magnitude from 
model predictions for this quantity (see discussion in Chapter \ref{sec5}).
      
A related asymmetry which can be used in the case when only one top decays
leptonically was also suggested in \cite{plb415p193}:

\beq
A_{cut}^{\mu}(E_{T cut})=
{\sigma^-(E_T^- > E_{T cut})-\sigma^+(E_T^+ > E_{T cut})
\over \sigma } \label{ppbartteq13}~.
\eeq

\ni The $\sigma$ in Eq.~\ref{ppbartteq13} denotes the integrated
cross-section with no cuts except for the standard experimental cuts. We
note that, with the CP-violating coupling $c_P$ and $c_D$ of the order of
0.1, which tends to be somewhat optimistic, this asymmetry in
Eq.~\ref{ppbartteq13} can also reach the $\sim 10\%$ level.


\subsubsection{Optimal observables \label{sssec814}}

It is useful to be able to experimentally separate CP violation
in the production from that in the decay. The optimization method
outlined in section \ref{ssec26} 
can provide for such a detection and was also
used in \cite{plb415p193}.

Consider the transverse lepton energy spectrum in the single leptonic
(say $\ell^+$) and the dileptonic final states

\beq
\frac{1}{\sigma} \frac{d \sigma}{dE_T^+}=f_1^+(E_T^+)+c_P f_P^+(E_T^+)
+c_D f_D^+(E_T^+) \label{ppbartteq14}~,
\eeq

\beq
\frac{1}{\sigma} \frac{d^2 \sigma}{dE_T^+\;dE_T^-}=
f_1^\pm(E_T^+,E_T^-)+c_P f_P^\pm(E_T^+,E_T^-)
+c_D f_D^\pm(E_T^+,E_T^-) \label{ppbartteq15}~,
\eeq

\ni where in Eqs.~\ref{ppbartteq14} and \ref{ppbartteq15}
$\sigma$ denotes the cross-section for the process $p \bar{p} \ra
\ell^+ \; +\hbox{jets}$ and $p \bar{p} \ra \ell^+ \; \ell^- \; +\hbox{jets}$,
respectively. Also, $c_P,c_D$ are defined in Eq.~\ref{cpcd} 
and $f^\pm$ 
are known functions
of $E_T^+$ and $E_T^-$.

As can be seen from Eqs.~\ref{ppbartteq14} and \ref{ppbartteq15}, the
transverse lepton energy spectrums, in both the single and double-leptonic
channels, are sensitive to $c_P$ and $c_D$.
Using the above transverse lepton energy spectrum, the optimal
weighting functions can be obtained. This was done in \cite{plb415p193}
where in both cases the statistical significances for the experimental
determination of $c_P$ and $c_D$, {\it i.e.}, $N_{SD}^{P,D}\equiv
|c_{P,D}|/\Delta
c_{P,D}$, were calculated. For the single leptonic events they obtained

\beq
N_{SD}^{P}=\frac{|c_P|}{2.37}\sqrt{N_\ell}~, \lsp
N_{SD}^{D}=\frac{|c_D|}{18.43}\sqrt{N_\ell} \label{ppbartteq16}~,
\eeq

\ni and for the dileptonic events

\beq
N_{SD}^P=\frac{|c_P|}{1.17}\sqrt{N_{\ell \ell}}~, \lsp
N_{SD}^D=\frac{|c_D|}{5.76}\sqrt{N_{\ell \ell}} \label{ppbartteq17}~,
\eeq

\ni where $N_\ell$ and $N_{\ell \ell}$ are the expected number of single
and double-leptonic events, respectively. $N_\ell \sim 1300(20,000)$ and
$N_{\ell \ell} \sim 80(1200)$ for $L=2(30)~ \fmbarn^{-1}$, respectively
\cite{hepph9704243}. The minimal values of $c_P$ and $c_D$ necessary to
observe a 3-$\sigma$ CP-violating effect with the optimization technique
are listed in Tables \ref{ppbartttab1} and \ref{ppbartttab2} for the
single and double-leptonic channels, respectively. We see that the
single-leptonic modes are more sensitive to the non-standard couplings.
Thus, for example, the Tevatron upgrade Run III with $L=30$ fb${-1}$ will
be able to probe $\Im{\rm m}d_t^g$ down to values of $\sim 5 \times
10^{-18}$ $g_s$-cm in the single-leptonic channel.  Note that, as
expected, this result is somewhat better than what can be achieved with
``naive'' observables such as $A_{cut}^\mu$ in Eq.~\ref{ppbartteq13}.

We note in passing that comparable limits for $\Im{\rm m}d_t^g$ but also 
for $\Re{\rm e}d_t^g$, i.e., $\Im{\rm m},\Re{\rm e}d_t^g \sim {\rm few} \times 
10^{-18}$ $g_s$-cm, were 
found also in \cite{hepph9805358} using optimal observables for 
the reaction $p\bar p \to t \bar t +X$ and 
with the Tevatron upgrade parameters.

To summarize,
as was noted in \cite{plb415p193}, it is not inconceivable that
non-standard CP-violating couplings of the top quark to a gluon may be
discovered at the Tevatron before precision measurements at the LHC
are done.


\begin{table}
\vspace*{-0.4cm}
\bce
\begin{tabular}{||c|c|c||}
\hline\hline
$L[\fmbarn^{-1}]$&2&30\\
\hline
$|c_P|$&0.20&0.05\\
\hline
$|c_D|$&1.50&0.40\\
\hline\hline
\end{tabular}\\
\vspace*{0.3cm}
\ece
\vspace*{-0.5cm}
\caption{\emph{The minimal values of $c_P$ and $c_D$ necessary to observe CP
violation in the single-lepton mode at the 3-$\sigma$ level for $L=$2, 30
fb$^{-1}$. As a reference value, recall 
that $c_P=1$ corresponds to $\Im{\rm m}d_t^g = 1.1 \times 
10^{-16}$ $g_s$-cm. Table taken from \cite{plb415p193}.}}
\label{ppbartttab1}
\end{table}

\begin{table}
\vspace*{-0.4cm}
\bce
\begin{tabular}{||c|c|c||}
\hline\hline
$L[\fmbarn^{-1}]$&2&30\\
\hline
$|c_P|$&0.39&0.10\\
\hline
$|c_D|$&1.93&0.50\\
\hline\hline
\end{tabular}\\
\vspace*{0.3cm}
\ece
\vspace*{-0.5cm}
\caption{\emph{The minimal values of $c_P$ and $c_D$ necessary to observe CP
violation in the dilepton mode at the 3-$\sigma$ level for $L=$2, 30
fb$^{-1}$. See also caption to Table \ref{ppbartttab1}.
 Table taken from \cite{plb415p193}.}}
\label{ppbartttab2}
\end{table}


\newpage
~

\newpage
~

\newpage
~

\newpage

\subsection{$p \bar p \to t \bar b + X$ \label{ssec82}}

\begin{figure}[ht]
\psfull
 \begin{center}
  \leavevmode
  \epsfig{file=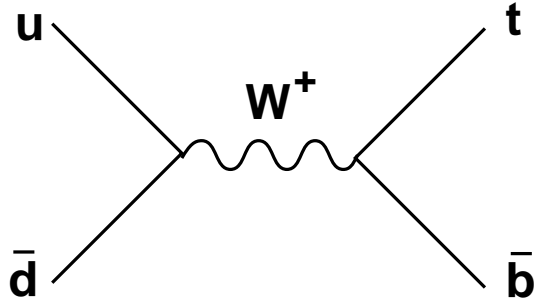
,height=4cm}
 \end{center}
\caption{\emph{The tree-level Feynman diagram contributing to $u \bar d \to t
\bar b$.}}
\label{ppbartbfig1}
\end{figure}

In spite of the fact that 
at the Tevatron $p \bar p$ collider, top quarks are  mainly
produced as $t \bar t$ pairs via an s-channel gluon exchange
~\cite{hepph9802305},
the subleading electroweak  production mechanism
of a single top forms a significant fraction of the $t \bar t$
pair production. It will therefore be closely scrutinized in
the next runs of the Tevatron \cite{stelzer}.  The
production rate of $t \bar b$ through
an s-channel off-shell $W$-boson, $p\bar{p} \to W^* \rightarrow t\bar{b}+X$,
(the corresponding partonic reaction, $u \bar d \to W^* \to t \bar b$, is
depicted in Fig.~\ref{ppbartbfig1}) is expected to yield about $10\%$ of
the $t \bar t$ production rate \cite{stelzer}.

In this section we examine CP violation asymmetries in top quark
production and its subsequent decay via the basic quark level reactions
\cite{prd57p1495,prd54p5412,sthesis}: 

\beq
u \bar d\to t \bar b \to b \,W^+ \bar b ~~, ~~ \bar u d\to \bar t b
\to \bar b \,W^- b \label{ppbartbeq1}~.
\eeq

\ni Indeed this reaction is  rich for CP violation studies as it
exhibits many different types of asymmetries. Some of these, which we
consider below, involve the top
spin. Therefore, the ability to track the top spin through its decays becomes
important and  top decays have to be examined as well 
(see e.g. \cite{hepph9811219}).

The asymmetries in $t\bar b$ production
can be appreciably larger than those in $t\bar t$ pair production 
wherein they tend to be about
a few tenths of percent (see Chapters \ref{sec6} and \ref{sec7}). 
Moreover, while in the SM,
CP-odd effects in $p\bar{p} \to W^* \rightarrow t\bar{b}+X$ are expected to be
extremely small  since they are severely
suppressed by the GIM mechanism (see e.g.,
\cite{prl67p1979,plb319p526}),
it is shown below that, in extensions of the SM, CP asymmetries can be
sizable - in some cases at the level of a few percent.
Therefore, since the number of events needed 
to observe an asymmetry
scales as (asymmetry)$^{-2}$, the enhanced CP-violating effects in
$t\bar b(\bar tb)$ may make up for the reduced production rates for
$t\bar b$ compared to $t\bar t$. In fact larger asymmetries
could  be essential as detector systematics can
be a serious limitation for asymmetries less than about $1 \%$.

Let us discuss now the asymmetries in the $u\bar d$ ($\bar ud$)
subprocess. We consider four types of asymmetries that may be present
\cite{prd54p5412}. First, is the CP-violating asymmetry in the
cross-section 

\beq
A_0 = (\sigma_{t \bar b} -\sigma_{\bar t b}) / (\sigma_{t \bar b} + 
\sigma_{\bar t b})
\label{ppbartbeq2} ~,
\eeq

\ni where $\sigma_{t \bar b}$ and $\sigma_{\bar t b}$ are the cross-sections
for $u\bar d\to t\bar b$ and $\bar ud\to \bar tb$, respectively, at
$\hat s = (p_t+p_{\bar b})^2$.

The spin of the top allows us to define three additional types of
CP-violating polarization asymmetries. For these it is convenient to
introduce the coordinate system in the top quark (or anti-top)
rest frame where the unit vectors are $\vec e_z\propto -\vec p_b$,
$\vec e_y\propto \vec p_u\times \vec p_b$ and $\vec e_x=\vec e_y\times
\vec e_z$. Here $\vec p_b$ and $\vec p_u$ are the 3-momenta of the
$\bar b$-quark and the initial $u$-quark in that frame. We denote  the
longitudinal polarization or helicity asymmetry as

 \beq
 A(\hat z) = (N_R-N_L -\bar N_R+\bar N_L) / 
(N_R + N_L + \bar N_R + \bar N_L) 
\label{ppbartbeq3} ~,
 \eeq

 \ni where $N_L$ is the number of left-handed top quarks produced
 in $u\bar d\to t\bar b$ and $\bar{N}_R$ is the number of right-handed
 $\bar t$ produced in $\bar u d\to \bar tb$, {\it etc}.   Therefore, in
 the  frame introduced above right-handed tops have spin up along the
 $z$-axis and left-handed ones spin down.
 We further define the CP-violating spin asymmetries in the $x$ and $y$
 directions as follows

\beqa
A (\hat x) & = & (N_{x+} - N_{x-} + \bar N_{x+} - \bar N_{x-}) / (
N_{x+} + N_{x-} + \bar N_{x+} + \bar N_{x-}), 
\nonumber\\
A (\hat y) & = & (N_{y+} - N_{y-} - \bar N_{y+} + \bar N_{y-}) / (
N_{y+} + N_{y-} + \bar N_{y+} + \bar N_{y-} ) \label{ppbartbeq5} ~,
\eeqa

 \ni where, for example, 
$N_{j+} (\bar N_{j+})$ represent the number of $t (\bar
 t)$ with spin up with respect to $\hat j$-axis, for $j=x$, $y$, etc.

While all these four asymmetries are manifestly CP-violating,
$A_0$, $A(\hat z)$ and $A(\hat x)$ are even under naive time
reversal ($T_N$) whereas $A(\hat y)$ is $T_N$-odd. So the first three,
unlike $A(\hat y)$, require a complex amplitude, i.e., absorptive phases.

These asymmetries are related to form factors arising from
radiative corrections of the $W^* \to tb$ 
production vertex due to non-standard
physics. To see this, it is useful to parameterize the 1-loop $t \bar b$ and
$\bar t b$ currents of the production amplitude as follows \cite{prd57p1495}:

\beqa
J^{\mu(t \bar b)} \equiv i \frac{g_W}{\sqrt 2} \sum_{P=L,R} \bar{u}_t
\left( \frac{{\cal P}_{1}^P p_b^{\mu}}{m_t} + {\cal P}_{2}^P \gamma^{\mu}
\right) Pv_b \label{ppbartbeq6}~, \\
J^{\mu(\bar t b)} \equiv  i \frac{g_W}{\sqrt 2} \sum_{P=L,R} \bar{u}_b
\left( \frac{{\bar {\cal P}}_{1}^P p_b^{\mu}}{m_t} + \bar{{\cal P}}_{2}^P
\gamma^{\mu} \right) Pv_t \label{ppbartbeq7}~,
\eeqa

\ni where $P=L~{\rm or}~R$ and $L(R) \equiv (1 -(+) \gamma_5)/2$.
${\cal P}_{1}^{L,R}$ and ${\cal P}_{2}^{L,R}$, defined
in Eqs.~\ref{ppbartbeq6} and \ref{ppbartbeq7}, contain the necessary
absorptive phases as well as the CP-violating phases in
a given model. It is easy to show that if one defines

\beqa
{\cal P}_{1}^L &\sim& e^{i \delta_s^{1}} \times e^{i \delta_w^{1}}
\label{ppbartbeq8}~,\\
{\cal P}_{2}^L &\sim& e^{i \delta_s^{2}} \times e^{i \delta_w^{2}}
\label{ppbartbeq9}~,
\eeqa

\ni where $\delta_s^{1},\delta_s^{2}$ are the CP-even absorptive
phases ({\it i.e.}, FSI phases) and
$\delta_w^{1},\delta_w^{2}$ are the CP-odd phases, then

\beqa
{\bar {\cal P}}_{1}^R &\sim& - e^{i \delta_s^{1}} \times e^{-i
\delta_w^{1}} \label{ppbartbeq10}~,\\
{\bar {\cal P}}_{2}^L &\sim& e^{i \delta_s^{2}} \times e^{-i
\delta_w^{2}} \label{ppbartbeq11}~.
\eeqa

\noindent In terms of these form factors, the two $T_N$-even asymmetries
$A_0$ and $A(\hat z)$ are given by \cite{prd54p5412}:

\beqa
A_0 = \frac{(x-1)}{2(x+2)} {\eRe}({\cal
P}_{1}^L + {\bar {\cal P}}_{1}^R) -  {\eRe}({\cal P}_{2}^L - {\bar
{\cal P}}_{2}^L) \label{ppbartbeq12}~, \\
A(\hat z) = \frac{(x-1)}{2(x+2)} {\eRe}({\cal
P}_{1}^L + {\bar {\cal P}}_{1}^R) - \frac{x-2}{x+2} {\eRe}({\cal P}_{2}^L -
{\bar
{\cal P}}_{2}^L) \label{ppbartbeq13}~,
\eeqa

\ni where $x \equiv m_t^2/\hat s$. Notice that the three quantities $\{
A_0,\ A(\hat z), A(\hat x) \}$ are linear combinations of only two form
factors. Thus one can show that 

\beqa
A(\hat x)= -3\pi x^{-{1\over 2}}
\left[ (2+x) A_0 +(2-x)A(\hat z) \right]/32 \label{ppbartbeq14}~.
\eeqa

\ni which will therefore hold if the new CP violating physics takes
place through such a vertex correction of $ W^* \to t b$. 

Similarly, the asymmetry $A(\hat y)$ is proportional 
to the real parts of the 1-loop
integrals and may therefore be obtained from the corresponding
imaginary parts through the use of dispersion relations.
In particular, since the $T_N$-even asymmetries are proportional
to the absorptive phases in the above form factors, one can express 
$A(\hat y)$ in terms of $A_0$ and $A(\hat z)$, 

 \beqa
 A(\hat y)[\hat s]  &=& -  {3\over 32}  \left[
 1-x\over (2+x)\sqrt{x}  \right]  {\eRe}  \left\{  \int_{0}^{\infty} 
 { 2\xi+x\over  (\xi-x)(\xi-1+i\epsilon)\xi} \right.  \nonumber\\
 &&\times  \left[(2\xi-x)A_0[\xi\hat s]  +  (2\xi+x)A(\hat z)[\xi\hat
 s] \right]  d\xi  \Biggr\}
 \label{ppbartbeq15} ~.
 \eeqa

\ni Note that the integrand is 0 if $\xi \hat s$ is below the
threshold to produce an  imaginary part since then $A_0$ and $A(\hat z)$ 
will vanish.

Let us now evaluate the form factors defined in Eqs.~\ref{ppbartbeq6}
and \ref{ppbartbeq7} in two extensions of the SM: the
2HDM of type II and the MSSM\null.
As was pointed above, once these form factors are calculated in a
given model, the asymmetries $A_0,A(\hat z),A(\hat x)$ and $A(\hat y)$
can be readily obtained.

\subsubsection{2HDM and the CP-violating asymmetries \label{sssec821}}

\begin{figure}[ht]
\psfull
 \begin{center}
  \leavevmode
  \epsfig{file=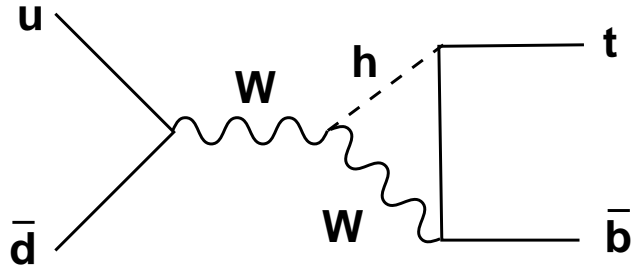
,height=3.5cm}
 \end{center}
\caption{\emph{The CP-violating 1-loop graph in 2HDM's with 
CP violation from neutral Higgs exchanges; $h$ denotes the lightest 
neutral Higgs.}}
\label{ppbartbfig2}
\end{figure}

As emphasized through out this article, 
in the 2HDM of type II, a CP-odd phase can reside in neutral
Higgs exchanges and there is only one Feynman diagram that
contributes to CP violation in $u \bar d \to W^* \to t \bar b$ 
to 1-loop order. This diagram is
shown in Fig.~\ref{ppbartbfig2}. The relevant Feynman rules for this
diagram, required for calculating the asymmetries of interest, can
be extracted from the parts of the 2HDM Lagrangian involving the
$t \bar t {\cal H}^k$ and $WW {\cal H}^k$ couplings in Eqs.~\ref{2hdmab}
and \ref{2hdmc}, with $k=1,2,3$ for the three neutral Higgs fields.
Recall again that the coupling constants,
$a_t^k$, $b_t^k$ and $c^k$ are functions of $\tan\beta$, which is
the ratio between the two VEV's in this
model, and of the three mixing angles
$\alpha_{1,2,3}$ which diagonalize the $3\times3$ Higgs mass matrix
(see section \ref{sssec323}).
As usual, for simplicity, we have assumed that two of the three neutral
Higgs-bosons are much heavier compared to the third one which we
denote by $h$. Thus, the CP-violating effect will be dominated by the
lightest neutral Higgs, $h$, and is proportional to $b_t^h c^h$.
Choosing the angles $\alpha_1=\alpha_2 =\pi/2$ and $\alpha_3=0$
which give maximal effects \cite{prd54p5412}, one obtains $b_t^h c^h
\propto \cos\beta  \cot\beta$ so the asymmetries are now
functions of $\tan\beta$ and $m_h$ only.

Using the Lagrangian in Eqs.~\ref{2hdmab} and \ref{2hdmc},
the form factors ${\eRe}({\cal
P}_{1}^L + {\bar {\cal P}}_{1}^R)$ and ${\eRe}({\cal P}_{2}^L - {\bar
{\cal P}}_{2}^L)$ can be readily calculated and we get

\beqa
{\eRe}({\cal P}_{1}^L + {\bar {\cal P}}_{1}^R) &=&
 - \left( \frac{\alpha}{\sqrt 2} \right) \frac{b_t^h c^h }
{2 \pi \sin^2 \theta_W}
\frac{m_t^2}{m_W^2} {\mIm} \left[ 2m_W^2 (C_{11} \right.
-C_{12}) \nonumber \\
&&\left. - m_t^2 (C_{12}+C_{23}) -
\tilde{C}_{0} -\tilde{C}_{11} \right] \label{ppbartbeq18} ~,\\
{\eRe}({\cal P}_{2}^L - {\bar
{\cal P}}_{2}^L) &=&
\left( \frac{\alpha}{\sqrt 2} \right) \frac{b_t^h c^h }{4 \pi \sin^2
\theta_W} \frac{m_t^2}{m_W^2}
 {\mIm} \left[ 2m_W^2 (C_0 + C_{12}) \right. \nonumber \\
&&\left. -2C_{24} \right] \label{ppbartbeq19}~,
\eeqa

\ni where the $C_x$ above, 
$x \in \left\{0,11,12,21,22,23,24 \right\}$, 
are the three-point form factor associated with the 1-loop diagram 
in Fig.~\ref{ppbartbfig2}, and 
are given via \cite{prd54p5412,sthesis}:

\bea
C_x &=& C_x(m_t^2,m_W^2,m_h^2,m_b^2,\hat s,m_t^2)~,
\eea

\ni where $\hat s=(p_t+p_{\bar b})^2$ and 
$C_x(m_1^2,m_2^2,m_3^2,p_1^2,p_2^2,p_3^2)$ is defined in appendix A.

\begin{figure}[htb]
\psfull
\begin{center}
\leavevmode
\epsfig{file=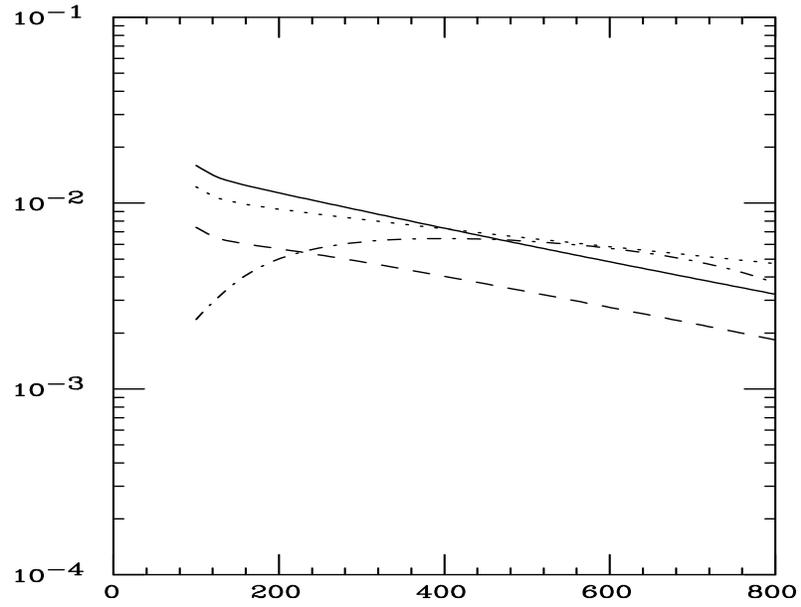
,height=9cm,width=12cm,bbllx=0cm,bblly=2cm,
bburx=20cm,bbury=25cm,angle=0}
 \end{center}
\caption{\emph{The CP-violating
asymmetries in $p \bar p \to t \bar b +X$ 
for the 2HDM case and in the $p \bar p$ c.m. frame for $\sqrt s=2$ TeV, 
as a function
of $m_h$ (horizontal axis); $A_0$ (solid),
$A(\hat z)$ (dashed), $A(\hat x)$ (dotted) and $A(\hat y)$ (dot-dashed).
Figure taken from \cite{prd54p5412}.}}
\label{ppbartbfig3}
\end{figure}

The quark level asymmetries of interest can be converted to the hadron
(i.e., $p \bar p$) level by folding in the structure functions in
the standard manner \cite{collider}. The results for the 2HDM case
are shown in Fig.~\ref{ppbartbfig3}, for $\tan\beta=0.3$ and as a
function of the lightest Higgs-boson mass, $m_h$.
For the asymmetry $A(\hat y)$ we apply a cut of
$\hat s > (m_h+m_W)^2$. We can see that $A_0$ and $A(\hat x)$
can reach above the percent level for $m_h \lsim 200$ GeV.
The measurable consequences for such an asymmetry are discussed 
in section \ref{sssec823} below.

It is interesting to note that in the 2HDM (to the 1-loop
order) the $T_N$-even asymmetries $A_0$, $A(\hat z)$ and $A(\hat x)$
do not receive any contribution from the decay vertex in $t \to b W$.
The only diagram that can potentially drive CP violation in $t \to b W$
is the same one as shown in Fig.~\ref{ppbartbfig2} with the momenta
of the $t$ and the $W$ reversed. Thus,
an important
necessary condition for $A_0,A(\hat z),A(\hat x) \neq 0$,
that there is an imaginary part in the decay amplitude, is not
satisfied. Moreover, as it turns out, the observed value of $A(\hat y)$ is not
affected by CP violation in the decay process. The key point is that 
the measurement of $A(\hat y)$ through the decay chain
%
%
$u(p_u)\ \bar d(p_d) \to \bar b (p_{b1})\  t(p_t)$ followed by
$t(p_t) \to b(p_{b2})\  e^+(p_e)\  \nu(p_\nu)$
is equivalent to a measurement of a term proportional to
$\epsilon(p_e,p_d,p_t,p_{b1})$ ($\epsilon$ being here the Levi-Civita tensor).
 On the other hand,
CP violation arising from the decay process is proportional to
$\epsilon(p_e,p_d,p_t,p_{b2})$.
It is easy to see that an observable related to the first of these will be
insensitive to the second. So, in this way asymmetries in the
production can be separated from those in the decay.

\newpage
~

\newpage

\subsubsection{SUSY and the 
CP-violating cross-section asymmetry \label{sssec822}}

The MSSM possesses several CP-odd phases that can give rise to
CP violation in $p \bar p \to t \bar b +X$ and in the
subsequent top decay $t \to b W$ (see section \ref{sssec514}). 
In \cite{prd57p1495} we have
constructed a plausible low-energy MSSM framework in which
there are only five relevant free parameters needed to evaluate
the cross-section asymmetry $A_0$ (see also sections \ref{sssec334} and 
\ref{sssec514}). These are: $M_S$ - a typical SUSY mass scale
that characterizes the 
mass of the heavy squarks, $m_G$ - the mass of
the gluino, $\mu$ - the Higgs mass parameter in the superpotential,
$m_l$ - the mass of the lighter stop and $\tan\beta$ - the ratio
between the VEV's of the two Higgs fields in the theory. In
this framework there are two sources of CP violation that can
potentially contribute to $A_0$ \cite{prd57p1495}. The
first may arise from the Higgs mass parameter $\mu$
which may be complex in general. The second CP-violating
phase arises from ${\tilde t}_L - {\tilde t}_R$ mixing 
(see e.g., Eq.~\ref{pra510}) 
and may be
parameterized by a single quantity $\xi_{CP}^t$ defined in 
Eq.~\ref{xicptchap5}.

\begin{figure}[ht]
\psfull
 \begin{center}
  \leavevmode
\epsfig{file=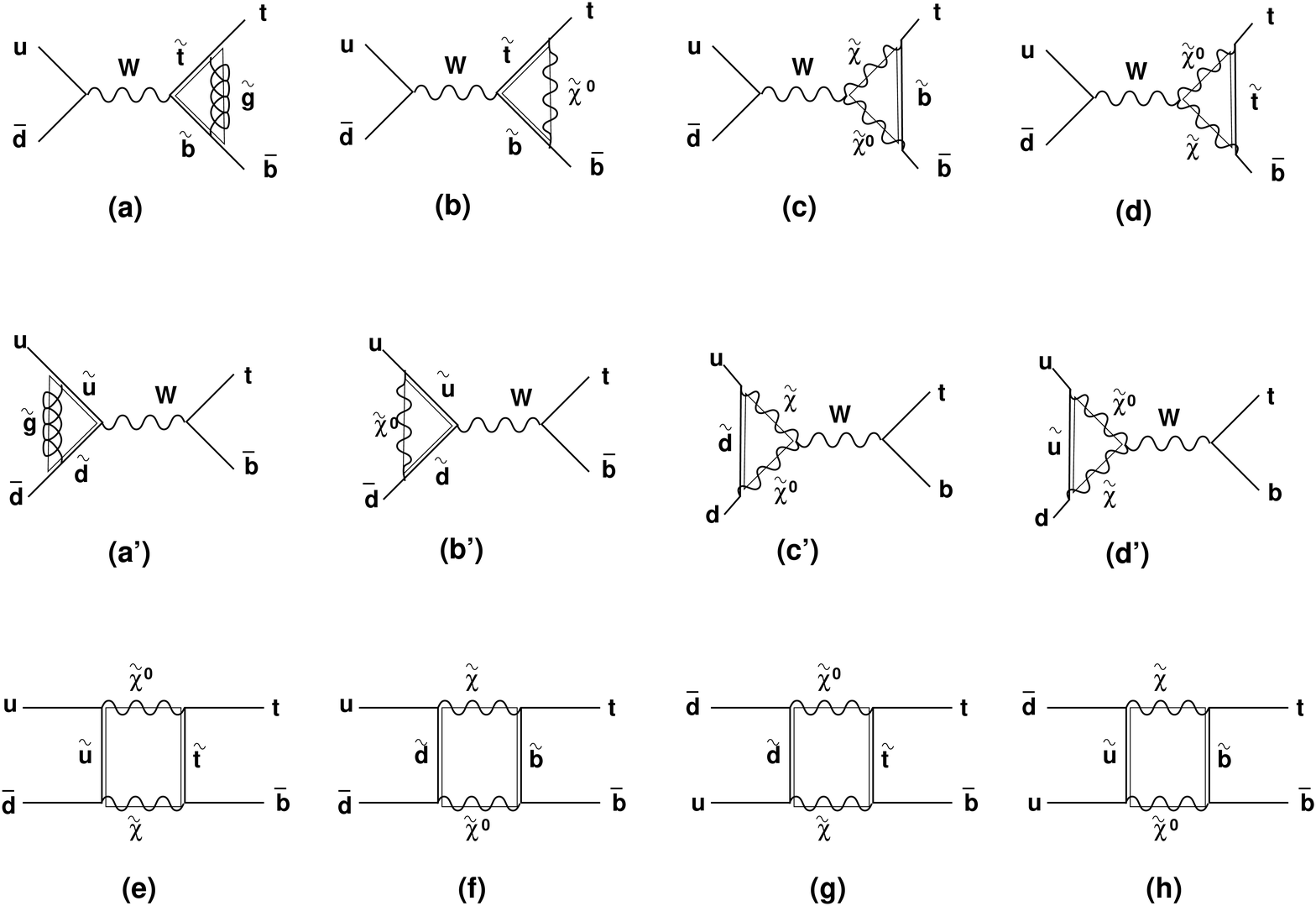
,height=9cm}
 \end{center}
\caption{\emph{CP-violating, SUSY induced, 1-loop diagrams for the processes
$u \bar d \to t \bar b$. $\tilde g$ is the gluino, 
${\tilde \chi}$ is a chargino, ${\tilde \chi}^0$ 
is a neutralino and ${\tilde t}$ and ${\tilde b}$ 
are top and bottom squarks, respectively.}}
\label{ppbartbfig4}
\end{figure}

In this scenario, when no further assumptions are made, there are
12 Feynman diagrams that can give rise to CP violation at the
parton level process $u \bar d \to t \bar b$ which are depicted in
Fig.~\ref{ppbartbfig4}. However, if we assume that
${\rm arg} (\mu) = 0$ as implied from the existing
experimental limit on the NEDM (see discussion in section \ref{sssec334}) 
and
take $m_u=m_d=m_b=0$, then only diagrams (a), (b), (d), (e) and
(g) have the necessary CP-violating phase, being proportional
to one quantity - $\xi_{CP}^t$. 
In \cite{prd57p1495} we have shown
that with $M_S$ and $m_G$ of about several hundred GeV, diagram (d)
is in fact responsible for $\sim 90\%$ of the total CP violation effect.

Let us, therefore, present the results for the asymmetry $A_0$
corresponding only to
diagram (d) in Fig.~\ref{ppbartbfig4}. The Feynman rules needed
to calculate $A_0$ can be derived from the following parts of
the SUSY Lagrangian  \cite{prd41p4464}:

\beqa
{\cal L}_{\tilde{u}_i d \tilde{\chi}_m} &=& \tilde{u}_i^+ \bar{d}   
\left[\left(-g_W Z^{1i}_u Z^{+*}_{1m} + \frac{\sqrt{2}m_u}{v_2}
Z^{2i}_u Z^{+*}_{2m}\right) \right.R \nonumber \\
&&+ \left. \frac{\sqrt{2}m_d}{v_1} Z^{1i}_u Z^-_{2m} L\right] V^{ud}
\tilde{\chi}_m^c + h.c. \label{ppbartbeq20}~,\\
{\cal L}_{\tilde{u}_i u\tilde{\chi}_n^0} &=&   \tilde{u}^-_i
\bar{\tilde{\chi}}^0_n
\left[\left(-\frac{g_W}{\sqrt{2}} Z^{1i*}_u L^+ -
\frac{\sqrt{2} m_u}{v_2} Z^{2i*}_u Z^{4n}_N \right) L\right. \nonumber \\
&&+\left. \left(\frac{2\sqrt{2}}{3} g_W \tan \theta_W Z^{2i*}_u Z^{1n*}_N -
\frac{\sqrt{2}m_u}{v_2}Z^{1i*}_u Z^{4n*}_N\right)R\right]u + h.c.
\label{ppbartbeq21}~, \nonumber \\
\\
{\cal L}_{W \tilde{\chi}_m\tilde{\chi}^0_n} &=& g_W \bar{\tilde{\chi}}_m
\gamma^\mu
\left( K^- L + K^+ R \right) \tilde{\chi}^0_n W^+_\mu + h.c.
\label{ppbartbeq22}~,
\eeqa

\ni where $L(R) = \frac{1}{2}(1-(+)\gamma_5)$, $\tilde{u}_i$
and $u$ ($\tilde{d}_j$ and $d$) stand for up squark and up quark (down squark
and down quark), respectively, $\tilde{\chi}_m$ ($m=1,2$) and 
$\tilde{\chi}^0_n$ ($n=1-4$)
are the charginos and neutralinos, respectively.
Also, in Eqs.~\ref{ppbartbeq22} and \ref{ppbartbeq21} we have defined

\beqa
L^{\pm} &\equiv& \frac{1}{3} \tan\theta_W Z_N^{1n} \pm Z_N^{2n}
\label{ppbartbeq23}~, \\
K^+ &\equiv& Z_N^{2n*} Z_{1m}^- + \frac{1}{\sqrt 2} Z_N^{3n*} Z_{2m}^-
\label{ppbartbeq24}
~, \\
K^- &\equiv& Z_N^{2n} Z_{1m}^{+*} - \frac{1}{\sqrt 2} Z_N^{4n} Z_{2m}^{+*}
\label{ppbartbeq25} ~,
\eeqa

\ni and the mixing matrices $Z_u, \ Z_d, \ Z_N, \ Z^-$ and $Z^+$ were
given in \cite{prd57p1495} (also given in section \ref{sssec332}).

Using the Lagrangian in Eqs.~\ref{ppbartbeq20} - \ref{ppbartbeq22},
the form factors ${\eRe}({\cal
P}_{1}^L + {\bar {\cal P}}_{1}^R)$ and ${\eRe}({\cal P}_{2}^L - {\bar
{\cal P}}_{2}^L)$ (corresponding to diagram (d) in
Fig.~\ref{ppbartbfig4}) can be evaluated within the MSSM 
and are given by

\beqa
{\eRe}({\cal P}_{1}^L + {\bar {\cal P}}_{1}^R) &=& - \frac{\alpha}{\pi
\sin^2\theta_W} m_t \left[  m_t {\cal O}_d^1 \Im{\rm m}\left( C^d_{12} + 
C^d_{23} \right)
+ m_{\tilde{\chi}_n^0} {\cal O}_d^2 \Im{\rm m} C^d_{12} \right. \nonumber \\
&& \left. - m_{\tilde{\chi}_m}
{\cal O}_d^3 \Im{\rm m} \left( C^d_{12} - C^d_{11}
\right) \right] \label{ppbartbeq26} ~, \\
{\eRe}({\cal P}_{2}^L - {\bar {\cal P}}_{2}^L) &=& \frac{1}{2}
\frac{\alpha}{\pi
\sin^2\theta_W} \left[{\cal O}_d^1 \left( (\hat s -m_t^2) 
\Im{\rm m} C^d_{23} -
\hat s \Im{\rm m} C^d_{22}  \right. \right. \nonumber \\
&& \left. \left. -2 \Im{\rm m}C^d_{24} -m_t^2 \Im{\rm m}C^d_{12} \right) 
\right. \nonumber \\
&& \left. - m_t m_{\tilde{\chi}_n^0} {\cal O}_d^2 \Im{\rm m} C^d_{12} + m_t
m_{\tilde{\chi}_m}
{\cal O}_d^3 \Im{\rm m} \left( C^d_0 + C^d_{12} \right) \right. \nonumber \\
&& \left. + m_{\tilde{\chi}_m} m_{\tilde{\chi}_n^0} {\cal O}_d^4 
\Im{\rm m} C^d_0 \right]
\label{ppbartbeq27} ~,
\eeqa

\ni where $\hat s=(p_t+p_{\bar b})^2$ and ${\cal O}_d^i$ contain the SUSY
CP-weak phases associated with diagram (d) in Fig.~\ref{ppbartbfig4}. In
fact, the same CP-violating phases occur also in the decay $t \to bW$ (see
section \ref{sssec514}) and, therefore, the ${\cal O}_d^i$ above are the
same as the ones given in Eqs.~\ref{od1chap5}-\ref{od4chap5}. The $\Im{\rm
m}C^d_x$, $x \in \left\{0,11,12,21,22,23,24 \right\}$, in
Eqs.~\ref{ppbartbeq26} and \ref{ppbartbeq27} are the imaginary parts
of the three-point form factors associated with
diagram (d) in Fig.~\ref{ppbartbfig4}.  Thus, $C^d_x$ are given via
\cite{prd57p1495}: 

\bea
C^d_x &=& C_x(m_{{\tilde t}_i}^2,m_{{\tilde\chi}_m}^2
,m_{{\tilde\chi}_n^0}^2,m_b^2,\hat s,m_t^2)~,
\eea

\ni and $C_x(m_1^2,m_2^2,m_3^2,p_1^2,p_2^2,p_3^2)$ is defined in appendix A.

\begin{figure}[htb]
\psfull
 \begin{center}
  \leavevmode
  \epsfig{file=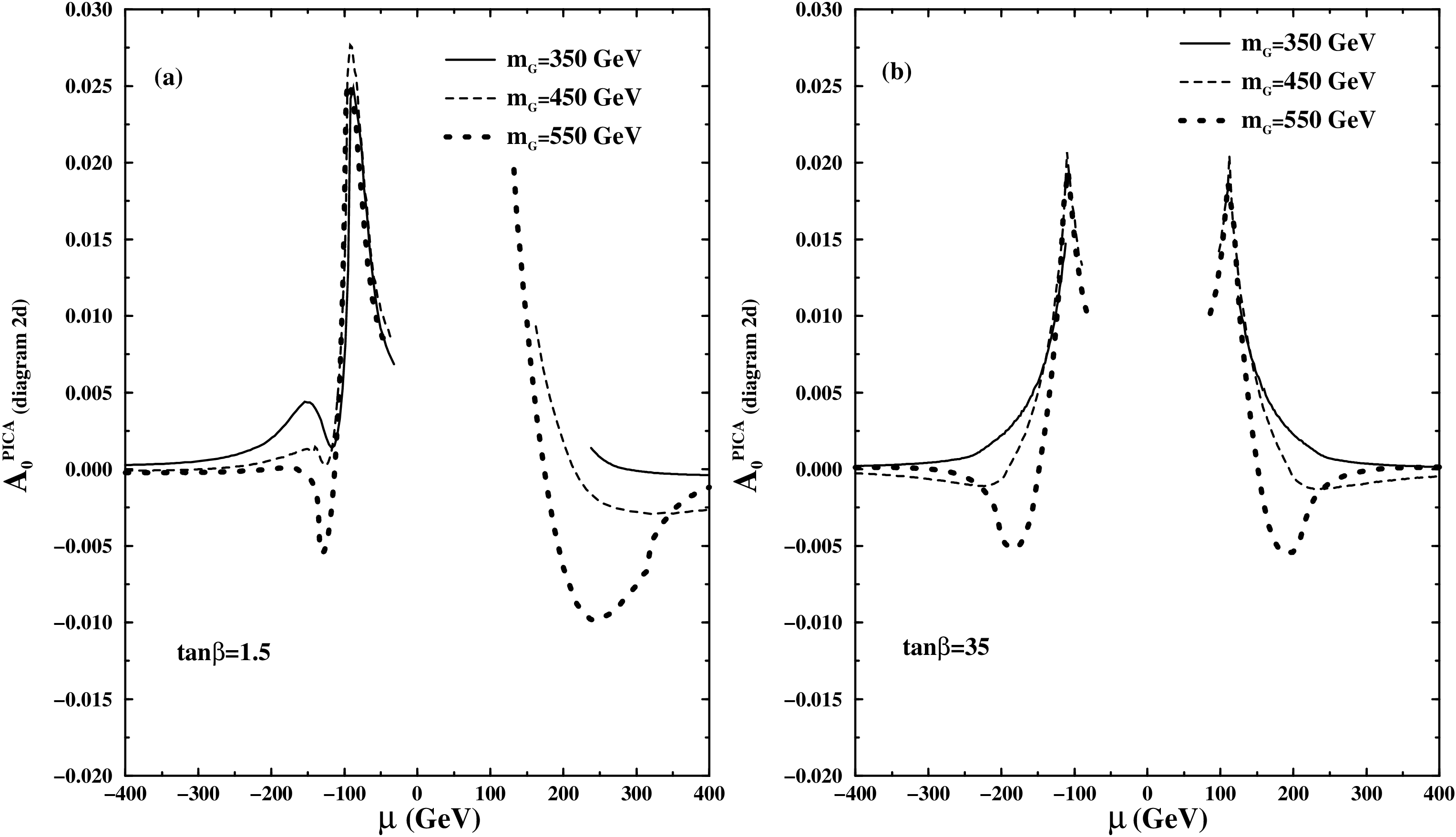
,height=6cm,width=6cm,bbllx=0cm,bblly=2cm,
bburx=20cm,bbury=25cm,angle=0}
 \end{center}
\caption{\emph{The SUSY induced partially integrated cross-section asymmetry 
$A_0^{\rm PICA}$ defined in the text, 
as a function of $\mu$, for $M_S=$400 GeV, $m_l=$50 GeV and for $\sqrt{s}=$2
TeV\null. With (a) $\tan\beta=$1.5 and (b) $\tan\beta=$35. Figure taken 
from \cite{prd57p1495}.}}
\label{ppbartbfig5}
\end{figure}

\begin{figure}[htb]
\psfull
 \begin{center}
  \leavevmode
  \epsfig{file=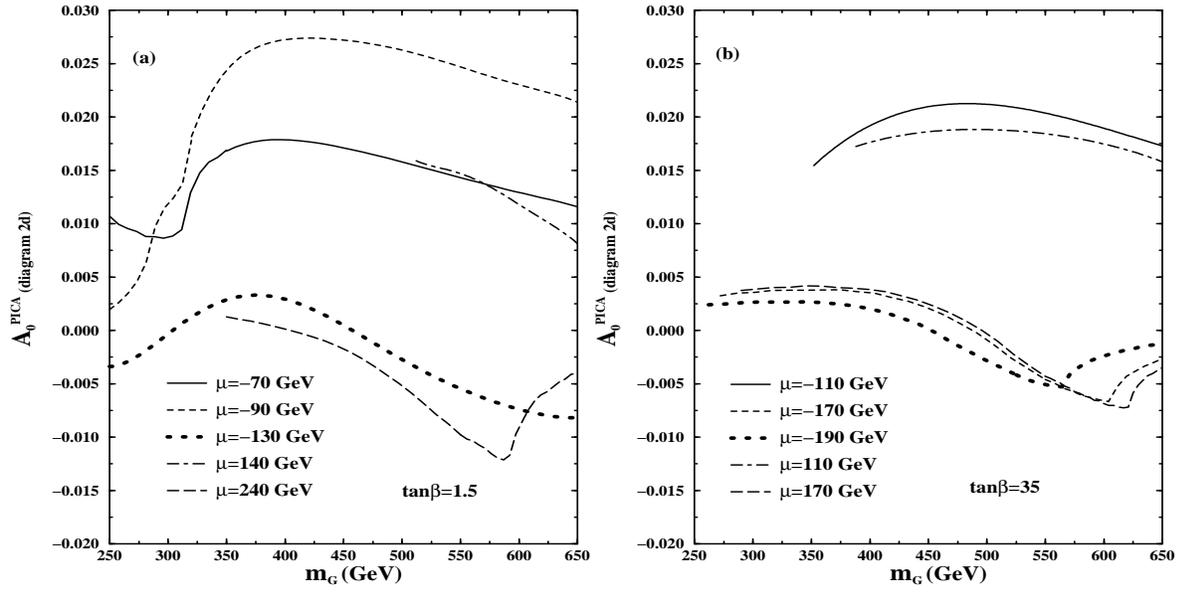
,height=6cm,width=6cm,bbllx=0cm,bblly=2cm,
bburx=20cm,bbury=25cm,angle=0}
 \end{center}
\caption{\emph{The SUSY induced partially integrated cross-section asymmetry
$A_0^{\rm PICA}$ defined in the text, 
as a function of $m_G$, for several values of $\mu$, $M_S=$400 GeV, $m_l=$50
GeV and for $\sqrt{s}=$2 TeV\null. With (a) $\tan\beta=$1.5 and (b)
$\tan\beta=$35. Figure taken 
from \cite{prd57p1495}.}}
\label{ppbartbfig6}
\end{figure}

In \cite{prd57p1495}, instead of calculating the 
cross-section asymmetry $A_0$, we
considered a partially integrated cross-section asymmetry, $A_0^{\rm
PICA}$, in which we have imposed a cut on the $tb$ invariant mass, 
$m_{tb} < 350$ GeV. Such a cut on $m_{tb}$ may help to remove the
$t \bar t$ ``background'' from a measurement of a cross-section
asymmetry in $p \bar p \to t \bar b +X$.
The results in the SUSY case are shown in Figs.~\ref{ppbartbfig5}
and \ref{ppbartbfig6} as a function of $\mu$ and $m_G$, respectively, 
for $M_S=400$
GeV, $m_l=50$ GeV ($m_l$ is the mass of the lighter stop particle) 
and for $\tan\beta=1.5,~35$. Maximal CP violation
was chosen in the sense that $\xi_{CP}^t=1$, thus the asymmetry 
plotted in Figs.~\ref{ppbartbfig5}
and \ref{ppbartbfig6}
is
in fact given in units of $\xi_{CP}^t$.
Evidently, for some values of $\mu$ around $100$ GeV
and with $m_G \sim 450$ GeV the asymmetry can almost reach the
$3\%$ level. The asymmetry is above the 1\% level for several
other choices of $\mu$. It was also shown in \cite{prd57p1495}
that, in general, in order for the asymmetry to be above the percent
level the mass of the lighter stop is required to be below $\sim 75$
GeV. Furthermore, the asymmetry tends to drop as $\tan\beta$ is increased in
the range $1 \lsim \tan\beta \lsim 10$, and it is almost insensitive
to $\tan\beta$ for $\tan\beta \gsim 10$.

In the MSSM, 1-loop radiative
corrections to the amplitude of top decay $t \to b W$ that can violate CP
are also present. In fact, disregarding the incoming $u$ and $d$ lines,
diagrams (a)-(d) in Fig.~\ref{ppbartbfig4} with the $t$ and $W$ momenta
reversed, can give rise to a CP-violating $tbW$ decay vertex. 
This was discussed in some detail in section \ref{sssec514}.
To 1-loop order in perturbation theory, where the CP-violating virtual
corrections enter only  either the production or the decay vertices
of the top in the overall reaction $p \bar p \to t \bar b +X \to W^+ b \bar
b +X$, and in the narrow width  approximation for the decaying top, an overall
CP asymmetry, $A$, can be broken into

\beq
A = A_P + A_D \label{ppbartbeq28}~.
\eeq

\ni In Eq.~\ref{ppbartbeq28} $A_P$ and $A_D$ are the CP
asymmetries emanating
from the production and decay of the top, respectively, and are defined
by

\beqa
&&A_P \equiv \frac{ \sigma(p \bar p \to t \bar b +X) - \bar\sigma(p \bar p
\to \bar t b +X) }{ \sigma(p \bar p \to t \bar b+X) + \bar\sigma(p \bar p
\to \bar t b+X) } \label{ppbartbeq29}~,\\
&&A_D \equiv \frac{ \Gamma(t \to W^+ b) - \bar\Gamma(\bar t \to W^- \bar b) }
{ \Gamma(t \to W^+ b) + \bar\Gamma(\bar t \to W^- \bar b) }
\label{ppbartbeq30}~.
\eeqa

\ni The PRA $A_D$, defined in Eq.~\ref{ppbartbeq30},
 does not depend on the specific production mechanism of the top
and was calculated in section \ref{sssec514}. We have shown there
that,
with the low-energy MSSM parameters described above, one gets
$|A_D| < 0.3 \%$ where, in most instances, it tends to be even 
smaller - $|A_D| < 0.1 \%$.
Therefore, it is about one order of magnitude smaller than the
asymmetry in the production of $tb$ and its relevance to the
overall asymmetry in $p \bar p \to t \bar b +X \to W^+ b \bar
b +X$ is negligible.

As a final remark here, let us recall 
that in the 2HDM case discussed before, 
the PRA $A_D$ in 
Eq.~\ref{ppbartbeq30} is 
forbidden at 1-loop order because of CPT invariance, i.e., 
no rescattering of final states as shown in section \ref{ssec23}. 
This was also discussed in section \ref{sssec512}.


\newpage
~

\newpage
~

\newpage
~

\newpage

\subsubsection{Feasibility of extraction from experiment \label{sssec823}}

To summarize the results of sections \ref{sssec821} and \ref{sssec822},
CP-violating asymmetries in single top production and decay at the
Tevatron through $p \bar p \to t \bar b +X \to W^+ b \bar b +X$ may
optimistically reach a few percent in extensions of the SM such as SUSY
and 2HDM's. In future upgrades of the Tevatron to $\sqrt s = 2$ TeV, the
cross-section for $p \bar p \to t \bar b +X$ is expected to be about 300
fb, if a cut of $m_{tb}<350$ GeV is applied on the invariant mass of the
$t \bar b$ \cite{prd57p1495}. Therefore, with an integrated luminosity of
${\cal L} = 30$ fb$^{-1}$ \cite{stelzer}, an asymmetry of $\sim 3\%$ can
be naively detected with a statistical significance of 3-$\sigma$. 
Therefore, a percent level CP-violating signal in the reaction $p \bar p
\to t \bar b +X \to W^+ b \bar b +X$ is especially notable as it may
become accessible at the near future 2 TeV $p \bar p$ collider. In
particular, based on the results presented in this section, such a
measurement may impose limits on the CP-violating parameters arg($A_t$)
and $b_t^h c^h$ of the MSSM and the 2HDM, respectively. However, it should
be noted that such a detection at the Tevatron will require the
identification of all $t\bar b$ pairs , which, in principle, can be
achieved only if the top can be reconstructed even when the $W$ decays
hadronically.

It will be also useful to explore SUSY or 2HDM mediated CP-violating
effects that can originate from the $W$-gluon fusion subprocess which
contributes to the same final state ({\it i.e.}, $W g \to t \bar b d$) and
which has a comparable production rate to that of the simple $u \bar d \to
t \bar b$ in the 2 TeV Tevatron. While in the MSSM various 1-loop triangle
and box corrections can give rise to CP nonconservation in the $W$-gluon
fusion subprocess, in the 2HDM CP-violating radiative corrections to $Wg$
fusion, at the 1-loop order, do not yield absorptive parts in the limit
$m_b=0$. Therefore, it will not contribute to CP asymmetries of the
$T_N$-even type in single top production.  Note, however, that the
$W$-gluon fusion subprocess has its own characteristics, e.g., the extra
light jet in the final state, which may be used in order to experimentally
distinguish it from the ``simple'' $ud$ fusion process (see e.g., Heinson
{\it et al.} in \cite{stelzer}).


\subsection{$p \bar p \to t \bar b h +X$, a case of tree-level CP
violation \label{ssec83}}

Motivated by the large, tree-level, CP-violating effects found in the the
reaction $e^+ e^- \to t \bar{t}h$ (see section \ref{ssec62}), we were led
to consider an analogous reaction in the Tevatron $p \bar{p}$ collider
with a $t\bar{b}h$ final state \cite{sthesis}. Thus, in this section we
focus on CP violation, driven by 2HDM in the process $p \bar{p} \to
t\bar{b}h + X$, where $h$ is the lightest neutral Higgs in the 2HDM of
type II\null. From the outset we remark that a statistically significant
CP study in the reaction $p \bar p \to t \bar{t}h +X$ in a future Tevatron
upgrade with c.m. energy of 2 TeV and even 4 TeV , is unlikely due to the
low $t\bar b h$  event rate. 

As in the case of $e^+ e^- \to t \bar{t}h$, a
very interesting feature of the reaction $u \bar{d} \to t \bar{b} h$
(at the parton level) is that it exhibits a CP asymmetry
 at the tree graph level. Such an effect arises from interference
of the Higgs emission from the $t$ (but not from the $\bar{b}$
in the limit $m_b \to 0$) with the Higgs emission from the
$W$-boson. Being a tree-level effect the resulting asymmetry is quite large.
This asymmetry may be measurable in principle, through
a CP-odd, $T_N$-odd observable.

\begin{figure}[ht]
\psfull
 \begin{center}
  \leavevmode
  \epsfig{file=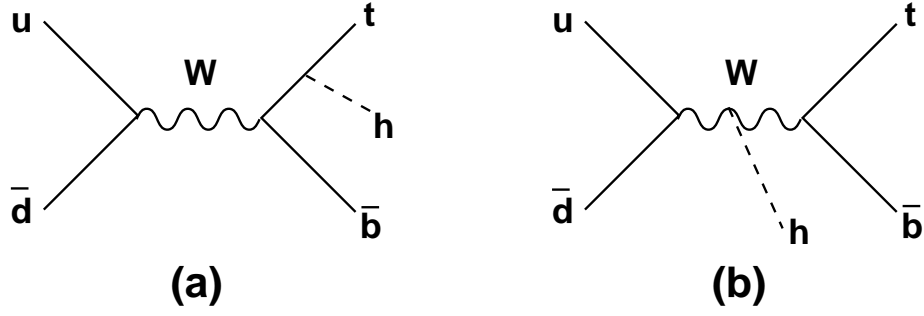
,height=4cm}
 \end{center}
\caption{\emph{Tree-level
Feynman diagrams contributing
to $u \bar d \to t\bar{b}h$ in the limit $m_d=m_u=m_b=0$;
$h$ stands for the lightest neutral Higgs in a 2HDM.}}
\label{ppbartbhfig1}
\end{figure}

Let us now discuss the tree-level cross-section and CP violation effects in our
reactions,

\beqa
u(p_u)\bar d(p_{\bar d}) \to t(p_t)  \bar b(p_{\bar b})  h(p_h) ~,~
\bar u(p_{\bar u})d(p_d) \to
\bar t(p_{\bar t})  b(p_b)  h(p_h) \ . \label{ppbartbheq1}
\eeqa

\ni In the limit $m_b = 0$ (and also $m_u=m_d=0$), and to lowest order,
the only two diagrams that can contribute to CP violation in the
reactions of Eq.~\ref{ppbartbheq1} are depicted in
Fig.~\ref{ppbartbhfig1}. The relevant Feynman rules needed to calculate
the tree-level CP asymmetry are extracted from the Lagrangian
in Eqs.~\ref{2hdmab} and \ref{2hdmc}.
Here also, we assume that two of the three neutral Higgs particles
are much heavier than the remaining one, {\it i.e.}\ $h$.
We, therefore, omit the index $k$ in Eqs.~\ref{2hdmab} and \ref{2hdmc},
and denote the couplings for the lightest neutral Higgs 
as: $a_t^h$, $b_t^h$ and $c^h$.  The
tree-level differential cross-section $\hat\Sigma^0$ at the parton level,
is a sum of two terms: the CP-even and CP-odd terms $\hat\Sigma^0_+$ and
$\hat\Sigma_-^0$\ , respectively,  

\beqa 
 \hat\Sigma^0\equiv \hat\Sigma_+^0 + \hat\Sigma^0_-
\label{ppbartbheq2}~,
\eeqa

\ni where $\hat\Sigma^0_\pm$ are
calculated from the tree-level diagrams in Fig.~\ref{ppbartbhfig1}.
The expression for the CP-even part, i.e.,
$\hat\Sigma^0_+$, can be parameterized as

\beqa
\hat\Sigma^0_+ =&&
\left( \frac{2 \pi \alpha }{\sin^2 \theta_W} \right)^3
\left( \frac{\Pi_W}{2} \right)^2 \frac{m_t^2}{m_W^2}
\times \left[ (a_t^h)^2 \Pi_t^2 {\cal A} +
(b_t^h)^2 \Pi_t^2 {\cal B} \right. \nonumber \\
&& \left.+ 2 (c^h)^2 \frac{m_W^2}{m_t^2} \Pi_{Wh}^2
{\cal C} + 2^{3/2} a_t^h c^h \Pi_t \Pi_{Wh} {\cal D} \right]
\label{ppbartbheq3}~,
\eeqa

\ni where the terms $\cal A,\cal B,\cal C$ and $\cal D$ are quite
involved and were calculated in \cite{sthesis}. $\Pi_W$ is the 
$W$-boson propagator, and together with $\Pi_t$ and $\Pi_{Wh}$ are given by

\beqa
&& \Pi_W \equiv \frac{1}{\hat s - m_W^2} \ , \
 \Pi_t\equiv \frac{1}{2p_t\cdot p_{h}+m^2_{h}} \nonumber \\
&&\Pi_{Wh}\equiv \frac{1}{m_t^2 - m_W^2 + 2 p_t \cdot p_{\bar b}}
\label{ppbartbheq4} ~.
\eeqa

\ni Furthermore, $p \equiv p_u + p_{\bar d}$ is
the $\hat s$-channel 4-momentum at the quark level,
and $\hat s$ is defined to be $\hat s = p^2$.

The CP-violating piece of the tree-level differential
cross-section is \cite{sthesis}:

\beq
\hat\Sigma^0_- = 2^{3/2}
\left( \frac{2 \pi \alpha }{\sin^2 \theta_W} \right)^3
\frac{m_t^2}{m_W^2} \Pi_W^2 \Pi_{Wh} \Pi_t b_t^h c^h \times
\epsilon(p_{\bar b},p_{\bar d},p_t,p_u) \times
(f - s_t +w) \label{ppbartbheq5} ~,
\eeq

\ni where $s_t\equiv(p_t+p_{\bar{b}})^2$, $f\equiv (p_u -
p_{\bar d})\cdot(p_t+p_{\bar{b}})$,
$w\equiv(p_u + p_{\bar d})\cdot(p_t+p_{\bar{b}})$ and $\epsilon$ is the 
Levi-Civita tensor.

\begin{figure}[htb]
\psfull
 \begin{center}
  \leavevmode
  \epsfig{file=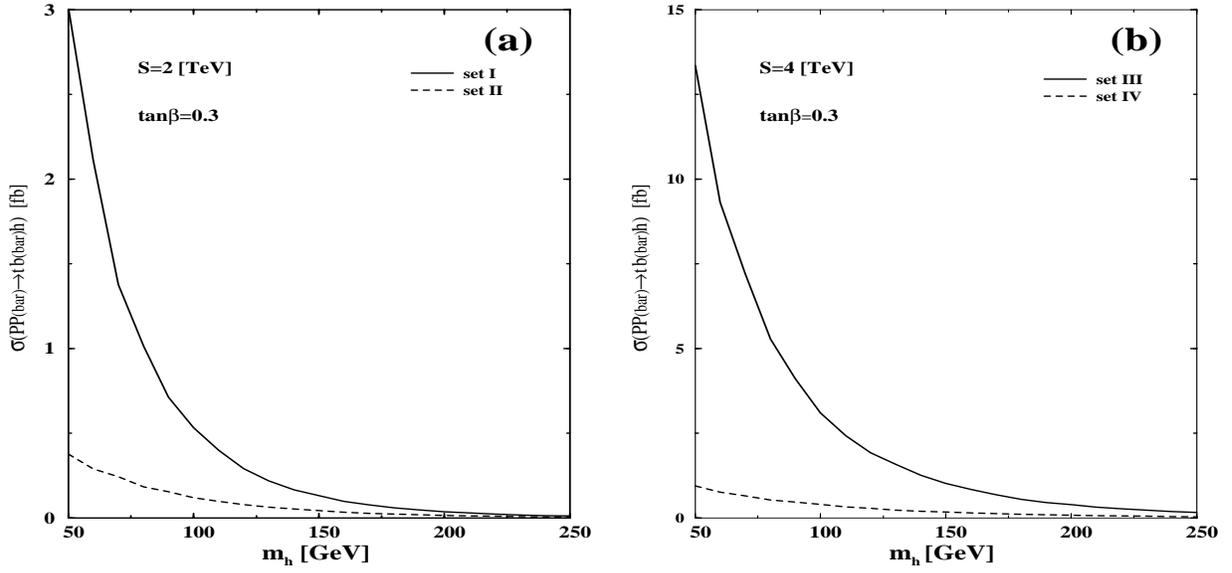
,height=8cm,width=9cm,bbllx=0cm,bblly=2cm,
bburx=20cm,bbury=25cm,angle=0}
 \end{center}
\caption{\emph{The cross-section for the reaction
$p \bar p \to t\bar{b}h + X$ (in fb), for (a): $\sqrt s= 2$ TeV and for
sets I (solid line) and II (dashed line), 
(b): $\sqrt s= 4$ TeV and for sets III
(solid line) and IV (dashed line). 
For the definition of the sets I,II,III and IV, see text.}}
\label{ppbartbhfig2}
\end{figure}

For illustration, we adopt here also the value tan$\beta=0.3$.
We fold in the structure functions of the $u$ and the $\bar d$
inside the $p$ and $\bar p$, respectively, and plot in Fig.~\ref{ppbartbhfig2}
the tree-level cross-section for $p \bar p \to t\bar{b}h +X$, with c.m.\
energies of $\sqrt s=2$ and $\sqrt s=4$ TeV\null. Four possible sets of
the Higgs coupling constants $a_t^h, \ b_t^h$ and $c^h$ were chosen.
For illustrative purposes,  the first two sets of parameters which are
chosen for a 2 TeV collider are: set I with $\tan\beta = 0.3$,
$\left\{\alpha_1,\alpha_2,\alpha_3 \right\} = 
\left\{\pi/4,\pi/2,3\pi/4 \right\}$
and set II with $\tan\beta = 0.3$,
$\left\{\alpha_1,\alpha_2,\alpha_3 \right\} = 
\left\{\pi/2,\pi/2,0 \right\}$.
The other two, chosen for a 4 TeV collider are: set III
with $\tan\beta = 0.3$,
$\left\{\alpha_1,\alpha_2,\alpha_3 \right\} = 
\left\{\pi/4,\pi/2,\pi/2 \right\}$
and set IV with $\tan\beta = 0.3$,
$\left\{\alpha_1,\alpha_2,\alpha_3 \right\} = 
\left\{\pi/2,3\pi/4,3\pi/4 \right\}$.
The general feature of these
sets is that sets I and III give rise to a large CP asymmetry but
``small'' cross-section, while sets II and IV increase the event rate
but decrease the asymmetry. Also, note that each set by itself is not
unique, as there are other values of the angles 
$\alpha_1,\alpha_2$ and $\alpha_3$
for each set which lead to the same effect.

%
%
%

As in the reaction $e^+ e^- \to t \bar t h$ discussed in section
\ref{ssec62}, we are dealing here with a tree-level CP-violating effect.
Thus the CP-violating term ${\hat {\Sigma}}^0_-$ can probe only CP
asymmetries of the $T_N$-odd type. In this case the final state is not a
CP eigenstate. Therefore, one has to construct a $T_N$-odd, triple
correlation product (or equivalently, a Levi-Civita) which takes into
account the conjugate reaction as well ($\bar u d \to \bar t b h$) , thus
endowing the observable with definite CP properties. This led us to
consider the following CP-odd, $T_N$-odd observable

\beq
O = (\epsilon(p_u,p_{\bar b},p_t,p_h) + \epsilon(p_{\bar u},p_b,p_{\bar t},
p_h))/s^2 \label{ppbartbheq6}~.
\eeq

\begin{figure}[htb]
\psfull
 \begin{center}
  \leavevmode
  \epsfig{file=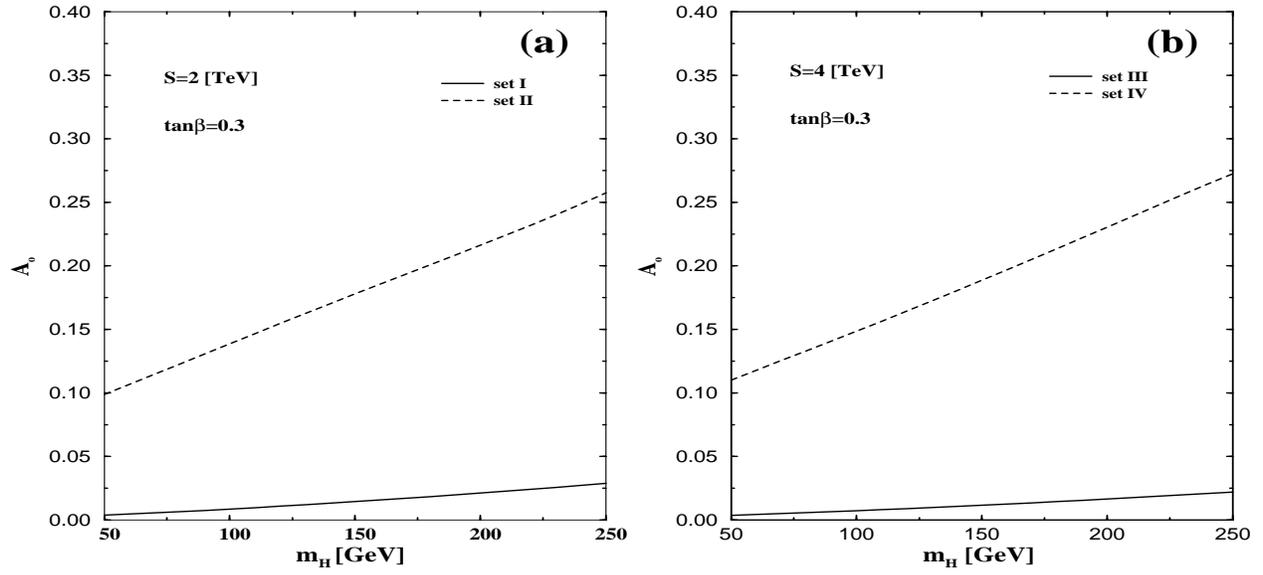,
height=8cm,width=9cm,bbllx=0cm,bblly=2cm,
bburx=20cm,bbury=25cm,angle=0}
 \end{center}
\caption{\emph{The asymmetry, $A_O$ (see text) as a
function of $m_h$ for (a): $\sqrt s=$2
TeV with sets I (solid line) and II (dashed line), (b):
$\sqrt s=$4 TeV with sets III (solid line) and IV (dashed line).  
For the definition of the sets I,II,III and IV,
see text.}}
\label{ppbartbhfig3}
\end{figure}

Fig.~\ref{ppbartbhfig3} shows the results for the signal to noise 
ratio, i.e., the asymmetry $A_O \equiv
<O>/ \sqrt {<O^2>}$, for $\sqrt s=2$ TeV with sets I and II and for $\sqrt
s=4$ TeV with sets III and IV\null. Evidently, 
the asymmetry $A_O$ is of the order of 10-15\%
for a light Higgs particle in the mass range 
$50 ~{\rm GeV} < m_h < 100$ GeV and of
the order of  20-30\% for a heavy Higgs particle with mass in the range 
$200 ~{\rm GeV} <
m_h < 250$ GeV, for both set II (which corresponds to a 2 TeV
collider) and set IV (which we chose for the 4 TeV collider). Sets I
and III give asymmetries of the order of a few percent. Although with
these sets, i.e., sets I and III, the cross-section can be 10
times larger than that corresponding  to sets II and IV, the
statistical significance of the CP-violating effect that can be
achieved when the free parameters of the 2HDM are chosen according to
sets I and III is much smaller than that with sets II and IV\null. This
is simply due to the fact that the number of events needed to detect
the CP-violating effect scale as (asymmetry)$^{-2}$. Therefore, the
enhanced effect for   sets II and IV makes up for the reduced
production rate in those scenarios. 

Let us proceed by analyzing the two scenarios which give the large
asymmetries. For the reaction at hand, 
the statistical significance $N_{SD}$ of
the CP-odd signal in the collider is

\beq
N_{SD} = \sqrt{{\cal L}} \sqrt {\sigma (p \bar p \to t\bar{b}h + X)}
\times A_O \label{ppbartbheq7}
\eeq

\ni where ${\cal L}$ is the collider luminosity. From
Fig.~\ref{ppbartbhfig2} we see that $\sigma (p \bar p \to t\bar{b}h +
X) \sim 0.1 -10$ fb, depending on the parameters $a_t^h,b_t^h$ and $c^h$
and the neutral Higgs mass. Thus, since $A_O \sim 0.1 -0.3$
in the best cases, it is evident from Eq.~\ref{ppbartbheq7} that,
typically, an integrated luminosity of about $\sim 100$ fb$^{-1}$
will be required to be able to observe a statistically significant
 effect in this reaction.
Therefore, although the CP asymmetry in this process
could reach the 10-30\% level, it is, unfortunately, 
not likely to be able to produce a CP-violating 
signal in the next runs of the Tevatron
with ${\cal L} =2$ fb$^{-1}$ and even with 30 fb$^{-1}$.

\newpage
~

\newpage
~

\pagebreak

\section{CP violation in \boldmath ${\gamma \gamma}$ 
collider experiments \label{sec9}}
\setcounter{equation}{0}

Future electron-positron colliders include the attractive option 
of a linear $\gamma \gamma$ collider, where each beam of photons 
is produced by Compton backscattering of laser light on an electron 
or positron beam. The peak energies and luminosities of the $\gamma
\gamma$   are expected to be slightly smaller than those 
of the corresponding $e^+e^-$ collider. 
The idea was originally suggested in \cite{ggreview1,ggreview2}; for 
recent reviews see \cite{ggreview3}.  
The attractive option of obtaining polarized photon beams is also being
considered.  Note also that $\gamma \gamma$ collisions
have been discussed in the context of heavy ion colliders (for a review
see  \cite{hepph9804348}) as well. Unfortunately, 
the invariant $\gamma \gamma$ mass reach
for the LHC (running in its heavy ion mode), will only be about 100
GeV, with lower values attainable at RHIC \cite{hepph9804348}. 
Therefore, this option will
not be discussed here.

\subsection{$\gamma \gamma \to X$: general comments \label{ssec91}}

In a $\gamma \gamma$ collider there are two distinguishable modes: 
unpolarized and polarized incoming photons. In the unpolarized case, 
in order to be able to detect CP violation in 
the reaction $\gamma \gamma \to t \bar t$ one needs information 
on the spins of the $t$ and the $\bar t$, or 
equivalently, one needs to construct asymmetries involving the 
decay products of the top quark. In this case, including 
the subsequent decays of the $t$ and $\bar t$, one 
can break the differential cross-section for the process 

\begin{eqnarray}
\gamma_1(k_1,\lambda_1) \gamma_2(k_2,\lambda_2) \to t(p_t,s_t) 
\bar t(p_{\bar t},s_{\bar t}) \label{ggtottgen}~,
\end{eqnarray}

\noindent to its CP-odd and CP-even parts as

\begin{eqnarray} 
&& d \sigma \equiv d \sigma^{odd} + d \sigma^{even} \label{ggtteq1}~.
\end{eqnarray}

\noindent In Eq.~\ref{ggtottgen} $\lambda_1,~\lambda_2$ denote 
the helicities of the incoming photons $\gamma_1,~\gamma_2$, respectively, 
and $s_t,~s_{\bar t}$ are the covariant spins of $t,~ \bar t$, 
respectively.   

In general, the CP-odd terms in Eq.~\ref{ggtteq1} has the form

\begin{eqnarray}
d \sigma^{odd} 
\propto 
\delta_- \vec {q} \cdot (\vec {s}_t \times \vec {s}_
{\bar t}) +
\delta_+ \vec {q} \cdot (\vec {s}_t - \vec {s}_{\bar t}) \label{ggtteq2} ~,
\end{eqnarray}

\noindent where $\vec {q}$ is a three momentum 
of any of the particles in the 
final or the initial state (i.e., $\vec {q}=\vec {k}_1,~
\vec {k}_2,~\vec {p}_t$ or 
$\vec {p}_{\bar t}$) and $\vec {s}_t(\vec {s}_{\bar t})$ is 
the spin three vectors of the $t(\bar t)$.
$\delta_-$ and $\delta_+$ are non-zero only if there is 
CP violation in the underlying dynamics of the process 
$\gamma_1 \gamma_2 \to t \bar t$. 
Furthermore, $\delta_-$ 
is proportional to the dispersive, CP-odd, $T_N$-odd 
contributions, while $\delta_+$ gets its contribution from 
absorptive, CP-odd, $T_N$-even terms.   
Of course, as mentioned above, the CP-odd polarization 
correlations of the top and the anti-top in Eq.~\ref{ggtteq2} 
will lead to CP-odd correlations among the momenta of 
the decay products of the $t$ and the $\bar t$. Asymmetries 
which involve the top decay products in the case of the unpolarized 
photons were investigated in \cite{hepph9602273,prd52p3803}.    

If the incoming photons are polarized, then one can 
construct CP-odd correlations by using linearly 
polarized photons where no information on the momenta 
and polarization of the top quark decay products are 
needed.  The amplitude squared for a general final state 
$X$, i.e., $\gamma \gamma \to X$, in the case where the 
two photons are linearly  polarized is given by \cite{hepph9704312}:

\begin{eqnarray}
d \Sigma (\eta,\bar{\eta};\chi,\phi)
  &=&\Sigma_{\rm unpol}
  -\frac{1}{2}[\eta\cos(\phi+\chi)+\bar{\eta}\cos(\phi-\chi)]
   \eRe(\Sigma_{02})\nonumber\\  
&+&\frac{1}{2}[\eta\sin(\phi+\chi)-\bar{\eta}\sin(\phi-\chi)]
   \mIm(\Sigma_{02})\nonumber\\  
&-&\frac{1}{2}[\eta\cos(\phi+\chi)-\bar{\eta}\cos(\phi-\chi)]
\eRe(\Delta_{02})\nonumber\\  
&+&\frac{1}{2}[\eta\sin(\phi+\chi)+\bar{\eta}\sin(\phi-\chi)]
   \mIm(\Delta_{02}) \nonumber\\  
&+&\eta\bar{\eta}\cos(2\phi)\eRe(\Sigma_{22})
+\eta\bar{\eta}\sin(2\phi)\mIm(\Sigma_{22}) \nonumber\\
&+&\eta\bar{\eta}\cos(2\chi)\eRe(\Sigma_{00})
+\eta\bar{\eta}\sin(2\chi)\mIm(\Sigma_{00}).
 \label{ggtteq3}
\end{eqnarray}

\noindent Here $\eta,\bar {\eta}$ are the degrees of linear
polarization of the  two initial photons, $\chi$ and $\phi$ are the
azimuthal  angle difference and sum, respectively, and the invariant 
functions are defined as \cite{hepph9704312}:

\begin{eqnarray}
\Sigma_{\rm unpol} &=& \frac{1}{4}\sum_X
              \left[|M_{++}|^2+|M_{+-}|^2
                   +|M_{-+}|^2+|M_{--}|^2\right]  \label{ggtteq004} ~,\\
\Sigma_{02} &=& \frac{1}{2}\sum_X\left[M_{++}(M^*_{+-}+M^*_{-+})
             +(M_{+-}+M_{-+})M^*_{--}\right] \label{ggtteq04}~,\\
\Delta_{02} &=& \frac{1}{2}\sum_X\left[M_{++}(M^*_{+-}-M^*_{-+})
             -(M_{+-}-M_{-+})M^*_{--}\right] ~,\\
\Sigma_{22} &=& \frac{1}{2}\sum_X(M_{+-}M^*_{-+}) ~,\\
\Sigma_{00} &=& \frac{1}{2}\sum_X(M_{++}M^*_{--}) ~.
\label{ggtteq4}
\end{eqnarray}
The subscripts $0$ and $2$ in Eqs.~\ref{ggtteq04} - \ref{ggtteq4} 
represent the magnitude of the 
sum of the initial photon helicities  and 
the notation for the helicity amplitudes for the reaction 
$\gamma_1 (\lambda_1) \gamma_2 (\lambda_2) \to X$  
using Eqs.~\ref{ggtteq004} - \ref{ggtteq4} is

\begin{eqnarray}
M_{\lambda_1\lambda_2}=\langle X|M|\lambda_1\lambda_2\rangle \label{ggtteq5}~.
\end{eqnarray}

\noindent Furthermore, the event rate of any final state 
production through $\gamma \gamma$ fusion can be written 
in general as \cite{plb294p361}: 

\begin{eqnarray}
dN=d{\cal L}_{\gamma \gamma} \sum_{i,j=0}^3 <\xi_1^{(i)} \xi_2^{(j)}> 
d \sigma^{ij} 
\label{ggtteq6}~,
\end{eqnarray}

\noindent where $\xi_1^{(i)}$($\xi_2^{(j)}$) 
are the so called Stokes polarization 
parameters for $\gamma_1$($\gamma_2$) with $\xi_1^{(0)}=\xi_2^{(0)}=1$. 
In particular, $\xi_1^{(2)}$ and $\xi_2^{(2)}$ are the mean helicities 
of $\gamma_1$ and $\gamma_2$, respectively, and 
$\sqrt {(\xi_i^{(1)})^2 + (\xi_i^{(3)})^2}$ are their degrees of 
linear polarization. Also, $d{\cal L}_{\gamma \gamma}$ 
is the luminosity of the two photons and $\sigma^{ij}$ are the 
corresponding cross-sections.

There are only three CP-odd functions out of the nine invariant 
functions in Eqs.~\ref{ggtteq004} - \ref{ggtteq4}: $\mIm(\Sigma_{02})$, 
$\mIm(\Sigma_{00})$ and $\eRe(\Delta_{02})$. While 
$\mIm(\Sigma_{02})$ and $\mIm(\Sigma_{00})$ are 
$T_N$-odd, $\eRe(\Delta_{02})$ is $T_N$-even. 
A CP-odd asymmetry can be formed at a $\gamma\gamma$ collider
if, for example, the $J=0$ amplitudes of two photons in the
CP-even and CP-odd states are both non-vanishing:

\begin{eqnarray}
&&{\cal M}\left[\gamma\gamma\rightarrow X({\rm CP}=+)\right]
 \propto \vec{\epsilon}_1\cdot\vec{\epsilon}_2~,  \nonumber \\
&&{\cal M}\left[\gamma\gamma\rightarrow X({\rm CP}=-)\right]
 \propto (\vec{\epsilon}_1\times\vec{\epsilon}_2)\cdot
         \vec{k}_1 \label{ggtteq7}~,
\end{eqnarray}

\noindent where $\vec{\epsilon}_1$ and $\vec{\epsilon}_2$ are 
the polarizations of
the two colliding photons  and $\vec{k}_1$ is the momentum
vector of one photon in the $\gamma\gamma$ c.m.\ frame. Such an asymmetry
can be constructed, for example, by taking the 
difference of distributions at $\chi=\pm\pi/4$
\cite{hepph9704312,plb359p369}:

\begin{eqnarray}
A_{00} \equiv \frac{\mIm(\Sigma_{00})}{\Sigma_{\rm unpol}}
= \frac{\int^{4\pi}_0{\rm d}\phi
\bigg[\left(\frac{{\rm d}\sigma}{{\rm d}\phi}\right)_{\chi=\frac{\pi}{4}}
    -\left(\frac{{\rm d}\sigma}{{\rm d}\phi}\right)_{\chi=-\frac{\pi}{4}}
\bigg]}{\int^{4\pi}_0{\rm d}\phi 
\bigg[\left(\frac{{\rm d}\sigma}{{\rm d}\phi}\right)_{\chi=\frac{\pi}{4}}
    +\left(\frac{{\rm d}\sigma}{{\rm d}\phi}\right)_{\chi=-\frac{\pi}{4}}
\bigg]} \label{ggtteq8}~.
\end{eqnarray}

\noindent Alternatively, in terms of event rates which correspond to 
the (0,0) and the unpolarized initial photon-photon states this reads

\begin{eqnarray}
A_{00} = \frac{N_{00}}{N_{\rm unpol}} \label{ggtteq9}~.
\end{eqnarray}


\subsection{$\gamma \gamma \to t \bar t$ and the top EDM \label{ssec92}}

Recall that a top EDM, i.e., $d_t^\gamma$, 
modifies the SM $t \bar t \gamma$ coupling to read 

\begin{eqnarray}
\Gamma_\mu\;= \;i e \f{2}{3}\,\gamma_\mu\;+
\;i d_t^\gamma\,\sigma_{\mu \nu}\,
\gamma_{\small 5}\,(p_t\,+\,p_{\bar{t}})^{\nu} 
\label{ggtteq10} ~,
\end{eqnarray}

\noindent and CP violation arising due to this EDM of the 
top can be studied in the reaction $\gamma_1 \gamma_2 \to t \bar t$ 
of Eq.~\ref{ggtottgen}. 

%
%
%
%

\begin{figure}[htb]
\psfull
\begin{center}
\leavevmode
\epsfig{file=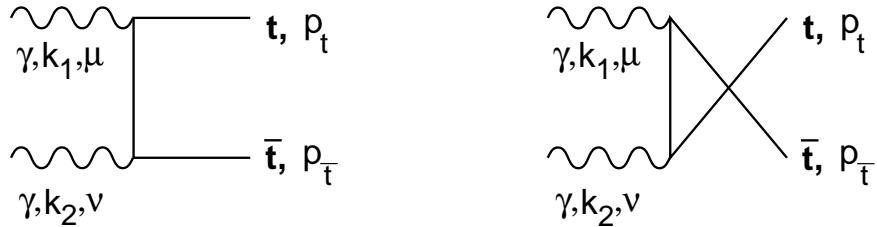,
height=3cm}
\end{center}
\caption{\emph{
Lowest order Feynman diagrams for the process 
$\gamma_1^\mu (k_1) \gamma_2^\nu(k_2) \to t(p_t) \bar t(p_{\bar t})$. 
$\mu$ and $\nu$ are Lorenz indices
}}
\label{ggttfig1}
\end{figure}

The relevant lowest order Feynman diagrams for $\gamma \gamma \to t \bar t$ 
are shown in Fig.~\ref{ggttfig1} wherein the top EDM 
can be folded into any of the $t \bar t \gamma$ vertices in 
those two diagrams. We note however again that, in general,
the CP-violating effects in $\gamma \gamma \to t \bar t$ can not
necessarily be all attributed to the top  quark EDM\null. 
For example, in a 2HDM (see next section)   
additional box diagrams can give rise to CP-nonconserving terms in the 
amplitude of the reaction $\gamma \gamma \to t \bar t$.

With the notation $M(\lambda_1,
\lambda_2,\lt,\ltb)$ for the amplitude,  where 
$\lambda_1,\,\,\lambda_2,\,\,\lt$ and 
$\ltb$ correspond to the helicities of the two incoming photons, the top quark 
and the top anti-quark, respectively, the non-vanishing helicity amplitudes 
for the process in Eq.~\ref{ggtottgen}, obtained 
for combinations such as $\lambda_2=\lambda_1=\lambda_\gamma$ 
or $\lambda_2=-\lambda_1=\lambda_\gamma$ and 
$\lambda_{\bar t} = \lambda_{t}$ or $\lambda_{\bar t} = - \lambda_{t}$, 
are given by \cite{hepph9704312,plb359p369,hepph9709225}:

\begin{eqnarray}
M(\lg,~\lg,~\lt,~\lt)&=& -
         4C_t{m_t\over \sqrt{s}}
        \left\{(\lg+\lt \bt) \right.
\nonumber\\ 
       && -i\,d_t^\gamma\;2m_t\left[2+\f{s}{4m_t^2}\bt
        (\bt-\lt \lg) \sin^2\theta_t\right] \non \\
        &&+(d_t^\gamma)^2\;\f{s\lg}{2}\,\left. 
        \left[\f{4 m_t^2}{s}+\bt(\bt-\lg \lt)
        \sin^2\theta_t\right]\right\} \non ~,\\
\\
M(\lg,~\lg,~\lt,-\lt)&=& -
        4m_t C_t
        \non \\
        &&\times \bt
        \sin\theta_t\;\cos\theta_t\;\left[ \lg\,i\,
        d_t^\gamma-m_t\,(d_t^\gamma)^2
        \right]~,\\
M(\lg,-\lg,~\lt,~\lt)&=& 
         4C_t{m_t\over \sqrt{s}}
        \non \\
        &&\times \left[ \lt\,\bt+i\,d_t^\gamma\;\f{s}{2m_t}\bt^2-
        (d_t^\gamma)^2\;\f{s}{2}\lt \bt \right]\;\sin^2\theta_t \non ~, \\
\\
M(\lg,-\lg,~\lt,-\lt)&=&
        2 C_t\bt
        \sin\theta_t
        \left\{\left( \lg \lt +\cos\theta_t\right.\right) \non \\
        &&\hspace{-1.5em} -(d_t^\gamma)^2\;\f{s}{2}\left. 
        \left[\f{4m_t^2}{s}\,\cos\tht+
        \lg \lt (1-\bt^2\cos^2\theta_t) \right]\right\} \non, \\
\label{ggtteq12}
\end{eqnarray} 

\noindent where $\theta_t$ is the scattering 
angle in the c.m.\ frame and $\beta_t$ is the 
top quark velocity and
$C_t=e^2Q_t^2/(1-\bt^2\cos^2\theta_t)$.

The CP-odd $T_N$-odd distribution 
$\mIm(\Sigma_{00})$ defined in Eq.~\ref{ggtteq4} 
depends linearly on $d_t^\gamma$ and is given by \cite{hepph9704312}:

\begin{eqnarray}
\mIm(\Sigma_{00}) = 24 \left( 1- \beta_t^2 \cos^2\theta_t \right) 
{\eRe}(d_t^\gamma) \label{ggtteq13} ~,
\end{eqnarray}

\noindent and the asymmetry $A_{00}$ defined in 
Eq.~\ref{ggtteq9} can then be calculated 
\cite{hepph9704312,plb359p369}. After extracting 
the top EDM from $A_{00}$ and defining 
$A_{00} \equiv {\eRe}(\tilde\kappa_\gamma) {\tilde A}_{00}$, 
where $\tilde\kappa_\gamma \equiv 2 m_t d_t^\gamma /e $ is a 
dimensionless EDM form factor, as in Eq.~\ref{dimdt}, 
one obtains the 
allowed sensitivity (i.e., $N_{SD}$ = number of standard deviations) to  
the dispersive part, $\eRe (\tilde\kappa_\gamma)$, in the case that 
no asymmetry is found 

\begin{eqnarray}
{\rm Max}(|{\eRe}(\tilde\kappa_\gamma)|)
   =\frac{\sqrt{2} N_{SD} }{|\tilde{A}_{00}\sqrt{\varepsilon N_{\rm unpol}}|}
\label{ggtteq14}~,
\end{eqnarray}
\noindent where $\varepsilon$ is the detection efficiency 
which was taken to be 10\% in \cite{hepph9704312}. 
The kinematics of the Compton backscattering process 
at hand is characterized in part by the dimensionless 
parameter $x \equiv 2p_e \cdot p_\gamma / m_e^2$. 
Larger $x$ values are favored to produce highly energetic 
photons but the degree of linear polarization is  
larger for smaller $x$ values (for more details see 
\cite{hepph9704312}). In particular, the denominator on the 
r.h.s. of Eq.~\ref{ggtteq14} depends on $x$, which for a 
given c.m. energy squared, $s$,  is bounded by

\begin{eqnarray}
\frac{2m_t}{\sqrt{s}-2m_t}\leq x \leq 2(1+\sqrt{2}) \label{ggtteq15}~,
\end{eqnarray} 

\noindent for the process $\gamma \gamma \to t \bar t$
where the upper bound is required to prevent $e^+e^-$ pair production in 
the scattering of the photon while the lower bound is required to have 
photons energetic enough to produce top pairs.

\begin{figure}[htb]
\psfull
\begin{center}
\leavevmode
\epsfig{file=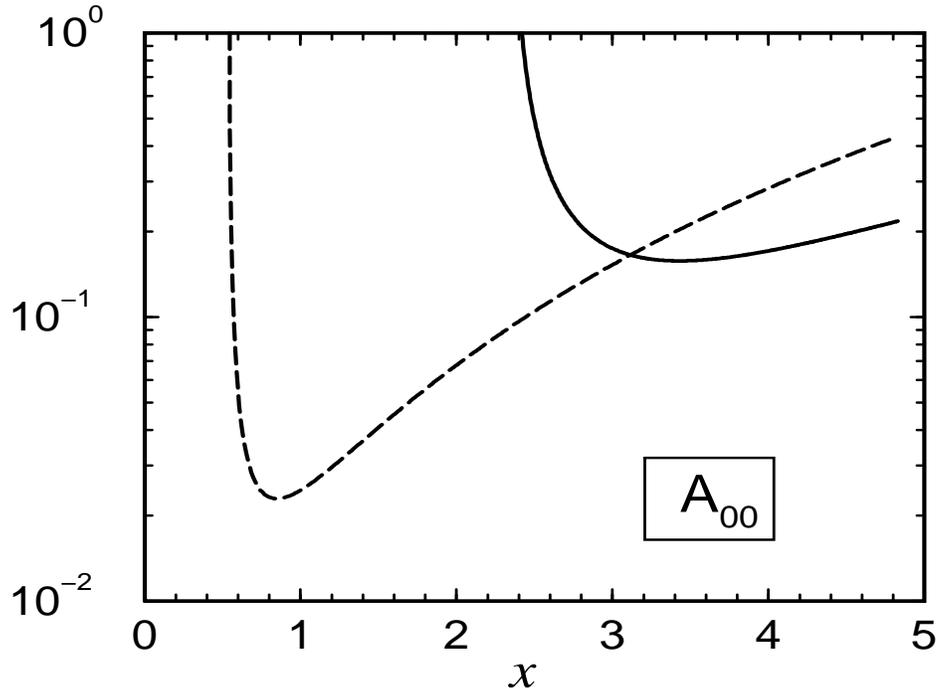,
height=10cm,width=10cm,bbllx=0cm,bblly=2cm,
bburx=20cm,bbury=25cm,angle=0}
 \end{center}
\caption{\emph{The $x$ dependence of the ${\eRe}(\tilde\kappa_\gamma)$ 
upper bound, i.e., 
${\rm Max}(|{\eRe}(\tilde\kappa_\gamma)|)$, 
at $\sqrt s=$0.5 TeV (solid line) and 1 TeV
(dashed line), 
from the asymmetry $A_{00}$. Figure taken from \cite{hepph9704312}.}}
\label{ggttfig2}
\end{figure}

In \cite{hepph9704312} the $x$ dependence of the
${\eRe}(\tilde\kappa_\gamma)$ upper bound, {\it i.e.} ${\rm
Max}(|{\eRe}(\tilde\kappa_\gamma)|)$, was given which is shown in
Fig.~\ref{ggttfig2} for two c.m.\ energies $\sqrt s=0.5$ and 1 TeV\null. 
We see that ${\rm Max}(|{\eRe}(\tilde\kappa_\gamma)|)$ can reach below 0.1
where the optimal sensitivities are obtained with $x=3.43$ and $x=0.85$
for $\sqrt s=0.5$ and 1 TeV, respectively. For these $x$ values, and to
1-$\sigma$, the upper bounds that can be achieved are: 
${\eRe}(\tilde\kappa_\gamma)=0.16$ and ${\eRe}(\tilde\kappa_\gamma)=0.02$
for $\sqrt s=0.5$ and 1 TeV, respectively.  This corresponds to
${\eRe}(d_t^\gamma) \approx 0.9 \times 10^{-17}$ and $0.1 \times 10^{-17}$
e-cm for $\sqrt s=0.5$ and 1 TeV, respectively.

Different type of asymmetries which involve  the 
polarization of both the initial photons beams and the 
decay products of the $t$ and $\bar t$ (e.g., 
$t \to b \,\ell \nu_{\ell}$) in 
the reaction $\gamma \gamma \to t \bar t$, were suggested 
in \cite{hepph9709225}. The first one is a charge asymmetry, 
$A_{{\rm ch}}$, which measures the difference between the number 
of leptons and anti-leptons produced as decay products of 
the top and anti-top, respectively. The second, $A_{{\rm FB}}$, 
is a sum of the forward-backward asymmetries of the leptons and anti-leptons 
and requires polarized laser beams.
In terms of the differential cross-sections these asymmetries are given by

\begin{eqnarray}
A_{{\rm ch}}(\theta_0)=\frac{
{\displaystyle          \int_{\theta_0}^{\pi-\theta_0}}d\theta_{\ell}
{\displaystyle          \left( \frac{d\sigma^+}{d\theta_{\ell}}
                -       \frac{d\sigma^-}{d\theta_{\ell}}\right)}}
{
{\displaystyle          \int_{\theta_0}^{\pi-\theta_0}}d\theta_{\ell}
{\displaystyle          \left( \frac{d\sigma^+}{d\theta_{\ell}} +
\frac{d\sigma^-}{d\theta_{\ell}}\right)}}
\label{ggtteq16}
\end{eqnarray}

\noindent and

\begin{eqnarray}
A_{{\rm FB}}(\theta_0)= \frac{ {\displaystyle
\int_{\theta_0}^{\frac{\pi}{2}}}d\theta_{\ell} {\displaystyle
\left( \frac{d\sigma^+}{d\theta_{\ell}} +
\frac{d\sigma^-}{d\theta_{\ell}}\right)} {\displaystyle
-\int^{\pi-\theta_0}_{\frac{\pi}{2}}}d\theta_{\ell} {\displaystyle
\left( \frac{d\sigma^+}{d\theta_{\ell}} +    \frac{d\sigma^-}{d\theta_{\ell}}
\right)}}
{
{\displaystyle          \int_{\theta_0}^{\pi-\theta_0}}d\theta_{\ell}
{\displaystyle          \left( \frac{d\sigma^+}{d\theta_{\ell}} +
\frac{d\sigma^-}{d\theta_{\ell}}\right)}}.
\label{ggtteq17}
\end{eqnarray}

\noindent where $\frac{d\sigma^+}{d\theta_{\ell}}$ and 
$\frac{d\sigma^-}{d\theta_{\ell}}$ refer to
the ${\ell}^+$ and ${\ell}^-$ distributions in the c.m. frame, respectively,
and $\theta_0$ is a cutoff on the polar angle of 
the lepton. It is important to note that if there is 
no CP violation in the top decays, then the charge 
asymmetry is zero in the absence of the cutoff 
$\theta_0$. 

%
%
%
%

Both $A_{{\rm ch}}$ and $A_{{\rm FB}}$ are $T_N$-even asymmetries, thus
they probe the imaginary part of the top EDM, ${\mIm}(d_t^\gamma)$. In
\cite{hepph9709225}, the case where one of the $t$ or $\bar t$ decays
leptonically and the other decays hadronically, was studied
\footnote{then the asymmetries correspond to samples $\cal{A}$ and $\bar
{{\cal A}}$ defined in Eqs.~\ref{eetthzeq31} and \ref{eetthzeq32} in
section \ref{sssec623}}.  Also, it was assumed that no CP violation enters
these top decays. The asymmetries $A_{{\rm ch}}$ and $A_{{\rm FB}}$ were
then evaluated in the $\gamma \gamma$ c.m. frame and 90\% C.L. limits on
the top EDM, in the case that no asymmetry is found in the experiment,
were obtained. The 90\% C.L. limits were evaluated for different electron
and laser beam energies as well as for different cutoff angles. Also,
different helicity combinations of the initial beam and different values
of the dimensionless parameter $x$ defined before (see Eq.~\ref{ggtteq15}
and the discussion above), were analyzed.  They found that for an electron
beam energy of 250 GeV, and for a suitable choice of circular
polarizations of the laser photons and longitudinal polarizations for the
electron beams, and assuming a luminosity of 20 fb$^{-1}$ for the electron
beam, in the best cases and in an ideal experiment, it is possible to
obtain limits on the imaginary part of the top EDM, again, of the order of
$10^{-17}$e-cm.  However, an order of magnitude improvement may be
possible if the beam energy is increased to 500 GeV \cite{hepph9709225}.

To conclude this section, the reaction $\gamma \gamma \to t \bar t$ 
can serve to limit both the real and the imaginary parts of the 
top EDM\null. However, it is worth mentioning that the limits 
that may be placed on the top EDM through 
a CP study in $\gamma \gamma \to t \bar t$ are 
roughly comparable to those which might be obtained through a 
study of the reaction $e^+e^- \to t \bar t$ at
the NLC (for comparison see section \ref{ssec61}).
Therefore, the motivation for
going to a $\gamma \gamma$ collider in order to study effects 
of the top EDM is somewhat arguable. 
On the other hand, model 
calculations of CP violation in $\gamma \gamma \to t \bar t$,
such as the one described below, i.e., a 2HDM, show that CP-nonconserving 
signals in this reaction, 
which are not necessarily associated with the top EDM, may be sizable; i.e.,  
at the detectable level in a future photon collider.

\newpage
~

\newpage

\subsection{$\gamma \gamma \to t \bar t$  
and $s$-channel Higgs exchange in a 2HDM \label{ssec93}} 

A $\gamma \gamma$ collider can also provide 
an interesting possibility for producing an s-channel 
neutral Higgs-boson, via $\gamma \gamma \to h$, which 
can then decay to a pair of fermions, $h \to f \bar f$ (recall that 
a related process was considered 
in the context of a $pp$ collider in section \ref{sssec732}). 
Once again,  
$h$ stands for the lightest neutral Higgs-boson in a MHDM and 
the other neutral Higgs particles are assumed to be much heavier,  
thus, neglecting their contribution in what follows.   

The decay of a neutral Higgs, for $m_h > 2 m_t$, to a pair of $t \bar t$ 
will inevitably dominate the other fermionic decays of 
the Higgs due to the largeness of the top mass.
CP violation in the reaction $\gamma \gamma \to t \bar t$ 
was investigated within a MHDM 
in \cite{prd52p3803} for unpolarized incoming photons, 
where the effects of the s-channel Higgs were 
included. In \cite{plb294p361} polarized laser beams were 
considered for the s-channel neutral Higgs production, 
$\gamma \gamma \to h$. 

\begin{figure}[htb]
\psfull
\begin{center}
\leavevmode
\epsfig{
file=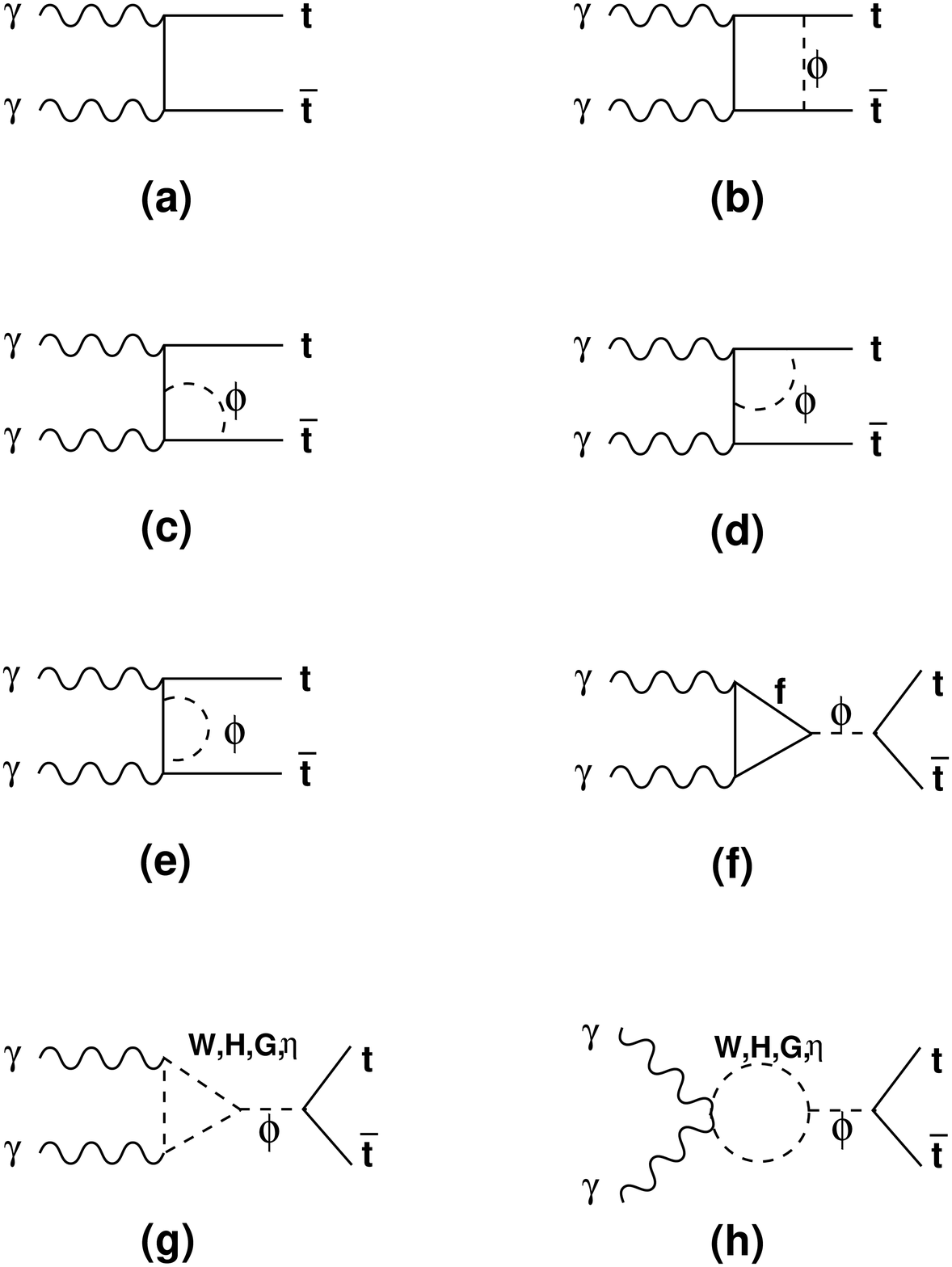
,height=11cm}
 \end{center}
\caption{\emph{Feynman diagrams for $\gamma \gamma \to t \bar t$. In (a) 
the Born diagram is shown (see also Fig.~\ref{ggttfig1}),
and in (b)-(h) the complete set of 1-loop 
diagrams that can violate CP are depicted. Diagrams with crossed lines 
are not shown.}}
\label{ggttfig3}
\end{figure}

In \cite{prd52p3803} the complete set of CP-nonconserving contributions to 
$\gamma \gamma \to t \bar t$, at the 1-loop order, were considered
within a 2HDM of type II\null. Recall that, in the SM, CP violation
cannot occur in this process at least to 2-loop order. This set of 
1-loop Feynman diagrams  is depicted in
Figs.~\ref{ggttfig3}(b)-\ref{ggttfig3}(h) and Fig.~\ref{ggttfig3}(a)
represents the  only tree-level diagram for this process;
this tree-level diagram and its permuted one 
are also shown in Fig.~\ref{ggttfig1}.

%
%

We define ${\cal M}_i$ to be the amplitude 
for a diagram $i$ in Fig.~\ref{ggttfig3} (i.e., $i=a,...,h$) where 
we further decompose ${\cal M}_i$ to its CP-odd part 
${\cal M}_i^{odd}$ and  CP-even part ${\cal M}_i^{even}$:
${\cal M}_i={\cal M}_i^{odd}+{\cal M}_i^{even}$. 
Then, to leading order, the CP-odd ($d\sigma^{odd}$) and the CP-even 
($d\sigma^{even}$) parts of the differential cross-section are given by

\begin{eqnarray} 
&&\hspace{-2.0em} 
d \sigma^{odd} = 2 {\eRe} \sum_{\gamma~ pol.} \sum_{i=b}^{h} {\cal M}_a 
{\cal M}_i^{odd~\dagger} \label{ggtteq18}~, \\    
&&\hspace{-2.0em} d \sigma^{even} = \sum_{\gamma~ pol.} \left[ {\cal M}_a
{\cal M}_a^{\dagger} + 
\sum_{i=f}^{h} 
2 {\eRe} {\cal M}_a {\cal M}_i^{even ~\dagger} 
\right]
\label{ggtteq19}.   
\end{eqnarray}

\noindent Note that the CP-even contribution from the 
interference of the $s$-channel Higgs graphs Fig.~\ref{ggttfig3}(f)-
\ref{ggttfig3}(h) with the Born amplitude of Fig.~\ref{ggttfig3}(a), which
is explicitly included in Eq.~\ref{ggtteq19}, can become 
important and was taken into account in 
\cite{prd52p3803} because of  the non-negligible width of the Higgs. 
Also, the CP-odd interference in Eq.~\ref{ggtteq18} 
will give rise to the simple form of $d \sigma^{odd}$ in Eq.~\ref{ggtteq2}. 

Again, to efficiently trace the CP-odd spin correlations in Eq.~\ref{ggtteq2}, 
one defines a $t \bar t$ decay scenario where 
the $t$ decays leptonically and the $\bar t$ decays hadronically and 
vice versa. 
As in Eqs.~\ref{eetthzeq31} and 
\ref{eetthzeq32} in section \ref{sssec623}, we denote by ${\cal A}$ 
the decay sample in the case that the top decays leptonically 
and the anti-top decays hadronically, and by 
$\bar {{\cal A}}$ the charged conjugate decay sample
\cite{hepph9602273,prd52p3803}. 

With these decay scenarios one can evaluate a 
few CP-odd asymmetries of both the $T_N$-odd and $T_N$-even 
type which may acquire a non-vanishing value only if $\delta_- \neq 0$
and $\delta_+ \neq 0$, respectively (see Eq.~\ref{ggtteq2}). 
To do so, let us define for sample ${\cal A}$, i.e., 
$t\to W^+ b \to {\ell}^+ \nu_{\ell} b$ and 
$\bar t \to W^- \bar b \to q \bar {q^{\prime}} \bar b$,
 the following operators

\begin{eqnarray}
&& {\cal O}_1= (\hat{\bf q}_{{\ell}^+} \times \hat {\bf q}_{W^-}^*) 
\cdot \hat {\bf p}_{\bar t} \label{ggtteq22} ~,\\
&& {\cal O}_2= E_{{\ell}^+} \label{ggtteq23}~,\\
&& {\cal O}_3= \hat {\bf q}_{{\ell}^+} \cdot \hat {\bf p}_{\bar t} 
\label{ggtteq24}~,
\end{eqnarray}

\noindent where the asterisk denotes the $t$($\bar t$) rest frame. 
The corresponding ones for the sample $\bar {{\cal A}}$ are

\begin{eqnarray}
&& \bar {{\cal O}}_1= (\hat {\bf q}_{{\ell}^-} \times \hat {\bf q}_{W^+}^*) 
\cdot \hat {\bf p}_t \label{ggtteq25}~,\\
&& \bar {{\cal O}}_2= E_{{\ell}^-} \label{ggtteq26}~,\\
&& \bar {{\cal O}}_3= \hat {\bf q}_{{\ell}^-} \cdot \hat {\bf p}_t 
\label{ggtteq27}~.
\end{eqnarray} 

\noindent Thus, the CP-odd asymmetries are constructed as
 \cite{hepph9602273,prd52p3803}:

\begin{eqnarray}
&& T_N-{\rm odd}:~~~ \alpha_1=<{\cal O}_1>+<\bar {{\cal O}}_1> 
\label{ggtteq28}~,\\
&& T_N-{\rm even}:~~~ \alpha_{2,3}=<{\cal O}_{2,3}>-<\bar {{\cal O}}_{2,3}> 
\label{ggtteq29}~.
\end{eqnarray} 

\noindent To calculate the above asymmetries one has to fold 
in the distribution functions of the backscattered laser 
photons (for more details see \cite{hepph9602273,prd52p3803}). 
Also, one has to choose a definite scheme for the type 
II 2HDM couplings $a_t^h$ and $b_t^h$ of a neutral Higgs particles 
to a pair of $t \bar t$ in Eq.~\ref{2hdmab}.
Recall that the scalar $a^t_h$ and pseudoscalar $b^t_h$ couplings are 
functions of the neutral Higgs mixing matrix $R_{ji}$ in Eq.~\ref{2hdmrij} 
and of the 
ratio between the two VEV's $\tan\beta$ (see section \ref{sssec323}). 
In particular, as in  
\cite{hepph9602273,prd52p3803}, assuming that
the other two neutral Higgs of the model are much heavier 
than $h$ such that their mass lies above the $\gamma \gamma$ 
c.m.\ energy, their contribution is neglected. Also, 
the mass of the charged Higgs-boson was taken as 
$m_{H^{\pm}}=500$ GeV and $R_{j1}=1/\sqrt 3$ 
for $j=1,2,3$ was assumed (see Eq.~\ref{2hdmrij}).

\begin{figure}[htb]
\psfull
\begin{center}
\leavevmode
\epsfig{file=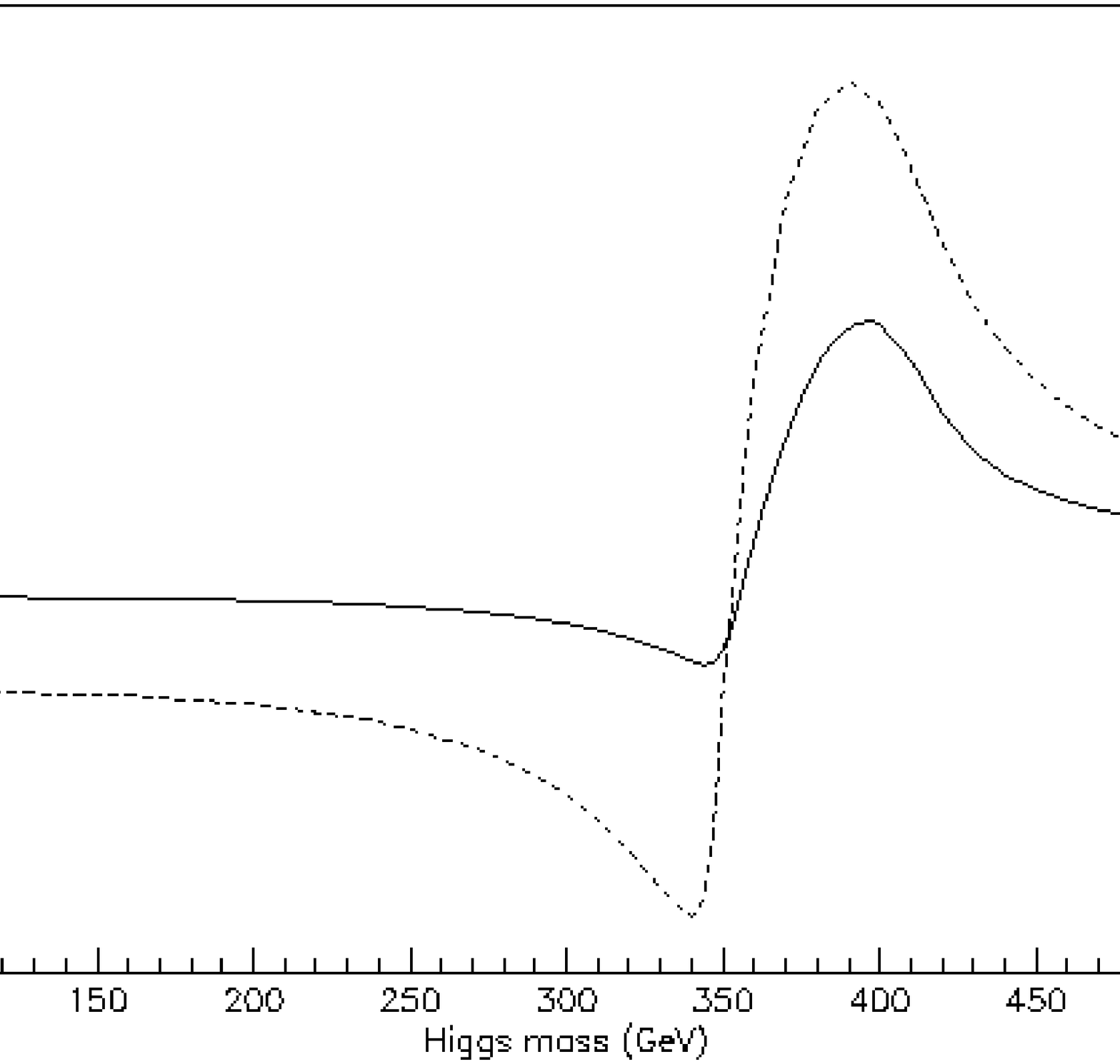,height=8cm,width=8cm,bbllx=0cm,bblly=2cm,
bburx=20cm,bbury=25cm,angle=0}
 \end{center}
\caption{\emph{The ratio $\alpha_1/\Delta \alpha_1$ 
as a function of the 
lightest Higgs-boson mass, $m_h$, 
at $\sqrt s=$0.5 TeV. The dashed line corresponds 
to $tan\,\beta=$0.5 and the solid line to $tan\,\beta=$1; $m_t=$175 GeV.
 Figure taken from \cite{prd52p3803}.}} 
\label{ggttfig4}
\end{figure}

\begin{figure}[htb]
\psfull
\begin{center}
\leavevmode
\epsfig{file=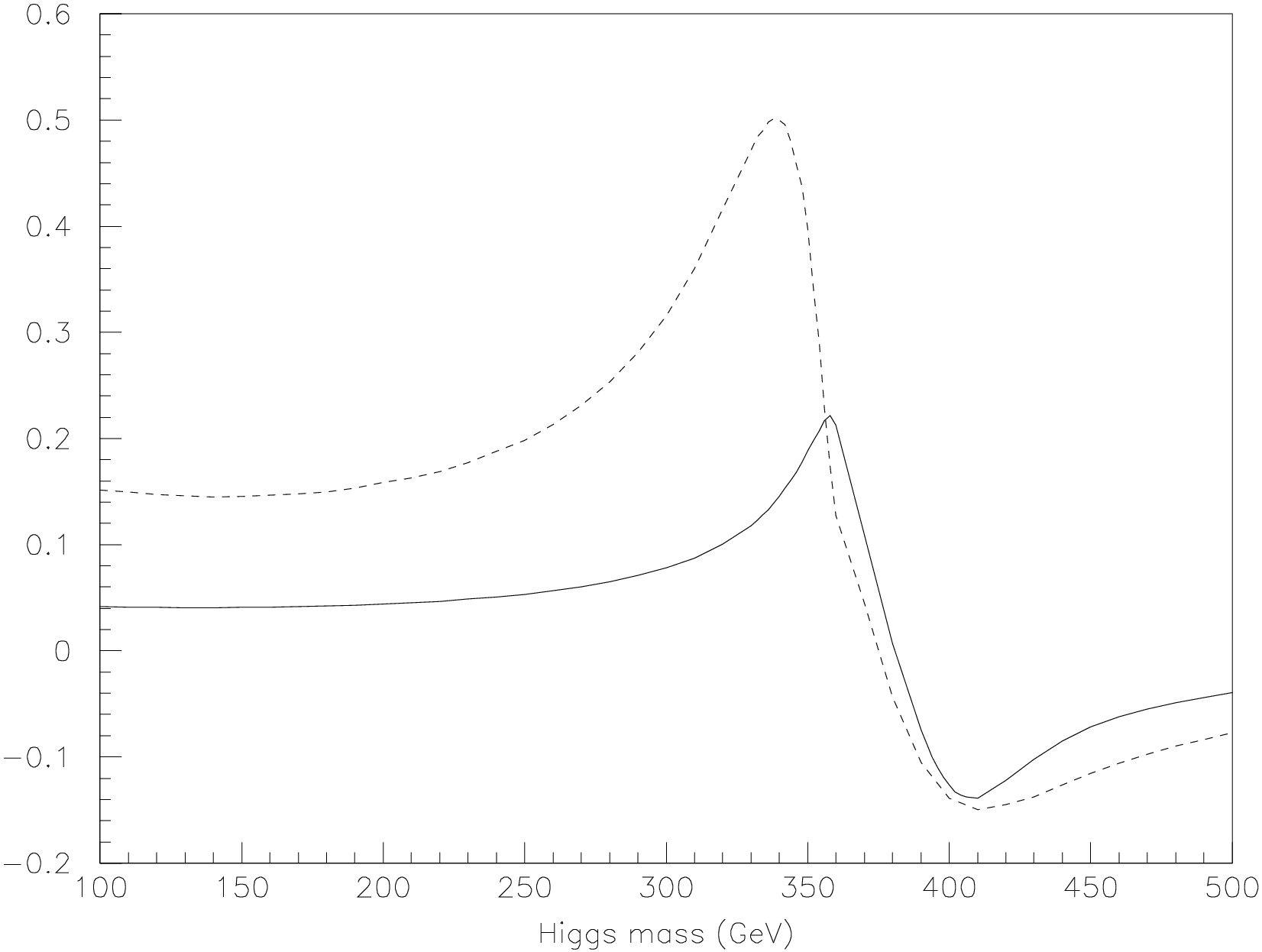,height=8cm,width=8cm,bbllx=0cm,bblly=2cm,
bburx=20cm,bbury=25cm,angle=0}
 \end{center}
\caption{\emph{The ratio $\alpha_3/\Delta \alpha_3$ 
as a function 
of the lightest Higgs-boson mass, $m_h$, at $\sqrt s=$0.5 TeV. The dashed 
line corresponds to $tan\,\beta=$0.3 and the solid line to $tan\,\beta=$0.627; 
$m_t=$180 GeV. Figure taken from \cite{hepph9602273}.}} 
\label{ggttfig5}
\end{figure}

Let us present a sample of the results that were obtained in 
\cite{hepph9602273,prd52p3803}. The $T_N$-odd 
asymmetry, i.e., the signal to noise ratio $\alpha_1/\Delta \alpha_1$,  
is shown in Fig.~\ref{ggttfig4} and the 
$T_N$-even asymmetry ratio $\alpha_3/\Delta 
\alpha_3$ is depicted in Fig.~\ref{ggttfig5}. The 
asymmetries are plotted for various values of $\tan\beta$. 
In Fig.~\ref{ggttfig4} $\tan\beta=0.5$ (dashed line) 
and $\tan\beta=1$ (solid line) were used, while in 
Fig.~\ref{ggttfig5} $\tan\beta=0.3$ (dashed line) and 
$\tan\beta=0.627$ (solid line) were chosen. 
In those figures an $e^+e^-$ collider energy of 
$\sqrt s=500$ GeV was taken, and the asymmetries 
were plotted for the above values of the 2HDM 
free parameters and as a function of the lightest Higgs mass.

We  see from Figs.~\ref{ggttfig4} and \ref{ggttfig5} 
that both the CP-odd $T_N$-odd asymmetry, $\alpha_1$, and the 
CP-odd $T_N$-even asymmetry, $\alpha_3$, peak twice. 
First when the mass of the Higgs is close to the 
$t \bar t$ threshold and then when it is close 
to the maximal $\gamma \gamma$ energy.   
With $m_h \approx 350~{\rm or} ~400$ GeV these 
asymmetries can reach above 10\%. In particular, 
$\alpha_3$ would lead to a somewhat higher CP-violating 
signal and we see from Fig.~\ref{ggttfig5} 
that, for $\tan\beta=0.3$, $|\alpha_3|/\Delta \alpha_3$ 
is above the 10\% level in almost the entire mass range
$100 ~{\rm GeV}<m_{h}<500 ~{\rm GeV}$, and peaks around 
$m_{h} \approx 2m_t$ at 50\%. 
Recall that the statistical significance $N^i_{SD}$ of the CP-violating 
signal that can be measured with a CP-odd asymmetry $\alpha_i$ is given by

\begin{eqnarray}
N^i_{SD} = \frac{|\alpha_i|}{\Delta \alpha_i} \sqrt {N_{\rm exp}} 
\label{ggtteq31}~,
\end{eqnarray}

\noindent where $N_{\rm exp} = R_{{\cal A},{\bar {\cal A}}} 
\times {\cal L} \sigma_0$ is the number of expected events, with
$R_{{\cal A},{\bar {\cal A}}}$ the branching ratios for the decay 
scenarios ${\cal A},{\bar {\cal A}}$, respectively, and
assuming a reconstruction efficiency of 1.
Furthermore, ${\cal L}$ is the collider 
integrated luminosity and $\sigma_0$ is the cross-section which, in 
the leading order, is calculated from $d \sigma^{even}$ 
in Eq.~\ref{ggtteq19}. With ${\cal L} = {\cal O}(10^2)$ fb$^{-1}$ 
the expected number of $\gamma \gamma \to t \bar t$ events is of 
the order of ${\rm few} \times 10^3$ for collider c.m. energies 
of $500 - 700$ GeV\null. Thus, for example, if we take $R_{{\cal A},
{\bar {\cal A}}} =4/27$ such that only leptonic top decays 
into electrons and muons are considered, then    
an asymmetry larger than $10\%$ will correspond to a 
signal-to-noise ratio above 
the 3-$\sigma$ level \cite{hepph9602273}. 

As previously discussed, if the polarization of the backscattered 
photons is adjustable then CP asymmetries involving these 
polarizations can be constructed and 
 they, in turn, can serve as an efficient tool for investigating the 
CP properties of the neutral Higgs-boson. Three such polarization 
asymmetries were suggested in \cite{plb294p361}:

\begin{eqnarray}
{\cal P}_1 \equiv \frac{|{\cal M}_{++}|^2 -|{\cal M}_{--}|^2}
{|{\cal M}_{++}|^2 +|{\cal M}_{--}|^2} 
\label{ggtteq32} ~, \\
{\cal P}_2 \equiv \frac{2{\mIm} ({\cal M}_{--}^* {\cal M}_{++})}
{|{\cal M}_{++}|^2 +|{\cal M}_{--}|^2} 
\label{ggtteq33} ~, \\   
{\cal P}_3 \equiv \frac{2{\eRe} ({\cal M}_{--}^* {\cal M}_{++})}{|
{\cal M}_{++}|^2 +|{\cal M}_{--}|^2} 
\label{ggtteq34}~.
\end{eqnarray}

\noindent In the helicity basis, one can choose the polarization of 
$\gamma_1$ (moving in the $+z$ direction) as: $\epsilon^{\pm}_1=\mp 
2^{-1/2} (0,1,\pm i,0)$, and that of $\gamma_2$ (moving in the 
$-z$ direction) as: $\epsilon^{\pm}_2=\mp 2^{-1/2} (0,-1,\pm i,0)$.
To understand how the above polarization asymmetries (${\cal P}_{1,2,3}$) 
trace the CP properties of the neutral Higgs, 
note that a CP-even (CP-odd) scalar couples to two photons via 
$F_{\mu\nu} F^{\mu\nu}$($F_{\mu\nu} {\tilde F}^{\mu\nu}$) 
(see \cite{plb294p361} and references therein). 
Therefore, as implied from Eq.~\ref{ggtteq7}, 
in the c.m.\ of the two photons
 this will yield a coupling proportional 
to $\vec{\epsilon}_1 \cdot \vec{\epsilon}_2$ 
($(\vec{\epsilon}_1 \times \vec{\epsilon}_2)_z$) 
for a CP-even (CP-odd) neutral Higgs to a $\gamma_1 \gamma_2$ pair. 

For the above convention of the polarizations 
one finds

\begin{eqnarray}
\vec{\epsilon}_1 \cdot \vec{\epsilon}_2 = -\frac{1}{2}(1
+\lambda_1\lambda_2)~~,~~ 
(\vec{\epsilon}_1 \times \vec{\epsilon_2})_z = \frac{i}{2}\lambda_1 (1
+\lambda_1\lambda_2) 
\label{ggtteq35} ~,
\end{eqnarray} 

\noindent where $\lambda_1,\lambda_2=\pm 1$ are the helicities of 
$\gamma_1,\gamma_2$, respectively. 
Now, for a mixed CP state, the general amplitude 
to couple to $\gamma_1 \gamma_2$ will have both the 
CP-even and the CP-odd pieces in Eq.~\ref{ggtteq7} and it can be 
written as

\begin{eqnarray}
{\cal M} = \delta_{even} \vec{\epsilon}_1 \cdot \vec{\epsilon}_2 +
\delta_{odd} (\vec{\epsilon}_1 \times \vec{\epsilon}_2)_z \label{ggtteq36} ~,
\end{eqnarray}

\noindent where $\delta_{even}$($\delta_{odd}$) is 
the CP-even (CP-odd) coupling strength of the neutral Higgs 
to the two photons. Using Eq.~\ref{ggtteq35}, the  squares of the helicity 
amplitudes which appear in ${\cal P}_{1,2,3}$ can be 
readily calculated \cite{plb294p361}:

\begin{eqnarray}
|{\cal M}_{++}|^2 +|{\cal M}_{--}|^2 = 2(|\delta_{even}|^2 + |\delta_{odd}|^2)
\label{ggtteq37} ~,\\
2{\eRe} ({\cal M}_{--}^* {\cal M}_{++})= 2(|\delta_{even}|^2 - 
|\delta_{odd}|^2) \label{ggtteq38} ~,\\     
|{\cal M}_{++}|^2 -|{\cal M}_{--}|^2 = -4{\mIm} (\delta_{even} 
\delta_{odd}^*) \label{ggtteq39} ~,\\
2{\mIm} ({\cal M}_{--}^* {\cal M}_{++})= -4{\eRe} 
(\delta_{even} \delta_{odd}^*) \label{ggtteq40} ~,
\end{eqnarray}

\noindent where $\delta_{even}$ and $\delta_{odd}$ are given 
in \cite{plb294p361} for a 2HDM with scalar and pseudoscalar 
couplings of a neutral Higgs to a pair of fermions.
It is then evident that ${\cal P}_1,{\cal P}_2 \neq 0$ 
and $|{\cal P}_3| < 1$ only if both $\delta_{even},\delta_{odd} \neq 0$. 
That is, only if both the CP-even and the CP-odd couplings are present. 

Using Eq.~\ref{ggtteq6}, for the Higgs-boson production of 
our interest, one gets \cite{plb294p361}:

\begin{eqnarray}
dN=&&d{\cal L}_{\gamma \gamma} d\Gamma \frac{1}{4} (|{\cal M}_{++}|^2 +
|{\cal M}_{--}|^2) \nonumber \\
&& \times \left[ (1+<\xi_1^{(2)} \xi_2^{(2)}>)+(<\xi_1^{(2)}>+<\xi_2^{(2)}>)
{\cal P}_1 \right.  \nonumber \\
&&\left. +(<\xi_1^{(3)} \xi_2^{(1)}>+<\xi_1^{(1)} \xi_2^{(3)}>){\cal P}_2 
 \right.  \nonumber \\
&&\left. + 
(<\xi_1^{(3)} \xi_2^{(3)}>-<\xi_1^{(1)} \xi_2^{(1)}>){\cal P}_3 \right] 
\label{ggtteq41} ~,
\end{eqnarray}

\noindent where $d\Gamma$ is the appropriate element of the 
final state phase space including the initial state flux 
factor. Note that the properties of $d{\cal L}_{\gamma \gamma}$ 
and of the various $\xi$'s (appearing in Eq.~\ref{ggtteq41}) 
as a function of the c.m.\ energy of the 
two photons are very important for this discussion as they depend strongly 
on the polarization of the incoming electrons and associated photons.
Instead of presenting a detailed analysis of those parameters
and the numerical results, we 
refer the reader to \cite{plb294p361}. We will only give their 
summary for a general 2HDM in which the CP properties of a single
neutral Higgs have to be determined.  
In particular, it was found in \cite{plb294p361}  
that out of the three  polarization asymmetries defined in
Eqs.~\ref{ggtteq32}-~\ref{ggtteq34}, ${\cal P}_1$  provides the 
best statistical significance for the task at hand.
A non-zero value for ${\cal P}_1$ requires that the $h \gamma\gamma$
coupling has an imaginary part, as well as both CP-even and CP-odd
contributions. For a mixed CP Higgs-boson with $m_h\lsim 2m_W$, a measurement
of ${\cal P}_1$ will be easiest if $\tan\beta$ is large since the 
$b$-quark loop, which makes the only large contribution
to the imaginary part for such $m_h$ values, will be enhanced.
For $m_h > 2m_W$, the required imaginary part is dominated by
the $W$-boson loop (or $t$-quark loop if $m_h$ is also $>2m_t$); 
large $\tan\beta$
makes detection more difficult since the dominant CP-odd contribution
originates from the $t$-quark loop, which will be suppressed. 

To summarize, the production of a neutral Higgs-boson by 
fusion of backscattered laser beams can provide a 
systematic analysis of the CP properties of the Higgs 
particle. In particular, $\gamma \gamma \to h \to t \bar t$ 
would be a promising channel for exploring CP-violating 
effects that can arise from an extended Higgs sector, as for 
quite a large range of the 2HDM parameter space this 
reaction can exhibit statistically significant CP-nonconserving 
signals in a high energy $\gamma \gamma$ collider running at c.m. 
energy of $\sqrt s \simeq 500$ GeV\null. Moreover, if the 
polarizations of the incoming photons are controlled, 
then detailed information on both the scalar and the pseudoscalar 
couplings of the neutral Higgs to a pair of fermions may be 
extracted by considering polarization asymmetries of the two 
colliding photons. If the neutral Higgs is a pure CP 
eigenstate, the polarization asymmetries ${\cal P}_1$ 
and ${\cal P}_2$ in Eqs.~\ref{ggtteq32} and \ref{ggtteq33} 
will vanish, while, in Eq.~\ref{ggtteq34}, ${\cal P}_3=1~
(-1)$ for a CP-even (odd) neutral Higgs. 
Therefore, a non-vanishing 
value for ${\cal P}_1$ and ${\cal P}_2$ and 
${\cal P}_3 < 1$ will imply the existence of an extended Higgs sector 
beyond the SM and of CP violation in the scalar potential.


\newpage
~

\newpage
~

\newpage
~

\newpage

\pagebreak

\section{CP violation in $\mu^+\mu^-$ collider experiments \label{sec10}}
\setcounter{equation}{0}

The idea to build a  high energy $\mu^+ \mu^-$ collider is more 
than 25 years old \cite{mumureview1}.
It has recently gained interest in part due to the interesting
possibility of using it as an ``$s$-channel Higgs factory''. Of course,
it may also be suitable for tackling some other physics issues,
e.g., SUSY\null. For recent reviews see \cite{mumureview2}.
The c.m.\ energies considered range from 100 GeV to 4 TeV or even 
more, with luminosity comparable or higher than in linear $e^+e^-$ 
colliders. 
The subject is still in its infancy compared with 
the more established technologies of 
linear $e^+e^-$ colliders, 
and of hadronic colliders such as the Tevatron or the LHC\null.   

\subsection{$\mu^+\mu^- \to t \bar t$ \label{ssec101}}

If there exists a Higgs-boson with mass of a few hundred GeV, a muon 
collider running at the Higgs resonance can
provide the fascinating and unique possibility of an in-depth study of
the Higgs particle in the $s$-channel. 
In particular, the CP-violating properties of its coupling to $t\bar t$ 
may be studied via the reaction 

\beq 
\mu^+\mu^-\to {\cal H} \to t \bar t ~,
\eeq

\noindent where we have generically denoted the neutral Higgs 
resonance under study by ${\cal H}$.

It is perhaps surprising that 
Higgs-bosons can be produced at an appreciable rate at a muon collider. 
Indeed, the ${\cal H}\mu\mu$ coupling is very small since it is 
proportional to the mass of the muon, $m_\mu$.
On the other hand, if the c.m. energy of
the accelerator can be tuned to be at $s=m^2_{\cal H}$, then the 
cross-section receives appreciable enhancement due to the resonant production 
of the Higgs. To see how this works, consider a collider tuned
precisely at the Higgs resonance, $s=m^2_{\cal H}$, then the
cross-section $\sigma_{\cal H} \equiv \sigma(\mu^+\mu^-\to{\cal H})$ for
neutral Higgs-boson production is given by

\beq
\sigma_{\cal H} = \frac{4\pi}{m^2_{\cal H}} B_\mu \label{eq102} ~,
\eeq

\noindent where $B_\mu$ is the branching ratio of ${\cal H} \to
\mu^+\mu^-$. It is useful to compare this with the point cross-section

\beq
\sigma_0 = \sigma(\mu^+\mu^-\to\gamma^\ast \to e^+e^-) ~.
\eeq

\noindent Thus

\beq
R({\cal H}) = \frac{\sigma_{\cal H}}{\sigma_0} = \frac{3}{\alpha^2}
B_\mu ~,
\eeq

\noindent where $\alpha$ is the fine-structure constant. 
Therefore, $\sigma_{\cal
H}$ and $R({\cal H})$ are enhanced if the neutral Higgs has a narrow width,
i.e., a relatively large 
$B_\mu=\Gamma({\cal H}\to \mu^+\mu^-)/\Gamma_{\cal H}$.

One simple way to study CP violation at a muon collider is via the decays
${\cal H}\to t\bar t$. CP-violating correlations can be studied in the
decays of the produced $t\bar t$ pair. Again, this is 
possible due to the fact that the 
weak decays of the top quark are very effective in analyzing the top
spin (see section \ref{ssec28}).

\subsubsection{A general model for the Higgs couplings \label{sssec1011}}

To keep the discussion completely general we will assume that a single
neutral Higgs-boson, ${\cal H}$, is under study although the underlying model
may contain several Higgs doublets. 
In practice, of course, the muon collider will only be tuned to one Higgs 
resonance at a time.  

In section \ref{sssec323} we have written an example of a useful 
parameterization for the ${\cal H} f \bar f$ ($f=$fermion) 
interaction (see Eq.~\ref{2hdmab}), taking into account 
possible CP violation in this vertex due to an extended Higgs sector.  
It is, however, also instructive to introduce a different notation - 
somewhat more compact - which is useful for the investigation at hand. 
Let us therefore parameterize the coupling of
${\cal H}$ to fermions with the Feynman rule \cite{prd52p6271}:

\beq
C_{{\cal H}ff} = i C^0_{ff} \chi_f e^{i\gamma_5\lambda_f} ~,
\label{fermicouple}
\eeq

\noindent where $C^0_{ff}=-(g_W/2)(m_f/m_W)$ is the coupling in the SM
and $\chi_f$ for each 
fermion $f$ (e.g. $f=l$ (i.e. a charged lepton), $u$, $d$) 
is a real constant which gives the magnitude of the coupling in relation 
to the standard model. 
The CP nature of the coupling is determined by 
the  value of $\lambda_f$. In particular, $\lambda_f$ which is not a 
multiple of $\pi/2$ is 
indicative of CP violation since the coupling will then contain both
scalar and pseudoscalar components.\footnote{We note that in the 
language of the interaction Lagrangian
in Eq.~\ref{2hdmab}, $\lambda_f \neq \pi/2$ corresponds to having 
$a_t^{\cal H},b_t^{\cal H} \neq 0$, where $a_t^{\cal H}$($b_t^{\cal H}$) 
is the scalar(pseudoscalar) ${\cal H}$ coupling to a 
$t\bar t$ 
pair.} 
CP violation is thus essential in any scalar coupling which is not either 
pure scalar or pseudoscalar and so learning about $\lambda$ is 
equivalent to investigating CP violation in ${\cal H}\to f\bar f$.

Moreover, in models with an extended Higgs sector the coupling of  ${\cal H}$  
to the boson sector of the theory may be characterized as either
scalar, $H$, or pseudoscalar, $A$. If ${\cal H} = A$ then it cannot
couple to gauge-bosons while if ${\cal H}=H$ we can parameterize its
coupling to two vector-bosons as

\beq
C_{{\cal H}VV} = C_V \cos\alphaX ~,
\eeq

\noindent where, again, 
$C_V$ 
is the coupling in the SM, $V=Z$ or
$W$, given by (see also Eq.~\ref{2hdmc})

\beqa
C_W=m_W g_W  ~~,~~C_Z=m_Z g_W/cos\theta_W ~,
\eeqa

\noindent and $\alphaX$ is the angle between the observed Higgs-boson ${\cal
H}$ and the orientation of the vacuum in the Higgs space. 

The mass eigenstate could also be a mixture of $H$ and $A$ which again
would violate CP. This aspect of CP violation in the Higgs sector 
can lead to enhanced CP violation if the two masses are close together
as discussed in section \ref{sssec1013}.
Such mixing could also lead to a CP-violating coupling to fermions
(Eq.~\ref{fermicouple}) whose implications  will be discussed in the
following pages. 

In the following sections we will consider a number of methods to
investigate CP-violating couplings of the s-channel neutral Higgs-boson 
to top
quarks at a muon collider. 
We first consider the reaction $\mu^+\mu^-\to t\bar t$. 
Note that in this 
reaction both initial and final states are CP eigenstates. Furthermore, 
there is no CP-odd observable that one can construct out of the total 
cross-section (such as a partial rate asymmetry); 
if one considers angular 
distributions of the final $t$-quark, such distributions depend only on 
the angle $\theta_{\mu t}$, i.e., the angle between the $\mu^-$ and the 
$t$-quark momenta in the c.m. frame. Since 
$\cos\theta_{\mu t}$ is a C-even P-even quantity
we clearly need more information if we are to observe CP violation. 
Indeed, if the dominant amplitude is mediated by scalar exchange the 
angular distribution will be isotropic in any case. 

To construct CP-violating observables we therefore need information about
the polarization of the fermions: either the final state $t\bar t$ or the
initial state $\mu^+\mu^-$. In section \ref{sssec1012} we consider the use of
correlations in the top polarization in $\mu^+\mu^-\to {\cal H} \to t\bar t$
to measure the CP-violating parameter of ${\cal H} t\bar t$ coupling, i.e., 
$\sin 2\lambda_t$, where $\lambda_t$ is the angle in
Eq.~\ref{fermicouple}. 

In 
section \ref{sssec1013} we  consider measurement of CP violation in
the same reaction ($\mu^+\mu^-\to {\cal H} \to t\bar t$) except this time
the asymmetry we construct is based on polarized muons. 
Clearly, to perform such experiments it is necessary to have a muon 
collider capable of producing muons with a significant polarization.

Finally in section \ref{ssec102} we consider the possibility of 
flavor changing
neutral Higgs couplings which could give rise to CP violation in the reaction
$\mu^+\mu^-\to t\bar c$ versus $\mu^+\mu^-\to \bar t c$.  Large couplings
of this sort may be expected in 2HDM of type III which is 
described in some detail in section \ref{sssec322}.  
Here again the use
of top and/or muon polarization is essential to obtain CP-violating
signals. 

\subsubsection{Decay correlation asymmetry \label{sssec1012}}

Let us now consider the case of a muon collider where unpolarized muons 
produce $t\bar t$ through
$\mu^+\mu^-\to {\cal H} \to t\bar t$. Thus,
in order to learn about the coupling in Eq.~\ref{fermicouple}, we can 
observe the polarization of the top quarks through their decays 
\cite{prd52p6271}, which is discussed extensively in section \ref{ssec28}.

Here we will just consider the determination of top polarization 
by its correlation with the momentum of a particle in various decay modes.
Thus, if $X$ is a decay product of a top decay, we define the 
``analyzing power'':

\beq
\epsilon^t_X \equiv 3 \langle\cos\theta_X\rangle ~,
\label{epsilondef}
\eeq

\noindent where $\theta_X$ is the angle between 
$\vec p_X$ and the spin of the top in the top rest frame.

Let us now extend this idea to study the correlations of the
polarizations of the top quarks where the polarizations are indicated by
the momenta of specific decay particles. As discussed in section 
\ref{ssec28}, for
the case of a single polarized top, some further optimization may follow
from going beyond this which we do not consider here.

Following~\cite{prd52p6271}, we work in the rest frame of the Higgs-boson
with the $t$ momentum along the $z$ axis.  Let each of the $t$-quarks
undergo a decay which can analyze the top polarization. Let $x_i$ and
$x_j$ be the outgoing particles which we wish to correlate with the $t$
and $\bar t$ polarization respectively; for instance, the lepton or
$W^\pm$. Also let $y_i$ and $y_j$ denote the rest of the decay products.
Thus the two decays are $t\to x_iy_i$ and $\bar t\to \bar x_j\bar y_j$. 
We can then define the azimuthal angle between $\vec p_{x_i}$ and $\vec
p_{\bar x_j}$ projected on to the $x,y$ plane as: 

\beq
\sin (\phi_{ij}) = \frac{\left( \vec p_{x_i} \times \vec p_{\bar x_j} \right)
 \cdot
\vec p_t}{|\vec p_{x_i}||\vec p_{\bar x_j}||\vec p_t|} ~.
\eeq

\noindent The azimuthal differential distribution 
of $t$ and $\bar t$ events is then given by

\beqa
\frac{d\Gamma}{\Gamma d\phi_{ij}} =  1 - \frac{\pi^2}{16}
\epsilon^t_i \epsilon^t_j \rho_t \cos 2\lambda_t \cos \phi_{ij}
 + \frac{\pi^2}{16} \epsilon^t_i\epsilon^t_j\eta_t
\sin2\lambda_t \sin\phi_{ij} ~.
\eeqa

\noindent The coupling $\lambda_t$ is defined in Eq.~\ref{fermicouple}
for $f=t$ and
$\epsilon^t_i,\epsilon^t_j$ are the analyzing power of
the decays (defined by Eq.~\ref{epsilondef}). Also, $\rho_t,\eta_t$ are 
phase space factors which approach 1 as $m_{\cal H} \muchbigger 2m_t$. 
They are given by

\beqa
\rho_t & = & -\frac{1-\beta^2_t - (1+\beta^2_t) \cos 2\lambda_t}{\cos
2\lambda_t [1+\beta^2_t - (1-\beta^2_t) \cos 2\lambda_t ]} ~, \\
\eta_t & = & \frac{\beta_t}{1- (1-\beta^2_t) \cos^2\lambda_t} ~,
\eeqa

\noindent where $\beta_t=\sqrt{1-4m_t^2/m_{\cal H}^2}$.

We may now define the following 
CP-violating, $T_N$-odd, $P$-odd, azimuthal asymmetry by 

\beq
A^t_{ij} = \frac{\Gamma (\sin\phi_{ij}>0) -
\Gamma(\sin\phi_{ij}<0)}{\Gamma (\sin\phi_{ij}>0) +
\Gamma(\sin\phi_{ij}<0)} \label{mumuat}~.
\eeq

\noindent From the distribution above, the value for this observable will be

\beq
A^t_{ij} = \frac{\pi}{8} \epsilon^t_i\epsilon^t_j \eta_t\sin
2\lambda_t ~. 
\eeq

\noindent For each pair $i$, $j$ of top decays that can be used to analyze 
the polarization of the $t$ and $\bar t$, one will 
obtain an experimental value of $A^t_{ij}$ from which one can infer 
the value of the quantity 
$\sin 2\lambda_t$. 
Clearly we would like to combine the information from 
all of the modes together in order to obtain 
better statistical significance for the determination of 
$\sin 2\lambda_t$. 
To 
combine the asymmetries from different pairs of modes
$\{i,j\}$ in an optimal fashion we
form a weighted average of the asymmetries 
defined by a set of 
weights $w_{ij}$ with the normalization defined by:

\beq
\sum w_{ij} B_iB_j = \sum B_iB_j ~,
\eeq

\noindent where the summation is over the modes under observation
and $B_i$ is the branching ratio of mode $i$.

The total weighted asymmetry is thus defined as

\beq
A^t = \sum w_{ij} A^t_{ij} B_iB_j ~,
\eeq

\noindent which is maximized by 
taking
$w_{ij}\propto A^t_{ij}$ \cite{prd45p2405}. 
With these weights, then, the 
maximal asymmetry is

\beq
A^t = \frac{\pi}{8} (\epsilon^t)^2 \eta_t \sin2\lambda_t ~,
\eeq

\noindent where

\beq
\epsilon^t = \left[ \sum B_i (\epsilon^t_i)^2 \right]^{1/2} \label{eq20} ~. 
\eeq

\noindent Now, for the top quark, the branching ratio into electron or muon,
$B_e=B_\mu\sim\frac{1}{9}$ and the analyzing power is 
$\epsilon^t_e\sim\epsilon^t_\mu=1$. For decays (of $W$) into hadrons
(i.e., jets), $B_h =\frac{2}{3}$ with $\epsilon^t_h=.39$. The decays
into $\tau$ have, $B_\tau=\frac{1}{9}$ and the 
analyzing power is taken to be the same as into the jet modes,
$\epsilon^t_\tau=.39$. Using these we can then deduce, via Eq.~\ref{eq20}

\beq
\epsilon^t\simeq .58 ~.
\eeq

In order to quantify how well such experiments might detect CP violation,
let us define $\hat y^{(3\sigma)}_j$ to be the number of years needed to
accumulate a $3\sigma$ signal for the CP asymmetry, $A_j$, in the final
state $j$. Then

\beq
\hat y^{(3\sigma)}_j = 9 \frac{R^0_j +R_{\cal H} B ({\cal H}\to
j)}{(A_j)^2 R^2_{\cal H} B ({\cal H}\to j)^2
\sigma_0 {\cal L}} ~,
\label{yhatdef}
\eeq

\noindent where ${\cal L}$ is the integrated luminosity per year,
$\sigma_0$ is the cross-section for $\mu^+\mu^- \to \gamma^* \to e^+e^-$
and for a final state $X$, $R_X=\sigma(\mu^+\mu^- \to X)/\sigma_0$.
In the above $R_j^0$ is $R_j$ from SM processes only, which needs 
to be included as it contributes to the background.

\begin{figure}
\psfull
 \begin{center}
  \leavevmode
   \epsfig{file=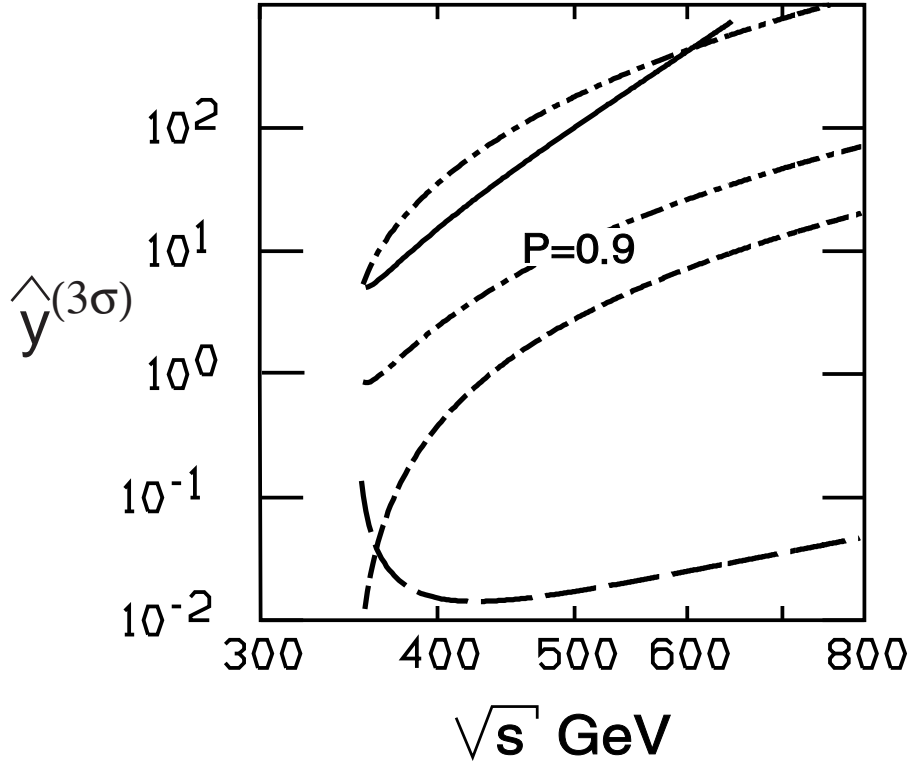,height=4 in}
 \end{center}
\caption{\emph{The values of 
$\hat y^{(3\sigma)}$ (i.e., the number of years required for a $3 \sigma$
effect)  
for the three scenarios discussed in the text
obtained in~\cite{prd52p6271}. The solid line is 
$\hat y^{(3\sigma)}_{t \bar t}$ 
for scenario (1) using 
top polarization correlation. 
The upper dash-dot 
line is 
$\hat y^{(3\sigma)}_{t \bar t}$ 
for scenario (2) using 
top polarization correlation
while the lower dash-dot line 
(also in scenario (2))is for the case where the initial muon 
beams have a longitudinal polarization $P=0.9$. 
The short dashed line is 
$\hat y^{(3\sigma)}_{init}$ 
obtained in scenario (2) using 
transverse polarization of the initial muon beams.
The long dash 
line is 
$\hat y^{(3\sigma)}_{t \bar t}$ 
for scenario (3) using 
top polarization correlation.
Here we take ${\cal L}=10^{34}$ cm$^{-2}$s$^{-1}$. 
}}
\label{newfig3}
\end{figure}

Note that the interference between the SM and the Higgs
exchange will be negligible; from helicity considerations, such an
interference term will be suppressed by $m_\mu/m_{\cal H}$.  The SM
does however contribute as a background, hence the term $R^0_j$ in
the numerator of Eq.~\ref{yhatdef}. We must also consider the effect of
all of the other decay modes of the Higgs taken together since $R_{\cal H}$ 
is proportional to $B_\mu$ and hence inversely proportional to the
total width (see Eq.~\ref{eq102}). 
To get an idea of how large the CP-violating effects can be, we consider 
$\hat y^{(3\sigma)}_j$ as a function of $m_{\cal H} \equiv \sqrt s$ in 
fig.~\ref{newfig3} 
in a number of different scenarios:

\begin{enumerate}
\item
${\cal H}=H$ with $\chi_f=1$ and $\lambda_f=45^\circ$ for all fermions
and $\alphaX=45^\circ$.
\item
${\cal H}=A$ with $\chi_f=1$ and $\lambda_f=45^\circ$ for all fermions.
\item
${\cal H}=A$ with $\chi_l=\chi_d=5$ and $\chi_u=1/5$, 
$\lambda_f=45^\circ$ for all fermions.
\end{enumerate}

\noindent We assume that 
$\chi_u=\chi_c=\chi_t$ 
and 
$\chi_d=\chi_s=\chi_b$.
The significance of ${\cal H}=H$ or $A$, as discussed above, is that only 
if ${\cal H}=H$ does ${\cal H}\to WW$, $ZZ$ contribute to the total 
width. Thus, in particular, the value of $\alphaX$ is not relevant to 
cases (2) and (3), since if ${\cal H}=A$, then no boson pairs are produced by 
the resonance at tree-level. 

In Fig.~\ref{newfig3} which shows the results from \cite{prd52p6271} we take
a nominal luminosity of $10^{34}cm^{-2}s^{-1}$ and a year of $10^7$ sec.
(i.e., with running efficiency of $1/3$). The solid line gives the result
in the case of scenario (1) while the upper dot-dash line is scenario (2).
In both of these cases, $\hat y_{t \bar t}^{3\sigma}$ starts at about 5 years near
threshold and increases thereafter.  The result for scenario (3) is shown
with the long dashed line and is considerably smaller, $.01-.1$ years, due
to the narrow width of the neutral Higgs ${\cal H}$ in this case.

One can enhance the signal with respect to the SM through the
use of longitudinally polarized beams. If both of the $\mu^+$ and $\mu^-$
beams are left polarized with polarization $P$, then the Higgs
production is multiplied by $(1+P^2)$ thus enhanced while the SM
backgrounds are multiplied by $(1-P^2)$ and thus reduced. 
More generally if the $\mu^+$ has polarization $P^+$ and the 
$\mu^-$ beam has polarization $P^-$ then the Higgs cross-section
gets multiplied by $( 1 + P^+ P^- )$ while the SM background
gets multiplied by $( 1 - P^+ P^-)$.
In the
lower dash-dot curve of fig.~\ref{newfig3} we consider the results for
scenario (2) where we have taken $P=0.9$ which gives a reduction of nearly
an order of magnitude. 

\subsubsection{Production 
asymmetry in $\mu^+\mu^-\to t\bar t$ via polarized muons \label{sssec1013}}

As discussed above, some knowledge about fermion polarization is 
required if information about CP violation in $\mu^+\mu^-\to t\bar t$ is 
to be obtained. Above we considered the case where we used the 
polarization of the top quarks to learn about $\sin 2\lambda_t$. 
Here we consider the case where the 
muons are polarized.

The initial production of muons results in a substantial longitudinal
polarization since the weak decay $\pi\to \mu \nu$ produces predominantly
left handed $\mu^-$ (and right handed $\mu^+$). If one constructed a
single pass colliding beam machine, it should not be too difficult to
preserve this polarization. On the other hand, in the case of muon storage
rings the polarization would have to be manipulated in some way since a
longitudinally polarized beam will precess at a rate proportional to
$g-2$. In a recent paper Grz\c{a}dkowski {\it et
al}~\cite{hepph0003091} discuss the details of how the polarization of
muons in a storage ring may be used to make measurements on the Higgs
resonance of the type considered in the following sections.  Here we will
assume that it is possible to prepare muons in a given initial state of
transverse or longitudinal polarization.

%
%

Let us first consider an experiment where the muon beams are polarized
transversely to the beam axis.  The cross-section is then measured as a
function of the angle between the polarizations. We can take the
$z$-axis in the c.m. frame to be in the direction of the $\mu^-$
beam and the $x$-axis to be its polarization while the $\mu^+$ beam is
polarized at an angle of $\phi_\mu$ to the $x$-axis, that is in the
direction $(\cos\phi_\mu,\sin\phi_\mu,0)$.

If the $\mu^\pm$ beams have polarization $P_\pm$ then the cross-section 
(e.g., for $\mu^+ \mu^- \to t \bar t$) as a 
function of $\phi_\mu$ is \cite{prd52p6271}:

\beq
\sigma(\phi_\mu)=
(1
-P_+P_- \cos 2\lambda_\mu \cos\phi_\mu + 
P_+P_-  \sin 2 \lambda_\mu\sin\phi_\mu)
\sigma_0 ~,
\eeq

\noindent
where $\sigma_0$ is the corresponding unpolarized cross-section.
We could therefore look for the presence of CP violation by comparing 
$\sigma(\phi_\mu=+90^\circ)$ 
with 
$\sigma(\phi_\mu = -90^\circ)$. Thus, we define 
the CP-odd (C-even and P-odd), $T_N$-odd  asymmetry

\beq
A_\mu=
{
\sigma(+90^\circ)-\sigma(-90^\circ)
\over
\sigma(+90^\circ)+\sigma(-90^\circ)
}=P_+P_-\sin 2\lambda_\mu \label{mumuamu}~.
\eeq

\noindent Clearly if appreciable polarizations are available and
$\sin2\lambda_\mu\approx 1$ these effects are dramatic.  

In this experiment, we are simply observing a change in Higgs production
as a function of $\phi_\mu$, so in the approximation that
the Higgs resonance is dominant, it would not matter in fact 
what the final state is. 

In practice, the SM effects will also produce the same final
states. Again using this asymmetry we can quantify the amount of run time
required to see a signal through Eq.~\ref{yhatdef}. In
fig.~\ref{newfig3} we show with
the short dashed line the value from~\cite{prd52p6271} of 
$\hat y^{(3\sigma)}_{init}$, which is 
the number of years required to obtain a $3 \sigma$ signal using the 
initial polarizations
for this asymmetry in scenario 2 taking $P_+=P_-=1$. 

It is also useful in some cases to consider asymmetries which make use of 
longitudinally polarized muons in the initial state. Such an asymmetry 
which is $T_N$-even, C-even and P-odd was considered in~\cite{prl77p4996}:

\beq
A_{CP} = \frac{\sigma(\mu^-_L\mu^+_L \to t\bar t) -
\sigma(\mu^-_R\mu^+_R\to t\bar t)}{\sigma (\mu^-_L\mu^+_L\to t\bar t) +
\sigma (\mu^-_R\mu^+_R\to t\bar t)} \label{papm1} ~.
\eeq

\noindent This asymmetry in the pair production cross-section clearly
requires longitudinally polarized muon beams. In principle, a similar
asymmetry is possible for other fermions as well.  Since this asymmetry is
$T_N$-even, some absorptive phase is required.  If the collider is running
near the Higgs resonance, this will be naturally provided by the complex
phase in the Higgs propagator.  Thus, a mechanism for generating this
asymmetry is the CP violation originating from the mixing between $H$ and
$Z$ and/or between the scalar ($H$) and pseudoscalar ($A$) Higgs that can
occur in extended models. In particular, the CP invariance of the Higgs
sector may be broken by the presence of heavy Majorana fermions. Such a
scenario can occur in the minimal SUSY model, in which heavy neutralinos
are Majorana fermions. E6 inspired theoretical scenarios offer another
possibility for heavy Majorana neutrinos at the TeV mass scale. 

The most interesting situation occurs when a CP-even $H$ mixes with a
CP-odd Higgs scalar, $A$, 
and, as is natural in SUSY models, for $M_A^2 \muchbigger M_Z^2$,
the two states are roughly degenerate,
$M_H\simeq M_A$. In particular, if 
$M_{A,H}>2M_Z$
the broadening of the $H$ due to the two vector decay channel 
can allow significant mixing between the two states.   
Consequently, Pilaftsis~\cite{prl77p4996}  finds

\beq
A_{CP} \sim \frac{-2\hat\Pi^{AH} [ \Im{\rm m} \hat \Pi^{HH} (m^2_H)
+ \Im{\rm m} \hat \Pi^{AA} (m^2_H)]}{(m^2_H - m^2_A)^2 +[ \Im{\rm m}\hat
\Pi^{AA} (m^2_H)]^2 + [\Im{\rm m}\hat\Pi^{HH} (m^2_H)]^2} \label{papm2}
\eeq

\noindent where $\Pi^{ij}$ are coupled channel propagators derived 
in that paper.
Note, in particular, the proportionality to the imaginary parts as expected 
since the asymmetry is CP$T_N$-even.

\begin{figure}
\psfull
\begin{center}
\leavevmode
\epsfig{file=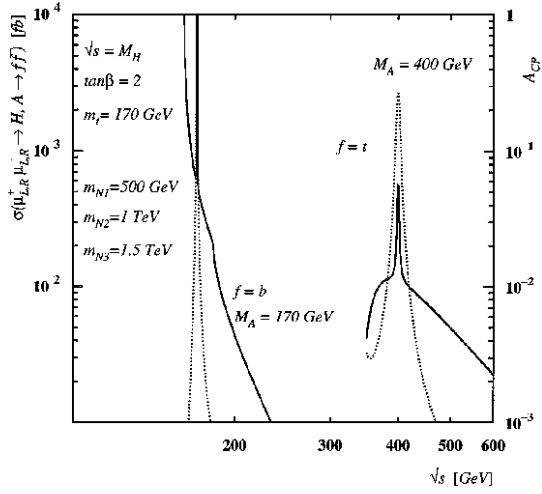,height=4.0 in}
\end{center}
\caption{\emph{ 
This figure shows the cross-section 
(solid curves)
and $A_{CP}$ in Eq.~\ref{papm2} (dotted curves) 
as a function of 
$\sqrt{s}=m_H$ in a model with $A-H$ mixing induced by heavy Majorana 
neutrinos with masses $M_N=.5$, $1.0$, and $1.5$ TeV and with 
$\chi_d=2=1/\chi_u$.
The two curves at 
the left are for  case (a) where $M_A=170$ GeV and $A_{CP}$ is 
observed in the $b\bar b$ channel (see text). 
The curves at the right are for case 
(b) where $M_A=400$ GeV and $A_{CP}$ is observed in the 
$t\bar t$ channel (see text).
Note that in both cases the curves are shown in the vicinity of 
$M_A\sim M_H$ where the mixing effects are likely to be most prominent.
 Figure taken from \cite{prl77p4996}.}}
\label{pilaf2}
\end{figure}

Fig.~\ref{pilaf2} shows the results from~\cite{prl77p4996}
in  a model where $\Pi^{HA}$ is generated by heavy Majorana neutrinos 
with masses $M_N=0.5$, $1.0$ and $1.5$ TeV. It is assumed that 
$\sqrt{s}=m_H$. Two scenarios for the masses and couplings of the 
Higgs-bosons are considered: 

\begin{itemize}
\item[a)]
$M_A=170$ GeV and $\cos^2\alphaX=1$, 
$\chi_d=2=1/\chi_u$
and the asymmetry is observed with a 
$b\bar b$ final state.
\item[b)]
$M_A=400$ GeV and $\cos^2\alphaX=0.1$,
$\chi_d=2=1/\chi_u$
and the asymmetry is observed with a $t\bar t$ final state.
\end{itemize}

\noindent The cross-section is shown with solid curves while the asymmetry
is shown with dotted curves. Scenario (a) is shown with the curves in the
region around $\sqrt{s}=170$ GeV where the final state is $b \bar b$ while
scenario (b) corresponds to the curves in the region $\sqrt{s}=400$ GeV,
with a $t \bar t$ final state. The enhancement of the asymmetry from the
imaginary part of the scalar propagators is apparent in the case where the
$A$ and $H$ masses are close together, within about 10\% of each other. 

\newpage

~

\newpage

\subsection{CP violation in the flavor changing reaction 
$\mu^+\mu^-\to t\bar c$ \label{ssec102}}

As mentioned before, one of the unique properties of a muon collider
is that, under favorable conditions, it may produce neutral Higgs states in
the $s$-channel. If the Higgs sector contains two or more doublets, then
the Higgs couplings may be FC (Flavor Changing)
\cite{prd55p3156,hou} (see also 
section \ref{sssec322}). 
This can lead to a dramatic
tree-level signature of $\mpmm\to t\bar c$ (or $c\bar t$) due to 
the neutral Higgs
resonance \cite{ars_mumu}. 
At the same time FC processes do occur at the
loop level in the SM and in practically all of its extensions, even if
they are forbidden at the tree-level. Thus a continuum of FCNC reactions
of the form $\mpmm\to Z^*,\gamma^* \to t \bar c,~\bar t c$ are expected.
Indeed such couplings 
with CP-violating phases may also naturally arise in 
R-parity violating SUSY models~\cite{hepph9910543}.
Of course such processes are 
GIM suppressed in the SM but for the purpose of this
discussion we are assuming that there is a FC 
Higgs sector as in 
section \ref{sssec322}, thus for the reactions
$Z^*, \gamma^* \to t \bar c, ~\bar t c$, rates
appreciably larger than the SM may be expected \cite{prd55p3156,ARS}.
Since many such
extensions of the SM contain a large number of unconstrained Yukawa
couplings, they will, in general, also contain CP-violating phases.
Therefore the interference between the resonant and the continuum
processes can lead to CP-odd observables; it is this possibility which we
wish to study in this section.

%
%
%
%
%
%
%

%
\begin{figure}
\psfull
 \begin{center}
  \leavevmode
   \epsfig{file=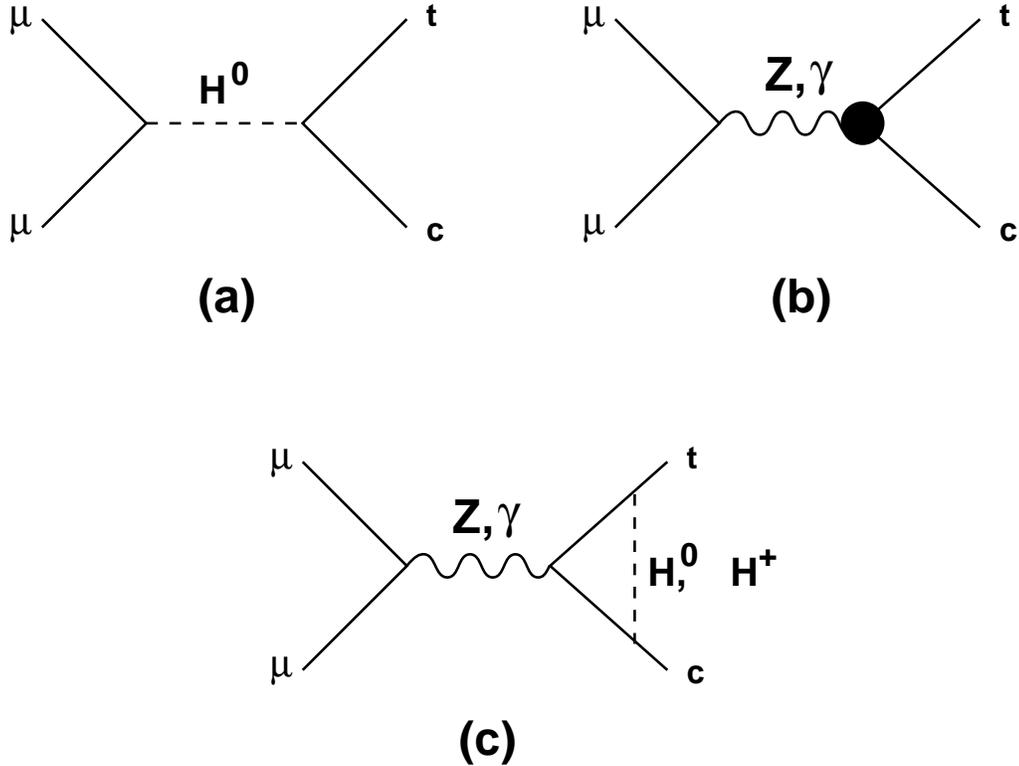
,height=4 in}
 \end{center}
\caption{\emph{
(a) Feynman diagram for $\mu^+\mu^-\to t c$ through $s$-channel 
neutral Higgs exchange.
(b) Feynman diagram for $\mu^+\mu^-\to t c$ through 
virtual $Z$ and $\gamma$  exchanges, 
where the circle indicates a vertex correction.
(c) An example of a vertex correction contributing to 
$\mu^+\mu^-\to tc$, where $H^0$ is a neutral Higgs with flavor changing
interactions to fermions and $H^\pm$ is a charged Higgs.}}
\label{tcfdia}
\end{figure}
 
Consider now the two processes shown in 
Fig.~\ref{tcfdia}(a) and \ref{tcfdia}(b). 
Since the 
Higgs 
flips the helicity of the $\mu$ while the $Z$ does not, for unpolarized or
longitudinally polarized beams the interference will be proportional to
the mass of the muon and consequently exceedingly small and uninteresting.
Such a suppression will not occur if the beams are transversely polarized
whence a large interference signal may be produced, especially if the
resonant 
(Fig.~\ref{tcfdia}(a)) 
and the continuum 
(Fig.~\ref{tcfdia}(b)) 
processes are of similar
strength. 

Bearing all this in mind we will thus proceed as follows: first we will
consider the general case of the resonance production of $t\bar c$
interfering with the continuum and then we will consider, more
specifically,  what signals are
produced in models similar to the ones discussed by~\cite{prd55p3156}. 

The process which produces $tc$ via s-channel Higgs exchange is 
controlled by the terms in the Lagrangian (for more details 
see section \ref{sssec322})

\beq
{\cal L}_H = [ \bar \mu \phi_{\mu} \mu + \bar t \phi_{tc} c +\cdots +
\mbox{h.c.}] H^0 \label{lh}
\eeq

\noindent where with the parameterization defined in Eq.~\ref{fermicouple}

\beqa
\phi_{\mu} 
=C^0_{\mu\mu}\chi_\mu e^{i\gamma_5\lambda_\mu}
\equiv  \alpha_{\mu} + \beta_{\mu} \gamma_5 ~.
\eeqa

\noindent Similarly, for $\phi_{tc}$ we can write 

\beqa
\phi_{tc} \equiv \alpha_{tc} + \beta_{tc} \gamma_5 \label{phimmtc} ~.
\eeqa

\noindent Note that unitarity implies that $\alpha_{\mu}$ is real while
$\beta_{\mu}$ is imaginary.

CP violation will then occur if there exists another mechanism for producing
$tc$ which the Higgs may interfere with. 
Here we take this  process
to be
$\mpmm\to\gamma^*\to t\bar c$ and/or $\mpmm\to Z^* \to t\bar c$ with the  
amplitude

\beq
{\cal M}_{Z,\gamma} = e^2\Pi_{Z,\gamma} \left( \bar \mu \gamma_\rho \eta_{\mu} \mu \right) 
\cdot \left( \bar t\gamma^\rho \eta_{tc} c \right) \label{mzg} ~,
\eeq

\noindent where $\bar\mu,~\mu,~\bar t$ and $c$ above are Dirac spinors
and  

\beqa
\Pi_Z = 
s \left [ (s-m_Z^2)s_W^2c_W^2\right]^{-1}
\quad , \quad 
\Pi_\gamma = 1 ~, \nonumber\\
\eta_{\mu} \equiv  A^{Z,\gamma}_{\mu} + B^{Z,\gamma}_{\mu}
\gamma_5 
\quad , \quad  
\eta_{tc}  \equiv  A^{Z,\gamma}_{tc} + B^{Z,\gamma}_{tc} \gamma_5 ~.
\label{chimmtc} 
\eeqa

\noindent Here $s_W = \sin\theta_W$, $c_W =\cos\theta_W$, where $\theta_W$
is the weak mixing angle. 
$A_{\mu}^{Z,\gamma}$, $B_{\mu}^{Z,\gamma}$ 
are real and they, of course,
occur in the tree-level SM Lagrangian. $A_{tc}^{Z,\gamma}$, 
$B_{tc}^{Z,\gamma}$ are form factors
which are induced at the loop level. 
As mentioned previously,
although small in the SM, 
they may be generated at reasonable levels in some extensions of the SM,
for instance, in multi Higgs scenarios which give $H^0 \to t\bar c$
\cite{hou,ars_mumu}.

Since we are interfering a vector continuum with a scalar resonance, 
this interference is naturally suppressed by $m_\mu$. This suppression, 
however, does not apply if at least one of the beams is transversely 
polarized.
To construct a quantity which is CP odd, we 
consider the 
case where the $\mu^-$ beam is polarized in the $+x$ direction 
and add it to the result where the $\mu^+$ is also 
polarized in the $+x$ direction.
We will also consider the case where the 
continuum is dominated by the 
$Z$ exchange.

Combining the 
transversely polarized 
$\mu^+$ and 
transversely polarized 
$\mu^-$ data as suggested 
above, we now 
consider some 
angular distributions of this combined data which
have specific properties under $CP$ and $T_N$.
Let us define the polar coordinates $(\theta,\phi)$ of $\vec p_c$; in
particular, $\phi$ is the azimuthal separation between the beam
polarization and $\vec p_c$. 
For each event of the form $\mu^+\mu^-\to t \bar c$ or $\bar t c$, let us 
also define, $L_t$ to be $+1$ for the $t\bar c$ final state and $-1$ for the 
$\bar t c$ final state. 
It is natural, therefore, to consider the following possible asymmetries

\beqa
x_1 = \langle \sigma(\cos\phi) \rangle & \qquad & 
x_2 = \langle L_t \sigma(\cos\phi) \rangle \nonumber \\
x_3 = \langle \sigma(\sin\phi) \rangle & \qquad & 
x_4 = \langle L_t \sigma(\sin\phi) \rangle
\label{x1234one} 
\eeqa

\noindent where $\sigma(x)=+1$ if $x\ge0$ and $-1$ if $x<0$.

These expectation values can be 
characterized in terms of their symmetry
properties. 
Thus $x_2$, $x_4$ are CP-odd; $x_2$ is
$T_N$-even and $x_4$ is $T_N$-odd. $x_1$ and $x_3$ are CP-even; $x_1$
is $T_N$-even and $x_3$ is $T_N$-odd, indicating that $x_2$ and $x_3$
require complex Feynman amplitudes, i.e., FSI phase(s). 
In the process at hand one source of this is the
Higgs propagator (see Fig.~\ref{tcfdia}(a)). 
In fact, since the Higgs is close to resonance, in the
experiments being envisioned here, these (CP$T_N$-odd) observables (i.e.,
$x_2$ and $x_3$) are likely to be the most 
prominent of the observables 
since they are enhanced by the resonant phase of the Higgs propagator.

In order to observe the signals suggested above one first requires a
muon collider which is able to deliver beams with a large transverse
polarization as well as an energy spread for the beam which is
comparable to or smaller than the Higgs peak.

Clearly, theories which produce detectable signals should, of course,
have fairly large FC $tc$ couplings. As a specific example
let us consider 2HDM of types III \cite{prd55p3156,hou}, also discussed 
in section \ref{sssec322}. In these scenarios a
2HDM is considered where the second doublet has
arbitrary Yukawa couplings. The popular Cheng-Sher Ansatz \cite{prd35p3484}:

\beqa
|\phi_{tc}| \simeq g_W \frac{\sqrt{m_tm_c}}{m_W} \lambda \label{lamcs}~, 
\eeqa

\noindent for $\phi_{tc}$ in Eq.~\ref{lh} is then 
imposed where $\lambda$ is a parameter that needs to
be extracted from experiment. It is perhaps natural to expect $\lambda$
to be of ${\cal O}(1)$. It is clear that the first obstacle to a large
signal is having the $Z^*$-exchange continuum in Fig.~\ref{tcfdia}(c), 
generated by loop corrections, to be sizable.

Let us define 
\beqa
R_H&=&
\sigma(\mu^+\mu^-\to H^0 \to t\bar c,~ \bar t c)/
\sigma(\mu^+\mu^-\to \gamma^* \to e^+e^-) ~,\\
R_Z&=&
\sigma(\mu^+\mu^-\to Z^* \to t\bar c,~ \bar t c)/
\sigma(\mu^+\mu^-\to \gamma^* \to e^+e^-) ~.
\eeqa

Clearly then, a necessary condition for there to be large ${\cal O}(1)$
asymmetries is that $R_H\approx R_Z$. For $m_{H^0} \sim 150 - 350$ GeV range,
typically $R_Z\sim10^{-3} - 10^{-4}$ \cite{ARS}. Such a signal would 
have a marginal
chance at a ${\cal L} = 10^{34}$ cm$^{-2}$s$^{-1}$ luminosity machine. The
Higgs signal in this case can also be of order 
$R_H\sim 10^{-3} - 10^{-4}$ 
in the
scenario where the Higgs decay to two vector-bosons is allowed. In that
case, $\Gamma_H\sim {\cal O}$(1 GeV), 
so the spread in beam energy needed to be in the resonance region of the 
Higgs  should be achievable at 
a muon collider. 
This could allow asymmetries of a few tens of percents provided that  
the CP-odd weak phase
difference were large.  Since $R_H \sim R_Z\sim 10^{-3}$, observing the
asymmetry would still require a $10^{34}$ cm$^{-2}$s$^{-1}$
collider. The situation, of course, would improve considerably if the
continuum were larger; for example, this happens if $\lambda>1$ in 
Eq.~\ref{lamcs}. Since
the continuum scales as $\lambda^4$ the integrated luminosity required to
observe these asymmetries scales as $\lambda^{-4}$. 

\newpage

\def \cp{{\rm CP}\!\!\!\!\!\!\!\!\!\not~~~~}
\def \cpint{{\cp\mid_t}}

\section{Summary and outlook}\label{summary_chapter}

There are only two known systems 
which have been shown
to violate the CP symmetry: the neutral kaon through the parameters
$\epsilon$ and $\epsilon^\prime$ and the entire universe through the
dominance of matter over antimatter.  It would be of great significance to
understand the relation between these two effects or trace them to a
common origin. Although experiments in the near term are likely to clarify
the source of CP violation in the kaon system, the mechanisms of
baryogenesis remain in the realm of theoretical speculation and may not be
directly tested in the lab for some time. The top quark, however, offers a
unique system where new CP-violating effects could be discovered which
could, in time, shed light on the processes which were important in the early
universe.

The immediate source of CP violation in the kaon is thought to be the CKM
phase in the SM. This will be tested in detail in the next few years 
through the study of the  
$B$ meson. Ironically, although the 
exchange of virtual top quarks
generates the large CP violation in the $B$ and $K$ mesons, the CKM phase will
not produce any signal in top quark systems that is large enough to be of
experimental interest. Instead, if a CP-violating signal is seen in the
top quark, it must be due to some inherently large, non-standard,
CP-odd phase which
becomes manifest 
only
at high energy scales. Since the effect of the CKM phase
is also thought to be too weak to explain baryogenesis, 
the required phase for this process is likely
to show up in top quark physics as well. 
Thus, the 
observation of CP
violation in top quark reactions 
is an unambiguous signal of physics beyond the
SM which may well shed light on baryogenesis. 

In this review, we have considered a number of laboratory tests of CP
violation in the top quark in the context of various 
non-standard
models for physics
beyond the SM.  In particular, we focus on two classes of models which are
described in some detail in Chapter \ref{sec3}:

\begin{enumerate} 

\item 
Multi Higgs models containing at least two Higgs doublets with
phases in the Yukawa couplings.  Here CP violation manifests
either in the neutral or in charged Higgs sectors.

\item 
SUSY models
wherein we consider in detail 
the MSSM
with the Yukawa couplings given by $N=1$
minimal SUGRA models. 
One manifestation of SUSY CP violation is  
through mixing in the sfermion sector.

\end{enumerate}

We chose to focus on these models because they seem to be representative
of models for physics beyond the SM which could give rise to CP violation
and are most often considered in the literature. It is likely that the
ability of a particular signal to detect CP violation in one of these
models is a good indication of its general utility.

In this review we highlight
some notable CP-violating phenomena which follow from these models:

\begin{itemize}

\item 
The transverse polarization of the $\tau$ in $ t \rightarrow b \tau \nu$
which follow from CP violation in the charged Higgs sector
(see Chapter \ref{sec5})

\item 
CP-violating correlations in  
$e^+ e^- \to t \bar t H^0$, $t \bar t Z$
and $e^+e^-\to t \bar t \nu_e \bar\nu_e$ at
high energy $e^+e^-$ colliders
generated by CP violation in the neutral Higgs sector. 
Since these 
effects arise through the interference of two tree-level graphs 
the resulting correlations can be very large indeed. 
(see Chapter \ref{sec6})

\item 
CP-violating correlations in hadronic top pair and single top production
which can 
originate from 
by CP-odd phase(s) in the neutral Higgs sector or
in the squark sector of SUSY models.
(see Chapters \ref{sec7} and \ref{sec8})

\item 
CP-violating top polarizations may arise 
in top production 
at 
muon and/or photon colliders.
In particular at such colliders
the neutral Higgs(es) can be 
produced in the $s$-channel, giving rise to a distinct resonant 
enhancement which in turn may magnify the CP-violating effect in 
reactions such as
$\mu^+ \mu^-,~\gamma \gamma \to t \bar t$ or even in the flavor changing 
channels $\mu^+ \mu^-,~\gamma \gamma \to t \bar c + \bar t c$.
(Chapters \ref{sec9} and \ref{sec10})

\item 
CP-violating moments of the top analogous to the electric dipole moment
which may be observed at an NLC from top polarimetry in the reaction 
$e^+e^- \to t \bar t$.  
Such moments can be generated in SUSY models 
as well as models with an extended
neutral 
Higgs sector. 
(Chapters \ref{sec6} and \ref{sec10})

\item

CP violation in the main top decay $t \to b W$.
In this case a CP-odd phase in the stop 
sector of the MSSM can cause a partial rate asymmetry in $t \to b W^+$ 
at the level of a ${\rm few} \times 0.1\%$.
(Chapter \ref{sec5})

\end{itemize}

A common feature of both the SM and the models mentioned
throughout this review is that CP violation is driven directly or
indirectly by Yukawa couplings in the scalar sector of the theory. In the
SM the CKM matrix which contains the CP-violating parameter results from
the Higgs-fermion coupling while with multi Higgs models additional CP
violation may result from the couplings between the various Higgs fields,
either from explicit CP violation in the Higgs potential (e.g., Model II)
and/or in the Yukawa interaction terms (Model III), or CP can be violated
spontaneously if there are more than two Higgs doublets. In SUSY models,
phases may be associated with 
the scalar Lagrangian as well, for
instance, from squarks and sleptons mixing.  As in the SM, the
amount of CP violation is proportional to the non-degeneracy of the mass
spectrum.  For instance, in MHDM's the non-degeneracy of the Higgs
particles is required while in SUSY it is the non-degeneracy of squarks or
sleptons of different helicities.

%
%
%
%

It is important to emphasize that, in many ways,
the phenomena of CP violation in top quark systems strongly relies on
the large mass of the top   
which, therefore, becomes the key property of the top
as far as CP violation is concerned:



\begin{itemize}
\item The large mass
of the top quark
allows its polarization to be determined by
its weak decays because, unlike the other 5 quark flavors, it decays
before it hadronizes and
so the information carried by its spin is not diluted.
As discussed in Chapter \ref{sec2},
this allows experiments
to consider CP-odd observables involving polarization
(i.e., top spin correlations) which is crucial
since in many settings no CP-violating observables could be constructed
without this information.
For example,
the CP-violating
transverse top polarization asymmetry, suggested in Chapter \ref{sec8},
may be used to probe
tree-level CP violation in $p \bar p \to t \bar b$ which otherwise
(i.e., without the use of top spins) cannot be observed.
\item In MHDM's, it is 
the enhancement in the Yukawa coupling of a neutral Higgs 
to the top quark 
that is 
responsible for the enhanced CP-violating effect. As discussed
in Chapter \ref{sec6}, this clearly manifests 
in e.g., $e^+e^- \to t \bar t Z$, where the only 
CP-violating diagram present, i.e., the one  
with a virtual neutral Higgs exchange, is comparable 
in size to the CP-even SM diagrams that contribute to the same final state,
due to the fact 
that
the CP-odd, 
$H^0t \bar t$ Yukawa coupling may be as large as the gauge coupling.
\item As mentioned above, 
in the CP-nonconserving effect associated with SUSY particles exchanges,
it is the large mass splitting between the two stop mass eigenstates of the 
theory that may be the cause for an enhanced CP-violating effect, again, due 
the corresponding large mass of their SM partner --
the top quark. As discussed
in Chapters \ref{sec7} and \ref{sec8},
this is the case for example in $pp \to t \bar t$ and 
$p \bar p \to t \bar b$ where the effect arises from CP-violating loop 
exchanges of stop particles. 
\item Large $m_t$ enables
the study of CP violation in 
cases where the CP-odd effect is driven by new thresholds (i.e., 
absorptive cuts across heavy particles of the underlying theory).
As discussed in Chapter \ref{sec5},
this is the case in 
e.g., PRA in $t \to b W$ within supersymmetry, where it is only viable 
if $m_t > m_{\tilde t} + m_{\tilde\chi^0}$ -- still allowed by present 
experimental data, basically, because of the heaviness of the top. 
\item The cases where the CP-odd effect is enhanced to the detectable level 
due only to an intermediate resonance are also a clear manifestation of 
the important role played by the large $m_t$ in CP violation studies.
Such is the case in e.g., CP violation in the decay 
$t \to b \tau \nu_\tau$, as discussed in Chapter \ref{sec5}, 
where the intermediate $W$-boson resonance provides
the necessary enhancement, of course, since $m_t > m_W$. 
\end{itemize}

It is therefore evident that, due to its large mass, the top is very
sensitive to new effects from possible new short distance theories.
This sensitivity of the top quark to short distance effects from
many models 
leads one to consider a more general approach
for such studies.  For instance, by parameterizing CP violation
in a model
independent way using CP-violating form factors.  Such form factors which
contain the information of the dynamics of some new physics scenarios at higher
energy scales are expected to be more pronounced in top quark
interactions.  This technique is a useful prescription for extracting
limits on various CP-violating couplings that may arise in new physics.
Examples of such effective form factors are the top dipole moments and the
CP-violating form factors in the top decays which were separately
discussed in Chapters \ref{sec4} and \ref{sec5} respectively.


In Chapter \ref{sec4} we find that in models with extra Higgs doublets as
well as in SUSY models, one can expect an EDM and ZEDM of the top on the
order of $ \sim 10^{-19}$ e-cm and, likewise, a CEDM of $ \sim 10^{-19}$
$g_s$-cm. These values are many orders of magnitude larger than the SM
prediction for these quantities. Thus, a discovery of such an effect in $t
\bar t$ production in leptonic or hadronic colliders and perhaps also in
photon and muon colliders, would be a clear signal of beyond the SM
dynamics. In Chapter \ref{sec6} we discuss the sensitivity of an $e^+ e^-$
NLC collider to these EDM and ZEDM. We find that optimal observable
techniques seem to indicate that high energy $e^+e^-$ colliders will be
sensitive, at best, to a top dipole moment at the level of $\sim 10^{-17}
- 10^{-18}$ e-cm, about one to two orders of magnitude larger than what is
expected in the models mentioned above.

In Chapter \ref{sec9} we find that a similar statement is true at
$\gamma\gamma$ colliders based on the backscattered laser light from an
NLC. On the other hand, hadron colliders are expected to be more sensitive
to the CEDM, in particular, in Chapter \ref{sec7} we find that a CEDM at
the level of $10^{-19}~-~10^{-20}~g_s-cm$ might be observable at the LHC. 
Although this sensitivity of a NLC to the top electric dipole moment may
seem a little discouraging, there is still very strong motivation to look
for this effect; the observation of a top dipole moment with this
magnitude (i.e., $\sim 10^{-17}$ e-cm), will clearly be a surprise, since
such a large dipole moment cannot be accounted for in the popular models
such as SUSY and MHDM's.

Thus, in spite of the very large enhancements expected in such beyond the
SM scenarios for the top dipole moments, it is evident that this type of
signal is useful only if the dipole moments are on the very large side of
the theoretical range. That being the case, one would like to search for
other alternatives for the observation of CP violation in top quark
systems.  It may, for example, be more promising to look for specific
signals of CP violation in the production or decay of top quarks which are
not related to the dipole moments.

In Chapter \ref{sec5} we consider the CP-violating effects which
might be present in the decay of top quarks. The simplest kind of signal
is a PRA in the decay $t\to bW$ (mentioned above) which in SUSY 
can have an asymmetry of
$\sim 10^{-3}$ and thus may be detectable at the LHC. Another promising
signal which is particularly applicable to 3HDM or other models with
charged scalars is $\tau$ polarization asymmetries in the decay $t\to \tau
\nu b$ which arises from the interference of the W pole with the charged
Higgs propagator and could result in asymmetries on the order of 
a few tens of percents.  

While CP-violating effects in the decay of top quarks may be searched for
at any experiment where top quarks are produced, there are a broader range
of signals where the CP violation occurs in the production of the top
quark. In this case one must consider each kind of top quark production
mechanism separately.
CP violation in the production mechanism of the top 
was discussed in the context of an 
$e^+e^-$ collider (in Chapter \ref{sec6}), hadronic colliders 
(in Chapters \ref{sec7} and \ref{sec8}), photon collider (in Chapter 
\ref{sec9}) and muon collider (in Chapter \ref{sec10}).

Each of these machines has its own characteristics and so special
attention needs to be given in constructing appropriate CP-violating
observables. For example, lepton and photon colliders have the advantage
of their relative cleanliness as far as background is concerned; it should
be easier to reconstruct the top quark in such colliders even when it
decays via purely hadronic modes.  On the other hand, it may be quite
challenging for such colliders, e.g., the NLC, to posses the necessary
luminosity for studying rare phenomena in top physics such as loop induced
CP-violating effects.  Hadron colliders such as the LHC may have an
advantage in this context, since top quarks will be more readily produced
there. However, the hadronic 
environment
requires more effort in disentangling the CP-violating signal both from
the experimental and the theoretical points of view.  In
addition for hadron colliders, it should be noted that one would prefer,
in principle, to always use a $p\bar p$ collider (such as the Tevatron)
for CP studies since then the initial state is a CP eigenstate.
Unfortunately, the LHC which is expected to produce a 
very large 
number
of top pairs is a $pp$ collider.  It turns out, however, that the initial
state at the LHC may not effect the CP studies considered here greatly,
primarily because 
the dominating $t \bar t$ production mechanism there is in
fact a CP eigenstate, i.e., gluon-gluon fusion.

Nonetheless, 
in such colliders it is important to use
clean CP-violating observables 
that can reduce the backgrounds. 
One such useful observable for the LHC, that was suggested by 
Schmidt and Peskin was discussed in Chapter \ref{sec7}.
Their CP-odd signal uses the difference between the energy of the 
positrons from $t \to b e^+ \nu_e$ and electrons from 
$\bar t \to \bar b e^- \bar\nu_e$  
in the overall rather complicated 
reaction, $gg \to t \bar t \to b W^+ \bar b W^- \to b e^+ \nu_e \bar b 
e^- \bar\nu_e$. 
Unfortunately, the expected asymmetry 
is unlikely to be larger than a few times $10^{-3}$. 
However, Bernreuther, Brandenburg and Flesch
have shown that a
considerable improvement may be achieved in this reaction by employing   
clever cuts on the $t \bar t$ invariant mass. By doing that they were able
to isolate the possible CP-violating contribution from an $s$-channel   
Higgs exchange in $gg \to t \bar t$. Thus, 
in their analysis, asymmetries at the  
level of a few percent may arise leading to a CP-odd signal well above the
$3-\sigma$ level in $pp \to t \bar t +X$ at the LHC.

%
%

Another useful CP-violating signal designed for the Tevatron 
setting  
was discussed in
Chapter \ref{sec8}. Specifically, it uses an apparent advantage 
of $p\bar p$ colliders: that there should be a high rate 
of virtual $W$ production
via $u\bar d$ annihilation. In this case a number of asymmetries involving
the transverse and longitudinal components of the top spin may be
constructed. In both MHDM's 
and SUSY models we find that asymmetries
around $1\%$
may thus occur in single top production at the Tevatron.

Loop induced CP-violating effects such as that of Schmidt and
Peskin as well as dipole moments tend to give asymmetries at the level of
$\sim 0.1 -1 \%$. Thus experimental detection of rare CP violation effects
in top physics, both in hadronic and leptonic colliders, leads to at least
two important challenges. 1) Can detector systematics be controlled to the
point that a CP asymmetry of ${\cal O}(0.1\%)$ can be observed?  2) Can CP
violation be studied with purely hadronic decay modes of the $ t \bar t$ pair.
That is, to
what extent will the experimentalists be able to distinguish between the
top and the anti-top via purely hadronic modes; if that can be done to a
significant level, then the increased statistics 
will improve the 
prospects for the observability of such rare CP-odd signals.

The small CP-asymmetries which arise from phenomena that occur at
one-loop may make most of those signals too small to be of great use in
putting bounds on models of new physics.  On the other hand, signals which
arise from the interference of tree graphs only are likely to give rise to
larger asymmetries. In Chapter \ref{sec6} we discuss some candidate
signals of this type such as $e^+ e^- \to t \bar t H^0, ~t \bar t Z$,
and $t \bar t \nu_e \bar\nu_e$.  In addition, the decay discussed in
Chapter \ref{sec5}, $t \to b \tau \nu_\tau$ falls into this category. In
these cases one finds that CP-asymmetries at the level of tens of
percents are possible in models with CP-odd phase(s)  in the Higgs sector. 
This makes that type of CP-violating mechanism quite robust, requiring
about a few thousands $t \bar t$ events per year in order to be
detected; such a number may well be within the reach of the
future colliders presently under consideration.

The two other exotic technologies which may be used in the future in this 
context (i.e., tree-level CP violation) are 
$\gamma\gamma$ colliders and muon storage rings. In Chapter \ref{sec9} 
we discuss 
reactions which can take place at a $\gamma\gamma$ collider 
constructed from an $e^+e^-$ 
NLC collider by backscattering laser light from the 
$e^\pm$ beams. As mentioned above, these machines 
can be used 
to produce $ t \bar t$ pairs through an 
intermediate Higgs state and interfere it with the born cross-section for 
top pair production. In this case, observables constructed by considering 
the top polarization can give asymmetries of up to $\sim 10\%$ in 2HDM. 
In Chapter \ref{sec10} we further 
discuss experiments at muon colliders. Clearly any 
experiment which can be performed at an $e^+e^-$ collider may also be 
performed at a muon collider. In addition, however, the larger mass of 
the muon allows us to contemplate the production of Higgs bosons in the 
s-channel. In such scenarios one can analyze the scalar versus 
pseudoscalar couplings of the Higgs to $t\bar t$ by studying the polarization 
correlations of the tops produced which, again,
could give $\sim 10\%$ asymmetries.

Now, since a large portion of the experimental effort in these future
colliders will be devoted to the search of supersymmetry, it is
particularly gratifying that for such studies of tree-level CP violation,
supersymmetry may play an important role in our understanding of the
underlying mechanism for CP violation. In particular, once SUSY is
discovered and SUSY particles are readily produced in high energy collider
experiments, the next step would clearly be to start scrutinizing the
basic ingredients of the SUSY Lagrangian, e.g., its CP-violating sector.
Indeed, due to the potential richness of CP-odd phases in SUSY theories,
tree-level CP violation can easily occur in production and decay of
SM+SUSY particles. A promising venue to investigate such tree-level SUSY
CP violation may be to search for reactions involving associated top
production in final states which contain additional SUSY particles and to
probe the CP-violating effect through top polarimetry, i.e., bypassing the
missing energy limitation (typical to SUSY signatures) by using the top
spins. A simple example may be CP violation in $e^+e^- \to t +X+{\rm
missing ~energy}$ vs.  $e^+e^- \to \bar t +\bar X+{\rm missing ~energy}$,
where $X$ is some non-SUSY hadronic final state.

Another interesting related venue in the context of tree-level CP
violation within SUSY models is to search for CP-odd signals in reactions
where, although involving SUSY particles, only SM particles are produced
in the final state. Indeed, if SUSY theories posses $R$-parity violating
interactions, CP may be violated at tree-level even in $2 \to 2$ processes
in which the initial and final states consist of SM particles only. In
particular, through SUSY scalar exchanges in which the CP-odd phases are
carried by the $R$-parity violating couplings in the interaction vertices
of a pair of SM particles to squarks and/or sleptons. Again, such
tree-level CP violation may be probed even in a $2 \to 2$ process if one
uses 
top spin asymmetries.  Consider, for example, single top
production at the Tevatron, $p \bar p \to t \bar b + \bar t b$. As was
discussed in Chapter \ref{sec8}, a transverse top polarization asymmetry
can 
probe 
potentially tree-level CP violation in this process. Indeed, since
$s$-channel exchanges of charged sleptons can mediate $p \bar p \to t \bar
b + \bar t b$ in $R$-parity violating SUSY, this transverse top
polarization asymmetry can
potentially
lead to 
large tree-level SUSY CP-violating signal
in this reaction. 
          
These types of tree-level CP violation in SUSY models were not discussed in
this review or anywhere else in the literature to date 
and could be useful 
to 
examine in the future, especially once SUSY is directly
observed. More generally, the subject of tree-level CP violation 
seems promising and requires additional effort from the theoretical point
of view.

In parting, the study of CP violation in top quark physics deserves to be one
of the main issues on the agenda of the future high energy colliders. The
expected high production rate of top quarks in these colliders turns these
machines into practically top factories enabling the
examination of what is presently considered rare phenomena in top physics.
In particular, these colliders provide a unique opportunity for the
study of CP violation -- a phenomenon that till now seems to be 
essentially confined
only to the kaon system -- and its relation to top quark dynamics.  The
manifestation of CP violation in heavy particles systems in general and in
the top quark system in particular, can shed light on new aspects of this
phenomena due to the high energy scales involved, possibly on new physics
related to the dynamics of our universe in its very early stages.

One, of course, should not forget the importance of the up coming CP
measurements in the $B$ system. On the other hand, it is also important to note
that there is a very interesting interplay between CP violation in
$b$ physics and in $t$ physics. In $b$ physics, one expects large
CP-violating signals due to the CKM phase alone. Therefore,
non-observation of CP violation in $B$ decays would, in fact, stand out as
a signal of new physics.  This is, of course, in complete contrast to the
situation in the top system in which one does not expect any CP-odd signal
with the CKM phase of the SM. Therefore, any signal of CP violation in top
reactions will unambiguously prove the existence of new physics. 
Moreover, in order to disentangle effects of new physics in
the $B$ system, one will need precision measurements and cross-checking of
the different available CP-violating $B$ decay channels. In top systems
the advantage is that no significant effort is needed in order to establish the
existence of new physics phenomena in CP-odd top correlations -- any
measured CP-nonconserving effect in top systems will suffice. 
  
Finally, in Tables~\ref{sumtab1} and \ref{sumtab2} 
we summarize the main features of some of the most interesting CP-violating 
signals that were discussed in this review. 

\begin{table}
\caption{The underlying source of CP-odd phase  and the 
mechanism responsible
for CP violation in the processes indicated.   
Note that: ${\tilde t}$ $\cp$ means $\cp$ from  
${\tilde t}_L - {\tilde t}_R$ mixing, $H^0$ $\cp$ means $\cp$ 
from scalar - pseudo-scalar mixing in the $H^0 t t$ vertex and 
$H^+$ $\cp$ means a $\cp$ phase in the $H^+ t b$ and/or  
$H^+ \tau \nu_\tau$ vertices. See also Chapter \ref{sec3}.} \label{sumtab1}
\begin{center}
\begin{tabular}{|c|c|c|c|} 
\hline
\hline
process & CP source & mechanism \\ 
\hline  
\hline
$t \to bW$ & MSSM (${\tilde t}$ $\cp$) & 1-loop \\
\hline  
& &  $W^+$-resonance \\
$t \to b \tau \nu_\tau$ & MHDM ($H^+$ $\cp$) & in tree-level \\ 
& & $W^+-H^+$ interference \\
\hline
& MSSM (${\tilde t}$ $\cp$) & \\
$p \bar p \to t \bar b$ &  $\&$  & 1-loop \\
& MHDM ($H^0$ $\cp$) & \\
\hline  
& MSSM (${\tilde t}$ $\cp$) & \\
$p \bar p \to t \bar t$ &  $\&$  & top - CEDM (1-loop) \\
& MHDM ($H^0$ $\cp$) & \\
\hline
& MSSM (${\tilde t}$ $\cp$) $\Longrightarrow$ & 1-loop \\
$pp \to t \bar t$ & $\&$ & \\
& MHDM ($H^0$ $\cp$) $\Longrightarrow$ & $s$-channel $H^0$ $\&$ 1-loop \\
\hline
& MSSM (${\tilde t}$ $\cp$) & \\
$e^+ e^- \to t \bar t$ &  $\&$  & top - EDM,ZEDM (1-loop) \\
& MHDM ($H^0$ $\cp$) & \\
\hline
$e^+e^- \to t \bar t H^0,~t \bar t Z$ & 
MHDM ($H^0$ $\cp$) & tree-level interference \\
\hline
& & $s$-channel $H^0$ \\
$e^+e^- \to t \bar t \nu_e \bar\nu_e$ & MHDM ($H^0$ $\cp$) & in  \\
& & tree-level interference \\
\hline
$\mu^+ \mu^- \to t \bar t$ & MHDM ($H^0$ $\cp$) & $H^0$-resonance \\
\hline
$\gamma \gamma \to t \bar t$ & MHDM ($H^0$ $\cp$) & 
$s$-channel $H^0$ $\&$ 1-loop \\
\hline
\hline
\end{tabular}
\end{center}
\end{table}

\begin{table}
\caption{The asymmetries that can probe   
CP violation in the processes indicated and their 
expected size. 
Also indicated 
is the place in which each asymmetry was discussed in the review, i.e., 
its equation number.
The size of the asymmetry given tends to be optimistic,
i.e., 
on the large side of its theoretical 
range.} \label{sumtab2}
\begin{center}
\begin{tabular}{|c|c|c|c|} 
\hline
\hline
process & type of asymmetry &  size \\ 
\hline  
\hline
$t \to bW$ & PRA - ${\cal A}_3$ (Eq.~\ref{GK4}) & 0.1\% \\
\hline  
$t \to b \tau \nu_\tau$ & $\tau$ pol. 
- e.g., (transverse) ${\cal A}_z^\prime$ (Eq.~\ref{eq585}) & $10\%$ \\ 
\hline
& cross-section -  $A_0$ (Eq.~\ref{ppbartbeq2}) & \\    
 $p \bar p \to t \bar b$ & & $1\%$ \\
& top pol. - e.g., (transverse) $A(\hat y)$ (Eq.~\ref{ppbartbeq5}) & \\ 
\hline  
$p \bar p \to t \bar t$ & lepton energy - 
e.g., (transverse) $A_T$ (Eq.~\ref{ppbartteq8}) & $0.1\%$ \\
\hline
& optimal observable - e.g., ${\cal O}^\prime$ (Eq.~\ref{opt2pptt}) & \\
$pp \to t \bar t$ & top pol./lepton momenta 
- e.g., $\Delta N_{LR}$ (Eq.~\ref{sp6})  & $0.1-1\%$ \\
& lepton energy - e.g., (transverse) $\Delta N(E_T)$ (Eq.~\ref{sp11}) & \\
\hline
& optimal observable - e.g., ${\cal O}_{R}$ (Eq.~\ref{eetteq7}) & \\
& top pol./lepton momenta - e.g., ${\hat T}_{ij}$ (Eq.~\ref{eetteq25}) & \\
$e^+ e^- \to t \bar t$ & & $0.1\%$ \\
 & angular distributions - e.g., $A_{ud}(\theta)$ (Eq.~\ref{eetteq42}) & \\ 
& energy distributions - e.g., $A_{\ell \ell}$ (Eq.~\ref{eetteq49}) & \\
\hline
$e^+e^- \to t \bar t H^0,~t \bar t Z$ & 
top momenta, optimal observable 
- ${\cal O},{\cal O}_{\rm opt}$ (Eq.~\ref{eetthzeq21}) & $10\%$ \\
\hline
$e^+e^- \to t \bar t \nu_e \bar\nu_e$ 
& top pol./lepton momenta - e.g., $A_y$ (Eq.~\ref{wwtteq1}) & $10\%$ \\
\hline 
& top pol./lepton momenta - $A^t$ (Eq.~\ref{mumuat}) & \\
$\mu^+ \mu^- \to t \bar t$ & & $10\%$ \\
& muon beam pol. - e.g., $A_\mu$ (Eq.~\ref{mumuamu}) & \\
\hline
 & top pol./lepton momenta - e.g., $\alpha_1$ 
(Eq.~\ref{ggtteq28}) & \\
$\gamma \gamma \to t \bar t$ & & $10\%$ \\
& photon pol. - e.g., ${\cal P}_1$ (Eq.~\ref{ggtteq32}) & \\
\hline
\hline
\end{tabular}
\end{center}
\end{table}

\newpage
\addcontentsline{toc}{section}{Notes}

\section*{\bf Notes}

\subsection*{Note on literature survey}

The literature survey for this review was primarily completed in Dec. 1999.

\vspace{0.5in}

\subsection*{Acknowledgments}

Two of us (GE and AS) are most grateful to the US-Israel Binational
Science Foundation for its support that proved very valuable during
the four year period that took to write this review. This work
was also supported in part by US DOE Contract Nos.
DE-FG02-94ER40817 (ISU), DE-AC02-98CH10886 (BNL) and DE-FG03-94ER40837 (UCR).

\newpage
\addcontentsline{toc}{section}{Appendix A}
\section*{\bf Appendix A}

In this appendix we define the coefficients ${\rm C}_x$,
($x=0,11,12,21,22,23,24$, see below) corresponding to one-loop integrals
with three internal propagators in the loop, i.e. ``triangle-like'' one
loop diagrams. In the review they appear in section \ref{ssec43}
(Eqs.~\ref{tdeq0018} - \ref{tdeq018}), in section \ref{ssec44}
(Eqs.~\ref{chargeloopform00} and \ref{chargeloopform}), in section
\ref{ssec45} (Eqs.~\ref{tdeq36}, \ref{tdeq39}, \ref{tdeq40}, \ref{tdeq47}
and \ref{tdeq48}), in section \ref{sssec514} (Eqs.~\ref{aalrd} -
\ref{dalrd}), in section \ref{sssec821} (Eqs.~\ref{ppbartbeq18} and
\ref{ppbartbeq19}) and in section \ref{sssec822} (Eqs.~\ref{ppbartbeq26}
and \ref{ppbartbeq27}). 

These three-point loop from factors which are functions of masses and momenta
are defined by the one-loop momentum integrals as follows \cite{olden}:
\newcounter{num}
\setcounter{num}{1}
\setcounter{equation}{0}
\def\theequation{\Alph{num}.\arabic{equation}}
\begin{eqnarray}
{\rm C}_0; {\rm C}_\mu; {\tilde{\rm C}}_\mu; 
{\rm C}_{\mu\nu}(m^2_1,m^2_2,m^2_3,p^2_1,p^2_2,p^2_3) \equiv  
\int \frac{d^4k}{i\pi^2}
\frac{1;k_\mu;k^2k_\mu;k_\mu k_\nu}{ {\cal D}_1 {\cal D}_2 {\cal D}_3 } ~,
\end{eqnarray}

\noindent where: 

\begin{eqnarray}
{\cal D}_1 &\equiv& k^2-m^2_1 ~,\\
{\cal D}_2 &\equiv& (k+p_1)^2-m^2_2 ~,\\
{\cal D}_3 &\equiv& (k-p_3)^2-m^2_3 ~,
\end{eqnarray}

\noindent and $\sum_i p_i = 0$, $i=1-3$, is to be understood above.

The three-point loop from factors are then given 
through the following relations \cite{veltman}:
\begin{eqnarray}
&&{\rm C}_\mu=p_{1\mu}{\rm C}_{11}+p_{2\mu}{\rm C}_{12} \ ,\\
&&{\tilde{\rm C}}_\mu=p_{1\mu} {\tilde{\rm C}}_{11}+
p_{2\mu}{\tilde{\rm C}}_{12} \ ,\\
&&{\rm C}_{\mu\nu} = p_{1\mu}p_{1\nu}{\rm C}_{21} 
+ p_{2\mu} p_{2\nu}{\rm C}_{22}
+\{p_1p_2\}_{\mu\nu}{\rm C}_{23}+g_{\mu\nu}{\rm C}_{24} \ ,
\end{eqnarray}
where $\{ab\}_{\mu\nu}\equiv a_\mu b_\nu +a_\nu b_\mu$.
The numerical evaluation of the above form factors can be performed 
using the algorithm developed in \cite{olden}.

\newpage
\addcontentsline{toc}{section}{Appendix B}
\section*{\bf Appendix B}

We list in this appendix all the 
abbreviations used throughout this review:

\begin{tabular}{ll}
SM & Standard Model \\
CKM & Cabibbo-Kobayashi-Maskawa \\
GIM & Glashow-Iliopoulos-Maiani \\
NLC & Next Linear Collider \\
LHC & Large Hadron Collider \\
PRA & Partial Rate Asymmetry \\
FSI & Final State Interactions \\
PIRA & Partially Integrated Rate Asymmetry \\
MHDM & Multi Higgs Doublet Models \\
SUSY & SUperSYmmetry or SUperSYmmetric\\
SSB & Spontaneous Symmetry Breaking\\
VEV & Vacuum Expectation Value\\
NLO & Next-to-Leading Order\\
2HDM & Two Higgs Doublet Model\\
3HDM & Three Higgs Doublet Model\\
MSSM & Minimal Supersymmetric Standard Model\\
FCNC & Flavor Changing Neutral Currents\\
FC & Flavor Changing\\
NFC & Natural Flavor Conservation \\
REWSB & Radiative ElectroWeak Symmetry Breaking\\
NEDM & Neutron Electric Dipole Moment\\
EDM & Electric Dipole Moment\\
RGE & Renormalization Group Equations\\
SUGRA & SUperGRAvity\\
EW & ElectroWeak\\
TDM & Top Dipole Moment\\
ZEDM & weak($Z$) - Dipole Moment\\
CEDM & Chromo - Electric Dipole Moment\\
FF & Form Factor \\
LSP & Lightest Supersymmetric Particle\\
DCS & Differential Cross-Section
\end{tabular}

\end{document}